%% file: PhDThesis.tex
\documentclass[a4paper,11pt]{book}
\usepackage[utf8]{inputenc}
\usepackage{{amsmath,amsfonts,latexsym, amstext, amssymb, amsthm}}
\usepackage{xspace} 
\usepackage{slashed}
\usepackage[bf]{caption}
\usepackage{subcaption}
\usepackage{comment}
\usepackage{cite}
\usepackage[bookmarks=true]{hyperref}
\usepackage{dsfont}
\usepackage[english]{babel}
\usepackage{placeins}
\usepackage[dvips]{graphicx}
\usepackage[dvips]{color}
\usepackage{hyphenat}

\usepackage{pstricks}
\usepackage{pst-node}
\usepackage{pst-coil}

\hypersetup{
	    pdfstartpage=1, 
	    pdfpagemode=UseOutlines,
	    bookmarksnumbered=true,
	    bookmarksopen=false,
	    pdflang=English,
            pdfauthor={Matthias Wilhelm},
            pdftitle={Form factors and the dilatation operator in N=4 super Yang-Mills theory and its deformations},
            pdfkeywords={Super Yang-Mills theory, Form factors, On-shell methods, Anomalous dimensions, Integrability, Renormalisation}
}

\allowdisplaybreaks

\makeatletter

\divide\@tempdima\baselineskip
\@tempcnta=\@tempdima
\makeatother

\def\ps@plain{\let\@mkboth\@gobbletwo%
     \let\@oddfoot\@empty\def\@oddhead{\reset@font\hfil\thepage}%
     \let\@evenfoot\@empty\def\@evenhead{\reset@font\thepage\hfil}}

\newcommand{\AdS}{\text{AdS}}

\newcommand{\lax}{\mathcal{L}}
\newcommand{\mono}{\mathcal{M}}
\newcommand{\trans}{\mathcal{T}}
\newcommand{\transosc}{\mathbf{T}}

\newcommand{\deriv}[2][]{\frac{\de #1}{\de #2}}
\newcommand{\parderiv}[2][]{\frac{\partial #1}{\partial #2}}

\newcommand{\Eulerbeta}{\text{B}}
\newcommand{\Bubbleop}{\mathbb{B}}

\newcommand{\rr}{R}

\newcommand{\dualmom}{y}
\newcommand{\dualsmom}{\vartheta}
\newcommand{\mtwistor}{\mathcal{Z}}

\newcommand{\mtwistoruu}{\smash{\underline{\mathcal{Z}}}}
\newcommand{\ptwistoruu}{\smash{\underline{\mathcal{W}}}}

\newcommand{\Zuu}{\smash{\underline{Z}}}

\newcommand{\NmaxMHV}{\ensuremath{\text{N}^{\text{max}}\MHV}}

\newcommand{\sigmapart}{\tilde{\sigma}}

\newcommand{\vlluu}{\underline{\smash{\lambda}}}
\newcommand{\vltuu}{\underline{\smash{\tilde{\lambda}}}}
\newcommand{\vleuu}{\underline{\smash{\tilde{\eta}}}}
\newcommand{\vletuu}{\underline{\smash{\eta}}}
\newcommand{\vmmuu}{\underline{\smash{\mu}}}

\newcommand{\etatt}{\tilde{\eta}}

\newcommand{\splus}{\! + \!}
\newcommand{\sminus}{\! - \!}

\newcommand{\refspina}{\xi_A}
\newcommand{\refspinb}{\xi_B}
\newcommand{\refspinal}{\xi_A}
\newcommand{\refspinbl}{\xi_B}

\newcommand{\ffratio}{\hat{r}}

\newcommand{\res}{\text{Res}}
\newcommand{\rest}{\widetilde{\res}}

\usepackage{feynmp}
\usepackage{ifpdf}

\ifpdf
\else
\fi

\usepackage{xkeyval}
\input{picturemacro}

\input{deformationpicturemacro}

\input{DoubleLineMakro}

\usepackage{tikz}
\usetikzlibrary{shapes}

\definecolor{grayn}{gray}{0.7}
\definecolor{lightgrayn}{gray}{0.8}

\def\bridgedistance{0.75}
\def\vacuumheight{1}
\def\ddist{0.75}
\def\hdist{\ddist*0.70710678118}
\def\labelvdist{0.3}
\def\labelhdist{0.3}
\def\labelddist{\labelvdist*0.70710678118}

\newlength{\vacuumradius}
\newlength{\onshellradius}
\tikzstyle{db}=[circle, black, fill=black, minimum width=\onshellradius, draw, inner sep=0pt]
\tikzstyle{dw}=[circle, black, fill=white, minimum width=\onshellradius, draw, inner sep=0pt]
\tikzstyle{dvac}=[circle, black, fill=lightgray, minimum width=\vacuumradius, inner sep=0pt]

\tikzstyle{dl}=[circle, black, fill=white, inner sep=2pt]

\newcommand{\drawminimalff}[1]{
        \draw[thick,double] (#1-0.5,-0) -- (#1-0.5,-0.5); 
	\draw (#1-0.5,-0.5) -- (#1-1,-\vacuumheight);  
	\draw (#1-0.5,-0.5) -- (#1,-\vacuumheight);}

\newcommand{\drawvacp}[1]{       
        \draw (#1-1,-.25) -- (#1-1,-\vacuumheight); 
        \node[dvac] at (#1-1,-0.25) {$+$};}
\newcommand{\drawvacm}[1]{       
        \draw (#1-1,-.25) -- (#1-1,-\vacuumheight); 
        \node[dvac] at (#1-1,-0.25) {$-$};}
\newcommand{\drawbridge}[2]{
        \draw (#1-1,-\vacuumheight -#2*\bridgedistance+\bridgedistance) -- (#1,-\vacuumheight -#2*\bridgedistance+\bridgedistance); 
	\node[dw] at (#1-1,-\vacuumheight -#2*\bridgedistance+\bridgedistance) {};
	\node[db] at (#1,-\vacuumheight -#2*\bridgedistance+\bridgedistance) {};}
\newcommand{\drawvline}[2]{
        \draw (#1-1,-\vacuumheight) -- (#1-1,-\vacuumheight -#2*\bridgedistance);}

\newcommand{\athreetwofthreetwo}[4]{
 \begin{aligned}
 \begin{tikzpicture}[scale=0.8]
 			\draw[thick,double] (0,-0) -- (0,0.5); 
  			\draw (0,0) -- (-\hdist,-\hdist) -- (0,-2*\hdist) -- (+\hdist,-\hdist) -- (0,0); 
  			\draw (0,-2*\hdist) -- (0,-2*\hdist-\ddist) -- (+\ddist,-2*\hdist-\ddist) -- (+\hdist+\ddist,-\hdist-\ddist) -- (+\hdist,-\hdist); 
 			\draw (+\hdist+\ddist,-\hdist-\ddist) -- (+2*\hdist+\ddist,-2*\hdist-\ddist);   
 			\draw (-\hdist,-\hdist) -- (-2*\hdist,-2*\hdist);   
 			\draw (0,-2*\hdist-\ddist) -- (0,-2*\hdist-2*\ddist); 
 			\draw (\ddist,-2*\hdist-\ddist) -- (\hdist+\ddist,-3*\hdist-\ddist);
                         \node[dw] at (+\hdist,-\hdist) {}; 
                         \node[db] at (0,-2*\hdist) {};
                         \node[dw] at (-\hdist,-\hdist) {}; 
                         \node[db] at (0,-2*\hdist-\ddist) {}; 
                         \node[dw] at (+\ddist,-2*\hdist-\ddist) {};
                         \node[db] at (+\hdist+\ddist,-\hdist-\ddist) {}; 
                         \node at (+2*\hdist +\ddist +\labelddist,-2*\hdist-\ddist-\labelddist) {#1};
                         \node at (-2*\hdist-\labelddist,-2*\hdist-\labelddist) {#4};
                         \node at (0,-2*\hdist-2*\ddist-\labelvdist) {#3};
                         \node at (\hdist +\ddist +\labelddist,-3*\hdist-\ddist-\labelddist) {#2};
         \end{tikzpicture}
\end{aligned}
}
\newcommand{\afourtwoftwotwo}[4]{
 \begin{aligned}
 \begin{tikzpicture}[scale=0.8]
			\draw[thick,double] (0,-0+\hdist) -- (0,0.5+\hdist); 
 			\draw (0,\hdist) -- (+\hdist,0);
 			\draw (0,\hdist) -- (-\hdist,0);
 			\draw (+\hdist,0) -- (1*\hdist,0) -- (2*\hdist,\hdist);
 			\draw (-\hdist,0) -- (-3*\hdist,0) -- (-4*\hdist,\hdist);
 			\draw (2*\hdist,-3*\hdist) -- (1*\hdist,-2*\hdist) -- (-3*\hdist,-2*\hdist) -- (-4*\hdist,-3*\hdist);
 			\draw (-\hdist,0) -- (-\hdist,-2*\hdist);
			\draw (-3*\hdist,0) -- (-3*\hdist,-2*\hdist);
 			\draw (\hdist,0) -- (\hdist,-2*\hdist);
                         \node[db] at (+\hdist,0) {}; 
                         \node[dw] at (-\hdist,0) {}; 
                         \node[db] at (-3*\hdist,0) {}; 
                         \node[dw] at (+\hdist,-2*\hdist) {}; 
                         \node[db] at (-\hdist,-2*\hdist) {}; 
                         \node[dw] at (-3*\hdist,-2*\hdist) {}; 
                         \node at (-4*\hdist-\labelddist,\hdist +\labelddist) {#4};
                         \node at (-4*\hdist-\labelddist,-3*\hdist-\labelddist) {#3};
                         \node at (2*\hdist +\labelddist,-3*\hdist-\labelddist) {#2};
                         \node at (2*\hdist +\labelddist,\hdist +\labelddist) {#1};
         \end{tikzpicture}
\end{aligned}
}
\newcommand{\fthreethreeathreeone}[4]{
 \begin{aligned}
 \begin{tikzpicture}[scale=0.8]
 			\draw[thick,double] (0,-0) -- (0,0.5); 
  			\draw (0,0) -- (\hdist,-\hdist) -- (0,-2*\hdist) -- (-\hdist,-\hdist) -- (0,0); 
  			\draw (0,-2*\hdist) -- (0,-2*\hdist-\ddist) -- (-\ddist,-2*\hdist-\ddist) -- (-\hdist-\ddist,-\hdist-\ddist) -- (-\hdist,-\hdist); 
 			\draw (-\hdist-\ddist,-\hdist-\ddist) -- (-2*\hdist-\ddist,-2*\hdist-\ddist);   
 			\draw (+\hdist,-\hdist) -- (+2*\hdist,-2*\hdist);   
 			\draw (0,-2*\hdist-\ddist) -- (0,-2*\hdist-2*\ddist); 
 			\draw (-\ddist,-2*\hdist-\ddist) -- (-\hdist-\ddist,-3*\hdist-\ddist);
                         \node[db] at (-\hdist,-\hdist) {}; 
                         \node[dw] at (0,-2*\hdist) {};
                         \node[db] at (+\hdist,-\hdist) {}; 
                         \node[dw] at (0,-2*\hdist-\ddist) {}; 
                         \node[db] at (-\ddist,-2*\hdist-\ddist) {};
                         \node[dw] at (-\hdist-\ddist,-\hdist-\ddist) {};  
                         \node at (-2*\hdist -\ddist -\labelddist,-2*\hdist-\ddist-\labelddist) {#2};
                         \node at (+2*\hdist+\labelddist,-2*\hdist-\labelddist) {#3};
                         \node at (0,-2*\hdist-2*\ddist-\labelvdist) {#4};
                         \node at (-\hdist -\ddist -\labelddist,-3*\hdist-\ddist-\labelddist) {#1};
         \end{tikzpicture}
\end{aligned}
}
\newcommand{\ftwotwoafourtwo}[4]{
 \begin{aligned}
 \begin{tikzpicture}[scale=0.8]
 			\draw[thick,double] (0,-0+\hdist) -- (0,0.5+\hdist); 
 			\draw (0,\hdist) -- (+\hdist,0);
 			\draw (0,\hdist) -- (-\hdist,0);
 			\draw (+\hdist,0) -- (3*\hdist,0) -- (4*\hdist,\hdist);
 			\draw (-\hdist,0) -- (-1*\hdist,0) -- (-2*\hdist,\hdist);
 			\draw (4*\hdist,-3*\hdist) -- (3*\hdist,-2*\hdist) -- (-1*\hdist,-2*\hdist) -- (-2*\hdist,-3*\hdist);
 			\draw (-\hdist,0) -- (-\hdist,-2*\hdist);
 			\draw (\hdist,0) -- (\hdist,-2*\hdist);
			\draw (3*\hdist,0) -- (3*\hdist,-2*\hdist);
                         \node[db] at (+\hdist,0) {}; 
                         \node[dw] at (+3*\hdist,0) {};
                         \node[dw] at (-\hdist,0) {}; 
                         \node[dw] at (+\hdist,-2*\hdist) {}; 
                         \node[db] at (+3*\hdist,-2*\hdist) {};
                         \node[db] at (-\hdist,-2*\hdist) {}; 
                         \node at (4*\hdist +\labelddist,\hdist +\labelddist) {#3};
                         \node at (4*\hdist +\labelddist,-3*\hdist-\labelddist) {#4};
                         \node at (-2*\hdist -\labelddist,-3*\hdist-\labelddist) {#1};
                         \node at (-2*\hdist -\labelddist,\hdist +\labelddist) {#2};
         \end{tikzpicture}
\end{aligned}
}

\def\blobhdist{0.4}
\def\blobvdist{0.1}
\def\extradist{0.12}
\def\blobheight{0.7}
\def\blobwidth{\blobhdist*1.5}

\tikzstyle{twoblob}=[ellipse, black, fill=white, minimum width=\blobwidth cm, minimum height=\blobheight cm, draw, inner sep=0pt]
\tikzstyle{threeblob}=[ellipse, black, fill=white, minimum width=\blobhdist cm + \blobwidth cm, minimum height=\blobheight cm, draw, inner sep=0pt]

\newcommand{\bloblabelsize}{\scriptsize}

\newcommand{\drawvlineblob}[2]{
        \draw (#1*\blobhdist-1*\blobhdist,0) -- (#1*\blobhdist-1*\blobhdist,#2*\blobvdist +#2*\blobheight +\blobvdist);
        \draw[white] (#1*\blobhdist-1*\blobhdist-\extradist,0) -- (#1*\blobhdist-1*\blobhdist +\extradist,0);}
\newcommand{\drawtwoblob}[3]{
\node[twoblob] at (#1*\blobhdist-1*\blobhdist+0.5*\blobhdist,#2*\blobvdist +#2*\blobheight-0.5*\blobheight) {\bloblabelsize #3};
}        
\newcommand{\drawthreeblob}[3]{
\node[threeblob] at (#1*\blobhdist-1*\blobhdist +1*\blobhdist,#2*\blobvdist +#2*\blobheight-0.5*\blobheight) {\bloblabelsize #3};
}

\input{Macros2.tex}

\newcommand{\DOT}[1]{\dot{#1}}

\newcommand{\deltap}[1]{\delta^+_{#1}}
\newcommand{\deltam}[1]{\delta^-_{#1}}

\definecolor{labelcolor}{rgb}{0,0,0}

\newcommand{\BPS}{{\scriptscriptstyle\text{BPS}}}

\newcommand{\dLIPS}{\de\text{LIPS}}
\newcommand{\dLIPSt}{\de\widetilde{\text{LIPS}}}
\newcommand{\twops}{\int\dLIPS_{2, \{l\}}}
\newcommand{\threeps}{\int \dLIPS_{3,\{l\}}}
\newcommand{\twopst}{\int\dLIPSt_{2, \{l\}}}
\newcommand{\threepst}{\int \dLIPSt_{3,\{l\}}}

\newcommand{\XYsize}{\scriptscriptstyle}

\newcommand{\Rem}{R^{(2)}}
\newcommand{\rem}{r^{(2)}}
\newcommand{\remi}{\rem_i}
\newcommand{\remif}[2]{(\remi)_{\XYsize #1}^{\XYsize #2}}

\newcommand{\interaction}{I}
\newcommand{\Interaction}{\cI}
\newcommand{\interactionr}{\underline{\interaction}}
\newcommand{\Interactionr}{\underline{\Interaction}}
\newcommand{\inttwo}[1][]{\interaction^{(2)}_{#1}}

\newcommand{\intone}[1][]{\interaction^{(1)}_{#1}}

\newcommand{\inttwoi}{\interaction^{(2)}_i}
\newcommand{\inttwoip}{\interaction^{(2)}_{i+1}}
\newcommand{\intonei}{\interaction^{(1)}_i}
\newcommand{\intoneip}{\interaction^{(1)}_{i+1}}

\newcommand{\intoneir}{\interactionr^{(1)}_i}
\newcommand{\intoneipr}{\interactionr^{(1)}_{i+1}}
\newcommand{\inttwoif}[2]{(\inttwoi)_{\XYsize #1}^{\XYsize #2}}

\newcommand{\intoneif}[2]{(\intonei)_{\XYsize #1}^{\XYsize #2}}

\newcommand{\atreef}[2]{(\ampco^{(0)}_6)_{\XYsize #1}^{\XYsize #2}}

\newcommand{\intonef}[3][]{({\interaction^{(1)}_{#1}})_{\XYsize #2}^{\XYsize #3}}

\newcommand{\zone}{\cZ^{(1)}}
\newcommand{\zonei}{\zone_i}
\newcommand{\zoneip}{\zone_{i+1}}
\newcommand{\ztwo}{\cZ^{(2)}}
\newcommand{\ztwoi}{\ztwo_i}

\newcommand{\symb}{\mathcal{S}}
\newcommand{\Dila}{\mathfrak{D}}
\newcommand{\dila}{\mathfrak{D}}
\newcommand{\loopDila}[1]{\Dila^{(#1)}}
\newcommand{\loopdila}[1]{\dila^{(#1)}}
\newcommand{\loopdilai}[2]{(\loopdila{#1})_{#2}}

\newcommand{\dilatwo}[1][]{\dila^{(2)}_{#1}}

\newcommand{\perm}{\mathbb{P}}

\newcommand{\Interactionamp}{\tilde{\Interaction}}

\newcommand{\peps}{\varepsilon}
\newcommand{\teps}{\epsilon}

\newcommand{\amp}{\cA}
\newcommand{\ampco}{\hat{\amp}}
\newcommand{\ff}{\cF}
\newcommand{\ffco}{\hat{\ff}}

\newcommand{\g}{g}
\newcommand{\gmod}{\tilde{g}}

\newcommand{\ren}{{\text{ren}}}
\newcommand{\bare}{{\text{bare}}}

\newcommand{\PSU}[1]{\operatorname{PSU}(#1)}
\renewcommand{\SU}[1]{\operatorname{SU}(#1)}
\newcommand{\SUbar}[1]{\overline{\operatorname{SU}}(#1)}
\renewcommand{\U}[1]{\operatorname{U}(#1)}
\renewcommand{\SO}[1]{\operatorname{SO}(#1)}
\newcommand{\SL}[1]{\operatorname{SL}(#1)}
\newcommand{\GL}[1]{\operatorname{GL}(#1)}

\newcommand{\GrassmannSymbol}{\operatorname{Gr}}

\newcommand{\deltaf}[1]{\delta^{\ff}_{#1}}

\newcommand{\abr}[1]{\langle #1 \rangle}
\newcommand{\sbr}[1]{\left[ #1 \right]}
\newcommand{\bigabr}[1]{\left\langle #1 \right\rangle}

\newcommand{\vll}{{\smash{\lambda}}}
\newcommand{\vlt}{{\smash{\tilde{\lambda}}}}

\newcommand{\vle}{{\smash{\tilde{\eta}}}}
\newcommand{\vlet}{{\smash{\eta}}}

\newcommand{\vmtuu}{\smash{\underline{{\smash{\tilde{\mu}}}}}}

\newcommand{\ssep}{\;}
\newcommand{\sssep}{\,}

\newcommand{\jj}{\tilde{\mathfrak{J}}}

\DeclareMathOperator{\str}{str}

\newcommand{\ffgen}{\hat{\tilde{\ff}}}

\numberwithin{equation}{chapter}

\usepackage{feynmp}
\makeatletter

\define@boolkey{FDiagram}{schannel}[true]{}
\define@boolkey{FDiagram}{tchannel}[true]{}
\define@boolkey{FDiagram}{xchannel}[true]{}
\define@boolkey{FDiagram}{leftSE}[true]{}
\define@boolkey{FDiagram}{rightSE}[true]{}
\define@boolkey{FDiagram}{long}[true]{}
\define@boolkey{FDiagram}{longup}[true]{}
\define@boolkey{FDiagram}{oldlabels}[true]{}
\define@key{FDiagram}{styleleftbottom}%
 {\def\FDiagramstyleleftbottom{#1}}
\define@key{FDiagram}{stylerightbottom}%
 {\def\FDiagramstylerightbottom{#1}}
\define@key{FDiagram}{stylelefttop}%
 {\def\FDiagramstylelefttop{#1}}
\define@key{FDiagram}{stylerighttop}%
 {\def\FDiagramstylerighttop{#1}}
\define@key{FDiagram}{stylemid}%
 {\def\FDiagramstylemid{#1}}
\define@key{FDiagram}{labelleftbottom}%
 {\def\FDiagramlabelleftbottom{#1}}
\define@key{FDiagram}{labelrightbottom}%
 {\def\FDiagramlabelrightbottom{#1}}
\define@key{FDiagram}{labellefttop}%
 {\def\FDiagramlabellefttop{#1}}
\define@key{FDiagram}{labelrighttop}%
 {\def\FDiagramlabelrighttop{#1}}
\define@key{FDiagram}{labelmid}%
 {\def\FDiagramlabelmid{#1}}
\presetkeys{FDiagram}{%
schannel=false,%
tchannel=false,%
xchannel=false,%
leftSE=false,%
rightSE=false,%
long=false,%
longup=false,%
oldlabels=false,%
labelleftbottom=$\phantom{0}$,%
labelrightbottom=$\phantom{0}$,%
labellefttop=$\phantom{0}$,%
labelrighttop=$\phantom{0}$,%
labelmid=$\phantom{0}$,%
styleleftbottom=plain,%
styleleftbottom=plain,%
stylerightbottom=plain,%
stylelefttop=plain,%
stylerighttop=plain,%
stylemid=plain}{}
\newcommand*\FDiagram[6][]{%
 \setkeys{FDiagram}{#1}%
\settoheight{\eqoff}{$\times$}%
\setlength{\eqoff}{0.5\eqoff}%
\addtolength{\eqoff}{-12.0\unitlength}%
\raisebox{\eqoff}{%
\fmfframe(2,2)(2,2){%
\begin{fmfchar*}(12,20)
\fmfbottom{vb1,vbb1,v1,vb2,vbb2,vb3,vb4,v2,vbb5,vb5}
\fmftop{vt1,vtt1,v3,vt2,vtt2,vt3,vt4,v4,vtt5,vt5}
\ifKV@FDiagram@schannel
\fmf{#2,left=0.3,tension=1}{v1,vc1}
\fmf{#3,right=0.3,tension=1}{v2,vc1}
\fmf{#4,tension=2}{vc1,vc2}
\fmf{#5,left=0.3,foreground=(0.65,,0.65,,0.65)}{vc2,v3}
\fmf{#6,right=0.3,foreground=(0.65,,0.65,,0.65)}{vc2,v4}
\else\fi
\ifKV@FDiagram@tchannel
\fmf{#2,left=0,tension=1}{v1,vc1}
\fmf{#3,right=0,tension=1}{v2,vc2}
\fmf{#4,tension=0}{vc1,vc2}
\fmf{#5,foreground=(0.65,,0.65,,0.65)}{vc1,v3}
\fmf{#6,foreground=(0.65,,0.65,,0.65)}{vc2,v4}
\else\fi
\ifKV@FDiagram@xchannel
\fmf{#2,left=0.25,tension=1}{v1,vc1}
\fmf{#3,right=0.25,tension=1}{v2,vc1}
\fmf{#5,left=0.25,foreground=(0.65,,0.65,,0.65)}{vc1,v3}
\fmf{#6,right=0.25,foreground=(0.65,,0.65,,0.65)}{vc1,v4}
\else\fi
\ifKV@FDiagram@leftSE
\fmf{#2,left=0,tension=1}{v1,vc1}
\fmf{#2,right=0,tension=1,foreground=(0.65,,0.65,,0.65)}{vc1,v3}
\fmf{#3,foreground=(0.65,,0.65,,0.65)}{v2,v4}
\fmfv{decor.shape=circle,decor.filled=full,decor.size=15}{vc1}
\else\fi
\ifKV@FDiagram@rightSE
\fmf{#3,left=0,tension=1}{v2,vc1}
\fmf{#3,right=0,tension=1,foreground=(0.65,,0.65,,0.65)}{vc1,v4}
\fmf{#2,foreground=(0.65,,0.65,,0.65)}{v1,v3}
\fmfv{decor.shape=circle,decor.filled=full,decor.size=15}{vc1}
\else\fi
\fmffreeze
\fmfposition
\fmfipath{p[]}
\fmfipair{vm[]}
\fmfcmd{pair verta, vertb, vertc, vertd, vertca, vertcb; verta = vloc(__v1); vertb = vloc(__v2); vertc = vloc(__v3); vertd = vloc(__v4); vertca = vloc(__vc1); vertcb = vloc(__vc2);}
\ifKV@FDiagram@oldlabels
\fmfiv{label=\FDiagramlabelleftbottom,l.a=+120,l.dist=0.07w}{verta}
\fmfiv{label=\FDiagramlabelrightbottom,l.a=+60,l.dist=0.07w}{vertb}
\ifKV@FDiagram@tchannel
\fmfiv{label=\FDiagramlabelmid,l.a=30,l.dist=0.18w}{vertca}
\else
\fmfiv{label=\FDiagramlabelmid,l.a=60,l.dist=0.20w}{vertca}
\fi
\fmfiv{label=\FDiagramlabellefttop,l.a=-120,l.dist=0.07w}{vertc}
\fmfiv{label=\FDiagramlabelrighttop,l.a=-60,l.dist=0.07w}{vertd}
\else
\fmfiv{label=\FDiagramlabelleftbottom,l.a=+120,l.dist=0.07w}{verta}
\fmfiv{label=\FDiagramlabelrightbottom,l.a=+60,l.dist=0.07w}{vertb}
\ifKV@FDiagram@tchannel
\fmfiv{label=\FDiagramlabellefttop,l.a=+120,l.dist=0.07w}{vertca}
\fmfiv{label=\FDiagramlabelrighttop,l.a=+60,l.dist=0.07w}{vertcb}
\else
\fi
\ifKV@FDiagram@schannel
\fmfiv{label=\FDiagramlabellefttop,l.a=+165,l.dist=0.22w}{vertcb}
\fmfiv{label=\FDiagramlabelrighttop,l.a=+15,l.dist=0.22w}{vertcb}
\else
\fi
\ifKV@FDiagram@xchannel
\fmfiv{label=\FDiagramlabellefttop,l.a=+158,l.dist=0.20w}{vertca+(0,0.025h)}
\fmfiv{label=\FDiagramlabelrighttop,l.a=+22,l.dist=0.20w}{vertca+(0,0.025h)}
\else
\fi
\ifKV@FDiagram@leftSE
\fmfiv{label=\FDiagramlabellefttop,l.a=+120,l.dist=0.07w}{verta+(0,0.6125h)}
\else
\fi
\ifKV@FDiagram@rightSE
\fmfiv{label=\FDiagramlabelrighttop,l.a=+60,l.dist=0.07w}{vertb+(0,0.6125h)}
\else
\fi
\fi
\fmfdraw
\ifKV@FDiagram@long
\fmf{plain,width=1mm}{vb1,vb5}
\else 
\fmf{plain,width=1mm}{v1,v2}
\fi
\ifKV@FDiagram@longup
\fmf{plain,width=1mm,foreground=(0.65,,0.65,,0.65)}{vt1,vt5}
\fi
\end{fmfchar*}%
}}%
}
\makeatother

\makeatletter
\DeclareRobustCommand*{\bfseries}{%
  \not@math@alphabet\bfseries\mathbf
  \fontseries\bfdefault\selectfont
  \boldmath
}

\makeatother

\newcommand{\beq}{\begin{equation}}
\newcommand{\eeq}{\end{equation}}
\newcommand{\beqa}{\begin{eqnarray}}
\newcommand{\eeqa}{\end{eqnarray}}
\newcommand{\bea}{\begin{aligned}}
\newcommand{\eea}{\end{aligned}}

\title{Form factors and the dilatation operator in $\mathcal{N}=4$ super Yang-Mills theory and its deformations}
\author{Matthias Wilhelm}

\begin{document}
\begin{fmffile}{diagrams}
\fmfcmd{%
thin := 1pt; 
thick := 2thin;
arrow_len := 3mm;
arrow_ang := 15;
curly_len := 3mm;
dash_len :=0.3; 
dot_len := 0.75mm; 
wiggly_len := 2mm; 
wiggly_slope := 60;
zigzag_len := 2mm;
zigzag_width := 2thick;
decor_size := 5mm;
dot_size := 2thick;
}
\fmfcmd{%
marksize=7mm;
def draw_cut(expr p,a) =
  begingroup
    save t,tip,dma,dmb; pair tip,dma,dmb;
    t=arctime a of p;
    tip =marksize*unitvector direction t of p;
    dma =marksize*unitvector direction t of p rotated -90;
    dmb =marksize*unitvector direction t of p rotated 90;
    linejoin:=beveled;
    drawoptions(dashed dashpattern(on 3bp off 3bp on 3bp));
    draw ((-.5dma.. -.5dmb) shifted point t of p);
    drawoptions();
  endgroup
enddef;
style_def phantom_cut expr p =
    save amid;
    amid=.5*arclength p;
    draw_cut(p, amid);
    draw p;
enddef;
}
\fmfcmd{%
smallmarksize=4mm;
def draw_smallcut(expr p,a) =
  begingroup
    save t,tip,dma,dmb; pair tip,dma,dmb;
    t=arctime a of p;
    tip =smallmarksize*unitvector direction t of p;
    dma =smallmarksize*unitvector direction t of p rotated -90;
    dmb =smallmarksize*unitvector direction t of p rotated 90;
    linejoin:=beveled;
    drawoptions(dashed dashpattern(on 2bp off 2bp on 2bp) withcolor red);
    draw ((-.5dma.. -.5dmb) shifted point t of p);
    drawoptions();
  endgroup
enddef;
style_def phantom_smallcut expr p =
    save amid;
    amid=.5*arclength p;
    draw_smallcut(p, amid);
    draw p;
enddef;
}

\fmfcmd{%
style_def plain_ar expr p =
  cdraw p;
  shrink (0.6);
  cfill (arrow p);
  endshrink;
enddef;
style_def plain_rar expr p =
  cdraw p; 
  shrink (0.6);
  cfill (arrow reverse(p));
  endshrink;
enddef;
style_def dashes_ar expr p =
  draw_dashes p;
  shrink (0.6);
  cfill (arrow p);
  endshrink;
enddef;
style_def dashes_rar expr p =
  draw_dashes p;
  shrink (0.6);
  cfill (arrow reverse(p));
  endshrink;
enddef;
style_def dots_ar expr p =
  draw_dots p;
  shrink (0.6);
  cfill (arrow p);
  endshrink;
enddef;
style_def dots_rar expr p =
  draw_dots p;
  shrink (0.6);
  cfill (arrow reverse(p));
  endshrink;
enddef;
}

\fmfcmd{%
style_def phantom_cross expr p =
    save amid,ang;
    amid=.5*length p;
    ang= angle direction amid of p;
    draw ((polycross 4) scaled 8 rotated ang) shifted point amid of p;
enddef;
}

\fmfcmd{%
style_def plain_sarrow expr p =
  cdraw p;
  shrink (0.55); 
  cfill (arrow p);
  endshrink;
enddef;
style_def dashes_sarrow expr p =
  draw_dashes p;
  shrink (0.55);
  cfill (arrow p);
  endshrink;
enddef;
style_def dots_sarrow expr p =
  draw_dots p;
  shrink (0.55);
  cfill (arrow p);
  endshrink;
enddef;
style_def plain_srarrow expr p =
  cdraw p;
  shrink (0.55);
  cfill (arrow (reverse p));
  endshrink;
enddef;
style_def dashes_srarrow expr p =
  draw_dashes p;
  shrink (0.55);
  cfill (arrow (reverse p));
  endshrink;
enddef;
style_def dots_srarrow expr p =
  draw_dots p;
  shrink (0.55);
  cfill (arrow (reverse p));
  endshrink;
enddef;
}

\feynmpdefinitionscaps

\begingroup\parindent0pt
\vspace*{4em}
\centering
\begingroup\LARGE
\bf Form factors and the dilatation operator in  \\ $\mathcal{N}=4$ super Yang-Mills theory and its deformations
\par\endgroup
\vspace{2.5em}
 \vspace{\baselineskip}
D i s s e r t a t i o n 
 \vspace{\baselineskip}

(Überarbeitete Fassung) 
 \vspace{2\baselineskip}

eingereicht an der 
 \vspace{\baselineskip}

Mathematisch-Naturwissenschaftlichen Fakultät

der Humboldt-Universität zu Berlin 
 \vspace{\baselineskip}

von 
 \vspace{\baselineskip}

{\bf Matthias Wilhelm}
 \vspace{\baselineskip}

\href{mailto:matthias.wilhelm@nbi.ku.dk}{matthias.wilhelm@nbi.ku.dk} 
 
 \vspace{2\baselineskip}
{\it 
Institut für Physik und Institut für Mathematik,Humboldt-Universität zu Berlin,\\
IRIS Gebäude,
Zum Großen Windkanal 6,
12489 Berlin
\vspace{\baselineskip}

Niels Bohr Institute, Copenhagen University,\\
Blegdamsvej 17, 2100 Copenhagen \O{}, Denmark
}

\endgroup

\thispagestyle{empty}

\cleardoublepage
\phantomsection
\addcontentsline{toc}{chapter}{Zusammenfassung}
\chapter*{Zusammenfassung}
\markboth{Zusammenfassung}{Zusammenfassung}

Seit mehr als einem halben Jahrhundert bietet die Quantenfeldtheorie (QFT) den genausten und erfolgreichsten 
 theoretischen Rahmen zur Beschreibung der fundamentalen Wechselwirkungen zwischen Elementarteilchen, wenn auch mit Ausnahme der Gravitation.
Dennoch sind QFTs im Allgemeinen weit davon entfernt, vollständig verstanden zu sein.
Dies liegt an einem Mangel an theoretischen Methoden zur Berechnung ihrer Observablen sowie an fehlendem Verständnis der auftretenden mathematischen Strukturen.
In den letzten anderthalb Jahrzehnten kam es zu bedeutendem Fortschritt im Verständnis von speziellen Aspekten einer bestimmten QFT, der maximal supersymmetrischen Yang-Mills-Theorie in vier Dimensionen, auch $\cN=4$ SYM-Theorie genannt. Diese haben die Hoffnung geweckt, dass die $\cN=4$ SYM-Theorie exakt lösbar ist.
Besonders bemerkenswert 
 war der Fortschritt auf den Gebiet der Streuamplituden auf Grund der Entwicklung sogenannter Masseschalen-Methoden und auf dem Gebiet der Korrelationsfunktionen zusammengesetzter Operatoren auf Grund von Integrabilität.
In dieser Dissertation gehen wir der Frage nach, ob und in welchem Umfang die in diesem Kontext gefunden Methoden und Strukturen 
auch zum Verständnis weitere Größen in dieser Theorie sowie zum Verständnis anderer Theorien beitragen können.

Formfaktoren beschreiben den quantenfeldtheoretischen Überlapp eines lokalen, eichinvarianten, zusammengesetzten Operators mit einem asymptotischen Streuzustand.
Als solche bilden sie eine Brücke zwischen der Welt der Streuamplituden, deren externe Impulse sich auf der Masseschale befinden, auf der einen Seite und der Welt der Korrelationsfunktionen von zusammengesetzten Operatoren, welche keine entsprechende Bedingung erfüllen, auf der anderen Seite.
Im ersten Teil dieser Arbeit berechnen wir Formfaktoren von allgemeinen, geschützten und ungeschützten Operatoren für verschiedene Schleifenordnungen und Multiplizitäten externer Teilchen in der 
 $\cN=4$ SYM-Theorie.
Dies gelingt durch Anwendung verschiedener Masseschalen-Methoden, die im Kontext von Streuamplituden entwickelt wurden und sehr erfolgreich angewandt werden konnten, wenn auch erst nach wichtigen Weiterentwicklungen.
Insbesondere zeigen wir, wie Formfaktoren und die zuvor genannten Methoden es ermöglichen, den Di\-la\-ta\-ti\-ons\-ope\-ra\-tor zu bestimmen. Dieser Operator liefert das Spektrum der anomalen Skalendimensionen der zusammengesetzten Operatoren und wirkt als Hamilton-Operator der integrablen Spin-Kette des Spektralproblems.
Auf Einschleifenordnung nutzen wir verallgemeinerte Unitarität, um den aus entsprechenden Schnitten rekonstruierbaren
 Teil des Formfaktors mit minimaler Multiplizität für beliebige zusammengesetzte Operatoren zu berechnen, von dem wir den vollständigen Dilatationsoperator auf Einschleifenordnung ablesen können.
Am Beispiel des Konishi-Operators und Operatoren des $\SU{2}$-Sektors auf Zweischleifenordnung zeigen wir, dass 
Masseschalen-Methoden und Formfaktoren auch auf höheren Schleifenordnungen zur Bestimmung des Dilatationsoperators eingesetzt werden können.
Die Rückstandsfunktionen letztgenannter Formfaktoren erfüllen interessante universelle Eigenschaften im Bezug auf ihre Transzendenz.
Auf Baumgraphenniveau konstruieren wir Formfaktoren über erweiterte Masseschalen-Diagramme, Graßmann-Integrale und die integrabilitätsinspirierte Technik der $\rr$-Operatoren. Letztere 
ermöglicht es, Formfaktoren als Eigenzustände der integrablen Transfermatrix zu konstruieren, was die Existenz eines Satzes erhaltener Ladungen impliziert.

Deformationen der $\cN=4$ SYM-Theorie erlauben es uns, andere Theorien mit den gleichen speziellen Eigenschaften zu finden und neue Erkenntnisse über den Ursprung von Integrabilität und der AdS/CFT-Korrespondenz zu gewinnen.
Im zweiten Teil dieser Arbeit untersuchen wir die $\cN=1$ supersymmetrische $\beta$-Deformation und die nichtsupersymmetrische $\gamma_i$-Deformation.
Beide teilen viele Eigenschaften mit der $\cN=4$ SYM-Theorie, speziell im planaren Limes. Sie zeigen jedoch auch neue Merkmale, insbesondere das Auftreten von Doppelspurtermen in ihrem Wirkungsfunktional. Zwar scheinen diese Terme im planaren Limes zu verschwinden, doch können sie durch einen neuen Effekt der endlichen Systemgröße, welchen wir Vorwickeln nennen, in führender Ordnung beitragen.
In der $\beta$-Deformation werden diese Terme für die konforme Invarianz benötigt und wir berechnen die durch sie entstehenden Korrekturen zum vollständigen planaren Di\-la\-ta\-ti\-ons\-ope\-ra\-tor auf Einschleifenordnung und dessen Spektrum.
In der $\gamma_i$-Deformation zeigen wir, dass Quantenkorrekturen rennende Doppelspurkopplungen ohne Fixpunkte induzieren, was die konforme Invarianz bricht.
Dann berechnen wir die planaren anomalen Skalendimensionen von Einspuroperatoren, die aus $L$ identischen Skalarfeldern bestehen, bei der kritischen Wickelordnung $\ell=L$ für alle $L\geq 2$. Für $L\geq3$ stimmen die Ergebnisse unser feldtheoretischen Rechnung exakt mit den durch Integrabilität gewonnenen Vorhersagen überein.
Für $L=2$, wo die Vorhersage durch Integrabilität divergiert, finden wir ein endliches, rationales Ergebnis. Dieses hängt jedoch von der rennenden Doppelspurkopplung und durch sie vom Renormierungsschema ab.

\cleardoublepage
\phantomsection
\addcontentsline{toc}{chapter}{Abstract}
\chapter*{Abstract}
\markboth{Abstract}{Abstract}

For more than half a century, quantum field theory (QFT) has been the most accurate and successful framework to describe the fundamental interactions among elementary particles, albeit with the notable exception of gravity.
Nevertheless, QFTs are in general far from being completely understood. This is due to a lack of calculational techniques and tools as well as our limited understanding of the mathematical structures that emerge 
 in them.
In the last one and a half decades, tremendous progress has been made in understanding certain aspects of a particular QFT, namely the maximally supersymmetric Yang-Mills theory in four dimensions, termed $\cN=4$ SYM theory, which has risen the hope that this theory could be exactly solvable.
In particular, this progress occurred for scattering amplitudes due to the development of on-shell methods and for correlation functions of gauge-invariant local composite operators due to integrability.
In this thesis, we address the question to which extend the methods and structures found there can be generalised to other quantities in the same theory and to other theories. 

Form factors 
describe the overlap between a gauge-invariant local composite operator on the one hand and an asymptotic on-shell scattering state on the other hand.
Thus, they form a bridge between the purely off-shell 
 correlation functions and the purely on-shell scattering amplitudes.
In the first part of this thesis, we calculate form factors of general, protected as well as non-protected, operators at various loop orders and numbers of external points in 
 $\cN=4$ SYM theory.
This is achieved using many of the successful on-shell methods that were developed in the context of scattering amplitudes, albeit
 after some important extensions.
In particular, we show how form factors and on-shell methods allow us to obtain the dilatation operator, which yields the spectrum of anomalous dimensions of composite operators and acts as Hamiltonian of the integrable spin chain of the spectral problem.
At one-loop level, we calculate the cut-constructible part of the form factor with minimal particle multiplicity 
 for any operator using generalised unitarity and obtain the complete one-loop dilatation operator from it.
We demonstrate that on-shell methods and form factors can be used to calculate the dilatation operator also at higher loop orders, using the Konishi operator and the $\SU{2}$ sector at two loops as examples.
Remarkably, the finite remainder functions of the latter form factors possess universal properties with respect to their transcendentality.
Moreover, form factors of non-protected operators share many features of scattering amplitudes in QCD, such as UV divergences and rational terms.
At tree level, we show how to construct form factors via extended on-shell diagrams, a Graßmannian integral as well as the integrability-based technique of $\rr$ operators.
Using the latter technique, form factors can be constructed as eigenstates of an integrable transfer matrix,
which implies the existence of a tower of conserved charges.

Deformations of $\cN=4$ SYM theory allow us to find further theories with its special properties 
 and to shed light on the origins of integrability and of the AdS/CFT correspondence.
In the second part of this thesis, we study the $\cN=1$ supersymmetric $\beta$-deformation and the non-supersymmetric $\gamma_i$-deformation. 
While they share many properties of their undeformed parent theory, in particular in the planar limit, also new features arise.
These new features are related to the 
 occurrence of double-trace terms in the action. 
Although apparently suppressed, double-trace terms can contribute at leading order in the planar limit via a new kind of finite-size effect, which we call prewrapping.
In the $\beta$-deformation, these double-trace terms are required for conformal symmetry, and we calculate the corresponding corrections to the complete planar one-loop dilatation operator and its spectrum.
In the $\gamma_i$-deformations, we show that running double-trace terms without fixed points are induced via quantum corrections, thus breaking conformal invariance.
We then calculate the planar anomalous dimensions of single-trace operators built from $L$ identical scalars at critical wrapping order $\ell=L$ for any $L\geq2$. 
At $L\geq3$, our field-theory results perfectly match the predictions from integrability.
At $L=2$, where the integrability-based prediction diverges, we find a finite rational result, which does however depend on the running double-trace coupling and thus on the renormalisation scheme.

\setcounter{tocdepth}{1}

\tableofcontents

\cleardoublepage
\phantomsection
\addcontentsline{toc}{chapter}{Publications}
\chapter*{Publications}
\markboth{Publications}{Publications}

This thesis is based on the following publications by the author:
\begin{itemize}
 \item[\cite{Fokken:2013aea}]
J.~Fokken, C.~Sieg, and M.~Wilhelm, 
``{Non-conformality of $\gamma_i$-deformed $\mathcal{N}=4$ SYM theory},''
\href{http://dx.doi.org/10.1088/1751-8113/47/45/455401}{{\em J. Phys. A: Math. Theor.} {\bfseries 47} (2014) 455401}, 
\href{http://arxiv.org/abs/1308.4420}{{\ttfamily arXiv:1308.4420 [hep-th]}}.
\item[\cite{Fokken:2013mza}]
J.~Fokken, C.~Sieg, and M.~Wilhelm, 
``{The complete one-loop dilatation operator of planar real $\beta$-deformed $ \mathcal{N}  = 4$ SYM theory},''
 \href{http://dx.doi.org/10.1007/JHEP07(2014)150}{{\em JHEP} {\bfseries 1407} (2014) 150},\\
\href{http://arxiv.org/abs/1312.2959}{{\ttfamily arXiv:1312.2959 [hep-th]}}.
\item[\cite{Fokken:2014soa}]
J.~Fokken, C.~Sieg, and M.~Wilhelm, 
``{A piece of cake: the ground-state energies in $\gamma_{i}$-deformed $ \mathcal{N} = 4$ SYM theory at leading wrapping order},'' \href{http://dx.doi.org/10.1007/JHEP09(2014)078}{{\em JHEP} {\bfseries 1409} (2014) 78},
\href{http://arxiv.org/abs/1405.6712}{{\ttfamily arXiv:1405.6712 [hep-th]}}.
\item[\cite{Wilhelm:2014qua}]
M.~Wilhelm, 
``{Amplitudes, Form Factors and the Dilatation Operator in $\mathcal{N}=4$ SYM Theory},'' 
\href{http://dx.doi.org/10.1007/JHEP02(2015)149}{{\em JHEP} {\bfseries 1502} (2015) 149},
\href{http://arxiv.org/abs/1410.6309}{{\ttfamily arXiv:1410.6309 [hep-th]}}.
\item[\cite{Nandan:2014oga}]
D.~Nandan, C.~Sieg, M.~Wilhelm, and G.~Yang, 
``{Cutting through form factors and cross sections of non-protected operators in $\mathcal{N}=4$ SYM},'' 
\href{http://dx.doi.org/10.1007/JHEP06(2015)156}{{\em JHEP} {\bfseries 1506} (2015) 156},
\href{http://arxiv.org/abs/1410.8485}{{\ttfamily arXiv:1410.8485 [hep-th]}}.
\item[\cite{Loebbert:2015ova}]
F.~Loebbert, D.~Nandan, C.~Sieg, M.~Wilhelm, and G.~Yang, 
``{On-Shell Methods for the Two-Loop Dilatation Operator and Finite Remainders },'' 
\href{http://dx.doi.org/10.1007/JHEP10(2015)012}{{\em JHEP} {\bfseries 1510} (2015) 012}, 
\href{http://arxiv.org/abs/1504.06323}{{\ttfamily arXiv:1504.06323 [hep-th]}}.
\item[\cite{Frassek:2015rka}]
R.~Frassek, D.~Meidinger, D.~Nandan, and M.~Wilhelm,
``On-shell Diagrams, Gra{\ss}mannians and Integrability for Form Factors,''
\href{http://dx.doi.org/10.1007/JHEP01(2016)182}{{\em JHEP} {\bfseries 1601} (2016) 182}, \\
\href{http://arxiv.org/abs/1506.08192}{{\ttfamily arXiv:1506.08192 [hep-th]}}.
\end{itemize}
The author has also contributed to the following publications:
\begin{itemize}
 \item[\cite{Schroers:2014dua}]
B.~Schroers, and M.~Wilhelm,
``{Towards Non-Commutative Deformations of Relativistic Wave Equations in 2+1 Dimensions},''
\href{http://dx.doi.org/10.3842/SIGMA.2014.053}{{\em SIGMA} {\bfseries 1410} (2014) 053}, \\
\href{http://arxiv.org/abs/1402.7039}{{\ttfamily arXiv:1402.7039 [hep-th]}}.
\item[\cite{Fokken:2014moa}]
J.~Fokken, and M.~Wilhelm, 
``{One-Loop Partition Functions in Deformed $\mathcal{N}=4$ SYM Theory},'' 
\href{http://dx.doi.org/10.1007/JHEP03(2015)018}{{\em JHEP} {\bfseries 1503} (2015) 018}, 
\href{http://arxiv.org/abs/1411.7695}{{\ttfamily arXiv:1411.7695 [hep-th]}}.
\end{itemize}

\cleardoublepage
\phantomsection
\addcontentsline{toc}{chapter}{Introduction}
\chapter*{Introduction}
\markboth{Introduction}{Introduction}
\label{chap: general introduction}

Quantum field theory (QFT) is arguably the most successful theoretical framework to describe and predict the fundamental interactions between the elementary particles, albeit 
 with the notable exception of gravity. 
In the form of the Standard Model of particle physics (SM), it describes three of the four known fundamental forces of nature: electromagnetism, the weak force and the strong force. 
A particle which is consistent with being the last missing piece to the Standard Model, a Higgs boson, was recently discovered at the Large Hadron Collider (LHC) \cite{Chatrchyan:2012xdj,Aad:2012tfa}. 
Using the Standard Model, theoretical predictions could be made that were confirmed by experiments with unprecedented precision.
The magnetic moment of the electron, for example, is known with an accuracy of $10^{-12}$, which is the equivalent of knowing the distance from New York to Moscow by the width of a hair.

Despite these successes, however, quantum field theory and in particular the Standard Model are far from being completely understood.
One reason for this is that many quantities 
 are currently only accessible via perturbation theory, in which the accuracy of the prediction decreases as the strength of the interaction increases. 
While processes involving only the electromagnetic force and the weak force are relatively well accounted for by considering only the first quantum correction, those involving the strong force require considerably more computational effort. 
For instance, very involved calculations \cite{Anastasiou:2015ema} are required to determine whether all properties of the discovered Higgs boson agree with the predictions of the Standard Model and where new physics might emerge.
Moreover, the strength of interactions is not constant but depends on the energy scale.
At low energies, the strong force, which is described by quantum chromodynamics (QCD), is so strong that perturbation theory becomes meaningless.
Hence, non-perturbative methods are required e.g.\ to answer why the elementary quarks are confined to hadrons such as protons and to calculate the mass of the latter composite particles.

In order to develop a qualitative understanding of quantum field theories in general as well as calculational techniques that can later be applied to the Standard Model, it is useful to look at the simplest non-trivial quantum field theory in four dimensions.
Arguably, this is the maximally supersymmetric Yang-Mills theory ($\mathcal{N}=4$ SYM theory) \cite{Brink:1976bc}, which is sometimes also called the harmonic oscillator of the $21^{\text{st}}$ century.
As the Standard Model, it is a non-Abelian gauge theory, but in contrast to the Standard Model, it enjoys many more symmetries.
Its field content also consists of 
gauge bosons, fermions and scalars, but these are all related by supersymmetry. 
Moreover, $\cN=4$ SYM theory is conformally invariant, which implies that the strength of the interactions is scale independent.
Together with Poincar\'{e} invariance, these symmetries combine to the superconformal invariance with the symmetry group $\PSU{2,2|4}$.

Remarkably, we cannot only learn something about gauge theories by studying $\cN=4$ SYM theory. 
Via the anti-de Sitter / conformal field theory (AdS/CFT) correspondence \cite{Maldacena:1997re,Gubser:1998bc,Witten:1998qj}, $\cN=4$ SYM theory is conjectured to be dual to a certain kind of string theory, namely type IIB superstring theory on the curved background $\AdS_5\times\text{S}^5$, which is the product of five-dimensional anti-de Sitter space and the five-sphere.%
\footnote{See \cite{Aharony:1999ti,D'Hoker:2002aw} for reviews.}
String theory is a candidate for quantum gravity, i.e.\ a quantum theory of gravity. Gravity, the fourth known fundamental force of nature, cannot be incorporated into the framework of perturbative quantum field theory. 
Using the AdS/CFT correspondence, we can hence learn something about string theory and thus gravity by studying gauge theory, and vice versa.

Both $\cN=4$ SYM theory and type IIB superstring theory can be further simplified by taking 't Hooft's planar limit \cite{'tHooft:1973jz}.
In the gauge theory with gauge group $\U{N}$ or $\SU{N}$, this amounts to taking the number of colours
 $N\to \infty$ and the Yang-Mills coupling $g_\YM\to0$ 
 while keeping the 't Hooft coupling $\lambda=g_\YM^2 N$ fixed. 
As a result, only Feynman diagrams that are planar with respect to their colour structure contribute.
The string theory, on the other hand, becomes free.

A surprising property of both theories in the 't Hooft limit is integrability; see \cite{Beisert:2010jr} for a review.
The concept 
of integrability goes back to Hans Bethe. In 1931, he solved the spectrum of the $(1+1)$-dimensional Heisenberg spin chain, a simple model for magnetism in solid states, with an ansatz that now bears his name \cite{Bethe:1931}. 
As some principles of integrability are fundamentally two-dimensional, its first occurrence in a four-dimensional theory, concretely in high-energy scattering in planar QCD as found by Lipatov \cite{Lipatov:1993yb}%
, came unexpected.  
Later, integrability was also found in $\cN=4$ SYM theory in the spectrum of anomalous dimensions of gauge-invariant local composite operators.
These operators are built from products of traces of elementary fields at the same point in spacetime.
Conformal symmetry significantly constrains the form 
 of their correlation functions,
which are 
 an important class of observables in a gauge theory. 
In particular, it guarantees that a basis of operators exists in which the two-point correlation functions are determined entirely by the operators' scaling dimensions. For so-called scalar conformal primary 
operators in this basis, the non-vanishing two-point functions read
\begin{equation}
 \ev{\cO(x)\cO(y)}=\frac{1}{(x-y)^{2\Delta}} \eqncom \qquad \Delta=\Delta_0+\gamma \eqncom
\end{equation}
where $\Delta$ is the scaling dimension of the operator $\cO$.
For operators saturating a Bogomolny-Prasad-Sommerfield-type bound \cite{Bogomolny:1975de,Prasad:1975kr}, called BPS operators, $\Delta$ is protected by supersymmetry and equals the classical scaling dimension $\Delta_0$; generically, however, $\Delta$ receives quantum corrections captured in terms of an anomalous part $\gamma$ that is added to $\Delta_0$.
The scaling dimensions can be measured as eigenvalues of the generator of dilatations, the dilatation operator, which is part of the conformal algebra. 
Diagonalising the dilatation operator, though, is a non-trivial problem which can be simplified by restricting to certain subsectors of the complete theory.
It was found by Minahan and Zarembo that the action of the one-loop dilatation operator of $\cN=4$ SYM theory on single-trace operators in the so-called $\SO{6}$ subsector maps to the action of the 
Hamiltonian of an integrable spin chain and that it can hence be diagonalised by a Bethe ansatz \cite{Minahan:2002ve}. 
This was later extended to the complete one-loop dilatation operator \cite{Beisert:2003yb}, which was found in \cite{Beisert:2003jj}.
Postulating integrability to be present also at higher loop orders, an all-loop asymptotic Bethe ansatz was formulated \cite{Beisert:2005fw}.
This ansatz is valid provided that the range of the interaction is smaller than the number $L$ of fields in the single-trace operator, which corresponds to the length of the spin chain, and hence for loop orders $\ell<L-1$.%
\footnote{Due to the structure of the interactions and the presence of supersymmetry, the asymptotic Bethe ansatz in $\cN=4$ SYM theory is in fact even valid for higher loop orders.
}

A further important step in the development of integrability in $\cN=4$ SYM theory is marked by finite-size effects.
As gauge-invariant local composite operators are colour singlets, Feynman diagrams which are non-planar in momentum space can still be planar with respect to their colour structure and hence contribute in the 't Hooft limit.
One mechanism giving rise to such diagrams is the so-called wrapping effect \cite{Sieg:2005kd},%
\footnote{See also \cite{Beisert:2004ry,Beisert:2004hm} for earlier discussions.}
which stems from interactions wrapping once around the operator.
The leading wrapping correction to the Konishi operator, which is the prime example of a non-protected 
operator, was calculated in \cite{Fiamberti:2007rj,Fiamberti:2008sh,Velizhanin:2008jd}.
The wrapping effect is incorporated into the framework of integrability in terms of Lüscher corrections \cite{Bajnok:2008bm} and the thermodynamic Bethe ansatz (TBA) \cite{Ambjorn:2005wa,Arutyunov:2007tc,Arutyunov:2009zu,Bombardelli:2009ns,Gromov:2009bc,Arutyunov:2009ur}.
After several reformulations as Y-system \cite{Gromov:2009tv}, T-system, Q-system and finite system of non-linear integral equations (FINLIE) \cite{Gromov:2011cx}, the present formulation as quantum spectral curve (QSC) \cite{Gromov:2013pga} is currently able to yield anomalous dimensions up to the tenth loop order \cite{Marboe:2014gma}.
Furthermore, numeric results at any value of the coupling are available \cite{Gromov:2015wca}. 
Thus, integrability opens up a window of quantitative non-perturbative understanding of gauge theories.

The latter successes in solving the spectral problem, however, are much closer in spirit to the string-theory description, where the classically integrable two-dimensional sigma model serves as a natural starting point. 
The field-theoretic origin of integrability is still largely unclear. %
One further complication is that the length of the spin chain in $\cN=4$ SYM theory
is not constant beyond one loop order, which makes it hard to describe using the solid-state-physics-inspired spin-chain techniques.
Moreover, the complete dilatation operator, and hence also the eigenstates that correspond to the anomalous dimensions, are 
still only known at one-loop order.

Although most insights of $\cN=4$ SYM theory into QCD are of qualitative nature, a surprising quantitative relation exists as well.
In \cite{Kotikov:2001sc}, it was argued that the anomalous dimensions of twist-two operators in $\cN=4$ SYM theory are of uniform transcendentality and given by the leading transcendental part of the corresponding expressions in QCD. This relation
 is known as principle of maximal transcendentality; see \cite{Kotikov:2004er, Kotikov:2006ts, Gehrmann:2011xn, Li:2014afw} for further discussions.

A further important advancement in understanding $\cN=4$ SYM theory, and gauge theories in general, was the development 
 of so-called 
 on-shell methods for scattering amplitudes; see e.g.\ \cite{Elvang:2013cua,Henn:2014yza} for reviews.
Scattering amplitudes describe the interaction of usually 
 two incoming elementary particles producing $n-2$ outgoing elementary particles.
They are the basic ingredients for cross sections, which are the observables determined experimentally at colliders. 
Using 
 crossing symmetry 
 to choose all $n$ elementary fields to be outgoing, the $n$-point scattering amplitude is given by the overlap of an outgoing $n$-particle on-shell state with the vacuum $\ket{0}$:
\begin{equation}
\label{eq: amplitude introduction}
\amp_n(1,2,\dots,n)=\ev{1,2,\dots,n|0}\eqndot
\end{equation}
Here, on-shell means that the external momenta $p_i^\mu$ satisfy the mass-shell condition $p_i^2=p_i^\mu p_{i,\mu}=m^2$, where $m^2=0$ in the case of $\cN=4$ SYM theory. 
Almost 30 years ago, Parke and Taylor succeeded in writing down a closed formula for the tree-level scattering amplitude of two polarised gluons of negative helicity with $n-2$ polarised gluons of positive helicity in any Yang-Mills theory \cite{Parke:1986gb}.
Proving this formula is greatly facilitated by choosing a set of variables in which the on-shell condition of the external fields in four dimensions is manifest, namely spinor-helicity variables $\lambda_i^\alpha$, $\lambdat_i^\alphadot$. 
In the maximally supersymmetric $\cN=4$ SYM theory, their fermionic analogues are given by the expansion parameters $\etatt_i^A$ of Nair's $\cN=4$ on-shell superspace \cite{Nair:1988bq}.

The main idea behind on-shell methods is to build amplitudes not via Feynman diagrams with virtual particles and gauge dependence. Instead, they are built from other amplitudes with a lower number of legs or a lower number of loops, which are manifestly gauge-invariant and whose external particles are real.

One important on-shell method is unitarity \cite{Bern:1994zx,Bern:1994cg}, which uses the fact that the scattering matrix is unitary and generalises the optical theorem.
Via unitarity, loop-level amplitudes can be reconstructed from their discontinuities, which are given by products of lower-loop and tree-level amplitudes.
These discontinuities can by calculated via so-called cuts, which impose the on-shell condition on internal propagators. 
In generalised unitarity \cite{Britto:2004nc}, also cuts are taken that do not correspond to discontinuities but still lead to a factorisation of the loop-level amplitude into lower-loop and tree-level amplitudes.

One problem in the calculation of amplitudes as well as other quantities is the occurrence of divergences, which need to be regularised. This can be achieved by continuing the dimension of spacetime from $D=4$ to $D=4-2\peps$.
Although $D$-dimensional unitarity exists, the on-shell unitarity method as well as other on-shell methods are most powerful in four dimensions, where spinor-helicity variables can be used.
Integrands that vanish in four dimensions can, however, integrate to expressions that are non-vanishing in four dimensions.
At one-loop level, they evaluate to rational terms, which have no discontinuities.
Hence, they cannot be reconstructed via four-dimensional unitarity.
In $\cN=4$ SYM theory, however, all one-loop amplitudes were proven to be cut-constructible \cite{Bern:1994zx}.

The structure of divergences in amplitudes is well understood. Since $\cN=4$ SYM theory is conformally invariant, no ultraviolet (UV) divergences arise in amplitudes,
 only infrared (IR) divergences.
Based on the universality and exponentiation properties of the latter, 
 Bern, Dixon and Smirnov (BDS) conjectured that the all-loop expression for the 
logarithm of the
amplitude is completely determined by the IR structure and the one-loop finite part \cite{Bern:2005iz}.%
\footnote{See also the previous studies \cite{Anastasiou:2003kj} including those in QCD \cite{Catani:1998bh, Sterman:2002qn}.}%
$^,$\footnote{The coefficient of the leading IR divergence, the so-called cusp anomalous dimension, was actually determined via integrability for all values of the 't Hooft coupling \cite{Beisert:2006ez}.
}
Although correct for four and five points, the BDS ansatz deviates from the complete amplitude at higher points \cite{Alday:2007he}.
The difference, which was termed remainder function, was first studied for six points in \cite{Bartels:2008ce, Bern:2008ap,Drummond:2008aq}. It exhibits uniform transcendentality and is composed of (generalised) polylogarithms, which can be simplified using the Hopf-algebraic structure of these functions, in particular the so-called symbol \cite{Goncharov09, Goncharov:2010jf}; see \cite{Duhr:2014woa} for a review.%
\footnote{Using the structure of the occurring transcendental functions, the six-point remainders can currently be bootstrapped up to four-loop order; see \cite{Dixon:2011pw,Dixon:2014xca} and references therein.}$^,$%
\footnote{For higher loops and points, examples of amplitudes are known that contain also elliptic functions \cite{CaronHuot:2012ab,ArkaniHamed:2012nw,Nandan:2013ip}.
}

Further important on-shell methods, namely Cachazo-Svrcek-Witten (CSW) \cite{Cachazo:2004kj} and Britto-Cachazo-Feng-Witten (BCFW) \cite{Britto:2004ap,Britto:2005fq} recursion relations, make use of the fact that tree-level amplitudes as well as their loop-level integrands are analytic functions of the external momenta. 
The poles of these functions correspond to propagators going on-shell, resulting in the factorisation of the amplitude into lower-point amplitudes or the forward limit of a lower-loop amplitude with two additional points.
Using these methods, all tree-level amplitudes of $\cN=4$ SYM theory could be calculated \cite{Drummond:2008cr} as well as the unregularised integrand of all loop-level amplitudes \cite{ArkaniHamed:2010kv}.
To understand the structure and the symmetries of these results, also the formulation in twistor \cite{Penrose:1967wn} 
and momentum-twistor \cite{Hodges:2009hk} variables has been very useful.

Tree-level scattering amplitudes and their unregularised loop-level integrands can also be represented by so-called on-shell diagrams \cite{ArkaniHamed:2012nw}, which furthermore yield the leading singularities of loop-level amplitudes.%
\footnote{More recently, on-shell diagrams were also studied for non-planar amplitudes \cite{Arkani-Hamed:2014via,Chen:2014ara,Arkani-Hamed:2014bca,Bern:2014kca,Franco:2015rma,Chen:2015qna} and planar amplitudes in less supersymmetric theories \cite{ArkaniHamed:2012nw,Benincasa:2015zna}.
}
Moreover, all tree-level amplitudes can be obtained as residues of integrating a certain on-shell form over the Graßmannian manifold $\GrassmannSymbol(n,k)$, i.e.\ the set of $k$-planes in $n$-dimensional space \cite{ArkaniHamed:2009dn,Mason:2009qx,ArkaniHamed:2009vw}. 
Here, $k$ is the maximally-helicity-violating (MHV) degree of the amplitude, which corresponds to a degree of $4k$ in the fermionic $\etatt$ variables.
Amplitudes with $k=2$ are denoted as MHV, amplitudes with $k=3$ as next-to-MHV (NMHV) and amplitudes with general $k$ as N$^{k-2}$MHV.
Furthermore, amplitudes can be understood geometrically as volumes of polytopes that triangulate the so-called amplituhedron \cite{Arkani-Hamed:2013jha,Arkani-Hamed:2013kca,ArkaniHamed:2010gg}, which also generalises to loop level.

In addition to providing qualitative understanding of scattering amplitudes and calculational techniques, the study of scattering amplitudes in $\cN=4$ SYM theory can also serve as an intermediate step in calculating scattering amplitudes in pure Yang-Mills theory or massless QCD. The latter theories share the computationally most challenging part, the gauge fields, with $\cN=4$ SYM theory. Therefore, the differences can be accounted for as corrections that are easier to calculate; see e.g.\ \cite{Bern:1994zx}.%
\footnote{At tree level, the contributions from the differing field content can even be projected out \cite{Dixon:2010ik}.}

For several years, the developments in scattering amplitudes and integrability proceeded independently.
In \cite{Drummond:2009fd},
however, it was found that the superconformal symmetry and the newly discovered dual superconformal symmetry \cite{Drummond:2008vq} of scattering amplitudes combine into a Yangian symmetry, which is a smoking gun of integrability.
Moreover, based on the fact that both objects are completely fixed by symmetry, Benjamin Zwiebel found a connection between the leading length-changing contributions to the  dilatation operator and all tree-level amplitudes \cite{Zwiebel:2011bx}. In particular, it connects the complete one-loop dilatation operator to the four-point tree-level amplitude.\footnote{For this special case, this connection goes back to Niklas Beisert.}
These findings inspired the study of the integrable structure of scattering amplitudes at weak coupling as well as their deformation with respect to the central-charge extension of $\PSU{2,2|4}$ \cite{Ferro:2012xw,Ferro:2013dga,Chicherin:2013ora,Frassek:2013xza,Beisert:2014qba,Kanning:2014maa,Broedel:2014pia,Broedel:2014hca,Bargheer:2014mxa,Ferro:2014gca,Kanning:2014cca}.
In particular, a spin chain appeared in this context as well, albeit a slightly different one. 
Via the duality between scattering amplitudes and Wilson loops \cite{Alday:2007hr}, the integrable structure of scattering amplitudes is currently better understood at strong coupling, where it can be mapped to a minimal surface problem \cite{Alday:2007hr} that can be solved via a Y-system \cite{Alday:2009dv,Alday:2010vh}.
Recently, much progress using the latter approach has also been made at finite coupling in certain kinematic regimes; see \cite{Basso:2013vsa,Basso:2014hfa} and references therein.
 \vspace{\baselineskip}

Given the success of on-shell methods for scattering amplitudes and the interesting structures found in them, it is an intriguing question 
whether 
 they may be generalised 
to quantities that include one or more composite operators.
An ideal starting point to answer this question is given by form factors.
Form factors describe the overlap of a state created by a composite operator $\cO$ from the vacuum with an $n$-particle on-shell state, i.e.\
\begin{equation}\label{eq: form factor introduction}
 \ff_{\cO,n}(1,\dots,n;x)=\bra{1,\dots,n}\cO(x)\ket{0}\eqndot
\end{equation}
In contrast to the elementary fields in the on-shell state, the momentum $q$ associated with the composite operator via a Fourier transformation 
 does not satisfy the on-shell condition, i.e.\ $q^2\neq0$, and we hence call it off-shell.
Containing $n$ on-shell fields and one off-shell composite operator, form factors form a bridge between the purely on-shell scattering amplitudes and the 
purely off-shell correlation functions.
Moreover, generalised form factors, which contain multiple operators, are the most general correlators composed of local objects alone.

Similar to scattering amplitudes, form factors occur in many physical applications including collider physics.
For instance, the composite operator can arise as part of a vertex in an effective Lagrangian.
A concrete example for this is the dominant Higgs production mechanism at the LHC, in which two gluons fuse to a Higgs boson via a top-quark loop.
As the top mass is much larger than the Higgs mass, the top-quark loop can be integrated out to obtain an effective dimension-five operator  $H\tr(F_{\mu\nu}F^{\mu\nu})$, see e.g.\ \cite{Schmidt:1997wr}.%
\footnote{In particular, this approximation is used in the calculation of \cite{Anastasiou:2015ema}.}
In $\cN=4$ SYM theory, 
 $\tr(F_{\mu\nu}F^{\mu\nu})$ is part of the stress-tensor supermultiplet, which contains $\tr(\phi_{14}\phi_{14})$ as its lowest component.
The operator can also be the (conserved) current describing a two-particle scattering such as $e^+ e^-$ annihilation into a virtual photon or Drell-Yan scattering. 
Moreover, form factors appear in the calculation of ‘event shapes’ such as energy or charge correlation functions \cite{Hofman:2008ar, Engelund:2012re, Belitsky:2013xxa, Belitsky:2013bja} as well as deep inelastic scattering in $\mathcal{N}=4$ SYM theory \cite{Bianchi:2013sta}. 
Form factors have also played an important role in understanding the exponentiation and universal structure of IR divergences, which in turn helped to understand scattering amplitudes \cite{Mueller:1979ih, Collins:1980ih, Sen:1981sd, Magnea:1990zb}.

Form factors in $\cN=4$ SYM theory were first studied 30 years ago by van Neerven \cite{vanNeerven:1985ja}.
Interest resurged when a description at strong coupling was found via the AdS/CFT correspondence as a minimal surface problem \cite{Alday:2007he}. This minimal surface problem is similar to the one of amplitudes and can also be solved via integrability techniques \cite{Maldacena:2010kp,Gao:2013dza}. 
Many studies at weak coupling followed \cite{Brandhuber:2010ad,Bork:2010wf,Brandhuber:2011tv,Bork:2011cj,Henn:2011by,Gehrmann:2011xn,Brandhuber:2012vm,Bork:2012tt,Engelund:2012re,Johansson:2012zv,Boels:2012ew,Penante:2014sza,Brandhuber:2014ica,Bork:2014eqa,Huang:2016bmv}.
In particular, it was shown that many of the successful on-shell techniques that were developed in the context of scattering amplitudes can also be applied to form factors.
Concretely, spinor-helicity variables \cite{Brandhuber:2010ad}, Nair's $\cN=4$ on-shell superspace \cite{Brandhuber:2011tv}, twistor \cite{Brandhuber:2010ad} and momentum-twistor \cite{Brandhuber:2011tv} variables,
BCFW \cite{Brandhuber:2010ad} and CSW recursion relations \cite{Brandhuber:2011tv} as well as 
(generalised) unitarity \cite{Brandhuber:2010ad,Johansson:2012zv} were shown to be applicable. 
In certain examples, also colour-kinematic duality \cite{Bern:2008qj} was found to be present \cite{Boels:2012ew}.
Furthermore, an interpretation of the tree-level expressions in terms of the volume of polytopes exists \cite{Bork:2014eqa}.
Interestingly, the remainder of the two-loop three-point form factor of $\tr(\phi_{14}\phi_{14})$ \cite{Brandhuber:2012vm} was found to match the highest transcendentality part of the remainder of the Higgs-to-three-gluon amplitude in QCD \cite{Gehrmann:2011aa}, thus extending the maximal transcendentality principle from numbers to functions of the kinematic variables.%
\footnote{A relation between the transcendental functions describing energy-energy correlation in $\cN=4$ SYM theory and QCD was also found in \cite{Belitsky:2013ofa}.}
Via generalised unitarity, also correlation functions can be built using amplitudes, form factors and generalised form factors 
 as building blocks \cite{Engelund:2012re}.
As for scattering amplitudes, the complexity of calculating form factors increases with the number of loops and external fields.
A form factor with the minimal number of external fields, namely as many as there are fields in the operator, is called a \emph{minimal} form factor. 

However, most previous studies have 
 focused on the form factors of the stress-tensor supermultiplet and its lowest component $\tr(\phi_{14}\phi_{14})$ as well as its generalisation to $\tr(\phi_{14}^L)$.
The minimal form factors of these operators have been calculated up to three-loop order \cite{Gehrmann:2011xn} and 
 two loop-order \cite{Brandhuber:2014ica}, respectively.%
\footnote{The integrand of the minimal form factor of $\tr(\phi_{14}\phi_{14})$ is even known up to four-loop order \cite{Boels:2012ew,Boels:2015yna}.}
The only exceptions are operators from the $\SU{2}$ and $\SL{2}$ subsectors, whose tree-level \MHV form factors were given in \cite{Engelund:2012re}, and the Konishi operator, whose minimal one-loop form factor was calculated in \cite{Bork:2010wf}.\footnote{We address an important subtlety occurring in the latter result further below.}
In fact, among experts, it has been 
 a vexing problem how to calculate the minimal two-loop Konishi form factor via unitarity.
Moreover, not all interesting structures that were discovered for scattering amplitudes have found a counterpart for form factors yet and the role of integrability for form factors at weak coupling has remained 
 unclear.

In the first part of this thesis, which is based on a series of papers \cite{Wilhelm:2014qua,Nandan:2014oga,Loebbert:2015ova,Frassek:2015rka} by the present author and collaborators, we focus on form factors.

We calculate form factors of general, protected as well as non-protected, operators at various loop orders and for various numbers of external points.
We show that the minimal tree-level form factor of a generic operator is essentially given by considering the operator in the oscillator representation \cite{Gunaydin:1981yq,Gunaydin:1998sw,Beisert:2003jj} of the spin-chain picture and replacing the oscillators by super-spinor-helicity variables. 
Moreover, the generators of the superconformal algebra in the corresponding representations are related by the same replacement.%
\footnote{This replacement was already studied in \cite{Beisert:2010jq} and also played an important role in \cite{Zwiebel:2011bx}. However, no connection to form factors was made in these works.}
In particular, this allows us to use on-shell techniques from the study of scattering amplitudes to determine the dilatation operator, which is the spin-chain Hamiltonian.
Hence, minimal form factors realise the spin chain of the spectral problem of $\cN=4$ SYM theory in the language of scattering amplitudes.

At one-loop level, we calculate the cut-constructible part of the minimal form factor of any operator via generalised unitarity and extract the complete one-loop dilatation operator from its UV divergence.
In particular, this yields a field-theoretic derivation of the connection between the one-loop dilatation operator and the four-point amplitude found in \cite{Zwiebel:2011bx}.
Furthermore, we calculate the minimal form factor of the Konishi primary operator up to two-loop order using unitarity and obtain the two-loop Konishi anomalous dimension from it,
thus solving this long known problem. 
The occurrence of general operators, such as the Konishi operator, requires an extension of the unitarity method to include the correct regularisation.
At one-loop order this extension leads to a new kind of rational terms, whereas from two-loop order on it affects also the divergent contributions and hence the dilatation operator. 
We also calculate the two-loop minimal form factors in the $\SU{2}$ sector and extract the corresponding dilatation operator.
In contrast to the aforementioned cases, this case involves both the mixing of UV and IR divergences and operator mixing,
such that the exponentiation of the divergences takes an operatorial form. 
Moreover, we calculate the two-loop remainder function, which is an operator in this case, via the BDS ansatz, which has to be promoted to an operatorial form as well.
Its matrix elements satisfy linear relations which are a consequence of Ward identities for the form factor.
For generic operators, the remainder function is not of uniform transcendentality.
However, its maximally transcendental part is universal and agrees with the remainder of the BPS operator $\tr(\phi_{14}^L)$ calculated in \cite{Brandhuber:2014ica}, thus extending the principle of maximal transcendentality even further.%
\footnote{Anomalous dimensions and the dilatation operator can also be determined via on-shell methods and correlation functions.
In \cite{Engelund:2012re}, certain matrix elements of the one-loop dilatation operator in the $\SL{2}$ sector were obtained via three-point functions and generalised unitarity.
In \cite{Koster:2014fva}, which appeared contemporaneously with \cite{Wilhelm:2014qua} by the present author, the one-loop dilatation operator in the $\SO{6}$ sector was calculated via two-point functions and the twistor action.
In \cite{Nandan:2014oga}, the present author and collaborators have calculated the two-loop Konishi anomalous dimension also via two-point functions and unitarity, where the same subtlety in the regularisation appears as for form factors.
The results of \cite{Koster:2014fva} were later reproduced using MHV rules and generalised unitarity in \cite{Brandhuber:2014pta} and \cite{Brandhuber:2015boa}, respectively.
For further on-shell approaches to correlation functions using a spacetime version of generalised unitarity and twistor-space Lagrangian-insertion techniques, see \cite{Engelund:2015cfa,Laenen:2015jia} and \cite{Chicherin:2014uca}, respectively.
}

Furthermore,
we study tree-level form factors for a generic number of external on-shell fields with a focus on the stress-tensor supermultiplet. We extend on-shell diagrams to describe form factors, which requires 
 to include the minimal form factor as an additional building block.
This allows us to find a Graßmannian integral representation of form factors in spinor-helicity variables, twistors and momentum twistors.  
Moreover, we introduce a central-charge deformation of form factors and show that they can be constructed via the integrability-based technique of $\rr$ operators.
In the non-minimal case, form factors embed the spin chain of the spectral problem in the one that appeared in the study of scattering amplitudes.
In particular, we find that form factors are eigenstates of the transfer matrix of the latter spin chain provided that the corresponding operators are eigenstates of the transfer matrix of the former spin chain.
This implies the existence of a tower of conserved charges and symmetry under the action of a part of the Yangian.%
\footnote{Some of the results presented in \cite{Frassek:2015rka} were also independently found in \cite{Bork:2015fla}.}

 \vspace{\baselineskip}
 
Given the success of integrability in $\cN=4$ SYM theory, in particular in the planar spectrum of anomalous dimensions, as well as its many remarkable properties, such as the existence of an AdS/CFT dual, the question arises whether more theories with these properties can be found that can be equally solved via integrability.
Moreover, one wonders how 
integrability is related to conformal symmetry and the high amount of supersymmetry and what its origin is.
These questions can be addressed by studying deformations of $\cN=4$ SYM theory in which the high amount of (super)symmetry is reduced in a controlled way.
The deformations fall into two classes: discrete orbifold theories and continuous deformations,
see \cite{Zoubos:2010kh,vanTongeren:2013gva} for reviews.

The prime example of a continuous deformation is the so-called $\beta$-deformation, which has one real deformation parameter $\beta$.
It is a special case of the $\cN=1$ supersymmetric exactly marginal deformations of $\cN=4$ SYM theory, which were classified by 
Leigh and Strassler \cite{Leigh:1995ep}.
In \cite{Lunin:2005jy},  Lunin and Maldacena conjectured the $\beta$-deformation to be dual to type IIB superstring theory on a certain deformed background. 
This background can be constructed by applying a sequence of a T duality, a shift (s) along an angular coordinate and another T duality to the $\text{S}^5$ factor of  $\AdS_5\times\text{S}^5$.
Applying three such TsT transformations instead,
Frolov generalised this setup to the non-supersymmetric three-parameter $\gamma_i$-deformation \cite{Frolov:2005dj}, which reduces to the $\beta$-deformation in the limit where all real deformation parameters $\gamma_i$, $i=1,2,3$, are equal.

Both the $\beta$- and the $\gamma_i$-deformation can be formulated in terms of a Moyal-like $\ast$-product, which replaces the usual product of fields in the action.
A similar $\ast$-product occurs in a certain type of spacetime non-commutative field theories, where the deformation parameter is related to the Planck constant $\hbar$; see \cite{Szabo:2001kg} for a review.
In the latter theories, planar diagrams of elementary interactions can be related to their undeformed counterparts via a theorem by Thomas Filk \cite{Filk:1996dm}.
This theorem can be adapted to planar single-trace 
diagrams in the $\beta$- and the $\gamma_i$-deformation, in particular to the diagrams that yield the asymptotic dilatation operator density.%
\footnote{For discussions in the context of orbifold theories, see \cite{Bershadsky:1998mb,Bershadsky:1998cb}.}
This was used in \cite{Beisert:2005if} to relate the asymptotic one-loop dilatation operator in the deformed theories to the one in $\cN=4$ SYM theory and to formulate an asymptotic Bethe ansatz, showing that the deformed theories are asymptotically integrable,
i.e.\ integrable in the absence of finite-size effects. 
Moreover, it was shown that the $\beta$- and the $\gamma_i$-deformation are the most general $\cN=1$ supersymmetric and non-supersymmetric continuous asymptotically integrable field-theory deformations of  
$\cN=4$ SYM theory, respectively.

Further checks of integrability in the deformed theories must hence go beyond the asymptotic level, to where finite-size effects contribute.
Their corresponding subdiagrams of elementary interactions are non-planar, and therefore, a priori, Filk's theorem is not applicable.
In \cite{Fiamberti:2008sn}, the anomalous dimensions of the so-called single-impurity states in the $\beta$-deformation were calculated via Feynman diagrams at leading wrapping order, yielding explicit results for $3\leq \ell= L \leq 11$.
These are single-trace states in the $\SU{2}$ sector which are composed of one complex scalar of one kind and $L-1$ complex scalars of a second kind, say $\tr(\phi_{14}^{L-1}\phi_{24}^{1})$.  
Being protected in the undeformed theory, their anomalous dimensions receive contributions only due to the presence of the deformation.
Using integrability, the results of \cite{Fiamberti:2008sn} have been reproduced in \cite{Gunnesson:2009nn} for $\beta=\frac{1}{2}$ and in \cite{Gromov:2010dy} and \cite{Arutyunov:2010gu} for generic $\beta$, based on Lüscher corrections, Y-system and TBA equations, respectively.
At $L=2$, however, the integrability-based predictions diverge. 
In the $\gamma_i$-deformation, also a state composed of only one kind of complex scalars, say $\tr(\phi_{14}^L)$, is not protected. 
In contrast to the single-impurity states in the $\beta$-deformation, which receive also corrections from deformed single-trace interactions, the anomalous dimensions of the vacuum states in the $\gamma_i$-deformation receive contributions only from finite-size effects.
This makes them particularly well suited for testing the non-trivial effects of the deformation on integrability.
For these states, which correspond to the vacuum of the spin chain, integrability-based predictions exist up to double-wrapping order $\ell=2L$ using Lüscher corrections, the TBA and the Y-system \cite{Ahn:2011xq}.%
\footnote{These results were also recently reproduced at single-wrapping order using the QSC \cite{Kazakov:2015efa}.
}
However, also in this case, the integrability-based prediction diverges for $L=2$.%
\footnote{A similar divergence for the vacuum state $\tr(\phi_{14}\phi_{14})$ has previously occurred in the undeformed theory \cite{Frolov:2009in} and in non-supersymmetric orbifold theories \cite{deLeeuw:2011rw}. In the undeformed theory, the divergence can be regularised using a twist in the $\AdS_5$ direction to show that the anomalous dimension of $\tr(\phi_{14}\phi_{14})$ vanishes \cite{deLeeuw:2012hp}. This regularisation extends to the vacuum state in the $\beta$-deformation \cite{FrolovPC}.}

An interesting property of the deformed theories which does not have a counterpart in $\cN=4$ SYM theory is related to the choice of $\U{N}$ or $\SU{N}$ as gauge group.
In $\cN=4$ SYM theory, all interactions are of commutator type, i.e.\ the interaction part of the action can be formulated such that the colour matrices of any given field only occur in a commutator.
Hence, the additional $\U{1}$ mode in the theory with gauge group $\U{N}$ decouples from all interaction and is thus free.
As a consequence, the undeformed theories with gauge group $\U{N}$ and $\SU{N}$ are essentially the same.

In the deformed theories, the commutators are replaced by $\ast$-commutators, from which the $\U{1}$ mode no longer decouples. 
The theories with gauge group $\U{N}$ and $\SU{N}$ are hence different \cite{Freedman:2005cg,Frolov:2005iq}.%
\footnote{See \cite{Frolov:2005iq} also for a discussion in the context of the AdS/CFT correspondence.}
Moreover, the $\beta$-deformed theory with gauge group $\U{N}$ is not even conformally invariant, as quantum corrections induce the running of a double-trace coupling in the component action \cite{Hollowood:2004ek}. 
In the conformally invariant $\beta$-deformation with gauge group $\SU{N}$, this coupling is at its non-vanishing IR fixed point; its fixed-point value can be obtained by integrating out the auxiliary fields in the deformed action in $\cN=1$ superspace, see e.g.\ \cite{Fokken:2013aea}.
Furthermore, this double-trace coupling is responsible for making the planar one-loop anomalous dimension of $\tr(\phi_{14}\phi_{24})$, the aforementioned single-impurity state with $L=2$, vanish for gauge group $\SU{N}$ while it is non-vanishing for gauge group $\U{N}$ \cite{Freedman:2005cg}.

In contrast to single-trace couplings, double-trace couplings are in general not restricted by Filk's theorem.
In particular, they are not covered by the proofs of conformal invariance of the planar deformed theories \cite{Ananth:2006ac,Ananth:2007px}, which only apply to the single-trace couplings.
In non-supersymmetric orbifold theories, running double-trace couplings without fixed point were found, which break conformal invariance \cite{Dymarsky:2005uh}. 
These findings amounted to a no-go theorem that no perturbatively accessible conformally invariant non-supersymmetric orbifold theory can exist \cite{Dymarsky:2005nc}.%
\footnote{The above arguments exclude fixed points of the double-trace coupling as a function of the Yang-Mills coupling, i.e.\ fixed lines. They cannot exclude Banks-Zaks fixed points \cite{Banks:1981nn} though, which are isolated fixed points at some finite but perturbatively accessible value of the Yang-Mills coupling.
}
Moreover, 
these running double-trace couplings were related to the occurrence of tachyons in the dual string theory \cite{Dymarsky:2005uh}, similar to the case for non-commutative field theories treated in \cite{Armoni:2003va}.%
\footnote{%
Non-supersymmetric orientifolds of type $0\,$B string theory can be tachyon free, see e.g.\ \cite{Blumenhagen:1999ns, Blumenhagen:1999uy,Angelantonj:1999qg}, and the corresponding gauge theory was shown to have no running double-trace couplings \cite{Liendo:2011da}. 
}

In the second part of this thesis, 
we discuss further developments in the field of deformations based on the series of papers \cite{Fokken:2013aea,Fokken:2013mza,Fokken:2014soa} by the present author and collaborators.
In particular, we study the influence of double-trace couplings and the double-trace structure in the $\SU{N}$ propagator on correlation functions of general operators.
It can be understood in terms of a new kind of finite-size effect, which starts to affect operators one loop order earlier than the wrapping effect and which we hence call prewrapping.
Based on the mechanism behind it, we classify which operators are potentially affected by prewrapping.
Moreover, we incorporate prewrapping and wrapping into the asymptotic one-loop dilatation operator of \cite{Beisert:2005if} to obtain the complete one-loop dilatation operator of the planar $\beta$-deformation.%
\footnote{At one-loop order, prewrapping affects operators of length two for gauge group $\SU{N}$ while wrapping affects operators of length one for gauge group $\U{N}$.}

We show that the $\gamma_i$-deformation in the form proposed in \cite{Frolov:2005dj} 
is not conformally invariant due to a running double-trace coupling without fixed point, neither for gauge group $\U{N}$ nor $\SU{N}$.
Furthermore, it cannot be rendered conformally invariant by including further multi-trace couplings that fulfil a set of minimal requirements.
We then calculate the anomalous dimension of the vacuum states $\tr(\phi^L_{14})$ in the $\gamma_i$-deformation at critical wrapping order $\ell=L$. 
For $L\geq3$, the calculation can be reduced to four Feynman diagrams which can be evaluated analytically for any $L$.
We find a perfect match with the prediction of integrability.
For $L=2$, the finite planar two-loop anomalous dimension depends on the running double-trace coupling and hence on the renormalisation scheme. 
This explicitly demonstrates that the theory is not conformally invariant, not even in the planar limit.
Interestingly, the (unresolved) divergences in the integrability-based description occur in the same cases in which the double-trace couplings contribute.

\cleardoublepage
\phantomsection
\addcontentsline{toc}{chapter}{Overview}
\chapter*{Overview}
\markboth{Overview}{Overview}

This work is structured as follows.

In chapter \ref{chap: N=4 SYM}, we give a short introduction to $\cN=4$ SYM theory and other concepts that will be important in both parts of this work. These include the spin-chain picture of composite operators, the 't Hooft limit and the complete one-loop dilatation operator of $\cN=4$ SYM theory.

The main body of this thesis in divided into two parts. 
The first part treats form factors in $\cN=4$ SYM theory and encompasses chapters \ref{chap: minimal tree-level form factors}, \ref{chap: minimal one-loop form factors}, \ref{chap: two-loop Konishi form factor}, \ref{chap: two-loop su(2) form factors} and \ref{chap: tree-level form factors}.

In the first section of chapter \ref{chap: minimal tree-level form factors}, we introduce important concepts for form factors as well as our conventions and notation. Based on \cite{Wilhelm:2014qua}, we then calculate the minimal tree-level form factors for generic composite operators.

In chapter \ref{chap: minimal one-loop form factors}, which is largely based on \cite{Wilhelm:2014qua}, we start to calculate loop corrections to the minimal form factors. In section \ref{sec: general structure of loop corrections}, we discuss the general structure of loop corrections to the minimal form factors and how one can read off the dilatation operator from them.
In section \ref{sec: one-loop unitarity}, we give a pedagogical example of using the on-shell unitarity method to calculate the minimal one-loop form factors in the $\SU{2}$ sector and the corresponding one-loop dilatation operator.
We then calculate the cut-constructible part of the one-loop correction to the minimal form factor of a generic 
 operator using generalised unitarity in section \ref{sec: one-loop generalised unitarity}. From its UV divergence, we can read off the complete one-loop dilatation operator of $\cN=4$ SYM theory.

In chapter \ref{chap: two-loop Konishi form factor}, which is based on \cite{Nandan:2014oga}, we demonstrate that on-shell methods and form factors can also be employed to calculate anomalous dimensions at two-loop level using the Konishi primary operator as an example.
After giving a short introduction to this operator in section \ref{sec: konishi}, we calculate its minimal one- and two-loop form factors via the unitarity method in section \ref{sec: konishi form factors}.
However, for operators 
 like the Konishi primary operator, important subtleties occur when using four-dimensional on-shell methods, 
which require the extension of these methods.
In section \ref{sec: subtleties}, we analyse these subtleties in detail and show how to treat them correctly. 
We give the resulting form factors in section \ref{sec: Konishi results}.

A further challenge at two-loop order, the non-trivial 
 exponentiation of UV and IR divergences due to operator mixing, is tackled in chapter \ref{chap: two-loop su(2) form factors}, where we treat the two-loop form factors in the $\SU{2}$ sector.
In section \ref{sec: two-loop form factors via unitarity}, we calculate the two-loop minimal form factors of all operators in the $\SU{2}$ sector.
We extract the two-loop dilatation operator in section \ref{sec: two-loop dilatation operator}.
In section \ref{sec: remainder}, we calculate the corresponding finite remainder functions via the BDS ansatz, which has to be promoted to an operatorial form, and find interesting universal behaviour with respect to their transcendentality. 

In chapter \ref{chap: tree-level form factors}, which is based on \cite{Frassek:2015rka}, we consider tree-level form factors with a focus on the stress-tensor supermultiplet.
After a short introduction to this supermultiplet and its form factors in section \ref{sec: stress-energy supermultiplet}, we briefly introduce on-shell diagrams and extend them to form factors in section \ref{sec: On-shell diagrams}.
We then define a central-charge deformation for form factors and show how to systematically construct them via the integrability-based method of $\rr$ operators in section \ref{sec: r operators and integrability}. 
In section \ref{sec: Grassmannian integrals}, we find a Graßmannian integral in spinor-helicity variables, twistors and momentum twistors, whose residues yield the form factors.

The second part of this work, which encompasses chapters \ref{chap: introduction to integrable deformations}, \ref{chap: prewrapping}, \ref{chap: nonconformality} and \ref{chap: anomalous dimensions}, treats deformations of $\cN=4$ SYM theory.
We give somewhat less details on the calculations in this part as compared to the more recent work on form factors.

In chapter \ref{chap: introduction to integrable deformations}, we give a short introduction to the $\beta$- and $\gamma_i$-deformation 
 of $\cN=4$ SYM theory.
Based on \cite{Fokken:2013mza}, we then discuss and extend the relation between the deformed theories and their undeformed parent theory in section \ref{sec: relation}.

In chapter \ref{chap: prewrapping}, which is largely based on \cite{Fokken:2013mza}, we analyse the effect of double-trace couplings in the $\beta$-deformation on two-point functions and the spectrum of planar anomalous dimensions.
In section \ref{sec: prewrapping}, we find that these couplings contribute at leading order in $N$ via a new kind of finite-size effect, which we call prewrapping, and we determine which states are potentially affected by it. 
In section \ref{sec: one-loop dilatation operator in beta deformation}, we calculate the corresponding finite-size corrections to the asymptotic dilatation operator to obtain the complete one-loop dilatation operator of the planar $\beta$-deformation.

In chapter \ref{chap: nonconformality}, based on \cite{Fokken:2013aea}, we show that the three-parameter non-supersymmetric $\gamma_i$-deformation proposed in \cite{Frolov:2005dj} is not conformally invariant.
In section \ref{sec: multi-trace couplings}, we formulate minimal requirements on multi-trace couplings that can be added to the single-trace part of the action and list all couplings that fulfil them.
In section \ref{sec: renormalisation}, we then show that for any choice of the tree-level values of these couplings, a particular double-trace is renormalised non-trivially.
Moreover, its beta function has no zeros such that it runs without fixed points, as is shown in section \ref{sec: beta function}.
Hence, conformal symmetry is broken. Moreover, this also affects the spectrum of planar anomalous dimensions, as is demonstrated in the subsequent chapter.

In chapter \ref{chap: anomalous dimensions}, which is based on \cite{Fokken:2014soa}, we calculate the planar anomalous dimensions of the operators $\tr(\phi_{14}^L)$ in the $\gamma_i$-deformation at the critical wrapping order $\ell=L$.
In section \ref{sec: classification}, we classify all diagrams contributing to the renormalisation of these operators with respect to their deformation dependence.
In the case $L\geq3$, which is covered in section \ref{sec: L geq 3}, this reduces the calculational effort to only four Feynman diagrams, which can be evaluated analytically for any $\ell=L$. We find perfect agreement with the integrability-based prediction of \cite{Ahn:2011xq}.
In the case $L=2$, treated in section \ref{sec: L eq 2}, also the previously discussed running double-trace coupling contributes, such that the anomalous dimension is finite but depends on the renormalisation scheme. 

We conclude with a summary of our results and an outlook on interesting directions for further research.
Moreover, several appendices are provided. Appendix \ref{appchap: integrals} contains our conventions and several explicit expressions for Feynman integrals. We give explicit expressions for scattering amplitudes in appendix \ref{app: scattering amplitudes}. In appendix \ref{app: deformed theories}, we give a short review on the renormalisation of fields, couplings and composite operators, which provides further details on the calculations in the second part of this work.

\chapter{\texorpdfstring{$\mathcal{N}=4$}{N=4} SYM theory}
\label{chap: N=4 SYM}

In this chapter, we give a short introduction to $\mathcal{N}=4$ SYM theory.
In particular, we introduce important concepts that will be required in both parts of this work.
For introductions to $\cN=4$ SYM theory that go beyond what is covered here, see \cite{Beisert:2004ry,Minahan:2010js}.

\section{Field content, action and symmetries}

The maximally supersymmetric Yang-Mills theory in four dimensions, termed $\cN=4$ SYM theory, was first constructed via dimensional reduction of $\cN=1$ SYM theory in ten dimensions by Brink, Schwarz and Scherk almost forty years ago \cite{Brink:1976bc}.
Although it is now argued to be the simplest quantum field theory \cite{ArkaniHamed:2008gz}, this is not manifest in its field content or action.

As follows from the dimensional reduction, the field content of $\mathcal{N}=4$ SYM theory consists of one gauge field $A_\mu$ with $\mu=0,1,2,3$, four fermions $\psi_{\alpha}^{A}$ with $\alpha=1,2$, $A=1,2,3,4$ transforming in the anti-fundamental representation of $\SU{4}$, four antifermions $\bar\psi_{\alphadot A}$ with $\alphadot=\DOT1,\DOT2$ transforming in the fundamental representation of $\SU{4}$ as well as six real scalars $\phi_I$ with $I=1,2,3,4,5,6$ transforming in the fundamental representation of $\SO{6}$.
Using the matrices $\sigma_{\alpha\dot\alpha}^\mu=(\idm,\sigma_1,\sigma_2,\sigma_3)_{\alpha\dot\alpha}$, where $\sigma_i$ are the Pauli matrices, we can exchange a Lorentz index $\mu$ for a pair of spinor indices $\alpha$, $\dot\alpha$, which exploits the isomorphism between (the algebras of) the Lorentz group and $\SU{2}\times\SUbar{2}$. 
For instance, we define $A_{\alpha\dot\alpha}=\sigma_{\alpha\dot\alpha}^\mu A_\mu$. Note that throughout this work we are using Einstein's summation convention, i.e.\ a pair of repeated indices is implicitly summed over.
Similarly, we can exploit the isomorphism between (the algebras of) $\SO{6}$ and $\SU{4}$ to define scalars $\phi_{AB}=\sigma_{AB}^I \phi_I$ via the 
corresponding 
 matrices $\sigma_{AB}^I$.
These scalars transform in the antisymmetric representation of $\SU{4}$, $\phi_{AB}=-\phi_{BA}$, and satisfy $(\phi_{AB})^*=\phi^{AB}=\frac{1}{2}\teps^{ABCD}\phi_{CD}$, where $\teps^{ABCD}$ is the completely antisymmetric tensor in four dimensions. 
Moreover, in particular in the context of the second part of this work, it is useful to define complex scalars $\phi_i=\phi_{i4}$, $\bar\phi^i=(\phi_i)^*$ with $i=1,2,3$, which transform in the fundamental and anti-fundamental representations of $\SU{3}\subset\SU{4}$, respectively.

All fields in $\mathcal{N}=4$ SYM theory transform in the adjoint representation of the gauge group.
We define the covariant derivative  
\begin{equation}
\label{eq: def covariant derivative}
 \cder_\mu=\partial_\mu-i g_\YM \comm{A_\mu}{\bullet} 
\end{equation}
and the field strength
\begin{equation}
\label{eq: field stengths definition 0}
 F_{\mu\nu}=\frac{i}{g_\YM}\comm{\cder_\mu}{\cder_\nu} \eqndot
\end{equation}

In Euclidean signature, the action of $\mathcal{N}=4$ is given by 
\begin{equation}\label{eq: action of N=4 SYM theory}
\begin{aligned}
S
 &=\int\de^4x\,\tr
  \bigg(
 -\frac{1}{4} F^{\mu\nu}F_{\mu\nu}- (\cder^{\mu}\bar\phi^j)\cder_{\mu}\phi_j
 +i \bar\psi^{\dot\alpha A} \cder_{\dot\alpha}{}^\alpha\psi_{\alpha A}\\
 &\hphantom{{}={}\int\de^4x\,\Bigl[\tr\Big({}}{}
 +g_\YM\Bigl(
 \frac{i}{2}\epsilon^{ijk}\phi_i \acomm{\psi^{\alpha}_{j}}{\psi_{\alpha k}}
 +\phi_j \acomm{\bar\psi^{\dot\alpha 4}}{\bar\psi_{\dot\alpha}^{j}}+\text{h.c.}
 \Bigr)\\
 &\phantom{{}={}\int\de^4x\,\Bigl[\tr\Big({}}{}
 -\frac{g^2_\YM}{4}
 \comm{\bar\phi^j}{\phi_j}\comm{\bar\phi^k}{\phi_k}
 +\frac{g^2_\YM}{2}
 \comm{\bar\phi^j}{\bar\phi^k}\comm{\phi_j}{\phi_k}
 \bigg)
\eqncom
\end{aligned}
\end{equation}
where h.c.\ denotes Hermitian conjugation and $\teps^{ijk}$ is the completely antisymmetric tensor in three dimensions. 
In fact, the extended $\cN=4$ supersymmetry fixes the action \eqref{eq: action of N=4 SYM theory} uniquely up to the choice of the gauge group.

Throughout this work, we consider the gauge group to be either $\SU{N}$ or $\U{N}$. 
We denote their generators as $(\T^a)^i_j$, where $i,j=1,\dots, N$ and $a=\colors,\dots,N^2-1$.
Here,
\begin{equation}\label{eq: color s}
 \colors=\begin{cases}
    0\quad\text{ for }\U{N}\eqncom \\
    1\quad\text{ for }\SU{N}\eqndot
   \end{cases}
\end{equation}
While immaterial for $\cN=4$ SYM theory, the difference between choosing either $\SU{N}$ or $\U{N}$ as gauge group plays a major role in its deformations, which are treated in the second part of this work.
We normalise the generators via
\begin{align}
 \tr(\T^a\T^b)=\delta^{ab} \eqncom
\end{align}
where $\delta^{ab}$ denotes the Kronecker delta.
They satisfy the completeness relation
\begin{equation}\label{eq: Ts summed over a}
\sum_{a=\colors}^{N^2-1}(\T^a)^i{}_j(\T^a)^k{}_l
=\delta^i_l\delta^k_j-\frac{\colors}{N}\delta^i_j\delta^k_l \eqndot
\end{equation}
We expand the elementary fields in terms of the gauge group generators as $A_\mu=A_\mu^a\T^a$, etc.

In addition to the $\cN=4$ super Poincar\'{e} group, $\mathcal{N}=4$ SYM theory is invariant under the conformal group. These two symmetry groups combine into the larger $\cN=4$ superconformal group $\PSU{2,2|4}$.
It is generated by the translations $\mathfrak{P}^{\alpha\dot\alpha}$, the super translations $\mathfrak{Q}^{\alpha A}$ and $\dot{\mathfrak{Q}}^{\dot\alpha}_A$, the dilatations $\mathfrak{D}$, the special conformal transformations $\mathfrak{K}_{\alpha\alphadot}$, the special superconformal transformations $\mathfrak{S}_{\alpha A}$ and $\dot{\mathfrak{S}}_{\dot\alpha}^A$ as well as the $\SU{2}$, $\SUbar{2}$ and $\SU{4}$ rotations $\mathfrak{L}^\alpha_\beta$, $\dot{\mathfrak{L}}^{\dot\alpha}_{\dot\beta}$ and $\mathfrak{R}^A_B$, respectively.
Moreover, we can add the central charge $\mathfrak{C}$ and the hypercharge $\mathfrak{B}$ in order to obtain $\U{2,2|4}$.
The commutation relations of these generators are rather lengthy but follow immediately from the oscillator representation given in \eqref{eq: oscillator algebra} and \eqref{eq: oscillator algebra extension} below.

\section{Composite operators}
\label{sec: composite operators}

Apart from the elementary fields, which can occur e.g.\ in asymptotic scattering states, an important class of objects are gauge-invariant local composite operators.

Local composite operators $\cO(x)$ contain products of fields evaluated at a common spacetime point $x$.
Using the momentum generators $\mathfrak{P}_\mu$, we can write
\begin{equation}
 \label{eq: O(x) shifted}
 \cO(x)=\e^{i x \mathfrak{P}}\cO(0)\e^{-i x \mathfrak{P}} \eqncom
\end{equation}
where $x \mathfrak{P}=x^\mu \mathfrak{P}_\mu$ is understood.
It follows that the action of any generator $\mathfrak{J}$ of $\PSU{2,2|4}$ on $\cO(x)$ is entirely determined by the action of $\PSU{2,2|4}$ on $\cO(0)$:
\begin{equation}
\begin{aligned}
\label{eq: J on O(x)}
 \mathfrak{J}\cO(x)&=\e^{i x \mathfrak{P}}\left(\e^{-i x \mathfrak{P}}\mathfrak{J}\e^{i x \mathfrak{P}}\right)\cO(0)\e^{-i x \mathfrak{P}}\\
 &=\e^{i x \mathfrak{P}}\left(\mathfrak{J}+(-i)x^\mu \comm{\mathfrak{P}_\mu}{\mathfrak{J}}+\frac{(-i)^2}{2!}x^\mu x^\nu \comm{\mathfrak{P}_\mu}{\comm{\mathfrak{P}_\nu}{\mathfrak{J}}} + \dots \right)\cO(0)\e^{-i x \mathfrak{P}}
 \eqncom
\end{aligned}
\end{equation}
where the sum actually terminates at the third term or before. 
Based on these arguments, it suffices to look at local composite operators $\cO(x)$ for $x=0$.

In order to obtain gauge-invariant expressions, we can take traces of products of fields that transform covariantly under gauge transformations.
These include the scalar fields $\phi_{AB}$, the antifermions $\bar{\psi}_{\dot\alpha A}$ and the fermions $\psi_{\alpha ABC}=\teps_{ABCD}\psi^{D}_\alpha$.%
\footnote{Hence, $\psi^A_\alpha=-\frac{1}{3!}\teps^{ABCD}\psi_{\alpha BCD}$.}
The gauge field $A_\mu$ itself does not transform covariantly under gauge transformations. It can, however, occur in the gauge-covariant combinations of the covariant derivative 
\begin{equation}
\label{eq: def covariant derivative in spinor indices}
 \cder_{\alpha\alphadot}=\cder_\mu(\sigma^\mu)_{\alpha\alphadot}
\end{equation}
and the field strength
\begin{equation}
\label{eq: def field strength slitting}
 F_{\alpha\beta\alphadot\betadot}=F_{\mu\nu}(\sigma^\mu)_{\alpha\alphadot}(\sigma^\nu)_{\beta\betadot}=-\sqrt{2}\epsilon_{\alphadot\betadot}\cfstrength_{\alpha\beta}-\sqrt{2}\epsilon_{\alpha\beta}\cantifstrength_{\alphadot\betadot} \eqncom
\end{equation}
which we have split into its self-dual part $\cfstrength_{\alpha\beta}$ and its anti-self-dual part $\cantifstrength_{\alphadot\betadot}$. We normalise the occurring antisymmetric tensors in two dimensions as $\epsilon^{21}=\epsilon_{12}=\epsilon^{\DOT2\DOT1}=\epsilon_{\DOT1\DOT2}=1$. 
The covariant derivatives can act on all fields that transform covariantly under gauge transformations to yield further fields that transform covariantly under gauge transformations.

Using the Bianchi identity%
\footnote{The brackets $[\dots]$ denote antisymmetrisation in the respective indices.}
\begin{equation}
 \label{eq: Bianchi identities}
 \cder_{[\mu} F_{\nu\rho]}=0 \eqncom
\end{equation}
the definition of the field strength \eqref{eq: field stengths definition 0} and \eqref{eq: def field strength slitting} as well as the equations of motion, 
 every field with antisymmetric $\alpha$ and $\alphadot$ indices can be replaced by a sum of fields that are individually symmetric under the exchange of all $\alpha$ and $\alphadot$ indices.
Thus, we arrive at irreducible fields composing the alphabet
\begin{equation}
\label{eq: alphabet of fields}
\begin{aligned}
 \cA=\{ 
 &\cder_{(\alpha_1\dot\alpha_1}\cdots \cder_{\alpha_k\dot\alpha_k}\cfstrength_{\alpha_{k+1}\alpha_{k+2})},\\
 &\cder_{(\alpha_1\dot\alpha_1}\cdots \cder_{\alpha_k\dot\alpha_k}\ferm_{\alpha_{k+1})ABC},\\
 &\cder_{(\alpha_1\dot\alpha_1}\cdots \cder_{\alpha_k\dot\alpha_k)}\phi_{AB},\\
 &\cder_{(\alpha_1\dot\alpha_1}\cdots \cder_{\alpha_k\dot\alpha_k}\antiferm_{\dot\alpha_{k+1})A},\\
 &\cder_{(\alpha_1\dot\alpha_1}\cdots \cder_{\alpha_k\dot\alpha_k}\cantifstrength_{\dot\alpha_{k+1}\dot\alpha_{k+2})} 
 \}\eqncom
\end{aligned}
\end{equation}
where $k\geq0$ and $(\dots)$ denotes symmetrisation in all $\alpha_i$ as well as all $\dot\alpha_i$.
 
The irreducible fields \eqref{eq: alphabet of fields} transform in the so-called singleton representation $\cV_{\text{S}}$ of $\PSU{2,2|4}$ and form the spin chain of $\cN=4$ SYM theory.
The singleton representation can be constructed by two sets of bosonic oscillators $\aosc_{i,\alpha}$, $\aosc_{i}^{\dagger \alpha}$ and $\bosc_{i,\alphadot}$, $\bosc_{i}^{\dagger \alphadot}$ as well as one set of fermionic oscillators $\dosc_{i,A}$, $\dosc_{i}^{\dagger A}$ \cite{Gunaydin:1981yq,Gunaydin:1998sw,Beisert:2003jj}. These satisfy the following non-vanishing commutation relations: 
\begin{equation}\label{eq: (anti)commutation relations}
\comm{\aosc_{i,\alpha}}{\aosc_j^{\dagger\beta}}=\delta_\alpha^\beta \delta_{i, j}\eqncom \qquad
\comm{\bosc_{i,\alphadot}}{\bosc_j^{\dagger\betadot}}=\delta_{\alphadot}^{\betadot}\delta_{i, j} \eqncom \qquad
\acomm{\dosc_{i,A}}{\dosc^{\dagger B}_j}=\delta^B_A\delta_{i, j}\eqncom  
\end{equation}
while all other (anti)commutators vanish. 
The fields \eqref{eq: alphabet of fields} can be obtained by acting with the creation operators on a Fock vacuum $\vac$:
\begin{equation}\label{eq: fields in oscillators}
\begin{aligned}
&\cder^k\cfstrength_{\phantom{ABC}} &\mathrel{\widehat{=}} \quad&
  (\aoscdag)^{k+2} 
  (\boscdag)^{k\phantom{+0}}
  \dosc^{\dagger 1} \dosc^{\dagger 2} \dosc^{\dagger 3} \dosc^{\dagger 4} 
  \vac \eqncom & &\\
&\cder^k\ferm_{ABC} &\mathrel{\widehat{=}}     \quad&
  (\aoscdag)^{k+1} 
  (\boscdag)^{k\phantom{+0}}
  \dosc^{\dagger A} \dosc^{\dagger B} \dosc^{\dagger C}
  \vac \eqncom & &\\
&\cder^k{}\phi_{AB{}\phantom{C}}{} &\mathrel{\widehat{=}}  \quad &  
  (\aoscdag)^{k\phantom{+0}} 
  (\boscdag)^{k\phantom{+0}} 
  \dosc^{\dagger A} \dosc^{\dagger B} 
  \vac \eqncom & &\\
&\cder^k\antiferm_{A\phantom{BC}} &\mathrel{\widehat{=}} \quad&
  (\aoscdag)^{k\phantom{+0}} 
  (\boscdag)^{k+1} 
  \dosc^{\dagger A}    
  \vac \eqncom & &\\
&\cder^k\cantifstrength_{\phantom{ABC}} &\mathrel{\widehat{=}} \quad  &
  (\aoscdag)^{k\phantom{+0}}
  (\boscdag)^{k+2} 
  \vac \eqncom & &
\end{aligned}
\end{equation} 
where we have suppressed all spinor indices.
We can characterise the irreducible fields in \eqref{eq: fields in oscillators} by vectors containing the occupation numbers of the eight oscillators,
\begin{equation}\label{eq: occupation number vector}
  \vec{n}_i=(\akindsite[1]{i},\akindsite[2]{i},\bkindsite[\DOT1]{i},\bkindsite[\DOT2]{i},\dkindsite[1]{i},\dkindsite[2]{i},\dkindsite[3]{i},\dkindsite[4]{i})\eqndot                                                                                                                                                                                         \end{equation}

In terms of the oscillators, the generators of $\PSU{2,2|4}$ can be written as
\begin{equation}\label{eq: oscillator algebra}
 \begin{aligned}
  \mathfrak{L}^\alpha_{i,\beta}&=\aosc_i^{\dagger \alpha}\aosc_{i,\beta}-\frac{1}{2}\delta^\alpha_\beta \aosc_i^{\dagger \gamma}\aosc_{i,\gamma} \eqncom &
  \quad\mathfrak{Q}_i^{\alpha A}&=\aosc_i^{\dagger\alpha}\dosc_i^{\dagger A} \eqncom \\
  \dot{\mathfrak{L}}^{\alphadot}_{i,\betadot}&=\bosc_i^{\dagger \alphadot}\bosc_{i,\betadot}-\frac{1}{2}\delta^{\alphadot}_{\betadot} \bosc_i^{\dagger \gammadot}\bosc_{i,\gammadot} \eqncom &
  \mathfrak{S}_{i,\alpha A}&=\aosc_{i,\alpha}\dosc_{i,A}  \eqncom\\
  \mathfrak{R}^A_{i,B}&=\dosc_i^{\dagger A}\dosc_{i,B}-\frac{1}{4}\delta^A_B \dosc_i^{\dagger C}\dosc_{i,C} \eqncom&
  \dot{\mathfrak{Q}}^{\alphadot}_{i,A}&=\bosc_i^{\dagger\alphadot}\dosc_{i,A}  \eqncom\\
  \mathfrak{D}_i&=\frac{1}{2}(\aosc_i^{\dagger \gamma}\aosc_{i,\gamma}+\bosc_i^{\dagger \gammadot}\bosc_{i,\gammadot}+2) \eqncom&
  \dot{\mathfrak{S}}^{ A}_{i,\alphadot}&=\bosc_{i,\alphadot}\dosc_i^{\dagger A}  \eqncom\\
  \mathfrak{P}_i^{\alpha \alphadot}&=\aosc_i^{\dagger\alpha}\bosc_i^{\dagger \alphadot} \eqncom &
  \mathfrak{K}_{i,\alpha \alphadot}&=\aosc_{i,\alpha}\bosc_{i,\alphadot} \eqndot
 \end{aligned}
\end{equation}
Moreover, the additional generators of $\U{2,2|4}$ are%
\footnote{Note that some authors also define the hypercharge as $\mathfrak{B}_i=\frac{1}{2}(\aosc_i^{\dagger \gamma}\aosc_{i,\gamma}-\bosc_i^{\dagger \gammadot}\bosc_{i,\gammadot}+\dosc_i^{\dagger C}\dosc_{i,C}+2)$.
Both definitions agree for vanishing central charge $\mathfrak{C}_i$.}
\begin{equation}\label{eq: oscillator algebra extension}
 \begin{aligned}
 \mathfrak{C}_i&=\frac{1}{2}(\aosc_i^{\dagger \gamma}\aosc_{i,\gamma}-\bosc_i^{\dagger \gammadot}\bosc_{i,\gammadot}-\dosc_i^{\dagger C}\dosc_{i,C}+2) \eqncom& \quad
  \mathfrak{B}_i&=\dosc_i^{\dagger C}\dosc_{i,C} \eqndot&
 \end{aligned}
\end{equation}
The central charge $\mathfrak{C}_i$ vanishes on all physical fields, cf.\ \eqref{eq: fields in oscillators}, whereas the hypercharge $\mathfrak{B}_i$ counts the fermionic oscillators. 

Single-trace operators containing $L$ irreducible fields can be described using the $L$-fold tensor product of the singleton representation. %
The different factors in the tensor product are labelled by $i=1,\dots,L$.
In the language of spin chains, they correspond to the sites and their number is referred to as length $L$.
Single-trace operators are invariant under graded cyclic symmetry, i.e.\ they are invariant under permuting a field from the first position of the trace to the last if the field and/or the rest of the operator is bosonic but obtain a sign if both are fermionic.
Hence, the spin-chain states describing them have to share this property.

\section{'t Hooft limit and finite-size effects}
\label{sec: 't Hooft limit}

In the perturbative expansion in terms of Feynman diagrams, different powers of the number of colours $N$ occur.
The $N$-power of a given contribution can be easily determined using the so-called fat or ribbon graphs, also known as double-line notation \cite{'tHooft:1973jz}.
In this notation, the flow of each fundamental gauge group index $i,j,k,l,\ldots=1,\dots, N$ is depicted by a line.
Each closed line yields a factor of $\delta_i^i=N$. 

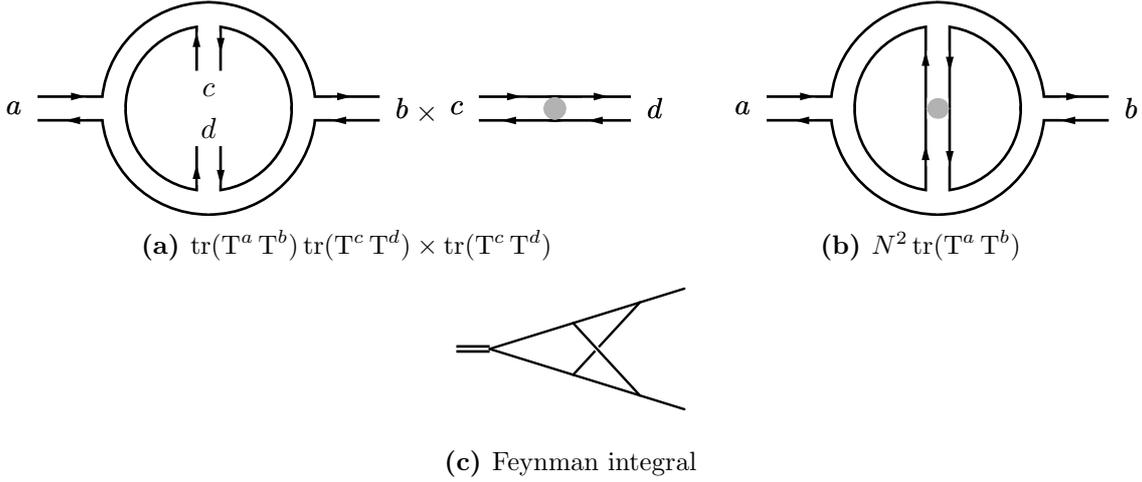
\begin{figure}[htbp]
\centering
 \begin{subfigure}[c]{0.55\textwidth}
 \centering
 $
  \begin{aligned}
 \begin{fmfgraph*}(45,25)
  \fmfpen{10}
  \fmfleft{in}
  \fmfright{out}
  \fmf{plain_h,foreground=black}{in,v1}
  \fmf{plain_t,foreground=black}{v2,out}
  \fmf{plain_ht,left,tension=0.2,foreground=black,tag=1}{v1,v2}
  \fmf{plain_ht,left,tension=0.2,foreground=black,tag=2}{v2,v1}
  \fmffreeze
  \fmfposition
\fmfipath{p[]}
\fmfiset{p1}{vpath1(__v1,__v2)}
\fmfiset{p2}{vpath2(__v2,__v1)}
\fmfi{plain_t,foreground=black}{(point length(p1)/2 of p1) -- (0.7*(point length(p1)/2 of p1)+0.3*(point length(p2)/2 of p2))}
\fmfi{plain_t,foreground=black}{(point length(p2)/2 of p2) -- (0.7*(point length(p2)/2 of p2)+0.3*(point length(p1)/2 of p1))}
  \fmfdraw
   \fmfpen{8}
   \fmf{iplain_h,foreground=white}{in,v1}
   \fmf{iplain_t,foreground=white}{v2,out}
   \fmf{plain,left,tension=0.3,foreground=white}{v1,v2}
   \fmf{plain,right,tension=0.3,foreground=white}{v1,v2}
\fmfi{iplain_t,foreground=white}{(point length(p1)/2 of p1) -- (0.7*(point length(p1)/2 of p1)+0.3*(point length(p2)/2 of p2))}
\fmfi{iplain_t,foreground=white}{(point length(p2)/2 of p2) -- (0.7*(point length(p2)/2 of p2)+0.3*(point length(p1)/2 of p1))}
   \fmfdraw
   \fmf{leftrightarrows,foreground=black}{in,v1}
   \fmf{leftrightarrows,foreground=black}{v2,out}
\fmfi{leftrightarrows,foreground=black}{(point length(p1)/2 of p1) -- (0.7*(point length(p1)/2 of p1)+0.3*(point length(p2)/2 of p2))}
\fmfi{leftrightarrows,foreground=black}{(point length(p2)/2 of p2) -- (0.7*(point length(p2)/2 of p2)+0.3*(point length(p1)/2 of p1))}
   \fmfpen{0}
   \fmfdraw
   \fmfv{label=$a$}{in}
   \fmfv{label=$b$}{out}
   \fmfiv{label=$c$,l.d=0}{(0.6*(point length(p1)/2 of p1)+0.4*(point length(p2)/2 of p2))}
   \fmfiv{label=$d$,l.d=0}{(0.6*(point length(p2)/2 of p2)+0.4*(point length(p1)/2 of p1))}
   \fmfdraw
 \end{fmfgraph*}
\end{aligned}
\quad\times\quad 
\begin{aligned}
 \begin{fmfgraph*}(20,5)
  \fmfpen{10}
  \fmfleft{in}
  \fmfright{out}
  \fmf{plain_h,foreground=black}{in,v1}
  \fmf{plain_t,foreground=black}{v1,out}
  \fmffreeze
  \fmfposition
\fmfipath{p[]}
  \fmfdraw
   \fmfpen{8}
   \fmf{iplain_h,foreground=white}{in,v1}
   \fmf{iplain_t,foreground=white}{v1,out}
   \fmfdraw
   \fmf{leftrightarrows,foreground=black}{in,v1}
   \fmf{leftrightarrows,foreground=black}{v1,out}
   \fmfpen{0}
  \fmfv{decor.shape=circle,decor.filled=full,decor.size=7.9,foreground=.7white}{v1}
   \fmfdraw
   \fmfv{label=$c$}{in}
   \fmfv{label=$d$}{out}
   \fmfdraw
 \end{fmfgraph*}
\end{aligned}
\quad$
\caption{$\tr(\T^a\T^b)\tr(\T^c\T^d)\times\tr(\T^c\T^d)$}
  \label{subfig: contraction}
 \end{subfigure}
 \qquad
 \begin{subfigure}[c]{0.33\textwidth}
  \centering
 $\quad
  \begin{aligned}
 \begin{fmfgraph*}(45,25)
  \fmfpen{10}
  \fmfleft{in}
  \fmfright{out}
  \fmf{plain_h,foreground=black}{in,v1}
  \fmf{plain_t,foreground=black}{v2,out}
  \fmf{plain_ht,left,tension=0.2,foreground=black,tag=1}{v1,v2}
  \fmf{plain_ht,left,tension=0.2,foreground=black,tag=2}{v2,v1}
  \fmffreeze
  \fmfposition
\fmfipath{p[]}
\fmfiset{p1}{vpath1(__v1,__v2)}
\fmfiset{p2}{vpath2(__v2,__v1)}
\fmfi{plain_t,foreground=black}{(point length(p1)/2 of p1) -- (0.5*(point length(p1)/2 of p1)+0.5*(point length(p2)/2 of p2))}
\fmfi{plain_t,foreground=black}{(point length(p2)/2 of p2) -- (0.5*(point length(p2)/2 of p2)+0.5*(point length(p1)/2 of p1))}
  \fmfdraw
   \fmfpen{8}
   \fmf{iplain_h,foreground=white}{in,v1}
   \fmf{iplain_t,foreground=white}{v2,out}
   \fmf{plain,left,tension=0.3,foreground=white}{v1,v2}
   \fmf{plain,right,tension=0.3,foreground=white}{v1,v2}
\fmfi{iplain_t,foreground=white}{(point length(p1)/2 of p1) -- (0.5*(point length(p1)/2 of p1)+0.5*(point length(p2)/2 of p2))}
\fmfi{iplain_t,foreground=white}{(point length(p2)/2 of p2) -- (0.5*(point length(p2)/2 of p2)+0.5*(point length(p1)/2 of p1))}
   \fmfdraw
   \fmf{leftrightarrows,foreground=black}{in,v1}
   \fmf{leftrightarrows,foreground=black}{v2,out}
\fmfi{leftrightarrows,foreground=black}{(point length(p1)/2 of p1) -- (0.5*(point length(p1)/2 of p1)+0.5*(point length(p2)/2 of p2))}
\fmfi{leftrightarrows,foreground=black}{(point length(p2)/2 of p2) -- (0.5*(point length(p2)/2 of p2)+0.5*(point length(p1)/2 of p1))}
   \fmfpen{0}
   \fmfdraw
   \fmfv{label=$a$}{in}
   \fmfv{label=$b$}{out}
   \fmfcmd{draw (0.5*(point length(p1)/2 of p1)+0.5*(point length(p2)/2 of p2)) withpen pencircle scaled 7.9 withcolor 0.7*white;}
   \fmfdraw
 \end{fmfgraph*}
\end{aligned}$
\caption{$N^2\tr(\T^a\T^b)$}
  \label{subfig: result}
 \end{subfigure}
 \\
 \begin{subfigure}[c]{0.35\textwidth}
 \centering
  $\FDinline[big,rainbownonplanar]$
  \caption{Feynman integral}
 \label{subfig: momentum space integral} 
 \end{subfigure}
\caption{Planarity and non-planarity in double-line notation and momentum space: (\subref{subfig: contraction}) contraction of a non-planar double-trace interaction with a composite operator, (\subref{subfig: result}) resulting planar diagram in double-line notation, (\subref{subfig: momentum space integral}) corresponding non-planar momentum space integral.
In double-line notation the operator is depicted as a grey blob, while it is depicted by a double line in momentum space.  
(We trust that the reader will not confuse both kinds of double lines.)
} 
\label{fig: 't Hooft limit}
\end{figure}

The perturbative expansion becomes particularly simple in the so-called 't Hooft, large $N$ or planar limit $N\to \infty$, $g_\YM\to0$ with the 't Hooft coupling $\lambda=g_\YM^2 N$ fixed \cite{'tHooft:1973jz}. 
In this limit, only Feynman diagrams contribute that are planar with respect to their double-line notation, i.e.\ with respect to their colour structure.
It should be stressed that this notion of planarity is in general different from planarity in momentum space, where all external momenta are understood to point outwards.
Both notions coincide for Feynman diagrams containing only elementary interactions and hence in particular for scattering amplitudes. 
However, this ceases to be the case in the presence of gauge-invariant local composite operators, as these are colour singlets but have a non-vanishing momentum.
Figure \ref{fig: 't Hooft limit} shows an example of a Feynman diagram that is planar in double-line notation (\subref{subfig: result}) but leads to a Feynman integral that is non-planar in momentum space (\subref{subfig: momentum space integral}).
Whether a diagram is planar or non-planar in the sense of the 't Hooft limit can in general only be determined via $N$ counting after all colour lines have been closed. Following \cite{Sieg:2005kd}, we call a diagram with $n$ open double lines planar if it contributes at the leading order in $N$ after a planar contraction of these lines with an $n$-point vertex at infinity. 
Of course, whether a diagram contributes at leading order to a particular process depends on the contraction of the colour lines in this process.
In particular, subdiagrams of elementary interactions which appear to be suppressed in the 't Hooft limit as they are non-planar and have a double-trace structure can contribute at the leading order when contracted with composite operators in a certain way. 
The reason for this is that the complete planar contraction of the composite operator with a trace factor in the double-trace interaction leads to an additional power of $N$ compared to the contraction with a single-trace interaction.
As the mechanism behind this enhancement of the $N$-power requires the interaction range to equal the number of fields in the single-trace operator, i.e.\ the length $L$ of the spin chain, it is called finite-size effect.
The right factor in figure  \ref{fig: 't Hooft limit} (\subref{subfig: contraction}) shows a non-planar double-trace diagram; the fact that it is non-planar can be seen when taking all external fields to point outwards. 
However, the depicted contraction of this double-trace interaction with a composite operator leads to the planar diagram in figure \ref{fig: 't Hooft limit} (\subref{subfig: result}).
One source of the double-trace structure of the interaction can be a sequence of fields wrapping around the composite operator \cite{Sieg:2005kd}.
We will encounter this wrapping effect in both parts of this work --- as well as a new finite-size effect in the second part.

\section{One-loop dilatation operator}
\label{sec: one-loop dilatation operator}

In this section, we give a short introduction to the quantum corrections to the dilatation operator, in particular at one-loop order.
These play a major role in both parts of this work.

Defining the effective planar coupling constant 
\begin{equation}
 \label{eq: effective planar coupling constant}
 \g^2=\frac{\lambda}{(4\pi)^2}=\frac{g_\YM^2 N}{(4\pi)^2}\eqncom
\end{equation}
the dilatation operator can be expanded as%
\footnote{Beyond one-loop order and certain subsectors, also odd powers of $\g$ can occur in the expansion \eqref{eq: expansion dilatation operator}. In this thesis, however, we restrict ourselves to cases where even powers suffice.}
\begin{equation}
 \label{eq: expansion dilatation operator}
 \Dila=\sum_{\ell=0}^\infty \g^{2\ell} \loopDila{\ell}\eqndot
\end{equation}
At $\ell$-loop order, connected interactions can involve at most $\ell+1$ fields of a composite operator of length $L$ at a time. 
Moreover, in the planar limit, these have to be neighbouring fields in the same trace factor.
Hence, in this limit the $\ell$-loop dilatation operator $\loopDila{\ell}$ can be written as sum of a density $\loopdilai{\ell}{i\dots i+\ell}$ that acts on $\ell+1$ neighbouring sites of the corresponding spin chain:
\begin{equation}
 \loopDila{\ell}=\sum_{i=1}^L \loopdilai{\ell}{i\dots i+\ell} \eqncom
\end{equation}
where cyclic identification $i+L\sim i$ is understood.
The study of the dilatation operator can be simplified by restricting to closed subsectors, which are defined via constraints on the various quantum numbers \cite{Beisert:2003jj}.

The complete one-loop dilatation operator density $\loopdilai{1}{i\,i+1}$ of $\cN=4$ SYM theory was first calculated by Niklas Beisert via a direct Feynman diagram calculation in the $\SL2$ sector that was then lifted to the complete theory via symmetry \cite{Beisert:2003jj}. It was later shown that
$\loopdilai{1}{i\,i+1}$ is completely fixed by symmetry apart from one global multiplicative constant \cite{Beisert:2004ry}.
Several different representations of $\loopdilai{1}{i\,i+1}$ exist.

The first kind of representation employs the following decomposition of the tensor product of two singleton representations \cite{Beisert:2003jj}:
\begin{equation}
 \cV_S\otimes\cV_S=\bigoplus_{i=0}^\infty \cV_j \eqndot
\end{equation}
Denoting the projection operator to the subspace $\cV_j$ as 
\begin{equation}
 \PP_j: \cV_S\otimes\cV_S \longrightarrow \cV_j \eqncom
\end{equation}
the one-loop dilatation operator density can be written as 
\begin{equation}
 \loopdilai{1}{i\,i+1}= 2\sum_{j=0}^\infty h(j)(\PP_j)_{i\, i+1} \eqncom
\end{equation}
where $h(j)=\sum_{i=1}^j\frac{1}{i}$ is the $j^{\text{th}}$ harmonic number.
Though quite compact, this representation is not very useful in direct calculations of anomalous dimensions due to the lack of handy expressions for $\PP_j$.

For direct calculations, a second kind of representation is advantageous, which is known as \emph{harmonic action}. It uses the oscillators defined in section \ref{sec: composite operators}, which can be combined into one superoscillator 
\begin{equation}\label{eq: def super oscillator}
\Aosc_i^\dagger=(\aosc_i^{\dagger 1},\aosc_i^{\dagger 2},\bosc_i^{\dagger \DOT1},\bosc_i^{\dagger \DOT2},\dosc_i^{\dagger 1},\dosc_i^{\dagger 2},\dosc_i^{\dagger 3},\dosc_i^{\dagger 4})\eqndot 
\end{equation} 
We specify the individual component oscillators of $\Aosc_i^\dagger$ by superscripts $A_i$ as $\Aosc_i^{\dagger A_i}$, i.e.\ $\Aosc_i^{\dagger1}=\aosc_i^{\dagger 1},\dots,\Aosc_i^{\dagger8}=\dosc_i^{\dagger 4}$.
Using these superoscillators, the one-loop dilatation operator density can be written as a weighted sum over all their reorderings
 \cite{Beisert:2003jj}:
\begin{equation}\label{eq: one-loop dila in oscillators beisert}
 \loopdilai{1}{1\,2} \, \Aosc_{s_1}^{\dagger A_1}\cdots\Aosc_{s_n}^{\dagger A_n} \vac = \sum_{s^\prime_1,\dots,s^\prime_n=1}^2 \delta_{C_2,0} \,c(n,n_{12},n_{21}) \Aosc_{s_1^\prime}^{\dagger A_1}\cdots\Aosc_{s_n^\prime}^{\dagger A_n} \vac \eqncom
\end{equation}
where $n$ denotes the total number of oscillators at both sites, $n_{12}$ ($n_{21}$) denotes the number of oscillators changing their site from 1 to 2 (2 to 1) and the Kronecker delta ensures that the resulting states fulfil the central charge constraint.
The coefficient is given by
\begin{equation}\label{eq: harmonic action coefficient}
 c(n,n_{12},n_{21})=\begin{cases}
 2 h\left(\frac12 n\right) \hfill \text{if } n_{12}=n_{21}=0 \eqncom& \\
 2 (-1)^{1+n_{12}n_{21}}  \Eulerbeta\left(\frac12(n_{12}+n_{21}),1+\frac12(n-n_{12}-n_{21})\right) \qquad \text{else,}&
                     \end{cases}             
\end{equation}
where $\Eulerbeta$ denotes the Euler beta function. 

An integral formulation of the latter representation was found 
 by Benjamin Zwiebel in \cite{Zwiebel:2007cpa}.\footnote{An alternative integral representation of the harmonic action can be found in \cite{Fokken:2014moa} and an operatorial form in \cite{Ferro:2013dga,Frassek:2013xza}.}
Defining 
\begin{equation}
 (\Aosc^\dagger_i)^{\vec{n}_i}=(\aosc_i^{\dagger 1})^{\akindsite[1]{i}}(\aosc_i^{\dagger 2})^{\akindsite[2]{i}}(\bosc_i^{\dagger \DOT1})^{\bkindsite[\DOT1]{i}}(\bosc_i^{\dagger \DOT2})^{\bkindsite[\DOT2]{i}}(\dosc_i^{\dagger 1})^{\dkindsite[1]{i}}(\dosc_i^{\dagger 2})^{\dkindsite[2]{i}}(\dosc_i^{\dagger 3})^{\dkindsite[3]{i}}(\dosc_i^{\dagger 4})^{\dkindsite[4]{i}}\eqncom
\end{equation}
the representation \eqref{eq: one-loop dila in oscillators beisert} can be recast into the form 
\begin{equation}\label{eq: one-loop dila in oscillators}
 \loopdilai{1}{1\,2} \, (\Aosc_{1}^{\dagger})^{ \vec{n}_{1}}(\Aosc_{2}^{\dagger})^{ \vec{n}_{2}}\vac = 4 \delta_{C_2,0} \int_0^{\frac{\pi}{2}} \de \theta \cot\theta \left((\Aosc_{1}^{\dagger})^{ \vec{n}_{1}}(\Aosc_{2}^{\dagger})^{ \vec{n}_{2}}-(\Aosc_{1}^{\prime\dagger})^{ \vec{n}_{1}}(\Aosc_{2}^{\prime \dagger})^{ \vec{n}_{2}} \right)\vac \eqncom
\end{equation}
where 
\begin{equation}
\left( \begin{array}{c}
\Aosc_{1}^{\prime \dagger}  \\
\Aosc_{2}^{\prime \dagger}   \end{array} \right)
=V(\theta)  \left( \begin{array}{c}
\Aosc_{1}^{\dagger}  \\
\Aosc_{2}^{\dagger}   \end{array} \right) \eqncom \qquad
V(\theta)=
\left( \begin{array}{cc}
\cos\theta & -\sin\theta \\
\sin\theta & \cos\theta \end{array} \right) \eqndot
\end{equation}
The equivalence of \eqref{eq: one-loop dila in oscillators} and \eqref{eq: one-loop dila in oscillators beisert} can be seen from the known integral representations 
\begin{equation}
 \begin{aligned}
\Eulerbeta(x,y)&=2\int_0^{\frac{\pi}{2}} \de \theta (\sin\theta)^{2x-1}(\cos\theta)^{2y-1} \eqncom  \\
h(y)&=2\int_0^{\frac{\pi}{2}} \de \theta \cot\theta  \left(1-(\cos \theta )^{2 y}\right) \eqndot
 \end{aligned}
\end{equation}
Note that the first summand in the integral on the right hand side of \eqref{eq: one-loop dila in oscillators} acts as regularisation, altering the divergent integral representation of $\Eulerbeta(0,y)$ such that it yields $h(y)$ instead.
In  chapter \ref{chap: minimal one-loop form factors}, we will find that the integral representation \eqref{eq: one-loop dila in oscillators} emerges naturally when deriving the complete one-loop dilatation operator via field theory alone.

Beyond one-loop order, several complications arise. 
The range of the interaction on a composite operator of length $L$ is naturally bounded by $L$. If this bound is saturated, the dilatation operator acts on the whole spin chain at once, invalidating the notion of an interaction density. 
At this point, finite-size effects set in, which will be a main topic of the second part of this thesis.%
\footnote{In fact, we will find a new kind of finite-size effect in the second part of this thesis, which already sets in at one-loop order in the deformed theories.}
Moreover, the length of a composite operator is not a conserved quantum number beyond one-loop order. 
The leading length-changing contributions to the dilatation operator are completely fixed by symmetry and were found in
\cite{Zwiebel:2011bx}.
In certain subsectors of the theory, such as the $\SU{2}$ sector, the length $L$ is connected to global charges of the theory, and hence preserved.
In this thesis, we will not consider cases where length-changing occurs.

\makeatletter
\def\toclevel@part{0}
\makeatother

\part{Form factors}
\label{part: form factors}

\chapter{Introduction to form factors}
\label{chap: minimal tree-level form factors}

In this chapter, we start our investigation of form factors in $\mathcal{N}=4$ SYM theory.
We review some important concepts that underlie the modern study of scattering amplitudes as well as form factors in section \ref{sec: generalities about form factors}. This also allows us to introduce our notation and conventions.
In section \ref{sec: minimal tree-level form factors for any operator}, we then calculate the minimal tree-level form factors of all operators via Feynman diagrams and relate them to the spin-chain picture.
We discuss the general problems arising for non-minimal and loop-level form factors in section \ref{sec: difficulties for non-minimal and loop-level form factors}.

While section \ref{sec: generalities about form factors} is a review of well known results, section \ref{sec: minimal tree-level form factors for any operator} is based on original work by the author first published in \cite{Wilhelm:2014qua}.

\section{Generalities}
\label{sec: generalities about form factors}

The physical quantities we are going to study are form factors of local gauge-invariant composite operators $\cO$.
For such an operator, the form factor is defined as the overlap between the state created by $\cO(x)$ from the vacuum $\ket{0}$ and an $n$-particle on-shell state $\bra{1,\dots,n}$,%
\footnote{The $n$ external fields are set on-shell using Lehmann-Symanzik-Zimmermann (LSZ) reduction \cite{Lehmann:1954rq}.}
i.e.\
\begin{equation}\label{eq: form factor definition}
 \ff_{\cO,n}(1,\dots,n;x)=\bra{1,\dots,n}\cO(x)\ket{0}\eqndot
\end{equation}
This definition reduces to the one for the scattering amplitude when setting $\cO=\idm$:
\begin{equation}\label{eq: amplitude definition}
 \amp_n(1,\dots,n)=\ev{1,\dots,n|0}\eqndot
\end{equation}
In both cases, the on-shell state is specified by the momenta $p_i$, the helicities $h_i$, the flavours and the gauge-group indices $a_i$ of the on-shell fields $i=1,\dots,n$.
We take all on-shell fields to be outgoing.

\begin{figure}[htbp]
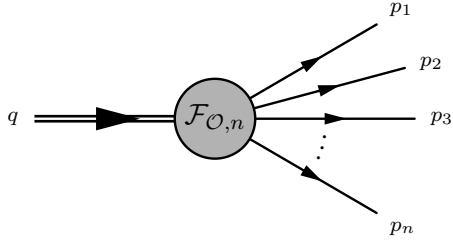

 \centering
\begin{equation*}
 \begin{aligned}
\settoheight{\eqoff}{$\times$}%
\setlength{\eqoff}{0.5\eqoff}%
\addtolength{\eqoff}{-14.5\unitlength}%
\raisebox{\eqoff}{%
\fmfframe(2,2)(6,2){%
\begin{fmfchar*}(50,25)
\fmfleft{vq}
\fmfright{vpL,vp,vp3,vp2,vp1}
\fmf{dbl_plain_arrow,tension=5}{vq,vqa}
\fmf{plain_arrow,tension=1}{vpLa,vpL}
\fmf{plain_arrow,tension=1}{vp3a,vp3}
\fmf{phantom,tension=1}{vpa,vp}
\fmf{plain_arrow}{vp1a,vp1}
\fmf{plain_arrow}{vp2a,vp2}
\fmf{dbl_plain_arrow,tension=25}{vqa,v1}
\fmf{plain_arrow,tension=5}{v1,vpLa}
\fmf{plain_arrow,tension=5}{v1,vp3a}
\fmf{phantom,tension=5}{v1,vpa}
\fmf{plain_arrow,tension=5}{v1,vp1a}
\fmf{plain_arrow,tension=5}{v1,vp2a}
\fmffreeze
\fmfdraw
 \fmfcmd{pair vertq, vertpone, vertptwo, vertpthree, vertpL, vertone, verttwo, vertp; vertone = vloc(__v1); verttwo = vloc(__v2); vertq = vloc(__vq); vertpone = vloc(__vp1); vertptwo = vloc(__vp2); vertpthree = vloc(__vp3);vertp = vloc(__vp);vertpL = vloc(__vpL);}
 \fmfiv{decor.shape=circle,decor.filled=30,decor.size=30}{vertone}
 \fmfiv{label=$ \cF_{\cO,,n}$,l.d=0,l.a=0}{vertone}
 \fmfiv{label=$\scriptstyle q$}{vertq}
 \fmfiv{label=$\scriptstyle p_1$}{vertpone}
 \fmfiv{label=$\scriptstyle p_2$}{vertptwo}
 \fmfiv{label=$\scriptstyle p_3$}{vertpthree}
 \fmfiv{label=$\scriptstyle p_n$}{vertpL}
 \fmfiv{label=$\cdot$,l.d=40,l.a=-10}{vertone}
 \fmfiv{label=$\cdot$,l.d=40,l.a=-16}{vertone}
 \fmfiv{label=$\cdot$,l.d=40,l.a=-22}{vertone}
\end{fmfchar*}%
}}%
\end{aligned}
\end{equation*}
\caption{The form factor of an operator $\cO$ in momentum space. The $n$ on-shell fields with momenta $p_i$ ($p_i^2=0$ for $i=1,\dots,n$) are depicted as single lines while the operator with off-shell momentum $q$ ($q^2\neq 0$) is depicted as double line. The direction of each momentum is indicated by an arrow.}
\label{fig: form factor}
\end{figure}

Most on-shell techniques that were developed for scattering amplitudes work in momentum space.%
\footnote{For recent reviews about on-shell techniques for scattering amplitudes, see \cite{Elvang:2013cua,Henn:2014yza}.}
Hence, as a first step, we Fourier transform \eqref{eq: form factor definition} from position space to momentum space:
\begin{equation}\label{eq: form factor momentum space}
\begin{aligned}
 \cF_{\cO}(1,\dots,n;q)&=\int \de^4x \e^{-iqx}\bra{1,\dots,n}\cO(x)\ket{0}
 =\int \de^4x \e^{-iqx}\bra{1,\dots,n}\e^{i x \mathfrak{P}}\cO(0)\e^{-i x \mathfrak{P}}\ket{0}\\
 &=(2\pi)^4\delta^4\left(q-\sum_{i=1}^n p_i\right)\bra{1,\dots,n}\cO(0)\ket{0} \eqncom
 \end{aligned}
\end{equation}
where we have used \eqref{eq: O(x) shifted}
 in the second line and the delta function in the third line guarantees momentum conservation.
We depict the Fourier-transformed form factor as shown in figure~\ref{fig: form factor}.

Using the 
 matrices $\sigma_{\mu}^{\alpha\alphadot}=(\idm,\sigma_1,\sigma_2,\sigma_3)^{\alpha\alphadot}$, where $\sigma_i$ are the usual Pauli matrices, the four-dimensional momenta $p_i^\mu$ can be written in terms of $2\times2$ matrices as $p_i^{\alpha\alphadot}=p_i^\mu \sigma_\mu^{\alpha\alphadot}$.
The on-shell condition $p_i^2=p_i^\mu p_{i,\mu}=0$ then translates to $\det p=0$.
Hence, on-shell momenta $p_i^{\alpha\alphadot}$ can be expressed as products of two two-dimensional spinors $\lambda_{p_i}^\alpha$ and $\lambdat_{p_i}^\alphadot$, which transform in the 
 anti-fundamental representations of $\SU{2}$ and $\SUbar{2}$, respectively:
\begin{equation}
\label{eq: spinor helicity variables}
 p_i^{\alpha\alphadot}=\lambda_{p_i}^\alpha \lambdat_{p_i}^\alphadot \eqndot
\end{equation}
These are known as spinor-helicity variables.
For real momenta and Minkowski signature, they are conjugate to each other, i.e.\  $\lambdat_{p_i}^\alphadot=\pm(\lambda_{p_i}^\alpha)^*$, where the $+$ ($-$) occurs for positive (negative) energy.
Moreover, multiplying $\lambda_{p_i}^\alpha$ by a phase factor $t\in\CC$, $|t|=1$, and $\lambdat_{p_i}^\alphadot$ by $t^{-1}$ 
leaves the momentum $p_i^{\alpha\alphadot}$ invariant. 
The behaviour of amplitudes and form factors under this transformation is called little group scaling.
In order to simplify notation, we will frequently abbreviate $\lambda_{p_i}^\alpha$ and $\lambdat_{p_i}^\alphadot$ as $\lambda_{i}^\alpha$ and $\lambdat_{i}^\alphadot$, respectively.
Contractions of the spinor-helicity variables are denoted as $\ab{i j}=\epsilon_{\alpha\beta}\lambda_i^\alpha\lambda_j^\beta$ and $\sb{i j}=-\epsilon_{\alphadot \betadot }\lambdat_i^{\alphadot}\lambdat_j^{\betadot}$.
They satisfy the so-called Schouten identities 
\begin{equation}
\label{eq: Schouten identities}
 \begin{aligned}
  \ab{ij}\ab{kl}+\ab{ik}\ab{lj}+\ab{il}\ab{jk}=0\eqncom \qquad
  \sb{ij}\sb{kl}+\sb{ik}\sb{lj}+\sb{il}\sb{jk}=0\eqndot
 \end{aligned}
\end{equation}

Furthermore, we can use Nair's $\mathcal{N}=4$ on-shell superfield \cite{Nair:1988bq} to describe external fields of all different flavours and helicities in a unified way:
\begin{equation}
\label{eq: N=4 superspace}
 \Phi(p_i,\etatt_i)=g_+(p_i) +  \etatt_i^A \, \bar\psi_A(p_i) + \frac{\etatt_i^A\etatt_i^B}{2!} \, \phi_{AB}(p_i) + \frac{ \teps_{ABCD} \etatt_i^A\etatt_i^B\etatt_i^C }{ 3!} \, \psi^D(p_i) + \etatt_i^1\etatt_i^2\etatt_i^3\etatt_i^4 \, g_-(p_i) \eqncom
\end{equation}
where $g_+$ ($g_-$) denotes the gluons of positive (negative) helicity and $\etatt_i^A$, $A=1,2,3,4$, are anticommuting Graßmann variables transforming in the 
 anti-fundamental representation of $\SU{4}$.
We can then combine all component amplitudes into one super amplitude and all component form factors into one super form factor.
The component expressions can be extracted from the super expressions by taking suitable derivatives with respect to the $\etatt_i^A$ variables. For instance,
\begin{equation}
\label{eq: def component expression}
 \begin{aligned}
 \ff_{\cO}(1^{g^+},2^{g^-},\dots,n^{\phi_{12}};q)=\left(\frac{\partial}{\partial \etatt_2^1}\frac{\partial}{\partial \etatt_2^2}\frac{\partial}{\partial \etatt_2^3}\frac{\partial}{\partial \etatt_2^4}\right)\cdots\left(-\frac{\partial}{\partial \etatt_n^1}\frac{\partial}{\partial \etatt_n^2}\right)\cF_{\cO}(1,2,\dots,n;q)\bigg|_{\etatt_i^A=0} \eqncom
 \end{aligned}
\end{equation}
where the superscripts specify the component fields and the sign takes into account that the $\etatt_i$ variables anticommute. 
Due to $\SU4$ invariance, the elementary interactions of $\mathcal{N}=4$ SYM theory can change the $\etatt$-degree of super amplitudes and super form factors only in units of four. 
The respective expressions with the minimal degree in $\etatt$ that allows for a non-vanishing result are called maximally-helicity-violating (MHV).
For super amplitudes, this minimal degree is eight, whereas it depends on the composite operator for super form factors.%
\footnote{Concretely, it is given by the eigenvalue of $\mathfrak{B}$ acting on the composite operators.}
Expressions with a Graßmann degree that is higher by four are called next-to-MHV (NMHV) and those with a Graßmann degree higher by $4\tilde{k}$ are called N$^{\tilde{k}}$MHV.

In addition to momenta, helicities and flavours, amplitudes and form factors also depend on the colour degrees of freedom of each on-shell field. 
We can split off this dependence by defining colour-ordered amplitudes $\ampco_n$ and colour-ordered form factors $\ffco_{\cO,n}$ as  
\begin{equation}\label{eq: def colour-ordered amplitude}
 \begin{aligned}
 \amp_n(1,\dots,n)&= g_\YM^{n-2}\sum_{\sigma\in \SS_n/\ZZ_n} \Tr(\T^{a_{\sigma(1)}}\cdots\T^{a_{\sigma(n)}}) \ampco_n(\sigma(1),\dots,\sigma(n))
  \\ &\phaneq+\text{multi-trace terms}
 \end{aligned}
\end{equation}
and
\begin{equation}\label{eq: def colour-ordered form factor}
 \begin{aligned}
 \ff_{\cO,n}(1,\dots,n;q)&= g_\YM^{n-L}\sum_{\sigma\in \SS_n/\ZZ_n} \Tr(\T^{a_{\sigma(1)}}\cdots\T^{a_{\sigma(n)}}) \ffco_{\cO,n}(\sigma(1),\dots,\sigma(n);q) 
 \\ &\phaneq+\text{multi-trace terms}
 \eqncom
 \end{aligned}
\end{equation}
where the sum is over all non-cyclic permutations.
Starting at one-loop order, 
also multi-trace terms can occur in \eqref{eq: def colour-ordered amplitude} and \eqref{eq: def colour-ordered form factor}. 
These multi-trace terms are formally suppressed in the planar limit. 
Hence, the double-trace part of amplitudes is irrelevant when studying the single-trace part of amplitudes and the double-trace part of form factors is irrelevant when studying the single-trace part of form factors. However, the double-trace part of amplitudes is relevant for calculating the single-trace part of form factors, as we will see in chapter \ref{chap: two-loop Konishi form factor}.%
\footnote{See also the discussion in section \ref{sec: 't Hooft limit}.}

Expressions for tree-level scattering amplitudes that will be used throughout the first part of this thesis are collected in appendix \ref{app: scattering amplitudes}.

\section{Minimal tree-level form factors for all operators}
\label{sec: minimal tree-level form factors for any operator}

Let us now calculate the form factors of all operators, starting in the free theory.
As every occurrence of the coupling constant $g_\YM$ either increases the number of loops or the number of legs, the form factor in the free theory coincides with the minimal tree-level form factor in the interacting theory.
It is sufficient to look at operators built from irreducible fields as reviewed in section \ref{sec: composite operators} since all operators are given by linear combinations of such operators and the form factor is linear in the operator.
We first look at single-trace operators.

\begin{figure}[htbp]
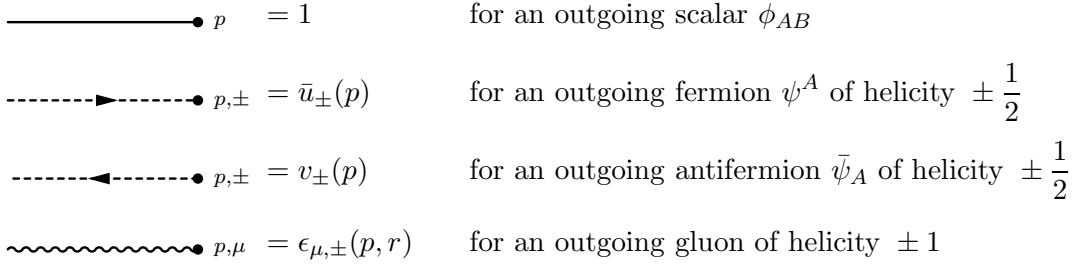
%
\begin{equation*}
 \begin{aligned}
\settoheight{\eqoff}{$\times$}%
\setlength{\eqoff}{0.5\eqoff}%
\addtolength{\eqoff}{-5\unitlength}%
\raisebox{\eqoff}{%
\fmfframe(2,0)(8,0){%
\begin{fmfchar*}(25,8)
\fmfleft{vq}
\fmfright{vp}
\fmf{plain,tension=1}{vq,vp}
\fmffreeze
\fmfdraw
 \fmfcmd{pair vertq, vertp;  vertq = vloc(__vq); vertp = vloc(__vp);}
 \fmfiv{decor.shape=circle,decor.filled=100,decor.size=3}{vertp}
 \fmfiv{label=$\scriptstyle p$}{vertp}
\end{fmfchar*}%
}}%
&=
1 \quad & &\text{for an outgoing scalar }\phi_{AB} \\
\settoheight{\eqoff}{$\times$}%
\setlength{\eqoff}{0.5\eqoff}%
\addtolength{\eqoff}{-5\unitlength}%
\raisebox{\eqoff}{%
\fmfframe(2,0)(8,0){%
\begin{fmfchar*}(25,8)
\fmfleft{vq}
\fmfright{vp}
\fmf{dashes_arrow,tension=1}{vq,vp}
\fmffreeze
\fmfdraw
 \fmfcmd{pair vertq, vertp;  vertq = vloc(__vq); vertp = vloc(__vp);}
 \fmfiv{decor.shape=circle,decor.filled=100,decor.size=3}{vertp}
 \fmfiv{label=$\scriptstyle p,,\pm$}{vertp}
\end{fmfchar*}%
}}%
&=
\bar{u}_\pm(p) \quad & & \text{for an outgoing fermion }\psi^A\text{ of helicity }\pm\frac{1}{2} \\
\settoheight{\eqoff}{$\times$}%
\setlength{\eqoff}{0.5\eqoff}%
\addtolength{\eqoff}{-5\unitlength}%
\raisebox{\eqoff}{%
\fmfframe(2,0)(8,0){%
\begin{fmfchar*}(25,8)
\fmfleft{vq}
\fmfright{vp}
\fmf{dashes_arrow,tension=1}{vp,vq}
\fmffreeze
\fmfdraw
 \fmfcmd{pair vertq, vertp;  vertq = vloc(__vq); vertp = vloc(__vp);}
 \fmfiv{decor.shape=circle,decor.filled=100,decor.size=3}{vertp}
 \fmfiv{label=$\scriptstyle p,,\pm$}{vertp}
\end{fmfchar*}%
}}%
&=
v_\pm(p) \quad & & \text{for an outgoing antifermion }\bar{\psi}_A\text{ of helicity }\pm\frac{1}{2}  \\
\settoheight{\eqoff}{$\times$}%
\setlength{\eqoff}{0.5\eqoff}%
\addtolength{\eqoff}{-5\unitlength}%
\raisebox{\eqoff}{%
\fmfframe(2,0)(8,0){%
\begin{fmfchar*}(25,8)
\fmfleft{vq}
\fmfright{vp}
\fmf{photon,tension=1}{vq,vp}
\fmffreeze
\fmfdraw
 \fmfcmd{pair vertq, vertp;  vertq = vloc(__vq); vertp = vloc(__vp);}
 \fmfiv{decor.shape=circle,decor.filled=100,decor.size=3}{vertp}
 \fmfiv{label=$\scriptstyle p,,\mu$}{vertp}
\end{fmfchar*}%
}}%
&=
\epsilon_{\mu,\pm}(p,r) \quad & & \text{for an outgoing gluon of helicity }\pm 1
\end{aligned}
\end{equation*}
\caption{Feynman rules for outgoing scalars, fermions, antifermions and gluons in momentum space; cf.\ for instance \cite{Henn:2014yza}.}%
\label{fig: Feynman rules}%
\end{figure}%

In the free theory, the form factors can be easily computed via Feynman diagrams.
As interactions are absent, the Feynman diagram reduces to the vertex for the composite operator and the collection of outgoing fields.
The required colour-ordered Feynman rules for outgoing fields are shown in figure \ref{fig: Feynman rules}.
For an outgoing scalar field $\phi_{AB}$, the momentum-space Feynman rule is simply $1$. As the scalar fields comes with a factor of $\etatt_i^A\etatt_i^B$ in Nair's $\mathcal{N}=4$ superfield \eqref{eq: N=4 superspace}, the total factor for a scalar field $\phi_{AB}$ in the composite operator that exits the diagram at leg $i$ is $\etatt_i^A\etatt_i^B$.
In the free theory, the covariant derivative $\cder_{\alpha\alphadot}$ reduces to the ordinary derivative $\partial_{\alpha\alphadot}$, which gives a factor of $p_i^{\alpha\alphadot}=\lambda_i^\alpha\lambdat_i^\alphadot$ in momentum space.%
\footnote{We have absorbed a factor of the imaginary unit $i$, which one would expect from the Fourier transformation, into the (covariant) derivative.
}
For instance, the total factor for an irreducible field $\cder_{\alpha\alphadot}\phi_{AB}$ in the composite operator that exits the diagram at leg $i$ is $\lambda_i^\alpha\lambdat_i^\alphadot\etatt_i^A\etatt_i^B$.
For an outgoing fermion $\psi_{\alpha ABC}=\teps_{ABCD}\psi^D_\alpha$ 
of helicity $-\frac{1}{2}$, the momentum space Feynman rule yields one of the solutions to the massless Dirac equation, namely $\bar{u}_-(p_i)=(\lambda_i^\alpha,0)$, and hence $\lambda_i^\alpha$. In order to obtain the super form factor, this has to be dressed by $\etatt_i^A\etatt_i^B\etatt_i^C$, so the total factor is $\lambda_i^\alpha \etatt_i^A\etatt_i^B\etatt_i^C$.
For an outgoing antifermion $\bar\psi_{\alphadot A}$ with helicity $+\frac{1}{2}$, the momentum space Feynman rule yields $v_+(p_i)=(0,\lambdat_i^\alphadot)^T$, which is another solution to the massless Dirac equation, and hence $\lambdat_i^\alphadot$. 
Taking the required Graßmann variables into account, the total factor is $\lambdat_i^\alphadot\etatt_i^A$.
For an outgoing gauge field of positive or negative helicity, the momentum space Feynman rules yield the polarisation vectors 
\begin{equation}\label{eq: def polarisation vectors}
 \epsilon_+^{\alpha\alphadot}(p_i;r_i)=\sqrt{2}\frac{\lambda_{r_i}^\alpha \lambdat_{p_i}^\alphadot}{\ab{ r_i p_i}}\eqncom \quad
 \epsilon_-^{\alpha\alphadot}(p_i;r_i)=\sqrt{2}\frac{\lambda_{p_i}^\alpha\lambdat_{r_i}^\alphadot}{\sb{p_{i} r_{i}}} \eqncom
\end{equation}
respectively, where $r_i$ is a light-like reference vector that can be chosen independently for each $i$.
As reviewed in section \ref{sec: composite operators}, gauge-invariant local composite operators contain gauge fields only in the gauge-covariant combinations of covariant derivatives and the self-dual and anti-self-dual parts of the fields strength. 
For vanishing coupling, the latter read
\begin{equation}\label{eq: field stengths definition}
 F_{\alpha\beta}=\frac{1}{2\sqrt{2}}\epsilon^{\alphadot\betadot}(\partial_{\alpha\alphadot}A_{\beta\betadot}-\partial_{\beta\betadot}A_{\alpha\alphadot})\eqncom \quad  
 \bar{F}_{\alphadot\betadot}=\frac{1}{2\sqrt{2}} \epsilon^{\alpha\beta}(\partial_{\alpha\alphadot}A_{\beta\betadot}-\partial_{\beta\betadot}A_{\alpha\alphadot})\eqndot
\end{equation}
Calculating the contribution of a field strength in the composite operator that exits the diagram at leg $i$ amounts to replacing the gauge fields in \eqref{eq: field stengths definition} by the polarisation vectors \eqref{eq: def polarisation vectors} and the derivatives by $\lambda_i\lambdat_i$. After a short calculation, we find  
\begin{equation}
 \begin{aligned}
   F_{\alpha\beta}&\overset{\epsilon_+}{\longrightarrow}
   -\frac{1}{2\sqrt{2}}\epsilon_{\alphadot\betadot}(\lambda_{p_i}^{\alpha}\lambdat_{p_i}^{\alphadot}\epsilon_+^{\beta\betadot}-\lambda_{p_i}^{\beta}\lambdat_{p_i}^{\betadot}\epsilon_+^{\alpha\alphadot})
   =0\eqncom\\
   F_{\alpha\beta}&\overset{\epsilon_-}{\longrightarrow}
   -\frac{1}{2\sqrt{2}}\epsilon_{\alphadot\betadot}(\lambda_{p_i}^{\alpha}\lambdat_{p_i}^{\alphadot}\epsilon_-^{\beta\betadot}-\lambda_{p_i}^{\beta}\lambdat_{p_i}^{\betadot}\epsilon_-^{\alpha\alphadot})
   =\lambda_{p_i}^\alpha\lambda_{p_i}^\beta\eqncom\\
   \bar{F}_{\alphadot\betadot}&\overset{\epsilon_+}{\longrightarrow}
   -\frac{1}{2\sqrt{2}}\epsilon_{\alpha\beta}(\lambda_{p_i}^{\alpha}\lambdat_{p_i}^{\alphadot}\epsilon_+^{\beta\betadot}-\lambda_{p_i}^{\beta}\lambdat_{p_i}^{\betadot}\epsilon_+^{\alpha\alphadot})
   =\lambdat_{p_i}^\alphadot\lambdat_{p_i}^\betadot\eqncom\\
   \bar{F}_{\alphadot\betadot}&\overset{\epsilon_-}{\longrightarrow}
   -\frac{1}{2\sqrt{2}}\epsilon_{\alpha\beta}(\lambda_{p_i}^{\alpha}\lambdat_{p_i}^{\alphadot}\epsilon_-^{\beta\betadot}-\lambda_{p_i}^{\beta}\lambdat_{p_i}^{\betadot}\epsilon_-^{\alpha\alphadot})
   =0 
 \end{aligned}
\end{equation}
for the different combinations of polarisation vectors and parts of the field strength; cf.\ also \cite{Witten:2003nn}.
The gauge fields of positive and negative helicity occur in Nair's $\cN=4$ on-shell superfield with zero and four factors of $\etatt_i$, respectively. Hence, the total contributions of the self-dual and anti-self-dual parts of the field strength are $\lambda_i^\alpha\lambda_i^\beta\etatt_i^1\etatt_i^2\etatt_i^3\etatt_i^4$ and $\lambdat_i^{\alphadot}\lambdat_i^{\betadot}$, respectively.
As in the scalar case, covariant derivatives $\cder^{\alpha\alphadot}$ that act on any of the fields in the composite operator yield an additional factor of $\lambda_i^\alpha\lambdat_i^\alphadot$, where $i$ is the leg at which the respective field exits the diagram. 
Note that the resulting expressions at leg $i$ are manifestly symmetric in all $\SU{2}$ and $\SUbar{2}$ indices and manifestly antisymmetric in all $\SU{4}$ indices, as are the corresponding expressions in the oscillator picture. 
Let us now summarise these results.

For a given gauge-invariant local composite single-trace operator $\cO$ characterised by $\{\vec{n}_i\}_{i=1,\dots,L}=\{(\akindsite[1]{i},\akindsite[2]{i},\bkindsite[\DOT1]{i},\bkindsite[\DOT2]{i},\dkindsite[1]{i},\dkindsite[2]{i},\dkindsite[3]{i},\dkindsite[4]{i})\}_{i=1,\dots,L}$, the minimal colour-ordered tree-level super form factor is 
\begin{multline}\label{eq: form factor from spin chain}
 \ffco_{\cO,L}(\Lambda_1,\dots,\Lambda_L;q)=(2\pi)^4\delta^4\left(q-\sum_{i=1}^L p_i\right) \sum_{\sigma\in\ZZ_L}\\ \prod_{i=1}^L (\lambda_{\sigma(i)}^1)^{\akindsite[1]{i}}(\lambda_{\sigma(i)}^2)^{\akindsite[2]{i}}(\lambdat_{\sigma(i)}^{\DOT1})^{\bkindsite[\DOT1]{i}}(\lambdat_{\sigma(i)}^{\DOT2})^{\bkindsite[\DOT2]{i}}(\etatt_{\sigma(i)}^1)^{\dkindsite[1]{i}}(\etatt_{\sigma(i)}^2)^{\dkindsite[2]{i}}(\etatt_{\sigma(i)}^3)^{\dkindsite[3]{i}}(\etatt_{\sigma(i)}^4)^{\dkindsite[4]{i}} \eqncom
 \end{multline}
where $\Lambda_i=(\lambda_i^\alpha,\lambdat_i^{\alphadot},\etatt_i^A)$. 
The sum over all cyclic permutations accounts for all possible colour-ordered contractions of the fields in the operator with the external legs and reflects the (graded) cyclic invariance of the single-trace operator.
The grading is implemented in the product in the second line of \eqref{eq: form factor from spin chain}, which should be understood as ordered with respect to $i$. Restoring the canonical order with respect to $\sigma(i)$ requires to anticommute the Graßmann variables, which can result in a total sign.
Form factors for composite operators that are 
 characterised by linear combinations of $\{\vec{n}_i\}_{i=1,\dots,L}$ 
 are given by the respective linear combinations of \eqref{eq: form factor from spin chain}.

Note that 
the expression \eqref{eq: form factor from spin chain} coincides with replacing the oscillators in the oscillator representation \eqref{eq: fields in oscillators} of the operator according to 
\begin{equation}\label{eq: oscillator replacements}
 \begin{aligned}
  \aosc_i^{\dagger \alpha} &\to \lambda_i^\alpha \eqncom & \bosc_i^{\dagger \alphadot}&\to \lambdat_i^{\alphadot}\eqncom & \dosc_i^{\dagger A}&\to\etatt_i^A\eqncom\\
  \aosc_{i,\alpha} &\to \partial_{i,\alpha}=\frac{\partial}{\partial\lambda_i^\alpha} \eqncom & 
  \bosc_{i,\alphadot}&\to \partial_{i,\alphadot}=\frac{\partial}{\partial\lambdat_i^{\alphadot}}\eqncom & \dosc_{i,A}&\to\partial_{i,A}=\frac{\partial}{\partial \etatt_i^A}
 \end{aligned}
\end{equation}
and multiplying the result with the momentum-conserving delta function and a normalisation factor of $L$:%
\footnote{The normalisation factor $L$ arises as we assume the states in the oscillator picture to be graded cyclic symmetric and normalised to unity. 
}
\begin{equation}
\label{eq: minimal tree-level form factor from oscillator replacement}
\ffco_{\cO,L}(\Lambda_1,\dots,\Lambda_L;q)=L (2\pi)^4\delta^4\left(q-\sum_{i=1}^L p_i\right) \times \cO \,\rule[-1.06cm]{0.1mm}{1.415cm} \!\phantom{|}_{\substack{\\
\begin{smallmatrix}
        \aosc_i^{\dagger \alpha} &\to& \vll_i^\alpha \\
  \bosc_i^{\dagger \dot\alpha} &\to& \vlt_i^{\dot\alpha}\\
  \dosc_i^{\dagger A} &\to& \vlet_i^A
    \end{smallmatrix}}}
\eqndot    
\end{equation}

As was already observed in \cite{Beisert:2010jq},%
\footnote{Moreover, the replacement \eqref{eq: oscillator algebra} plays an important role in the algebraic considerations of \cite{Zwiebel:2011bx}, which connect the one-loop dilatation operator to the four-point tree-level scattering amplitude. This connection will be the subject of section \ref{sec: one-loop generalised unitarity}. However, the connection between the replacement \eqref{eq: oscillator replacements} and form factors was made neither in \cite{Beisert:2010jq} nor in \cite{Zwiebel:2011bx}.
}
 the replacement \eqref{eq: oscillator replacements} relates the generators \eqref{eq: oscillator algebra} of $\PSU{2,2|4}$ in the oscillator representation 
 to their well known form 
 on on-shell superfields in scattering amplitudes: 
\begin{equation}\label{eq: onshell algebra}
 \begin{aligned}
  \mathfrak{L}^\alpha_{i,\beta}&=\lambda^{ \alpha}_i\partial_{i,\beta}-\frac{1}{2}\delta^\alpha_\beta \lambda^{ \gamma}_i\partial_{i,\gamma} \eqncom &
  \mathfrak{Q}^{\alpha A}_i&=\lambda^{\alpha}_i\etatt^{ A}_i \eqncom \\
  \dot{\mathfrak{L}}^{\alphadot}_{i,\betadot}&=\lambdat^{ \alphadot}_i\partial_{i,\betadot}-\frac{1}{2}\delta^{\alphadot}_{\betadot} \lambdat^{ \gammadot}_i\partial_{i,\gammadot} \eqncom &
  \mathfrak{S}_{i,\alpha A}&=\partial_{i,\alpha}\partial_{i,A}  \eqncom\\
  \mathfrak{R}^A_{i,B}&=\etatt^{ A}_i\partial_{i,B}-\frac{1}{4}\delta^A_B \etatt^{ C}_i\partial_{i,C} \eqncom&
  \dot{\mathfrak{Q}}^{\alphadot}_{i,A}&=\lambdat^{\alphadot}_i\partial_{i,A}  \eqncom\\
  \mathfrak{D}_i&=\frac{1}{2}(\lambda^{ \gamma}_i\partial_{i,\gamma}+\lambdat^{ \gammadot}_i\partial_{i,\gammadot}+2) \eqncom&
  \dot{\mathfrak{S}}^{ A}_{i,\alphadot}&=\partial_{i,\alphadot}\etatt^{ A}_i  \eqncom\\
  \mathfrak{C}_i&=\frac{1}{2}(\lambda^{ \gamma}_i\partial_{i,\gamma}-\lambdat^{ \gammadot}_i\partial_{i,\gammadot}-\etatt^{ C}_i\partial_{i,C}+2) \eqncom&
  \mathfrak{P}^{\alpha \alphadot}_i&=\lambda^{\alpha}_i\lambdat^{ \alphadot}_i \eqncom \\
  \mathfrak{B}&=\etatt^{ C}_i\partial_{i,C} \eqncom&
  \mathfrak{K}_{i,\alpha \alphadot}&=\partial_{i,\alpha}\partial_{i,\alphadot} \eqnsem
 \end{aligned}
\end{equation}
cf.\ \cite{Witten:2003nn}.

The action of any generator $\mathfrak{J}$ of $\PSU{2,2|4}$ on the on-shell part of the form factor \eqref{eq: form factor from spin chain} is given by 
\begin{equation}
 \sum_{i=1}^n\mathfrak{J}_i \ffco_{\cO,n}(1,\dots,n;q)\eqndot
\end{equation}
Note that some of the generators $\mathfrak{J}_i$ contain differential operators, which act both on the polynomial in the super-spinor-helicity variables and on the momentum-conserving delta function in \eqref{eq: form factor from spin chain}. 
The action on the polynomial is precisely given by the action of the corresponding generator written in terms of oscillators \eqref{eq: oscillator algebra} on the fields \eqref{eq: fields in oscillators} in the oscillator representation. Moreover, the terms arising from the action on the momentum-conserving delta function agree with the spacetime-dependent terms in \eqref{eq: J on O(x)}, i.e.\ those vanishing for $x=0$, after the Fourier transformation \eqref{eq: form factor momentum space}.
Hence,
\begin{equation}\label{eq: action on form factor}
 \sum_{i=1}^L\mathfrak{J}_i\ffco_{\cO,L}(1,\dots,L;q)=\ffco_{\mathfrak{J}\cO,L}(1,\dots,L;q) \eqncom
\end{equation}
where $\mathfrak{J}\cO$ is given in \eqref{eq: J on O(x)} and \eqref{eq: oscillator algebra}.

We have seen that, apart from a normalisation factor $L$ and the momentum-conserving delta function, the minimal tree-level for factors can be obtained by replacing the superoscillators of the oscillator picture by super-spinor-helicity variables. 
Moreover, the corresponding generators of $\PSU{2,2|4}$ are related by the same replacement.
Hence, minimal tree-level form factors translate the spin-chain of free $\mathcal{N}=4$ SYM theory into the language of scattering amplitudes.

In fact, \eqref{eq: action on form factor} is a special case of a superconformal Ward identity for form factors derived in \cite{Brandhuber:2011tv}.
In principle, this Ward identity should also hold in the interacting theory.
In practise, however, the generators are known to receive quantum corrections for both scattering amplitudes and composite operators, and in the former case also anomalies occur; see \cite{Beisert:2010jq,Bargheer:2011mm} for reviews. 
For form factors, both the corrections for scattering amplitudes and the corrections for composite operators contribute.
In the following chapters, we calculate the corrections to the action of a particular generator on composite operators via form factors, namely the dilatation operator. 
We leave the study of corrections to further generators on composite operators via form factors for future work.

Finally, let us mention that for multi-trace operators the single-trace minimal tree-level form factors naturally vanish.
The respective multi-trace minimal tree-level form factors are obtained by performing the replacement \eqref{eq: minimal tree-level form factor from oscillator replacement} for each trace factor and multiplying by the length of each of them.

\section{Difficulties for non-minimal and loop-level form factors}
\label{sec: difficulties for non-minimal and loop-level form factors}

Let us conclude this chapter with some discussion on non-minimal and loop-level form factors.
The minimal tree-level form factors coincide with the form factors in the free theory.
For non-minimal and loop-level form factors, interactions play a role.
In principle, these interactions can be calculated via Feynman diagrams.
In practise, however, this becomes intractable after the first few loop orders and additional legs due to the large growth in the number of contributing Feynman diagrams.%
\footnote{Moreover, the individual Feynman diagrams depend on the gauge choice, which only drops out at the end. Thus, the final result is often much shorter than each of the intermediate steps.}
In the case of amplitudes, 
efficient on-shell methods have been developed to overcome this limitation.
Previous studies have shown that these are at least partially also applicable to form factors.
However, these studies have largely focused on the form factors of the stress-tensor supermultiplet and its lowest component $\tr(\phi_{14}\phi_{14})$ as well as its generalisation to $\tr(\phi_{14}^L)$.

At tree level, for instance BCFW recursion relations can be used to calculate general $n$-point form factors. These were applied to the stress-tensor supermultiplet at general MHV degree \cite{Brandhuber:2010ad,Brandhuber:2011tv,Bork:2014eqa}. Moreover, they were applied to the supermultiplet of $\tr(\phi_{14}^L)$ \cite{Penante:2014sza} and to operators from the $\SU{2}$ and $\SL{2}$ subsectors \cite{Engelund:2012re} at MHV level.
In this thesis, we will restrict our study of non-minimal tree-level form factors to the stress-tensor supermultiplet.
In chapter \ref{chap: tree-level form factors}, we will investigate the structure and symmetries of these form factors and present alternative ways to construct them.
Non-minimal tree-level form factors of general operators will be studied in \cite{KMSW}.

At loop level, an additional complication apart from the growing number of Feynman diagrams is the occurrence of loop integrals, which need to be evaluated. 
Moreover, these can contain divergences, such that the theory needs to be regularised.
A natural starting point to calculate loop corrections is given by the minimal form factors.
In previous studies, the minimal form factors of the stress-tensor supermultiplet and its generalisation to $\tr(\phi_{14}^L)$ have been calculated via unitarity 
 up to three-loop order \cite{Gehrmann:2011xn} and 
 two loop-order \cite{Brandhuber:2014ica}, respectively.%
\footnote{The integrand of the minimal form factor of $\tr(\phi_{14}\phi_{14})$ is even known up to four loops \cite{Boels:2012ew,Boels:2015yna}.}
However, among experts, it has been 
 a vexing problem how to calculate the minimal two-loop Konishi form factor via unitarity.
In chapter \ref{chap: minimal one-loop form factors} of this thesis, we study one-loop corrections to minimal form factors of general operators.
In chapter \ref{chap: two-loop Konishi form factor}, we address the problem of calculating the minimal two-loop Konishi form factor via unitarity, finding that it is related to a subtlety in the regularisation.
We then study minimal two-loop form factors of operators in the $\SU{2}$ sector in chapter \ref{chap: two-loop su(2) form factors}.

Let us mention that also non-minimal loop-level form factors have been studied, namely two-loop three-point and one-loop $n$-point MHV and NMHV form factors of the stress-tensor supermultiplet \cite{Brandhuber:2010ad,Brandhuber:2011tv,Brandhuber:2012vm,Bork:2012tt}, one-loop $n$-point MHV form factors of its generalisation to $\tr(\phi_{14}^L)$ \cite{Penante:2014sza} as well as one-loop three-point form factors of the Konishi primary operator \cite{Nandan:2014oga}.

Finally, as already alluded to in the end of the last section, a further complication for non-minimal and loop-level form factors lies in the fact that the symmetries of these expressions are obscured by corrections and anomalies in the symmetry generators.

\chapter{Minimal one-loop form factors}
\label{chap: minimal one-loop form factors}

In this chapter, we compute one-loop corrections to minimal form factors.
We begin by discussing the general structure of loop corrections to form factors in section \ref{sec: general structure of loop corrections}.
In particular, we show how to obtain the dilatation operator from them.
In section \ref{sec: one-loop unitarity}, we introduce the on-shell method of unitarity by calculating the one-loop corrections to the minimal form factors for composite operators in the $\SU2$ sector. 
After this warm-up exercise, we use generalised unitarity to calculate the cut\hyp constructible part of the one-loop correction to the minimal form factor of any 
 composite operator in section \ref{sec: one-loop generalised unitarity}. This allows us to (re)derive the complete one-loop dilatation operator.

This chapter is based on results first published in \cite{Wilhelm:2014qua} and partially adapts a presentation later developed in \cite{Loebbert:2015ova}.

\section{General structure of loop corrections and the dilatation operator}
\label{sec: general structure of loop corrections}

Before starting to compute loop corrections to minimal form factors, let us first discuss their general structure.

A general property of loop calculations in QFTs is the appearance of divergences in the occurring Feynman integrals. 
Ultraviolet (UV) divergences stem from integration regions where the energy of a virtual particle is very large, while infrared (IR) divergences stem from integration regions where the energy of the virtual particle is very low and/or it is collinear to an external particle.
In order to perform calculations, the divergences have to be regularised. 
This can be achieved by continuing the dimension of spacetime from $D=4$ to $D=4-2\peps$. At the same time, also the fields have to be continued, which leads to some subtleties when applying on-shell methods. We will postpone their discussion to section \ref{sec: subtleties} in the next chapter.

Infrared divergences occur in theories with massless fields. In all observables, the IR divergences from virtual loop corrections are cancelled by contributions from the emission of soft and collinear real particles according to the Kinoshita-Lee-Nauenberg (KLN) theorem \cite{Kinoshita:1962ur,Lee:1964is}.

Ultraviolet divergences signal that certain quantities appearing in the 
 formulation of the theory (fields, masses, coupling constants, etc.) depend on the energy scale due to quantum effects. 
They require renormalisation, i.e.\ the absorption of divergences into a redefinition of these quantities. 
In a conformal field theory like $\mathcal{N}=4$ SYM theory, all beta functions are zero, i.e.\ the strength of the interactions is independent of the energy scale.
Hence, scattering amplitudes in $\mathcal{N}=4$ SYM theory are UV finite and require no renormalisation.

Loop corrections to scattering amplitudes can be written as 
\begin{equation}
 \ampco_n(\gmod^2,\peps)=\Interactionamp(\gmod^2,\peps) \ampco^{(0)}_n=\left(1+\sum_{\ell=1}^\infty \gmod^{2\ell} \Interactionamp^{(\ell)}(\peps) \right) \ampco^{(0)}_n \eqncom
\end{equation}
where $\Interactionamp^{(\ell)}$ is the ratio between the $\ell$-loop and tree-level amplitude.
The modified effective planar coupling constant $\gmod^2$ is defined as  
\begin{equation}
\label{eq: modified effective planar coupling constant}
 \gmod^2=\left(4\pi\e^{-\gamma_{\text{E}}}\right)^\peps \g^2=\left(4\pi\e^{-\gamma_{\text{E}}}\right)^\peps \frac{g_\YM^2 N}{(4\pi)^2}\eqncom
\end{equation}
where $\gamma_{\text{E}}$ is the Euler-Mascheroni constant.%
\footnote{The rescaling of $\g^2$ defined in \eqref{eq: effective planar coupling constant} with $\left(4\pi\e^{-\gamma_{\text{E}}}\right)^\peps$ absorbs terms containing $\gamma_{\text{E}}$ and $\log(4\pi)$, which would otherwise appear in the Laurent expansion of $\Interactionamp^{(\ell)}(\peps)$ in $\peps$. In particular, this makes certain number-theoretic properties of $\Interactionamp^{(\ell)}(\peps)$ manifest, see e.g.\ \cite{Duhr:2014woa}.}
The structure of the IR divergences occurring in $\Interactionamp(\gmod^2,\peps)$ is universal and well understood \cite{Mueller:1979ih, Collins:1980ih, Sen:1981sd, Magnea:1990zb}:
\begin{equation}\label{eq: universal IR divergences}
  \log\left(\Interactionamp(\gmod^2,\peps)\right)= \sum_{\ell=1}^\infty \gmod^{2\ell}\left[ -\frac{\gamma^{(\ell)}_{\text{cusp}}}{8(\ell\peps)^2}-\frac{\cG_0^{(\ell)}}{4\ell\peps}\right]\sum_{i=1}^n\left(-\frac{s_{i\,i+1}}{\mu^2}\right)^{-\ell\peps} + \text{Fin}(\gmod^2) +\cO(\peps)
  \eqncom
\end{equation}
where $s_{i\,i+1}=(p_i+p_{i+1})^2$
\begin{equation}
\label{eq: cusp anomalous dimension}
 \gamma_{\text{cusp}}(\gmod^2)=\sum_{\ell=1}^\infty \gmod^{2\ell}\gamma^{(\ell)}_{\text{cusp}}= 8 \gmod^2-\frac{8 \pi ^2 }{3}\gmod^4+\frac{88 \pi^4 }{45}\gmod^6+\cO(\gmod^8)
\end{equation}
is the cusp anomalous dimension and
\begin{equation}
\label{eq: collinear anomalous dimension}
 \cG_0(\gmod^2)=\sum_{\ell=1}^\infty \gmod^{2\ell}\cG^{(\ell)}_0= -4  \zeta_3\gmod^4+8  \left(\frac{5 \pi ^2 \zeta_3}{9}+4 \zeta_5\right)\gmod^6+\cO(\gmod^8)
\end{equation}
is the collinear anomalous dimension.
The 't Hooft mass $\mu$, which sets the renormalisation scale, originates from a rescaling of $g_\YM$ in order to render it dimensionless in $D=4-2\peps$ dimensions \cite{'tHooft:1973mm}.%
\footnote{The occurrence of $\mu$ in \eqref{eq: universal IR divergences} signals in particular that the collinear anomalous dimension $\cG_0$ depends on the renormalisation scheme and is hence not an observable. The expansion \eqref{eq: collinear anomalous dimension} for $\cG_0$ is valid in the scheme which is given by minimal subtraction in the coupling $\gmod^2$ shown in \eqref{eq: modified effective planar coupling constant}.}
$\text{Fin}(\gmod^2)$ denotes a finite part in the $\peps$ expansion, which can also be a function of the coupling constant \eqref{eq: modified effective planar coupling constant}.

The above form of the loop corrections and their IR divergences is also shared by form factors of protected operators \cite{vanNeerven:1985ja,Brandhuber:2010ad,Bork:2010wf,Henn:2011by,Brandhuber:2012vm,Penante:2014sza,Brandhuber:2014ica}.%
\footnote{In fact, form factors have played a central role in the developments leading to \eqref{eq: universal IR divergences} \cite{Mueller:1979ih, Collins:1980ih, Sen:1981sd}.
}
The situation becomes more complicated for form factors of general operators, as analysed in \cite{Wilhelm:2014qua,Nandan:2014oga,Loebbert:2015ova}. Although $\mathcal{N}=4$ SYM theory is conformally invariant, these form factors are UV divergent due to the presence of the composite operator. The UV divergences can be absorbed by renormalising this operator. 
Moreover, general operators are not eigenstates under renormalisation but mix with other operators that have the same quantum numbers. 
We define the renormalised operators $\cO_{\ren}^a$ in terms of the bare operators $\cO_{\bare}^a$ as
\begin{equation}\label{eq: def renomalisation constant}
 \cO_{\ren}^a=\cZ^a{}_b \cO_{\bare}^b \eqncom
\end{equation}
where the indices $a$ and $b$ range over the set of operators and the matrix-valued renormalisation constant has the following loop expansion:%
\footnote{Beyond one-loop order and certain subsectors of the theory, also odd powers of $\gmod$ appear in \eqref{eq: expansion of Z}. These correspond to mixing between operators with different lengths. In this work, however, we are not treating cases where length-changing occurs and hence disregard the corresponding terms to simplify the presentation.}
\begin{equation}\label{eq: expansion of Z}
 \cZ^a{}_b=\delta^a{}_b+\sum_{\ell=1}^\infty \gmod^{2\ell} (\cZ^{(\ell)})^a{}_b \eqndot
\end{equation}
The renormalisation constant is connected to the dilatation operator as%
\footnote{As in \eqref{eq: expansion of Z}, we have neglected terms with odd power of $\gmod$, which correspond to length-changing contributions to the dilatation operator.}$^,$%
\footnote{Note that the expansion of the dilatation operator in $\gmod$ coincides with its expansion in $\g$ shown in \eqref{eq: expansion dilatation operator}.
}
\begin{equation}
\label{eq: Z in terms of D}
 \cZ=\exp\sum_{\ell=1}^\infty \gmod^{2\ell}\frac{\loopDila{\ell}}{2\ell\peps} \eqndot
\end{equation}
The renormalised form factor is then given as 
\begin{equation}
\label{eq: renormalised form factor}
 \ffco_{\cO_{\ren}^a,n}
 =\cZ^a{}_b \ffco_{ \cO_{\bare}^b,n}\eqndot
\end{equation}
Due to the mixing of the operators, the loop correction to the form factor of one operator is no longer proportional to the tree-level form factor of that operator. However, we can still write 
\begin{equation}
\label{eq: loop correction form factor}
 \ffco_{\cO,L}(1,\dots,L;q)=\Interaction \,\ffco_{\cO,L}^{(0)}(1,\dots,L;q)=\left( 1+ \sum_{\ell=1}^\infty \gmod^{2\ell} \Interaction^{(\ell)}\right) \ffco_{\cO,L}^{(0)}(1,\dots,L;q) \eqncom
\end{equation}
if we promote $\Interaction^{(\ell)}$ to \emph{operators} that act on the minimal tree-level form factor. 
We will give concrete examples of these interaction operators in the next sections.
Due to the close connection \eqref{eq: minimal tree-level form factor from oscillator replacement} between the composite operators and their minimal form factors, we can equally write the renormalisation constant as an operator acting on the minimal form factor:
\begin{equation}
\label{eq: renormalisation constant as operator}
 \cZ^a{}_b \ffco_{ \cO_{\bare}^b,L}(1,\dots,L;q)=\cZ \ffco_{ \cO_{\bare}^a,L}(1,\dots,L;q)\eqndot
\end{equation}
The renormalised form factors can then be obtained by acting with the renormalised interaction operators 
\begin{equation}
\label{eq: def renormalised interaction operators}
 \Interactionr=\Interaction \cZ\eqncom \qquad \Interactionr^{(\ell)}=\sum_{l=0}^\ell \Interaction^{(l)} \cZ^{(\ell-l)} \eqncom
\end{equation}
on the minimal tree-level form factor.
As the IR divergences are universal, the renormalised interaction operators have to satisfy \eqref{eq: universal IR divergences}. Inserting \eqref{eq: Z in terms of D}, we find that the bare interaction operators satisfy:
\begin{equation}
\label{eq: general divergences of form factor}
 \begin{aligned}
  \log\left(\cI\right)&= \sum_{\ell=1}^\infty \gmod^{2\ell}\left[ -\frac{\gamma^{(\ell)}_{\text{cusp}}}{8(\ell\peps)^2}-\frac{\cG_0^{(\ell)}}{4\ell\peps}
  \right]\sum_{i=1}^L\left(-\frac{s_{i\,i+1}}{\mu^2}\right)^{-\ell\peps}
 -\sum_{\ell=1}^\infty \gmod^{2\ell}\frac{\loopDila{\ell}}{2\ell\peps} + \text{Fin}(\gmod^2) +\cO(\peps) \eqndot
\end{aligned}
\end{equation}
Thus, we can determine the dilatation operator $\Dila$ via $\cI$. 
Let us now show how to calculate $\cI$ via on-shell methods.

\section{One-loop corrections in the \texorpdfstring{$\SU2$}{SU(2)} sector via unitarity}
\label{sec: one-loop unitarity}

In this section, we demonstrate in detail how to calculate one-loop corrections to minimal form factors via the on-shell method of unitarity.%
\footnote{For reviews of unitarity for scattering amplitudes, see e.g.\ \cite{Dixon:1996wi,Bern:2007dw,Elvang:2013cua}.} We focus on composite operators from the $\SU2$ sector for explicitness.

Operators in the $\SU2$ sector are formed from two kinds of scalar fields with one common $\SU4$ index, say $X=\phi_{14}$ and $Y=\phi_{24}$.
According to \eqref{eq: form factor from spin chain}, the colour-ordered minimal tree-level super form factor of such operators is given by a polynomial in $\etatt_i^1\etatt_i^4$ and $\etatt_i^2\etatt_i^4$ multiplied by a momentum-conserving delta function; e.g.\
 for $\cO=\tr(XXYX\dots)$, we have
\begin{equation}
  \ffco^{(0)}_{\cO,L}(1,\dots,L;q)=(2\pi)^4\delta^4(q-\sum_{i=1}^L\lambda_i\lambdat_i)\left(\etatt_1^1\etatt_1^4\etatt_2^1\etatt_2^4\etatt_3^2\etatt_3^4\etatt_4^1\etatt_4^4\dots +\text{cyclic permutations}\right) \eqndot
\end{equation}

We encode the one-loop corrections in the interaction operator $\Interaction^{(1)}$ defined in \eqref{eq: loop correction form factor}.
At this loop order, only two fields in the composite operator can interact at a time, and those have to be neighbouring in order to produce a single-trace structure.
Hence, we can write $\Interaction^{(1)}$ in terms of its interaction density $\intone[i\,i+1]$ as 
\begin{equation}
\label{eq: one-loop interaction}
 \Interaction^{(1)}=\sum_{i=1}^L \intone[i\,i+1]\eqncom
\end{equation}
where $\intone[i\,i+1]$ acts on the fields $i$ and $i+1$ and cyclic identification $i+L\sim i$ is understood.
We depict $\intone[i\,i+1]$ as 
\begin{equation}
 \intone[i\,i+1]=  
 \begin{aligned}
 \begin{tikzpicture}
  \drawvlineblob{1}{1}
  \drawvlineblob{2}{1}
  \drawtwoblob{1}{1}{$\intonei$}
 \end{tikzpicture}
 \end{aligned}
 \eqncom
\end{equation}
where we in general only specify the first field the density acts on in the case that this determines all fields it acts on unambiguously.

In the $\SU2$ sector, we can write $\intone[i\,i+1]$ explicitly as a differential operator in the fermionic variables:
\begin{equation}
\label{eq: one-loop int expansion}
 \interaction^{(1)}_{i\,i+1}=\sum_{A,B,C,D=1}^2 \intoneif{Z_AZ_B}{Z_CZ_D} \etatt_{i}^C\frac{\partial}{\partial \etatt_{i}^A} \etatt_{i+1}^D\frac{\partial}{\partial \etatt_{i+1}^B}\eqncom
\end{equation}
where $Z_1=X$ and $Z_2=Y$. The matrix element $\intoneif{Z_AZ_B}{Z_CZ_D}$ encodes the contributions of all interactions that transform the fields $Z_A$, $Z_B$ in the operator to external fields $Z_C$, $Z_D$.

As the global charges are conserved, the only non-vanishing matrix elements are 
\begin{equation}
\intoneif{XX}{XX}\eqncom \quad\intoneif{XY}{XY}\eqncom \quad\intoneif{XY}{YX}\eqncom \quad\intoneif{YX}{YX}\eqncom \quad\intoneif{YX}{XY} \quad \text{and} \quad \intoneif{YY}{YY} \eqndot 
\end{equation}
Moreover, matrix elements that are connected by a simple relabelling of $X$ and $Y$ coincide, i.e.\ 
\begin{equation}
 \intoneif{XX}{XX}=\intoneif{YY}{YY}\eqncom \quad\intoneif{XY}{XY}=\intoneif{YX}{YX}\eqncom \quad \intoneif{XY}{YX}=\intoneif{YX}{XY}\eqndot
\end{equation}
Hence, we only have to calculate $\intoneif{XX}{XX}$, $\intoneif{XY}{XY}$ and $\intoneif{XY}{YX}$. As already mentioned, this can be achieved via the on-shell method of unitarity.

The general idea behind unitarity \cite{Bern:1994zx,Bern:1994cg} and generalised unitarity \cite{Britto:2004nc} is to reconstruct processes, such as scattering amplitudes, form factors or correlation functions, at loop order by applying cuts.
Here, a cut denotes replacing one or more propagators according to
\begin{equation}
\label{eq: cut replacement}
 \frac{i}{l_i^2}\to 2\pi \delta^+(l_i^2)=2\pi \delta(l_i^2)\Theta(l_i^0) \eqncom
\end{equation}
where $\delta^+(l_i^2)$ denotes the delta function picking the positive-energy branch of the on-shell condition $l_i^2=0$.
This delta function can be explicitly written using the Heaviside step function $\Theta$, as shown on the right hand side of \eqref{eq: cut replacement}.
On such a cut, the process factorises into the product of one or more tree-level or lower-loop processes.
In the case of unitarity, the cut has to result in exactly two factors and corresponds to a discontinuity in one of the kinematic variables. 
For generalised unitarity, more general cuts are possible.%
\footnote{See \cite{Abreu:2014cla} for a discussion of the relation between cuts and discontinuities across the corresponding branch cuts in generalised unitarity.}
Throughout this work, we are using four-dimensional (generalised) unitarity, i.e.\ we evaluate the expressions on the cut in $D=4$ dimensions.%
\footnote{For a review of $D$-dimensional unitarity, see \cite{Britto:2010xq}.
}
This allows us to use the simpler building blocks in four dimensions but requires that the results are lifted to $D=4-2\peps$, leading to some subtleties as discussed in the next chapter.

In this section, we apply unitarity at the level of the integrand, i.e.\ we aim to reconstruct the integrand of the minimal one-loop form factors via cuts. As each interaction density $\intone[i\,i+1]$ depends only on a single scale, namely $s_{i\,i+1}=(p_i+p_{i+1})^2$, it is sufficient to consider the double cut corresponding to the discontinuity in $s_{i\,i+1}$. For convenience, we set $i=1$. On this cut, the minimal one-loop form factor $\ffco^{(1)}_{\cO,L}$ factorises into the product of the minimal tree-level form factor $\ffco^{(0)}_{\cO,L}$ and the four-point tree-level amplitude $\ampco^{(0)}_4$, as shown in figure \ref{fig: one-loop double cut}:%
\footnote{Here and throughout the first part of this work, we are splitting off the dependence on the gauge group generators $\T^a$ to work with colour-ordered objects as defined in \eqref{eq: def colour-ordered amplitude} and \eqref{eq: def colour-ordered form factor}. The contractions of the generators $\T^a$ in the trace factors of \eqref{eq: def colour-ordered amplitude} and \eqref{eq: def colour-ordered form factor} can be performed via \eqref{eq: Ts summed over a}, which is trivially done in the case of planar cuts.}
\begin{equation}\label{eq: one-loop double cut}
\begin{aligned}
 \ffco_{\cO,L}^{(1)}(1,2,3,\dots,L)\Big|_{s_{12}} \! =\int \dLIPS_{2,\{l\}} \de^4 \etatt_{l_{1}}\de^4 \etatt_{l_{2}}  \ffco^{(0)}_{\cO,L}(l_1,l_2,{3},\dots,{L};q) \ampco^{(0)}_4(-l_2,-l_1,{1},{2})\eqndot
\end{aligned}
\end{equation}
The loop integral $\int \frac{\de^Dl}{(2\pi)^D}$ is reduced to a phase-space integral, which for a general number of particles $n$ and general $D$ is given by 
\begin{equation}
\label{eq: def phase space integral}
\int\dLIPS_{n,\{l\}} = \int\bigg(\prod_{i=1}^n {\de^D l_i \over (2\pi)^D}  2\pi  \delta_+(l_i^2) \bigg)  \eqndot
\end{equation}
In four dimensions, we can alternatively write the integration as 
\begin{equation}
\label{eq: dLambda}
 \int \de^4 l_i \delta_+(l_i^2)\de^4 \etatt_{l_{i}} = \int \frac{\de^2\lambda_{l_i}\de^2\lambdat_{l_i}}{\U{1}}\de^4\etatt_{l_i} \equiv \int \de \Lambda_{l_i} \eqncom
\end{equation}
where the factor of $\U{1}$ refers to the fact that $\lambda_{l_i}$ and $\lambdat_{l_i}$ are defined only up to a phase, which is not integrated over; cf.\ the discussion below \eqref{eq: spinor helicity variables}. 
In our conventions, the super-spinor-helicity variables corresponding to $-l_i$ are related to those corresponding to $l_i$ as $\lambda_{-l_i}=
-\lambda_{l_i}$, $\lambdat_{-l_i}=\lambdat_{l_i}$, $\etatt_{-l_i}=\etatt_{l_i}$.

\begin{figure}[htbp]
 \centering
$
\settoheight{\eqoff}{$\times$}%
\setlength{\eqoff}{0.5\eqoff}%
\addtolength{\eqoff}{-12.0\unitlength}%
\raisebox{\eqoff}{%
\fmfframe(2,2)(2,2){%
\begin{fmfchar*}(80,30)
\fmfleft{vp3,vp,vpL,vq}
\fmfright{vp2,vp1}
\fmf{dbl_plain_arrow,tension=1.2}{vq,v1}
\fmf{plain_arrow,tension=0}{v1,vpL}
\fmf{plain_arrow,tension=1.2}{v1,vp3}
\fmf{plain_arrow,left=0.7,label=$\scriptstyle l_1\,,$,l.d=15}{v1,v2}
\fmf{plain_arrow,right=0.7,label=$\scriptstyle l_2\,,$,l.d=15}{v1,v2}
\fmf{phantom_smallcut,left=0.7,tension=0}{v1,v2}
\fmf{phantom_smallcut,right=0.7,tension=0}{v1,v2}
\fmf{plain_arrow}{v2,vp1}
\fmf{plain_arrow}{v2,vp2}
\fmfv{decor.shape=circle,decor.filled=30,decor.size=30,label=$\ffco_{\cO,,L}$,label.dist=0}{v1}
\fmfv{decor.shape=circle,decor.filled=10,decor.size=30,label=$\ampco_{4}$,label.dist=0}{v2}
\fmffreeze
 \fmfcmd{pair vertq, vertpone, vertptwo, vertpthree, vertpL, vertone, verttwo; vertone = vloc(__v1); verttwo = vloc(__v2); vertq = vloc(__vq); vertpone = vloc(__vp1); vertptwo = vloc(__vp2); vertpthree = vloc(__vp3);vertpL = vloc(__vpL);}
 \fmfiv{label=$\scriptstyle q$}{vertq}
 \fmfiv{label=$\scriptstyle p_1$}{vertpone}
 \fmfiv{label=$\scriptstyle p_2$}{vertptwo}
 \fmfiv{label=$\scriptstyle p_3$}{vertpthree}
 \fmfiv{label=$\scriptstyle p_L$}{vertpL}
 \fmfiv{label=$\cdot$,l.d=20,l.a=-150}{vertone}
 \fmfiv{label=$\cdot$,l.d=20,l.a=-165}{vertone}
 \fmfiv{label=$\cdot$,l.d=20,l.a=-180}{vertone}
\end{fmfchar*}%
}}%
$
\caption{The double cut of the minimal one-loop form factor $\ffco^{(1)}_{\cO,L}$ in the channel $(p_1+p_2)^2$.}
\label{fig: one-loop double cut}
\end{figure}

We look at $\intonef[1]{XX}{XX}$ in detail.%
\footnote{This case was already treated in \cite{Brandhuber:2010ad}.} 
We have 
\begin{equation}\label{eq: su2 XXXX}
 \begin{aligned}
  &\ffco^{(1)}_{\cO,L}(1,2,3,\dots,L)\Big|_{\intonef[1]{XX}{XX},s_{12}}\\
  &=\int \dLIPS_{2,\{l\}} \de^4 \etatt_{l_{1}}\de^4 \etatt_{l_{2}}  \ffco^{(0)}_{\cO,L}(l_1|^X,l_2|^X,{3},\dots,{L};q) \ampco_4^{(0)}(-l_2,-l_1,{1}|^X,{2}|^X)
\eqncom
 \end{aligned}
\end{equation} 
where we denote by the superscript and the vertical bar the specified component defined in \eqref{eq: def component expression} dressed with the corresponding $\etatt$ factors. For instance,
\begin{equation}
\ampco^{(0)}_4(-l_2,-l_1,{1}|^X,{2}|^X)=\ampco^{(0)}_4(-l_2,-l_1,{1}^X,{2}^X) \etatt_1^1\etatt_1^4\etatt_2^1\etatt_2^4 \eqndot                                                                  \end{equation}
Inserting the expression \eqref{eq: colour-ordered MHV super amplitude} for the super amplitude and performing the integration over the fermionic variables, we find 
\begin{equation}\label{eq: su2 XXXX 0}
 \begin{aligned}
  \ffco^{(1)}_{\cO,L}(1,2,3,\dots,L)\Big|_{\intonef[1]{XX}{XX},s_{12}}
  =\etatt_1^1\etatt_1^4\etatt_2^1\etatt_2^4 \ffco^{(0)}_{\cO,L}(1^X,2^X,3,\dots,L)\, i\int \dLIPSt_{2,\{l\}}\frac{\ab{1 2}\ab{ l_1l_2}}{\ab{1 l_1}\ab{2 l_2}}\eqndot
 \end{aligned}
\end{equation} 
Here, we have defined $\dLIPSt$ as $\dLIPS$ multiplied by 
 the momentum-conserving delta function from the amplitude. In general,
\begin{equation}
 \label{eq: dLIPS tilde}
 \dLIPSt_{n,\{l\}}=\dLIPS_{n,\{l\}}(2\pi)^D \delta^{D} \Big(p_{i,j} -\sum_{i=1}^n l_i \Big)\eqncom
\end{equation}
where $p_{i,j}=\sum_{k=i}^j p_k$ is the total external momentum traversing the cut, which is $p_{1,2}=p_1+p_2$ in the case under consideration.
The Schouten identity \eqref{eq: Schouten identities} and momentum conservation yield 
\begin{equation}\label{eq: triangle propagator identity}
 \frac{\ab{ 1 2}\ab{ l_1 l_2}}{\ab{1 l_1}\ab{2 l_2}}=-\frac{(p_1+p_2)^2}{(p_1-l_1)^2}\eqndot
\end{equation}
Thus, the cut of this one-loop form factor is proportional to the cut of the triangle integral:%
\footnote{Here, the depicted cut integral denotes the double cut of \eqref{eq: oneloopint:1masstriangle} but with measure $\int \frac{\de^Dl}{(2\pi)^D}$ instead of $\e^{\gamma_{\text{E}} \peps}\int \frac{\de^Dl}{i\pi^{\frac{D}{2}}}$.}
\begin{equation}\label{eq: su2 XXXX 2}
 \begin{aligned}
  \ffco^{(1)}_{\cO,L}(1,2,3,\dots,L)\Big|_{\intonef[1]{XX}{XX},s_{12}}=\etatt_1^1\etatt_1^4\etatt_2^1\etatt_2^4 \ffco^{(0)}_{\cO,L}(1^X,2^X,3,\dots,L)\,i\left(-s_{12} \FDinline[triangle,doublecut,cutlabels,twolabels, labelone=\scriptscriptstyle p_1,labeltwo=\scriptscriptstyle p_2] \right)\!\eqndot
 \end{aligned}
\end{equation} 
Now, we lift the result to the $D$-dimensional uncut expressions, i.e.\ we conclude that the same proportionality exists between the uncut one-loop form factor and the uncut triangle integral.
The precise rules for this lifting procedure, which in particular removes the factor $i$ in \eqref{eq: su2 XXXX 2}, are explained in appendix \ref{appsec: lifting}.
We find that 
\begin{equation}
 \intonef[1]{XX}{XX}=-s_{12} \FDinline[triangle,twolabels,labelone=\scriptscriptstyle p_1,labeltwo=\scriptscriptstyle p_2]\eqndot
\end{equation}

A similar calculation for $\intonef[1]{XY}{YX}$ shows that
\begin{equation}\label{eq: su2 XYYX 2}
 \begin{aligned}
  \ffco^{(1)}_{\cO,L}(1,2,3,\dots,L)\Big|_{\intonef[1]{XY}{YX},s_{12}}=\etatt_1^2\etatt_1^4\etatt_2^1\etatt_2^4 \ffco^{(0)}_{\cO,L}(1^X,2^Y,3,\dots,L)\,i\left( \FDinline[bubble,doublecut,cutlabels,twolabels, labelone=\scriptscriptstyle p_1,labeltwo=\scriptscriptstyle p_2] \right)\eqncom
 \end{aligned}
\end{equation} 
and hence 
\begin{equation}
 \intonef[1]{XY}{YX}= \FDinline[bubble,twolabels,labelone=\scriptscriptstyle p_1,labeltwo=\scriptscriptstyle p_2]\eqndot
\end{equation}
This is an explicit example where the one-loop form factor is not proportional to the tree-level form factor of the same composite operator, making it necessary to promote $\Interaction^{(\ell)}$ to operators as done in \eqref{eq: loop correction form factor}.

The case $\intonef[1]{XY}{XY}$ can be calculated in complete analogy to the previous cases. We have summarised the results of the three calculations in table \ref{tab: 1-loop su 2}.
\begin{table}[tbp]
\begin{center}
\begin{tabular}{|l|c|c|c|}\hline
\qquad\qquad$\intoneif{}{}$&${}^{\XYsize XX}_{\XYsize XX}$&${}_{\XYsize XY}^{\XYsize XY}$&${}_{\XYsize XY}^{\XYsize YX}$
\\\hline
\FDinline[triangle,twolabels,labelone=\scriptscriptstyle i,labeltwo=\scriptscriptstyle i+1]
$s_{i\, i+1} $
&-1&-1&0
\\\hline
\FDinline[bubble,twolabels,labelone=\scriptscriptstyle i,labeltwo=\scriptscriptstyle i+1]
&0&-1&+1
\\\hline
\end{tabular}
\end{center}
\caption{Linear combinations of Feynman integrals forming the matrix elements for the minimal one-loop form factors in the $\SU2$ sector.}
\label{tab: 1-loop su 2}
\end{table}
Note that the different matrix elements satisfy
\begin{equation}
\label{eq: one-loop identity}
 \intoneif{XY}{XY}+\intoneif{XY}{YX}=\intoneif{XX}{XX}\eqndot
\end{equation}
This identity is in fact a consequence of the Ward identity \eqref{eq: action on form factor} for the generators 
\begin{equation}
\mathfrak{J}_i^1=\etatt_i^1\frac{\partial}{\partial\etatt_i^2}+\etatt_i^2\frac{\partial}{\partial\etatt_i^1}\eqncom \quad
\mathfrak{J}_i^2=-i\etatt_i^1\frac{\partial}{\partial\etatt_i^2}+i\etatt_i^2\frac{\partial}{\partial\etatt_i^1}\eqncom \quad
\mathfrak{J}_i^3=\etatt_i^1\frac{\partial}{\partial\etatt_i^1}-\etatt_i^2\frac{\partial}{\partial\etatt_i^2}
\end{equation}
of $\SU2$.
Applying \eqref{eq: action on form factor} once for the tree-level form factor in \eqref{eq: loop correction form factor} and once for its one-loop correction, we find
\begin{equation}
\label{eq: symmetry commutator one-loop}
 [\mathfrak{J}^A,\Interaction^{(1)}]=0\eqncom
\end{equation}
which implies \eqref{eq: one-loop identity}. 

Defining the identity operator 
\begin{equation}
\label{eq: def identity operator}
 \idm_{i\,i+1}= \sum_{A,B=1}^2\etatt_i^A\frac{\partial}{\partial\etatt_i^A}\etatt_{i+1}^B\frac{\partial}{\partial\etatt_{i+1}^B} 
\end{equation}
and the permutation operator
\begin{equation}
\label{eq: def permutation operator}
 \PP_{i\,i+1}= \sum_{A,B=1}^2\etatt_i^B\frac{\partial}{\partial\etatt_i^A}\etatt_{i+1}^A\frac{\partial}{\partial\etatt_{i+1}^B} \eqncom
\end{equation}
we can recast the one-loop corrections into the following form:
\begin{equation}
\label{eq: one-loop int su2}
 \interaction^{(1)}_{i\,i+1}= -\ s_{i\, i+1}\FDinline[triangle,twolabels,labelone=\scriptscriptstyle i,labeltwo=\scriptscriptstyle i+1]\times
\idm_{i\,i+1} \,-\, \FDinline[bubble,twolabels,labelone=\scriptscriptstyle i,labeltwo=\scriptscriptstyle i+1] \times (\idm-\PP)_{i\,i+1} 
 \eqndot
\end{equation}

Explicit expressions for the one-mass triangle integral and the bubble integral are given in \eqref{eq: oneloopint:1masstriangle} and \eqref{eq: oneloopint:bubble}, respectively.
The divergence of the triangle integral is  
\begin{equation}
 -s_{i\, i+1}\FDinline[triangle,twolabels,labelone=\scriptscriptstyle i,labeltwo=\scriptscriptstyle i+1]
 =-\frac{1}{\peps^2}(-s_{i\, i+1})^{-\peps} +\cO(\peps^0)
 =\left[ -\frac{\gamma^{(1)}_{\text{cusp}}}{8 \peps^2}-\frac{\cG_0^{(1)}}{4\peps}
  \right](-s_{i\,i+1})^{-\peps}+\cO(\peps^0)
\end{equation}
and the one of the bubble integral is 
\begin{equation}
 \FDinline[bubble,twolabels,labelone=\scriptscriptstyle i,labeltwo=\scriptscriptstyle i+1]
 =\frac{1}{\peps}+\cO(\peps^0) \eqncom
\end{equation}
where $\gamma^{(1)}_{\text{cusp}}$ and $\cG_0^{(1)}$ were given in \eqref{eq: cusp anomalous dimension} and \eqref{eq: collinear anomalous dimension}, respectively. 
Comparing \eqref{eq: one-loop int su2} to the general form \eqref{eq: general divergences of form factor}, we can immediately read off the one-loop dilatation operator in the $\SU2$ sector as 
\begin{equation}
 \loopDila{1}=\sum_{i=1}^{L} \loopDila{1}_{i\,i+1}
\end{equation}
with density
\begin{equation}
 \loopDila{1}_{i\,i+1}=2(\idm-\PP)_{i\,i+1}\eqndot
\end{equation}
This is exactly the Hamiltonian density of the integrable Heisenberg XXX spin chain and perfectly agrees with the result first obtained in \cite{Minahan:2002ve}.

After this warm-up exercise, let us now derive the complete one-loop dilatation operator via generalised unitarity.

\section{One-loop corrections for all operators via generalised unitarity}
\label{sec: one-loop generalised unitarity}

In this section, we use four-dimensional generalised unitarity to calculate the cut\hyp constructible part of the one-loop correction to the minimal form factor of any 
 composite operator. As demonstrated in the previous section in the $\SU2$ sector, this immediately yields the one-loop dilatation operator.

Whereas we have worked at the level of the \emph{integrand} in the previous section, we now work at the level of the \emph{integral}.
In non-compact subsectors, such as the $\SL2$ subsector, as well as in the complete theory, minimal form factors can have arbitrarily high powers of $\lambda_i$ and $\lambdat_i$. Via \eqref{eq: one-loop double cut}, they lead to one-loop integrands with arbitrarily high powers of the loop momentum in the numerator, which can be expressed in a countably infinite basis of tensor structures. The resulting integrals, however, satisfy considerably more identities than the integrands, as non-vanishing integrands can integrate to zero. They can thus be reduced further. Instead of working at the level of the integrand and performing this reduction, we can simply work at the level of the integral.

It is well known that every one-loop Feynman integral in strictly four dimensions can be written as a linear combination of box integrals, triangle integrals, bubble integrals, tadpole integrals and rational terms \cite{Passarino:1978jh}.%
\footnote{In order to also obtain all terms in $D=4-2\peps$ dimensions that vanish for $D=4$, the pentagon integral has to be included; see e.g.\ \cite{Henn:2014yza}.} 
In massless theories such as $\mathcal{N}=4$ SYM theory, the tadpole integrals vanish. 
Hence, we can make a general ansatz for the one-loop form factor, which is shown in figure \ref{fig: ansatz for generalised unitarity}.
\begin{figure}[htbp]
 \centering
\begin{equation*}
 \begin{aligned}
\settoheight{\eqoff}{$\times$}%
\setlength{\eqoff}{0.5\eqoff}%
\addtolength{\eqoff}{-14.5\unitlength}%
\raisebox{\eqoff}{%
\fmfframe(2,2)(6,2){%
\begin{fmfchar*}(25,25)
\fmfsurround{vp3,vp2,vp1,vq,vpL,vp}
\fmf{dbl_plain_arrow,tension=1}{vq,vqa}
\fmf{plain_arrow,tension=1}{vpLa,vpL}
\fmf{plain_arrow,tension=1}{vp3a,vp3}
\fmf{phantom,tension=1}{vpa,vp}
\fmf{plain_arrow}{vp1a,vp1}
\fmf{plain_arrow}{vp2a,vp2}
\fmf{dbl_plain_arrow,tension=2}{vqa,v1}
\fmf{plain_arrow,tension=2}{v1,vpLa}
\fmf{plain_arrow,tension=2}{v1,vp3a}
\fmf{phantom,tension=2}{v1,vpa}
\fmf{plain_arrow,tension=2}{v1,vp1a}
\fmf{plain_arrow,tension=2}{v1,vp2a}
\fmffreeze
\fmfdraw
 \fmfcmd{pair vertq, vertpone, vertptwo, vertpthree, vertpL, vertone, verttwo, vertp; vertone = vloc(__v1); verttwo = vloc(__v2); vertq = vloc(__vq); vertpone = vloc(__vp1); vertptwo = vloc(__vp2); vertpthree = vloc(__vp3);vertp = vloc(__vp);vertpL = vloc(__vpL);}
 \fmfiv{decor.shape=circle,decor.filled=30,decor.size=30}{vertone}
 \fmfiv{decor.shape=circle,decor.filled=0,decor.size=20}{vertone}
 \fmfiv{label=$\scriptstyle  \ffco_{\cO,,n}$,l.d=0,l.a=0}{vertone}
 \fmfiv{label=$\scriptstyle q$}{vertq}
 \fmfiv{label=$\scriptstyle p_1$}{vertpone}
 \fmfiv{label=$\scriptstyle p_2$}{vertptwo}
 \fmfiv{label=$\scriptstyle p_3$}{vertpthree}
 \fmfiv{label=$\scriptstyle p_n$}{vertpL}
 \fmfiv{label=$\cdot$,l.d=20,l.a=-45}{vertone}
 \fmfiv{label=$\cdot$,l.d=20,l.a=-60}{vertone}
 \fmfiv{label=$\cdot$,l.d=20,l.a=-75}{vertone}
\end{fmfchar*}%
}}%
&=
\sum_{i,j,k,l}c_{\text{box}}^{(i,j,k,l)}
\quad
\settoheight{\eqoff}{$\times$}%
\setlength{\eqoff}{0.5\eqoff}%
\addtolength{\eqoff}{-12\unitlength}%
\raisebox{\eqoff}{%
\fmfframe(2,2)(6,2){%
\begin{fmfchar*}(20,20)
\fmfsurround{vpjjj,vpj1,vpjj,vpj,vpiii,vpi1,vpii,vpi,vq,vpl1,vpll,vpl,vpkkk,vpk1,vpkk,vpk}
\fmf{dbl_plain_arrow,tension=0}{vq,vi}
\fmf{plain_arrow,tension=1}{vi,vpi}
\fmf{plain_arrow,tension=1}{vi,vpl1}
\fmf{plain_arrow,tension=1}{vj,vpj}
\fmf{plain_arrow,tension=1}{vj,vpi1}
\fmf{plain_arrow,tension=1}{vk,vpk}
\fmf{plain_arrow,tension=1}{vk,vpj1}
\fmf{plain_arrow,tension=1}{vl,vpl}
\fmf{plain_arrow,tension=1}{vl,vpk1}
\fmf{plain_arrow,tension=1}{vi,vj}
\fmf{plain_arrow,tension=1}{vi,vl}
\fmf{plain_arrow,tension=1}{vj,vk}
\fmf{plain_arrow,tension=1}{vl,vk}
\fmffreeze
\fmfdraw
 \fmfcmd{pair vertq, verti, vertj, vertk, vertl, vertpi, vertpj, vertpk, vertpl, vertpii, vertpjj, vertpkk, vertpll; vertq = vloc(__vq); verti = vloc(__vi); vertj = vloc(__vj); vertk = vloc(__vk); vertl = vloc(__vl);vertpi = vloc(__vpi); vertpj = vloc(__vpj); vertpk = vloc(__vpk); vertpl = vloc(__vpl);vertpii = vloc(__vpi1); vertpjj = vloc(__vpj1); vertpkk = vloc(__vpk1); vertpll = vloc(__vpl1);}
 \fmfiv{label=$\scriptstyle q$}{vertq}
 \fmfiv{label=$\scriptstyle p_i$}{vertpi}
 \fmfiv{label=$\scriptstyle p_{i+1}$}{vertpii}
 \fmfiv{label=$\scriptstyle p_j$}{vertpj}
 \fmfiv{label=$\scriptstyle p_{j+1}$}{vertpjj}
 \fmfiv{label=$\scriptstyle p_k$}{vertpk}
 \fmfiv{label=$\scriptstyle p_{k+1}$}{vertpkk}
 \fmfiv{label=$\scriptstyle p_l$}{vertpl}
 \fmfiv{label=$\scriptstyle p_{l+1}$}{vertpll}
 \fmfiv{label=$\cdot$,l.d=0,l.a=0}{verti+(12*sind(243),12*cosd(243))}
 \fmfiv{label=$\cdot$,l.d=0,l.a=0}{verti+(12*sind(-67),12*cosd(-67))}
 \fmfiv{label=$\cdot$,l.d=10,l.a=105}{vertj}
 \fmfiv{label=$\cdot$,l.d=10,l.a=90}{vertj}
 \fmfiv{label=$\cdot$,l.d=10,l.a=75}{vertj}
 \fmfiv{label=$\cdot$,l.d=10,l.a=-15}{vertk}
 \fmfiv{label=$\cdot$,l.d=10,l.a=-0}{vertk}
 \fmfiv{label=$\cdot$,l.d=10,l.a=+15}{vertk}
 \fmfiv{label=$\cdot$,l.d=10,l.a=-105}{vertl}
 \fmfiv{label=$\cdot$,l.d=10,l.a=-90}{vertl}
 \fmfiv{label=$\cdot$,l.d=10,l.a=-75}{vertl}
\end{fmfchar*}%
}}%
+\sum_{i,j,k}c_{\text{triangle}}^{(i,j,k)}
\quad
\settoheight{\eqoff}{$\times$}%
\setlength{\eqoff}{0.5\eqoff}%
\addtolength{\eqoff}{-12\unitlength}%
\raisebox{\eqoff}{%
\fmfframe(2,2)(6,2){%
\begin{fmfchar*}(20,20)
\fmfsurround{vpjj,vpj,vpiii,vpi1,vpii,vpiiii,vpi,vq,vpk1,vpkk,vpkkkk,vpk,vpjjj,vpj1}
\fmf{dbl_plain_arrow,tension=0}{vq,vi}
\fmf{plain_arrow,tension=1}{vi,vpi}
\fmf{plain_arrow,tension=1}{vi,vpk1}
\fmf{plain_arrow,tension=1}{vj,vpj}
\fmf{plain_arrow,tension=1}{vj,vpi1}
\fmf{plain_arrow,tension=1}{vk,vpk}
\fmf{plain_arrow,tension=1}{vk,vpj1}
\fmf{plain_arrow,tension=0.66}{vi,vj}
\fmf{plain_arrow,tension=0.66}{vi,vk}
\fmf{plain_arrow,tension=0.66}{vj,vk}
\fmffreeze
\fmfdraw
 \fmfcmd{pair vertq, verti, vertj, vertk, vertpi, vertpj, vertpk, vertpii, vertpjj, vertpkk; vertq = vloc(__vq); verti = vloc(__vi); vertj = vloc(__vj); vertk = vloc(__vk); vertpi = vloc(__vpi); vertpj = vloc(__vpj); vertpk = vloc(__vpk); vertpii = vloc(__vpi1); vertpjj = vloc(__vpj1); vertpkk = vloc(__vpk1); }
 \fmfiv{label=$\scriptstyle q$}{vertq}
 \fmfiv{label=$\scriptstyle p_i$}{vertpi}
 \fmfiv{label=$\scriptstyle p_{i+1}$}{vertpii}
 \fmfiv{label=$\scriptstyle p_j$}{vertpj}
 \fmfiv{label=$\scriptstyle p_{j+1}$}{vertpjj}
 \fmfiv{label=$\scriptstyle p_k$}{vertpk}
 \fmfiv{label=$\scriptstyle p_{k+1}$}{vertpkk}
 \fmfiv{label=$\cdot$,l.d=0,l.a=0}{verti+(12*sind(243),12*cosd(243))}
 \fmfiv{label=$\cdot$,l.d=0,l.a=0}{verti+(12*sind(-67),12*cosd(-67))}
 \fmfiv{label=$\cdot$,l.d=10,l.a=35}{vertj}
 \fmfiv{label=$\cdot$,l.d=10,l.a=50}{vertj}
 \fmfiv{label=$\cdot$,l.d=10,l.a=65}{vertj}
 \fmfiv{label=$\cdot$,l.d=10,l.a=-35}{vertk}
 \fmfiv{label=$\cdot$,l.d=10,l.a=-50}{vertk}
 \fmfiv{label=$\cdot$,l.d=10,l.a=-65}{vertk}
\end{fmfchar*}%
}}%
\\ &\phaneq
+\sum_{i,j}c_{\text{bubble}}^{(i,j)}
\quad
\settoheight{\eqoff}{$\times$}%
\setlength{\eqoff}{0.5\eqoff}%
\addtolength{\eqoff}{-12\unitlength}%
\raisebox{\eqoff}{%
\fmfframe(2,2)(6,2){%
\begin{fmfchar*}(20,20)
\fmfsurround{vpiii,vpi1,vpii,vpiiii,vpiiiii,vpiiiiii,vpi,vq,vpj1,vpjj,vpjjjj,vpjjjjj,vpjjjjjj,vpj}
\fmf{dbl_plain_arrow,tension=0}{vq,vi}
\fmf{plain_arrow,tension=1}{vi,vpi}
\fmf{plain_arrow,tension=1}{vi,vpj1}
\fmf{plain_arrow,tension=1}{vj,vpj}
\fmf{plain_arrow,tension=1}{vj,vpi1}
\fmf{plain_arrow,tension=0.5,right=0.66}{vi,vj}
\fmf{plain_arrow,tension=0.5,left=0.66}{vi,vj}
\fmffreeze
\fmfdraw
 \fmfcmd{pair vertq, verti, vertj,  vertpi, vertpj, vertpii, vertpjj; vertq = vloc(__vq); verti = vloc(__vi); vertj = vloc(__vj);  vertpi = vloc(__vpi); vertpj = vloc(__vpj);  vertpii = vloc(__vpi1); vertpjj = vloc(__vpj1);  }
 \fmfiv{label=$\scriptstyle q$}{vertq}
 \fmfiv{label=$\scriptstyle p_i$}{vertpi}
 \fmfiv{label=$\scriptstyle p_{i+1}$}{vertpii}
 \fmfiv{label=$\scriptstyle p_j$}{vertpj}
 \fmfiv{label=$\scriptstyle p_{j+1}$}{vertpjj}
 \fmfiv{label=$\cdot$,l.d=0,l.a=0}{verti+(12*sind(243),12*cosd(243))}
 \fmfiv{label=$\cdot$,l.d=0,l.a=0}{verti+(12*sind(-67),12*cosd(-67))}
 \fmfiv{label=$\cdot$,l.d=10,l.a=15}{vertj}
 \fmfiv{label=$\cdot$,l.d=10,l.a=0}{vertj}
 \fmfiv{label=$\cdot$,l.d=10,l.a=-15}{vertj}
\end{fmfchar*}%
}}%
+\text{rational terms}
\end{aligned}
\end{equation*}
\caption{The $n$-point one-loop form factor $\ffco^{(1)}_{\cO,n}$ of a generic single-trace operator $\cO$ can be written as a linear combination of box integrals, triangle integrals, bubble integrals and rational terms. The coefficients of these integrals are labelled by the different combinations of momenta flowing out of their corners.
}
\label{fig: ansatz for generalised unitarity}
\end{figure}

The coefficients in this ansatz can be fixed by applying cuts to both sides in figure \ref{fig: ansatz for generalised unitarity} and integrating over all remaining degrees of freedom. First, the maximal cuts are taken, which are the quadruple cuts. They isolate the box integrals and hence fix their coefficients. Next, the triple cuts are taken, which isolate the box integrals and triangle integrals. As the box coefficients are already known, this fixes the triangle coefficients. Finally, the double cuts are taken, which have contributions from box integrals, triangle integrals and bubble integrals. Knowing the coefficients of the former, this fixes the bubble coefficients.
The finite rational terms vanish in all cuts and can hence not be obtained via this variant of four-dimensional generalised unitarity.
We refer to the expression in figure \ref{fig: ansatz for generalised unitarity} without the rational terms as the cut-constructible part.%
\footnote{Note that at least some of the rational terms in this ansatz can be constructed by applying the double cut at the level of the integrand and using PV reduction, see \cite{Wilhelm:2014qua} for an example.}

The above method shares many features with other variants of generalised unitarity from the literature on scattering amplitudes, but there are also important differences which are designed to make it well suited for form factors of general operators.
As the method of Ossola, Papadopoulos and Pittau (OPP) \cite{Ossola:2006us}, it first fixes the coefficients in an ansatz that correspond to many propagators and then uses them to determine the coefficients corresponding to less propagators. 
In contrast to OPP, however, it works at the level of the integral and not the integrand. 
Moreover, the integral over all unfixed degrees of freedom in the cut is taken in several other methods, including ones for the direct extraction of integral coefficients \cite{Forde:2007mi,ArkaniHamed:2008gz}. Our way to perform the integration, though, is different.

For minimal one-loop form factors, the general ansatz shown in figure \ref{fig: ansatz for generalised unitarity} simplifies even further.
By definition, minimal form factors have as many external fields as there are fields in the composite operator. 
The box integral involves two fields of the composite operator, which enter its left corner; see figure \ref{fig: ansatz for generalised unitarity}.
However, at least one external field is connected to each of the three other corners of the box integral. 
Hence, it can only contribute to form factors which have at least one more external field than fields in the operator, i.e.\ to at least next-to-minimal form factors.
For the same reason, exactly one external field has to be connected to each of the two corners of the triangle integral that are not connected to the composite operator, and exactly two external fields have to be connected to the right corner of the bubble integral.
Thus, only one-mass triangle integrals, bubble integrals and rational terms can contribute to the minimal one-loop form factor.
This simplified ansatz is shown in figure \ref{fig: simplified ansatz for generalised unitarity}.
\begin{figure}[tbp]
 \centering
\begin{equation*}
 \begin{aligned}
\settoheight{\eqoff}{$\times$}%
\setlength{\eqoff}{0.5\eqoff}%
\addtolength{\eqoff}{-14.5\unitlength}%
\raisebox{\eqoff}{%
\fmfframe(2,2)(6,2){%
\begin{fmfchar*}(25,25)
\fmfsurround{vp3,vp2,vp1,vq,vpL,vp}
\fmf{dbl_plain_arrow,tension=1}{vq,vqa}
\fmf{plain_arrow,tension=1}{vpLa,vpL}
\fmf{plain_arrow,tension=1}{vp3a,vp3}
\fmf{phantom,tension=1}{vpa,vp}
\fmf{plain_arrow}{vp1a,vp1}
\fmf{plain_arrow}{vp2a,vp2}
\fmf{dbl_plain_arrow,tension=2}{vqa,v1}
\fmf{plain_arrow,tension=2}{v1,vpLa}
\fmf{plain_arrow,tension=2}{v1,vp3a}
\fmf{phantom,tension=2}{v1,vpa}
\fmf{plain_arrow,tension=2}{v1,vp1a}
\fmf{plain_arrow,tension=2}{v1,vp2a}
\fmffreeze
\fmfdraw
 \fmfcmd{pair vertq, vertpone, vertptwo, vertpthree, vertpL, vertone, verttwo, vertp; vertone = vloc(__v1); verttwo = vloc(__v2); vertq = vloc(__vq); vertpone = vloc(__vp1); vertptwo = vloc(__vp2); vertpthree = vloc(__vp3);vertp = vloc(__vp);vertpL = vloc(__vpL);}
 \fmfiv{decor.shape=circle,decor.filled=30,decor.size=30}{vertone}
 \fmfiv{decor.shape=circle,decor.filled=0,decor.size=20}{vertone}
 \fmfiv{label=$\scriptstyle  \ffco_{\cO,,L}$,l.d=0,l.a=0}{vertone}
 \fmfiv{label=$\scriptstyle q$}{vertq}
 \fmfiv{label=$\scriptstyle p_1$}{vertpone}
 \fmfiv{label=$\scriptstyle p_2$}{vertptwo}
 \fmfiv{label=$\scriptstyle p_3$}{vertpthree}
 \fmfiv{label=$\scriptstyle p_L$}{vertpL}
 \fmfiv{label=$\cdot$,l.d=20,l.a=-45}{vertone}
 \fmfiv{label=$\cdot$,l.d=20,l.a=-60}{vertone}
 \fmfiv{label=$\cdot$,l.d=20,l.a=-75}{vertone}
\end{fmfchar*}%
}}%
&=
\sum_{i}c_{\text{triangle}}^{i,i+1}
\quad \,\,
\settoheight{\eqoff}{$\times$}%
\setlength{\eqoff}{0.5\eqoff}%
\addtolength{\eqoff}{-12\unitlength}%
\raisebox{\eqoff}{%
\fmfframe(2,2)(6,2){%
\begin{fmfchar*}(20,20)
\fmfsurround{vpjj,vpiii,vpj,vpi1,vpii,vpiiii,vpi,vq,vpk1,vpkk,vpkkkk,vpjjj,vpk,vpj1}
\fmf{dbl_plain_arrow,tension=0}{vq,vi}
\fmf{plain_arrow,tension=1}{vi,vpi}
\fmf{plain_arrow,tension=1}{vi,vpk1}
 \fmf{plain_arrow,tension=2}{vj,vpj}
 \fmf{plain_arrow,tension=2}{vk,vpk}
\fmf{plain_arrow,tension=0.66}{vi,vj}
\fmf{plain_arrow,tension=0.66}{vi,vk}
\fmf{plain_arrow,tension=0.66}{vj,vk}
\fmffreeze
\fmfdraw
 \fmfcmd{pair vertq, verti, vertj, vertk, vertpi, vertpj, vertpk, vertpii, vertpjj, vertpkk; vertq = vloc(__vq); verti = vloc(__vi); vertj = vloc(__vj); vertk = vloc(__vk); vertpi = vloc(__vpi); vertpj = vloc(__vpj); vertpk = vloc(__vpk); vertpii = vloc(__vpi1); vertpjj = vloc(__vpj1); vertpkk = vloc(__vpk1); }
 \fmfiv{label=$\scriptstyle q$}{vertq}
 \fmfiv{label=$\scriptstyle p_{i-1}$}{vertpi}
 \fmfiv{label=$\scriptstyle p_i$}{vertpj}
 \fmfiv{label=$\scriptstyle p_{i+1}$}{vertpk}
 \fmfiv{label=$\scriptstyle p_{i+2}$}{vertpkk}
 \fmfiv{label=$\cdot$,l.d=0,l.a=0}{verti+(12*sind(243),12*cosd(243))}
 \fmfiv{label=$\cdot$,l.d=0,l.a=0}{verti+(12*sind(-67),12*cosd(-67))}
\end{fmfchar*}%
}}%
\!\!\!\!+\sum_{i}c_{\text{bubble}}^{i,i+1}
\quad \,\,
\settoheight{\eqoff}{$\times$}%
\setlength{\eqoff}{0.5\eqoff}%
\addtolength{\eqoff}{-12\unitlength}%
\raisebox{\eqoff}{%
\fmfframe(2,2)(6,2){%
\begin{fmfchar*}(20,20)
\fmfsurround{vpiii,vpi1,vpii,vpiiii,vpiiiii,vpiiiiii,vpi,vq,vpj1,vpjj,vpjjjj,vpjjjjj,vpjjjjjj,vpj}
\fmf{dbl_plain_arrow,tension=0}{vq,vi}
\fmf{plain_arrow,tension=1}{vi,vpi}
\fmf{plain_arrow,tension=1}{vi,vpj1}
\fmf{plain_arrow,tension=1}{vj,vpj}
\fmf{plain_arrow,tension=1}{vj,vpi1}
\fmf{plain_arrow,tension=0.5,right=0.66}{vi,vj}
\fmf{plain_arrow,tension=0.5,left=0.66}{vi,vj}
\fmffreeze
\fmfdraw
 \fmfcmd{pair vertq, verti, vertj,  vertpi, vertpj, vertpii, vertpjj; vertq = vloc(__vq); verti = vloc(__vi); vertj = vloc(__vj);  vertpi = vloc(__vpi); vertpj = vloc(__vpj);  vertpii = vloc(__vpi1); vertpjj = vloc(__vpj1);  }
 \fmfiv{label=$\scriptstyle q$}{vertq}
 \fmfiv{label=$\scriptstyle p_{i-1}$}{vertpi}
 \fmfiv{label=$\scriptstyle p_{i}$}{vertpii}
 \fmfiv{label=$\scriptstyle p_{i+1}$}{vertpj}
 \fmfiv{label=$\scriptstyle p_{i+2}$}{vertpjj}
 \fmfiv{label=$\cdot$,l.d=0,l.a=0}{verti+(12*sind(243),12*cosd(243))}
 \fmfiv{label=$\cdot$,l.d=0,l.a=0}{verti+(12*sind(-67),12*cosd(-67))}
\end{fmfchar*}%
}}%
\\ &\phaneq
+\text{rational terms}
\end{aligned}
\end{equation*}
\caption{The minimal one-loop form factor $\ffco^{(1)}_{\cO,L}$ of a generic single-trace operator $\cO$ can be written as a linear combination of one-mass triangle integrals, bubble integrals and rational terms. Here, the coefficients of the integrals are labelled by the two momenta that flow through the integrals and thus set their scale.}
\label{fig: simplified ansatz for generalised unitarity}
\end{figure}

\begin{figure}[tbp]
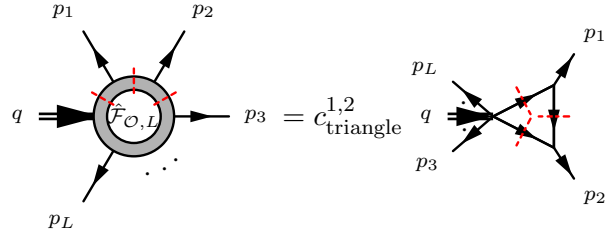

 \centering
\begin{equation*}
 \begin{aligned}
\settoheight{\eqoff}{$\times$}%
\setlength{\eqoff}{0.5\eqoff}%
\addtolength{\eqoff}{-14.5\unitlength}%
\raisebox{\eqoff}{%
\fmfframe(2,2)(6,2){%
\begin{fmfchar*}(25,25)
\fmfsurround{vp3,vp2,vp1,vq,vpL,vp}
\fmf{dbl_plain_arrow,tension=1}{vq,vqa}
\fmf{plain_arrow,tension=1}{vpLa,vpL}
\fmf{plain_arrow,tension=1}{vp3a,vp3}
\fmf{phantom,tension=1}{vpa,vp}
\fmf{plain_arrow}{vp1a,vp1}
\fmf{plain_arrow}{vp2a,vp2}
\fmf{dbl_plain_arrow,tension=2}{vqa,v1}
\fmf{plain_arrow,tension=2}{v1,vpLa}
\fmf{plain_arrow,tension=2}{v1,vp3a}
\fmf{phantom,tension=2}{v1,vpa}
\fmf{plain_arrow,tension=2}{v1,vp1a}
\fmf{plain_arrow,tension=2}{v1,vp2a}
\fmffreeze
\fmfdraw
 \fmfcmd{pair vertq, vertpone, vertptwo, vertpthree, vertpL, vertone, verttwo, vertp; vertone = vloc(__v1); verttwo = vloc(__v2); vertq = vloc(__vq); vertpone = vloc(__vp1); vertptwo = vloc(__vp2); vertpthree = vloc(__vp3);vertp = vloc(__vp);vertpL = vloc(__vpL);}
 \fmfiv{decor.shape=circle,decor.filled=30,decor.size=30}{vertone}
 \fmfiv{decor.shape=circle,decor.filled=0,decor.size=20}{vertone}
 \fmfiv{label=$\scriptstyle  \ffco_{\cO,,L}$,l.d=0,l.a=0}{vertone}
 \fmfdraw
 \fmfi{dashes,fore=red}{(vertone+(-18*sqrt(0.75),18*sqrt(0.25)))--(vertone+(-7*sqrt(0.75),7*sqrt(0.25)))}
 \fmfi{dashes,fore=red}{(vertone+(0,18))--(vertone+(0,7))}
 \fmfi{dashes,fore=red}{(vertone+(+18*sqrt(0.75),18*sqrt(0.25)))--(vertone+(+7*sqrt(0.75),7*sqrt(0.25)))}
 \fmfiv{label=$\scriptstyle q$}{vertq}
 \fmfiv{label=$\scriptstyle p_1$}{vertpone}
 \fmfiv{label=$\scriptstyle p_2$}{vertptwo}
 \fmfiv{label=$\scriptstyle p_3$}{vertpthree}
 \fmfiv{label=$\scriptstyle p_L$}{vertpL}
 \fmfiv{label=$\cdot$,l.d=20,l.a=-45}{vertone}
 \fmfiv{label=$\cdot$,l.d=20,l.a=-60}{vertone}
 \fmfiv{label=$\cdot$,l.d=20,l.a=-75}{vertone}
\end{fmfchar*}%
}}%
&=
c_{\text{triangle}}^{1,2}
\quad
\settoheight{\eqoff}{$\times$}%
\setlength{\eqoff}{0.5\eqoff}%
\addtolength{\eqoff}{-12\unitlength}%
\raisebox{\eqoff}{%
\fmfframe(2,2)(6,2){%
\begin{fmfchar*}(20,20)
\fmfsurround{vpjj,vpiii,vpj,vpi1,vpii,vpiiii,vpi,vq,vpk1,vpkk,vpkkkk,vpjjj,vpk,vpj1}
\fmf{dbl_plain_arrow,tension=0}{vq,vi}
\fmf{plain_arrow,tension=1}{vi,vpi}
\fmf{plain_arrow,tension=1}{vi,vpk1}
 \fmf{plain_arrow,tension=2}{vj,vpj}
 \fmf{plain_arrow,tension=2}{vk,vpk}
\fmf{plain_arrow,tension=0.66}{vi,vj}
\fmf{plain_arrow,tension=0.66}{vi,vk}
\fmf{plain_arrow,tension=0.66}{vj,vk}
\fmf{phantom_smallcut,tension=0}{vi,vj}
\fmf{phantom_smallcut,tension=0}{vi,vk}
\fmf{phantom_smallcut,tension=0}{vj,vk}
\fmffreeze
\fmfdraw
 \fmfcmd{pair vertq, verti, vertj, vertk, vertpi, vertpj, vertpk, vertpii, vertpjj, vertpkk; vertq = vloc(__vq); verti = vloc(__vi); vertj = vloc(__vj); vertk = vloc(__vk); vertpi = vloc(__vpi); vertpj = vloc(__vpj); vertpk = vloc(__vpk); vertpii = vloc(__vpi1); vertpjj = vloc(__vpj1); vertpkk = vloc(__vpk1); }
 \fmfiv{label=$\scriptstyle q$}{vertq}
 \fmfiv{label=$\scriptstyle p_{L}$}{vertpi}
 \fmfiv{label=$\scriptstyle p_1$}{vertpj}
 \fmfiv{label=$\scriptstyle p_{2}$}{vertpk}
 \fmfiv{label=$\scriptstyle p_{3}$}{vertpkk}
 \fmfiv{label=$\cdot$,l.d=0,l.a=0}{verti+(12*sind(243),12*cosd(243))}
 \fmfiv{label=$\cdot$,l.d=0,l.a=0}{verti+(12*sind(-67),12*cosd(-67))}
\end{fmfchar*}%
}}%
\end{aligned}
\end{equation*}
\caption{Three-particle cut of the ansatz for the minimal one-loop form factor $\ffco^{(1)}_{\cO,L}$ of a generic single-trace operator $\cO$. Taken between $p_1$, $p_2$ and the rest of the diagram, this cut isolates the triangle integral with external on-shell legs $p_1$ and $p_2$ and its  coefficient $c_{\text{triangle}}^{1,2}$.}
\label{fig: triple cut of simplified ansatz for generalised unitarity}
\end{figure}

\begin{figure}[tbp]
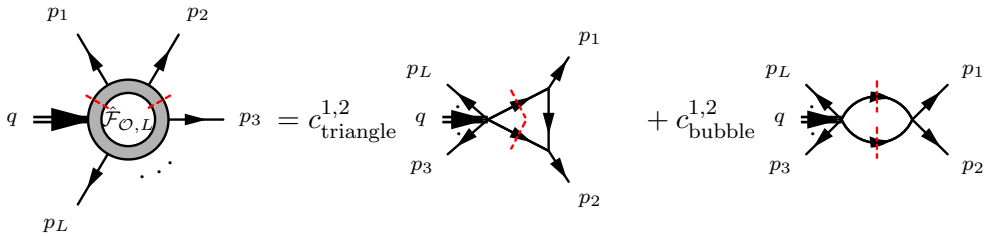

 \centering
\begin{equation*}
 \begin{aligned}
\settoheight{\eqoff}{$\times$}%
\setlength{\eqoff}{0.5\eqoff}%
\addtolength{\eqoff}{-14.5\unitlength}%
\raisebox{\eqoff}{%
\fmfframe(2,2)(6,2){%
\begin{fmfchar*}(25,25)
\fmfsurround{vp3,vp2,vp1,vq,vpL,vp}
\fmf{dbl_plain_arrow,tension=1}{vq,vqa}
\fmf{plain_arrow,tension=1}{vpLa,vpL}
\fmf{plain_arrow,tension=1}{vp3a,vp3}
\fmf{phantom,tension=1}{vpa,vp}
\fmf{plain_arrow}{vp1a,vp1}
\fmf{plain_arrow}{vp2a,vp2}
\fmf{dbl_plain_arrow,tension=2}{vqa,v1}
\fmf{plain_arrow,tension=2}{v1,vpLa}
\fmf{plain_arrow,tension=2}{v1,vp3a}
\fmf{phantom,tension=2}{v1,vpa}
\fmf{plain_arrow,tension=2}{v1,vp1a}
\fmf{plain_arrow,tension=2}{v1,vp2a}
\fmffreeze
\fmfdraw
 \fmfcmd{pair vertq, vertpone, vertptwo, vertpthree, vertpL, vertone, verttwo, vertp; vertone = vloc(__v1); verttwo = vloc(__v2); vertq = vloc(__vq); vertpone = vloc(__vp1); vertptwo = vloc(__vp2); vertpthree = vloc(__vp3);vertp = vloc(__vp);vertpL = vloc(__vpL);}
 \fmfiv{decor.shape=circle,decor.filled=30,decor.size=30}{vertone}
 \fmfiv{decor.shape=circle,decor.filled=0,decor.size=20}{vertone}
 \fmfiv{label=$\scriptstyle  \ffco_{\cO,,L}$,l.d=0,l.a=0}{vertone}
 \fmfdraw
 \fmfi{dashes,fore=red}{(vertone+(-18*sqrt(0.75),18*sqrt(0.25)))--(vertone+(-7*sqrt(0.75),7*sqrt(0.25)))}
 \fmfi{dashes,fore=red}{(vertone+(+18*sqrt(0.75),18*sqrt(0.25)))--(vertone+(+7*sqrt(0.75),7*sqrt(0.25)))}
 \fmfiv{label=$\scriptstyle q$}{vertq}
 \fmfiv{label=$\scriptstyle p_1$}{vertpone}
 \fmfiv{label=$\scriptstyle p_2$}{vertptwo}
 \fmfiv{label=$\scriptstyle p_3$}{vertpthree}
 \fmfiv{label=$\scriptstyle p_L$}{vertpL}
 \fmfiv{label=$\cdot$,l.d=20,l.a=-45}{vertone}
 \fmfiv{label=$\cdot$,l.d=20,l.a=-60}{vertone}
 \fmfiv{label=$\cdot$,l.d=20,l.a=-75}{vertone}
\end{fmfchar*}%
}}%
&=
c_{\text{triangle}}^{1,2}
\quad
\settoheight{\eqoff}{$\times$}%
\setlength{\eqoff}{0.5\eqoff}%
\addtolength{\eqoff}{-12\unitlength}%
\raisebox{\eqoff}{%
\fmfframe(2,2)(6,2){%
\begin{fmfchar*}(20,20)
\fmfsurround{vpjj,vpiii,vpj,vpi1,vpii,vpiiii,vpi,vq,vpk1,vpkk,vpkkkk,vpjjj,vpk,vpj1}
\fmf{dbl_plain_arrow,tension=0}{vq,vi}
\fmf{plain_arrow,tension=1}{vi,vpi}
\fmf{plain_arrow,tension=1}{vi,vpk1}
 \fmf{plain_arrow,tension=2}{vj,vpj}
 \fmf{plain_arrow,tension=2}{vk,vpk}
\fmf{plain_arrow,tension=0.66}{vi,vj}
\fmf{plain_arrow,tension=0.66}{vi,vk}
\fmf{plain_arrow,tension=0.66}{vj,vk}
\fmf{phantom_smallcut,tension=0}{vi,vj}
\fmf{phantom_smallcut,tension=0}{vi,vk}
\fmffreeze
\fmfdraw
 \fmfcmd{pair vertq, verti, vertj, vertk, vertpi, vertpj, vertpk, vertpii, vertpjj, vertpkk; vertq = vloc(__vq); verti = vloc(__vi); vertj = vloc(__vj); vertk = vloc(__vk); vertpi = vloc(__vpi); vertpj = vloc(__vpj); vertpk = vloc(__vpk); vertpii = vloc(__vpi1); vertpjj = vloc(__vpj1); vertpkk = vloc(__vpk1); }
 \fmfiv{label=$\scriptstyle q$}{vertq}
 \fmfiv{label=$\scriptstyle p_{L}$}{vertpi}
 \fmfiv{label=$\scriptstyle p_1$}{vertpj}
 \fmfiv{label=$\scriptstyle p_{2}$}{vertpk}
 \fmfiv{label=$\scriptstyle p_{3}$}{vertpkk}
 \fmfiv{label=$\cdot$,l.d=0,l.a=0}{verti+(12*sind(243),12*cosd(243))}
 \fmfiv{label=$\cdot$,l.d=0,l.a=0}{verti+(12*sind(-67),12*cosd(-67))}
\end{fmfchar*}%
}}%
+
c_{\text{bubble}}^{1,2}
\quad
\settoheight{\eqoff}{$\times$}%
\setlength{\eqoff}{0.5\eqoff}%
\addtolength{\eqoff}{-12\unitlength}%
\raisebox{\eqoff}{%
\fmfframe(2,2)(6,2){%
\begin{fmfchar*}(20,20)
\fmfsurround{vpiii,vpi1,vpii,vpiiii,vpiiiii,vpiiiiii,vpi,vq,vpj1,vpjj,vpjjjj,vpjjjjj,vpjjjjjj,vpj}
\fmf{dbl_plain_arrow,tension=0}{vq,vi}
\fmf{plain_arrow,tension=1}{vi,vpi}
\fmf{plain_arrow,tension=1}{vi,vpj1}
\fmf{plain_arrow,tension=1}{vj,vpj}
\fmf{plain_arrow,tension=1}{vj,vpi1}
\fmf{plain_arrow,tension=0.5,right=0.66}{vi,vj}
\fmf{plain_arrow,tension=0.5,left=0.66}{vi,vj}
\fmf{phantom_smallcut,tension=0,right=0.66}{vi,vj}
\fmf{phantom_smallcut,tension=0,left=0.66}{vi,vj}
\fmffreeze
\fmfdraw
 \fmfcmd{pair vertq, verti, vertj,  vertpi, vertpj, vertpii, vertpjj; vertq = vloc(__vq); verti = vloc(__vi); vertj = vloc(__vj);  vertpi = vloc(__vpi); vertpj = vloc(__vpj);  vertpii = vloc(__vpi1); vertpjj = vloc(__vpj1);  }
 \fmfiv{label=$\scriptstyle q$}{vertq}
 \fmfiv{label=$\scriptstyle p_{L}$}{vertpi}
 \fmfiv{label=$\scriptstyle p_{1}$}{vertpii}
 \fmfiv{label=$\scriptstyle p_{2}$}{vertpj}
 \fmfiv{label=$\scriptstyle p_{3}$}{vertpjj}
 \fmfiv{label=$\cdot$,l.d=0,l.a=0}{verti+(12*sind(243),12*cosd(243))}
 \fmfiv{label=$\cdot$,l.d=0,l.a=0}{verti+(12*sind(-67),12*cosd(-67))}
\end{fmfchar*}%
}}%
\end{aligned}
\end{equation*}
\caption{Two-particle cut of the ansatz for the minimal one-loop form factor $\ffco^{(1)}_{\cO,L}$ of a generic single-trace operator $\cO$. Taken between $p_1$, $p_2$ and the rest of the diagram, this cut isolates the triangle integral and the bubble integral with external on-shell legs $p_1$ and $p_2$ and their respective coefficients  $c_{\text{triangle}}^{1,2}$ and $c_{\text{bubble}}^{1,2}$.}
\label{fig: double cut of simplified ansatz for generalised unitarity}
\end{figure}

As the box integral is absent, the maximal possible cut of the minimal one-loop form factor is the triple cut. As shown in figure \ref{fig: triple cut of simplified ansatz for generalised unitarity}, it isolates the triangle integral and hence fixes the triangle coefficient.
We explicitly compute the triangle coefficient from the triple cut in the next subsection.
Next, we take the double cut. It has contributions from the triangle integral and the bubble integral, as shown in figure \ref{fig: double cut of simplified ansatz for generalised unitarity}. Knowing the triangle coefficient, we can calculate the bubble coefficient from this cut. We perform this calculation in subsection \ref{subsec: double cut and bubble coefficient}. 
In subsection \ref{subsec result and one-loop dilatation operator}, we summarise our result and extract the complete one-loop dilatation operator from it. Moreover, we give a short discussion of the rational terms.

\subsection{Triple cut and triangle coefficient}
\label{subsec: triple cut and triangle coefficient}

\begin{figure}[h]
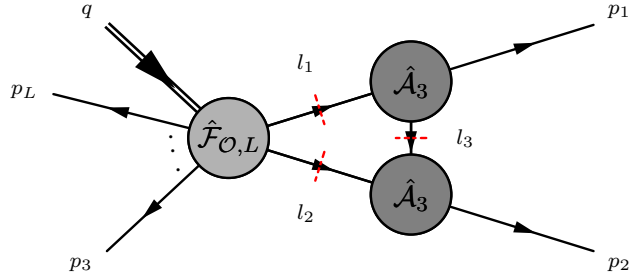

 \centering
$
\settoheight{\eqoff}{$\times$}%
\setlength{\eqoff}{0.5\eqoff}%
\addtolength{\eqoff}{-12.0\unitlength}%
\raisebox{\eqoff}{%
\fmfframe(2,2)(2,2){%
\begin{fmfchar*}(80,30)
\fmfleft{vp3,vp,vpL,vq}
\fmfright{vp2,vp1}
\fmf{dbl_plain_arrow,tension=1.5}{vq,v1}
\fmf{plain_arrow,tension=0}{v1,vpL}
\fmf{plain_arrow,tension=1.5}{v1,vp3}
\fmf{plain_arrow,left=0,l.s=left,label=$\scriptstyle l_1\,,$,l.d=15}{v1,v2}
\fmf{plain_arrow,right=0,l.s=right,label=$\scriptstyle l_2\,,$,l.d=15}{v1,v3}
\fmf{plain_arrow,right=0,tension=0,label=$\scriptstyle l_3\,,$,l.d=-25}{v2,v3}
\fmf{phantom_smallcut,tension=0}{v1,v2}
\fmf{phantom_smallcut,tension=0}{v1,v3}
\fmf{phantom_smallcut,tension=0}{v2,v3}
\fmf{plain_arrow}{v2,vp1}
\fmf{plain_arrow}{v3,vp2}
\fmfv{decor.shape=circle,decor.filled=30,decor.size=30,label=$\ffco_{\cO,,L}$,label.dist=0}{v1}
\fmfv{decor.shape=circle,decor.filled=50,decor.size=30,label=$\ampco_{3}$,label.dist=0}{v2}
\fmfv{decor.shape=circle,decor.filled=50,decor.size=30,label=$\ampco_{3}$,label.dist=0}{v3}
\fmffreeze
\fmfdraw
 \fmfcmd{pair vertq, vertpone, vertptwo, vertpthree, vertpL, vertone, verttwo; vertone = vloc(__v1); verttwo = vloc(__v2); vertq = vloc(__vq); vertpone = vloc(__vp1); vertptwo = vloc(__vp2); vertpthree = vloc(__vp3);vertpL = vloc(__vpL);}
 \fmfiv{label=$\scriptstyle q$}{vertq}
 \fmfiv{label=$\scriptstyle p_1$}{vertpone}
 \fmfiv{label=$\scriptstyle p_2$}{vertptwo}
 \fmfiv{label=$\scriptstyle p_3$}{vertpthree}
 \fmfiv{label=$\scriptstyle p_L$}{vertpL}
 \fmfiv{label=$\cdot$,l.d=20,l.a=-150}{vertone}
 \fmfiv{label=$\cdot$,l.d=20,l.a=-165}{vertone}
 \fmfiv{label=$\cdot$,l.d=20,l.a=-180}{vertone}
\end{fmfchar*}%
}}%
$
\caption{On the three-particle cut, the minimal one-loop form factor $\ffco^{(1)}_{\cO,L}$ shown on the left hand side of figure \ref{fig: triple cut of simplified ansatz for generalised unitarity} factorises into the product of the minimal tree-level form factor $\ffco^{(0)}_{\cO,L}$ and two three-point tree-level amplitudes $\ampco^{(0)}_{3}$.
}
\label{fig: triple cut}
\end{figure}

In this subsection, we calculate the triangle coefficient of the minimal one-loop form factor shown in figure \ref{fig: simplified ansatz for generalised unitarity}.
To this end, we study the triple cut between two neighbouring external fields $i$ and $i+1$ and the rest of the diagram. We set $i=1$ to simplify the notation.
As shown in figure \ref{fig: triple cut of simplified ansatz for generalised unitarity}, this cut isolates the one-mass triangle integral with external momenta $p_1$ and $p_2$, which is multiplied by the coefficient $c_{\text{triangle}}^{1,2}$. On the cut, the minimal one-loop form factor $\ffco_{\cO,L}^{(1)}$ on the left hand side of figure \ref{fig: triple cut of simplified ansatz for generalised unitarity} factorises into the product of the minimal tree-level form factor $\ffco_{\cO,L}^{(0)}$ and two three-point tree-level amplitudes $\ampco_{3}^{(0)}$, as shown in figure \ref{fig: triple cut}.
The corresponding four-dimensional phase-space integral is 
\begin{equation}
 \label{eq: triple cut phase space integral}
 \frac{1}{(2\pi)^9}\int \de\Lambda_{l_1}\de\Lambda_{l_2}\de\Lambda_{l_3} \ffco_{\cO,L}^{(0)}(\Lambda_{l_1},\Lambda_{l_2},\Lambda_3,\dots,\Lambda_L;q)\ampco_3^{(0)}(\Lambda_1,\Lambda_{l_3},\Lambda_{-l_1}) 
 \ampco_3^{(0)}(\Lambda_{-l_3},\Lambda_{2},\Lambda_{-l_2}) \eqndot
\end{equation}

The triple cut imposes the following three constraints:
\begin{equation}
\begin{aligned}
  l_1^2&=l^2=0 \eqncom\\
  l_2^2&=(p_1+p_2+l)^2=l^2+(p_1+p_2)^2+2(p_1+p_2)\cdot l=0\eqncom\\
  l_3^2&=(p_1+l)^2=l^2+2l\cdot p_1=0 \eqncom
\end{aligned}
\end{equation}
where the loop momentum $l$ is chosen as $-l_1$. 
Since $l$ has four components, one might expect a one-parameter real solution.
Instead, the real solution for $l_1$ and $l_2$ is unique:
\begin{equation}\label{eq: condition}
 l_1=p_1\eqncom \quad l_2=p_2 \eqndot
\end{equation}
This is a consequence of $p_1^2=p_2^2=0$, cf.\ for example \cite{ArkaniHamed:2008gz}.
In terms of spinor-helicity variables, \eqref{eq: condition} reads
\begin{equation}\label{eq: condition in spinors}
 \begin{aligned}
  \lambda_{l_1}^\alpha&=\e^{i\phi_1} \lambda_{1}^\alpha\eqncom & \lambda_{l_2}^\alpha&=\e^{i\phi_2} \lambda_{2}^\alpha\eqncom \\
  \lambdat_{l_1}^{\alphadot}&=\e^{-i\phi_1} \lambdat_{1}^{\alphadot}\eqncom & \lambdat_{l_2}^{\alphadot}&=\e^{-i\phi_2} \lambdat_{2}^{\alphadot}\eqndot
 \end{aligned}
\end{equation}
where $\phi_1$ and $\phi_2$ parametrise the $\U1$ freedom in defining a pair of spinor-helicity variables that corresponds to a given vector.

Up to now, we have neglected the momentum of the third cut propagator $l_3$. Together with momentum conservation at the three-point amplitudes, the on-shell conditions for $p_1$ and $p_2$ impose the constraints
\begin{equation}
\begin{aligned}
  p_1^2=(l_1-l_3)^2=0 \eqncom\qquad
  p_2^2=(l_2+l_3)^2=0 \eqncom
\end{aligned}
\end{equation}
which, in terms of spinor-helicity variables, read
\begin{equation}
\label{eq: on-shell l3 in spinors}
\begin{aligned}
  \ab{l_1 l_3}\sb{l_1 l_3}=0 \eqncom\qquad
  \ab{l_2 l_3}\sb{l_2 l_3}=0 \eqndot
\end{aligned}
\end{equation}
For real momenta and Minkowski signature, the angle and square brackets in \eqref{eq: on-shell l3 in spinors} are negative conjugates of each other.%
\footnote{Recall that $l_1$, $l_2$ and $l_3$ have positive energy due to the Heaviside step function in \eqref{eq: cut replacement}.}
Hence, we have to allow complex momenta to obtain a non-trivial solution, as is usual for massless three-particle kinematics. We then consider the limit of the complex solutions where the momenta become real and \eqref{eq: condition}, \eqref{eq: condition in spinors} are satisfied.
For complex momenta, the constraints \eqref{eq: on-shell l3 in spinors} imply
$\lambda_{l_3} \propto \lambda_{l_1}$ or $\lambdat_{l_3} \propto \lambdat_{l_1}$ and
$\lambda_{l_3} \propto \lambda_{l_2}$ or $\lambdat_{l_3} \propto \lambdat_{l_2}$, respectively.
Choosing $\lambda_{l_1}\propto\lambda_{l_2}\propto\lambda_{l_3}$ would imply $\lambda_{p_1}\propto\lambda_{p_2}$, which is incompatible with having generic external momenta $p_1$ and $p_2$. Analogously, not all $\lambdat_{l_i}$ can be proportional.
This leaves us with two possibilities: either 
\begin{align}
\text{(i) } \qquad &\lambdat_{l_3} \propto \lambdat_{l_1} \quad\text{ and }\quad\lambda_{l_3} \propto \lambda_{l_2}
\intertext{or}
\text{(ii)} \qquad &\lambda_{l_3} \propto \lambda_{l_1} \quad\text{ and }\quad\lambdat_{l_3} \propto \lambdat_{l_2}\eqndot
\end{align}
The resulting contributions for both 
 solutions 
 should be averaged over, i.e.\ summed with a prefactor of $\frac{1}{2}$, cf.\ \cite{Britto:2004nc,Kosower:2011ty}.%
\footnote{The above procedure has an equivalent description in terms of a sequence of four generalised cuts as follows. 
On the double cut in $l_1$ and $l_2$, the squared momentum in the third propagator factors to $l_3^2=(l_1-p_1)^2=\abr{l_1p_1}\sbr{l_1p_1}$. 
Hence, further generalised cuts can be taken in $\abr{l_1p_1}$ and $\sbr{l_1p_1}$ individually.
}

The three-point amplitudes in the three-particle cut can be either of MHV type or of \MHVb type, and we have to sum both contributions.
However, only the combination of an upper MHV amplitude with a lower \MHVb amplitude is non-vanishing on the support of solution (i) while only the opposite combination is non-vanishing  on the support of solution (ii). 
Let us look at solution (i) first.

The product of the two colour-ordered tree-level three-point amplitudes is given by 
\begin{equation}
 \begin{aligned}
 &\ampco_3^{\MHV\,(0)}(\Lambda_1,\Lambda_{l_3},\Lambda_{-l_1}) 
 \ampco_3^{\MHVb\,(0)}(\Lambda_{-l_3},\Lambda_{2},\Lambda_{-l_2}) \\
 &=\frac{i(2\pi)^4\delta^4(p_1+l_3-l_1)}{\ab{1 l_3}\ab{l_3 l_1}\ab{l_1 1}}\prod_{A=1}^4(\ab{1l_3}\etatt_1^A\etatt_{l_3}^A-\ab{l_1l_3}\etatt_{l_1}^A\etatt_{l_3}^A-\ab{1l_1}\etatt_1^A\etatt_{l_1}^A)\\
 &\phaneq
 \frac{(-i)(2\pi)^4\delta^4(p_2-l_2-l_3)}{\sb{ 2l_2}\sb{l_2 l_3}\sb{l_3 2}}\prod_{A=1}^4(\sb{l_2l_3}\etatt_2^A+\sb{2l_2}\etatt_{l_3}^A+\sb{l_3 2}\etatt_{l_2}^A)
\\
&=(2\pi)^8\delta^4(p_1+l_3-l_1)\delta^4(p_2-l_2-l_3)\frac{\ab{12}\e^{2i(\phi_1+\phi_2)}}{\sb{12}^3\ab{1l_1}^4}\\
&\phaneq\prod_{A=1}^4 \Big(\ab{1l_3}(\etatt_1^A-\e^{i\phi_1}\etatt_{l_1}^A)\etatt_{l_3}^A-\ab{1l_1}\etatt_1^A\etatt_{l_1}^A\Big)
\prod_{A=1}^4 \Big(\sb{2l_3}(\e^{-i\phi_2}\etatt_2^A-\etatt_{l_2}^A)+\sb{2l_2}\etatt_{l_3}^A\Big) \eqncom
\label{eq: product of three-point amplitudes}
 \end{aligned}
\end{equation}
where we have used 
the identities
\begin{equation}\label{eq: identities}
 \begin{aligned}
 \ab{1l_3}\sb{l_3 2}&= \asb{1}{l_3}{2}= \asb{1}{l_1}{2}=\ab{1 l_1}\sb{l_12}=\ab{1l_1}\sb{12}\e^{-i\phi_1} \eqncom\\
 \ab{l_3l_1}\sb{l_2l_3}&=\sab{l_2}{l_3}{l_1}=-\sab{l_2}{p_1}{l_1}=-\ab{1l_1}\sb{l_21}=\ab{1l_1}\sb{12}\e^{-i\phi_2}\eqncom\\
 \ab{12}\sb{2l_2}&=\asb{1}{p_2}{l_2}=\asb{1}{l_1}{l_2}=\ab{1l_1}\sb{l_1l_2}=\ab{1l_1}\sb{12}\e^{-i(\phi_1+\phi_2)}\eqncom
\end{aligned}
\end{equation}
which follow from momentum conservation.
Moreover, we have dropped terms that are subleading in the limit where the momenta become real and satisfy \eqref{eq: condition in spinors}.
Integrating out the fermionic $\etatt_{l_3}^A$ variables, we find
\begin{multline}\label{eq: intermediate step in triple cut}
(2\pi)^8\delta^4(p_1+l_3-l_1)\delta^4(p_2-l_2-l_3)\frac{\ab{12}\e^{2i(\phi_1+\phi_2)}}{\sb{12}^3\ab{1l_1}^4}\\
\prod_{A=1}^4 \Big(\ab{1l_3}(\etatt_1^A-\e^{i\phi_1}\etatt_{l_1}^A)\sb{2l_3}(\e^{-i\phi_2}\etatt_2^A-\etatt_{l_2}^A)+\ab{1l_1}\etatt_1^A\etatt_{l_1}^A\sb{2l_2}\Big)
\eqndot
 \end{multline}
Using \eqref{eq: identities}, we can cancel all angle and square brackets in the denominator such that the expression is manifestly finite in the limit where the momenta are real and satisfy \eqref{eq: condition in spinors}. 
Furthermore, the second term in the parenthesis in \eqref{eq: intermediate step in triple cut} is found to vanish in this limit, 
such that we have  
\begin{multline}
-(p_1+p_2)^2 (2\pi)^8 \delta^4(p_1+l_3-l_1)\delta^4(p_2-l_2-l_3) \e^{2i(\phi_1+\phi_2)} \\
\prod_{A=1}^4\!\Big((\e^{-i\phi_1}\etatt_1^A-\etatt_{l_1}^A)(\e^{-i\phi_2}\etatt_2^A-\etatt_{l_2}^A)\Big)\eqndot
\end{multline}
 The subsequent integration over the fermionic variables $\etatt_{l_1}^A$ and $\etatt_{l_2}^A$ replaces
 \begin{equation}
 \etatt_{l_1}^A\to\e^{-i\phi_1}\etatt_1^A\eqncom \quad \etatt_{l_2}^A\to \e^{-i\phi_2} \etatt_2^A 
\end{equation}
in the tree-level form factor, while the phase-space integral leads to similar replacements in the bosonic variables via \eqref{eq: condition in spinors}. The resulting total phase factor is $\e^{2 i\phi_1 C_{l_1}}\e^{2 i\phi_2 C_{l_2}}$, which equals unity as the central charges $C_{l_1}$ and $C_{l_2}$ corresponding to the tree-level form factor vanish.
Thus, the phase-space integral \eqref{eq: triple cut phase space integral} evaluates to
\begin{equation}
 -\frac{1}{2\pi}(p_1+p_2)^2 \ffco^{(0)}_{\cO,L}(\Lambda_1,\Lambda_2,\Lambda_3,\dots,\Lambda_L;q)
\end{equation}
on the support of solution (i). 
A completely analogous calculation yields the same result for solution (ii), which cancels the prefactor of $\frac{1}{2}$.%
\footnote{In fact, we could hence also have taken only one of the two solution. The crucial point is that the same procedure is applied to both sides of the ansatz in figure \ref{fig: simplified ansatz for generalised unitarity}.
}
In comparison, the phase-space integral of the triple-cut triangle integral yields%
\footnote{Note that we are considering the cut triangle integral based on the measure factor $\frac{1}{i(2\pi)^4}$ here, which is a mixture of both sides of \eqref{eq: integral-measure-relation}. This is a consequence of the fact that $\gmod^2$ is factored out in the ansatz for $\ffco_{\cO,L}^{(1)}$ in figure \ref{fig: simplified ansatz for generalised unitarity}.}
\begin{equation}\label{eq: triple cut of triangle}
\begin{aligned}
\FDinline[triangle,triplecut,cutlabels,twolabels,labelone=\scriptscriptstyle p_1,labeltwo=\scriptscriptstyle p_2] 
&=\frac{1}{2\pi} \eqndot
\end{aligned}
\end{equation}
Thus, we find
\begin{equation}
 c^{1,2}_{\text{triangle}}=-(p_1+p_2)^2 \ffco^{(0)}_{\cO,L}(\Lambda_1,\Lambda_2,\Lambda_3,\dots,\Lambda_L;q)\eqndot
\end{equation}

\subsection{Double cut and bubble coefficient}
\label{subsec: double cut and bubble coefficient}

In this subsection, we calculate the bubble coefficient of the minimal one-loop form factor shown in figure \ref{fig: simplified ansatz for generalised unitarity}. 
We study the double cut between the two external fields $1$ and $2$ and the rest of the diagram. 
As shown in figure \ref{fig: double cut of simplified ansatz for generalised unitarity}, this cut is the sum of two contributions. 
The first contribution is the double-cut one-mass triangle integral with external momenta $p_1$ and $p_2$ multiplied by the triangle coefficient $c_{\text{triangle}}^{1,2}$:%
\footnote{In this section, as in the previous subsection, we are considering the cut triangle integral and the cut bubble integral based on the measure factor $\frac{1}{i(2\pi)^4}$.}
\begin{equation}
\begin{aligned}
\label{eq: double cut triangle}
 &c^{1,2}_{\text{triangle}} \FDinline[triangle,doublecut,cutlabels,twolabels, labelone=\scriptscriptstyle p_1,labeltwo=\scriptscriptstyle p_2]\\
 &= - \frac{i}{(2\pi)^2}\int \de\Lambda_{l_1}\de\Lambda_{l_2}\delta^4(p_1+p_2-l_1-l_2)\frac{(p_1+p_2)^2}{(l_1-p_1)^2}\ffco^{(0)}_{\cO,L}(\Lambda_1,\Lambda_2,\Lambda_3,\dots,\Lambda_L;q)\eqndot
 \end{aligned}
\end{equation}
The second contribution is the double-cut bubble integral with external momenta $p_1$ and $p_2$ multiplied by the bubble coefficient $c_{\text{bubble}}^{1,2}$:
\begin{equation}
\label{eq: double cut bubble}
 c^{1,2}_{\text{bubble}} \FDinline[bubble,doublecut,cutlabels,twolabels, labelone=\scriptscriptstyle p_1,labeltwo=\scriptscriptstyle p_2]
 = c^{1,2}_{\text{bubble}}  \frac{i}{(2\pi)^2}\int \de\Lambda_{l_1}\de\Lambda_{l_2} \delta^4(p_1+p_2-l_1-l_2)\eqndot
\end{equation}
On the cut, the minimal one-loop form factor $\ffco^{(1)}_{\cO,L}$ on the right side of figure \ref{fig: double cut of simplified ansatz for generalised unitarity} factorises into the product of the minimal tree-level form factor $\ffco^{(0)}_{\cO,L}$ and the four-point tree-level amplitude $\ampco^{(0)}_{4}$, as shown in figure \ref{fig: one-loop double cut}. This yields
\begin{equation}\label{eq: minimal form factor times four point tree amplitude}
 \begin{aligned}
  \frac{1}{(2\pi)^6}\int \de \Lambda_{l_1}\de \Lambda_{l_2}  \ffco_{\cO,L}^{(0)}(\Lambda_{l_1},\Lambda_{l_2},\Lambda_{3},\dots,\Lambda_{L};q) \ampco_4^{(0)}(\Lambda_{-l_2},\Lambda_{-l_1},\Lambda_{1},\Lambda_{2})\eqncom
 \end{aligned}
\end{equation}
cf.\ \eqref{eq: one-loop double cut}.

To obtain the bubble coefficient $c_{\text{bubble}}^{1,2}$, we have to evaluate the four-dimensional phase-space integrals in \eqref{eq: double cut triangle}, \eqref{eq: double cut bubble} and \eqref{eq: minimal form factor times four point tree amplitude}. 
This can be achieved via an explicit parametrisation. 
We use the parametrisation of \cite{Zwiebel:2011bx} in order to make contact with the observation on the connection between the four-point amplitude and the one-loop dilatation operator presented in that paper:
\begin{equation}\label{eq: parametrisation}
\begin{aligned}
 \left( \begin{array}{c}
\lambda_{l_1}^1  \\
\lambda_{l_2}^1    \end{array} \right)
&=r_1 \e^{i\sigma_1} U  \left( \begin{array}{c}
\lambda_{1}^1  \\
\lambda_{2}^1    \end{array} \right)\eqncom
\qquad
 \left( \begin{array}{c}
\lambda_{l_1}^2  \\
\lambda_{l_2}^2    \end{array} \right)
&=r_2  \,U\,  V(\sigma_2) \left( \begin{array}{c}
\lambda_{1}^2  \\
\lambda_{2}^2    \end{array} \right) \eqncom
\end{aligned} 
\end{equation}
where
\begin{equation}
U=\text{diag}(\e^{i\phi_2},\e^{i\phi_3})\,V(\theta)\,\text{diag}(1,\e^{i\phi_1})\eqncom\quad 
V(\theta)=
\left( \begin{array}{cc}
\cos\theta & -\sin\theta \\
\sin\theta & \cos\theta \end{array} \right) \eqndot
\end{equation}
The spinors $\lambdat_{l_1}^{\DOT1}$, $\lambdat_{l_1}^{\dot2}$, $\lambdat_{l_2}^{\DOT1}$ and $\lambdat_{l_2}^{\DOT2}$ are obtained from \eqref{eq: parametrisation} by complex conjugation. The parameters are $r_1,r_2\in(0,\infty)$, $\theta,\sigma_2\in(0,\frac{\pi}{2})$ and $\sigma_1,\phi_1,\phi_2,\phi_3\in(0,2\pi)$. 
Different values of the phases $\phi_2$ and $\phi_3$ yield the same momenta $l_1$ and $l_2$, and hence we do not need to integrate over them. Below, we will explicitly show that the phase-space integrals are independent of $\phi_2$ and $\phi_3$, as required by consistency.%
\footnote{In fact, the phases $\phi_2$ and $\phi_3$ exactly parametrise the $\U{1}$ in \eqref{eq: dLambda} for $i=1,2$.}

In \cite{Zwiebel:2011bx}, the parametrisation \eqref{eq: parametrisation} was used to obtain a compact expression for \eqref{eq: minimal form factor times four point tree amplitude}.
We briefly review the corresponding calculation below.

Under the change of variables, the momentum-conserving delta function transforms as
\begin{equation}\label{eq: delta function change}
\begin{aligned}
 \delta^4(P)&=\delta^4(p_1+p_2-l_1-l_2)=\prod_{\alpha=1}^2\prod_{\alphadot=\DOT1}^{\DOT2}\delta(\lambda_1^\alpha\lambdat_1^{\alphadot}+\lambda_2^\alpha\lambdat_2^{\alphadot}-\lambda_{l_1}^\alpha\lambdat_{l_1}^{\alphadot}-\lambda_{l_2}^\alpha\lambdat_{l_2}^{\alphadot})\\
 &=\frac{i \delta(1-r_1)\delta(1-r_2)\delta(\sigma_1)\delta(\sigma_2)}{4(\lambda_1^1\lambdat_1^{\DOT1}+\lambda_2^1\lambdat_2^{\DOT1})(\lambda_1^2\lambdat_1^{\DOT2}+\lambda_2^2\lambdat_2^{\DOT2})\left(
 \ab{12}(\lambdat_1^{\DOT1}\lambdat_1^{\DOT2}+\lambdat_2^{\DOT1}\lambdat_2^{\DOT2})
 -\sb{12}(\lambda_1^1\lambda_1^{2}+\lambda_2^1\lambda_2^{2})\right)} \eqndot
\end{aligned}
\end{equation}
Thus, momentum conservation localises the integrals over $r_1$, $r_2$ and $\sigma_1$, $\sigma_2$ at $1$ and $0$, respectively.
The Jacobian corresponding to the change of variables equals $2 \cos\theta \sin\theta$ times the denominator of the second line in \eqref{eq: delta function change} when evaluated at these values.
Hence, 
\begin{equation}
\label{eq: transformation of delta times Jacobian}
 \frac{\de^2\lambda_{l_1}\de^2\lambdat_{l_1}}{\U1}\frac{\de^2\lambda_{l_2}\de^2\lambdat_{l_2}}{\U1} \, \delta^4(P) \to \de\phi_1
\de\theta \, 2i\cos\theta\sin\theta\eqndot
\end{equation}
The MHV denominator of the four-point amplitude can then be simplified to 
\begin{equation}
 \ab{12}\ab{2l_2}\ab{l_2 l_1}\ab{l_1 1}=   \ab{12}^4 \e^{2 i (\phi_1+\phi_2+\phi_3)} \sin^2\theta \eqncom
\end{equation}
and the supermomentum-conserving delta function becomes
\begin{equation}\label{eq: expansion of super momentum conserving delta function}
\begin{aligned}
 \delta^8(Q)
 &=\prod_{A=1}^4 \Big( \ab{12}\etatt_{1}^A\etatt_{2}^A-\ab{1l_1}\etatt_{1}^A\etatt_{l_1}^A-\ab{1l_2}\etatt_{1}^A\etatt_{l_2}^A
 -\ab{2l_1}\etatt_{2}^A\etatt_{l_1}^A-\ab{2l_2}\etatt_{2}^A\etatt_{l_2}^A+\ab{l_1l_2}\etatt_{l_1}^A\etatt_{l_2}^A\Big)\\
 &=\ab{12}^4 \e^{4 i (\phi_1+\phi_2+\phi_3)} \prod_{A=1}^4 \Big(\e^{-i (\phi_1+\phi_2+\phi_3)} \, \etatt_{1}^A\etatt_{2}^A 
 +\e^{-i\phi_3}(\sin\theta   \, \etatt_{1}^A+\e^{-i \phi_1} \cos \theta \, \etatt_{2}^A)\etatt_{l_1}^A\\
 &\phaneq\phantom{\ab{12}^4 \e^{4 i (\phi_1+\phi_2+\phi_3)} \prod_{A=1}^4 \Big(}
  +\e^{-i \phi_2}(\e^{-i \phi_1} \sin \theta \, \etatt_{2}^A-  \cos \theta \, \etatt_{1}^A)\etatt_{l_2}^A 
 +\etatt_{l_1}^A\etatt_{l_2}^A\Big)\eqndot
 \end{aligned}
\end{equation}
Integrating over the fermionic variables $\etatt_{l_1}^{A}$, $\etatt_{l_2}^{A}$ amounts to the following replacements in $\ffco_{\cO,L}^{(0)}$:
\begin{equation}
\begin{aligned}
   \left( \begin{array}{c}
\etatt_{l_1}^{A}  \\
\etatt_{l_2}^{A}  \end{array} \right)
= U^*
\left( \begin{array}{c}
\etatt^A_{1}  \\
\etatt^A_{2}    \end{array} \right) \eqncom
\end{aligned} 
\end{equation}
while
\begin{equation}\label{eq: bosonic loop spinor variables}
\begin{aligned}
  \left( \begin{array}{c}
\lambda_{l_1}^{\alpha}  \\
\lambda_{l_2}^{\alpha}  \end{array} \right)=U
\left( \begin{array}{c}
\lambda_{1}^\alpha  \\
\lambda_{2}^\alpha    \end{array} \right)\eqncom\quad
  \left( \begin{array}{c}
\lambdat_{l_1}^{\alphadot}  \\
\lambdat_{l_2}^{\alphadot}  \end{array} \right)=U^*
\left( \begin{array}{c}
\lambdat^{\alphadot}_{1}  \\
\lambdat^{\alphadot}_{2}    \end{array} \right)\eqndot
\end{aligned} 
\end{equation}
Assembling all previous steps, \eqref{eq: minimal form factor times four point tree amplitude} can be simplified to%
\footnote{Note that there is an additional factor of $i$ in the amplitude \eqref{eq: colour-ordered MHV super amplitude} combining with the one in \eqref{eq: transformation of delta times Jacobian} to make the total prefactor real. Similarly, the additional factor of $(2\pi)^4$ in \eqref{eq: colour-ordered MHV super amplitude} combines with the corresponding factor in \eqref{eq: minimal form factor times four point tree amplitude}.}
\begin{equation}
\begin{aligned}
 -\frac{2}{(2\pi)^2} \int_0^{2\pi}\de \phi_1 \e^{2i\phi_1 C_{p_2}}  \e^{2i\phi_2 C_{l_1}}   \e^{2i\phi_3 C_{l_2}} \int_0^{\frac{\pi}{2}} \de \theta \cot\theta 
 \ffco^{(0)}_{\cO,L}(\Lambda_{1}^\prime,\Lambda_{2}^\prime,\Lambda_{3},\dots,\Lambda_{L};q)\eqncom
\end{aligned}
\end{equation}
where 
\begin{equation}
 \begin{aligned}
  \left( \begin{array}{c}
\lambda_{1}^{\prime \alpha}  \\
\lambda_{2}^{\prime \alpha}  \end{array} \right)=V(\theta)
\left( \begin{array}{c}
\lambda_{1}^\alpha  \\
\lambda_{2}^\alpha    \end{array} \right)\eqncom\quad
  \left( \begin{array}{c}
\lambdat_{1}^{\prime \alphadot}  \\
\lambdat_{2}^{\prime \alphadot}  \end{array} \right)=V(\theta)
\left( \begin{array}{c}
\lambdat^{\alphadot}_{1}  \\
\lambdat^{\alphadot}_{2}    \end{array} \right)\eqncom\quad
   \left( \begin{array}{c}
\etatt_{1}^{\prime A}  \\
\etatt_{2}^{\prime A}  \end{array} \right)
=V(\theta)
\left( \begin{array}{c}
\etatt^A_{1}  \\
\etatt^A_{2}    \end{array} \right)\eqndot
 \end{aligned}
\end{equation}
The central charges $C_{l_1}$ and $C_{l_2}$ vanish as the fields at $l_1$ and $l_2$ correspond to the minimal tree-level form factor and hence satisfy the required little group scaling. 
Thus, the dependence on $\phi_2$ and $\phi_3$ drops out as expected.
The integral over $\phi_1$ yields a Kronecker delta that ensures that the central charge vanishes also for $p_2$:
\begin{equation}
\label{eq: phase integral}
\int_0^{2\pi}\de \phi_1 \e^{2i\phi_1 C_{p_2}} = 2\pi \delta_{C_{p_2},0} \eqndot
\end{equation}
As the amplitude conserves the central charge, this ensures that both $p_1$ and $p_2$ have the correct little group scaling.
In total, we have
\begin{equation}
\label{eq: minimal form factor times amplitude final}
\begin{aligned}
&\frac{1}{(2\pi)^6}\int \de \Lambda_{l_1}\de \Lambda_{l_2}  \Big(\ffco^{(0)}_{\cO,L}(\Lambda_{l_1},\Lambda_{l_2},\Lambda_{3},\dots,\Lambda_{L};q) \ampco^{(0)}_4(\Lambda_{-l_2},\Lambda_{-l_1},\Lambda_{1},\Lambda_{2})\Big)\\ 
&=-\frac{1}{\pi}  \delta_{C_{p_2},0} \int_0^{\frac{\pi}{2}} \de \theta \cot\theta \ffco^{(0)}_{\cO,L}(\Lambda_{1}^\prime,\Lambda_{2}^\prime,\Lambda_{3},\dots,\Lambda_{L};q)\eqndot
\end{aligned}
\end{equation}

As a next step, we evaluate the phase-space integral \eqref{eq: double cut triangle}. 
In terms of spinor-helicity variables, the denominator of \eqref{eq: double cut triangle} can be written as 
\begin{equation}
 (l_1-p_1)^2=\ab{l_1 1}\sb{l_1 1}\eqndot
\end{equation}
Inserting the parametrisation \eqref{eq: parametrisation} localised by momentum conservation, we find
\begin{equation}
 \ab{l_1 1}= \ab{ 1 2} \e^{i(\phi_1+\phi_2)} \sin\theta \eqncom
\end{equation}
and hence
\begin{equation}
 (l_1-p_1)^2=\ab{ 1 2}\sb{ 1 2}\sin^2\theta=-(p_1+p_2)^2\sin^2\theta\eqndot
\end{equation}
Combining this with $c^{1,2}_{\text{triangle}}$, \eqref{eq: transformation of delta times Jacobian} and \eqref{eq: phase integral}, we have
\begin{equation}
\label{eq: evaluation of cut triangle}
 \begin{aligned}
 c^{1,2}_{\text{triangle}} \FDinline[triangle,doublecut,cutlabels,twolabels, labelone=\scriptscriptstyle p_1,labeltwo=\scriptscriptstyle p_2]=
 -\frac{1}{\pi} \delta_{C_{p_2},0} \int_0^{\frac{\pi}{2}} \de\theta \cot\theta \ffco^{(0)}_{\cO,L}(\Lambda_{1},\Lambda_{2},\Lambda_{3},\dots,\Lambda_{L};q)
  \eqncom
 \end{aligned}
\end{equation}
where we have included $\delta_{C_{p_2},0}$ to match the prefactor of \eqref{eq: minimal form factor times amplitude final}, which is possible since the central charge $C_{p_2}$ belongs to the tree-level form factor in this case and hence vanishes automatically.

Via \eqref{eq: transformation of delta times Jacobian} and explicit integration, the cut bubble integral yields 
\begin{equation}\label{eq: cut of bubble}
\begin{aligned}
\FDinline[bubble,doublecut,cutlabels,twolabels,labelone=\scriptscriptstyle p_1,labeltwo=\scriptscriptstyle p_2] 
=\frac{i}{(2\pi)^2}\int \de\Lambda_{l_1}\de\Lambda_{l_2}   \delta^4(P)  
=-\frac{1}{\pi}  \int_0^{\frac{\pi}{2}} \de \theta \sin\theta \cos\theta =-\frac{1}{2\pi} \eqndot
\end{aligned}
\end{equation}

As shown in figure \ref{fig: double cut of simplified ansatz for generalised unitarity}, the bubble coefficient can be obtained by subtracting \eqref{eq: evaluation of cut triangle} from \eqref{eq: minimal form factor times amplitude final} and dividing by \eqref{eq: cut of bubble}. This gives
\begin{equation}\label{eq: bubble coefficient final}
\begin{aligned}
c_{\text{bubble}}^{1,2}=-2 \delta_{C_{p_2},0} \int_0^{\frac{\pi}{2}} \de \theta \cot\theta \Big(& \ffco^{(0)}_{\cO,L}(\Lambda_{1},\Lambda_{2},\Lambda_{3},\dots,\Lambda_{L};q)\\&-\ffco^{(0)}_{\cO,L}(\Lambda_{1}^\prime,\Lambda_{2}^\prime,\Lambda_{3},\dots,\Lambda_{L};q) \Big)\eqndot
\end{aligned}
\end{equation}
Note that the integral of each summand in \eqref{eq: bubble coefficient final} diverges when taken individually. This divergence occurs in the integral region where $\theta=0$, i.e.\ where the uncut propagator in \eqref{eq: double cut triangle} goes on-shell. Hence, it is the collinear divergence of the tree-level four-point amplitude. This divergence in \eqref{eq: minimal form factor times four point tree amplitude} is precisely cancelled by \eqref{eq: double cut triangle}.

\subsection{Results and the complete one-loop dilatation operator}
\label{subsec result and one-loop dilatation operator}

Let us summarise our results from the previous two subsections in terms of the interaction density defined in \eqref{eq: one-loop interaction}. The cut-constructible part of the one-loop correction to the minimal form factor of a generic single-trace operator is obtained by acting with the density 
\begin{equation}\label{eq: summary}
 \begin{aligned}
  \intone[i\,i+1]&= - s_{i\, i+1} \FDinline[triangle,twolabels,labelone=\scriptscriptstyle i,labeltwo=\scriptscriptstyle i+1] \times \idm_{i\,i+1}
  \,+\, \FDinline[bubble,twolabels,labelone=\scriptscriptstyle i,labeltwo=\scriptscriptstyle i+1] \times \Bubbleop_{i\,i+1} 
  \,+\,\text{rational terms} 
 \end{aligned}
 \end{equation}
on the minimal tree-level form factor, where $s_{i\,i+1}=(p_i+p_{i+1})^2$, $\idm_{i\,i+1}$ is the identity operator and the integral operator $\Bubbleop_{i\,i+1}$ acts as
\begin{equation}\label{eq: bubble coefficient summary}
\begin{aligned}
\Bubbleop_{i\,i+1}\ffco^{(0)}_{\cO,L}(\Lambda_{1},\dots,\Lambda_{L};q)=-2 \delta_{C_{i+1},0} \int_0^{\frac{\pi}{2}} \de \theta \cot\theta \Big( &\ffco^{(0)}_{\cO,L}(\Lambda_{1},\dots,\Lambda_{i},\Lambda_{i+1},\dots,\Lambda_{L};q)\\
&-\ffco^{(0)}_{\cO,L}(\Lambda_{1},\dots,\Lambda_{i}^\prime,\Lambda_{i+1}^\prime,\dots,\Lambda_{L};q) \Big)\eqncom
\end{aligned}
\end{equation}
with
\begin{equation}
 \begin{aligned}
   \left( \begin{array}{c}
\Lambda_{i\phantom{+1}}^{\prime}  \\
\Lambda_{i+1}^{\prime }  \end{array} \right)
=V(\theta)
\left( \begin{array}{c}
\Lambda_{i\phantom{+1}}  \\
\Lambda_{i+1}    \end{array} \right)\eqncom \qquad V(\theta)=
\left( \begin{array}{cc}
\cos\theta & -\sin\theta \\
\sin\theta & \cos\theta \end{array} \right)\eqndot
 \end{aligned}
\end{equation} 

Note that the finite rational terms in \eqref{eq: summary} are absent in the $\SU2$ sector discussed in the last section. For general operators, however, they are non-vanishing; see \cite{Wilhelm:2014qua} for a simple example in the $\SL2$ sector. It would be interesting to determine these rational terms in general.%
\footnote{Methods to determine rational terms were developed in the context of scattering amplitudes in QCD, see e.g.\ \cite{Bern:2007dw} for a review. These methods might also be applicable here.}

As discussed in detail in the end of the last section, we can read off the dilatation operator from \eqref{eq: summary} by comparison with the general form \eqref{eq: general divergences of form factor}. Accordingly, the complete one-loop dilatation operator of $\mathcal{N}=4$ SYM theory is given by the density 
\begin{equation}
\loopdilai{1}{i\,i+1}= -2 \Bubbleop_{i\,i+1} \eqndot
\end{equation}
Note that this precisely agrees with \eqref{eq: one-loop dila in oscillators} after replacing all superoscillators by super-spinor-helicity variables according to \eqref{eq: oscillator replacements}.

In \cite{Zwiebel:2011bx}, a connection between the leading length-changing part of the complete $\ell$-loop dilatation operator and the $n$-point tree-level scattering amplitudes was derived via symmetry considerations, which was based on the fact that both objects are completely determined by $\PSU{2,2|4}$. 
In particular, the second part in \eqref{eq: one-loop dila in oscillators} was obtained from the four-point tree-level amplitude.
The first part in \eqref{eq: one-loop dila in oscillators} was added as a regularisation. The author of \cite{Zwiebel:2011bx} has numerically shown that it is uniquely fixed by commutation relations with certain leading length-changing algebra generators but states that a (more) physical argument would be desirable.
Above, we have given this physical argument and, moreover, derived the complete result via field theory.

Let us mention that our result for the minimal one-loop form factor is not limited to the planar theory.
It can be immediately generalised to the non-planar case by acting with the interaction density \eqref{eq: summary} on pairs of non-neighbouring legs and performing the occurring contractions of traces via \eqref{eq: Ts summed over a}.%
\footnote{This is a straightforward generalisation of the way the complete one-loop dilatation operator acts in the non-planar case, which is described in detail in \cite{Beisert:2003jj}. It is similar to the situation of one-loop amplitudes as well, whose non-planar double-trace contributions are also completely determined by the planar single-trace contributions \cite{Bern:1994zx}.}

In the next chapters, we will proceed to two-loop order. 
In contrast to the situation at one-loop order, the basis of integrals is not known at two-loop order; see e.g.\ \cite{Gluza:2010ws} for an approach in this direction, though.
Hence, at two-loop order, we cannot apply exactly the same method as used in this section. 
We will work with unitarity at the level of the integrand as used in the previous section instead.
Moreover, there are some conceptual problems which need to be solved before proceeding to treat all operators.
These problems concern the non-trivial mixing between IR and UV divergences and operator mixing beyond one-loop order and the long known problem how to calculate the minimal two-loop Konishi form factor via unitarity. We will address them in the following two chapters.

\chapter{Minimal two-loop Konishi form factor}
\label{chap: two-loop Konishi form factor}

Having derived the complete one-loop dilatation operator via form factors and on-shell methods in the previous chapter, we now proceed to two-loop order.  
Concretely, we calculate the minimal two-loop form factor of the Konishi primary operator and obtain the Konishi anomalous dimension up to two-loop order from it.
The Konishi operator is the best-studied example of a non-protected operator in $\mathcal{N}=4$ SYM theory. 
Calculating its form factors and correlation functions via on-shell methods involves some important subtleties concerning the regularisation. 
These subtleties also occur for a wide class of other operators and  require an extension of the on-shell method of unitarity.

We introduce the Konishi primary operator in section \ref{sec: konishi}. 
In section \ref{sec: konishi form factors}, we present the calculation of its two-loop form factor via unitarity. 
Subsequently, we analyse the occurring subtleties and show how to treat them correctly in section \ref{sec: subtleties}.
Finally, we summarise our results in section \ref{sec: Konishi results}.

The results presented in this chapter were first published in \cite{Nandan:2014oga}.

\section{Konishi operator}
\label{sec: konishi}

The best-studied composite operators in $\mathcal{N}=4$ SYM theory arise from taking the trace of two scalar fields, where we are now considering real scalars transforming in the fundamental representation of $\SO6$.

One of them is the traceless symmetric part
\begin{equation}
\label{eq: BPS in so6}
\cO_\BPS=\tr(\phi_{(I}\phi_{J)})\eqncom \qquad I,J=1,\dots,6\eqncom
\end{equation}
which transforms in the 20' of $\SO6$. It is part of the stress-tensor supermultiplet, which also contains the on-shell Lagrangian. Moreover, it is half BPS, and hence protected.

The other one is the trace part
\begin{equation}
\label{eq: K in so6}
\cK=\delta^{IJ}\tr(\phi_I\phi_J)
\end{equation}
and known as the Konishi primary operator. 
It is the primary operator of a long supermultiplet of $\PSU{2,2|4}$.
Although not protected, the Konishi primary is an eigenstate under renormalisation, i.e.\
\begin{equation}
 \label{eq: K renormalisation constant}
 \cK_\ren=\cZ_\cK \cK_\bare\eqncom \qquad \cZ_\cK(\gmod,\peps)=\exp\sum_{\ell=1}^\infty \gmod^{2\ell}\frac{\gamma_{\cK}^{(\ell)}}{2\ell\peps} \eqndot
\end{equation}
In the planar limit, the anomalous dimension of the Konishi operator is%
\footnote{In a conformal field theory like $\cN=4$ SYM theory, the anomalous dimensions are scheme independent. Hence, the expansions of $\gamma_{\cK}$ in $\g$ and $\gmod$ coincide.}
\begin{equation}
\begin{aligned}
\label{eq: gammaK}
\gamma_{\cK}=\sum_{\ell=1}^\infty \gmod^{2\ell} \gamma_{\cK}^{(\ell)}&=12\gmod^2-48\gmod^4+336\gmod^6-96(26-6\zeta_3+15\zeta_5)\gmod^8\\
&\phantom{{}={}6[}
+96(158+72\zeta_3-54\zeta_3^2-90\zeta_5+315\zeta_7)\gmod^{10}
+\cO(\gmod^{12}) \eqndot
\end{aligned}
\end{equation}
Non-planar corrections to \eqref{eq: gammaK} start to occur at the fourth loop order \cite{Velizhanin:2009gv}.
The first two orders in \eqref{eq: gammaK}, which we reproduce in this chapter, were first calculated via Feynman diagrams in \cite{Anselmi:1996mq,Anselmi:1996dd} and
\cite{Bianchi:1999ge,Bianchi:2000hn,Eden:2000mv}, respectively.%
\footnote{%
Currently, the anomalous dimension $\gamma_{\cK}$ of the Konishi operator is known up to $\ell=5$ from field theory \cite{Kotikov:2004er,Eden:2004ua,Sieg:2010tz,Fiamberti:2007rj,Fiamberti:2008sh,Velizhanin:2008jd,Eden:2012fe} and up to $\ell=10$ from integrability \cite{Beisert:2003tq,Bajnok:2008bm,Bajnok:2009vm,Arutyunov:2010gb,Balog:2010xa,Bajnok:2012bz,Leurent:2013mr,Volin:IGST,Marboe:2014gma}.
}

In order to apply four-dimensional supersymmetric on-shell methods, we need to express the operators \eqref{eq: BPS in so6} and \eqref{eq: K in so6} in terms of the antisymmetric scalars appearing in Nair's $\mathcal{N}=4$ on-shell superfield \eqref{eq: N=4 superspace}.
In terms of these fields, \eqref{eq: BPS in so6} can be written as 
\begin{equation}
\label{eq: BPS in su4 general}
\cO_\BPS=\tr(\phi_{AB}\phi_{CD}) - \frac{1}{12} \teps_{ABCD} \, \tr(\phi^{EF} \phi_{EF})\eqndot
\end{equation}
Without loss of generality, we focus on the particular component
\begin{equation}
\label{eq: BPS in su4}
\cO_\BPS=\tr(\phi_{AB}\phi_{AB})\eqncom
\end{equation}
where $A$ and $B$ are not summed over.
Similarly, the Konishi primary operator \eqref{eq: K in so6} can be expressed as 
\begin{equation}
\label{eq: K in su4}
\cK_6=\frac{1}{8}\teps^{ABCD}\tr(\phi_{AB}\phi_{CD})
=\tr(\phi_{12}\phi_{34})-\tr(\phi_{13}\phi_{24})+\tr(\phi_{14}\phi_{23}) \eqndot
\end{equation} 
Note that \eqref{eq: K in su4} as well as the superfield \eqref{eq: N=4 superspace} manifestly require $N_\phi=6$ scalars. Hence, we denote the operator defined in \eqref{eq: K in su4} as $\cK_6$. The expression \eqref{eq: K in so6}, however, is meaningful for any $N_\phi$. 
In fact, supersymmetric regularisation in $D=4-2\peps$ dimensions requires to continue the number of scalars to $N_\phi=10-D=6+2\peps$. Hence, $\cK_6 \neq \cK$.
We come back to this subtlety in section \ref{sec: subtleties}.

\section{Calculation of form factors}
\label{sec: konishi form factors}

Let us now calculate the one- and two-loop corrections to the Konishi two-point form factor via the on-shell method of unitarity in four dimensions. 
As already mentioned, this yields direct results only for $\cK_6\neq \cK$.%
\footnote{We postpone a detailed discussion of this subtlety to the next section.}

The Konishi primary operator is an eigenstate under renormalisation. Thus, all loop corrections to its form factors are proportional to the tree-level expressions and the operators $\Interaction^{(\ell)}$ defined in \eqref{eq: loop correction form factor} reduce to simple functions, as in the case of amplitudes and form factors of the BPS operator \eqref{eq: BPS in so6}.
We denote these functions as $f^{(\ell)}$.

According to section \ref{sec: minimal tree-level form factors for any operator}, the colour-ordered minimal tree-level super form factor of $\cK_6$ is given by 
\begin{equation}
\label{eq: K6 0-loop 2-pnt}
\begin{aligned}
 \ffco_{\cK_6,2}^{(0)}(1,2)&=\frac{1}{4} \teps_{ABCD}
\etatt_1^A \etatt_1^B \etatt_2^C \etatt_2^D \, (2\pi)^4\delta^4\big(\lambda_1 \lambdat_1+\lambda_2 \lambdat_2 -q \big)
\\ &= \big(
+\etatt_1^1 \etatt_1^2 \etatt_2^3 \etatt_2^4 - \etatt_1^1 \etatt_1^3 \etatt_2^2 \etatt_2^4 + \etatt_1^1 \etatt_1^4 \etatt_2^2 \etatt_2^3 
\\ &\phaneq \phantom{\big(}
+\etatt_1^3 \etatt_1^4 \etatt_2^1 \etatt_2^2 - \etatt_1^2 \etatt_1^4 \etatt_2^1 \etatt_2^3 + \etatt_1^2 \etatt_1^3 \etatt_2^1 \etatt_2^4 \big) (2\pi)^4\delta^4\big(\lambda_1 \lambdat_1+\lambda_2 \lambdat_2 -q \big)\eqndot
\end{aligned}
\end{equation}

\begin{figure}[tbp]
 \centering
$
\settoheight{\eqoff}{$\times$}%
\setlength{\eqoff}{0.5\eqoff}%
\addtolength{\eqoff}{-12.0\unitlength}%
\raisebox{\eqoff}{%
\fmfframe(2,2)(2,2){%
\begin{fmfchar*}(80,30)
\fmfleft{vq}
\fmfright{vp2,vp1}
\fmf{dbl_plain_arrow,tension=2}{vq,v1}
\fmf{plain_arrow,tension=0.5,left=0.7,label=$\scriptstyle l_2$,l.s=left,l.d=8}{v2,v1}
\fmf{plain_arrow,tension=0.5,right=0.7,label=$\scriptstyle l_1$,l.s=right,l.d=8}{v2,v1}
\fmf{phantom_smallcut,left=0.7,tension=0}{v1,v2}
\fmf{phantom_smallcut,right=0.7,tension=0}{v1,v2}
\fmf{plain_arrow}{v2,vp1}
\fmf{plain_arrow}{v2,vp2}
\fmfv{decor.shape=circle,decor.filled=30,decor.size=30,label=$\ffco_{\cK,,2}$,label.dist=0}{v1}
\fmfv{decor.shape=circle,decor.filled=50,decor.size=30,label=$\ampco_{4}$,label.dist=0}{v2}
\fmffreeze
 \fmfcmd{pair vertq, vertpone, vertptwo, vertone, verttwo; vertone = vloc(__v1); verttwo = vloc(__v2); vertq = vloc(__vq); vertpone = vloc(__vp1); vertptwo = vloc(__vp2); }
 \fmfiv{label=$\scriptstyle q$}{vertq}
 \fmfiv{label=$\scriptstyle p_1$}{vertpone}
 \fmfiv{label=$\scriptstyle p_2$}{vertptwo}
\end{fmfchar*}%
}}%
$
\caption{The double cut of the minimal one-loop Konishi form factor $\ffco_{\cK,2}^{(1)}$ in the channel $(p_1+p_2)^2$.}
\label{fig: one-loop Konishi double cut}
\end{figure}

The one-loop correction to \eqref{eq: K6 0-loop 2-pnt} can be calculated via unitarity in analogy to section \ref{sec: one-loop unitarity}.
We only have to consider the double cut given in \eqref{eq: one-loop double cut} and depicted in figure \ref{fig: one-loop double cut}.
Specialising to $L=2$ and $\cO=\cK_6$, figure \ref{fig: one-loop double cut} reduces to figure \ref{fig: one-loop Konishi double cut} and we find%
\footnote{Note that we have reversed the momenta $l_1$ and $l_2$ with respect to the last chapter.}
\begin{equation}
\label{eq: f 1-loop double cut}
\begin{aligned}
\ffco_{\cK_6,2}^{(1)} (1,2) \Big|_{s_{12}}\!&=\twops \de^4 \etatt_{l_{1}}\de^4 \etatt_{l_{2}}\ffco_{\cK_6,2}^{(0)}(-l_1,-l_2)\times \ampco_4^{(0)}(p_1,p_2,l_2,l_1)
\\
&=\ffco_{\cK_6,2}^{(0)}(1,2) \underbrace{i \!\!\twopst \frac{\abr{l_1 2}^2\abr{l_2 1}^2+4\abr{l_1 1}\abr{l_1 2}\abr{l_2 1}\abr{l_2 2}+\abr{l_1 1}^2\abr{l_2 2}^2}{\abr{l_1 1}\abr{1 2}\abr{2 l_2}\abr{l_2 l_1}}}_{f^{(1)}_{\cK_6}\big|_{s_{12}}} \!\eqndot
\end{aligned}
\end{equation}
Applying the Schouten identity \eqref{eq: Schouten identities}, this yields
\begin{equation}
\label{eq: f  1-loop double cut further simplified}
\begin{aligned}
f^{(1)}_{\cK_6}\Big|_{s_{12}}&= i\twopst \left(\frac{\abr{l_1 l_2}\abr{1 2}}{\abr{l_1 1}\abr{l_2 2}}+6\frac{\abr{l_1 2}\abr{l_2 1}}{\abr{1 2}\abr{l_1 l_2}} \right)\\
&= i\twopst\left(-\frac{s_{12}}{(l_1+p_1)^2} +6 \frac{(l_1+p_2)^2}{s_{12}}\right)\\
&=- i\, s_{12}  \FDinline[triangle, doublecut,cutlabels, twolabels, labelone=p_1, labeltwo=p_2] + i\, 6{}{}\frac{(l_1+p_2)^2}{s_{12}} \FDinline[bubble, doublecut,cutlabels, twolabels, labelone=p_1, labeltwo=p_2] \eqncom
\end{aligned}
\end{equation}
where the loop-momentum dependent prefactor $(l_1+p_2)^2$ is understood to appear in the numerator of the depicted integral.

As in the previous chapter, we have to sum the contributions from cuts in all pairs of neighbouring legs. In the case of $L=2$, those are $p_1$ and $p_2$ as well as $p_2$ and $p_1$, which are inequivalent when considering colour-ordered quantities. The contributions from both cuts, however, do agree, resulting in a total prefactor of $2$.
In total, we find 
\begin{equation}
\label{eq: Konishi 1-loop full}
\begin{aligned}
f_{\cK_6,2}^{(1)} &=2 \left(-s_{12} 
\FDinline[triangle, twolabels, labelone=p_1, labeltwo=p_2] +6 \frac{s_{2 l}}{s_{12}} 
\FDinline[bubble, twolabels, labelone=p_1, labeltwo=p_2,momentum]
\right) \eqndot
\end{aligned}
\end{equation}

In contrast to the one-loop results in section \ref{sec: one-loop unitarity}, not all integrals appearing in the result \eqref{eq: Konishi 1-loop full} are scalar. 
Instead, also a linear tensor integral occurs. 
Via Passarino-Veltmann (PV) reduction \cite{Passarino:1978jh}, however, it can be reduced to a scalar integral, as shown in appendix \ref{appsec: PV reduction}. 
Using \eqref{eq: linear bubble PV}, we have 
\begin{equation}
\begin{aligned}
f_{\cK_6,2}^{(1)} (1,2) &= \underbrace{-2s_{12} 
\FDinline[triangle, twolabels, labelone=p_1, labeltwo=p_2] }_{f_{\BPS,2}^{(1)}(1,2)} 
\,-\, 6 
\FDinline[bubble, twolabels, labelone=p_1, labeltwo=p_2]
\eqncom
\label{eq: K6 FF 2pt 1loop final}
\end{aligned}
\end{equation}
where the first summand is equal to the corresponding result for $\cO_\BPS$.

\begin{figure}[tbp]
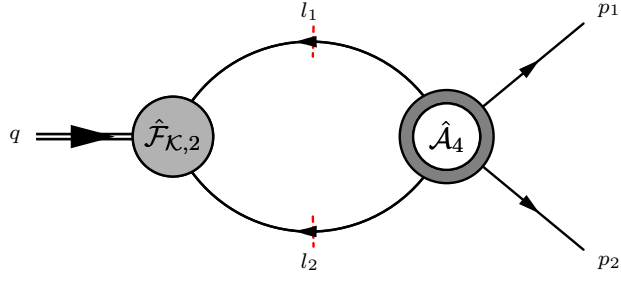

 \centering
$
\settoheight{\eqoff}{$\times$}%
\setlength{\eqoff}{0.5\eqoff}%
\addtolength{\eqoff}{-12.0\unitlength}%
\raisebox{\eqoff}{%
\fmfframe(2,2)(2,2){%
\begin{fmfchar*}(80,30)
\fmfleft{vq}
\fmfright{vp2,vp1}
\fmf{dbl_plain_arrow,tension=2}{vq,v1}
\fmf{plain_arrow,tension=0.5,left=0.7,label=$\scriptstyle l_2$,l.s=left,l.d=8}{v2,v1}
\fmf{plain_arrow,tension=0.5,right=0.7,label=$\scriptstyle l_1$,l.s=right,l.d=8}{v2,v1}
\fmf{phantom_smallcut,left=0.7,tension=0}{v1,v2}
\fmf{phantom_smallcut,right=0.7,tension=0}{v1,v2}
\fmf{plain_arrow}{v2,vp1}
\fmf{plain_arrow}{v2,vp2}
\fmfv{decor.shape=circle,decor.filled=30,decor.size=30,label=$\ffco_{\cK,,2}$,label.dist=0}{v1}
\fmfv{decor.shape=circle,decor.filled=0,decor.size=25}{v2}
\fmffreeze
\fmfdraw
 \fmfcmd{pair vertq, vertpone, vertptwo, vertone, verttwo; vertone = vloc(__v1); verttwo = vloc(__v2); vertq = vloc(__vq); vertpone = vloc(__vp1); vertptwo = vloc(__vp2); }
 \fmfiv{label=$\scriptstyle q$}{vertq}
 \fmfiv{label=$\scriptstyle p_1$}{vertpone}
 \fmfiv{label=$\scriptstyle p_2$}{vertptwo}
\fmfiv{decor.shape=circle,decor.filled=50,decor.size=35,label=$\ampco_{4}$,label.dist=0}{verttwo}
\fmfiv{decor.shape=circle,decor.filled=0,decor.size=25,label=$\ampco_{4}$,label.dist=0}{verttwo}
\end{fmfchar*}%
}}%
$
\caption{The planar two-particle cut of the two-loop Konishi form factor $\ffco_{\cK,2}^{(2)}$ in the channel $(p_1+p_2)^2$.}
\label{fig: planar double cut at two loop}
\end{figure}

\begin{figure}[tbp]
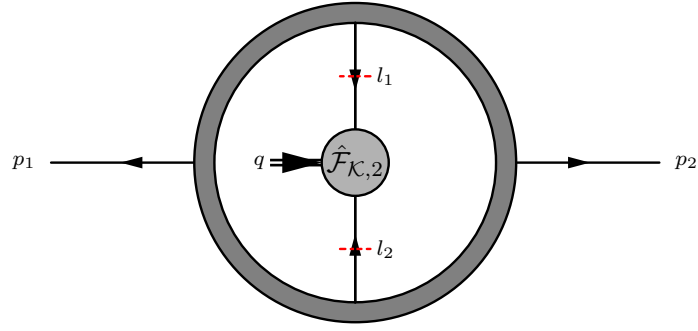

\centering
$
 \settoheight{\eqoff}{$\times$}%
 \setlength{\eqoff}{0.5\eqoff}%
 \addtolength{\eqoff}{-21.0\unitlength}%
 \raisebox{\eqoff}{%
 \fmfframe(2,6)(2,6){%
 \begin{fmfchar*}(80,30)
 \fmfleft{vp1}
 \fmfright{vp2}
 \fmftop{va}
 \fmf{plain_arrow,tension=2}{vl,vp1}
 \fmf{plain_arrow,tension=2}{vr,vp2}
 \fmf{plain,tension=0.5}{v1,vl}
 \fmf{plain,tension=0.5}{v1,vr}
 \fmf{phantom_smallcut,left=0,tension=0}{v1,vl}
 \fmf{phantom_smallcut,right=0,tension=0}{v1,vr}
 \fmf{phantom,tension=0.5,left=1}{vl,vr}
 \fmf{phantom,tension=0.5,right=1}{vl,vr}
 \fmffreeze
 \fmf{phantom,tension=3}{va,vq}
 \fmf{dbl_plain_arrow}{vq,v1}
 \fmffreeze
 \fmfdraw
   \fmfcmd{pair vertq, vertpone, vertptwo, vertone, vertl, vertr;
vertone = vloc(__v1); vertl = vloc(__vl); vertr = vloc(__vr); vertq =
vloc(__vq); vertpone = vloc(__vp1); vertptwo = vloc(__vp2);}
  \fmfiv{label=$\scriptstyle q$,l.d=2,l.a=180}{vertone+0.352778(-90,0)}
   \fmfiv{label=$\scriptstyle p_1$}{vertpone}
   \fmfiv{label=$\scriptstyle p_2$}{vertptwo}
\fmfiv{decor.shape=circle,decor.filled=50,decor.size=120}{vertone}
\fmfiv{decor.shape=circle,decor.filled=0,decor.size=105}{vertone}
\fmfi{dbl_plain_arrow}{(vertone+0.352778(-90,0))--(vertone+0.352778(-35,0))}
 \fmfi{plain_arrow,label=$\scriptstyle l_2$,l.s=right,l.d=8}{(vertone-0.352778(0,149))--(vertone-0.352778(0,35))}
 \fmfi{plain_arrow,label=$\scriptstyle l_1$,l.s=left,l.d=8}{(vertone+0.352778(0,149))--(vertone+0.352778(0,35))}
 \fmfi{phantom_smallcut}{(vertone-0.352778(0,149))--(vertone-0.352778(0,35))}
 \fmfi{phantom_smallcut}{(vertone+0.352778(0,149))--(vertone+0.352778(0,35))}
    \fmfiv{decor.shape=circle,decor.filled=30,decor.size=25,label=$\ffco_{\cK,,2}$,label.dist=0}{vertone}
\end{fmfchar*}%
}}%
$
\caption{The non-planar two-particle cut of two-loop Konishi form factor $\ffco_{\cK,2}^{(2)}$ in the channel $(p_1+p_2)^2$.}
\label{fig: nonplanar double cut two loop}
\end{figure}

At two-loop order, several different cuts have to be considered, which are depicted in figures \ref{fig: planar double cut at two loop}, \ref{fig: nonplanar double cut two loop} and \ref{fig: two-loop Konishi triple cut}. We treat these cuts one after the other.
%
The first cut is the planar double cut depicted in figure \ref{fig: planar double cut at two loop}, on which the minimal colour-ordered two-loop form factor $\ffco_{\cK_6,2}^{(2)}$ factorises into the product of the minimal colour-ordered tree-level form factor $\ffco_{\cK_6,2}^{(0)}$ and the colour-ordered one-loop four-point amplitude $\ampco_4^{(1)}$.
The latter is given by \cite{Bern:1996je}%
\footnote{The sign in \eqref{eq: one-loop four-point amplitude} is related to our conventions for the box integral \eqref{eq: one-loop massless box integral}.}
\begin{equation}
\label{eq: one-loop four-point amplitude}
\begin{aligned}
 \ampco_4^{(1)}(p_1,p_2, p_3, p_4)&= \ampco_4^{(0)}(p_1,p_2, p_3, p_4)(-\, s_{12} s_{23}) I_4^{(1)}(p_1,p_2,p_3, p_4) \eqncom
\end{aligned}
\end{equation}
where the box integral is 
\begin{equation}
\label{eq: one-loop massless box integral}
 I_4^{(1)}(p_1,p_2,p_3, p_4)= (\e^{\gamma_{\text{E}}}\mu^2)^\peps \int \frac{\de l^D}{i\pi^{\frac{D}{2}}}\frac{1}{l^2(l+p_1)^2(l+p_1+p_2)^2(l+p_1+p_2+p_3)}\eqndot
\end{equation}
Hence, we have 
\begin{equation}
\label{eq: Konishi two-loop first cut}
\begin{aligned}
\ffco_{\cK_6,2}^{(2)}\Big|_{s_{12}}^{\text{I}}
&= \twops \de^4 \etatt_{l_{1}}\de^4 \etatt_{l_{2}} \ffco_{\cK_6,2}^{(0)}(-l_1,-l_2)\ampco_4^{(0)}(p_1,p_2,l_2,l_1)  \\
&\hphantom{{}={}\twops \de^4 \etatt_{l_{1}}\de^4 \etatt_{l_{2}}}\times \, (-) s_{12} s_{1 l_1} I_4^{(1)}(p_1,p_2,l_2,l_1) \eqndot
\end{aligned}
\end{equation}
The first line in \eqref{eq: Konishi two-loop first cut} can now be simplified in complete analogy to the one-loop case in \eqref{eq: f  1-loop double cut further simplified} such that we find 
\begin{equation}
\label{eq: Konishi two-loop first cut simplified}
\begin{aligned}
f_{\cK_6,2}^{(2)}\Big|_{s_{12}}^{\text{I}}
& =  \twopst
\Big(\frac{s_{12}}{s_{1 l_1}}-6\frac{s_{2 l_1}}{s_{12}}\Big)
s_{12} s_{1 l_1}  I_4^{(1)}(p_1,p_2,l_2,l_1)\\
& =\Big( s_{12}^2   - 6  s_{1 l_1} s_{2 l_1} \Big)
\FDinline[rainbow, doublecut,cutlabels, twolabels, labelone=p_1, labeltwo=p_2]  \eqndot
\end{aligned}
\end{equation}
We denote the resulting contribution to the two-loop form factor of $\cK_6$ as
\begin{equation}
\label{eq: Konishi two-loop first cut result}
\begin{aligned}
f_{\cK_6,2}^{(2),\text{I}}
& =\Big( s_{12}^2   - 6  s_{1 l_1} s_{2 l_1} \Big)
\FDinline[rainbow, momentum, twolabels,labelone=p_1, labeltwo=p_2]  \eqndot
\end{aligned}
\end{equation}
As in the one-loop case, this cut can be taken between the legs $p_1$ and $p_2$ as well as $p_2$ and $p_1$, which are inequivalent configurations for colour-ordered objects.

In addition to the planar double cut, also a non-planar double cut contributes, as shown in figure \ref{fig: nonplanar double cut two loop}.
On this cut, the colour-ordered two-loop minimal form factor $\ffco_{\cK_6,2}^{(2)}$ factorises into the product of the colour-ordered minimal tree-level form factor $\ffco_{\cK_6,2}^{(2)}$ and the double-trace part $\hat{\hat{\amp}}_4^{(1)}$ of the one-loop four-point amplitude $\amp_4^{(1)}$. The latter appears with trace structure
\begin{equation}
\label{eq: double-trace structure of double-trace amplitude}
 \hat{\hat{\amp}}_4^{(1)}(l_1,l_2;p_1,p_2)\frac{1}{N}\tr(\T^{a_{l_1}}\T^{a_{l_2}})\tr(\T^{a_{p_1}}\T^{a_{p_2}})\eqncom
\end{equation}
where the momenta in different traces are separated by a semicolon.
Although \eqref{eq: double-trace structure of double-trace amplitude} is apparently suppressed in $\frac{1}{N}$, it contributes at leading order in $N$ in this cut due to the wrapping effect \cite{Sieg:2005kd}, cf.\ the discussion in section \ref{sec: 't Hooft limit}. 
The double-trace part $\hat{\hat{\amp}}_4^{(1)}$ can be expressed in terms of colour-ordered amplitudes as \cite{Bern:1994zx}
\begin{equation}
\label{eq: double-trace one-loop four-point amplitude}
\begin{aligned}
 \hat{\hat{\amp}}_{4}^{(1)}(l_1,l_2;p_1,p_2)&= \ampco_{4}^{(1)}(p_1,p_2,l_1,l_2)+\ampco_{4}^{(1)}(p_1,l_1,p_2,l_2)+\ampco_{4}^{(1)}(p_1,l_1,l_2,p_2)
\\ &\phaneq +\ampco_{4}^{(1)}(p_1,p_2,l_2,l_1)+\ampco_{4}^{(1)}(p_1,l_2,p_2,l_1)+\ampco_{4}^{(1)}(p_1,l_2,l_1,p_2)\eqncom
\end{aligned} 
\end{equation}
The two lines in \eqref{eq: double-trace one-loop four-point amplitude} are related by relabelling $l_1 \leftrightarrow l_2$. 
Since we are working with full amplitudes at this point, we have to include a prefactor of $\frac{1}{2}$ in the phase-space integral to compensate for the freedom to relabel $l_1 \leftrightarrow l_2$, which effectively reduces \eqref{eq: double-trace one-loop four-point amplitude} to its first line.
The first and the last term in the first line of \eqref{eq: double-trace one-loop four-point amplitude} both contribute the same integral as the previous cut, such that the total prefactor of $f_{\cK_6,2}^{(2),\text{I}}$  becomes $4$.
The second term in the first line of \eqref{eq: double-trace one-loop four-point amplitude} yields%
\footnote{See \cite{Brandhuber:2012vm} for an alternative derivation of the contributions \eqref{eq: Konishi two-loop first cut} and \eqref{eq: Konishi two-loop second cut} of the (planar and non-planar) double cut to the two-point two-loop form factor of an operator of length two, which uses fundamental and adjoint gauge-group indices.}
\begin{equation}
\label{eq: Konishi two-loop second cut}
\begin{aligned}
\ffco_{\cK_6,2}^{(2)}\Big|_{s_{12}}^{\text{II}}
&= \twops \de^4 \etatt_{l_{1}}\de^4 \etatt_{l_{2}}\ffco_{\cK_6 , 2}^{(0)}(-l_1,-l_2)\ampco_4^{(0)}(p_1,l_1,p_2,l_2)\\
&\hphantom{{}={}\twops \de^4 \etatt_{l_{1}}\de^4 \etatt_{l_{2}}}\times (-)s_{12} s_{1 l_1} I_4^{(1)}(p_1,l_1,p_2,l_2) \eqncom
\end{aligned}
\end{equation}
such that 
\begin{equation}
\label{eq: Konishi two-loop second cut simplified}
\begin{aligned}
f_{\cK_6,2}^{(2)}\Big|_{s_{12}}^{\text{II}}=\Big( s_{12}^2  - 6 s_{1 l_1} s_{2 l_1} \Big) \FDinline[fliprainbownonplanar, doublecut, big, cutlabels,twolabels,labelone=p_1,labeltwo=p_2, splitlabels]
\eqndot
\end{aligned}
\end{equation}
Its contribution to the two-loop form factor of $\cK_6$ is 
\begin{equation}
\label{eq: Konishi two-loop second cut result}
\begin{aligned}
f_{\cK_6,2}^{(2),{\text{II}}} =  \Big( s^2_{12} - 6 s_{1 l} s_{2 l} \Big)
\FDinline[rainbownonplanar,momentum,twolabels,labelone=p_1,labeltwo=p_2] \eqndot
\end{aligned}
\end{equation}

\begin{figure}[tbp]
 \centering
$
\settoheight{\eqoff}{$\times$}%
\setlength{\eqoff}{0.5\eqoff}%
\addtolength{\eqoff}{-12.0\unitlength}%
\raisebox{\eqoff}{%
\fmfframe(2,2)(2,2){%
\begin{fmfchar*}(80,30)
\fmfleft{vq}
\fmfright{vp2,vp1}
\fmf{dbl_plain_arrow,tension=2}{vq,v1}
\fmf{plain_arrow,tension=0.333,left=0.7,label=$\scriptstyle l_3$,l.s=left,l.d=8}{v2,v1}
\fmf{plain_arrow,tension=0.333,right=0.7,label=$\scriptstyle l_1$,l.s=right,l.d=8}{v2,v1}
\fmf{plain_arrow,tension=0.333,right=0.0,label=$\scriptstyle l_2$,l.s=right,l.d=8}{v2,v1}
\fmf{phantom_smallcut,left=0.7,tension=0}{v1,v2}
\fmf{phantom_smallcut,right=0.7,tension=0}{v1,v2}
\fmf{phantom_smallcut,right=0,tension=0}{v1,v2}
\fmf{plain_arrow}{v2,vp1}
\fmf{plain_arrow}{v2,vp2}
\fmfv{decor.shape=circle,decor.filled=30,decor.size=30,label=$\ffco_{\cK,,3}$,label.dist=0}{v1}
\fmfv{decor.shape=circle,decor.filled=50,decor.size=30,label=$\ampco_{5}$,label.dist=0}{v2}
\fmffreeze
 \fmfcmd{pair vertq, vertpone, vertptwo, vertone, verttwo; vertone = vloc(__v1); verttwo = vloc(__v2); vertq = vloc(__vq); vertpone = vloc(__vp1); vertptwo = vloc(__vp2); }
 \fmfiv{label=$\scriptstyle q$}{vertq}
 \fmfiv{label=$\scriptstyle p_1$}{vertpone}
 \fmfiv{label=$\scriptstyle p_2$}{vertptwo}
\end{fmfchar*}%
}}%
$
\caption{The three-particle cut of the two-loop Konishi form factor $\ffco_{\cK,2}^{(2)}$ in the channel $(p_1+p_2)^2$.}
\label{fig: two-loop Konishi triple cut}
\end{figure}

The three-particle cut, or triple cut (TC), of the minimal two-loop form factor $\ffco_{\cK_6,2}^{(2)}$ is shown in figure \ref{fig: two-loop Konishi triple cut}.
On this cut, $\ffco_{\cK_6,2}^{(2)}$ factorises into the product of the tree-level three-point form factor $\ffco_{\cK_6,3}^{(0)}$ and the tree-level five-point amplitude $\ampco_{5}^{(0)}$:
\begin{equation}
\label{eq: Konishi two-loop triple cut}
\begin{aligned}
\ffco_{\cK_6,2}^{(2)}(1,2)\Big|_{\text{TC}} & = \threeps \prod_{i=1}^3\de^4 \etatt_{l_{i}} \Big(\ffco_{\cK_6,3}^{(0),\text{MHV}}(-l_1, -l_2,-l_3) \ampco_5^{(0),\text{NMHV}}(p_1,p_2, l_3, l_2, l_1)\\
&\hphantom{\threeps  } 
+\ffco_{\cK_6,3}^{(0),\text{NMHV}}(-l_1, -l_2, -l_3) \ampco_5^{(0),\text{MHV}}(p_1,p_2, l_3, l_2, l_1)\Big)
\\
&=\ffco_{\cK_6,2}^{(0)}(1,2) f_{\cK_6,2}^{(2)}\Big|_\text{TC} \eqncom
\end{aligned}
\end{equation}
where two summands arise due to the different possibilities to distribute the MHV degree between the amplitude and the form factor. In fact, those two summands are conjugates of each other.

For any composite operator built from $L$ scalar fields, the next-to-minimal tree-level form factors can be easily obtained via Feynman diagrams or from the component expansion of the super form factors of $\tr(\phi_{14}^L)$ given in \cite{Penante:2014sza}.\footnote{Since no interactions among the different scalar fields in the composite operator can occur at tree level, the corresponding components of the tree-level form factors are insensitive to the flavours of the scalars.} They can be of MHV or NMHV type.
For MHV, two different cases can occur.
In the first case, a $g^+$ can be emitted between two neighbouring scalars $\phi_{AB}$ and $\phi_{CD}$ at positions $i$ and $i+1$. This leads to the replacement
\begin{equation}
\label{eq: Nminimal ff MHV gluon}
 \cdots \etatt_i^A\etatt_i^B\etatt_{i+1}^C\etatt_{i+1}^D \cdots \longrightarrow \cdots \etatt_i^A\etatt_i^B \frac{\ab{i\,i\textrm{+}2}}{\ab{i\,i\textrm{+}1}\ab{i\textrm{+}1\,i\textrm{+}2}}  \etatt_{i+2}^C\etatt_{i+2}^D \cdots 
\end{equation}
in the minimal tree-level form factor.
In the second case, a scalar field $\phi_{CD}$ at position $i$ splits into two antifermions $\bar\psi_C$ and $\bar\psi_D$, leading to  
\begin{equation}
\label{eq: Nminimal ff MHV fermions}
\cdots \etatt_{i-1}^A\etatt_{i-1}^B \etatt_i^C\etatt_i^D  \etatt_{i+1}^E\etatt_{i+1}^F \cdots \longrightarrow \cdots \etatt_{i-1}^A\etatt_{i-1}^B \frac{1}{\ab{i\,i\textrm{+}1}} (\etatt_i^C\etatt_{i+1}^D - \etatt_i^D\etatt_{i+1}^C)   \etatt_{i+2}^E\etatt_{i+2}^F \cdots \eqndot
\end{equation}
For NMHV, two different cases can occur as well, which are in fact conjugates of the above cases.%
\footnote{They can be obtained using the conjugation rule described in appendix \ref{app: scattering amplitudes}.
}
In the first case, a $g^-$ can be emitted between two neighbouring scalars $\phi_{AB}$ and $\phi_{CD}$ at positions $i$ and $i+1$, leading to 
\begin{equation}
\label{eq: Nminimal ff NMHV gluon}
\cdots \etatt_i^A\etatt_i^B\etatt_{i+1}^C\etatt_{i+1}^D  \cdots \longrightarrow \cdots \etatt_i^A\etatt_i^B \frac{-\sb{i\,i\textrm{+}2}}{\sb{i\,i\textrm{+}1}\sb{i\textrm{+}1\,i\textrm{+}2}} \etatt_{i+1}^1\etatt_{i+1}^2\etatt_{i+1}^3\etatt_{i+1}^4\etatt_{i+2}^C\etatt_{i+2}^D \cdots \eqndot
\end{equation}
In the second case, the scalar field $\phi_{CD}$ at position $i$ splits into two fermions $\psi^{C'}$, $\psi^{D'}$ with $\teps_{CDC'D'}=1$, leading to  
\begin{equation}
\label{eq: Nminimal ff NMHV fermions}
\cdots \etatt_{i-1}^A\etatt_{i-1}^B \etatt_i^C\etatt_i^D  \etatt_{i+1}^E\etatt_{i+1}^F \cdots \longrightarrow \cdots \etatt_{i-1}^A\etatt_{i-1}^B \frac{-1}{\sb{i \, i\textrm{+}1}} (\bar\etatt_{i,C'} \bar\etatt_{i+1,D'} - \bar\etatt_{i,D'} \bar\etatt_{i+1,C'})   \etatt_{i+2}^E\etatt_{i+2}^F \cdots \eqncom
\end{equation}
where $\bar\etatt_{j,A}=\frac{1}{3!}\teps_{ABCD}\etatt_j^B\etatt_j^C\etatt_j^D$.
These replacements have to be summed over all possible insertion points.

\begin{figure}[tbp]
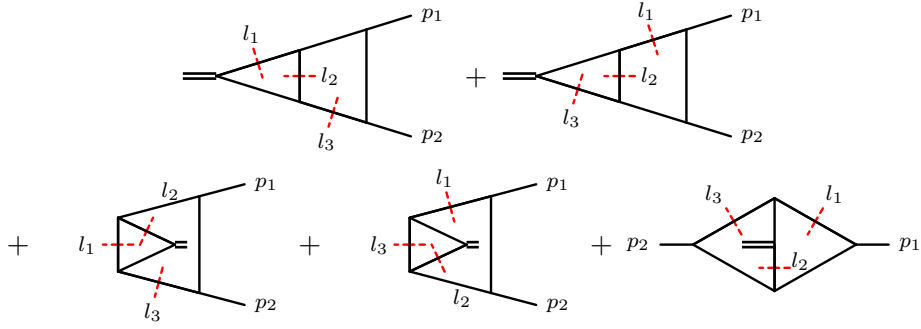

 \centering
\FDinline[rainbow,triplecut,big,cutlabels,twolabels,labelone=p_1,labeltwo=p_2]+\FDinline[rainbow,alttriplecut,big,cutlabels,twolabels,labelone=p_1,labeltwo=p_2]\\
+ \hskip -0.5cm \FDinline[fliprainbow,triplecut,big,cutlabels,twolabels,labelone=p_1,labeltwo=p_2]+
 \hskip -0.5cm \FDinline[fliprainbow,alttriplecut,big,cutlabels,twolabels,labelone=p_1,labeltwo=p_2]+\FDinline[splitlabels,fliprainbownonplanar,alttriplecut,big,cutlabels,twolabels,labelone=p_2,labeltwo=p_1]
\caption{Triple cuts of the integrals in \eqref{eq: Konishi two-loop first cut result} and \eqref{eq: Konishi two-loop second cut result}, which yield the first five terms in \eqref{eq: Konishi two-loop triple cut simplified}. Here, we have suppressed the numerator factors occurring in \eqref{eq: Konishi two-loop triple cut simplified}.}
\label{fig: triple cuts of previously known integrals}
\end{figure}

Inserting the above expressions for the next-to-minimal form factors as well as those for the five-point amplitudes into \eqref{eq: Konishi two-loop triple cut}, we find
\begin{equation}
\label{eq: Konishi two-loop triple cut simplified}
\begin{aligned}
f_{\cK_6,2}^{(2)}\Big|_\text{TC}&= i\threepst  \left(
\frac{s_{12}^2 - 6 s_{1l_1} s_{2l_1}}{s_{2l_3} s_{l_1l_2} s_{l_2l_3}}
   +\frac{s_{12}^2 - 6 s_{1l_3} s_{2l_3}}{s_{1l_1} s_{l_1l_2} s_{l_2l_3}}
   +\frac{s_{12}^2 - 6 s_{1l_2} s_{2l_2}}{s_{2l_3} s_{l_1l_2} s_{l_1l_3}}
 \right.
   \\
   & \hphantom{{}={} \threeps \bigg(} \left.    +\frac{s_{12}^2 - 6 s_{1l_2} s_{2l_2} }{s_{1l_1} s_{l_1l_3}
   s_{l_2l_3}} +\frac{s_{12}^2 - 6 s_{1l_2} s_{2l_2} }{s_{1l_1} s_{2l_3} s_{l_1l_3}}
   +\frac{18}{s_{12}}
   -\frac{18 s_{1l_3}}{s_{12} s_{l_1l_2}}-\frac{18 s_{2l_1}}{s_{12} s_{l_2l_3}} \right) \!\eqndot
\end{aligned}
\end{equation}
The first five terms in \eqref{eq: Konishi two-loop triple cut simplified} stem from triple cuts of the integrals in \eqref{eq: Konishi two-loop first cut result} and \eqref{eq: Konishi two-loop second cut result}, as shown in figure \ref{fig: triple cuts of previously known integrals}.
   
The remaining three terms correspond to integrals that could not be detected in the previous cuts:
\begin{equation}
\label{eq: Konishi two-loop triple cut new cut diagrams}
\begin{aligned}
f_{\cK_6,2}^{(2)} \Big|_{\text{TC}}^{\text{III}} = i 18 \left(
\frac{1}{s_{12}}\FDinline[sunrise,triplecut,cutlabels,twolabels,labelone=p_1,labeltwo=p_2] 
- \frac{s_{1 l_3} }{s_{12}}\FDinline[fishtop,triplecut,cutlabels,twolabels,labelone=p_1,labeltwo=p_2] 
- \frac{s_{2 l_1} }{s_{12}}\FDinline[fishbottom,triplecut,cutlabels,twolabels,labelone=p_1,labeltwo=p_2]
\right) \eqndot
\end{aligned}
\end{equation}
Hence,
\begin{equation}
\label{eq: Konishi two-loop triple cut integrals}
\begin{aligned}
f_{\cK_6,2}^{(2),{\text{III}}} &=18 \left(
\frac{1}{s_{12}}\FDinline[sunrise,twolabels,labelone=p_1,labeltwo=p_2] 
- \frac{s_{1 l} }{s_{12}}\FDinline[fishtop,momentum,twolabels,labelone=p_1,labeltwo=p_2] 
- \frac{s_{2 l} }{s_{12}}\FDinline[fishbottom,momentum,twolabels,labelone=p_1,labeltwo=p_2]
\right)
 \\ &
=18\FDinline[fishtop,twolabels,labelone=p_1,labeltwo=p_2] \eqncom
\end{aligned}
\end{equation}
where the last step is valid at the level of the integral, i.e.\ up to terms that integrate to zero.
Similarly to the previous cases, we have to add the result from the triple cut in the legs $p_2$ and $p_1$, which yields a factor of two.

Note that there is also a fourth cut, namely the double cut on which the minimal two-loop form factor $\ffco_{\cK_6,2}^{(2)}$ factorises into the product of the minimal one-loop form factor $\ffco_{\cK_6,2}^{(1)}$ and the tree-level four-point amplitude $\ampco_{4}^{(0)}$. This cut is consistent with the previous cuts and contributes no new integrals; see \cite{Nandan:2014oga} for details. 

Assembling all pieces, the total result for the two-loop minimal form factor of $\cK_6$ is\footnote{This result agrees with the unpublished notes of Boucher-Veronneau, Dixon and Pennington \cite{BDP-notes}. We thank Camille Boucher-Veronneau, Lance Dixon and Jeffrey Pennington for sharing these notes.
}
\begin{equation}
\label{eq: K6 minimal two-loop form factor complete}
\begin{aligned}
f_{\cK_6,2}^{(2)} & =   4 f_{\cK_6,2}^{(2),{\text{I}}}+f_{\cK_6,2}^{(2),{\text{II}}} + 2 f_{\cK_6,2}^{(2),{\text{III}}} \\
& =
-\ 6  (l+p_1)^2(l+p_2)^2
\bigg(4\FDinline[rainbow,momentum,twolabels,labelone=p_1,labeltwo=p_2]+\FDinline[rainbownonplanar,momentum,twolabels,labelone=p_1,labeltwo=p_2]
\bigg)\\
&\phantom{{}={}}+ 36\FDinline[fishtop,twolabels,labelone=p_1,labeltwo=p_2] 
+ \underbrace{s_{12}^2\bigg(4
\FDinline[rainbow, twolabels, labelone=p_1,labeltwo=p_2] +
\FDinline[rainbownonplanar, twolabels, labelone=p_1,labeltwo=p_2]\bigg)}_{f_{\BPS,2}^{(2)}}\eqndot
\end{aligned}
\end{equation}
The integrals occurring in \eqref{eq: K6 minimal two-loop form factor complete} are given in appendix \ref{appsec: selected integrals}. 

Employing \eqref{eq: general divergences of form factor}, however, we find a mismatch with the known two-loop Konishi anomalous dimension \eqref{eq: gammaK}. As already mentioned, this is because $\cK_6$ does not coincide with the Konishi primary operator $\cK$ when regulating the theory in $D=4-2\peps$ dimensions. This subtlety is the subject of the following section.

\section{Subtleties in the regularisation}
\label{sec: subtleties}

In this section, we discuss some important subtleties that arise when applying on-shell methods that were developed for scattering amplitudes to the computation of form factors and correlation functions. These subtleties are related to regularisation and require an extension of the on-shell methods.
For concreteness, we focus on the calculation of the Konishi form factor via unitarity as an example.

As discussed in section \ref{sec: general structure of loop corrections}, the occurrence of divergences in loop calculations requires us to regularise the theory.
In general, such a regularisation should be compatible with the symmetries of the theory.
For gauge theories, the regularisation of choice is dimensional regularisation, i.e.\ continuing the dimension of spacetime from $D=4$ to $D=4-2\peps$.
Conventional dimensional regularisation (CDR) \cite{Collins:1984xc} and the 't Hooft Veltmann (HV) scheme \cite{'tHooft:1972fi}, however, break supersymmetry as the number of vector degrees of freedom is changed --- since the polarisation vectors $\teps_\mu^\pm$ are in $D=4-2\peps$ dimensions --- while the number of scalars $N_\phi=6$ and fermions $N_\psi=4$ stays the same.
A way to regularise the theory while preserving supersymmetry is dimensional reduction (DR) from ten dimensions \cite{Siegel:1979wq,Capper:1979ns}.
It exploits the fact that four-dimensional $\mathcal{N}=4$ SYM theory is the dimensional reduction of ten-dimensional $\mathcal{N}=1$ SYM theory to $D=4$. 
Dimensionally reducing to $D=4-2\peps$ instead, one obtains a regularised supersymmetric theory with $N_\psi=4$ and $N_\phi=10-D=6+2\peps$.
In the DR scheme, the ten-dimensional metric $g^{MN}$, $M,N=0,\dots,9$, is split into the $(4-2\peps)$-dimensional metric $g^{\mu\nu}$, $\mu,\nu=0,\dots,3-2\peps$, and the metric $\delta^{IJ}$, $I,J=1,\dots,6+2\peps$, of the scalar field flavours.
The ten-dimensional gauge field $A_M$ splits into the $(4-2\peps)$-dimensional gauge field $A_\mu$ and $N_\phi=6+2\peps$ scalars $\phi_I$. 

A modification of the DR scheme is the so-called four-dimensional-helicity (FDH) scheme \cite{Bern:1991aq,Bern:2002zk}.
In this scheme, the additional $2\peps$ scalars are absorbed into the vector bosons such that the polarisation vectors are in four dimensions.
As the DR scheme, the FDH scheme apparently preserves supersymmetry, as the number of bosonic and fermionic degrees of freedom match.
Moreover, it allows to use spinor-helicity variables \eqref{eq: spinor helicity variables} and Nair's $\cN=4$ on-shell superfield \eqref{eq: N=4 superspace}.
Most on-shell methods implicitly use the FDH scheme, which has been successful for amplitudes and form factors of BPS operators.
However, as we will argue below, it is incompatible with the occurrence of operators that are sensitive to the reorganisation of the $2\peps$ scalars,
such as the Konishi primary operator $\cK$.

In order to investigate the differences between working in four dimensions, the DR scheme or the FDH scheme, we study the underlying Feynman diagrams. 
In Feynman diagrams, factors of $D=g^\mu_\mu$ and $N_\phi=\delta_I^I$ arise from gauge fields and scalar fields that circulate in a loop in such a way that also their indices form a loop. 
Moreover, such an index loop can exist even though the loop in the field flavours is interrupted e.g.\ by a self-energy insertion.
We call an index loop internally closed if it involves only the elementary vertices of the theory and externally closed if it involves also a composite operator.

Let us consider internally closed index loops first. 
Both vector fields and scalar fields in $\cN=4$ SYM theory originate from the ten-dimensional vector field in $\cN=1$ SYM theory, and so do their elementary interaction vertices.
Hence, for every Feynman diagram with an internally closed vector index loop, there exists an accompanying diagram with an internally closed scalar index loop. 
The sum of both contributions is proportional to $N_\phi+D=10$, which is independent of $\peps$.
As far as internally closed index loops are concerned, one is thus free to work in strictly four dimensions, the FDH scheme, or the DR scheme.
Scattering amplitudes and form factors of the BPS operators $\tr(\phi^L_{14})$ contain only internally closed index loops.
This explains their successful calculation via on-shell methods.

The situation changes for externally closed index loops.
Composite operators in $\mathcal{N}=4$ SYM theory do not in general arise from the dimensional reduction of ten-dimensional composite operators in $\cN=1$ SYM theory. 
In particular, there are composite operators that contain only scalar fields and no vector fields.
Hence, a factor of $N_\phi$ from an externally closed scalar index loop is in general not accompanied by a factor of $D$ from a closed vector index loop and vice versa.
For externally closed index loops, the result thus depends on working in the FDH scheme, in four dimensions or in the DR scheme.
For instance, the $\delta^{IJ}$ in the Konishi primary $\cK=\delta^{IJ}\phi_I\phi_J$ in \eqref{eq: K in so6} can give rise to externally closed index loops.
Via its tensor structure, this operator explicitly depends on the dimension of spacetime.%
\footnote{The difference $\cK-\cK_6$, which vanishes for $D=4$, is an example of a so-called evanescent operator, which also occur in QCD \cite{Buras:1989xd}. For a textbook treatment, see \cite{Collins:1984xc}.}

\begin{figure}[tbp]
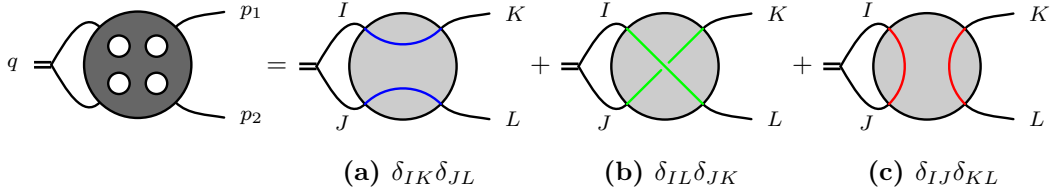

\centering
 $\quad\begin{subfigure}[c]{29\unitlength}
 \FDflow[black,spacetime]
 \end{subfigure}
   \,=\, 
   \begin{subfigure}[c]{29\unitlength}
   \FDflow[blue,flavour]\rule[0.585cm]{0cm}{1.2cm}
   \caption{$\delta_{IK}\delta_{JL}$}
   \label{subfig: a}
   \end{subfigure}
   \,+\,
   \begin{subfigure}[c]{29\unitlength}
   \FDflow[green,flavour]\rule[0.585cm]{0cm}{1.2cm}
   \caption{$\delta_{IL}\delta_{JK}$}
   \label{subfig: b}
   \end{subfigure}
   \,+\,
   \begin{subfigure}[c]{29\unitlength}
   \FDflow[red,flavour]\rule[0.585cm]{0.0cm}{1.2cm}
   \caption{$\delta_{IJ}\delta_{KL}$}
   \label{subfig: c}
   \end{subfigure}$ 
\caption{As a consequence of R-charge conservation, only three different tensor structures can occur in a Feynman diagram that contributes to the minimal $\ell$-loop form factor of the operator $\tr(\phi_I\phi_J)$ with external fields $\phi_K$ and $\phi_L$: (\subref{subfig: a}) $ \delta_{IK}\delta_{JL}$, (\subref{subfig: b}) $\delta_{IL}\delta_{JK}$ and (\subref{subfig: c}) $\delta_{IJ}\delta_{KL}$.}
 \label{fig: flavour flows}
\end{figure}

Let us consider a Feynman diagram that contributes to the $\ell$-loop minimal form factor of the operator $\tr(\phi_I\phi_J)$ with external fields $\phi_K$ and $\phi_L$.
As a consequence of R-charge conservation, only three different tensor structures can occur, as shown in figure \ref{fig: flavour flows}: (\subref{subfig: a}) $\delta_{IK}\delta_{JL}$, (\subref{subfig: b}) $\delta_{IL}\delta_{JK}$ and (\subref{subfig: c}) $\delta_{IJ}\delta_{KL}$. 
They are denoted as identity, permutation and trace, respectively.
In the cases (\subref{subfig: a}) and (\subref{subfig: b}), no externally closed index loop occurs. 
In the case (\subref{subfig: c}), one externally closed index loop occurs.
In strictly four dimensions or the FDH scheme, it yields $N_\phi=6$, while it yields $N_\phi=6+2\peps$ in the DR scheme.
We can hence multiply the tensor structure (\subref{subfig: c}), which is the trace, by a factor 
\begin{equation}
 \label{eq: def r phi}
 r_\phi=\frac{6+2\peps}{6}
\end{equation}
to account for the difference.

The contributions of the tensor structures (\subref{subfig: a}), (\subref{subfig: b}) and (\subref{subfig: c}) to the BPS operator \eqref{eq: BPS in so6} and the Konishi primary operator \eqref{eq: K in so6} can be obtained by contraction with the respective tensor structures in \eqref{eq: BPS in so6} and \eqref{eq: K in so6}.
We find that the sum of (\subref{subfig: a}) and (\subref{subfig: b}) contributes to the BPS operator \eqref{eq: BPS in so6} while the sum of (\subref{subfig: a}), (\subref{subfig: b}) and (\subref{subfig: c}) contributes to the Konishi primary operator \eqref{eq: K in so6}.
Hence, we can single out the tensor structure (\subref{subfig: c}) as difference between the form factors of the Konishi operator and the BPS operator.

In all results of the last section, we have already written the form factor ratios of $\cK_6$ as
\begin{equation}
\label{eq: K6 decomposition}
 f_{\cK_6,2}^{(\ell)}
=f_{{\BPS},2}^{(\ell)}+{\tilde f}_{\cK_6,2}^{(\ell)}
\eqncom
\end{equation}
where $f_{{\BPS},2}^{(\ell)}$ coincides with the form factor ratio of the BPS operator \eqref{eq: BPS in su4} and ${\tilde f}_{\cK_6,2}^{(\ell)}$ is the difference between the form factor ratios of $\cK_6$ and the BPS operator. 
Writing the form factor ratios of $\cK$ as 
\begin{equation}
\label{eq: K decomposition}
 f_{\cK,2}^{(\ell)}
=f_{{\BPS},2}^{(\ell)}+{\tilde f}_{\cK,2}^{(\ell)}
\eqncom
\end{equation}
we can obtain them by the simple replacement rule
\begin{equation}
\label{eq: ftilde-correction-rule}
{\tilde f}_{\cK_6,2}^{(\ell)}
\quad \xrightarrow{r_\phi} \quad 
r_\phi{\tilde f}_{\cK_6,2}^{(\ell)}={\tilde f}_{\cK,2}^{(\ell)}
\eqndot
\end{equation}

The above arguments do not depend on the loop order $\ell$ and should hence be valid for generic $\ell$.%
\footnote{They do, however, rely on the DR scheme, which has known inconsistencies at higher loop orders \cite{Siegel:1980qs,Avdeev:1981vf,Avdeev:1982np,Avdeev:1982xy}.}
Here, we have looked only at the minimal form factor. The above replacement \eqref{eq: ftilde-correction-rule}, however, is valid for general $n$-point form factors of $\cK$, cf.\ \cite{Nandan:2014oga}. 

Moreover, the subtleties analysed in this section for the Konishi form factor also occur for form factors, generalised form factors and correlation functions of other operators that depend on the spacetime dimension via contracted indices.
Our analysis can be straightforwardly generalised to these cases and so should our solution. 
The latter relies on the possibility to decompose the different contributions to these quantities with respect to tensor structures and to calculate the contributions to the different tensor structures independently.

Let us look at some further examples of spacetime-dependent operators. 
Another scalar example is the operator $\tr(\phi_I\phi_I\phi_K)$, which is an eigenstate of the one-loop dilatation operator with one-loop anomalous dimension $8$. 
An example with contracted vector indices instead of scalar indices is $\tr(\cder^\mu\phi_{12}\cder_\mu\phi_{12})$, which is an eigenstate of the one-loop dilatation operator with one-loop anomalous dimension $0$. 
In one-loop diagrams, this contraction of covariant derivatives leads to a contraction of the loop momentum with itself. 
The difference between performing this contraction in $D=4$ or $D=4-2\peps$ dimensions is given by the integral in the Passarino-Veltman reduction discussed in appendix \ref{appsec: PV reduction} that contains $l_\peps^2=l_{(4)}^2 - l^2$ in the numerator. This integral evaluates to a rational term.
Similarly, the replacement \eqref{eq: ftilde-correction-rule} leads to a rational term in $f_{\cK,2}^{(1)}$; it arises when multiplying the term in $r_\phi$ that is linear in $\peps$ with the $\frac{1}{\peps}$ pole in $\tilde{f}_{\cK,2}^{(1)}$. 
Both rational terms arise from contributions to the integrand that vanish for $D=4$ but integrate to a finite value in $D=4-2\peps$.%
\footnote{A complementary view on this subtlety is as follows. Using four-dimensional unitarity, the results have to be lifted to $D$ dimensions, i.e.\ the occurring functions have to be continued from $D=4$ to $D=4-2\peps$. 
This concerns the loop momenta but also the factors arising from flavours, at least if we are interested in the $D$-dimensional theory that is the dimensional reduction of ten-dimensional $\cN=1$ SYM theory. 
The continuation to $D=4-2\peps$ is not unique if terms contribute that vanish for $D=4$, such as $l_\peps^2=l_{(4)}^2 - l^2$ or $\cK-\cK_6$.
However, we can make the lifting unique by decomposing the four-dimensional result in terms of tensor structures that are meaningful for general $D$, such as identity, permutation and trace.
}

\section{Final result and Konishi anomalous dimension}
\label{sec: Konishi results}

Using the modification rule \eqref{eq: ftilde-correction-rule} and the integrals provided in appendix \ref{appsec: selected integrals}, the minimal one-loop form factor of the Konishi primary $\cK$ is given by 
\begin{equation}
\label{eq: final Konishi ff 1-loop}
\begin{aligned}
f_{{\BPS},2}^{(1)} &= \Big(-\frac{q^2}{\mu^2} \Big)^{-\peps}  \left[- \frac{2}{\peps^2} +  \frac{\pi^2}{6} + \frac{14}{3} \zeta_3 \peps + \frac{47}{720} \pi^4 \, \peps^2 \right] + {\cal O} (\peps^3) \eqncom 
\\
{\tilde f}_{\cK,2}^{(1)} & = \Big(-\frac{q^2}{\mu^2} \Big)^{-\peps} \left[- \frac{6}{\peps} - 14 - \Big(28 - \frac{\pi^2}{2} \Big) \peps - \Big( 56 - \frac{7\pi^2}{6} - 14 \zeta_3 \Big) \peps^2 \right] + {\cal O} (\peps^3) \eqncom
\end{aligned}
\end{equation}
where the decomposition into $f_{{\BPS}}$ and $\tilde{f}_{{\cK}}$ was defined in \eqref{eq: K decomposition}.
Similarly, the minimal two-loop form factor of the Konishi primary $\cK$ is given by 
\begin{equation}
\label{eq: final Konishi ff 2-loop}
\begin{aligned}
f_{{\BPS},2}^{(2)} &= \Big(-\frac{q^2}{\mu^2}\Big)^{-2\peps} \left[ \frac{2}{\peps^4} -  \frac{\pi^2}{6 \peps^2} - \frac{25  \zeta_3}{3 \peps} - \frac{7 \pi^4}{60} \right] + {\cal O} (\peps) \eqncom
\\
{\tilde f}_{\cK,2}^{(2)} &= \Big(-\frac{q^2}{\mu^2}\Big)^{-2\peps}  \left[ \frac{12}{\peps^3} +  \frac{46}{\peps^2} + \frac{152 - 2\pi^2 }{ \peps} + \Big( 484 - \frac{35 \pi^2 }{ 3} - 56 \zeta_3\Big) \right] + {\cal O} (\peps) \eqndot
\end{aligned}
\end{equation}
These results match with the results of a direct Feynman diagram calculation, which is presented in \cite{Nandan:2014oga}. 

Comparing \eqref{eq: final Konishi ff 1-loop} and \eqref{eq: final Konishi ff 2-loop} with the general form of loop corrections \eqref{eq: general divergences of form factor}, we find the anomalous dimensions 
\begin{equation}
 \gamma_\cK^{(1)}=12\eqncom \qquad \gamma_\cK^{(2)}=-48 \eqncom
\end{equation}
in perfect agreement with the known values \eqref{eq: gammaK}.

\chapter{Minimal two-loop \texorpdfstring{$\SU2$}{SU(2)} form factors}
\label{chap: two-loop su(2) form factors}

Next, we look at minimal two-loop form factors in the $\SU2$ sector.
The subtleties discussed in the last chapter do not occur in this case.
However, a non-trivial mixing of UV and IR divergences occurs, which demonstrates that the exponentiation of divergences is indeed in terms of interaction \emph{operators} as given in \eqref{eq: general divergences of form factor}.
Furthermore, we study the finite part of the form factors, or, more precisely, the remainder functions.

We calculate the minimal two-loop form factors for generic operators in the $\SU2$ sector via unitarity in section \ref{sec: two-loop form factors via unitarity}.
In section \ref{sec: two-loop dilatation operator}, we extract the two-loop dilatation operator in the $\SU2$ sector from them and, in section \ref{sec: remainder}, the finite remainder function.

This chapter is based on results first published in \cite{Loebbert:2015ova}.

\section{Two-loop form factors via unitarity}
\label{sec: two-loop form factors via unitarity}

In section \ref{sec: one-loop unitarity}, we have calculated the minimal one-loop form factors for operators in the $\SU2$ sector.
Now, we proceed to the next loop order.

Connected interactions at two-loop order can involve either two or three neighbouring fields of the composite operator at a time.%
\footnote{Recall that interactions involving only one field lead to integrals that can only depend on the vanishing scale $p_i^2=0$. These integrals vanish themselves.}
We denote them as having range two or range three, respectively. Moreover, disconnected interactions can occur, which are products of two one-loop interactions.
Hence, we can write the two-loop interaction operator $\Interaction^{(2)}$ in terms of corresponding densities as%
\footnote{We restrict ourselves to $L\geq3$ here. All operators in the $\SU{2}$ sector with $L=2$ are related via $\SU{2}$ symmetry to the BPS operator $\tr(\phi_{14}\phi_{14})$, whose minimal two-loop form factor was first calculated in \cite{vanNeerven:1985ja}. In particular, no finite-size effects contribute here.}
\begin{equation}
\label{eq: two-loop interaction}
 \Interaction^{(2)}=\sum_{i=1}^L \Big( \inttwo[i\,i+1\,i+2] + \inttwo[i\,i+1] + \frac{1}{2} \sum_{j=i+2}^{L+i-2} \intone[i\,i+1] \intone[j\,j+1] \Big) \eqndot
\end{equation}
Similarly to the one-loop case, the densities are operators and can be expressed in terms of their matrix elements as%
\footnote{We are using the notation introduced in section \ref{sec: one-loop unitarity}.}
\begin{equation}
\begin{aligned}
\label{eq: two-loop int expansion}
 \interaction^{(2)}_{i\,i+1} & 
 = \begin{aligned}
 \begin{tikzpicture}
  \drawvlineblob{1}{1}
  \drawvlineblob{2}{1}
  \drawtwoblob{1}{1}{$\inttwoi$}
 \end{tikzpicture}
 \end{aligned}
 &&=\sum_{A,B,C,D=1}^2 \inttwoif{Z_AZ_B}{Z_CZ_D} \etatt_{i}^C\frac{\partial}{\partial \etatt_{i}^A} \etatt_{i+1}^D\frac{\partial}{\partial \etatt_{i+1}^B}\eqncom \\
  \interaction^{(2)}_{i\,i+1\,i+2} & =
  \begin{aligned}
 \begin{tikzpicture}
  \drawvlineblob{1}{1}
  \drawvlineblob{2}{1}
  \drawvlineblob{3}{1}
  \drawthreeblob{1}{1}{$\inttwoi$}
 \end{tikzpicture}
 \end{aligned}
  &&=\sum_{A,B,C,D,E,F=1}^2 \inttwoif{Z_AZ_BZ_C}{Z_DZ_EZ_F} \etatt_{i}^D\frac{\partial}{\partial \etatt_{i}^A} \etatt_{i+1}^E\frac{\partial}{\partial \etatt_{i+1}^B}  \etatt_{i+2}^F\frac{\partial}{\partial \etatt_{i+2}^C}\eqndot &&
  \end{aligned}
\end{equation}
Several of these matrix elements vanish due to $\SU4$ charge conservation and further matrix elements are trivially related to each other via relabelling $X\leftrightarrow Y$ or inverting the order of the fields.
Hence, we only need to calculate $\inttwoif{XX}{XX}$, $\inttwoif{XY}{XY}$ and $\inttwoif{XY}{YX}$ at range two as well as $\inttwoif{XXX}{XXX}$, $\inttwoif{XXY}{XXY}$, $\inttwoif{XXY}{XYX}$, $\inttwoif{XXY}{YXX}$, $\inttwoif{XYX}{XYX}$ and $\inttwoif{XYX}{XXY}$ at range three.

The matrix elements can be calculated via the on-shell unitarity method as in the last two chapters. The required cuts are shown in figure \ref{fig: two-loop cuts SU(2)}. 
The cuts in figures \ref{fig: one two-loop double cut}, \ref{fig: two-loop p1+p2 triple cut} and \ref{fig: other two-loop double cut} have already been discussed in the case of the Konishi two-loop form factor treated in the previous chapter and they can be calculated analogously in the present case.
In addition to these cuts, also the triple cut in the three-particle channel is required, which is shown in figure \ref{fig: two-loop p1+p2+p3 triple cut}. On this cut, the minimal two-loop form factor $\ffco_{\cO,L}^{(2)}$ factorises into the product of the minimal tree-level form factor $\ffco_{\cO,L}^{(0)}$ and the tree-level six-point amplitude $\ampco_{6}^{(0)}$. The required amplitudes are of NMHV type; we give explicit expressions for them in \eqref{eq: six- point amplitudes} of appendix \ref{appsec: scalar nmhv six-point amplitudes}. Via the triple cut in the three-particle channel, each term in \eqref{eq: six- point amplitudes} maps to one of the integrals occurring in the interactions of range three.
\def\middletension{0.8}
\begin{figure}[tbp]	
\centering
\begin{subfigure}[t]{0.49\textwidth}
 \centering
$
\settoheight{\eqoff}{$\times$}%
\setlength{\eqoff}{0.5\eqoff}%
\addtolength{\eqoff}{-12.0\unitlength}%
\raisebox{\eqoff}{%
\fmfframe(2,2)(2,2){%
\begin{fmfchar*}(60,25)
\fmfleft{vp3,vp,vpL,vq}
\fmfright{vp2,vp1}
\fmf{dbl_plain_arrow,tension=1.2}{vq,v1}
\fmf{plain_arrow,tension=0}{v1,vpL}
\fmf{plain_arrow,tension=1.2}{v1,vp3}
\fmf{plain_arrow,left=0.7,label=$\scriptstyle l_1$,l.d=8,tension=0.4}{v1,v2}
\fmf{plain_arrow,right=0.7,label=$\scriptstyle l_2$,l.d=8,tension=0.4}{v1,v2}
\fmf{phantom_smallcut,left=0.7,tension=0.4}{v1,v2}
\fmf{phantom_smallcut,right=0.7,tension=0.4}{v1,v2}
\fmf{plain_arrow,tension=1.2}{v2,vp1}
\fmf{plain_arrow,tension=1.2}{v2,vp2}
\fmfv{decor.shape=circle,decor.filled=30,decor.size=30,label=$\scriptstyle \ffco_{\cO,,L}$,label.dist=0}{v1}
\fmfv{decor.shape=circle,decor.filled=0,decor.size=25}{v2}
\fmffreeze
\fmfdraw
 \fmfcmd{pair vertq, vertpone, vertptwo, vertpthree, vertpL, vertone, verttwo; vertone = vloc(__v1); verttwo = vloc(__v2); vertq = vloc(__vq); vertpone = vloc(__vp1); vertptwo = vloc(__vp2); vertpthree = vloc(__vp3);vertpL = vloc(__vpL);}
\fmfiv{decor.shape=circle,decor.filled=50,decor.size=35,label=$\scriptstyle \ampco_{4}$,label.dist=0}{verttwo}
\fmfiv{decor.shape=circle,decor.filled=0,decor.size=25,label=$\scriptstyle \ampco_{4}$,label.dist=0}{verttwo}
 \fmfiv{label=$\scriptstyle q$}{vertq}
 \fmfiv{label=$\scriptstyle p_1$}{vertpone}
 \fmfiv{label=$\scriptstyle p_2$}{vertptwo}
 \fmfiv{label=$\scriptstyle p_3$}{vertpthree}
 \fmfiv{label=$\scriptstyle p_L$}{vertpL}
 \fmfiv{label=$\cdot$,l.d=20,l.a=-150}{vertone}
 \fmfiv{label=$\cdot$,l.d=20,l.a=-165}{vertone}
 \fmfiv{label=$\cdot$,l.d=20,l.a=-180}{vertone}
\end{fmfchar*}%
}}%
$
\caption{First double cut of $\ffco_{\cO,L}^{(2)}$ in the two-particle channel $(p_1+p_2)^2$.}
\label{fig: one two-loop double cut}
\end{subfigure}
\begin{subfigure}[t]{0.49\textwidth}
 \centering
$
\settoheight{\eqoff}{$\times$}%
\setlength{\eqoff}{0.5\eqoff}%
\addtolength{\eqoff}{-12.0\unitlength}%
\raisebox{\eqoff}{%
\fmfframe(2,2)(2,2){%
\begin{fmfchar*}(60,25)
\fmfleft{vp4,vp,vpL,vq}
\fmfright{vp3,vp2,vp1}
\fmf{dbl_plain_arrow,tension=1.2}{vq,v1}
\fmf{plain_arrow,tension=0}{v1,vpL}
\fmf{plain_arrow,tension=1.2}{v1,vp4}
\fmf{plain_arrow,left=0.7,label=$\scriptstyle l_1$,l.d=8,tension=0.8}{v1,v2}
\fmf{plain_arrow,right=0.7,label=$\scriptstyle l_3$,l.d=8,tension=0.8}{v1,v2}
\fmf{phantom_smallcut,left=0.7,tension=0}{v1,v2}
\fmf{phantom_smallcut,right=0.7,tension=0}{v1,v2}
\fmf{plain_arrow,tension=0,tag=1}{v1,v2}
\fmf{phantom_smallcut,tension=0}{v1,v2}
\fmf{plain_arrow,tension=1.2}{v2,vp1}
\fmf{plain_arrow,tension=0}{v2,vp2}
\fmf{plain_arrow,tension=1.2}{v2,vp3}
\fmfv{decor.shape=circle,decor.filled=30,decor.size=30,label=$\scriptstyle \ffco_{\cO,,L}$,label.dist=0}{v1}
\fmfv{decor.shape=circle,decor.filled=50,decor.size=30,label=$\scriptstyle \ampco_{6}$,label.dist=0}{v2}
\fmffreeze
 \fmfcmd{pair vertq, vertpone, vertptwo, vertpthree, vertpfour, vertpL, vertone, verttwo; vertone = vloc(__v1); verttwo = vloc(__v2); vertq = vloc(__vq); vertpone = vloc(__vp1); vertptwo = vloc(__vp2); vertpthree = vloc(__vp3); vertpfour = vloc(__vp4);vertpL = vloc(__vpL);}
 \fmfiv{label=$\scriptstyle q$}{vertq}
 \fmfiv{label=$\scriptstyle p_1$}{vertpone}
 \fmfiv{label=$\scriptstyle p_2$}{vertptwo}
 \fmfiv{label=$\scriptstyle p_3$}{vertpthree}
 \fmfiv{label=$\scriptstyle p_4$}{vertpfour}
 \fmfiv{label=$\scriptstyle p_L$}{vertpL}
 \fmfiv{label=$\cdot$,l.d=20,l.a=-150}{vertone}
 \fmfiv{label=$\cdot$,l.d=20,l.a=-165}{vertone}
 \fmfiv{label=$\cdot$,l.d=20,l.a=-180}{vertone}
\fmfipath{p[]}
\fmfiset{p1}{vpath1(__v1,__v2)}
\fmfiv{label=$\scriptstyle l_2\,,$,l.d=5,l.a=-115}{point length(p1)/2 of p1}
\end{fmfchar*}%
}}%
$
\caption{Triple cut of $\ffco_{\cO,L}^{(2)}$ in the three-particle channel $(p_1+p_2+p_3)^2$.}
\label{fig: two-loop p1+p2+p3 triple cut}
\end{subfigure}
\\[0.5\baselineskip]
\begin{subfigure}[t]{0.49\textwidth}
 \centering
$
\settoheight{\eqoff}{$\times$}%
\setlength{\eqoff}{0.5\eqoff}%
\addtolength{\eqoff}{-12.0\unitlength}%
\raisebox{\eqoff}{%
\fmfframe(2,2)(2,2){%
\begin{fmfchar*}(60,25)
\fmfleft{vp3,vp,vpL,vq}
\fmfright{vp2,vp1}
\fmf{dbl_plain_arrow,tension=1.2}{vq,v1}
\fmf{plain_arrow,tension=0}{v1,vpL}
\fmf{plain_arrow,tension=1.2}{v1,vp3}
\fmf{plain_arrow,left=0.7,label=$\scriptstyle l_1$,l.d=8,tension=0.4}{v1,v2}
\fmf{plain_arrow,right=0.7,label=$\scriptstyle l_3$,l.d=8,tension=0.4}{v1,v2}
\fmf{phantom_smallcut,left=0.7,tension=0.4}{v1,v2}
\fmf{phantom_smallcut,right=0.7,tension=0.4}{v1,v2}
\fmf{plain_arrow,tension=0,tag=1}{v1,v2}
\fmf{phantom_smallcut,tension=0}{v1,v2}
\fmf{plain_arrow,tension=1.2}{v2,vp1}
\fmf{plain_arrow,tension=1.2}{v2,vp2}
\fmfv{decor.shape=circle,decor.filled=30,decor.size=30,label=$\scriptstyle \ffco_{\cO,,L+1}$,label.dist=0}{v1}
\fmfv{decor.shape=circle,decor.filled=50,decor.size=30,label=$\scriptstyle \ampco_{5}$,label.dist=0}{v2}
\fmffreeze
 \fmfcmd{pair vertq, vertpone, vertptwo, vertpthree, vertpL, vertone, verttwo; vertone = vloc(__v1); verttwo = vloc(__v2); vertq = vloc(__vq); vertpone = vloc(__vp1); vertptwo = vloc(__vp2); vertpthree = vloc(__vp3);vertpL = vloc(__vpL);}
 \fmfiv{label=$\scriptstyle q$}{vertq}
 \fmfiv{label=$\scriptstyle p_1$}{vertpone}
 \fmfiv{label=$\scriptstyle p_2$}{vertptwo}
 \fmfiv{label=$\scriptstyle p_3$}{vertpthree}
 \fmfiv{label=$\scriptstyle p_L$}{vertpL}
 \fmfiv{label=$\cdot$,l.d=20,l.a=-150}{vertone}
 \fmfiv{label=$\cdot$,l.d=20,l.a=-165}{vertone}
 \fmfiv{label=$\cdot$,l.d=20,l.a=-180}{vertone}
\fmfipath{p[]}
\fmfiset{p1}{vpath1(__v1,__v2)}
\fmfiv{label=$\scriptstyle l_2\,,$,l.d=5,l.a=-115}{point length(p1)/2 of p1}
\end{fmfchar*}%
}}%
$
\caption{Triple cut of $\ffco_{\cO,L}^{(2)}$ in the two-particle channel $(p_1+p_2)^2$.}
\label{fig: two-loop p1+p2 triple cut}
\end{subfigure}
\begin{subfigure}[t]{0.49\textwidth}
 \centering
$
\settoheight{\eqoff}{$\times$}%
\setlength{\eqoff}{0.5\eqoff}%
\addtolength{\eqoff}{-12.0\unitlength}%
\raisebox{\eqoff}{%
\fmfframe(2,2)(2,2){%
\begin{fmfchar*}(60,25)
\fmfleft{vp3,vp,vpL,vq}
\fmfright{vp2,vp1}
\fmf{dbl_plain_arrow,tension=1.2}{vq,v1}
\fmf{plain_arrow,tension=0}{v1,vpL}
\fmf{plain_arrow,tension=1.2}{v1,vp3}
\fmf{plain_arrow,left=0.7,label=$\scriptstyle l_1$,l.d=8,tension=0.4}{v1,v2}
\fmf{plain_arrow,right=0.7,label=$\scriptstyle l_2$,l.d=8,tension=0.4}{v1,v2}
\fmf{phantom_smallcut,left=0.7,tension=0.4}{v1,v2}
\fmf{phantom_smallcut,right=0.7,tension=0.4}{v1,v2}
\fmf{plain_arrow,tension=1.2}{v2,vp1}
\fmf{plain_arrow,tension=1.2}{v2,vp2}
\fmfv{decor.shape=circle,decor.filled=0,decor.size=25}{v1}
\fmfv{decor.shape=circle,decor.filled=50,decor.size=30,label=$\scriptstyle \ampco_{4}$,label.dist=0}{v2}
\fmffreeze
\fmfdraw
 \fmfcmd{pair vertq, vertpone, vertptwo, vertpthree, vertpL, vertone, verttwo; vertone = vloc(__v1); verttwo = vloc(__v2); vertq = vloc(__vq); vertpone = vloc(__vp1); vertptwo = vloc(__vp2); vertpthree = vloc(__vp3);vertpL = vloc(__vpL);}
\fmfiv{decor.shape=circle,decor.filled=30,decor.size=35,label=$\scriptstyle \ffco_{\cO,,L}$,label.dist=0}{vertone}
\fmfiv{decor.shape=circle,decor.filled=0,decor.size=25,label=$\scriptstyle \ffco_{\cO,,L}$,label.dist=0}{vertone}
 \fmfiv{label=$\scriptstyle q$}{vertq}
 \fmfiv{label=$\scriptstyle p_1$}{vertpone}
 \fmfiv{label=$\scriptstyle p_2$}{vertptwo}
 \fmfiv{label=$\scriptstyle p_3$}{vertpthree}
 \fmfiv{label=$\scriptstyle p_L$}{vertpL}
 \fmfiv{label=$\cdot$,l.d=20,l.a=-150}{vertone}
 \fmfiv{label=$\cdot$,l.d=20,l.a=-165}{vertone}
 \fmfiv{label=$\cdot$,l.d=20,l.a=-180}{vertone}
\end{fmfchar*}%
}}%
$
\caption{Second double cut of $\ffco_{\cO,L}^{(2)}$ in the two-particle channel $(p_1+p_2)^2$.}
\label{fig: other two-loop double cut}
\end{subfigure}
\caption{Unitary cuts of the minimal two-loop form factor $\ffco_{\cO,L}^{(2)}$ in the $\SU2$ sector.
}
\label{fig: two-loop cuts SU(2)}
\end{figure}

The resulting matrix elements are shown in table \ref{tab: 2-loop range 2} for range two and table \ref{tab: 2-loop range 3} for range three.%
\footnote{The matrix elements $\inttwoif{XX}{XX}$ and $\inttwoif{XXX}{XXX}$ also occur for the BPS vacuum treated in \cite{Brandhuber:2014ica} and our results agree with the ones found there.
}
\begin{table}[tbp]
\begin{center}
\begin{tabular}{|l|c|c|c|}\hline
\qquad\qquad$\inttwoif{}{}$&${}^{\XYsize XX}_{\XYsize XX}$&${}_{\XYsize XY}^{\XYsize XY}$&${}_{\XYsize XY}^{\XYsize YX}$
\\\hline
\FDinline[rainbow,twolabels,labelone=\scriptscriptstyle i,labeltwo=\scriptscriptstyle i+1]
$s_{i\,i+1}^2$
&+1&+1&0
\\
\FDinline[doubletriangletwo,threelabels,labelone=\scriptscriptstyle i,labeltwo=\scriptscriptstyle \phantom{i+1},labelthree=\scriptscriptstyle i+1]
$s_{i\,i+1}$
&+1&+1&0
\\\hline
\FDinline[rainbow,momentum,twolabels,labelone=\scriptscriptstyle i,labeltwo=\scriptscriptstyle i+1] 
$ s_{i\, i+1}s_{i\, l}$
&0&+1&-1
\\
\FDinline[fishtop,twolabels,labelone=\scriptscriptstyle i,labeltwo=\scriptscriptstyle i+1]
&0&+1&-1
\\
\FDinline[fishbottom,twolabels,labelone=\scriptscriptstyle i,labeltwo=\scriptscriptstyle i+1]
&0&+1&-1
\\\hline
\end{tabular}
\end{center}
\caption{Linear combinations of Feynman integrals forming the matrix elements of range two for the minimal two-loop form factors in the $\SU2$ sector.
Integrals between horizontal lines occur in fixed combinations.}
\label{tab: 2-loop range 2}
\end{table}
\begin{table}[tbp]
\begin{center}
\begin{tabular}{|l|c|c|c|c|c|c|c|}\hline
  \qquad\qquad$\inttwoif{}{}$&${}_{\XYsize XXX}^{\XYsize XXX}$&${}_{\XYsize XXY}^{\XYsize XXY}$&${}_{\XYsize XYX}^{\XYsize XYX}$&${}_{\XYsize XXY}^{\XYsize XYX}$&${}_{\XYsize XYX}^{\XYsize XXY}$&${}_{\XYsize XXY}^{\XYsize YXX}$
\\\hline
\FDinline[trianglebox,momentum,threelabels,labelone=\scriptscriptstyle i,labeltwo=\scriptscriptstyle i+1,labelthree=\scriptscriptstyle i+2]
$s_{il}s_{i+1i+2}$
&+1&+1&+1&0&0&0
\\
\FDinline[boxtriangle,momentum,threelabels,labelone=\scriptscriptstyle i,labeltwo=\scriptscriptstyle i+1,labelthree=\scriptscriptstyle i+2] 
$s_{ii+1}s_{i+2l}$
&+1&+1&+1&0&0&0
\\
\FDinline[doubletrianglethree,threelabels,labelone=\scriptscriptstyle i,labeltwo=\scriptscriptstyle i+1,labelthree=\scriptscriptstyle i+2]
$s_{ii+1i+2}$
&-1&-1&-1&0&0&0
\\\hline
\FDinline[boxbubble,threelabels,labelone=\scriptscriptstyle i,labeltwo=\scriptscriptstyle i+1,labelthree=\scriptscriptstyle i+2]
$s_{ii+1}$
&0&+1&+1&-1&-1&0
\\
\FDinline[triangleshiftedtriangle,momentum,threelabels,labelone=\scriptscriptstyle i,labeltwo=\scriptscriptstyle i+1,labelthree=\scriptscriptstyle i+2]
$s_{il}$
&0&+1&+1&-1&-1&0
\\
\FDinline[trianglebubblethree,threelabels,labelone=\scriptscriptstyle i,labeltwo=\scriptscriptstyle i+1,labelthree=\scriptscriptstyle i+2]
&0&-1&-1&+1&+1&0
\\\hline
\FDinline[shiftedtrianglebubble,threelabels,labelone=\scriptscriptstyle i,labeltwo=\scriptscriptstyle i+1,labelthree=\scriptscriptstyle i+2]
&0&0&+1&-1&0&+1
\\\hline
\FDinline[bubblebox,threelabels,labelone=\scriptscriptstyle i,labeltwo=\scriptscriptstyle i+1,labelthree=\scriptscriptstyle i+2]
$s_{i+1i+2}$
&0&0&+1&0&0&0
\\
\FDinline[shiftedtriangletriangle,momentum,threelabels,labelone=\scriptscriptstyle i,labeltwo=\scriptscriptstyle i+1,labelthree=\scriptscriptstyle i+2]
$s_{i+2l}$
&0&0&+1&0&0&0
\\
\FDinline[bubbletrianglethree,threelabels,labelone=\scriptscriptstyle i,labeltwo=\scriptscriptstyle i+1,labelthree=\scriptscriptstyle i+2]
&0&0&-1&0&0&0
\\\hline
\FDinline[bubbleshiftedtriangle,threelabels,labelone=\scriptscriptstyle i,labeltwo=\scriptscriptstyle i+1,labelthree=\scriptscriptstyle i+2]
&0&0&+1&0&-1&0
\\\hline
\end{tabular}
\end{center}
\caption{Linear combinations of Feynman integrals forming the matrix elements of range three for the minimal two-loop form factors in the $\SU2$ sector.
Integrals between horizontal lines occur in fixed combinations.}
\label{tab: 2-loop range 3}
\end{table}
They 
satisfy the linear relations
\begin{equation}
 \inttwoif{XY}{XY}+\inttwoif{XY}{YX}=\inttwoif{XX}{XX}
\end{equation}
and
\begin{equation}
\label{eq: I2 identities}
\begin{aligned}
 \inttwoif{XXY}{YXX}+\inttwoif{XXY}{XYX}+\inttwoif{XXY}{XXY}=\inttwoif{XXX}{XXX}\eqncom \\
 \inttwoif{XYX}{XYX}+\inttwoif{XYX}{YXX}+\inttwoif{XYX}{XXY}=\inttwoif{XXX}{XXX}\eqncom \\
 \inttwoif{XXY}{XYX}+\inttwoif{XXY}{YXX}=\inttwoif{XYX}{XXY}+\inttwoif{YXX}{XXY}\eqndot
\end{aligned}
\end{equation}
These identities are a consequence of 
\begin{equation}
\label{eq: symmetry commutator two-loop}
 [\mathfrak{J}^A,\Interaction^{(2)}]=0\eqncom
\end{equation}
which follows from the Ward identity \eqref{eq: action on form factor} as in the one-loop case.
Accordingly, we can bring the two-loop interaction density into the form
\begin{equation}
\label{eq: two-loop range two interaction operator}
\begin{aligned}
\inttwo[i\,i+1]&= \big(\inttwoif{XX}{XX}-\inttwoif{XY}{YX}\big) \idm_{i\,i+1} + \inttwoif{XY}{YX} \PP_{i\,i+1} 
\end{aligned}
\end{equation}
and 
\begin{equation}
\label{eq: two-loop range three interaction operator}
 \begin{aligned}
  \inttwo[i\,i+1\,i+2]&= 
  \big(\inttwoif{XXX}{XXX}-\inttwoif{YXX}{XYX}-\inttwoif{XXY}{XYX}\big) \idm_{i\,i+1\,i+2}\\
  &\phaneq
  +\big(\inttwoif{YXX}{XYX}-\inttwoif{XXY}{YXX}\big)\perm_{i\,i+1}%
  +\big(\inttwoif{XXY}{XYX}-\inttwoif{YXX}{XXY}\big)%
  \perm_{i+1\,i+2}\\
  &\phaneq
  +\inttwoif{XXY}{YXX}\perm_{i\,i+1}\perm_{i+1\,i+2}
  +\inttwoif{YXX}{XXY}\perm_{i+1\,i+2}\perm_{i\,i+1} 
  \eqncom
 \end{aligned}
\end{equation}
where $\idm_{i\,i+1\,i+2}$ is defined in analogy to \eqref{eq: def identity operator}.

The Feynman integrals occurring in tables \ref{tab: 2-loop range 2} and \ref{tab: 2-loop range 3} can be reduced to master integrals via IBP reduction, which is implemented e.g.\ in the \texttt{Mathematica} package \texttt{LiteRed}\cite{Lee:2013mka}. Expressions for the required master integrals can be found in \cite{Gehrmann:2000zt}.

\section{Two-loop dilatation operator}
\label{sec: two-loop dilatation operator}

As in the previous chapters, we can extract the dilatation operator by comparing the results \eqref{eq: one-loop int su2} and \eqref{eq: two-loop range two interaction operator}, \eqref{eq: two-loop range three interaction operator} for the one-loop and two-loop interaction operators \eqref{eq: one-loop int su2} and \eqref{eq: two-loop interaction} with the general form \eqref{eq: general divergences of form factor}.
Inserting the explicit expressions for the occurring Feynman integrals, we find%
\footnote{While there is in general a freedom to define the dilatation operator density \eqref{eq: dilatation operator density SU(2) two-loop}, the requirement of having a finite density effectively eliminates this freedom for the remainder function; see the discussion below. Here, we give the dilatation operator density that corresponds to the finite remainder function density.}
\begin{equation}
\label{eq: dilatation operator density SU(2) two-loop}
 \dilatwo[i\,i+1\,i+2]=-2(4\idm_{i\,i+1\,1+2}-3\perm_{i\,i+1}-3\perm_{i+1\,i+2}+\perm_{i\,i+1}\perm_{i+1\,i+2}+\perm_{i+1\,i+2}\perm_{i\,i+1})\eqndot
\end{equation}
This matches the known result of \cite{Beisert:2003tq}.

Having dealt with the divergent terms in \eqref{eq: general divergences of form factor}, let us now turn to the finite terms and in particular to the remainder function.

\section{Remainder}
\label{sec: remainder}

For scattering amplitudes, the BDS ansatz \cite{Anastasiou:2003kj,Bern:2005iz} states that the finite part of the logarithm of the loop correction \eqref{eq: universal IR divergences} is entirely determined by the 
one-loop result. 
While this ansatz gives correct predictions for four and five points, it misses certain contributions starting at six points \cite{Alday:2007he,Bartels:2008ce, Bern:2008ap,Drummond:2008aq}.
The contributions missed by the BDS ansatz are known as remainder functions.
In addition to scattering amplitudes, remainder functions were also studied for form factors of BPS operators \cite{Brandhuber:2012vm, Brandhuber:2014ica}.
In both cases, they were found to be of maximal uniform transcendentality and could be vastly simplified using the so-called symbol techniques \cite{Goncharov09, Goncharov:2010jf}.

For form factors of non-protected operators that renormalise non-diagonally, the definition of the remainder function has to be generalised to
\begin{equation}
\label{eq: remainder}
 \Rem=\Interactionr^{(2)}(\peps) - \frac{1}{2}\left(\Interactionr^{(1)}(\peps)\right)^2 -  f^{(2)}(\peps)\Interactionr^{(1)}(2\peps)  
 + \cO (\peps ) \eqncom
 \end{equation}
where 
\begin{equation}
 f^{(2)} (\peps)= - 2 \zeta_2  - 2 \zeta_3 \peps - 2 \zeta_4 \peps^2 
\end{equation}
as in the amplitude case \cite{Bern:2005iz}
and the renormalised interaction operators $\Interactionr^{(\ell)}$ have been defined in \eqref{eq: def renormalised interaction operators}.
In particular, $\Rem$ is an operator itself.
The term $f^{(2)}(\peps)\Interactionr^{(1)}(2\peps)$ subtracts the result expected due to the universality and exponentiation of the IR divergences.  
In general, $f^{(\ell)}$ is connected to the cusp and collinear anomalous dimensions as $f^{(\ell)}=\frac{1}{8}\gamma_{\text{cusp}}^{(\ell)}+\peps\frac{\ell}{4}\cG_0^{(\ell)}+\cO(\peps^2)$ \cite{Bern:2005iz}.

As all disconnected contributions in $\Interactionr^{(2)}$ are cancelled by the square of $\Interactionr^{(1)}$ in \eqref{eq: remainder}, the remainder can be written in terms of a density of range three:
\begin{equation}
\label{eq: sum of remainder densities}
\Rem = 
\sum_{i=1}^L \rem_{i\,i+1\,i+2} \eqndot
\end{equation}
This density is explicitly given by
\begin{equation}
\begin{aligned}
 &\rem_{i\,i+1\,i+2}\\&= \frac{1}{2} 
 \begin{aligned}
 \begin{tikzpicture}
  \drawvlineblob{1}{1}
  \drawvlineblob{2}{1}
  \drawvlineblob{3}{1}
  \drawtwoblob{1}{1}{$\inttwoi$}
 \end{tikzpicture}
 \end{aligned}
 +
 \begin{aligned}
 \begin{tikzpicture}
  \drawvlineblob{1}{1}
  \drawvlineblob{2}{1}
  \drawvlineblob{3}{1}
  \drawthreeblob{1}{1}{$\inttwoi$}
 \end{tikzpicture}
 \end{aligned}
 +
 \frac{1}{2} 
 \begin{aligned}
 \begin{tikzpicture}
  \drawvlineblob{1}{1}
  \drawvlineblob{2}{1}
  \drawvlineblob{3}{1}
  \drawtwoblob{2}{1}{$\inttwoip$}
 \end{tikzpicture}
 \end{aligned}
 + 
 \begin{aligned}
 \begin{tikzpicture}
  \drawvlineblob{1}{1}
  \drawvlineblob{2}{1}
  \drawvlineblob{3}{1}
  \drawthreeblob{1}{1}{$\ztwoi$}
 \end{tikzpicture}
 \end{aligned}
 +
 \frac{1}{2} 
 \begin{aligned}
 \begin{tikzpicture}
  \drawvlineblob{1}{2}
  \drawvlineblob{2}{2}
  \drawvlineblob{3}{2}
  \drawtwoblob{1}{1}{$\zonei$}
  \drawtwoblob{1}{2}{$\intonei$}
 \end{tikzpicture}
 \end{aligned}
 +
 \begin{aligned}
 \begin{tikzpicture}
  \drawvlineblob{1}{2}
  \drawvlineblob{2}{2}
  \drawvlineblob{3}{2}
  \drawtwoblob{1}{1}{$\zonei$}
  \drawtwoblob{2}{2}{$\intoneip$}
 \end{tikzpicture}
 \end{aligned}
 +
 \begin{aligned}
 \begin{tikzpicture}
  \drawvlineblob{1}{2}
  \drawvlineblob{2}{2}
  \drawvlineblob{3}{2}
  \drawtwoblob{1}{2}{$\intonei$}
  \drawtwoblob{2}{1}{$\zoneip$}
 \end{tikzpicture}
 \end{aligned}
 + 
 \frac{1}{2} 
 \begin{aligned}
 \begin{tikzpicture}
  \drawvlineblob{1}{2}
  \drawvlineblob{2}{2}
  \drawvlineblob{3}{2}
  \drawtwoblob{2}{1}{$\zoneip$}
  \drawtwoblob{2}{2}{$\intoneip$}
 \end{tikzpicture}
 \end{aligned}
 \\
 &\phaneq-\frac{1}{2}
 \left(
  \frac{1}{2} 
 \begin{aligned}
 \begin{tikzpicture}
  \drawvlineblob{1}{2}
  \drawvlineblob{2}{2}
  \drawvlineblob{3}{2}
  \drawtwoblob{1}{1}{$\intoneir$}
  \drawtwoblob{1}{2}{$\intoneir$}
 \end{tikzpicture}
 \end{aligned}
 +
 \begin{aligned}
 \begin{tikzpicture}
  \drawvlineblob{1}{2}
  \drawvlineblob{2}{2}
  \drawvlineblob{3}{2}
  \drawtwoblob{1}{1}{$\intoneir$}
  \drawtwoblob{2}{2}{$\intoneipr$}
 \end{tikzpicture}
 \end{aligned}
 +
 \begin{aligned}
 \begin{tikzpicture}
  \drawvlineblob{1}{2}
  \drawvlineblob{2}{2}
  \drawvlineblob{3}{2}
  \drawtwoblob{1}{2}{$\intoneir$}
  \drawtwoblob{2}{1}{$\intoneipr$}
 \end{tikzpicture}
 \end{aligned}
 + 
 \frac{1}{2} 
 \begin{aligned}
 \begin{tikzpicture}
  \drawvlineblob{1}{2}
  \drawvlineblob{2}{2}
  \drawvlineblob{3}{2}
  \drawtwoblob{2}{1}{$\intoneipr$}
  \drawtwoblob{2}{2}{$\intoneipr$}
 \end{tikzpicture}
 \end{aligned}
 \right)
 - f^{(2)}
 \left(
 \frac{1}{2} 
 \begin{aligned}
 \begin{tikzpicture}
  \drawvlineblob{1}{1}
  \drawvlineblob{2}{1}
  \drawvlineblob{3}{1}
  \drawtwoblob{1}{1}{$\intoneir$}
 \end{tikzpicture}
 \end{aligned}
 +
 \frac{1}{2} 
 \begin{aligned}
 \begin{tikzpicture}
  \drawvlineblob{1}{1}
  \drawvlineblob{2}{1}
  \drawvlineblob{3}{1}
  \drawtwoblob{2}{1}{$\intoneipr$}
 \end{tikzpicture}
 \end{aligned}
 \right)_{\peps\rightarrow2\peps}
 \eqndot
 \end{aligned}
 \label{eq: remainder density as graphs}
\end{equation}
Note that there is a freedom of defining the density $\rem_{i\,i+1\,i+2}$, which is related to distributing the contributions with effective range two  between the first and second pair of neighbouring legs.
In order to obtain a finite density, however, these contributions have to be distributed equally; hence the prefactors of $\frac{1}{2}$ in front of the respective terms in \eqref{eq: remainder density as graphs}.%
\footnote{For other choices that lead to divergent densities, the divergent contributions cancel in the sum \eqref{eq: sum of remainder densities}.}
The occurring renormalisation constants are depicted in analogy to the interaction operators. They can be obtained from the dilatation operator via \eqref{eq: Z in terms of D}.

The different matrix elements of the remainder density satisfy analogous identities to those of $\inttwo[i\,i+1\,i+2]$: 
\begin{equation}
\label{eq: remi identities}
\begin{aligned}
 \remif{XXY}{YXX}+\remif{XXY}{XYX}+\remif{XXY}{XXY}=\remif{XXX}{XXX}\eqncom \\
 \remif{XYX}{XYX}+\remif{XYX}{YXX}+\remif{XYX}{XXY}=\remif{XXX}{XXX}\eqncom \\
 \remif{XXY}{XYX}+\remif{XXY}{YXX}=\remif{XYX}{XXY}+\remif{YXX}{XXY}\eqndot
\end{aligned}
\end{equation}
These identities are a consequence of $\SU2$ symmetry and follow from 
\begin{equation}
\label{eq: symmetry commutator remainder}
 [\mathfrak{J}^A,\Rem]=0\eqncom
\end{equation}
which can be derived from \eqref{eq: symmetry commutator one-loop} and \eqref{eq: symmetry commutator two-loop}.
Accordingly, we can write 
\begin{equation}
\label{eq: remainder operator}
 \begin{aligned}
  \rem_{i\,i+1\,i+2}&= 
  \big(\remif{XXX}{XXX}-\remif{YXX}{XYX}-\remif{XXY}{XYX}\big) \idm_{i\,i+1\,i+2}\\
  &\phaneq
  +\big(\remif{YXX}{XYX}-\remif{XXY}{YXX}\big)\perm_{i\,i+1}%
  +\big(\remif{XXY}{XYX}-\remif{YXX}{XXY}\big)%
  \perm_{i+1\,i+2}\\
  &\phaneq
  +\remif{XXY}{YXX}\perm_{i\,i+1}\perm_{i+1\,i+2}
  +\remif{YXX}{XXY}\perm_{i+1\,i+2}\perm_{i\,i+1} 
  \eqndot
 \end{aligned}
\end{equation}
Taking into account the symmetry under inverting the order of the fields, it is hence sufficient to calculate $\remif{XXX}{XXX}$, $\remif{XXY}{XYX}$ and $\remif{XXY}{YXX}$.

Each of these matrix elements depends on the Mandelstam variables $s_{i\,i+1}$, $s_{i+1\,i+2}$ and $s_{i+2\,i}$ mainly through the ratios%
\footnote{The BPS remainder studied in \cite{Brandhuber:2014ica} depends only on the rations $u_i$, $v_i$ and $w_i$.
The fact that also $\frac{s_{i\,i+1}}{\mu^2}$, $\frac{s_{i+1\,i+2}}{\mu^2}$ and $\frac{s_{i+2\,i}}{\mu^2}$ occur in \eqref{eq: second matric element trans three}, \eqref{eq: second matrix element} and \eqref{eq: third matrix element} reflects the fact that we are considering form factors of composite operators that are in general non-protected. As required, the latter terms cancel for the BPS case.}
\begin{equation}
\label{eq: uvw}
u_i = \frac{s_{i\,i+1}}{s_{i\,i+1 \, i+2}}\eqncom \quad
v_i = \frac{s_{i+1\,i+2}}{s_{i\,i+1 \, i+2}}\eqncom \quad
w_i = \frac{s_{i+2\,i}}{s_{i\,i+1 \, i+2}} \eqncom
\end{equation}
where $s_{i\,i+1 \, i+2}=s_{i\,i+1}+s_{i+1\,i+2}+s_{i+2\,i}$ and $u_i+v_i+w_i=1$.

The first matrix element was already calculated in \cite{Brandhuber:2014ica} and is of uniform transcendentality four:
\begin{equation}
 \remif{XXX}{XXX}=\remif{XXX}{XXX}\Big|_4\eqncom
\end{equation}
where we denote the restriction to functions with a specific degree of transcendentality by a vertical bar.
Inserting the respective expressions for the Feynman integrals into \eqref{eq: remainder density as graphs}, one finds a result filling several pages. One can, however, simplify it using the symbol \cite{Goncharov09, Goncharov:2010jf}.%
\footnote{The symbol is implemented e.g.\ in the \texttt{Mathematica} code \cite{VerguCode}.}
The symbol of $\remif{XXX}{XXX}$ is given by \cite{Brandhuber:2014ica}%
\begin{equation}
\begin{aligned} 
\symb\left( \remif{XXX}{XXX}\Big|_4 \right) &= 
- u_i\otimes (1-u_i)\otimes \left[ \frac{1-u_i}{u_i} \otimes \frac{v_i}{w_i}+ \frac{v_i}{w_i}\otimes \frac{w_i^2}{u_i v_i} \right]  - u_i\otimes u_i\otimes \frac{1-u_i}{v_i} \otimes \frac{w_i}{v_i}  \\ 
&\phaneq
 - u_i\otimes v_i\otimes  \frac{v_i}{w_i} \otimes \frac{u_i}{w_i} 
 - u_i\otimes v_i\otimes  \frac{u_i}{w_i} \otimes \frac{v_i}{w_i} 
 + (u_i \leftrightarrow v_i ) \eqndot
\end{aligned}
\end{equation}
In \cite{Brandhuber:2014ica}, it was integrated to the relatively compact expression%
\begin{equation}
 \begin{aligned}
 &\remif{XXX}{XXX}|_4\\
 &=  -    \text{Li}_4(1-u_i)-\text{Li}_4(u_i)+\text{Li}_4\left(\frac{u_i - 1}{u_i}\right)
-  \log \left( \frac{1-u_i}{w_i }\right) 
\left[  \text{Li}_3\left(\frac{u_i - 1}{u_i}\right) - \text{Li}_3\left(1-u_i\right) \right] \\
 &\phaneq- \log \left(u_i\right) \left[\text{Li}_3\left(\frac{v_i}{1-u_i}\right)+\text{Li}_3\left(-\frac{w_i}{v_i}\right) + \text{Li}_3\left(\frac{v_i-1}{v_i}\right)
 -\frac{1}{3}  \log ^3\left(v_i\right) -\frac{1}{3} \log ^3\left(1-u_i\right)  \right]
 \\   
&\phaneq- \text{Li}_2\left(\frac{u_i-1}{u_i}\right) \text{Li}_2\left(\frac{v_i}{1-u_i}\right)+  \text{Li}_2\left(u_i\right) \left[
   \log \left(\frac{1-u_i}{ w_i }\right) \log \left(v_i \right) +\frac{1}{2} \log ^2\left( \frac{ 1-u_i }{ w_i }\right) \right] 
\\
&\phaneq+  \frac{1}{24} \log ^4\left(u_i\right)
-\frac{1}{8} \log ^2\left(u_i\right) \log ^2\left(v_i\right)  - \frac{1}{2} \log ^2\left(1-u_i\right) \log \left(u_i\right) \log \left( \frac{ w_i }{ v_i}\right)\\
&\phaneq- \frac{1}{2} \log \left(1-u_i\right) \log ^2\left(u_i\right) \log \left(v_i\right)
- \frac{1}{6} \log ^3\left(u_i\right) \log \left(w_i\right)  
  \\
  &\phaneq- \zeta_2 \Big[ \log \left(u_i\right) \log \left(\frac{1-v_i}{ v_i} \right)
+ \frac{1}{2}\log ^2\left( \frac{ 1-u_i }{ w_i }\right) - \frac{1}{2}\log ^2\left(u_i\right)   \Big]
\\
&\phaneq + \zeta_3 \log (u_i) + \frac{\zeta_4}{ 2}  +G\left(\left\{1-u_i,1-u_i,1,0\right\},v_i\right) +  (u_i\, \leftrightarrow\, v_i)\eqndot
 \end{aligned}
\end{equation}
where the Goncharov polylogarithm $G\left(\left\{1-u_i,1-u_i,1,0\right\},v_i\right)$ in the last line is the only term that cannot be expressed in terms of classical polylogarithms.

The matrix element $\remif{XXY}{XYX}$ has mixed transcendentality of degree three to zero.
Its maximally transcendental part has the symbol
\begin{equation}
\begin{aligned}
 \symb\left( \remif{XXY}{XYX}\Big|_3 \right)
&= v_i\otimes \frac{v_i}{1-v_i}\otimes \frac{u_i}{1-v_i} -v_i\otimes \frac{1-v_i}{u_i}\otimes \frac{v_i}{w_i}-u_i\otimes \frac{1-u_i}{v_i}\otimes \frac{v_i}{w_i}
\eqncom
\end{aligned}
\end{equation}
which can be integrated to 
\begin{equation}
\label{eq: second matric element trans three}
\begin{aligned}
\remif{XXY}{XYX}\Big|_3
&= \left[ \text{Li}_3\left(-\frac{u_i}{w_i}\right) - \log \left(u_i\right) \text{Li}_2\left(\frac{v_i}{1-u_i}\right)
+{1\over2} \log \left(1-u_i\right) \log \left(u_i\right) \log \left({w_i^2 \over 1-u_i} \right) \right.
\\
&\phaneq \left. -\ {1\over2} \text{Li}_3\left(-\frac{u_i v_i}{w_i}\right) - {1\over2}\log \left(u_i\right) \log \left(v_i\right) \log \left(w_i\right) - \frac{1}{12} \log ^3\left(w_i\right) + (u_i\, \leftrightarrow\, v_i) \right]
\\
&\phaneq -\ \text{Li}_3\left(1-v_i\right)+\text{Li}_3\left(u_i\right)-\frac{1}{2} \log ^2\left(v_i\right) \log \left(\frac{1-v_i}{u_i}\right)
+\frac{1}{6} \pi ^2 \log \left(\frac{v_i}{w_i}\right) \\
& \phaneq -\ \frac{1}{6} \pi ^2 \log \left(-\frac{s_{i\,i+1\,i+2}}{\mu^2}\right) 
 \eqndot
\end{aligned}
\end{equation}
Together with the less transcendental parts, the complete second matrix element reads
\begin{equation}
\label{eq: second matrix element}
\begin{aligned}
 \remif{XXY}{XYX}&=\remif{XXY}{XYX}\Big|_3 +\text{Li}_2\left(1-u_i\right)+\text{Li}_2\left(1-v_i\right)\\
 &\phaneq+\log \left(u_i\right) \log \left(v_i\right)-\frac{1}{2} \log \left(-\frac{s_{i+1\,i+2}}{\mu^2}\right) \log \left(\frac{u_i}{v_i}\right)+2 \log \left(-\frac{s_{i\,i+1}}{\mu^2}\right) +\frac{\pi ^2}{3} -7
 \eqndot
\end{aligned}
\end{equation}
 
The matrix element $\remif{XXY}{YXX}$ has mixed transcendentality with degree two to zero. It is given by 
\begin{equation}
\label{eq: third matrix element}
\begin{aligned}
 \remif{XXY}{YXX}
 &= \frac{1}{2} \log \left(-\frac{s_{i+1\,i+2}}{\mu^2}\right) \log \left(\frac{u_i}{v_i}\right)-\text{Li}_2\left(1-u_i\right)-\log \left(u_i\right) \log \left(v_i\right)+\frac{1}{2} \log ^2\left(v_i\right)\\
 &\phaneq+\log \left(-\frac{s_{i+1\,i+2}}{\mu^2}\right)-2 \log \left(-\frac{s_{i\,i+1}}{\mu^2}\right)+\frac{7}{2} \eqndot
\end{aligned}
\end{equation}

From the above results and the form \eqref{eq: remainder operator}, we find that the remainder of any 
 operator in the $\SU{2}$ sector is given by a linear combination of one function of transcendentality four, one function of transcendentality three and two functions of transcendentality two and less.
In particular, the transcendentality-four contribution to the remainder is the same for any operator in the $\SU{2}$ sector and agrees with the one of the BPS operator $\tr(\phi_{14}^L)$ studied in \cite{Brandhuber:2014ica}.
This generalises the principle of maximal transcendentality to non-protected operators in $\cN=4$ SYM theory.
The difference with respect to the BPS remainder is of transcendentality three or less and can be written entirely in terms of classical polylogarithms.

Given the above results, it is tempting to conjecture that the universality of the leading transcendental part extends to the remainder functions of the minimal form factors of all operators, also beyond the $\SU{2}$ sector.
Moreover, this suggests that the leading transcendental part of the three-point form factor remainder of any length-two operator matches the corresponding BPS remainder calculated in \cite{Brandhuber:2012vm}, which also matches the leading transcendental part of the Higgs-to-three-gluons amplitude calculated in \cite{Gehrmann:2011aa}.
It would be desirable to check this conjecture in further examples or even to prove it.%
\footnote{%
One approach to such a proof might be to show the universality of the leading singularities.
The latter are closely related to the maximally transcendental functions in the dlog form, see e.g.\ \cite{ArkaniHamed:2012nw}.
Moreover, leading singularities are related to on-shell diagrams, which will be one of the subjects of the next chapter.}

Furthermore, note that the maximal degree of transcendentality $t$ in a given matrix element is related to the shuffle number $s$ as $t=4-s$, where $s$ is the number of sites by which a single $Y$ among two $X$'s or vice versa is displaced.
Interestingly, we find that the degree-zero contributions to the remainder function are related to the two-loop dilatation operator as
\begin{equation}
 \dilatwo[i\,i+1\,i+2]=-\frac{4}{7}\,\rem_{i\,i+1\,i+2}\Big|_0 \eqndot
\end{equation}

An important property of scattering amplitudes and form factors is their behaviour under soft and collinear limits.
In these limits, they in general reduce to lower-point scattering amplitudes and form factors multiplied by some universal function. 
Similarly, the corresponding remainders reduce to lower-point remainders.
This poses important constraints on the remainders, which, together with other constraints, even made it possible to bootstrap them in several cases; see for instance \cite{Dixon:2011pw,Dixon:2014xca} and references therein.

As already observed for the BPS remainder in \cite{Brandhuber:2014ica}, also the soft and collinear limits of the remainders in the $\SU{2}$ sector considered here yield non-vanishing results. 
This is interesting as these remainders correspond to the minimal form factors,
i.e.\ no non-vanishing form factors with less legs exist to which they could be proportional.
In fact, a similar behaviour occurs also for scattering amplitudes. For example, several of the six-point NMHV amplitudes listed in appendix \ref{appsec: scalar nmhv six-point amplitudes} have non-vanishing soft and collinear limits that do not correspond to physical amplitudes. Via the three-particle unitarity cut, this behaviour of the amplitudes can be related to the one of the minimal form factor remainders.  
A better understanding of theses limits is clearly desirable.

This chapter concludes the treatment of minimal loop-level form factors in this thesis --- although many interesting questions remain unanswered. Most notably, it would be very interesting to extend the methods presented here to obtain the minimal two-loop form factor of a generic operator and from it the complete two-loop dilatation operator.
These questions are currently under active investigation.
In the next chapter, we will turn to non-minimal ($n$-point) form factors at tree-level.

\chapter{Tree-level form factors}
\label{chap: tree-level form factors}

In the previous chapters, we have studied loop-level form factors with the minimal number of external legs.
In this chapter, we turn to the opposite configuration --- $n$-point form factors at tree level.
We will mostly focus on the form factors of the chiral part of the stress-tensor supermultiplet, which we will briefly introduce in section \ref{sec: stress-energy supermultiplet}.
In the subsequent sections, we extend several powerful techniques that were developed in the context of scattering amplitudes to form factors, in particular on-shell diagrams, central-charge deformations, an integrability-based construction via $\rr$ operators and a formulation in terms of Graßmannian integrals. 
For each of these techniques, we first give a very short description in the case of scattering amplitudes and then show how to generalise them to form factors.
We introduce on-shell diagrams, corresponding permutations, their construction and the extension to top-cell diagrams in section \ref{sec: On-shell diagrams}.
In section \ref{sec: r operators and integrability}, we introduce central-charge deformations for form factors, show how form factors can be constructed via the integrability-based method of $\rr$ operators and demonstrate how they can thus be found as solutions to an eigenvalue equation of the transfer matrix, which also (partially) generalises to generic operators.
In the final section \ref{sec: Grassmannian integrals}, we find a Graßmannian integral representation of form factors in spinor-helicity variables, which we moreover translate to twistor and momentum-twistor variables.
As we are only discussing tree-level expressions, we will be dropping the superscript that indicates the loop order throughout this chapter.

The results presented in this chapter were first published in \cite{Frassek:2015rka}.

\section{Stress-tensor supermultiplet}
\label{sec: stress-energy supermultiplet}

As already mentioned in section \ref{sec: konishi}, one of the two best studied examples of composite operators in $\mathcal{N}=4$ SYM theory is the BPS operator $\tr(\phi_{14}\phi_{14})$, which is the lowest component of the stress-tensor supermultiplet.
In the previous chapters, we have studied (super) form factors of individual operators.
In this chapter, we consider the (super) form factors of the supermultiplet the operator belongs to, or, more precisely, its chiral half.
A convenient way to do so is $\cN=4$ harmonic superspace \cite{Hartwell:1994rp}, see also \cite{Eden:2011yp,Brandhuber:2011tv,Bork:2014eqa}. 

Harmonic superspace introduces the variables $u_A^{+a}$ and $u_A^{-a'}$ as well as their conjugates $\bar{u}_A^{+a}$ and $\bar{u}_A^{-a'}$, where the indices $a$, $a'$ and $\pm$ correspond to the respective factors in $\SU{2}\times\SU{2}'\times\U{1} \subset \SU{4}$.
Using these variables, the $\SU{4}$ charge index of the chiral superspace coordinates $\theta^{A}_\alpha$ can be decomposed as 
$\theta^{ + a}_\alpha=\theta^{ A}_\alpha u_A^{+ a}$, 
$\theta^{ - a'}_\alpha=\theta^{ A}_\alpha u_A^{- a'}$.
Similarly, we define $\phi^{++}=\frac{1}{2}\teps_{ab}u_A^{+a}u_B^{+b}\phi^{AB}$. 
The chiral part of the stress-tensor supermultiplet is then given by 
\begin{equation}
\label{eq: def stress tensor multiplet}
  T(x,\theta^+)=
  \tr(\phi^{++}\phi^{++})(x) + \dots + \frac{1}{3}(\theta^+)^4 \cL (x) \eqncom
\end{equation}
where the highest component $\cL$ is the on-shell Lagrangian.

In addition to the bosonic Fourier transformation \eqref{eq: form factor momentum space}, we can also take the Fourier transformation in the fermionic variables and define%
\footnote{Following the literature on on-shell diagrams, in this chapter we are suppressing a factor of $(2\pi)^4$ in each form factor and each amplitude.}
\begin{equation}
\begin{aligned}
 \ff_{T,n}(1,\dots,n;q,\gamma^-)&=\int \de^4x\de^4\theta^+ \e^{-iqx-i\theta^{+a}_\alpha \gamma_{a}^{-\alpha}}\bra{1,\dots,n}T(x,\theta^+)\ket{0}\\
&=\delta^4(P)\delta^4(Q^+)\delta^4(Q^-)\bra{1,\dots,n}T(0,0)\ket{0}
 \eqncom
 \end{aligned}
\end{equation}
where $\gamma^{- \alpha a }$ is the supermomentum of the supermultiplet,
\begin{equation}
\label{eq: deformed momenta and supermomenta intro}
        P=\sum_{i=1}^n \vll_i\vlt_i - q
        \eqncom \qquad
        Q^{+}=\sum_{i=1}^n \vll_i\vle_{i}^+ 
        \eqncom \qquad
        Q^{-}=\sum_{i=1}^n \vll_i\vle_{i}^- - \gamma^-
\end{equation}
and $\etat^{+ a}=  \bar{u}_A^{+ a} \etat^A$, $\etat^{- a'}=  \bar{u}_A^{- a'}\etat^A$.
Hence, supermomentum conservation is manifest in addition to momentum conservation, which allows the use of many supersymmetric methods that were developed in the context of amplitudes, such as the supersymmetric form of BCFW recursion relations \cite{Britto:2004ap,Britto:2005fq}.

The tree-level colour-ordered MHV super form factor of $T$ is \cite{Brandhuber:2011tv}%
\footnote{Following the conventions in the literature on form factors of $T$, we have absorbed a sign here. Hence, $\ffco_{T,2,2}(1^{\phi^{++}},2^{\phi^{++}};q,\gamma^-)=-1$ in contrast to \eqref{eq: minimal tree-level form factor from oscillator replacement}.
}
\begin{equation}
\label{eq: ff n,2 intro}
        \ffco_{T,n,2}(1,\dots,n;q,\gamma^-)
        =\frac{
                \delta^4(P)\delta^4(Q^+)\delta^4(Q^{-})
        }{
          \abr{12}\abr{23}\cdots\abr{n \sminus 1 \ssep n}\abr{n1}
        } \eqndot
\end{equation}
As for amplitudes, we can write the N$^{k-2}$MHV form factors of $T$ as
\begin{equation}
 \ffco_{T,n,k}(1,\dots,n;q,\gamma^-)=\ffco_{T,n,2}(1,\dots,n;q,\gamma^-)\times \ffratio(1,\dots,n;q,\gamma^-) \eqncom
\end{equation}
where $\ffratio$ is of Graßmann degree $4(k-2)$.
Expressions for $\ffratio$ of $T$ in momentum-twistor space were given in \cite{Brandhuber:2011tv} and \cite{Bork:2014eqa} at NMHV level and N$^{k-2}$MHV level, respectively.
Results for certain components can be found in \cite{Brandhuber:2010ad,Brandhuber:2011tv}.

\section{On-shell diagrams}
\label{sec: On-shell diagrams}

In the following section, we introduce on-shell diagrams for form factors. 

\subsection{On-shell diagrams}

On-shell diagrams for scattering amplitudes were intensively studied in \cite{ArkaniHamed:2012nw}.
They can be used to represent tree-level amplitudes and their unregularised loop-level integrands and furthermore yield the leading singularities of loop-level amplitudes.%
\footnote{For extensions to non-planar amplitudes, see 
\cite{Arkani-Hamed:2014via,Chen:2014ara,Arkani-Hamed:2014bca,Bern:2014kca,Franco:2015rma,Chen:2015qna}.} 
As already mentioned, we will restrict ourselves to tree level.
Note that we are using slightly different conventions than \cite{ArkaniHamed:2012nw}.

For scattering amplitudes, on-shell diagrams are built from two different building blocks, namely the tree-level three-point MHV and \MHVb amplitudes, which are depicted as black and white vertices, respectively:%
\footnote{In accordance with the literature on on-shell diagrams, we suppress the factor of $\pm i(2\pi)^4$ occurring in the amplitudes throughout this chapter.}
\begin{equation}
\label{eq: amplitude building blocks for on-shell diagrams}
 \begin{aligned}
  \begin{aligned}
        \begin{tikzpicture}[scale=0.8]
        \draw (1,1) -- (1,1+0.65);
        \draw (1,1) -- (1-0.5,1-0.5);
        \draw (1,1) -- (1+0.5,1-0.5);
        \node [] at (1,1+0.65+\labelvdist) {1};
        \node [] at (1-0.5-\labelddist,1-0.5-\labelddist) {3};
        \node [] at (1+0.5+\labelddist,1-0.5-\labelddist) {2};
        \node[db] at (1,1) {};
        \end{tikzpicture}
        \end{aligned}
        &=
        \ampco_{3,2}(1,2,3)=\frac{
                \delta^4(\vll_1\vlt_1+\vll_2\vlt_2+\vll_3\vlt_3)
                \delta^8(\vll_1\vle_1+\vll_2\vle_2+\vll_3\vle_3)
        }{\abr{12}\abr{23}\abr{31}} \eqncom\\
   \begin{aligned}
        \begin{tikzpicture}[scale=0.8]
        \draw (1,1) -- (1,1+0.65);
        \draw (1,1) -- (1-0.5,1-0.5);
        \draw (1,1) -- (1+0.5,1-0.5);
        \node [] at (1,1+0.65+\labelvdist) {1};
        \node [] at (1-0.5-\labelddist,1-0.5-\labelddist) {3};
        \node [] at (1+0.5+\labelddist,1-0.5-\labelddist) {2};
        \node[dw] at (1,1) {};
        \end{tikzpicture}
        \end{aligned}
        &=
        \ampco_{3,1}(1,2,3)
        =\frac{
                \delta^4(\vll_1\vlt_1+\vll_2\vlt_2+\vll_3\vlt_3)
                \delta^4(\sbr{12}\vle_3+\sbr{23}\vle_1+\sbr{31}\vle_2)
        }{\sbr{12}\sbr{23}\sbr{31}} \eqndot
 \end{aligned}
\end{equation}

One way to obtain the on-shell graph for a given scattering amplitude is by constructing the latter via the BCFW recursion relation \cite{Britto:2004ap,Britto:2005fq}, which can be depicted as \cite{ArkaniHamed:2012nw}%
\footnote{Note that we are using BCFW bridges with the opposite assignment of black and white vertices compared to \cite{ArkaniHamed:2012nw}.}$^,$%
\footnote{A more efficient way will be described further below.}
\begin{equation}
 \label{eq: BCFW for amplitudes}
 \ampco_{n,k}= \sum_{\substack{n',n'',k',k''\\ n'+n''=n+2\\ k'+k''=k+1}}
 \begin{aligned}
 \begin{tikzpicture}[scale=0.8]
 \draw (0,-0) -- (2,-0); 
 \draw (0,-1.5) -- (2,-1.5); 
 \draw (0,-0) -- (0,-2.25); 
 \draw (2,-0) -- (2,-2.25); 
 \draw (0,-0) -- (-1,-0);
 \node[] at (-1-\labelhdist,-0) {$3$};
 \draw (0,-0) -- (-0,+1);
 \node[] at (-0,+1+\labelvdist) {$n'$};
 \draw (2,-0) -- (+3,-0);
 \node[] at (+3+\labelhdist,-0) {$n$};
 \draw (2,-0) -- (+2,+1);
 \node[] at (+2,+1+\labelvdist) {$n'+1$};
 \node[] at (-0.7,+0.7) {\rotatebox{45}{$\cdots$}};
 \node[] at (+2.7,+0.7) {\rotatebox{-45}{$\cdots$}};
 \node[dw] at (0,-1.5) {};
 \node[db] at (2,-1.5) {};
 \node[circle, black, fill=gray, minimum width=4*\onshellradius, draw, inner sep=0pt] at (0,0) {$\scriptstyle \ampco_{n',k'}$};
 \node[circle, black, fill=gray, minimum width=4*\onshellradius, draw, inner sep=0pt] at (2,0) {$\scriptstyle \ampco_{n'',k''}$};
 \node[] at (0,-2.25-\labelvdist) {$2$};
 \node[] at (2,-2.25-\labelvdist) {$1$};
\end{tikzpicture}
\end{aligned} \eqndot
\end{equation}

The representation of an amplitude in terms of BCFW terms, however, is not unique, and neither is the one in terms of on-shell diagrams. 
Instead, several equivalent representations exist.
For planar on-shell diagrams, the set of equivalence relations is generated by two different \emph{moves}: the merge/unmerge move and the square move \cite{ArkaniHamed:2012nw}. These are depicted in figure \ref{fig: amplitude moves}. 

\begin{figure}[htbp]
\begin{subfigure}[t]{0.54\textwidth}
 \begin{equation*} 
 \begin{aligned}
   \begin{tikzpicture}[scale=0.8,rotate=0]
                         \draw (1,1) -- (1,2);
                         \draw (0.5,0.5) -- (1,1);
                         \draw (1.5,0.5) -- (1,1);
                         \draw (1.5,2.5) -- (1,2);
                         \draw (0.5,2.5) -- (1,2);
                         \node[db] at (1,1) {};
                         \node[db] at (1,2) {};
                         \node at (1.5+\labelddist,0.5-\labelddist) {1};
                         \node at (0.5-\labelddist,2.5+\labelddist) {3};
                         \node at (0.5-\labelddist,0.5-\labelddist) {2};
                         \node at (1.5+\labelddist,2.5+\labelddist) {4};
   \end{tikzpicture}
   \end{aligned}
   =
         \begin{aligned}
   \begin{tikzpicture}[scale=0.8,rotate=90]
                         \draw (0.5,0.5) -- (1,1);
                         \draw (1.5,0.5) -- (1,1);
                         \draw (1.5,1.5) -- (1,1);
                         \draw (0.5,1.5) -- (1,1);
                         \node[db] at (1,1) {};
                         \node at (1.5+\labelddist,0.5-\labelddist) {4};
                         \node at (0.5-\labelddist,1.5+\labelddist) {2};
                         \node at (0.5-\labelddist,0.5-\labelddist) {1};
                         \node at (1.5+\labelddist,1.5+\labelddist) {3};
                         \end{tikzpicture}
   \end{aligned}
   =
         \begin{aligned}
   \begin{tikzpicture}[scale=0.8,rotate=90]
                         \draw (1,1) -- (1,2);
                         \draw (0.5,0.5) -- (1,1);
                         \draw (1.5,0.5) -- (1,1);
                         \draw (1.5,2.5) -- (1,2);
                         \draw (0.5,2.5) -- (1,2);
                         \node[db] at (1,1) {};
                         \node[db] at (1,2) {};
                         \node at (1.5+\labelddist,0.5-\labelddist) {4};
                         \node at (0.5-\labelddist,2.5+\labelddist) {2};
                         \node at (0.5-\labelddist,0.5-\labelddist) {1};
                         \node at (1.5+\labelddist,2.5+\labelddist) {3};
   \end{tikzpicture}
   \end{aligned}
 \end{equation*} 
\caption{Merge/unmerge move for black vertices.}
\label{fig: merge move}
\end{subfigure}
\begin{subfigure}[t]{0.44\textwidth}
 \begin{equation*}
  \begin{aligned}
   \begin{tikzpicture}[scale=0.8,rotate=0]
                         \draw (1,1) -- (1,2) -- (2,2) -- (2,1) -- (1,1);
                         \draw (0.5,0.5) -- (1,1);
                         \draw (2.5,0.5) -- (2,1);
                         \draw (2.5,2.5) -- (2,2);
                         \draw (0.5,2.5) -- (1,2);
                         \node[dw] at (2,1) {};
                         \node[dw] at (1,2) {};
                         \node[db] at (1,1) {};
                         \node[db] at (2,2) {};
                         \node at (2.5+\labelddist,0.5-\labelddist) {1};
                         \node at (0.5-\labelddist,2.5+\labelddist) {3};
                         \node at (0.5-\labelddist,0.5-\labelddist) {2};
                         \node at (2.5+\labelddist,2.5+\labelddist) {4};
   \end{tikzpicture}
   \end{aligned}
   =
     \begin{aligned}
   \begin{tikzpicture}[scale=0.8,rotate=0]
                         \draw (1,1) -- (1,2) -- (2,2) -- (2,1) -- (1,1);
                         \draw (0.5,0.5) -- (1,1);
                         \draw (2.5,0.5) -- (2,1);
                         \draw (2.5,2.5) -- (2,2);
                         \draw (0.5,2.5) -- (1,2);
                         \node[db] at (2,1) {};
                         \node[db] at (1,2) {};
                         \node[dw] at (1,1) {};
                         \node[dw] at (2,2) {};
                         \node at (2.5+\labelddist,0.5-\labelddist) {1};
                         \node at (0.5-\labelddist,2.5+\labelddist) {3};
                         \node at (0.5-\labelddist,0.5-\labelddist) {2};
                         \node at (2.5+\labelddist,2.5+\labelddist) {4};
   \end{tikzpicture}
   \end{aligned}
 \end{equation*} 
\caption{Square move.}
\label{fig: square move}
\end{subfigure}
\caption{Equivalence moves between on-shell diagrams for scattering amplitudes. An analogous version of the merge/unmerge move also exists for white vertices.}
\label{fig: amplitude moves}
\end{figure}

For form factors of the stress-tensor supermultiplet, a similar construction via BCFW recursion relations exists \cite{Brandhuber:2010ad,Brandhuber:2011tv}, 
which we can depict as:
\begin{equation}
\label{eq: BCFW for form factors}
\ffco_{T,n,k}= \sum_{\substack{n',n'',k',k''\\ n'+n''=n+2\\ k'+k''=k+1}}
 \begin{aligned}
 \begin{tikzpicture}[scale=0.8]
 \draw (0,-0) -- (2,-0); 
 \draw (0,-1.5) -- (2,-1.5); 
 \draw (0,-0) -- (0,-2.25); 
 \draw (2,-0) -- (2,-2.25); 
 \draw (0,-0) -- (-1,-0);
 \node[] at (-1-\labelhdist,-0) {$3$};
 \draw (0,-0) -- (-0,+1);
 \node[] at (-0,+1+\labelvdist) {$n'$};
 \draw (2,-0) -- (+3,-0);
 \node[] at (+3+\labelhdist,-0) {$n$};
 \draw (2,-0) -- (+2,+1);
 \node[] at (+2,+1+\labelvdist) {$n'+1$};
 \node[] at (-0.7,+0.7) {\rotatebox{45}{$\cdots$}};
 \node[] at (+2.7,+0.7) {\rotatebox{-45}{$\cdots$}};
 \node[dw] at (0,-1.5) {};
 \node[db] at (2,-1.5) {};
 \draw [thick,double] (0,0) -- (-1,-1);
 \node[circle, black, fill=gray, minimum width=4*\onshellradius, draw, inner sep=0pt] at (0,0) {$\scriptstyle \ffco_{T,n',k'}$};
 \node[circle, black, fill=gray, minimum width=4*\onshellradius, draw, inner sep=0pt] at (2,0) {$\scriptstyle \ampco_{n'',k''}$};
 \node[] at (0,-2.25-\labelvdist) {$2$};
 \node[] at (2,-2.25-\labelvdist) {$1$};
\end{tikzpicture}
\end{aligned}
+
\begin{aligned}
 \begin{tikzpicture}[scale=0.8]
 \draw (0,-0) -- (2,-0); 
 \draw (0,-1.5) -- (2,-1.5); 
 \draw (0,-0) -- (0,-2.25); 
 \draw (2,-0) -- (2,-2.25); 
 \draw (0,-0) -- (-1,-0);
 \node[] at (-1-\labelhdist,-0) {$3$};
 \draw (0,-0) -- (-0,+1);
 \node[] at (-0,+1+\labelvdist) {$n'$};
 \draw (2,-0) -- (+3,-0);
 \node[] at (+3+\labelhdist,-0) {$n$};
 \draw (2,-0) -- (+2,+1);
 \node[] at (+2,+1+\labelvdist) {$n'+1$};
 \node[] at (-0.7,+0.7) {\rotatebox{45}{$\cdots$}};
 \node[] at (+2.7,+0.7) {\rotatebox{-45}{$\cdots$}};
 \node[dw] at (0,-1.5) {};
 \node[db] at (2,-1.5) {};
 \draw [thick,double] (2,0) -- (+3,-1);
 \node[circle, black, fill=gray, minimum width=4*\onshellradius, draw, inner sep=0pt] at (0,0) {$\scriptstyle \ampco_{n',k'}$};
 \node[circle, black, fill=gray, minimum width=4*\onshellradius, draw, inner sep=0pt] at (2,0) {$\scriptstyle \ffco_{T,n'',k''}$};
 \node[] at (0,-2.25-\labelvdist) {$2$};
 \node[] at (2,-2.25-\labelvdist) {$1$};
\end{tikzpicture}
\end{aligned}
\eqndot
\end{equation}
Via these recursion relations, all $\ffco_{T,n,k}$ can be written in terms of the three-point amplitudes \eqref{eq: amplitude building blocks for on-shell diagrams} and the minimal form factor $\ffco_{T,2,2}$.

In order to introduce on-shell diagrams for form factors, we have to add the minimal form factor as a further building block:%
\footnote{While the three-point amplitudes can be either MHV or $\overline{\text{MHV}}$, the minimal form factor is both MHV and \NmaxMHV. Hence, we only have one form-factor vertex \eqref{eq: form factor building block for on-shell diagrams} as building block, while we have two amplitude vertices \eqref{eq: amplitude building blocks for on-shell diagrams}.}
\begin{equation}
\label{eq: form factor building block for on-shell diagrams}
 \begin{aligned}
          \begin{aligned}
        \begin{tikzpicture}[scale=0.8]
\drawminimalff{1} 
        \node [] at (-\labelddist,-\vacuumheight-\labelddist) {2};
        \node [] at (1+\labelddist,-\vacuumheight-\labelddist) {1};
        \end{tikzpicture}
        \end{aligned}
        = \begin{aligned}
        \begin{tikzpicture}[scale=0.8]
\drawminimalff{1} 
        \node [] at (-\labelddist,-\vacuumheight-\labelddist) {1};
        \node [] at (1+\labelddist,-\vacuumheight-\labelddist) {2};
        \end{tikzpicture}
        \end{aligned}
        &=
        \ffco_{T,2,2}(1,2)\\&=\frac{
                \delta^4(\vll_1\vlt_1 + \vll_2\vlt_2 - q)
                \delta^4(\vll_1\vle_1^+ + \vll_2\vle_2^+)
                \delta^4(\vll_1\vle_1^- + \vll_2\vle_2^- - \gamma^-)
        }{\abr{12}\abr{21}} \eqndot
 \end{aligned}
\end{equation}

Thus, on-shell diagrams can be used to represent all tree-level form factors --- at least for the stress-tensor supermultiplet.
It remains to characterise these on-shell diagrams, to find the equivalence relations among them and to find a more direct way to construct them than via BCFW recursion relations.

\subsection{Inverse soft limits}

Another way to construct amplitudes is via the so-called inverse soft limit \cite{ArkaniHamed:2009si,ArkaniHamed:2010kv,Bullimore:2010pa}.
This construction amounts to gluing the following structures to two adjacent legs of an on-shell diagram:
\begin{equation}
\label{eq: inverse soft limit}
        \begin{aligned}
        \begin{tikzpicture}[scale=0.8,rotate=180]
        \draw (1,1) -- (0,0.7);
        \draw (1,1) -- (2,0.7);
        \draw (1,1) -- (1,1+0.65);
        \draw (0,0.7) -- (0-0.3,0.7+0.6);
        \draw (2,0.7) -- (2+0.3,0.7+0.6);
        \draw (0,0.7) -- (0+0.3,0.7-0.6);
        \draw (2,0.7) -- (2-0.3,0.7-0.6);
        \node[db] at (1,1) {};
        \node[dw] at (0,0.7) {};
        \node[dw] at (2,0.7) {};
        \end{tikzpicture}
        \end{aligned}
\,\eqncom \qquad       
        \begin{aligned}
        \begin{tikzpicture}[scale=0.8,rotate=180]
        \draw (1,1) -- (0,0.7);
        \draw (1,1) -- (2,0.7);
        \draw (1,1) -- (1,1+0.65);
        \draw (0,0.7) -- (0-0.3,0.7+0.6);
        \draw (2,0.7) -- (2+0.3,0.7+0.6);
        \draw (0,0.7) -- (0+0.3,0.7-0.6);
        \draw (2,0.7) -- (2-0.3,0.7-0.6);
        \node[dw] at (1,1) {};
        \node[db] at (0,0.7) {};
        \node[db] at (2,0.7) {};
        \end{tikzpicture}
        \end{aligned}
\,\eqncom        
\end{equation}
which respectively preserve the MHV degree $k$ or increase it by one.
In principle, all tree-level scattering amplitudes can be constructed via the inverse soft limit \cite{Nandan:2012rk}.
Adding $k$-preserving and $k$-increasing structures, however, does not commute. 
Hence, the inverse soft limit is most powerful for minimal and maximal MHV degree, where only one of the two structures in \eqref{eq: inverse soft limit} occurs and the order of applying them is irrelevant.
Moreover, the inverse soft limit can also be applied to construct tree-level form factors of $T$ \cite{Nandan:2012rk}.

Via the $k$-preserving inverse soft limit, the four-point MHV amplitude $\ampco_{4,2}$ can be constructed from the three-point MHV amplitude $\ampco_{3,2}$ as  
\begin{equation}
\label{eq: on-shell diagram amp 4,2 from inverse soft limit}
   \begin{aligned}
        \begin{tikzpicture}[scale=0.8]
        \draw (1,1) -- (1,1+0.65);
        \draw (1,1) -- (1-0.5,1-0.5);
        \draw (1,1) -- (1+0.5,1-0.5);
        \node [] at (1,1+0.65+\labelvdist) {2};
        \node [] at (1-0.5-\labelddist,1-0.5-\labelddist) {1};
        \node [] at (1+0.5+\labelddist,1-0.5-\labelddist) {3};
        \node[db] at (1,1) {};
        \end{tikzpicture}
        \end{aligned}
        \quad
   \xrightarrow{%
   \resizebox{1cm}{!} {%
   \begin{tikzpicture}[scale=0.8,rotate=180]
        \draw (1,1) -- (0,0.7);
        \draw (1,1) -- (2,0.7);
        \draw (1,1) -- (1,1+0.65);
        \draw (0,0.7) -- (0-0.3,0.7+0.6);
        \draw (2,0.7) -- (2+0.3,0.7+0.6);
        \draw (0,0.7) -- (0+0.3,0.7-0.6);
        \draw (2,0.7) -- (2-0.3,0.7-0.6);
        \node[db] at (1,1) {};
        \node[dw] at (0,0.7) {};
        \node[dw] at (2,0.7) {};
        \end{tikzpicture}
        }%
   }
   \quad
   \begin{aligned}
   \begin{tikzpicture}[scale=0.8,rotate=45]
                         \draw (1,1) -- (1,2) -- (2,2) -- (2,1) -- (1,1);
                         \draw (0.5,0.5) -- (1,1);
                         \draw (2.5,0.5) -- (2,1);
                         \draw (2.5,2.5) -- (2,2);
                         \draw (0.5,2.5) -- (1,2);
        \node [] at (2.5+\labelddist,2.5+\labelddist) {2};
        \node [] at (2.5+\labelddist,0.5-\labelddist) {3};
        \node [] at (0.5-\labelddist,0.5-\labelddist) {4};
        \node [] at (0.5-\labelddist,2.5+\labelddist) {1};
                         \node[dw] at (2,1) {};
                         \node[dw] at (1,2) {};
                         \node[db] at (1,1) {};
                         \node[db] at (2,2) {};
   \end{tikzpicture}
   \end{aligned}
   \,\eqncom
\end{equation}
which agrees with the result from the BCFW recursion relation \eqref{eq: BCFW for amplitudes}.%
\footnote{Throughout this chapter, we disregard terms in which external legs are connected by a chain of vertices of the same colour, as these do not contribute for generic external momenta; cf.\ also the discussion in subsection \ref{subsec: triple cut and triangle coefficient}.}

Similarly, the three-point MHV form factor $\ffco_{T,3,2}$ can be built from the minimal form factor $\ffco_{T,2,2}$ as
\begin{equation}
\label{eq: on-shell diagram ff 3,2 from inverse soft limit}
        \begin{aligned}
        \begin{tikzpicture}[scale=0.8]
\drawminimalff{1} 
        \node [] at (-\labelddist,-\vacuumheight-\labelddist) {1};
        \node [] at (1+\labelddist,-\vacuumheight-\labelddist) {2};
        \end{tikzpicture}
        \end{aligned}
        \quad
   \xrightarrow{%
   \resizebox{1cm}{!} {%
   \begin{tikzpicture}[scale=0.8,rotate=180]
        \draw (1,1) -- (0,0.7);
        \draw (1,1) -- (2,0.7);
        \draw (1,1) -- (1,1+0.65);
        \draw (0,0.7) -- (0-0.3,0.7+0.6);
        \draw (2,0.7) -- (2+0.3,0.7+0.6);
        \draw (0,0.7) -- (0+0.3,0.7-0.6);
        \draw (2,0.7) -- (2-0.3,0.7-0.6);
        \node[db] at (1,1) {};
        \node[dw] at (0,0.7) {};
        \node[dw] at (2,0.7) {};
        \end{tikzpicture}
        }%
   }
   \quad
\begin{aligned}
 \begin{tikzpicture}[scale=0.8]
 			\draw[thick,double] (0,-0) -- (0,0.5); 
  			\draw (0,0) -- (-\hdist,-\hdist) -- (0,-2*\hdist) -- (+\hdist,-\hdist) -- (0,0); 
 			\draw (+\hdist,-\hdist) -- (+2*\hdist,-2*\hdist);   
 			\draw (-\hdist,-\hdist) -- (-2*\hdist,-2*\hdist);   
 			\draw (0,-2*\hdist) -- (0,-2*\hdist-\ddist);  
                         \node[dw] at (+\hdist,-\hdist) {}; 
                         \node[db] at (0,-2*\hdist) {};
                         \node[dw] at (-\hdist,-\hdist) {}; 
                         \node at (+2*\hdist  +\labelddist,-2*\hdist-\labelddist) {2};
                         \node at (0,-2*\hdist-\ddist-\labelvdist) {3};
                         \node at (-2*\hdist -\labelddist,-2*\hdist-\labelddist) {1};
         \end{tikzpicture}
\end{aligned}
   \,\eqncom
\end{equation}
which agrees with the result from the BCFW recursion relation \eqref{eq: BCFW for form factors}.

Note that the on-shell diagram \eqref{eq: on-shell diagram ff 3,2 from inverse soft limit} for $\ffco_{T,3,2}$ is not manifestly invariant under cyclic permutations of its three on-shell legs, while $\ffco_{T,3,2}$ is.
Similarly, the on-shell diagram \eqref{eq: on-shell diagram amp 4,2 from inverse soft limit} for $\ampco_{4,2}$ is not manifestly invariant under cyclic permutations of its four on-shell legs, although $\ampco_{4,2}$ is.
The equivalence of both cyclic orderings of the on-shell diagram \eqref{eq: on-shell diagram amp 4,2 from inverse soft limit} is precisely the statement of the square move depicted in figure \ref{fig: square move}. 
The cyclic invariance of the on-shell diagrams for  all other MHV amplitudes, however, follows from its combination with the merge/unmerge move.
Hence, we have to add an additional equivalence move for on-shell diagrams of form factors, which reflects the cyclic invariance of the three-point form factor.
This move is depicted in figure \ref{fig: move for minimal ff} and we call it \emph{rotation move}.
As the square move for amplitudes, it implies the cyclic invariance of all other MHV form factors when combined with the moves in figure \ref{fig: amplitude moves}. 

\begin{figure}[htbp]
\begin{equation*}
\begin{aligned}
 \begin{tikzpicture}[scale=0.8]
 			\draw[thick,double] (0,-0) -- (0,0.5); 
  			\draw (0,0) -- (-\hdist,-\hdist) -- (0,-2*\hdist) -- (+\hdist,-\hdist) -- (0,0); 
 			\draw (+\hdist,-\hdist) -- (+2*\hdist,-2*\hdist);   
 			\draw (-\hdist,-\hdist) -- (-2*\hdist,-2*\hdist);   
 			\draw (0,-2*\hdist) -- (0,-2*\hdist-\ddist);  
                         \node[dw] at (+\hdist,-\hdist) {}; 
                         \node[db] at (0,-2*\hdist) {};
                         \node[dw] at (-\hdist,-\hdist) {}; 
                         \node at (+2*\hdist  +\labelddist,-2*\hdist-\labelddist) {1};
                         \node at (0,-2*\hdist-\ddist-\labelvdist) {2};
                         \node at (-2*\hdist -\labelddist,-2*\hdist-\labelddist) {3};
         \end{tikzpicture}
\end{aligned}
=
\begin{aligned}
 \begin{tikzpicture}[scale=0.8]
 			\draw[thick,double] (0,-0) -- (0,0.5); 
  			\draw (0,0) -- (-\hdist,-\hdist) -- (0,-2*\hdist) -- (+\hdist,-\hdist) -- (0,0); 
 			\draw (+\hdist,-\hdist) -- (+2*\hdist,-2*\hdist);   
 			\draw (-\hdist,-\hdist) -- (-2*\hdist,-2*\hdist);   
 			\draw (0,-2*\hdist) -- (0,-2*\hdist-\ddist);  
                         \node[dw] at (+\hdist,-\hdist) {}; 
                         \node[db] at (0,-2*\hdist) {};
                         \node[dw] at (-\hdist,-\hdist) {}; 
                         \node at (+2*\hdist  +\labelddist,-2*\hdist-\labelddist) {2};
                         \node at (0,-2*\hdist-\ddist-\labelvdist) {3};
                         \node at (-2*\hdist -\labelddist,-2*\hdist-\labelddist) {1};
         \end{tikzpicture}
\end{aligned}
=
\begin{aligned}
 \begin{tikzpicture}[scale=0.8]
 			\draw[thick,double] (0,-0) -- (0,0.5); 
  			\draw (0,0) -- (-\hdist,-\hdist) -- (0,-2*\hdist) -- (+\hdist,-\hdist) -- (0,0); 
 			\draw (+\hdist,-\hdist) -- (+2*\hdist,-2*\hdist);   
 			\draw (-\hdist,-\hdist) -- (-2*\hdist,-2*\hdist);   
 			\draw (0,-2*\hdist) -- (0,-2*\hdist-\ddist);  
                         \node[dw] at (+\hdist,-\hdist) {}; 
                         \node[db] at (0,-2*\hdist) {};
                         \node[dw] at (-\hdist,-\hdist) {}; 
                         \node at (+2*\hdist  +\labelddist,-2*\hdist-\labelddist) {3};
                         \node at (0,-2*\hdist-\ddist-\labelvdist) {1};
                         \node at (-2*\hdist -\labelddist,-2*\hdist-\labelddist) {2};
         \end{tikzpicture}
\end{aligned}
\end{equation*}
 \caption{Rotation move for on-shell diagrams that involve the minimal form factor. In addition to the depicted version with one black and two white vertices, an analogous version with one white and two black vertices exists.
 Similar to the other moves, this move can be applied to any subdiagram of a given on-shell diagram.
 }
 \label{fig: move for minimal ff}
\end{figure}

The three-point NMHV form factor $\ffco_{T,3,3}$ can be built from the minimal form factor $\ffco_{T,2,2}$ by the $k$-increasing inverse soft limit. The resulting on-shell diagram is related to the on-shell diagram \eqref{eq: on-shell diagram ff 3,2 from inverse soft limit} by inverting the colour of the vertices. As in the MHV case, its cyclic invariance gives rise to another equivalence move: the rotation move with inverted colours.

\subsection{Permutations}
\label{subsec: permutations}

For scattering amplitudes, a permutation $\sigma$ can be associated with every on-shell diagram \cite{Postnikov:2006kva,ArkaniHamed:2012nw}.
For brevity, we write permutations 
\begin{equation}
 \sigma=\begin{pmatrix}
         1&2&3&\dots&n\\
         \downarrow&\downarrow&\downarrow&\dots &\downarrow\\
         \sigma(1)&\sigma(2)&\sigma(3)&\dots&\sigma(n)
         \end{pmatrix}
\end{equation}
as $\sigma=(\sigma(1),\sigma(2),\sigma(3),\dots,\sigma(n))$. 
The association is as follows.
Entering the on-shell diagram at an external leg $i$, turn left at every white vertex and right at every black vertex until arriving at an external leg again, which is then identified as $\sigma(i)$.
For example,
\begin{equation}
  \begin{aligned}
        \begin{tikzpicture}[scale=0.8]
        \draw (1,1) -- (1,1+0.65);
        \draw (1,1) -- (1-0.5,1-0.5);
        \draw (1,1) -- (1+0.5,1-0.5);
        \node[db] at (1,1) {};
        \node [] at (1,1+0.65+\labelvdist) {1};
        \node [] at (1-0.5-0.2,1-0.5-0.2) {3};
        \node [] at (1+0.5+0.2,1-0.5-0.2) {2};
        \draw[blue,->] (1-0.15,1+0.65) to[out=270,in=45] (1-0.5-0.1,1-0.5+0.1);
        \draw[blue,<-] (1+0.15,1+0.65) to[out=270,in=135] (1+0.5+0.1,1-0.5+0.1); 
        \draw[blue,<-] (1+0.5-0.1,1-0.5-0.1) to[in=45,out=135] (1-0.5+0.1,1-0.5-0.1);
        \end{tikzpicture}
  \end{aligned}
  \rightarrow \sigma=(3,1,2)\eqncom
  \qquad
  \begin{aligned}
        \begin{tikzpicture}[scale=0.8]
        \draw (1,1) -- (1,1+0.65);
        \draw (1,1) -- (1-0.5,1-0.5);
        \draw (1,1) -- (1+0.5,1-0.5);
        \node[dw] at (1,1) {};
        \node [] at (1,1+0.65+\labelvdist) {1};
        \node [] at (1-0.5-0.2,1-0.5-0.2) {3};
        \node [] at (1+0.5+0.2,1-0.5-0.2) {2};
        \draw[blue,<-] (1-0.15,1+0.65) to[out=270,in=45] (1-0.5-0.1,1-0.5+0.1);
        \draw[blue,->] (1+0.15,1+0.65) to[out=270,in=135] (1+0.5+0.1,1-0.5+0.1); 
        \draw[blue,->] (1+0.5-0.1,1-0.5-0.1) to[in=45,out=135] (1-0.5+0.1,1-0.5-0.1);
        \end{tikzpicture}
  \end{aligned}
  \rightarrow \sigma=(2,3,1)\eqndot
\end{equation}
Note that the permutation associated with an on-shell diagram is invariant under the equivalence moves in figure \ref{fig: amplitude moves}.%
\footnote{Note that, in contrast to \cite{ArkaniHamed:2012nw}, we are not using decorated permutations.}

We can define a permutation for on-shell diagrams of form factors by the additional prescription to turn back at the minimal form factor, i.e.\
\begin{equation}
\label{eq: permutation minimal form factor}
 \begin{aligned}
        \begin{tikzpicture}[scale=0.8]
\drawminimalff{1} 
        \node [] at (-\labelddist,-\vacuumheight-\labelddist) {2};
        \node [] at (1+\labelddist,-\vacuumheight-\labelddist) {1};
        \draw[blue] (0-0.1,-\vacuumheight+0.1) to[in=135,out=45] (0.25,-0.75\vacuumheight);
        \draw[blue,<-] (0+0.1,-\vacuumheight-0.1) to[in=-45,out=45] (0.25,-0.75\vacuumheight);
        \draw[blue] (1-0.1,-\vacuumheight-0.1) to[in=-135,out=135] (0.75,-0.75\vacuumheight);
        \draw[blue,<-] (1+0.1,-\vacuumheight+0.1) to[in=45,out=135] (0.75,-0.75\vacuumheight);
        \end{tikzpicture}
        \end{aligned}
        \rightarrow \sigma=(1,2)
        \eqndot
\end{equation}
The resulting permutation is invariant under the equivalence moves in figure \ref{fig: amplitude moves} as well as figure \ref{fig: move for minimal ff}.

For $n$-point MHV and \MHVb scattering amplitudes, the permutations associated with the on-shell diagrams are well known to be $\sigma=(3,\dots,n,1,2)$ and $\sigma=(n-1,n,1\dots,n-2)$ \cite{ArkaniHamed:2012nw}, respectively.
From the construction via inverse soft limits, we find that the permutations associated with the on-shell diagrams of $n$-point MHV and $\NmaxMHV$ form factors are respectively given by $\sigma=(3,\dots,n,1,2)$ and $\sigma=(n-1,n,1\dots,n-2)$ as well.

\subsection{Systematic construction for MHV and \texorpdfstring{$\NmaxMHV$}{NmaxMHV}}
\label{subsec: systematic construction}

Using the permutations, the corresponding on-shell diagrams for MHV and \MHVb amplitudes can be reconstructed in a systematic way \cite{ArkaniHamed:2012nw}.
To this end, the permutation $\sigma$ is decomposed into a sequence of transpositions of minimal length. Note that multiplication of permutations is understood in the sense of the right action, i.e.\ $\sigma_1 \sigma_2 = (\sigma_2(\sigma_1(1)),\dots,\sigma_2(\sigma_1(n)))$.%
\footnote{This is different with respect to \cite{ArkaniHamed:2012nw} and related to the fact that we are using the opposite colour assignment for BCFW bridges.}
Each transposition $(i,j)$ is associated with a BCFW bridge between legs $i$ and $j$. 
Starting from an empty diagram with $n$ vacua,%
\footnote{These vacua will be given a specific meaning below. For now, they can be understood in a purely symbolic way.} 
the on-shell diagram can be built by acting with BCFW bridges, where the order of applying the bridges is the inverse of the order of the multiplication among the transpositions.
Then, all vacua are removed from the diagram as well as all lines that are directly connected to them.
In the final step, all vertices that have become two-valent by removing these lines are removed while connecting the two lines that enter the vertices.
For the three-point MHV amplitude $\ampco_{3,2}$, this construction is illustrated in figure \ref{fig: A3,2}.

\begin{figure}[htbp]
 \begin{equation*}
 \sigma=(3,1,2)=(2,3)(1,2)\longrightarrow
\begin{aligned}
 \begin{tikzpicture}[scale=0.8]
\drawvline{1}{2}
\drawvline{2}{2}
\drawvline{3}{2}
\drawvacm{1} 
\drawvacm{2} 
\drawvacp{3}
\drawbridge{2}{1}
\drawbridge{1}{2}
\node[dl] at (0,-\vacuumheight-2*\bridgedistance-\labelvdist) {3};
\node[dl] at (1,-\vacuumheight-2*\bridgedistance-\labelvdist) {2};
\node[dl] at (2,-\vacuumheight-2*\bridgedistance-\labelvdist) {1};
\end{tikzpicture}
\end{aligned}
\quad\longrightarrow\quad 
        \begin{aligned}
        \begin{tikzpicture}[scale=0.8]
        \draw (1,1) -- (1,1+0.65);
        \draw (1,1) -- (1-0.5,1-0.5);
        \draw (1,1) -- (1+0.5,1-0.5);
        \node[db] at (1,1) {};
        \node [] at (1,1+0.65+\labelvdist) {1};
        \node [] at (1-0.5-\labelddist,1-0.5-\labelddist) {3};
        \node [] at (1+0.5+\labelddist,1-0.5-\labelddist) {2};
        \end{tikzpicture}
        \end{aligned}
\end{equation*}
\caption{Permutation, construction via BCFW bridges and on-shell diagram for $\ampco_{3,2}$.}
\label{fig: A3,2}
\end{figure}

For MHV and $\NmaxMHV$ form factors, we can use an analogous construction. The only difference is that the two left-most vacua or the two right-most vacua have to be replaced by the minimal form factor, respectively.
We illustrate this construction for $\ffco_{T,3,2}$, $\ffco_{T,4,2}$, $\ffco_{T,5,2}$ and $\ffco_{T,3,3}$ in figures \ref{fig: F3,2}, \ref{fig: F4,2}, \ref{fig: F5,2} and \ref{fig: F3,3}, respectively.

\begin{figure}[htbp]
\begin{equation*}
 \sigma=(3,1,2)=(2,3)(1,2)  \longrightarrow 
\begin{aligned}
 \begin{tikzpicture}[scale=0.8]
\drawvline{1}{2}
\drawvline{2}{2}
\drawvline{3}{2}
\drawminimalff{1} 
\drawvacp{3}
\drawbridge{2}{1}
\drawbridge{1}{2}
\node[dl] at (0,-\vacuumheight-2*\bridgedistance-\labelvdist) {3};
\node[dl] at (1,-\vacuumheight-2*\bridgedistance-\labelvdist) {2};
\node[dl] at (2,-\vacuumheight-2*\bridgedistance-\labelvdist) {1};
\end{tikzpicture}
\end{aligned}
\quad\longrightarrow\quad 
\begin{aligned}
 \begin{tikzpicture}[scale=0.8]
 			\draw[thick,double] (0,-0) -- (0,0.5); 
  			\draw (0,0) -- (-\hdist,-\hdist) -- (0,-2*\hdist) -- (+\hdist,-\hdist) -- (0,0); 
 			\draw (+\hdist,-\hdist) -- (+2*\hdist,-2*\hdist);   
 			\draw (-\hdist,-\hdist) -- (-2*\hdist,-2*\hdist);   
 			\draw (0,-2*\hdist) -- (0,-2*\hdist-\ddist);  
                         \node[dw] at (+\hdist,-\hdist) {}; 
                         \node[db] at (0,-2*\hdist) {};
                         \node[dw] at (-\hdist,-\hdist) {}; 
                         \node at (+2*\hdist  +\labelddist,-2*\hdist-\labelddist) {1};
                         \node at (0,-2*\hdist-\ddist-\labelvdist) {2};
                         \node at (-2*\hdist -\labelddist,-2*\hdist-\labelddist) {3};
         \end{tikzpicture}
\end{aligned}
\end{equation*}
\caption{Permutation, construction via BCFW bridges and on-shell diagram for $\ffco_{T,3,2}$.}
\label{fig: F3,2}
\end{figure}

\begin{figure}[htbp]
\begin{equation*}
 \sigma=(3,4,1,2)=(2,3)(3,4)(1,2)(2,3) \longrightarrow
\begin{aligned}
 \begin{tikzpicture}[scale=0.8]
\drawvline{1}{3}
\drawvline{2}{3}
\drawvline{3}{3}
\drawvline{4}{3}
\drawminimalff{1} 
\drawvacp{3}
\drawvacp{4}
\drawbridge{2}{1}
\drawbridge{1}{2}
\drawbridge{3}{2}
\drawbridge{2}{3}
\node[dl] at (0,-\vacuumheight-3*\bridgedistance-\labelvdist) {4};
\node[dl] at (1,-\vacuumheight-3*\bridgedistance-\labelvdist) {3};
\node[dl] at (2,-\vacuumheight-3*\bridgedistance-\labelvdist) {2};
\node[dl] at (3,-\vacuumheight-3*\bridgedistance-\labelvdist) {1};
\end{tikzpicture}
\end{aligned}
\quad\longrightarrow\quad 
\begin{aligned}
 \begin{tikzpicture}[scale=0.8]
 			\draw[thick,double] (0,-0) -- (0,0.5); 
  			\draw (0,0) -- (-\hdist,-\hdist) -- (0,-2*\hdist) -- (+\hdist,-\hdist) -- (0,0); 
  			\draw (0,-2*\hdist) -- (0,-2*\hdist-\ddist) -- (\ddist,-2*\hdist-\ddist) -- (\hdist+\ddist,-\hdist-\ddist) -- (+\hdist,-\hdist); 
 			\draw (+\hdist+\ddist,-\hdist-\ddist) -- (+2*\hdist+\ddist,-2*\hdist-\ddist);   
 			\draw (-\hdist,-\hdist) -- (-2*\hdist,-2*\hdist);   
 			\draw (0,-2*\hdist-\ddist) -- (0,-2*\hdist-2*\ddist); 
 			\draw (\ddist,-2*\hdist-\ddist) -- (+\hdist+\ddist,-3*\hdist-\ddist);
                         \node[dw] at (+\hdist,-\hdist) {}; 
                         \node[db] at (0,-2*\hdist) {};
                         \node[dw] at (-\hdist,-\hdist) {}; 
                         \node[dw] at (0,-2*\hdist-\ddist) {}; 
                         \node[db] at (\ddist,-2*\hdist-\ddist) {};
                         \node[dw] at (\hdist+\ddist,-\hdist-\ddist) {}; 
                         \node at (+2*\hdist +\ddist +\labelddist,-2*\hdist-\ddist-\labelddist) {1};
                         \node at (-2*\hdist-\labelddist,-2*\hdist-\labelddist) {4};
                         \node at (0,-2*\hdist-2*\ddist-\labelvdist) {3};
                         \node at (\hdist +\ddist +\labelddist,-3*\hdist-\ddist-\labelddist) {2};
         \end{tikzpicture}
\end{aligned}
\end{equation*}
\caption{Permutation, construction via BCFW bridges and on-shell diagram for $\ffco_{T,4,2}$.}
\label{fig: F4,2}
\end{figure}

\begin{figure}[htbp]
\begin{equation*}
\begin{aligned}
\sigma=(3,4,5,1,2)=(2,3)(3,4)(4,5)(1,2)(2,3)(3,4)&\longrightarrow \\[0.25\baselineskip]
 \begin{aligned}
 \begin{tikzpicture}[scale=0.8]
\drawvline{1}{4}
\drawvline{2}{4}
\drawvline{3}{4}
\drawvline{4}{4}
\drawvline{5}{4}
\drawminimalff{1} 
\drawvacp{3}
\drawvacp{4}
\drawvacp{5}
\drawbridge{2}{1}
\drawbridge{3}{2}
\drawbridge{4}{3}
\drawbridge{1}{2}
\drawbridge{2}{3}
\drawbridge{3}{4}
\node[dl] at (0,-\vacuumheight-4*\bridgedistance-\labelvdist) {5};
\node[dl] at (1,-\vacuumheight-4*\bridgedistance-\labelvdist) {4};
\node[dl] at (2,-\vacuumheight-4*\bridgedistance-\labelvdist) {3};
\node[dl] at (3,-\vacuumheight-4*\bridgedistance-\labelvdist) {2};
\node[dl] at (4,-\vacuumheight-4*\bridgedistance-\labelvdist) {1};
\end{tikzpicture}
\end{aligned}
\qquad&\longrightarrow\qquad 
\begin{aligned}
 \begin{tikzpicture}[scale=0.8]
 			\draw[thick,double] (0,-0) -- (0,0.5); 
  			\draw (0,0) -- (-\hdist,-\hdist) -- (0,-2*\hdist) -- (+\hdist,-\hdist) -- (0,0); 
  			\draw (0,-2*\hdist) -- (0,-2*\hdist-\ddist) -- (\ddist,-2*\hdist-\ddist) -- (\hdist+\ddist,-\hdist-\ddist) -- (+\hdist,-\hdist); 
  			\draw (0+\ddist,-2*\hdist-\ddist) -- (0+\ddist,-2*\hdist-\ddist-\ddist) -- (\ddist+\ddist,-2*\hdist-\ddist-\ddist) -- (\hdist+\ddist+\ddist,-\hdist-\ddist-\ddist) -- (+\hdist+\ddist,-\hdist-\ddist); 
 			\draw (+\hdist+2*\ddist,-\hdist-2*\ddist) -- (+2*\hdist+2*\ddist,-2*\hdist-2*\ddist);   
 			\draw (-\hdist,-\hdist) -- (-2*\hdist,-2*\hdist);   
 			\draw (0,-2*\hdist-\ddist) -- (0,-2*\hdist-2*\ddist); 
 			\draw (\ddist,-2*\hdist-2*\ddist) -- (\ddist,-2*\hdist-3*\ddist); 
 			\draw (2*\ddist,-2*\hdist-2*\ddist) -- (+\hdist+2*\ddist,-3*\hdist-2*\ddist);
                         \node[dw] at (+\hdist,-\hdist) {}; 
                         \node[db] at (0,-2*\hdist) {};
                         \node[dw] at (-\hdist,-\hdist) {}; 
                         \node[dw] at (0,-2*\hdist-\ddist) {}; 
                         \node[db] at (\ddist,-2*\hdist-\ddist) {};
                         \node[dw] at (\hdist+\ddist,-\hdist-\ddist) {}; 
                         \node[dw] at (\ddist,-2*\hdist-2*\ddist) {}; 
                         \node[db] at (2*\ddist,-2*\hdist-2*\ddist) {};
                         \node[dw] at (\hdist+2*\ddist,-\hdist-2*\ddist) {}; 
                         \node at (+2*\hdist +2*\ddist +\labelddist,-2*\hdist-2*\ddist-\labelddist) {1};
                         \node at (-2*\hdist-\labelddist,-2*\hdist-\labelddist) {5};
                         \node at (0,-2*\hdist-2*\ddist-\labelvdist) {4};
                         \node at (\ddist ,-2*\hdist-3*\ddist-\labelvdist) {3};
                         \node at (\hdist +2*\ddist +\labelddist,-3*\hdist-2*\ddist-\labelddist) {2};
         \end{tikzpicture}
\end{aligned}
\end{aligned}
\end{equation*}
\caption{Permutation, construction via BCFW bridges and on-shell diagram for $\ffco_{T,5,2}$.}
\label{fig: F5,2}
\end{figure}

\begin{figure}[htbp]
\begin{equation*}
\sigma=(2,3,1)=(1,2)(2,3)\longrightarrow
\begin{aligned}
 \begin{tikzpicture}[scale=0.8]
\drawvline{1}{2}
\drawvline{2}{2}
\drawvline{3}{2}
\drawminimalff{2} 
\drawvacm{1}
\drawbridge{1}{1}
\drawbridge{2}{2}
\node[dl] at (0,-\vacuumheight-2*\bridgedistance-\labelvdist) {3};
\node[dl] at (1,-\vacuumheight-2*\bridgedistance-\labelvdist) {2};
\node[dl] at (2,-\vacuumheight-2*\bridgedistance-\labelvdist) {1};
\end{tikzpicture}
\end{aligned}
\quad\longrightarrow\quad 
\begin{aligned}
 \begin{tikzpicture}[scale=0.8]
 			\draw[thick,double] (0,-0) -- (0,0.5); 
  			\draw (0,0) -- (-\hdist,-\hdist) -- (0,-2*\hdist) -- (+\hdist,-\hdist) -- (0,0); 
 			\draw (+\hdist,-\hdist) -- (+2*\hdist,-2*\hdist);   
 			\draw (-\hdist,-\hdist) -- (-2*\hdist,-2*\hdist);   
 			\draw (0,-2*\hdist) -- (0,-2*\hdist-\ddist);  
                         \node[db] at (+\hdist,-\hdist) {}; 
                         \node[dw] at (0,-2*\hdist) {};
                         \node[db] at (-\hdist,-\hdist) {}; 
                         \node at (+2*\hdist  +\labelddist,-2*\hdist-\labelddist) {1};
                         \node at (0,-2*\hdist-\ddist-\labelvdist) {2};
                         \node at (-2*\hdist -\labelddist,-2*\hdist-\labelddist) {3};
         \end{tikzpicture}
\end{aligned}
\end{equation*}
\caption{Permutation, construction via BCFW bridges and on-shell diagram for $\ffco_{T,3,3}$.}
\label{fig: F3,3}
\end{figure}

\subsection{On-shell diagrams at \texorpdfstring{N$^{k-2}$MHV}{NkMHV} and top-cell diagrams}

One important difference between MHV, $\NmaxMHV$ and general N$^{k-2}$MHV is that sums of different BCFW terms occur in the latter case, whereas only one non-vanishing BCFW term contributes in the former cases. 
As the BCFW terms are represented by
 on-shell diagrams, this leads to a sum of different on-shell diagrams.
For a given amplitude, all these on-shell diagrams can be obtained from a so-called top-cell diagram by deleting edges, which corresponds to taking residues at the level of BCFW bridges.\footnote{Note that not all edges are removable. A criterion for removability is given in \cite{Postnikov:2006kva,ArkaniHamed:2012nw}.
}
The top-cell diagram for amplitudes can be obtained from the permutation 
\begin{equation}
 \sigma=(k+1,\dots,n,1,\dots,k)\eqndot
\end{equation}
For amplitudes, the MHV degree $k$ ranges from $2$ to $n-2$. Hence, the simplest amplitude which is neither MHV nor \MHVb is the NMHV six-point amplitude $\ampco_{6,3}$.

For form factors of the stress-tensor supermultiplet, the MHV degree $k$ ranges from $2$ to $n$. Hence, the simplest form factor which is neither MHV nor $\NmaxMHV$ is the NMHV four-point form factor $\ffco_{T,4,3}$. We will look at this example in some detail.
All BCFW terms with adjacent shifts that contribute to $\ffco_{T,4,3}$ are shown in figure \ref{fig: BCFW ff 4,3}.
\begin{figure}[htbp]
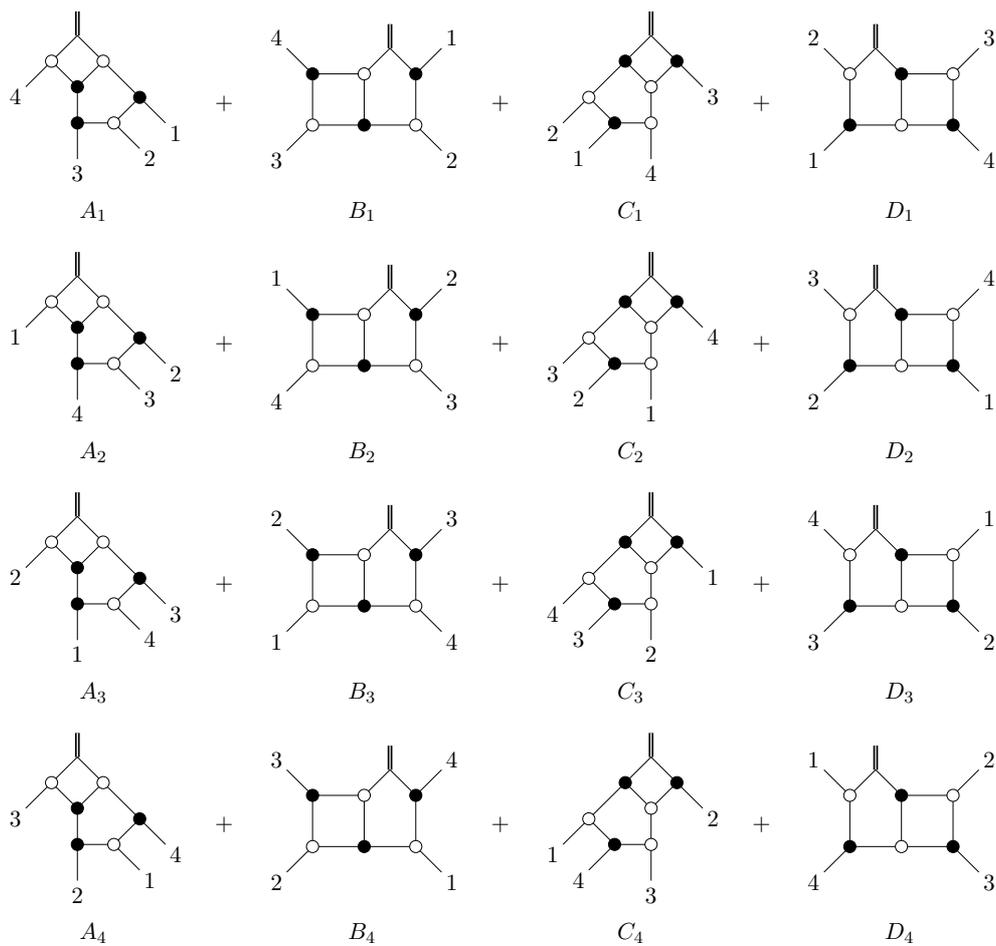

  \centering
  \scalebox{0.8}{
  \begin{tabular}{ccccccc}
 $\athreetwofthreetwo{1}{2}{3}{4}$   & $+$ &
 $\afourtwoftwotwo{1}{2}{3}{4}$      & $+$ &
 $\fthreethreeathreeone{1}{2}{3}{4}$ & $+$ &
 $\ftwotwoafourtwo{1}{2}{3}{4}$      \\
 $A_1$ && $B_1$ && $C_1$ && $D_1$    \\[11pt]
 $\athreetwofthreetwo{2}{3}{4}{1}$   & $+$ &
 $\afourtwoftwotwo{2}{3}{4}{1}$      & $+$ &
 $\fthreethreeathreeone{2}{3}{4}{1}$ & $+$ &
 $\ftwotwoafourtwo{2}{3}{4}{1}$      \\
 $A_2$ && $B_2$ && $C_2$ && $D_2$    \\[11pt]
 $\athreetwofthreetwo{3}{4}{1}{2}$   & $+$ &
 $\afourtwoftwotwo{3}{4}{1}{2}$      & $+$ &
 $\fthreethreeathreeone{3}{4}{1}{2}$ & $+$ &
 $\ftwotwoafourtwo{3}{4}{1}{2}$      \\
 $A_3$ && $B_3$ && $C_3$ && $D_3$    \\[11pt]
 $\athreetwofthreetwo{4}{1}{2}{3}$   & $+$ &
 $\afourtwoftwotwo{4}{1}{2}{3}$      & $+$ &
 $\fthreethreeathreeone{4}{1}{2}{3}$ & $+$ &
 $\ftwotwoafourtwo{4}{1}{2}{3}$      \\
 $A_4$ && $B_4$ && $C_4$ && $D_4$    
\end{tabular}
}
\caption{All BCFW terms of $\ffco_{T,4,3}$ that arise from adjacent shift. The terms are grouped such that the $i^{\text{th}}$ line stems from a shift in the legs $i$ and $i+1$.}
\label{fig: BCFW ff 4,3}
\end{figure}
Via the equivalence moves in figures \ref{fig: amplitude moves} and \ref{fig: move for minimal ff}, the different terms can be shown to satisfy the following identities:
\begin{equation}
  A_i=D_{(i+2) \text{ mod } 4}\eqncom \qquad B_i=C_{(i+2) \text{ mod } 4}\eqndot
\end{equation}
Hence, only eight different BCFW terms occur.
These BCFW terms can be obtained from the top-cell diagram depicted in figure \ref{fig: F4,3} as well as its image under a cyclic shift of the external on-shell legs by two.
Note that several differences occur with respect to the amplitude case. 
While one top-cell diagram suffices to generate all BCFW terms in the case of scattering amplitudes, this is no longer the case for form factors.
This can also be seen from the permutation associated with the top-cell diagram of $\ffco_{T,4,3}$, which is not cyclic.
Moreover, the decomposition of this permutation into transpositions that is required for the construction of the top-cell diagram in terms of BCFW bridges as discussed in the last subsection is not minimal in the sense it is for amplitudes, cf.\ figure \ref{fig: F4,3}.%
\footnote{It would be interesting to find a refined version of the construction in the amplitude case that yields the top-cell diagram directly from the permutation. 
}
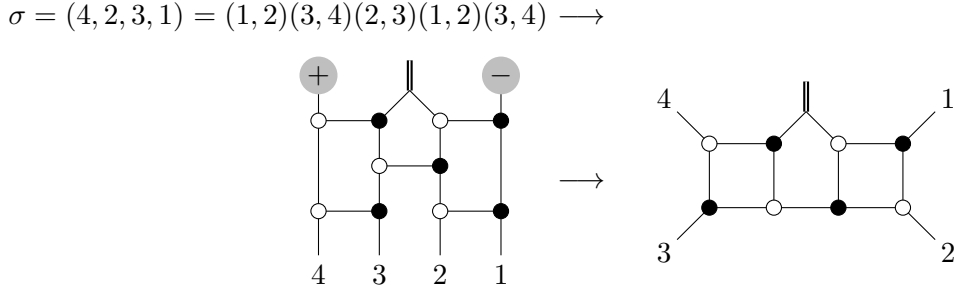
\begin{figure}[htbp]
\begin{equation*}
\begin{aligned}
\sigma=(4,2,3,1)=(1,2)(3,4)(2,3)(1,2)(3,4) &\longrightarrow  \\[0.25\baselineskip]
\begin{aligned}
 \begin{tikzpicture}[scale=0.8]
\drawvline{1}{3}
\drawvline{2}{3}
\drawvline{3}{3}
\drawvline{4}{3}
\drawvacp{1}
\drawminimalff{2} 
\drawvacm{4}
\drawbridge{1}{1}
\drawbridge{3}{1}
\drawbridge{2}{2}
\drawbridge{1}{3}
\drawbridge{3}{3}
\node[dl] at (0,-\vacuumheight-3*\bridgedistance-\labelvdist) {4};
\node[dl] at (1,-\vacuumheight-3*\bridgedistance-\labelvdist) {3};
\node[dl] at (2,-\vacuumheight-3*\bridgedistance-\labelvdist) {2};
\node[dl] at (3,-\vacuumheight-3*\bridgedistance-\labelvdist) {1};
\end{tikzpicture}
\end{aligned}
\quad&\longrightarrow\quad
\begin{aligned}
 \begin{tikzpicture}[scale=0.8]
 			\draw[thick,double] (0,-0+\hdist) -- (0,0.5+\hdist); 
 			\draw (0,\hdist) -- (+\hdist,0);
 			\draw (0,\hdist) -- (-\hdist,0);
 			\draw (+\hdist,0) -- (3*\hdist,0) -- (4*\hdist,\hdist);
 			\draw (-\hdist,0) -- (-3*\hdist,0) -- (-4*\hdist,\hdist);
 			\draw (4*\hdist,-3*\hdist) -- (3*\hdist,-2*\hdist) -- (-3*\hdist,-2*\hdist) -- (-4*\hdist,-3*\hdist);
 			\draw (-\hdist,0) -- (-\hdist,-2*\hdist);
			\draw (-3*\hdist,0) -- (-3*\hdist,-2*\hdist);
 			\draw (\hdist,0) -- (\hdist,-2*\hdist);
			\draw (3*\hdist,0) -- (3*\hdist,-2*\hdist);
                         \node[dw] at (+\hdist,0) {}; 
                         \node[db] at (+3*\hdist,0) {};
                         \node[db] at (-\hdist,0) {}; 
                         \node[dw] at (-3*\hdist,0) {}; 
                         \node[db] at (+\hdist,-2*\hdist) {}; 
                         \node[dw] at (+3*\hdist,-2*\hdist) {};
                         \node[dw] at (-\hdist,-2*\hdist) {}; 
                         \node[db] at (-3*\hdist,-2*\hdist) {}; 
                         \node at (4*\hdist +\labelddist,1*\hdist +\labelddist) {1};
                         \node at (4*\hdist +\labelddist,-3*\hdist-\labelddist) {2};
                         \node at (-4*\hdist -\labelddist,-3*\hdist-\labelddist) {3};
                         \node at (-4*\hdist -\labelddist,1*\hdist +\labelddist) {4};
\end{tikzpicture}
\end{aligned}
\end{aligned}
\end{equation*}
\caption{Permutation, construction via BCFW bridges and top-cell diagram for $\ffco_{T,4,3}$.}
\label{fig: F4,3}
\end{figure}

For higher $n$ and $k$, more than one edge has to be deleted to obtain the BCFW terms from the top-cell diagram, which makes an explicit construction of the latter via BCFW terms increasingly tedious.
Instead, we employ the following observation.
In all cases considered in this chapter, the on-shell diagrams encoding the BCFW terms for the form factor as well as the top-cell diagrams for the form factor can be obtained from their counterparts in the amplitude case with two more legs as follows. We use the moves in figure \ref{fig: amplitude moves} to expose a box at the boundary of the corresponding on-shell diagram in the amplitude case with two more legs and replace this box by the minimal form factor:
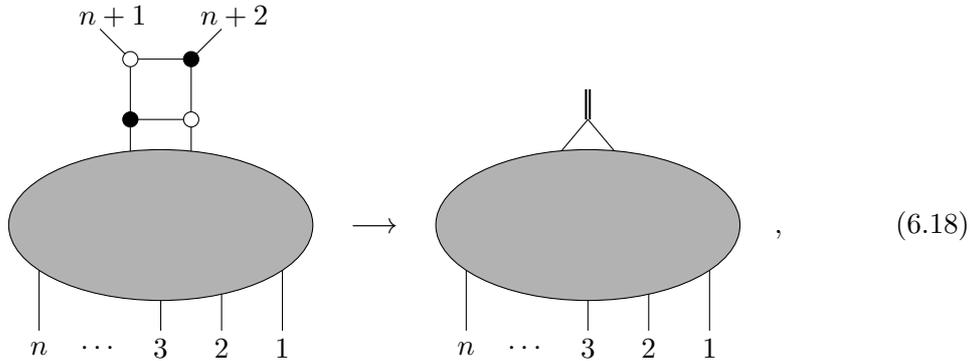
\begin{equation}
\label{eq: box eater}
 \begin{tikzpicture}[scale=0.8, baseline=-0.7cm]
                        \draw (1,1) -- (1,2) -- (2,2) -- (2,1) -- (1,1);
                        \draw (1,0) -- (1,1);
                        \draw (2,0) -- (2,1);
                        \draw (2.5,2.5) -- (2,2);
                        \draw (0.5,2.5) -- (1,2);
                        \draw (-0.5,-0.75) -- (-0.5,-2.5);
                        \draw (1.5,-0.75) -- (1.5,-2.5);
                        \draw (2.5,-0.75) -- (2.5,-2.5);
                        \draw (3.5,-0.75) -- (3.5,-2.5);
                        \node at (-0.5,-2.5-\labelvdist) {$n$};
                        \node at (0.5,-2.5-\labelvdist) {$\cdots$};
                        \node at (1.5,-2.5-\labelvdist) {$3$};
                        \node at (2.5,-2.5-\labelvdist) {$2$};
                        \node at (3.5,-2.5-\labelvdist) {$1$};
                        \node at (2.5+\labelddist,2.5+\labelddist) {$n+2$};
                        \node at (0.5+\labelddist,2.5+\labelddist) {$n+1$};
                        \node[dw] at (2,1) {};
                        \node[dw] at (1,2) {};
                        \node[db] at (1,1) {};
                        \node[db] at (2,2) {};
                        \node[ellipse, black, fill=grayn, minimum width=4 cm, minimum height=2 cm, draw, inner sep=0pt] at (1.5,-0.75) {};
        \end{tikzpicture}
        \quad
        \longrightarrow
        \quad
        \begin{tikzpicture}[scale=0.8, baseline=-0.7cm]
        \draw[thick,double] (1.5,-0+1.5) -- (1.5,-0.5+1.5); 
	\draw (1.5,-0.5+1.5) -- (2,-\vacuumheight+1.4);  
	\draw (1.5,-0.5+1.5) -- (1,-\vacuumheight+1.4);
                        \draw (-0.5,-0.75) -- (-0.5,-2.5);
                        \draw (1.5,-0.75) -- (1.5,-2.5);
                        \draw (2.5,-0.75) -- (2.5,-2.5);
                        \draw (3.5,-0.75) -- (3.5,-2.5);
                        \node at (-0.5,-2.5-\labelvdist) {$n$};
                        \node at (0.5,-2.5-\labelvdist) {$\cdots$};
                        \node at (1.5,-2.5-\labelvdist) {$3$};
                        \node at (2.5,-2.5-\labelvdist) {$2$};
                        \node at (3.5,-2.5-\labelvdist) {$1$};
                        \node[ellipse, black, fill=grayn, minimum width=4 cm, minimum height=2 cm, draw, inner sep=0pt] at (1.5,-0.75) {};
        \end{tikzpicture}
        \quad
      \eqncom
\end{equation}
where the grey area denotes the rest of the on-shell diagram and we have replaced a box at the external legs $n+1$ and $n+2$ for concreteness.

At the level of the BCFW terms, the above relation can be proven as follows. The on-shell diagram for $\ampco_{4,2}$ is nothing but a box and the on-shell diagram of $\ffco_{T,2,2}$ can indeed be obtained by replacing this box with the minimal form factor. Recursively constructing $\ampco_{n+2,k}$ via \eqref{eq: BCFW for amplitudes}, boxes can only occur at the boundary of the on-shell diagram. Comparing this to the recursive construction of $\ffco_{T,n,k}$ via \eqref{eq: BCFW for form factors}, we find that each BCFW term in \eqref{eq: BCFW for form factors} can be obtained from one term in \eqref{eq: BCFW for amplitudes} by replacing a box with the minimal form factor.
It would be interesting to prove relation \eqref{eq: box eater} also at the level of the top-cell diagrams. 
In the following, we will assume that it is valid, which will also yield further examples supporting this conjecture.

The permutation corresponding to the top-cell diagram for the amplitude $\ampco_{n+2,k}$ is given by 
\begin{equation}
 \ampco_{n+2,k}:\qquad \sigma=(k+1,\dots, n,n+1, n+2, 1, 2, \dots, k)\eqndot
\end{equation}
As the permutation corresponding to an additional box is a transposition, the permutation encoded in the top-cell diagram after removing a box at the legs $n+1$ and $n+2$ is
\begin{equation}
 \ffco_{T,n,k} \text{ without }\ffco_{T,2,2}:\qquad \sigmapart=(k+1,\dots, n, n+2, n+1, 1, 2, \dots,k-2, k,k-1)\eqndot
 \label{eq: permutation diagram to be glued}
\end{equation}
Gluing in the minimal form factor at the legs $n+1$ and $n+2$ according to \eqref{eq: permutation minimal form factor}, the permutation for the form factor top-cell diagram becomes
\begin{equation}
 \ffco_{T,n,k}:\qquad \sigma=(k+1,\dots,n,k-1, k, 1, 2, \dots, k-2)\eqndot
\end{equation}

Note that, in contrast to the situation for tree-level amplitudes, the permutation $\sigma$ cannot directly be used to construct on-shell diagrams for tree-level form factors. Instead, we can construct the on-shell diagrams without the minimal form factor via $\sigmapart$, e.g.\ using the \texttt{Mathematica} package \texttt{positroid.m} \cite{Bourjaily:2012gy}, and glue in the minimal form factor. 
This is similar to the situation for on-shell diagrams of one-loop amplitudes, which are not directly constructed from their permutations either, but via corresponding tree-level amplitudes with two additional legs and the forward limit.

The different top-cell diagrams can be obtained from the top-cell diagram with the box replaced by the minimal form factor at legs $n+1$ and $n+2$ via a cyclic permutation of the legs $1,\dots,n$. 
In general, one might expect that all copies of the top-cell diagram under this cyclic permutation are required to generate all BCFW terms. 
For the example of $\ffco_{T,4,3}$, we have seen that two out of four suffice.
It would be interesting to find a pattern for general $n$ and $k$.

\section{\texorpdfstring{$\rr$}{R} operators and integrability}
\label{sec: r operators and integrability}

In this section, we extend the integrability-based construction of scattering amplitude via $\rr$ operators \cite{Frassek:2013xza,Chicherin:2013ora,Broedel:2014pia,Kanning:2014maa} to form factors of the stress-tensor supermultiplet. In particular, this allows us to introduce a central-charge deformation to form factors in analogy to the amplitude case.
While amplitudes are Yangian invariant and hence eigenstates of the spin-chain monodromy matrix, we find that form factors can be constructed as solutions to an eigenvalue equation of the spin-chain transfer matrix.
In particular, the latter statement also generalises to minimal tree-level form factors of generic operators.

\subsection{Construction for amplitudes}

In the integrability-based construction, it is convenient to work with $\GL{4|4}$, which is the (centrally) extended and complexified version of $\PSU{2,2|4}$.
Its generators in what is conventionally called the quantum space can be given in Jordan-Schwinger form $\jj^{\cA\cB}=\hat{x}^\cA\, \hat{p}^\cB$ with the Heisenberg pairs
\begin{equation}
        \label{eq: def xp}
    \hat{x}^{\cA}= \left(\vll^\alpha,-\frac{\partial}{\partial\vlt^{\dot\alpha}} ,\frac{\partial}{\partial\vle^A}\right)
        \eqncom \qquad
        \hat{p}^{\cA}= \left(\frac{\partial}{\partial\vll^\alpha},\vlt^{\dot\alpha} ,\vle^A \right)
        \eqncom 
\end{equation}
which satisfy $[\hat{x}^\cA,\hat{p}^\cB] = (-1)^{|\cA|}\delta^{\cA\cB}$; see e.g.\ \cite{Ferro:2013dga}.%
\footnote{Note that this form of the generators differs from the one in \eqref{eq: onshell algebra} by a slight reorganisation.}
Here, we have combined the indices $\alpha$, $\dot\alpha$ and $A$ into $\cA$, $|\cdot|$ denotes the grading and $[\cdot,\cdot]$ the graded commutator.

One central object in integrability is the Lax operator $\lax$. It depends on the spectral parameter $u$ and acts on one copy of the quantum space associated with the external leg $i$ as well as an auxiliary space with generators $(e^{\cA\cB})_{\cC\cD}=\delta_\cC^\cA\delta_\cD^\cB$:
\begin{equation}
        \label{eq: def lax}
    \lax_i(u) =
  \begin{tikzpicture}[scale=0.8,baseline=-34pt]
  \drawvline{1}{1}   
  \draw[dashed] (-0.5, -1*\vacuumheight-0.5*\bridgedistance) -- (0.5, -1*\vacuumheight-0.5*\bridgedistance);
\node[dl] at (0,-\vacuumheight-\bridgedistance-\labelvdist) {$i$};
  \end{tikzpicture}
    = u + (-1)^{|\cB|}e^{\cA\cB} \;  \hat{x}_i^{\cB}\hat{p}_i^\cA
    \eqncom
\end{equation} 
where we have depicted the quantum space by a solid line and the auxiliary space by a dashed line.

The Lax operators can be used to define the spin-chain monodromy matrix by taking the $n$-fold tensor product with respect to the quantum space and the ordinary product in the auxiliary space:
\begin{equation}
    \label{eq: def monodromy}
\begin{aligned}
    \mono_n(u,\{v_i\})
    =
  \begin{tikzpicture}[scale=0.8,baseline=-34pt]
  \drawvline{1}{1}   
  \drawvline{3}{1}   
  \drawvline{4}{1}   
  \draw[dashed] (-0.5, -1*\vacuumheight-0.5*\bridgedistance) -- (3.5, -1*\vacuumheight-0.5*\bridgedistance);
  \node at (1, -1*\vacuumheight-0.25*\bridgedistance) {$\cdots$};
  \node at (1, -1*\vacuumheight-0.85*\bridgedistance) {$\cdots$};
\node[dl] at (0,-\vacuumheight-\bridgedistance-\labelvdist) {$n$};
\node[dl] at (2,-\vacuumheight-\bridgedistance-\labelvdist) {$2$};
\node[dl] at (3,-\vacuumheight-\bridgedistance-\labelvdist) {$1$};
  \end{tikzpicture}
    = \lax_n(u-v_n) \cdots \lax_2(u-v_2)\lax_1(u-v_1)\eqncom
\end{aligned}    
\end{equation}
where the $v_i$ are inhomogeneities associated with each quantum space. The inhomogeneities correspond to local shifts of the spectral parameter $u$.

The Yangian invariance of tree-level scattering amplitudes \cite{Drummond:2009fd} can be expressed as an eigenvalue equation for the monodromy matrix, cf.\ \cite{Frassek:2013xza,Chicherin:2013ora}:
\begin{equation}
    \label{eq: def monodom}
\mono_n(u,\{v_i\})   \mathcal{A}  \propto \idm \mathcal{A}  \eqndot
\end{equation}
In \cite{Chicherin:2013ora,Broedel:2014pia,Kanning:2014maa}, it was shown how to construct tree-level amplitudes as solutions to this equation.%
\footnote{See also the alternative construction \cite{Frassek:2013xza} via Bethe-ansatz methods.}
This construction is based on reinterpreting and generalising the BCFW bridges and vacua encountered in the previous section.

The BCFW bridges are interpreted as $\rr$ operators \cite{Chicherin:2013ora}, which can be formally written as
\begin{equation}\label{eq: rop}
    \rr_{ij}(u)
    = 
 \begin{tikzpicture}[scale=0.8,baseline=-43pt]
\drawvline{1}{2}
\drawvline{2}{2}
\drawbridge{1}{2}
\node[dl] at (0,-\vacuumheight-2*\bridgedistance-\labelvdist) {$j$};
\node[dl] at (1,-\vacuumheight-2*\bridgedistance-\labelvdist) {$i$};
\end{tikzpicture}
= \int \frac{\de\alpha}{\alpha^{1+u}}\e^{-\alpha (\hat{x}_j \cdot \hat{p}_i)}
   \eqncom
\end{equation}
where $u$ is the spectral parameter.
These operators satisfy the Yang-Baxter equation
\begin{equation}
        \label{eq: rll}
        \rr_{ij}(u_j-u_i)
        \lax_j(u_j)\lax_i(u_i)
        =
        \lax_j(u_i)\lax_i(u_j)
        \rr_{ij}(u_j-u_i)
        \eqncom
\end{equation}
which can be depicted as 
\begin{equation} \label{eq: property of r operators in picture} 
 \begin{tikzpicture}[scale=0.8, baseline=-43pt]
 	\drawvline{1}{2}   
 	\drawvline{2}{2}   
 	\drawbridge{1}{2}
	\draw[dashed] (-0.5, -1*\vacuumheight-1.5*\bridgedistance) -- (1.5, -1*\vacuumheight-1.5*\bridgedistance);
\node[dl] at (1,-\vacuumheight-2*\bridgedistance-\labelvdist) {$i$};
\node[dl] at (0,-\vacuumheight-2*\bridgedistance-\labelvdist) {$j$};
         \end{tikzpicture}
         =
           \begin{tikzpicture}[scale=0.8, baseline=-43pt]
	\drawvline{1}{2}   
 	\drawvline{2}{2}   
 	\drawbridge{1}{2}
	\draw[dashed] (-0.5, -1*\vacuumheight-0.5*\bridgedistance) -- (1.5, -1*\vacuumheight-0.5*\bridgedistance);
\node[dl] at (1,-\vacuumheight-2*\bridgedistance-\labelvdist) {$i$};
\node[dl] at (0,-\vacuumheight-2*\bridgedistance-\labelvdist) {$j$};
         \end{tikzpicture}
         \eqndot
\end{equation}
Note that the arguments of the Lax operators on the right hand side of \eqref{eq: rll} are exchanged with respect to the left hand side.

The vacua are given by 
\begin{equation}
  \begin{aligned}\begin{tikzpicture}[scale=0.8]
\drawvacp{1}
\node[dl] at (0,-\vacuumheight-0*\bridgedistance-\labelvdist) {$i$};
    \end{tikzpicture}\end{aligned} 
  =
      \deltap{i}=\delta^{2}(\vll_i)\eqncom\qquad
  \begin{aligned}\begin{tikzpicture}[scale=0.8]
\drawvacm{1}
\node[dl] at (0,-\vacuumheight-0*\bridgedistance-\labelvdist) {$i$};
    \end{tikzpicture}\end{aligned} 
  =
        \deltam{i}=\delta^{2}(\vlt_i)\delta^{4}(\vle_i)
  \eqndot
  \label{eq: vacua}
\end{equation}
They are eigenstates of the Lax operators, i.e.\ they satisfy 
\begin{equation}
        \label{eq: lax on vacua}
        \lax_i(u)\,\delta_{i}^+
        = (u-1)\; \idm \; \delta_{i}^+\eqncom
        \qquad
        \lax_i(u)\, \delta_{i}^-
        = u \; \idm \; \delta_{i}^-
        \eqncom
\end{equation}
which is depicted as
\begin{equation}
  \begin{tikzpicture}[scale=0.8,baseline=-20pt]
\drawvacp{1}
\draw[dashed] (-0.5, -0.75*\vacuumheight) -- (0.5, -0.75*\vacuumheight);
\node[dl] at (0,-\vacuumheight-0*\bridgedistance-\labelvdist) {$i$};
 \end{tikzpicture}
=(u-1)
  \begin{tikzpicture}[scale=0.8,baseline=-20pt]
\drawvacp{1}
\draw[dashed] (-0.5, 0.3*\vacuumheight) -- (0.5, 0.3*\vacuumheight);
\node[dl] at (0,-\vacuumheight-0*\bridgedistance-\labelvdist) {$i$};
 \end{tikzpicture}
\eqncom \qquad
  \begin{tikzpicture}[scale=0.8,baseline=-20pt]
\drawvacm{1}
\draw[dashed] (-0.5, -0.75*\vacuumheight) -- (0.5, -0.75*\vacuumheight);
\node[dl] at (0,-\vacuumheight-0*\bridgedistance-\labelvdist) {$i$};
 \end{tikzpicture}
=u
  \begin{tikzpicture}[scale=0.8,baseline=-20pt]
\drawvacm{1}
\draw[dashed] (-0.5, 0.3*\vacuumheight) -- (0.5, 0.3*\vacuumheight);
\node[dl] at (0,-\vacuumheight-0*\bridgedistance-\labelvdist) {$i$};
 \end{tikzpicture}
\eqndot
\end{equation}

Deformed tree-level scattering amplitudes can then be constructed by acting with a chain of $\rr$ operators on the vacua \cite{Chicherin:2013ora,Broedel:2014pia,Kanning:2014maa} following the procedure discussed in subsection \ref{subsec: systematic construction} for the undeformed case.
The resulting on-shell diagram has to be planar, which imposes certain constraints on the $\rr$ operators. In particular, we assume that $i<j$ for each $\rr_{ij}$. 
In order to yield a Yangian invariant, 
the different spectral parameters and inhomogeneities have to be related via the permutation $\sigma$ that is associated with the on-shell diagram as discussed in subsection \ref{subsec: permutations}.
Concretely, as a consequence of \eqref{eq: rll}, the monodromy matrix can be pulled through a sequence of $\rr$ operators, 
\begin{equation}
  \label{eq: rchain com}
\mono(u,\{v_i\})\,
    \rr_{i_1j_1}(z_1)\cdots \rr_{i_mj_m}(z_m)=\rr_{i_1j_1}(z_1)\cdots \rr_{i_mj_m}(z_m) \,
    \mono(u,\{v_{\sigma(i)}\}) \eqncom
\end{equation}
provided that 
we have the following relation among the inhomogeneities $v_i$ and the spectral parameters $z_i$: 
\begin{equation}
    z_\ell = v_{\tau_\ell(i_\ell)}-v_{\tau_\ell(j_\ell)}\quad\quad\text{with}\quad\quad     \tau_\ell= (i_1,j_1)\cdots(i_\ell,j_\ell)\eqncom \qquad \ell=1,\ldots,m\eqncom
\end{equation}
see \cite{Chicherin:2013ora,Broedel:2014pia,Kanning:2014maa} for details. 
The inhomogeneities $v_i$ are related to the central charges $c_i$ via \cite{Beisert:2014qba}
\begin{equation}
\label{eq: central charges from inhomogenities}
    c_i=v_i-v_{\sigma(i)}\eqndot
\end{equation}

For example, the deformed three-point tree-level MHV amplitude $\ampco_{3,2}$ can be constructed as
\begin{equation}
\label{eq: A 3,2 from r operators}
  \rr_{23}(v_{32})\rr_{12}(v_{31})\deltap{1}\deltam{2}\deltam{3}
  =
        \frac{
                \delta^4(\sum_{i=1}^3\vll_i\vlt_i)
                \delta^4(\sum_{i=1}^3\vll_i\vle_i^+)
                \delta^4(\sum_{i=1}^3\vll_i\vle_i^-)
              }{\abr{12}^{1-v_{23}}\abr{23}^{1-v_{31}}\abr{31}^{1-v_{12}}}
              \eqncom
\end{equation}
where 
\begin{equation}
 \label{eq: def uij}
 v_{ij}=v_i-v_j \eqndot 
\end{equation}
Its Yangian invariance is diagrammatically shown in figure \ref{fig: monodromy on amp 3,2}.

\begin{figure}[htbp]
 \begin{equation*}
\begin{aligned}
 \begin{tikzpicture}[scale=0.8]
\drawvline{1}{2}
\drawvline{2}{2}
\drawvline{3}{2}
\drawvacm{1} 
\drawvacm{2} 
\drawvacp{3}
\drawbridge{2}{1}
\drawbridge{1}{2}
\node[dl] at (0,-\vacuumheight-2*\bridgedistance-\labelvdist) {3};
\node[dl] at (1,-\vacuumheight-2*\bridgedistance-\labelvdist) {2};
\node[dl] at (2,-\vacuumheight-2*\bridgedistance-\labelvdist) {1};
\draw[dashed] (-0.5, -1*\vacuumheight-1.5*\bridgedistance) -- (2.5, -1*\vacuumheight-1.5*\bridgedistance);
\end{tikzpicture}
\end{aligned}
\;\;=\;\;
\begin{aligned}
 \begin{tikzpicture}[scale=0.8]
\drawvline{1}{2}
\drawvline{2}{2}
\drawvline{3}{2}
\drawvacm{1} 
\drawvacm{2} 
\drawvacp{3}
\drawbridge{2}{1}
\drawbridge{1}{2}
\node[dl] at (0,-\vacuumheight-2*\bridgedistance-\labelvdist) {3};
\node[dl] at (1,-\vacuumheight-2*\bridgedistance-\labelvdist) {2};
\node[dl] at (2,-\vacuumheight-2*\bridgedistance-\labelvdist) {1};
\draw[dashed] (-0.5, -1*\vacuumheight+0.45*\bridgedistance) -- (2.5, -1*\vacuumheight+0.45*\bridgedistance);
\end{tikzpicture}
\end{aligned}
=(u-v_3-1)(u-v_1)(u-v_2)
\begin{aligned}
 \begin{tikzpicture}[scale=0.8]
\drawvline{1}{2}
\drawvline{2}{2}
\drawvline{3}{2}
\drawvacm{1} 
\drawvacm{2} 
\drawvacp{3}
\drawbridge{2}{1}
\drawbridge{1}{2}
\node[dl] at (0,-\vacuumheight-2*\bridgedistance-\labelvdist) {3};
\node[dl] at (1,-\vacuumheight-2*\bridgedistance-\labelvdist) {2};
\node[dl] at (2,-\vacuumheight-2*\bridgedistance-\labelvdist) {1};
\draw[dashed] (-0.5, -1*\vacuumheight+1.75*\bridgedistance) -- (2.5, -1*\vacuumheight+1.75*\bridgedistance);
\end{tikzpicture}
\end{aligned}
\end{equation*}
\caption{Action of the monodromy matrix on $\ampco_{3,2}$.}
\label{fig: monodromy on amp 3,2}
\end{figure}

\subsection{Construction for form factors of the stress-tensor supermultiplet}

As for amplitudes, we can also construct on-shell diagrams for form factors of the chiral half of the stress-tensor supermultiplet via the $\rr$ operators \eqref{eq: rop} and the vacua \eqref{eq: vacua} if we include the minimal form factor as an additional vacuum.
We denote such on-shell diagrams, which can in particular encode top-cell diagrams, BCFW terms and factorisation channels, by $\ffgen_{T,n}$:
\begin{align}
\label{eq: ff function}
&\ffgen_{T,n}= \rr_{i_1j_1}(z_1)\cdots \rr_{i_mj_m}(z_m) \;\; \delta^+_1 \cdots \delta^+_{k-2} \;
    \ffco_{T,2,2}(k-1,k)\;
    \delta^-_{k+1}\cdots\delta^-_{n}
\eqncom
\end{align}
where $m$ is the number of $\rr$ operators.

\newcommand{\ffactor}{\mathfrak{f}}
Using the above steps, we find that 
\begin{equation}
    \label{eq: monodromy on ff}
    \begin{split}
    \mono_n(u,\{v_{i}\})
    \ffgen_{T,n}
    &=\ffactor(u,\{v_{\sigma(i)}\})\; \rr_{i_1j_1}(z_1)\cdots \rr_{i_mj_m}(z_m)\\
    &\quad\times\delta^+_1 \cdots \delta^+_{k-2} \;
        \left[
        \mono_2(u,\{v_{\sigma(i)}\})
        \ffco_{T,2,2}(k-1,k) \;
    \right]\;
    \delta^-_{k+1}\cdots\delta^-_{n} \eqncom
    \end{split}
\end{equation} 
where
\begin{equation}
\label{eq: length two monodromy}
\mono_2(u,\{v_{\sigma(i)}\})=\lax_{k}(u-v_{\sigma(k)})\lax_{k-1}(u-v_{\sigma(k-1)})
\end{equation}
is the reduced monodromy matrix of length two acting on sites $k-1$ and $k$ and 
\begin{equation}
\label{eq: ffactor}
 \ffactor(u,\{v_{\sigma(i)}\})=\prod_{i=1}^{k-2}(u-v_{\sigma(i)}-1)\prod_{i=k+1}^{n}(u-v_{\sigma(i)}) 
\end{equation}
arises from the action \eqref{eq: lax on vacua} of the Lax operators on the vacua $\deltap{}$ and $\deltam{}$. 
This procedure is illustrated in figure \ref{fig: monodromy on ff n,2} for the case of $\ffco_{T,n,2}$.

\begin{figure}[htbp]
\begin{equation*}
  \begin{aligned}
 \begin{tikzpicture}[scale=0.8]
\drawvline{1}{4}
\drawvline{2}{4}
\drawvline{4}{4}
\drawvline{5}{4}
\drawminimalff{1} 
\drawvacp{4}
\drawvacp{5}
\node at (0,-\vacuumheight-4*\bridgedistance-\labelvdist) {\strut $n$};
\node at (1,-\vacuumheight-4*\bridgedistance-\labelvdist) {$n \sminus 1$};
\node at (3,-\vacuumheight-4*\bridgedistance-\labelvdist) {$2$};
\node at (4,-\vacuumheight-4*\bridgedistance-\labelvdist) {$1$};
\node at (2,-0.25) {$\cdots$};
\node at (2,-\vacuumheight-4*\bridgedistance-\labelvdist) {$\cdots$};
\draw[dashed] (-0.5, -1*\vacuumheight-3.5*\bridgedistance) -- (4.5, -1*\vacuumheight-3.5*\bridgedistance);
\node[rectangle, rounded corners=\vacuumradius, black, fill=grayn, minimum width=4 cm, minimum height=1.5 cm, draw, inner sep=0pt] at (2,-\vacuumheight-1.5*\bridgedistance) {$\text{BCFW bridges}$};
\end{tikzpicture}
\end{aligned}
\;\;=\;\;
  \begin{aligned}
 \begin{tikzpicture}[scale=0.8]
\drawvline{1}{4}
\drawvline{2}{4}
\drawvline{4}{4}
\drawvline{5}{4}
\drawminimalff{1} 
\drawvacp{4}
\drawvacp{5}
\node at (0,-\vacuumheight-4*\bridgedistance-\labelvdist) {\strut $n$};
\node at (1,-\vacuumheight-4*\bridgedistance-\labelvdist) {$n \sminus 1$};
\node at (3,-\vacuumheight-4*\bridgedistance-\labelvdist) {$2$};
\node at (4,-\vacuumheight-4*\bridgedistance-\labelvdist) {$1$};
\node at (2,-0.25) {$\cdots$};
\node at (2,-\vacuumheight-4*\bridgedistance-\labelvdist) {$\cdots$};
\draw[dashed] (-0.5, -1*\vacuumheight+0.25*\bridgedistance) -- (4.5, -1*\vacuumheight+0.25*\bridgedistance);
\node[rectangle, rounded corners=\vacuumradius, black, fill=grayn, minimum width=4 cm, minimum height=1.5 cm, draw, inner sep=0pt] at (2,-\vacuumheight-1.5*\bridgedistance) {$\text{BCFW bridges}$};
\end{tikzpicture}
\end{aligned}
\;\;=\;\; \ffactor(u) \;
  \begin{aligned}
 \begin{tikzpicture}[scale=0.8]
\drawvline{1}{4}
\drawvline{2}{4}
\drawvline{4}{4}
\drawvline{5}{4}
\drawminimalff{1} 
\drawvacp{4}
\drawvacp{5}
\node at (0,-\vacuumheight-4*\bridgedistance-\labelvdist) {\strut $n$};
\node at (1,-\vacuumheight-4*\bridgedistance-\labelvdist) {$n \sminus 1$};
\node at (3,-\vacuumheight-4*\bridgedistance-\labelvdist) {$2$};
\node at (4,-\vacuumheight-4*\bridgedistance-\labelvdist) {$1$};
\node at (2,-0.25) {$\cdots$};
\node at (2,-\vacuumheight-4*\bridgedistance-\labelvdist) {$\cdots$};
\draw[dashed] (-0.5, -1*\vacuumheight+0.25*\bridgedistance) -- (1.5, -1*\vacuumheight+0.25*\bridgedistance);
\node[rectangle, rounded corners=\vacuumradius, black, fill=grayn, minimum width=4 cm, minimum height=1.5 cm, draw, inner sep=0pt] at (2,-\vacuumheight-1.5*\bridgedistance) {$\text{BCFW bridges}$};
\end{tikzpicture}
\end{aligned}
\end{equation*}
\caption{Action of the monodromy matrix on $\ffco_{T,n,2}$.}
\label{fig: monodromy on ff n,2}
\end{figure}

In contrast to the vacua $\deltap{}$ and $\deltam{}$, the minimal form factor is not an eigenstate of the monodromy matrix $\mono_2$ as would be required for Yangian invariance in analogy to the amplitude case.
For instance, the off-diagonal generator $\jj^{\alpha\dot\alpha}$ acts as 
\begin{equation}
 \sum_{i=k-1}^k \jj^{\alpha\dot\alpha}_i \ffco_{T,2,2}(k-1,k)= (\lambda_{k-1}^\alpha\lambdat_{k-1}^{\dot\alpha}+\lambda_{k}^\alpha\lambdat_{k}^{\dot\alpha})\ffco_{T,2,2}(k-1,k)
 = q^{\alpha\dot\alpha}\ffco_{T,2,2}(k-1,k)\eqncom
\end{equation}
which is non-vanishing; cf.\ \eqref{eq: action on form factor}.

\subsection{Transfer matrix}

Instead of considering the monodromy matrix $\mono_n$, we can take its supertrace  
to obtain the transfer matrix $\trans_n$:
 \begin{equation}
    \label{eq: def transfer}
   \trans_n(u,\{v_i\})
   =
  \begin{tikzpicture}[scale=0.8,baseline=-33.5pt]
  \drawvline{1}{1}   
  \drawvline{3}{1}   
  \drawvline{4}{1}   
  \draw[rounded corners=2mm, dashed] (-0.5, -1*\vacuumheight-0.5*\bridgedistance) rectangle (3.5, -1*\vacuumheight+0.25*\bridgedistance);
  \node at (1, -1*\vacuumheight-0.25*\bridgedistance) {$\cdots$};
  \node at (1, -1*\vacuumheight-0.85*\bridgedistance) {$\cdots$};
\node[dl] at (0,-\vacuumheight-\bridgedistance-\labelvdist) {$n$};
\node[dl] at (2,-\vacuumheight-\bridgedistance-\labelvdist) {$2$};
\node[dl] at (3,-\vacuumheight-\bridgedistance-\labelvdist) {$1$};
  \end{tikzpicture}
    =\str\mono_n(u,\{v_i\})  \eqndot  
\end{equation}
As the monodromy matrix, the transfer matrix can be pulled through the sequence of $\rr$ operators to obtain the action of the reduced transfer matrix $\trans_2(u,\{v_{\sigma(i)}\})$ on the minimal form factor.
Moreover, we require the reduced transfer matrix to be homogeneous, i.e.\ the two inhomogenities occurring in it have to be equal:
\begin{equation}
 \label{eq: trans two hom}
    \trans_2(u-v)
    =\str\lax_{k}(u-v_{\sigma(k)})
     \lax_{k-1}(u-v_{\sigma(k-1)})
    \eqncom    \qquad
    \text{with}\quad
    v_{\sigma(k-1)}=
    v_{\sigma(k)} \equiv v 
    \eqndot
\end{equation}
In contrast to the reduced monodromy matrix $\mono_2(u,\{v_{\sigma(i)}\})$, the reduced transfer matrix $\trans_2(u-v)$ satisfies an eigenvalue equation with respect to the minimal form factor, as we will see in the following.
Further below, we will show that this also generalises to the case of generic single-trace operators.

An important property of the transfer matrix is its $\GL{4|4}$-invariance:
\begin{equation}
    \label{eq: transfer matrix invariance}
    \left[\trans_n(u,\{ v_i\}), \sum_{i=1}^n \jj_i^{\cA \cB}\right]=0
    \eqndot
\end{equation} 
In particular, it commutes with any function of $\sum_{i=1}^n \jj_i^{\alpha \dot\alpha}=\sum_{i=1}^n \lambda^\alpha_i\lambdat^{\dot\alpha}_i$ and $\sum_{i=1}^n \jj_i^{\alpha A}=\sum_{i=1}^n \lambda^\alpha_i\etat^{A}_i$, and thus also with the momentum- and supermomentum-conserving delta functions.

Using this property, we find that 
\begin{equation}
    \trans_2(u-v)\ffco_{T,2,2}= \delta^4(P)\delta^4(Q^+)\delta^4(Q^{-})\trans_2(u-v)\frac{1}{\abr{12}\abr{21}} =0
    \eqncom
\end{equation}
where the last step is the consequence of a straightforward evaluation using \eqref{eq: trans two hom}, \eqref{eq: def lax} and \eqref{eq: def xp}.  
This means the minimal tree-level form factor of the stress-tensor supermultiplet is an eigenstate of the homogeneous transfer matrix $\trans_2(u-v)$ with eigenvalue zero.
Moreover, it follows from the previous discussion that also all planar on-shell diagrams containing an insertion of the minimal form factor $\ffco_{T,2,2}$ are annihilated by $\trans_n$:
\begin{equation}
 \trans_n(u,\{ v_i\}|_{v_{\sigma(k-1)}=
    v_{\sigma(k)}})\ffgen_{T,n}=0\eqndot
\end{equation}
In particular, this is the case for the (undeformed) tree-level form factors $\ffco_{T,n,k}$ themselves: 
\begin{equation}
\label{eq: eigenvalue equation ff T,n,k}
 \trans_n(u)\ffco_{T,n,k}=0\eqndot
\end{equation}

Using the construction above, we can also build deformed form factors as solutions to \eqref{eq: eigenvalue equation ff T,n,k}.
For instance, we find 
\begin{equation} 
\label{eq: deformed ff n,2}
\ffco_{T,n,2}(1,\dots,n;q,\gamma^-)=\frac{
  \delta^4(\sum_{i=1}^n\vll_i\vlt_i-q)
  \delta^4(\sum_{i=1}^n\vll_i\vle_i^+)
  \delta^4(\sum_{i=1}^n\vll_i\vle_i^- - \gamma^-)
}{
\prod_{i=1}^n \abr{i \ssep i \splus 1}^{1-v_{i+1 \sssep i+2}}
}\eqncom
\end{equation}
where $v_{ij}$ was defined in \eqref{eq: def uij} and $v_3=v_4$ has to be satisfied for the reduced transfer matrix to be homogeneous.
In the limit of vanishing deformation parameters, \eqref{eq: deformed ff n,2} reduces to \eqref{eq: ff n,2 intro} as required.

\subsection{Generic operators}

Let us now look at form factors of generic operators $\cO$ of length $L$, starting at the minimal case.
As the homogeneous transfer matrix of length $L$ commutes with the momentum-conserving delta function, it only acts on the last factor in \eqref{eq: minimal tree-level form factor from oscillator replacement}, which contains the operator in the spin-chain picture with oscillators replaced by super-spinor-helicity variables according to \eqref{eq: oscillator replacements}.
We thus have
\begin{equation}
\label{eq: transfer matrix pulled through form factor}
    \trans_L(u) \ffco_{\cO,L}
    =
    \ffco_{\transosc_L(u)\cO,L}
    \eqncom
\end{equation}
where 
\begin{equation}
 \transosc_L(u)=  \trans_L(u)\,\rule[-1.06cm]{0.1mm}{1.415cm} \!\phantom{|}_{\substack{\\
\begin{smallmatrix}
        \aosc_{i,\alpha},\aosc_i^{\dagger \alpha} &\to& \partial_{i,\alpha},\vll_i^\alpha \\
  \bosc_{i,\dot\alpha},\bosc_i^{\dagger \dot\alpha} &\to& \partial_{i,\dot\alpha},\vlt_i^{\dot\alpha}\\
  \dosc_{i,A},\dosc_i^{\dagger A} &\to& \partial_{i,A},\vle_i^A
    \end{smallmatrix}}}
    \label{eq: minimal form factor oscillatorls}
\end{equation} 
is the homogeneous transfer matrix in the oscillator representation, which also arises in the spectral problem.
Although the transfer matrix $\transosc_L(u)$ does not contain the spin-chain Hamiltonian, i.e.\ the one-loop dilatation operator $\loopDila{1}$ of \eqref{eq: one-loop dila in oscillators}, it can be used to diagonalise $\loopDila{1}$ and the transfer matrix that does contain $\loopDila{1}$;
see e.g.\ \cite{Faddeev:1996iy,PhysRevLett.19.103,Sklyanin:1991ss,Ferro:2013dga,Kazama:2015iua}.
Hence, an operator $\cO$ is an eigenstate of $\transosc_L(u)$,
\newcommand{\eigenval}{\mathfrak{t}}
\begin{equation}\label{eq: transfer matrix eigenvalue eq}
 \transosc_L(u) \cO=\eigenval(u) \cO \eqncom
\end{equation} 
provided that it is an eigenstate of the transfer matrix that contains $\loopDila{1}$.
In this case, \eqref{eq: transfer matrix pulled through form factor} yields
\begin{equation}
\label{eq: transfer matrix minimal form factor eigenvalue equation}
    \trans_L(u) \ffco_{\cO,L}
    =
    \eigenval(u)\ffco_{\cO,L}
    \eqncom
\end{equation}
i.e.\ the minimal form factor is an eigenstate of the transfer matrix occurring in the study of amplitudes provided that the corresponding operator is an eigenstate of the one-loop spectral problem.

\begin{figure}[htbp]
\begin{equation*}
        \begin{tikzpicture}[baseline=-1cm,scale=0.8]
                \node[draw,inner sep=0](ff) at (1,1) {};
                \draw[double] (1,1.4)--(ff);
                \node[] (onshell) at (1, -1) {};
                \draw (ff) to[in=65,out=305] ([shift={(0.4,0)}] onshell);
                \draw (ff) to[in=90,out=270] ([shift={(0,0)}] onshell);
                \draw (ff) to[in=115,out=235] ([shift={(-0.4,0)}] onshell);
                \draw ([shift={(-0.8,0)}] onshell) to[out=-140, in=90] (-1,-3);
                \draw ([shift={(-0.4,0)}] onshell) to[out=-115, in=90] (0,-3);
                \draw ([shift={(0.0,0)}] onshell) to[out=-90, in=90] (1,-3);
                \draw ([shift={(0.4,0)}] onshell) to[out=-65, in=90] (2,-3);
                \draw ([shift={(0.8,0)}] onshell) to[out=-40, in=90] (3,-3);
                \node[rectangle, rounded corners=15pt, draw, inner sep=10pt,  minimum height=55pt, scale=0.8, fill=lightgrayn,align=center] at (1, -1) {BCFW bridges \& $\delta^\pm$s};
                \draw[rounded corners=3pt, dashed] (-1.5,-2.7) rectangle (3.5,-2.4);
        \end{tikzpicture}
        \;\;
=
        \;\; \ffactor(u)\;
        \begin{tikzpicture}[baseline=-1cm,scale=0.8]
                \node[draw,inner sep=0](ff) at (1,1) {};
                \draw[ double] (1,1.4)--(ff);
                \node[] (onshell) at (1, -1) {};
                \draw (ff) to[in=65,out=305] ([shift={(0.4,0)}] onshell);
                \draw (ff) to[in=90,out=270] ([shift={(0,0)}] onshell);
                \draw (ff) to[in=115,out=235] ([shift={(-0.4,0)}] onshell);
                \draw ([shift={(-0.8,0)}] onshell) to[out=-140, in=90] (-1,-3);
                \draw ([shift={(-0.4,0)}] onshell) to[out=-115, in=90] (0,-3);
                \draw ([shift={(0.0,0)}] onshell) to[out=-90, in=90] (1,-3);
                \draw ([shift={(0.4,0)}] onshell) to[out=-65, in=90] (2,-3);
                \draw ([shift={(0.8,0)}] onshell) to[out=-40, in=90] (3,-3);
                \node[rectangle, rounded corners=15pt, draw, inner sep=10pt,  minimum height=55pt, scale=0.8, fill=lightgrayn] at (1, -1) {BCFW bridges \& $\delta^\pm$s};
                \draw[rounded corners=3pt, dashed] (0.2,0.15) rectangle (1.8,0.45);
        \end{tikzpicture}
        \;\;
        =
        \;\;
        \ffactor(u)\eigenval(u)\;
        \begin{tikzpicture}[baseline=-1cm,scale=0.8]
                \node[draw,inner sep=0](ff) at (1,1) {};
                \draw[ double] (1,1.4)--(ff);
                \node[] (onshell) at (1, -1) {};
                \draw (ff) to[in=65,out=305] ([shift={(0.4,0)}] onshell);
                \draw (ff) to[in=90,out=270] ([shift={(0,0)}] onshell);
                \draw (ff) to[in=115,out=235] ([shift={(-0.4,0)}] onshell);
                \draw ([shift={(-0.8,0)}] onshell) to[out=-140, in=90] (-1,-3);
                \draw ([shift={(-0.4,0)}] onshell) to[out=-115, in=90] (0,-3);
                \draw ([shift={(0.0,0)}] onshell) to[out=-90, in=90] (1,-3);
                \draw ([shift={(0.4,0)}] onshell) to[out=-65, in=90] (2,-3);
                \draw ([shift={(0.8,0)}] onshell) to[out=-40, in=90] (3,-3);
                \node[rectangle, rounded corners=15pt, draw, inner sep=10pt,  minimum height=55pt, scale=0.8, fill=lightgrayn] at (1, -1) {BCFW bridges \& $\delta^\pm$s};
        \end{tikzpicture}
\end{equation*}
\caption{Action of the transfer matrix on a planar on-shell diagram containing the minimal form factor of a generic operator.}
\label{fig: transfer}
\end{figure}
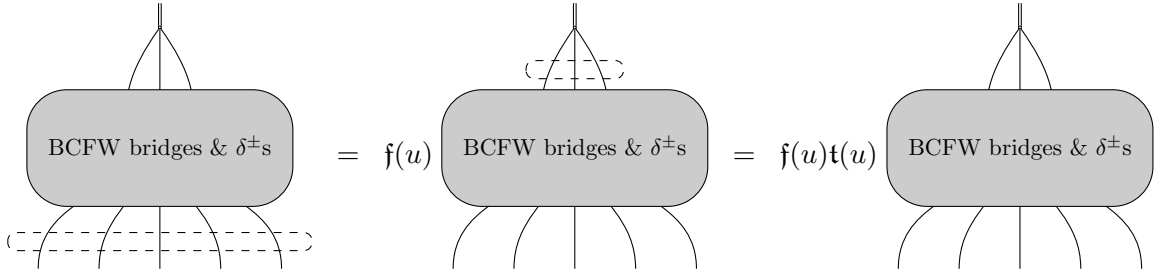

Moreover, we can build planar on-shell diagrams containing $\ffco_{\cO,L}$ in analogy to \eqref{eq: ff function}. They satisfy 
\begin{equation}
    \label{eq: eigenvalue equation2}
  \trans_n(u)\; \ffgen_{\cO,n}
=
\ffactor(u)
    \ffgen_{\transosc_L\cO,n}
  =\ffactor(u)
    \eigenval(u)\;\ffgen_{\cO,n}
    \eqncom
\end{equation}
where $\ffactor(u)$ denotes the factor arising from the action of the Lax operators on the amplitude vacua in generalisation of \eqref{eq: ffactor} and the last equality holds for operators satisfying \eqref{eq: transfer matrix eigenvalue eq}.
This equation is illustrated in figure \ref{fig: transfer}.%
\footnote{Actually, the second step in figure \ref{fig: transfer} shows the generalisation of \eqref{eq: monodromy on ff} to the transfer matrix and to generic operators, which coincides with the second step in \eqref{eq: eigenvalue equation2} via \eqref{eq: transfer matrix pulled through form factor}.}

Note that the on-shell diagrams $\ffgen_{\cO,n}$ containing an insertion of the minimal form factor  $\ffco_{\cO,L}$ do not necessarily correspond to tree-level form factors of the operators $\cO$.
However, they yield certain leading singularities of loop-level form factors of $\cO$.
It would be very interesting to clarify the relation between these on-shell diagrams and the general tree-level form factors of the operators.%
\footnote{In contrast to the BCFW recursion relation \eqref{eq: BCFW for form factors} for $T$, an analogous recursion relation for the MHV form factors of the supermultiplet of $\tr(\phi_{14}^L)$ contains non-vanishing residues at infinity \cite{Penante:2014sza}. Nevertheless, also the latter recursion relation can be solved and could be used as a basis for constructing on-shell diagrams for these operators. 
Other BCFW recursion relations for MHV form factors, which involve also shifts of the off-shell momentum $q$, were studied in the 
$\SU{2}$ and $\SL{2}$ sectors in \cite{Engelund:2012re}; these might also be suitable to construct on-shell diagrams.
In \cite{KMSW}, we will provide explicit expressions for non-minimal tree-level form factors for general operators.}
 We leave a detailed investigation for future work.

\section{Graßmannian integrals}
\label{sec: Grassmannian integrals}

An amazing discovery in the study of scattering amplitudes in $\cN=4$ SYM theory was that these objects can be written as integrals of a certain on-shell form on the Graßmannian manifold $\GrassmannSymbol(k,n)$, i.e.\ on the space of all $k$-dimensional planes in $n$-dimensional space \cite{ArkaniHamed:2009dn,Mason:2009qx,ArkaniHamed:2009vw,ArkaniHamed:2012nw}.
In the following, we will show that a similar formulation also exists for form factors.
Again, we focus on the stress-tensor supermultiplet $T$.

\subsection{Geometry of (super)momentum conservation}

One important idea behind the Graßmannian integral formulation for scattering amplitudes is to realise momentum conservation and supermomentum conservation geometrically.
To this end, the collection of $\lambda_i^\alpha$ variables is considered as a two-dimensional plane in an $n$-dimensional space. 
Similarly, the $\lambdat_i^\alphadot$ variables are considered as another two-dimensional plane, and the $\etat_i^A$ variables are considered as a four-dimensional plane. 

Momentum conservation is the statement that the $\lambda$-plane and the $\lambdat$-plane are orthogonal to each other:
\begin{equation}
\label{eq: momentum conservation as orthogonality of planes}
 \vll\cdot\vlt \equiv \sum_{i=1}^n\vll_i\vlt_i=0 \eqndot
\end{equation}
Similarly, supermomentum conservation is the statement that the $\lambda$-plane and the $\etat$-plane are orthogonal to each other:
\begin{equation}
\label{eq: super momentum conservation as orthogonality of planes}
 \vll\cdot\vle \equiv \sum_{i=1}^n\vll_i\vle_i=0 \eqndot
\end{equation}
These constraints can be linearised by introducing the auxiliary plane $C\in \GrassmannSymbol(k,n)$, which contains the $\lambda$-plane and is orthogonal to the $\lambdat$- and the $\etat$-plane:
\begin{equation}
\begin{aligned}
  (C\cdot\vlt)_I=\sum_{i=1}^nC_{Ii}\vlt_i=0 \quad\text{and}\quad (C^\perp\cdot\vll)_J=\sum_{i=1}^nC^\perp_{Ji}\vll_i=0 \quad\implies\quad
  \vll\cdot\vlt=0
  \eqncom\\
    (C\cdot\vle)_I=\sum_{i=1}^nC_{Ii}\vle_i=0 \quad\text{and}\quad (C^\perp\cdot\vll)_J=\sum_{i=1}^nC^\perp_{Ji}\vll_i=0 \quad\implies\quad
  \vll\cdot\vle=0
  \eqncom
\end{aligned}  
\end{equation}
with $I=1,\dots,k$, $J=1,\dots,n-k$.
Here, the requirement that $C$ contains the $\lambda$-plane is written as the $\lambda$-plane being orthogonal to the orthogonal complement $C^\perp$ of $C$, which satisfies
\begin{equation}
\label{eq: def Cperp}
 C(C^\perp)^T=0\eqndot
\end{equation}

We would like to proceed in a similar way for form factors. The Graßmannian $\GrassmannSymbol(n,k)$, however, is too small for this purpose; in particular, $k$ ranges from $2$ to $n$ for form factors but $\GrassmannSymbol(n,n)$ is just a point. The relation \eqref{eq: box eater} between the top-cell diagrams of $\ampco_{n+2,k}$ and $\ffco_{T,n,k}$ suggests that the correct Graßmannian is $\GrassmannSymbol(n+2,k)$. Moreover, this is consistent with the fact that one off-shell momentum can be parametrised by two on-shell momenta.
Concretely, we can define the new set of variables
\begin{equation}
 \label{eq: def underunderscore variables}
 \begin{aligned}
\vlluu_k&=\vll_k \eqncom \quad k=1,\dots,n \eqncom &
\vlluu_{n+1}&=\refspina\eqncom&
\vlluu_{n+2}&=\refspinb\eqncom 
\\
\vltuu_k&=\vlt_k \eqncom \quad k=1,\dots,n \eqncom &
\vltuu_{n+1}&=-\frac{\bra{\refspinbl}q}{\abr{\refspinbl\refspinal}}\eqncom&
\vltuu_{n+2}&=-\frac{\bra{\refspinal}q}{\abr{\refspinal\refspinbl}}\eqncom
\\
\vleuu_k^+&=\vle_k^+ \eqncom \quad k=1,\dots,n \eqncom &
\vleuu_{n+1}^+&=0\eqncom&
\vleuu_{n+2}^+&=0\eqncom 
\\
\vleuu_k^-&=\vle_k^- \eqncom \quad k=1,\dots,n \eqncom &
\vleuu_{n+1}^-&=-\frac{\bra{\refspinbl}\gamma^-}{\abr{\refspinbl\refspinal}}\eqncom&
\vleuu_{n+2}^-&=-\frac{\bra{\refspinal}\gamma^-}{\abr{\refspinal\refspinbl}}\eqncom
\end{aligned}
\end{equation}
where $\refspinal$ and $\refspinbl$ are arbitrary reference spinors which account for the fact that two on-shell momenta together have two more degrees of freedom than one off-shell momentum.
The two additional on-shell legs satisfy 
\begin{equation}
 \vlluu_{n+1}\vltuu_{n+1}+\vlluu_{n+2}\vltuu_{n+2}=-q \eqncom \qquad
 \vlluu_{n+1}\vleuu^\pm_{n+1}+\vlluu_{n+2}\vleuu^\pm_{n+2}=-\gamma^\pm \eqncom
\end{equation}
where $\gamma^+=0$.
Hence, momentum and supermomentum conservation can be written as $\vlluu \cdot \vltuu=0$ and $\vlluu \cdot \vleuu=0$, respectively. 
These constraints can be linearised via an auxiliary plane $C'\in \GrassmannSymbol(k,n+2)$ by requiring 
\begin{equation}
 C'\cdot\vltuu=0\eqncom \qquad C'\cdot\vleuu=0\eqncom \qquad C'^\perp\cdot\vlluu=0\eqndot
\end{equation}

For scattering amplitudes, the Graßmannian integral is given by \cite{ArkaniHamed:2009dn}%
\footnote{Two $n\times k$ matrices that differ by a $\GL{k}$ transformation still parametrise the same $k$-plane in $n$-dimensional space; hence, the measure factor in \eqref{eq: grassmannian amplitudes} is divided by the action of $\GL{k}$. Note that we are also relaxing the reality conditions on the momenta by allowing complex $n\times k$ matrices.
}
\begin{equation}
\label{eq: grassmannian amplitudes}
    \int\frac{\de^{k\times n}C}{\GL{k} }\;\,
  \Omega_{n,k}\,\;
  \delta^{2\times k}(C \cdot\vlt) \, \delta^{4\times k}(C\cdot\vle) \, \delta^{2\times(n-k)}(C^\perp\cdot\vll)\eqncom
\end{equation}
where the on-shell form is specified by 
\begin{equation}
\label{eq: grassmannian form amplitudes}
  \begin{aligned}
    \Omega_{n,k} &=
  \frac{1}{(1\cdots k)(2\cdots k \splus 1)\cdots(n \cdots k \sminus 1)}
\end{aligned}
\end{equation}
and $(1\cdots k)$ denotes the minor built from the first $k$ columns of $C$, etc.
By abuse of notation, we will also directly refer to $\Omega_{n,k}$ as the on-shell form.

From the above arguments, the Graßmannian integral for form factors is given by 
\begin{equation}
\label{eq: general grassmannian abstract}
  \int\frac{\de^{k\times(n+2)}C'}{\GL{k}}\;\,
  \Omega_{n,k}'\,\;
  \delta^{2\times k}(C'\cdot\vltuu) \, \delta^{4\times k}(C'\cdot\vleuu) \, \delta^{2\times(n+2-k)}(C'^\perp\cdot\vlluu)\eqncom
\end{equation}
where the on-shell form $\Omega_{n,k}'$ remains to be determined.
In order to find $\Omega_{n,k}'$, we use the on-shell diagrams of section \ref{sec: On-shell diagrams}.

\subsection{On-shell gluing}

Each of the amplitude vertices has a representation in terms of a Graßmannian integral. Moreover, these representations can be combined into the Graßmannian for an on-shell diagram by gluing the individual expressions together, i.e.\ by integrating over all degrees of freedom in the intermediate legs. 
Hence, we can obtain the on-shell form in the Graßmannian integral \eqref{eq: general grassmannian abstract} by assembling it from the individual expressions in the respective top-cell diagram obtained in section \ref{sec: On-shell diagrams}. 
For practical purposes, however, it is more convenient to start with the minimal form factor and the on-shell subdiagram that is obtained by excising the minimal form factor from the top-cell diagram. A Graßmannian integral representation for the latter can be readily obtained e.g.\ from the \texttt{Mathematica} package \texttt{positroid.m} \cite{Bourjaily:2012gy} via the permutation $\sigmapart$ given in \eqref{eq: permutation diagram to be glued}.
It can be written as 
\begin{equation}
  I=
        \int \frac{\de\alpha_1}{\alpha_1}\cdots\frac{\de\alpha_m}{\alpha_m} \;
        \delta^{k\times 2}(C(\alpha_i)\cdot\vlt) \,
        \delta^{k\times 4}(C(\alpha_i)\cdot\vle) \,
        \delta^{(n+2-k)\times 2}(C^\perp(\alpha_i)\cdot\vll)
        \eqncom
\end{equation}
where $C(\alpha_i)\in \GrassmannSymbol(k,n+2)$ and $m$ is the dimension of the corresponding cell in the Graßmannian.%
\footnote{In particular, the parametrisation by the $\alpha_i$'s fixes the $\GL{k}$ gauge freedom that appears in \eqref{eq: grassmannian amplitudes}.}

Writing the minimal form factor as 
\begin{equation}
 \label{eq: minimal form factor as vacuum}
 \begin{aligned}
        \deltaf{ij} \equiv \ffco_{2,2}(i,j)
        =\,
        &\delta^2\left(\vlt_i-\frac{\bra{j}q}{\abr{ji}}\right)\delta^2\left(\vle_i^--\frac{\bra{j}\gamma^-}{\abr{ji}}\right)\delta^2\big(\vle_i^+\big)\\
        &\delta^2\left(\vlt_j-\frac{\bra{i}q}{\abr{ij}}\right)\delta^2\left(\vle_j^--\frac{\bra{i}\gamma^-}{\abr{ij}}\right)\delta^2\big(\vle_j^+\big)
        \eqncom
 \end{aligned}       
\end{equation}
the Graßmannian integral corresponding to the top-cell diagram is given by%
\footnote{As is conventional in the literature on on-shell diagrams and Graßmannian integrals, we divide by the volume of $\GL{1}$ instead of $\U{1}$ here. This is related to relaxing the reality condition $\lambdat_{p_i}^\alphadot=\pm(\lambda_{p_i}^\alpha)^*$. Moreover, we have suppressed all factors of $(2\pi)$.}
\begin{equation}
\label{eq: gluing minimal ff to on-shell piece}
\begin{aligned}
        \int 
        \frac{\de^2\vll_{n+1}}{\GL{1}}
        \,
        \de^2\vlt_{n+1}
        \,
        \de^4\vle_{n+1}
        \,
        \frac{\de^2\vll_{n+2}}{\GL{1}}
        \,
        \de^2\vlt_{n+2}
        \,
        \de^4\vle_{n+2}
        \;
        \;
       \deltaf{n+1\; n+2} \Big\vert_{\vll\to -\vll}
        \;
        \;
        I(1,\ldots,n+2) \eqncom
\end{aligned}
\end{equation}
where the momentum flow in the minimal form factor is inverted as discussed below \eqref{eq: dLambda}.
Performing the integrations over $\vlt_{n+1}$, $\vlt_{n+2}$, $\vle_{n+1}$ and $\vle_{n+2}$ via the delta functions in $\deltaf{n+1\; n+2}$ leads to the following replacements in $I$:
\begin{equation}
\label{eq: replacement n+1 n+2}
 \begin{aligned}
  \vlt_{n+1} &\to -  \frac{\bra{n+2}q}{\abr{n \splus 2 \ssep n \splus 1}}\eqncom&
  \vle_{n+1}^+ &\to 0\eqncom&
  \vle_{n+1}^- &\to - \frac{\bra{n+2}\gamma^-}{\abr{n \splus 2 \ssep n \splus 1}}\eqncom\\
  \vlt_{n+2} &\to - \frac{\bra{n+1}q}{\abr{n \splus 1 \ssep n \splus 2}}\eqncom&
  \vle_{n+2}^+ &\to 0\eqncom&
  \vle_{n+2}^- &\to - \frac{\bra{n+1}\gamma^-}{\abr{ n \splus 1 \ssep n \splus 2}}\eqndot
\end{aligned}
\end{equation}
We can eliminate the $\GL{1}$ gauge freedom in \eqref{eq: gluing minimal ff to on-shell piece} by parametrising 
\begin{equation}
  \vll_{n+1}=\refspina-\beta_1 \refspinb
        \eqncom\qquad
        \vll_{n+2}=\refspinb-\beta_2 \refspina
        \eqncom
\end{equation}
where $\refspina$ and $\refspinb$ are arbitrary reference spinors that will be identified with the ones appearing in \eqref{eq: def underunderscore variables} shortly. 
Hence, $\abr{n \splus 1 \ssep n \splus 2 }=(\beta_1\beta_2-1)\abr{\refspinbl\refspinal}$  and the replacement \eqref{eq: replacement n+1 n+2} becomes
\begin{equation}
\label{eq: last to spinors in terms of reference spinors}
\begin{aligned}
  \vlt_{n+1} &\to  
                \frac{1}{\beta_1\beta_2-1}\,\frac{\bra{\refspinbl}q}{\abr{\refspinbl\refspinal}}
                +\frac{\beta_2}{\beta_1\beta_2-1}\,\frac{\bra{\refspinal}q}{\abr{\refspinal\refspinbl}}\eqncom
                \\
                \vle_{n+1}^{-} &\to  
                \frac{1}{\beta_1\beta_2-1}\,\frac{\bra{\refspinbl}\gamma^{-}}{\abr{\refspinbl\refspinal}}
                +\frac{\beta_2}{\beta_1\beta_2-1}\,\frac{\bra{\refspinal}\gamma^{-}}{\abr{\refspinal\refspinbl}}\eqncom
                \\
                \vlt_{n+2} &\to  
                \frac{1}{\beta_1\beta_2-1}\,\frac{\bra{\refspinal}q}{\abr{\refspinal\refspinbl}}
                +\frac{\beta_1}{\beta_1\beta_2-1}\,\frac{\bra{\refspinbl}q}{\abr{\refspinbl\refspinal}}\eqncom
                \\
                \vle_{n+2}^{-} &\to  
                \frac{1}{\beta_1\beta_2-1}\,\frac{\bra{\refspinal}\gamma^{-}}{\abr{\refspinal\refspinbl}}
                +\frac{\beta_1}{\beta_1\beta_2-1}\,\frac{\bra{\refspinbl}\gamma^{-}}{\abr{\refspinbl\refspinal}}
                \eqncom
\end{aligned}
\end{equation}
while the integration changes to 
\begin{equation}
         \int 
        \frac{\de^2\vll_{n+1}}{\GL{1}}
        \frac{\de^2\vll_{n+2}}{\GL{1}}
        =
        \abr{\refspinal\refspinbl}\abr{\refspinbl\refspinal}
        \int\de\beta_1\de\beta_2
        \eqndot
\end{equation}

We define a matrix $C'\in \GrassmannSymbol(n+2,k)$ with columns  
\begin{equation}
  C'=(C'_1 \cdots C'_{n+2})
  \eqncom
\end{equation}
by $C'_i=C_i$ for $i=1,\dots,n$ and modifying the last two columns 
to
\begin{equation}
\begin{aligned}
  C'_{n+1} & = \frac{1}{1-\beta_1\beta_2}C_{n+1}+\frac{\beta_1}{1-\beta_1\beta_2} C_{n+2} \eqncom  \qquad
  C'_{n+2} & = \frac{1}{1-\beta_1\beta_2}C_{n+2}+\frac{\beta_2}{1-\beta_1\beta_2} C_{n+1} \eqndot 
\end{aligned}
\end{equation}
In its orthogonal matrix $C'^\perp$, we have 
\begin{equation}
\begin{aligned}
  C'^\perp_{n+1} &= C^\perp_{n+1}-\beta_2 C^\perp_{n+2} \eqncom \qquad
  C'^\perp_{n+2} &= C^\perp_{n+2}-\beta_1 C^{\perp}_{n+1} \eqndot
\end{aligned}
\end{equation}
Note that in the delta function involving $C'^\perp$, which is defined as 
\begin{equation}
\label{eq: delta function C perp alternative for Jacobian}
        \delta^{(n+2-k)\times 2}(C'^\perp\cdot\vlluu) =
        \prod_{K=1}^k
        \int \de^{2}\rho_K  \;\,
        \delta^{(n+2)\times 2}\left(\vlluu_i-\rho_{L}C'_{Li}\right)
\end{equation}
with auxiliary variables $\rho_K^\alpha$, $K=1,\dots,k$, this leads to a Jacobian factor of $(1-\beta_1\beta_2)^2$; see \cite{Frassek:2015rka} for details.

In total, the Graßmannian integral obtained from gluing the top-cell diagram together as in \eqref{eq: gluing minimal ff to on-shell piece} is 
\begin{multline}
\label{eq: general Grassmannian gluing formula}
  I_{\ff}=
  \abr{\refspinal\refspinbl}\abr{\refspinbl\refspinal}
  \int \frac{\de\alpha_1}{\alpha_1}\cdots\frac{\de\alpha_m}{\alpha_m}\;\frac{\de\beta_1\,\de\beta_2}{(1-\beta_1\beta_2)^2} 
    \\ \times\;
        \delta^{k\times 2}(C'(\alpha_i,\beta_i)\cdot\vltuu) \,
        \delta^{k\times 4}(C'(\alpha_i,\beta_i)\cdot\vleuu)\,
        \delta^{(n+2-k)\times 2}(C'^\perp(\alpha_i,\beta_i)\cdot\vlluu)
        \eqncom
\end{multline}
where the variables $\vlluu_i$, $\vltuu_i$ and $\vleuu_i$ have been defined in \eqref{eq: def underunderscore variables}.

\subsection{Graßmannian integral}

Using the gluing procedure described in the last subsection and the on-shell subdiagrams $I$ obtained from the permutation $\sigmapart$ in \eqref{eq: permutation diagram to be glued}, we have found the following form of the Graßmannian integral:%
\footnote{We have explicitly checked this for all $\ffco_{T,n,k}$ with $k\leq n\leq 6$.}
\begin{equation}
\label{eq: general grassmannian spinor helicity}
    \int\frac{\de^{k\times(n+2)}C'}{\GL{k}}\;\,
  \Omega_{n,k}'\,\;
  \delta^{2\times k}(C'\cdot\vltuu) \, \delta^{4\times k}(C'\cdot\vleuu) \, \delta^{2\times(n+2-k)}(C'^\perp\cdot\vlluu)
\end{equation}
with
\begin{equation}
\label{eq: general grassmannian form spinor helicity}
  \begin{aligned}
    \Omega_{n,k}' &=
  \frac{\abr{\refspinal\refspinbl}^2
  Y(1-Y)^{-1}}{(1\cdots k)(2\cdots k \splus 1)\cdots(n \cdots k \sminus 3)(n \splus 1 \cdots k \sminus 2)(n \splus 2\cdots k \sminus 1)} \eqncom
  \\[0.5em]
  Y&=\frac{
  (n \sminus k \splus 2 \cdots n \ssep n \splus 1)(n \splus 2 \ssep 1 \cdots k \sminus 1)
}{
  (n \sminus k \splus 2 \cdots n \ssep n \splus 2)(n \splus 1 \ssep 1 \cdots k \sminus 1)
  } 
  \eqndot
\end{aligned}
\end{equation}

In contrast to the on-shell form \eqref{eq: grassmannian form amplitudes} in the case of planar amplitudes, the on-shell form \eqref{eq: general grassmannian form spinor helicity} contains consecutive as well as non-consecutive minors. The latter
 also appear for non-planar amplitudes \cite{Arkani-Hamed:2014bca,Chen:2014ara,Franco:2015rma,Chen:2015qna}.%
\footnote{As we have explicitly seen in chapter \ref{chap: two-loop Konishi form factor}, the diagrams contributing to loop-level form factors can become non-planar when removing the minimal form factor. At least at the level of leading singularities, it might hence not surprise to see features of non-planar amplitudes appear for form factors.}

Note that the permutation $\sigmapart$ in \eqref{eq: permutation diagram to be glued}, and hence also the on-shell form in \eqref{eq: general grassmannian form spinor helicity}, correspond
to 
the top-cell diagram with the minimal form factor glued in at positions $n+1$ and $n+2$. 
In addition, we have to consider the cyclic permutations of this on-shell diagram with respect to the $n$ on-shell legs.%
\footnote{Recall that the top-cell diagram is not cyclically symmetric unless $k=2$ or $k=n$.}
This can be achieved by permuting the super-spinor-helicity variables in the delta functions of \eqref{eq: general grassmannian spinor helicity} or, equivalently, by permuting the entries of the minors in \eqref{eq: general grassmannian form spinor helicity}.

Before evaluating \eqref{eq: general grassmannian spinor helicity}, \eqref{eq: general grassmannian form spinor helicity} explicitly, we bring it into an equivalent form that makes its evaluation easier.%
\footnote{For explicit evaluations of the Graßmannian integral in super-spinor-helicity variables, see \cite{Frassek:2015rka}.}

\subsection{Graßmannian integral in twistor space}

An alternative formulation of the Graßmannian integral \eqref{eq: grassmannian amplitudes} for scattering amplitudes, which was first given in \cite{ArkaniHamed:2009dn}, uses 
twistor variables \cite{Penrose:1967wn} instead of spinor-helicity variables.
The former formulation is related to the latter by Witten's half Fourier transform \cite{Witten:2003nn}:
\begin{equation}
\label{eq: half Fourier transformation}
 f(\vlluu_j)\longrightarrow\int \de^2 \vlluu_j \exp(-i\vmtuu^\alpha_j \vlluu_{j\alpha}) f(\vlluu_j) \eqndot
\end{equation}
In contrast to spinor-helicity variables, twistors 
\begin{equation}
 \label{eq: def position twistors}
 \ptwistoruu_i=\begin{pmatrix}
             \vmtuu_i^\alpha \\
             \vltuu_i^{\alphadot} \\
             \vleuu_i^A 
             \end{pmatrix}
\end{equation}
transform under the little group as $\ptwistoruu\to t^{-1}\ptwistoruu$.
Hence, they can be defined projectively.

In analogy to the amplitude case, we can also transform \eqref{eq: general grassmannian spinor helicity} and \eqref{eq: general grassmannian form spinor helicity} to twistor space.
The on-shell form \eqref{eq: general grassmannian form spinor helicity} depends on the $\vlluu_i$ only through the reference spinors $\refspina=\vlluu_{n+1}$ and $\refspinb=\vlluu_{n+2}$. Using 
\eqref{eq: half Fourier transformation}, the respective factor in \eqref{eq: general grassmannian form spinor helicity} becomes
\begin{equation}
 \abr{\refspinal\refspinbl}^2
 =\bigabr{\frac{\partial}{\partial \vmtuu_{n+1}}\frac{\partial}{\partial \vmtuu_{n+2}}}^2 \eqndot
\end{equation}
Writing the delta function $\delta^{2\times(n+2-k)}(C'^\perp\cdot\vlluu)$ as in \eqref{eq: delta function C perp alternative for Jacobian}, 
the integrals \eqref{eq: half Fourier transformation} over $\vlluu_i$ yield
\begin{equation}
 \prod_{K=1}^k \int \de^2\rho_K \exp\bigg(-i\sum_{j=1}^{n+2}\sum_{L=1}^k\rho^\alpha_{L}C'_{Lj}\vmtuu_{\alpha j}\bigg)=\delta^{2k}(C'\cdot\vmtuu)\eqndot
\end{equation}

In total, the Graßmannian integral \eqref{eq: general grassmannian spinor helicity} can be written as  
\begin{equation}
\label{eq: general grassmannian position twistor}
  \bigabr{\frac{\partial}{\partial \vmtuu_{n+1}}\frac{\partial}{\partial \vmtuu_{n+2}}}^2
  \int\frac{\de^{k\times(n+2)}C'}{\GL{k}}\;\,
  \Omega_{n,k}'\,\;
  \delta^{4 k| 4 k }(C'\cdot\ptwistoruu)\eqncom
\end{equation}
where the on-shell form is now given by 
\begin{equation}
\label{eq: general grassmannian form ptwistors}
  \begin{aligned}
    \Omega_{n,k}' &=
  \frac{ Y(1-Y)^{-1}}{(1\cdots k)(2\cdots k \splus 1)\cdots(n \cdots k \sminus 3)(n \splus 1 \cdots k \sminus 2)(n \splus 2\cdots k \sminus 1)} \eqncom
  \\[0.5em]
  Y&=\frac{
  (n \sminus k \splus 2 \cdots n \ssep n \splus 1)(n \splus 2 \ssep 1 \cdots k \sminus 1)
}{
  (n \sminus k \splus 2 \cdots n \ssep n \splus 2)(n \splus 1 \ssep 1 \cdots k \sminus 1)
  } 
  \eqndot
\end{aligned}
\end{equation}
It would be interesting to explore this representation of the Graßmannian integral in terms of twistors further, but this is beyond the scope of this work.

\subsection{Graßmannian integral in momentum-twistor space}

In \cite{Mason:2009qx}, a formulation of the Graßmannian integral \eqref{eq: grassmannian amplitudes}, \eqref{eq: grassmannian form amplitudes} in terms of different variables was given, namely in terms of Hodges's momentum-twistor variables \cite{Hodges:2009hk}.
A derivation of this momentum-twistor Graßmannian from \eqref{eq: grassmannian amplitudes}, \eqref{eq: grassmannian form amplitudes} was later provided in \cite{ArkaniHamed:2009vw}.
After a brief introduction of momentum twistors, we will show that the Graßmannian integral representation of form factors can be equally formulated in terms of momentum twistors.
This can be achieved by adapting the strategy of \cite{ArkaniHamed:2009vw} in the refined version of \cite{Elvang:2014fja}.

The definition of momentum-twistor variables \cite{Hodges:2009hk} is based on dual or regional (super)momenta,
which have also played a major role in showing dual superconformal invariance of scattering amplitudes \cite{Drummond:2008vq}.
For scattering amplitudes, the dual momenta $\dualmom_i$ and dual supermomenta $\dualsmom_i$ are defined by%
\footnote{Definitions in the literature differ by a sign and/or a cyclic relabelling.}
\begin{equation}
\label{eq: def dual (super) momenta}
\begin{aligned}
 \vll_i^{\alpha}\vlt_i^{\alphadot}=\dualmom_i^{\alpha\alphadot}-\dualmom_{i+1}^{\alpha\alphadot}\eqncom \qquad
 \vll_i^\alpha \vle_i^{A}=\dualsmom_i^{\alpha A}-\dualsmom_{i+1}^{\alpha A}\eqncom
\end{aligned}
\end{equation}
which determines them up to a global shift. 
As a consequence of (super)momentum conservation, they form a closed contour. 
Momentum twistor variables are then defined as 
\begin{equation}
\label{eq: def momentum twistors}
 \mtwistor_i=\begin{pmatrix}
             \lambda_i^\alpha \\
             \mu_i^{\alphadot} \\
             \eta_i^A 
             \end{pmatrix}
             =\begin{pmatrix}
             \lambda_i^\alpha \\
             \dualmom_{i}^{\alpha\alphadot} \lambda_{i,\alpha}\\
             \dualsmom_{i}^{\alpha A} \lambda_{i,\alpha}
             \end{pmatrix}
             \eqncom
\end{equation}
which transform homogeneously as $\mtwistor \to t \mtwistor$ under the little group. 

\begin{figure}[htbp]
\begin{equation*}
        \begin{tikzpicture}[scale=0.8]
        
        \draw[draw=none,fill=black!20, fill opacity=0.15] (-0.4,-1.7) rectangle (6.4,3.7);
        \draw[draw=none,fill=black!20, fill opacity=0.15] (2.2,-0.4) rectangle (8.9,4.4);
        
        \draw[dotted] (-0.8,1.2) -- (0,0);
        \draw[<-] (0,0) -- (1.7,3);
        \draw[<-] (1.7,3) -- (2.6,1.9);
        \draw[<-] (2.6,1.9) -- (3.9,3.3);
        \draw[<-] (3.9,3.3) -- (6,0);
        \draw[<-] (6,0) -- (7.7,3);
        \draw[<-] (7.7,3) -- (8.6,1.9);
        \draw[<-] (8.6,1.9) -- (9.9,3.3);
        \draw[<-] (9.9,3.3) -- (12,0);
        \draw[dotted,<-] (12,0) -- (12.8,1.2);
        
        \draw[dashed,<-] (2.2,-1.3) -- (0,0);
        \draw[dashed,<-] (6,0) -- (2.2,-1.3);

        \draw[dashed,<-] (4.8,0.6) -- (2.6,1.9);
        \draw[dashed,<-] (8.6,1.9) -- (4.8,0.6);

        \node[label={[label distance=-7pt]135:$p_1$}] at (0.85,1.5) {};
        \node[label={[label distance=-8pt]45:$p_2$}] at (2.15,2.45) {};
        \node[label={[label distance=-7pt]340:$p_3$}] at (3.25,2.6) {};
        \node[label={[label distance=-2pt]360:$p_4$}] at (4.95,1.65) {};
        \node[label={[label distance=-2pt]354:$p_5$}] at (4.1,-0.65) {};
        \node[label={[label distance=-2pt]270:$p_6$}] at (1.1,-0.65) {};

        \node[label={[label distance=-6pt]220:$y_1$}] at (0,0) {};
        \node[label={[label distance=-2pt]90:$y_2$}] at (1.7,3) {};
        \node[label={[label distance=-2pt]270:$y_3$}] at (2.6,1.9) {};
        \node[label={[label distance=-2pt]90:$y_4$}] at (3.9,3.3) {};
        \node[label={[label distance=-4pt]330:$y_5$}] at (6,0) {};
        \node[label={[label distance=-2pt]270:$y_6$}] at (2.2,-1.3) {};
        
    \end{tikzpicture}
\end{equation*}
\caption{%
For form factors, the momenta $p_i$ with $1\leq i \leq n$ form a periodic contour with period $q$. 
Using two auxiliary momenta $p_{n+1}$ and $p_{n+2}$ with $p_{n+1}+p_{n+2}=-q$, this contour can be closed in different ways to define dual momenta $y_i$.
The periodic contour is depicted in solid lines for $n=4$, the auxiliary momenta are depicted as dashed lines and two different ways of inserting the latter into the former are shown in grey boxes.}
\label{fig: dual contour}
\end{figure}
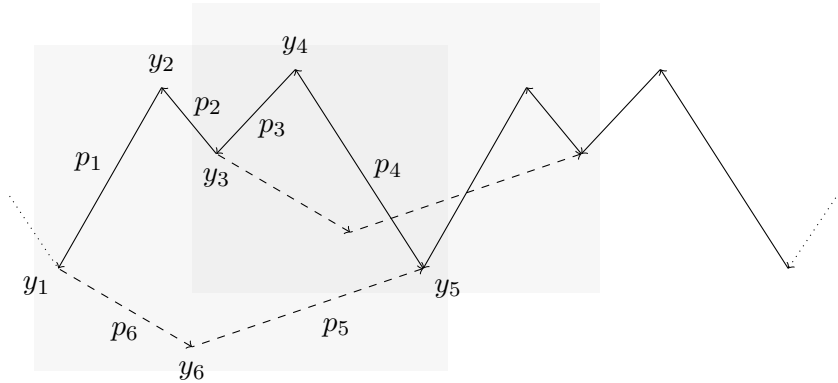

For form factors, the $n$ on-shell momenta do not add up to zero but to the off-shell momentum $q$. 
The contour is hence not closed but periodic with period $q$ \cite{Alday:2007he}.
We can close it in several ways using the two auxiliary on-shell momenta $p_{n+1}$ and $p_{n+2}$ with $p_{n+1}+p_{n+2}=-q$, see figure \ref{fig: dual contour}.
The construction of the momentum twistors proceeds as for amplitudes but using the variables with underscores defined in \eqref{eq: def underunderscore variables}.
In the following, only these variables occur, and we will hence drop the underscore in contractions of the spinor and twistor variables.

Now, let us transform the Graßmannian integral representation \eqref{eq: general grassmannian spinor helicity} and \eqref{eq: general grassmannian form spinor helicity} to momentum-twistor variables. 
In a first step, we can use part of the $\GL{k}$ freedom to choose  
\begin{equation}
 \rho=\begin{pmatrix}
       0 & \cdots & 0 & 1 & 0 \\
       0 & \cdots & 0 & 0 & 1 \\ 
      \end{pmatrix}
      \eqndot
\end{equation}
The integration over the auxiliary variables $\rho$ in \eqref{eq: delta function C perp alternative for Jacobian} then fixes 
\begin{equation}
 C_{k\sminus1 \, i}=\vlluu^1_i\eqncom \qquad C_{k \, i}=\vlluu^2_i\eqndot 
\end{equation}
As a consequence, the momentum- and supermomentum-conserving delta functions are split from the remaining delta functions and we can write \eqref{eq: general grassmannian spinor helicity} as  
\begin{equation}
\label{eq: momentum swistor Grassmann intermediate step}
   \delta^{4}(\vlluu\cdot\vltuu) \, \delta^{8}(\vlluu\cdot\vleuu) 
  \int\frac{\de^{(k-2)\times(n+2)}C'}{\GL{k-2}\ltimes T_{k-2}}\;\,
  \Omega_{n,k}'\,\;
  \delta^{2\times (k-2)}(C'\cdot\vltuu) \, \delta^{4\times (k-2)}(C'\cdot\vleuu)\eqncom
\end{equation}
where the shift symmetry $T_{k-2}$ as part of the remaining gauge freedom acts on the first $k-2$ rows of $C'$ as
\begin{equation}
 C'_{Ii}\longrightarrow C'_{Ii}+r_{1I}\vlluu_i^1+r_{2I}\vlluu_i^2\eqncom \quad I=1,\dots, k-2\eqncom
\end{equation}
with $r_{1I}$ and $r_{2I}$ arbitrary.

In a second step, we replace the super-spinor-helicity variables within the integral in \eqref{eq: momentum swistor Grassmann intermediate step} by momentum twistors \eqref{eq: def momentum twistors}. In terms of the former, the latter are explicitly given as 
\begin{equation}
 \begin{aligned}
  \vltuu_i&=\frac{\abr{i\splus1\,i}\vmmuu_{i-1}+\abr{i\,i\sminus1}\vmmuu_{i+1}+\abr{i\sminus1\, i\splus1}\vmmuu_{i}}{\abr{i\sminus1\,i}\abr{i\,i\splus1}}\eqncom \\
  \vleuu_i&=\frac{\abr{i\splus1\,i}\vletuu_{i-1}+\abr{i\,i\sminus1}\vletuu_{i+1}+\abr{i\sminus1\, i\splus1}\vletuu_{i}}{\abr{i\sminus1\,i}\abr{i\,i\splus1}}\eqndot
 \end{aligned}
\end{equation}
Defining the matrices $D$ as 
\begin{equation}
\label{eq: def D matrix}
 D_{Ii}=\frac{\abr{i\, i\splus1}C'_{I\,i\sminus1}+\abr{i\sminus1\,i}C'_{I\,i\splus1}+\abr{i\splus1\,i\sminus1}C'_{I\,i}}{\abr{i\sminus1\,i}\abr{i\,i\splus1}}\eqncom
\end{equation}
we have
\begin{equation}
 \sum_{i=1}^{n+2}C'_{Ii}\vltuu_i=-\sum_{i=1}^{n+2}D_{Ii}\vmmuu_i \eqncom \qquad
 \sum_{i=1}^{n+2}C'_{Ii}\vleuu_i=-\sum_{i=1}^{n+2}D_{Ii}\vletuu_i \eqncom \qquad I=1,\dots, k-2\eqndot
\end{equation}

In a third step, we express the minors of $C'$ in terms of minors of $D$.
In the case of planar amplitudes, only consecutive minors occur, which are related as 
\begin{equation}
\label{eq: old relation}
 (C'_1\cdots C'_{k})=-\abr{1\,2}\cdots\abr{k\sminus1\,k}(D_2\cdots D_{k\sminus1})
\end{equation}
and its cyclic permutations.
In our case, also non-consecutive minors occur in the on-shell form \eqref{eq: general grassmannian form spinor helicity}. 
These are related as 
\begin{equation}
\label{eq: new relation}
\begin{aligned}
 (C'_1\cdots C'_{k-1}C'_{k+1})&=
 -\abr{1\,2}\cdots\abr{k\sminus2\,k\sminus1}\abr{k\sminus1\,k\splus1}(D_2\cdots D_{k\sminus1})\\
 &\phaneq-\abr{1\,2}\cdots\abr{k\sminus2\,k\sminus1}\abr{k\,k\splus1}(D_2\cdots D_{k\sminus2}D_{k})\eqncom\\
  (C'_1C'_3 \cdots C'_{k+1})&=
 -\abr{1\,3}\abr{3\,4}\cdots\abr{k\,k\splus1}(D_3\cdots D_{k})\\
 &\phaneq-\abr{1\,2}\abr{3\,4}\cdots\abr{k\,k\splus1}(D_2D_4\cdots D_{k})\eqncom
\end{aligned}
\end{equation}
etc. Using these relations, we find 
\begin{equation}
 (1\cdots k)_{C'}\cdots(n \splus 2\cdots k \sminus 1)_{C'}=(-1)^{n+2}(\abr{1\,2}\cdots\abr{n\splus2\,1})^{k-1}(1\cdots k)_{D}\cdots(n \splus 2\cdots k \sminus 1)_{D}
\end{equation}
and 
\begin{equation}
\label{eq: Y mtwistor}
\begin{aligned}
 Y&=\frac{
  (n \sminus k \splus 2 \cdots n \ssep n \splus 1)_{C'}(n \splus 2 \ssep 1 \cdots k \sminus 1)_{C'}
}{
  (n \sminus k \splus 2 \cdots n \ssep n \splus 2)_{C'}(n \splus 1 \ssep 1 \cdots k \sminus 1)_{C'}
  } \\
&=\frac{
  \abr{n\,n\splus1}(n \sminus k \splus 3 \cdots n )_{D}
}{
  \abr{n\,n\splus2}(n \sminus k \splus 3 \cdots n )_{D}+\abr{n\splus1\,n\splus2}(n \sminus k \splus 3 \cdots n\sminus1 \, n\splus1)_{D}
  } \\
  &\phaneq
  \frac{
  \abr{n\splus2\,1}(1 \cdots k \sminus 2)_{D}
  }{
    \abr{n\splus1\,1}(1 \cdots k \sminus 2)_{D}
    +\abr{n\splus1\,n\splus2}(n\splus2\, 2 \cdots k \sminus 2)_{D}
  }
  \eqndot
\end{aligned}
\end{equation}

The remaining steps are again completely analogous to \cite{Elvang:2014fja}.
In a fourth step, the $T_{k-2}$ shift symmetry is used to set the first two columns of $C'$ to zero: $C'_{I1}=C'_{I2}=0$. 
The measure transforms under this change of variables as 
\begin{equation}
 \frac{\de^{(k-2)\times(n+2)}C'}{\GL{k-2}\ltimes T_{k-2}} = \abr{12}^{k-2}\frac{\de^{(k-2)\times(n)}C'}{\GL{k-2}} \eqndot
\end{equation}
In a fifth step, the integration variable is changed from $C'$ to $D$:
\begin{equation}
 \frac{\de^{(k-2)\times(n)}C'}{\GL{k-2}}=\left(\frac{\abr{12} \cdots \abr{n\splus2\,1}}{\abr{12}^2}\right)^{k-2}\frac{\de^{(k-2)\times(n)}D}{\GL{k-2}}\eqndot
\end{equation}
In the sixth and final step, the integration over the first two columns of $D$, which are fixed by fixing the first two columns of $C'$, is formally restored by introducing 
\begin{equation}
 \abr{12}\delta^2(D_{Ii}\vlluu_i)
\end{equation}
for each row $I=1,\dots,k-2$. For the details of these steps, we refer the interested reader to \cite{Elvang:2014fja}.

Assembling all pieces, we find 
\begin{equation}
\label{eq: general grassmannian mtwistor}
 \ffco_{T,n,2}
  \int\frac{\de^{(k-2)\times(n+2)}D}{\GL{k-2}}\;\,
  \Omega_{n,k}'\,\;
  \delta^{4(k-2)|4(k-2)}(D\cdot\mtwistoruu) \eqncom
\end{equation}
where
\begin{equation}
\label{eq: general grassmannian form mtwistor}
  \begin{aligned}
    \Omega_{n,k}' &=
  \frac{\abr{n\,1}\abr{n\splus1\,n\splus2}}{ \abr{n\,n\splus1}  \abr{n\splus2\,1}}
  \frac{Y(1-Y)^{-1}}{(1\cdots k\sminus2)(2\cdots k \sminus 1)\cdots(n \cdots k \sminus 5)(n \splus 1 \cdots k \sminus 4)(n \splus 2\cdots k \sminus 3)} 
\end{aligned}
\end{equation}
with $Y$ given in \eqref{eq: Y mtwistor}.

Note that we still have the freedom to choose the reference spinors $\refspina$ and $\refspinb$. A convenient choice is $\vlluu_{n+1}\equiv \refspinal = \vlluu_1$, $\vlluu_{n+2}\equiv \refspinbl = \vlluu_n$.
Using this choice, \eqref{eq: general grassmannian form mtwistor} becomes
\begin{equation}
\label{eq: form factor form gauged}
 \begin{aligned}
\Omega_{n,k}'=\frac{-\tilde{Y}(1-\tilde{Y})^{-1}}{(1\cdots k\sminus 2)\cdots(n \cdots k \sminus 5)(n \splus 1 \cdots k \sminus 4)(n \splus 2\cdots k \sminus 3)} 
   \end{aligned}
\end{equation}
with
\begin{equation}
\label{eq: form factor form Y gauged}
    \tilde{Y}=\frac{(n\sminus k \splus 3 \cdots n)(1\cdots k \sminus 2)}{(n \sminus k \splus 3 \cdots n \sminus 1 \ssep n \splus 1)(n\splus 2\ssep 2 \cdots k\sminus 2)}\eqndot
\end{equation}

The contributions from the other top-cell diagrams can be obtained by shifting the position at which the legs $n+1$ and $n+2$ are inserted into the contour from between $(n,1)$ to between $(n+s \mod n,1+s \mod n)$. This is illustrated in figure \ref{fig: dual contour} for $n=4$.

\subsection{Examples}

Let us now evaluate the momentum-twistor Graßmannian integral \eqref{eq: general grassmannian mtwistor}, \eqref{eq: form factor form gauged}, \eqref{eq: form factor form Y gauged} for certain examples. 

In the case $k=2$, the integral \eqref{eq: general grassmannian mtwistor} is zero-dimensional. 
All consecutive minors are equal to $1$ whereas all non-consecutive minors vanish. 
Inserting this into \eqref{eq: form factor form gauged} and \eqref{eq: form factor form Y gauged}, we find that $\Omega_{n,2}'=1$.%
\footnote{Note that $\tilde{Y}$ is formally divergent when inserting the values of the consecutive and non-consecutive minors that correspond to $k=2$. Hence, we have to expand $\frac{-\tilde{Y}}{1-\tilde{Y}}=\frac{-(n\sminus k \splus 3 \cdots n)(1\cdots k \sminus 2)}{(n \sminus k \splus 3 \cdots n \sminus 1 \ssep n \splus 1)(n\splus 2\ssep 2 \cdots k\sminus 2)-(n\sminus k \splus 3 \cdots n)(1\cdots k \sminus 2)}$ before setting $k=2$, in which case we find $\frac{-\tilde{Y}}{1-\tilde{Y}}=1$. 
} 
Hence, the only contribution comes from the prefactor $\ffco_{T,n,2}$ in \eqref{eq: general grassmannian mtwistor}, which is the correct result. This explicitly shows that our Graßmannian integral representation works at MHV level.

In the case $k=3$, 
\begin{equation}
D=
\begin{pmatrix}
 d_1 & d_2 & \cdots & d_{n+2}
\end{pmatrix}
\eqndot
\end{equation}
The consecutive minors $(i)$ of $D$ are equal to the $d_i$'s, whereas the non-consecutive minors of length one are by definition equal to the corresponding consecutive minors $(i)=d_i$, where $i$ is the index that stands alone in the non-consecutive minor for general $k$.
Thus, the Graßmannian integral \eqref{eq: general grassmannian mtwistor} becomes
 \begin{equation}
 \label{eq: NMHV Grassmann integral gauged}
 \begin{aligned}
  \ffco_{T,n,2}
   \int\frac{\de^{1\times(n+2)}D}{\GL{1}}\;\,
    \frac{1}{1-\frac{d_{n+1}d_{n+2}}{d_1 d_n}}
  \frac{1}{d_1\cdots d_{n}}\frac{1}{d_{n+1}d_{n+2}}
   \delta^{4|4}(D\cdot\mtwistoruu) \eqndot
   \end{aligned}
 \end{equation}
Its integrand diverges for $d_i=0$ with $i=2,\dots,n-1,n+1,n+2$ as well as for $d_1d_n=d_{n+1}d_{n+2}$. 
In fact, only the former poles will be relevant for computing form factors as residues of \eqref{eq: NMHV Grassmann integral gauged}.
For $k=3$ and general $n$, $n-3$ consecutive residues have to be taken.
As for scattering amplitudes \cite{Elvang:2014fja}, they can be characterised by a list of the five $d_i$'s with respect to which no residues are taken.
Considering only residues of the type $d_i=0$ with $i=2,\dots,n-1,n+1,n+2$, it follows that both $d_1$ and $d_n$ have to be included in the list of these five variables. 
This allows for two cases. In the first case, both $d_{n+1}$ and $d_{n+2}$ are included in this list as well.
Suppressing 
 a possible global sign,%
\footnote{See \cite{Elvang:2014fja} for a method to determine this sign in the case of amplitudes.}
the corresponding residue reads 
 \begin{multline}
 \label{eq: NMHV Grassmann integral gauged residue}
\ffco_{T,n,2}
   \int\frac{\de d_1\de d_i\de d_n\de d_{n+1}\de d_{n+2}}{\GL{1}}\;\,
    \frac{1}{1-\frac{d_{n+1}d_{n+2}}{d_1 d_n}}
  \frac{1}{d_1 d_i d_{n} d_{n+1}d_{n+2}} \\
   \delta^{4|4}(d_1\mtwistoruu_1+ d_i\mtwistoruu_i +d_{n}\mtwistoruu_n+ d_{n+1}\mtwistoruu_{n+1}+d_{n+2}\mtwistoruu_{n+2}) \eqndot
 \end{multline}
We can use the $\GL{1}$ redundancy to fix $d_1=\abr{i\,n\,n+1\,n+2}$,
where the four-bracket is defined as 
\begin{equation}
 \abr{i\,j\,k\,l}=\det(\Zuu_i\Zuu_j\Zuu_k\Zuu_l)=\epsilon_{ABCD}\Zuu_i^A\Zuu_j^B\Zuu_k^C\Zuu_l^D 
\end{equation}
and $\Zuu_\bullet$ denotes the four bosonic components of $\mtwistoruu_\bullet$.
The delta function then fixes the four remaining integration variables to
\begin{equation}
 d_i=\abr{n\,n\splus1\,n\splus2\,1}\eqncom \quad d_n=\abr{n\splus1\,n\splus2\,1\,i}\eqncom \quad d_{n+1}=\abr{n\splus2\,1\,i\,n}\eqncom \quad d_{n+2}=\abr{1\,i\,n\,n\splus1}\eqndot
\end{equation}
Hence, \eqref{eq: NMHV Grassmann integral gauged residue} reduces to
\begin{equation}
\res_i= \ffco_{T,n,2} \frac{1}{1-\frac{\abr{n\splus2\,1\,n\,i}\abr{1\,n\,i\,n\splus1}}{\abr{n\,i\,n\splus1\,n\splus2} \abr{i\,n\splus1\,n\splus2\,1}}}
  \sbr{i\,n\splus1\,n\splus2\,1\,n}\eqncom
\end{equation}
where $i\in\{2,\dots,n-1\}$ and the five bracket is defined as
\begin{equation}
 \sbr{i\,j\,k\,l\,m}=\frac{\delta^4(\abr{i\,j\,k\,l}\vletuu_m+\text{cyclic permutations})}{\abr{i\,j\,k\,l}\abr{j\,k\,l\,m}\abr{k\,l\,m\,i}\abr{l\,m\,i\,j}\abr{m\,i\,j\,k}}\eqndot
\end{equation}
In the second case, at least one of  $d_{n+1}$, $d_{n+2}$ is not part of the aforementioned list. The resulting residue can be evaluated in complete analogy to the previous case. This yields
\begin{equation}
\rest_{i,j,k}=\ffco_{T,n,2}\sbr{i\,j\,k\,1\,n}\eqncom
\end{equation}
where $i,j,k\in\{2,\dots,n-1,n+1,n+2\}$.

In the case $n=3$, no residues have to be taken and we obtain 
\begin{equation}
 \ffco_{T,3,3}=\ffco_{T,3,2} \,\res_2\eqndot
\end{equation}
We find a perfect numeric match between this expression and the known results \cite{Brandhuber:2011tv,Bork:2014eqa}.

For $n\geq4$, two complications occur. Residues have to be taken and --- due to residue theorems --- the sum of all residues vanishes such that only a particular combination yields the correct result for the form factor. Moreover, several top-cell diagrams are required. The first complication can be solved by a numeric comparison to the known results, whereas the second one can be solved by shifting the position at which the legs $n+1$ and $n+2$ are inserted into the contour from $(n,1)$ to $(n+s \mod n,1+s \mod n)$ as discussed above.

Numerically comparing with the results of \cite{Bork:2014eqa}, we find%
\footnote{Note that, due to residue theorems, the decomposition in \eqref{eq: 4,3 and 5,4 results} is not unique. In particular, some of the terms can be obtained from different top-cell diagrams.}
\begin{equation}
\label{eq: 4,3 and 5,4 results}
\begin{aligned}
 \ffco_{T,4,3}&=\ffco_{T,4,2} (\res_3+\rest_{2,3,5}+\res_3^{s=2}+\rest_{2,3,5}^{s=2})\eqncom \\
 \ffco_{T,5,3}&=\ffco_{T,5,2} (\res_4+\rest_{3,4,6}+\rest_{2,3,6}^{s=3}+\res_3^{s=3}-\rest_{2,3,4} \\
 &\qquad\qquad\qquad\,\,+\rest_{2,3,6}+\rest_{3,4,7}^{s=3}-\rest_{2,3,4}^{s=3}+\res_5^{s=1})\eqncom
 \end{aligned}
\end{equation}
where the superscript $s$ specifies the shift.

Finally, let us look at the simplest case for $k=4$, namely $n=4$.
Using the $\GL{2}$ redundancy, the $D$ matrix becomes
\begin{equation}
 D=\begin{pmatrix}
    1&0&d_{13}&d_{14}&d_{15}&d_{16}\\ 
    0&1&d_{23}&d_{24}&d_{25}&d_{26}
   \end{pmatrix}
\eqncom
\end{equation}
and the delta functions completely fix its remaining entries to 
\begin{equation}
 \begin{aligned}
  d_{i3}=-\frac{\abr{i\,4\,5\,6}}{\abr{3\,4\,5\,6}}\eqncom \quad
  d_{i4}=+\frac{\abr{i\,3\,5\,6}}{\abr{3\,4\,5\,6}}\eqncom \quad
  d_{i5}=-\frac{\abr{i\,3\,4\,6}}{\abr{3\,4\,5\,6}}\eqncom \quad
  d_{i6}=+\frac{\abr{i\,3\,4\,5}}{\abr{3\,4\,5\,6}}\eqncom
 \end{aligned}
\end{equation}
for $i=1,2$.
After using the generalised Schouten identity 
\begin{equation}
 \abr{i\,j\,k\,l}\abr{i\,j\,m\,n}+\abr{i\,j\,k\,m}\abr{i\,j\,n\,l}+\abr{i\,j\,k\,n}\abr{i\,j\,l\,m}=0
 \eqncom
\end{equation}
we find%
\footnote{Note that a subtle global sign occurs in the evaluation of this momentum-twistor Graßmannian integral; this sign is also present in the case of the related amplitude $\ampco_{6,4}$.}
\begin{equation}
\begin{aligned}
 \ffco_{T,4,4}
 =\ffco_{T,4,2} \frac{\abr{1\,3\,4\,5} \abr{1\,3\,4\,6} \abr{1\,3\,5\,6} \abr{2\,3\,4\,6} \abr{2\,3\,5\,6} \abr{2\,4\,5\,6}\sbr{1\,3\,4\,5\,6}\sbr{2\,3\,4\,5\,6}}{\abr{1\,2\,3\,4} \abr{1\,2\,3\,6} \abr{3\,4\,5\,6}^2 (\abr{1\,2\,4\,6} \abr{1\,3\,4\,5}+\abr{1\,2\,5\,6} \abr{3\,4\,5\,6})}
 \eqndot
 \end{aligned}
\end{equation}
We have numerically checked this against component results of \cite{Brandhuber:2011tv} and found perfect agreement.

The above results give strong evidence for the conjectured top-cell diagram for form factors \eqref{eq: box eater} and the resulting on-shell form \eqref{eq: general grassmannian form spinor helicity} in the Graßmannian integral  \eqref{eq: general grassmannian spinor helicity}.
Nevertheless, a proof of the relation \eqref{eq: box eater} at the level of the top-cell diagrams and a proof of \eqref{eq: general grassmannian form spinor helicity} would be desirable. 

\subsection*{Note on central-charge deformations}

In \cite{Bargheer:2014mxa,Ferro:2014gca}, a central-charge deformation of the Graßmannian integral formulation for scattering amplitudes was proposed.
For the case of MHV form factors of the stress-tensor supermultiplet, we have shown in \cite{Frassek:2015rka} that a similar deformation also exists for the Graßmannian integral formulation for form factors. 
The deformed Graßmannian integral is obtained immediately when constructing these form factors via deformed $\rr$ operators, see \cite{Frassek:2015rka} for details.
In the construction of the Graßmannian integral for N$^{k-2}$MHV form factors in this section, though, we have used the approach of on-shell gluing. 
It should also be possible to construct a central-charge deformed version of the Graßmannian integral for N$^{k-2}$MHV form factors via $\rr$ operators. However, this is beyond the scope of this work.

\part{Deformations}

\chapter{Introduction to integrable deformations}
\label{chap: introduction to integrable deformations}

In the previous chapters, we have studied the maximally supersymmetric $\mathcal{N}=4$ SYM theory.
Let us now turn to deformations of this theory in which this high amount of symmetry is reduced.%
\footnote{We give somewhat less details on the calculations in this part compared to the first part. Further and partially complementary details can be found in the Ph.D.\ thesis \cite{FokkenPhDthesis} of Jan Fokken, with whom I coauthored \cite{Fokken:2013aea,Fokken:2013mza,Fokken:2014soa}.
} 
Concretely, we look at the $\beta$- and $\gamma_i$-deformation. They were respectively shown to be the most general 
$\cN=1$ supersymmetric and non-supersymmetric field-theory deformations of $\mathcal{N}=4$ SYM theory that are 
integrable in the planar limit at the level of the asymptotic Bethe ansatz \cite{Beisert:2005if}, i.e.\
asymptotically integrable. 
In this chapter, we introduce these theories as well as some of their properties.

We give the single-trace part of the action of the deformed theories in section \ref{sec: single-trace action}. In section \ref{sec: relation}, we discuss the similarities and differences between the deformed theories and certain non-commutative field theories, in particular with respect to the notion of planarity.
This last discussion, which is based on \cite{Fokken:2013aea,Fokken:2013mza,Fokken:2014soa}, yields important relations between the deformed theories and their undeformed parent theory.

\section{Single-trace action}
\label{sec: single-trace action}

The single-trace part of the action of the deformed theories can be obtained from the action \eqref{eq: action of N=4 SYM theory} of $\mathcal{N}=4$ SYM theory via a certain type of non-commutative Moyal-like $\ast$-product \cite{Lunin:2005jy,Frolov:2005dj}.
For two fields $A$ and $B$, the $\ast$-product is defined as 
\begin{equation}
\label{eq: def star product}
 A\cstar B= A B \e^{{}\frac{\complexi}{2}\left(\mathbf{q}_{A}{}\wedge{}\mathbf{q}_{B}\right){}}\eqncom
\end{equation}
where $\mathbf{q}_{A}=(q_A^1,q_A^2,q_A^3)$ and $\mathbf{q}_{B}=(q_B^1,q_B^2,q_B^3)$ are the $\SU{4}$ Cartan charge vectors of the fields, which are given in table \ref{tab: su(4) charges}. 
\begin{table}[tbp]
\centering
$\begin{array}{l|c|ccc|ccccc}
	B&A_\mu, \cder_\mu, F_{\mu\nu} &\phi_1&\phi_2&\phi_3 & \psi_{\alpha1}&\psi_{\alpha2}&\psi_{\alpha3}&\psi_{\alpha4}\\
	& & & & & & & &\\[-0.4cm]
	\hline
	& & & & & & & &\\[-0.4cm]
	q^1_B & 0 & 1 & 0 & 0 &  +\frac12 & -\frac12 & -\frac12 & +\frac12 \\
	& & & & & & & &\\[-0.4cm]
	q^2_B &  0 & 0 & 1 & 0 &-\frac12 & +\frac12 & -\frac12 & +\frac12\\
	& & & & & & & &\\[-0.4cm]
	q^3_B &  0 & 0 & 0 & 1 &-\frac12 & -\frac12 & +\frac12 & +\frac12\\
\end{array}$
\caption{$\SU{4}$ Cartan charges of the different fields \cite{Beisert:2005if}.
Their respective antifields have the opposite charges.}
\label{tab: su(4) charges}
\end{table}
The antisymmetric product of the charge vectors is defined as  
\begin{equation}
\label{eq: def antisymmetric product}
\mathbf{q}_A\wedge \mathbf{q}_B=(\mathbf{q}_A)^{\Top}\mathbf{C}\,\mathbf{q}_B
\eqncom\qquad
\mathbf{C}=\begin{pmatrix}
0 & -\gamma_3 & \gamma_2 \\
\gamma_3 & 0 & -\gamma_1 \\
-\gamma_2 & \gamma_1 & 0 
\end{pmatrix}
\eqndot
\end{equation} 
The three real deformation parameters $\gamma_1$, $\gamma_2$ and $\gamma_3$ frequently occur in the linear combinations
\begin{equation}\label{eq: gamma pm def}
\gamma_1^\pm=\pm\frac{1}{2}(\gamma_2\pm\gamma_3)
\eqncom\qquad
\gamma_2^\pm=\pm\frac{1}{2}(\gamma_3\pm\gamma_1)
\eqncom\qquad
\gamma_3^\pm=\pm\frac{1}{2}(\gamma_1\pm\gamma_2)
\eqndot
\end{equation}
In the limit of the $\beta$-deformation, these assume the values $\gamma_i^+=\beta$ and $\gamma_i^-=0$.
Note that, although non-commutative, the $\ast$-product \eqref{eq: def star product} is associative.

We can then obtain the single-trace part of the action of the deformed theories by replacing all products of fields in \eqref{eq: action of N=4 SYM theory} by their $\ast$-products. This yields
\begin{equation}
\label{eq: deformed single-trace action}
\begin{aligned}
S
 &=\int\de^4x\,\tr\Big(
 -\frac{1}{4} F^{\mu\nu}F_{\mu\nu}- (\D^{\mu}\bar\phi^j)\D_{\mu}\phi_j
 +i \bar\psi^{\dot\alpha A} \D_{\dot\alpha}{}^\alpha\psi_{\alpha A}\\
 &\hphantom{{}={}\int\de^4x\,\tr\Big({}}{}
 +g_\YM\Bigl(
 \frac{i}{2}\epsilon^{ijk}\phi_i \staracomm{\psi^{\alpha}_{j}}{\psi_{\alpha k}}
 +\phi_j \staracomm{\bar\psi^{\dot\alpha4}}{
 \bar\psi_{\dot\alpha}^{j}}+\text{h.c.}
 \Bigr)\\
 &\phantom{{}={}\int\de^4x\,\tr\Big({}}{}
 -\frac{g^2_\YM}{4}
 \comm{\bar\phi^j}{\phi_j}\comm{\bar\phi^k}{\phi_k}
 +\frac{g^2_\YM}{2}
 \starcomm{\bar\phi^j}{\bar\phi^k}\starcomm{\phi_j}{\phi_k}
 \Big) 
 \eqncom
\end{aligned}
\end{equation}
where $\starcomm{\cdot}{\cdot}$ and $\staracomm{\cdot}{\cdot}$ are the $\ast$-deformed (anti)commutators defined via \eqref{eq: def star product}.
We have dropped the $\cstar$ in cases where the $\cstar$-products trivially reduce to the usual products in the case of the $\gamma_i$-deformation.
In the $\beta$-deformation, also the interactions of the gluino $\psi_{\alpha4}$ and the antigluino $\bar\psi^{4}_{\dot\alpha}$ are undeformed.

\section{Relation to the undeformed theory}
\label{sec: relation}

The non-commutative $\ast$-product \eqref{eq: def star product} is similar to a $\ast$-product appearing in certain types of non-commutative field theories, see \cite{Szabo:2001kg} for a review of the latter.
This allows to adapt a theorem developed in the context of these theories by Thomas Filk \cite{Filk:1996dm}.
This theorem relates planar single-trace diagrams built from elementary interactions in the deformed and undeformed theory. 
In the formulation of \cite{Khoze:2005nd}, it reads
\begin{equation}
\label{eq: diagram relation}
 \begin{aligned}
\settoheight{\eqoff}{$+$}%
\setlength{\eqoff}{0.5\eqoff}%
\addtolength{\eqoff}{-9\unit}%
\raisebox{\eqoff}{%
\begin{pspicture}(-2,1)(11,-17)
\rput(4.5,-8){%
\rotatebox{90}{%
\begin{pspicture}(-1,-2)(17,11)
\uinex{2}{9}%
\iinex{2}{6}%
\dinex{2}{0}%
\setlength{\ya}{9\unit}
\addtolength{\ya}{0.5\dlinewidth}
\setlength{\yb}{0\unit}
\addtolength{\yb}{-0.5\dlinewidth}
\setlength{\xc}{7.5\unit}
\setlength{\yc}{4.5\unit}
\addtolength{\yc}{-0.5\dlinewidth}
\setlength{\xd}{8.5\unit}
\setlength{\yd}{4.5\unit}
\addtolength{\yd}{0.5\dlinewidth}
\psline[linewidth=\blacklinew](3.5,\ya)(12.5,\ya)
\psline[linestyle=dotted,linewidth=\blacklinew](3.5,4)(3.5,2)
\psline[linewidth=\blacklinew](12.5,\yb)(3.5,\yb)
\doutex{14}{0}%
\ioutex{14}{6}%
\uoutex{14}{9}%
\psline[linestyle=dotted,linewidth=\blacklinew](12.5,4)(12.5,2)
\setlength{\xa}{3.5\unit}
\addtolength{\xa}{\dlinewidth}
\setlength{\xb}{12.5\unit}
\addtolength{\xb}{-\dlinewidth}
\setlength{\ya}{9\unit}
\addtolength{\ya}{-0.5\dlinewidth}
\setlength{\yb}{0\unit}
\addtolength{\yb}{0.5\dlinewidth}
\pscustom[linecolor=gray,fillstyle=solid,fillcolor=gray,linearc=\linearc]{%
\psline(\xa,4.5)(\xa,\ya)(\xb,\ya)(\xb,4.5)
\psline[liftpen=2](\xb,4.5)(\xb,\yb)(\xa,\yb)(\xa,4.5)
}
\end{pspicture}}
\rput(-2,1.25){%
\rput[b](-9,14.5){$\scriptstyle A_{1}$}
\rput[b](-6,14.5){$\scriptstyle A_{2}$}
\rput[b](0,14.5){$\scriptstyle A_{i}$}
\rput[t](0,1){$\scriptstyle A_{i+1}$}
\rput[t](-6,1){$\scriptstyle A_{n-1}$}
\rput[t](-9,1){$\scriptstyle A_{n}$}}
}
\rput(4.5,-6.75){planar}
\rput(4.5,-9.25){$\beta$ or $\gamma_i$}
\end{pspicture}}
\,
&=\,
\settoheight{\eqoff}{$+$}%
\setlength{\eqoff}{0.5\eqoff}%
\addtolength{\eqoff}{-9\unit}%
\raisebox{\eqoff}{%
\begin{pspicture}(-2,1)(11,-17)
\rput(4.5,-8){%
\rotatebox{90}{%
\begin{pspicture}(-1,-2)(17,11)
\uinex{2}{9}%
\iinex{2}{6}%
\dinex{2}{0}%
\setlength{\ya}{9\unit}
\addtolength{\ya}{0.5\dlinewidth}
\setlength{\yb}{0\unit}
\addtolength{\yb}{-0.5\dlinewidth}
\setlength{\xc}{7.5\unit}
\setlength{\yc}{4.5\unit}
\addtolength{\yc}{-0.5\dlinewidth}
\setlength{\xd}{8.5\unit}
\setlength{\yd}{4.5\unit}
\addtolength{\yd}{0.5\dlinewidth}
\psline[linewidth=\blacklinew](3.5,\ya)(12.5,\ya)
\psline[linestyle=dotted,linewidth=\blacklinew](3.5,4)(3.5,2)
\psline[linewidth=\blacklinew](12.5,\yb)(3.5,\yb)
\doutex{14}{0}%
\ioutex{14}{6}%
\uoutex{14}{9}%
\psline[linestyle=dotted,linewidth=\blacklinew](12.5,4)(12.5,2)
\setlength{\xa}{3.5\unit}
\addtolength{\xa}{\dlinewidth}
\setlength{\xb}{12.5\unit}
\addtolength{\xb}{-\dlinewidth}
\setlength{\ya}{9\unit}
\addtolength{\ya}{-0.5\dlinewidth}
\setlength{\yb}{0\unit}
\addtolength{\yb}{0.5\dlinewidth}
\pscustom[linecolor=gray,fillstyle=solid,fillcolor=gray,linearc=\linearc]{%
\psline(\xa,4.5)(\xa,\ya)(\xb,\ya)(\xb,4.5)
\psline[liftpen=2](\xb,4.5)(\xb,\yb)(\xa,\yb)(\xa,4.5)
}
\end{pspicture}}
\rput(-2,1.25){%
\rput[b](-9,14.5){$\scriptstyle A_{1}$}
\rput[b](-6,14.5){$\scriptstyle A_{2}$}
\rput[b](0,14.5){$\scriptstyle A_{i}$}
\rput[t](0,1){$\scriptstyle A_{i+1}$}
\rput[t](-6,1){$\scriptstyle A_{n-1}$}
\rput[t](-9,1){$\scriptstyle A_{n}$}}
}
\rput(4.5,-6.75){planar}
\rput(4.5,-9.25){$\mathcal{N}=4$}
\end{pspicture}}
\times
\,\,\,
\Phi\left(%
A_{1}{}\ast{}A_{2}
\ast{}\dots{}\ast{}A_{n}
\right)\eqncom
\end{aligned}
\end{equation}
where $A_i$, $i=1,\dots,n$, are elementary fields of the theory, the grey area depicts planar elementary interactions between them and $\Phi(A_{1}\ast A_{2}\ast \dots \ast A_{n})$ denotes the phase factor of the $\ast$-product $A_{1}\ast A_{2}\ast \dots \ast A_{n}$.
As the single-trace couplings are not renormalised in $\cN=4$ SYM theory, this relation could be used in \cite{Ananth:2006ac} and \cite{Ananth:2007px}, respectively, to show that the single-trace couplings in the $\beta$- and $\gamma_i$-deformation are not renormalised either.
Furthermore, it was used to relate planar colour-ordered scattering amplitudes in the deformed theories to their undeformed counterparts \cite{Khoze:2005nd}.

However, when applying relation \eqref{eq: diagram relation} to Feynman diagrams that contain composite operators, such as those contributing to correlation functions, form factors or operator renormalisation, one has to be very careful.
A priori, relation \eqref{eq: diagram relation} is only valid for planar single-trace diagrams built from the elementary interactions of the theory.
In order to apply it to a planar single-trace diagram that contains a composite operator, one needs to remove the operator from the diagram.
The resulting subdiagram of elementary interactions is either of single-trace type or of double-trace type, the latter occurring in the presence of finite-size effects \cite{Sieg:2005kd}. Relation \eqref{eq: diagram relation}, however, can only be applied in the former case.

In the context of non-commutative field theories, Filk's theorem can easily be extended to include composite operators.
They can simply be added to the action with appropriate source terms and can be deformed in analogy to the elementary interactions.
Problems arise when adapting this extension to the $\beta$- and $\gamma_i$-deformation.
In the non-commutative field theories, the ordering principle at each vertex 
refers to the positions in spacetime, or, equivalently, to the momenta.
The phase factor introduced by the corresponding $\ast$-product is also defined in terms of the momenta.
In particular, momentum conservation is satisfied at every vertex and composite operator such that the phase factor is invariant under a cyclic relabelling of the fields.
In the $\beta$- and $\gamma_i$-deformation, the ordering principle 
refers to colour, whereas the phase factor is defined by the $\SU{4}$ Cartan charges. 
Colour singlets such as traces can, however, be charged under the Cartan subgroup of $\SU{4}$. 
In this case, the definition of the phase factor in the $\ast$-product \eqref{eq: def star product} is incompatible with the (graded) cyclic invariance of the trace.
For example, $\tr(\phi_i\phi_j)=\tr(\phi_j\phi_i)$ but $\tr(\phi_i \ast \phi_j)\neq \tr(\phi_j \ast \phi_i)$ unless $i=j$.
Thus, the extension of Filk's theorem can only be adapted for composite operators whose trace factors are neutral with respect to the $\SU{4}$ charges that define the phase factor in the $\ast$-product \eqref{eq: def star product}.%
\footnote{Note that the phase factor in the $\beta$-deformation depends only on two linear combinations of the three Cartan charges of $\SU{4}$.}

One example of an operator for which we can extend Filk's theorem is the Konishi primary operator studied in chapter \ref{chap: two-loop Konishi form factor}. 
According to the above arguments, its planar anomalous dimension and minimal form factors are independent of the deformation parameters, as well as all planar correlation functions that contain only this operator.%
\footnote{Non-minimal and generalised Konishi planar form factors in the deformed theories are related to their undeformed counterparts via \eqref{eq: diagram relation}.}
In the $\beta$-deformation, another example is the chiral primary operator
\begin{equation}
\label{eq: tilde O_k}
 \tilde\cO_k= \tr\big((\phi_1)^k(\phi_2)^k(\phi_3)^k\big)+\text{permutations}\eqncom
\end{equation}
where each permutation is weighted by $\frac{S}{3k}$ with $S$ being the smallest cyclic shift that maps the operator to itself.
Its $\SU{4}$ Cartan charge is $(q^1,q^2,q^3)=(k,k,k)$, which vanishes in all antisymmetric products \eqref{eq: def antisymmetric product} in the $\beta$-deformation.
In contrast to the Konishi primary operator, the operators $\tilde\cO_k$ are altered when replacing all products in \eqref{eq: tilde O_k} by $\ast$-products.%
\footnote{In \cite{Freedman:2005cg}, an alternative prescription to obtain the deformed chiral primary operators $\tilde\cO_k$ is given; the result using this prescription differs only by a global phase factor.
}
By the above considerations, however, all their planar correlation functions have to be deformation independent.
This is consistent with the explicit results available in the literature. 
Concretely, the planar anomalous dimensions of the operators $\tilde\cO_k$ were argued to vanish in \cite{Frolov:2005iq}, generalising an argument of \cite{Berenstein:2000ux,Berenstein:2000hy} for rational $\beta$.
Moreover, three-point functions $\langle \tilde{\cO}_k \tilde{\cO}_{k^\prime} \tilde{\cO}_{k^{\prime\prime}} \rangle$ were studied in \cite{David:2013oha} at one-loop order in the planar gauge theory as well as at strong coupling via the Lunin-Maldacena background and found to be independent of $\beta$.

Considering the subdiagram of elementary interactions that is obtained from a planar single-trace diagram that contains a single uncharged operator by removing this operator, we find the following relation for double-trace diagrams:%
\footnote{This relation was derived earlier for the leading double-trace part of scattering amplitudes in the $\beta$-deformation in \cite{Jin:2012np}.}
\begin{equation}\label{eq: relation for double-trace diagrams}
 \begin{aligned}
\settoheight{\eqoff}{$+$}%
\setlength{\eqoff}{0.5\eqoff}%
\addtolength{\eqoff}{-9\unit}%
\raisebox{\eqoff}{%
\begin{pspicture}(-2,1)(11,-17)
\rput(4.5,-8){%
\rotatebox{90}{%
\begin{pspicture}(-1,-2)(17,11)
\uinex{2}{9}%
\iinex{2}{6}%
\dinex{2}{0}%
\setlength{\ya}{9\unit}
\addtolength{\ya}{0.5\dlinewidth}
\setlength{\yb}{0\unit}
\addtolength{\yb}{-0.5\dlinewidth}
\setlength{\xc}{7.0\unit}
\setlength{\yc}{4.5\unit}
\addtolength{\yc}{-0.5\dlinewidth}
\setlength{\xd}{9.0\unit}
\setlength{\yd}{4.5\unit}
\addtolength{\yd}{0.5\dlinewidth}
\psline[linearc=\linearc](3.5,\ya)(\xc,\ya)(\xc,\yb)(3.5,\yb)
\psline[linestyle=dotted](3.5,4.5)(3.5,1.5)
\psline[linearc=\linearc](12.5,\ya)(\xd,\ya)(\xd,\yb)(12.5,\yb)
\doutex{14}{0}%
\ioutex{14}{6}%
\uoutex{14}{9}%
\psline[linestyle=dotted](12.5,4.5)(12.5,1.5)
\setlength{\xa}{3.5\unit}
\addtolength{\xa}{\dlinewidth}
\setlength{\xb}{7.0\unit}
\addtolength{\xb}{-\dlinewidth}
\setlength{\ya}{9\unit}
\addtolength{\ya}{-0.5\dlinewidth}
\setlength{\yb}{0\unit}
\addtolength{\yb}{0.5\dlinewidth}
\pscustom[linecolor=gray,fillstyle=solid,fillcolor=gray,linearc=\linearc]{%
\psline(\xa,4.5)(\xa,\ya)(\xb,\ya)(\xb,4.5)
\psline[liftpen=2](\xb,4.5)(\xb,\yb)(\xa,\yb)(\xa,4.5)
}
\setlength{\xa}{12.5\unit}
\addtolength{\xa}{-\dlinewidth}
\setlength{\xb}{9.0\unit}
\addtolength{\xb}{\dlinewidth}
\setlength{\ya}{9\unit}
\addtolength{\ya}{-0.5\dlinewidth}
\setlength{\yb}{0\unit}
\addtolength{\yb}{0.5\dlinewidth}
\pscustom[linecolor=gray,fillstyle=solid,fillcolor=gray,linearc=\linearc]{%
\psline(\xa,4.5)(\xa,\ya)(\xb,\ya)(\xb,4.5)
\psline[liftpen=2](\xb,4.5)(\xb,\yb)(\xa,\yb)(\xa,4.5)
}
\pnode(7.25,4.5){left}
\pnode(8.75,4.5){right}
\ncline[arrows=->,linecolor=black]{left}{right}
\end{pspicture}}
\rput(-2,1.25){%
\rput[b](-9,14.5){$\scriptstyle A_{1}$}
\rput[b](-6,14.5){$\scriptstyle A_{2}$}
\rput[b](0,14.5){$\scriptstyle A_{i}$}
\rput[t](0,1){$\scriptstyle A_{i+1}$}
\rput[t](-6,1){$\scriptstyle A_{n-1}$}
\rput[t](-9,1){$\scriptstyle A_{n}$}}
}
\rput(4.5,-5.25){$\beta$ or $\gamma_i$}
\rput(4.5,-10.75){$\beta$ or $\gamma_i$}
 \rput[l](5.0,-8.125){$\scriptstyle \delta \mathbf{q} =0$}
\end{pspicture}}
\,
&=\,
\settoheight{\eqoff}{$+$}%
\setlength{\eqoff}{0.5\eqoff}%
\addtolength{\eqoff}{-9\unit}%
\raisebox{\eqoff}{%
\begin{pspicture}(-2,1)(11,-17)
\rput(4.5,-8){%
\rotatebox{90}{%
\begin{pspicture}(-1,-2)(17,11)
\uinex{2}{9}%
\iinex{2}{6}%
\dinex{2}{0}%
\setlength{\ya}{9\unit}
\addtolength{\ya}{0.5\dlinewidth}
\setlength{\yb}{0\unit}
\addtolength{\yb}{-0.5\dlinewidth}
\setlength{\xc}{7.0\unit}
\setlength{\yc}{4.5\unit}
\addtolength{\yc}{-0.5\dlinewidth}
\setlength{\xd}{9.0\unit}
\setlength{\yd}{4.5\unit}
\addtolength{\yd}{0.5\dlinewidth}
\psline[linearc=\linearc](3.5,\ya)(\xc,\ya)(\xc,\yb)(3.5,\yb)
\psline[linestyle=dotted](3.5,4.5)(3.5,1.5)
\psline[linearc=\linearc](12.5,\ya)(\xd,\ya)(\xd,\yb)(12.5,\yb)
\doutex{14}{0}%
\ioutex{14}{6}%
\uoutex{14}{9}%
\psline[linestyle=dotted](12.5,4.5)(12.5,1.5)
\setlength{\xa}{3.5\unit}
\addtolength{\xa}{\dlinewidth}
\setlength{\xb}{7.0\unit}
\addtolength{\xb}{-\dlinewidth}
\setlength{\ya}{9\unit}
\addtolength{\ya}{-0.5\dlinewidth}
\setlength{\yb}{0\unit}
\addtolength{\yb}{0.5\dlinewidth}
\pscustom[linecolor=gray,fillstyle=solid,fillcolor=gray,linearc=\linearc]{%
\psline(\xa,4.5)(\xa,\ya)(\xb,\ya)(\xb,4.5)
\psline[liftpen=2](\xb,4.5)(\xb,\yb)(\xa,\yb)(\xa,4.5)
}
\setlength{\xa}{12.5\unit}
\addtolength{\xa}{-\dlinewidth}
\setlength{\xb}{9.0\unit}
\addtolength{\xb}{\dlinewidth}
\setlength{\ya}{9\unit}
\addtolength{\ya}{-0.5\dlinewidth}
\setlength{\yb}{0\unit}
\addtolength{\yb}{0.5\dlinewidth}
\pscustom[linecolor=gray,fillstyle=solid,fillcolor=gray,linearc=\linearc]{%
\psline(\xa,4.5)(\xa,\ya)(\xb,\ya)(\xb,4.5)
\psline[liftpen=2](\xb,4.5)(\xb,\yb)(\xa,\yb)(\xa,4.5)
}
\pnode(7.25,4.5){left}
\pnode(8.75,4.5){right}
\ncline[arrows=->,linecolor=black]{left}{right}
\end{pspicture}}
\rput(-2,1.25){%
\rput[b](-9,14.5){$\scriptstyle A_{1}$}
\rput[b](-6,14.5){$\scriptstyle A_{2}$}
\rput[b](0,14.5){$\scriptstyle A_{i}$}
\rput[t](0,1){$\scriptstyle A_{i+1}$}
\rput[t](-6,1){$\scriptstyle A_{n-1}$}
\rput[t](-9,1){$\scriptstyle A_{n}$}}
}
\rput(4.5,-5.25){$\cN=4$}
\rput(4.5,-10.75){$\cN=4$}
 \rput[l](5.0,-8.125){$\scriptstyle \delta \mathbf{q} =0$}
\end{pspicture}}
\times
\,\,\,
\underbrace{\Phi\left(
A_{1}{}
\ast{}\dots{}\ast{}A_{i}
\right)
\Phi\left(
A_{i+1}{}%
\ast{}\dots{}\ast{}A_{n}
\right)\rule[-0.25cm]{0pt}{0.5cm}}_{\rule[0pt]{0pt}{0.5cm}\displaystyle  \Phi\left(
A_{1}{}
\ast{}\dots{}\ast{}A_{i}{}\ast{}
A_{i+1}{}
\ast{}\dots{}\ast{}A_{n}
\right)}\eqncom
\end{aligned}
\end{equation}
where $\delta \mathbf{q}$ denotes the flow of the relevant charges.

The above extension is very powerful for planar (generalised) form factors and correlation functions of composite operators that are neutral with respect to the charges which define the deformation.
For planar (generalised) form factors and correlation functions of charged operators, however, it is not applicable.
As already mentioned, the composite operators have to be removed from the diagrams in this case, resulting in a subdiagram of elementary interactions which can be either of single-trace type or of multi-trace type.
If this subdiagram is of single-trace type, relation \eqref{eq: diagram relation} can be applied.
In particular, it can be applied to the subdiagrams that yield the asymptotic dilatation operator.
At one-loop order, this gives \cite{Beisert:2005if}
\begin{equation}
\label{eq: deformation of D_2}
\begin{aligned}
(\loopdila{1}_{\beta,\gamma_i})_{A_iA_j}^{A_kA_l}&= \Phi(A_k\cstar A_l\cstar A_j\cstar A_i)(\loopdila{1}_{\cN=4})_{A_iA_j}^{A_kA_l}
=\e^{\frac{\complexi}{2} (\mathbf{q}_{A_k} \wedge \mathbf{q}_{A_l}- \mathbf{q}_{A_i} \wedge \mathbf{q}_{A_j})}(\loopdila{1}_{\cN=4})_{A_iA_j}^{A_kA_l} \eqndot
\end{aligned}
\end{equation} 
Moreover, relation \eqref{eq: diagram relation} was used to derive asymptotic integrability of the deformed theories from the assumption that $\mathcal{N}=4$ SYM theory is asymptotically integrable \cite{Beisert:2005if}.
However, if the subdiagram is of multi-trace type, which occurs for finite-size effects, relation \eqref{eq: diagram relation} cannot be applied to relate the deformed result to the undeformed one.
Hence, finite-size effects have to be studied for checks of integrability and the AdS/CFT correspondence that go beyond the undeformed case.
These will be the subject of the next three chapters.

\chapter{Prewrapping in the \texorpdfstring{$\beta$}{beta}-deformation}
\label{chap: prewrapping}

In this chapter, we study double-trace couplings in the planar $\beta$\hyp deformation.
We analyse the influence of these couplings on two-point correlation functions and anomalous dimensions in section \ref{sec: prewrapping}. We find that, although apparently suppressed by a factor of $\frac{1}{N}$, they can contribute at planar level via a new kind of finite-size effect. As this finite-size effect starts to contribute one loop order before the finite-size effect of wrapping, we call it prewrapping. Moreover, we classify which composite operators are potentially affected by prewrapping.
Finally, we obtain the complete one-loop dilatation operator of the planar $\beta$-deformation by incorporating this finite-size effect into the asymptotic dilatation operator in section \ref{sec: one-loop dilatation operator in beta deformation}.

The results presented in this chapter were first published in \cite{Fokken:2013mza}.

\section{Prewrapping}
\label{sec: prewrapping}

As reviewed in section \ref{sec: 't Hooft limit}, subdiagrams of elementary interactions with a multi-trace structure, and in particular a double-trace structure, can contribute at leading order in the planar limit if each trace factor in the subdiagram is planarly contracted with a trace factor of matching length in a composite operator \cite{Sieg:2005kd}.
This can only happen if the range of the interaction equals the length of the operator, and it is hence known as finite-size effect.
In the well known finite-size effect of wrapping \cite{Sieg:2005kd},
the double-trace structure is built from single-trace interactions that wrap around the operator.
In this section, we study the effect of double-trace structures whose origin is also of 
 double-trace type.
In particular, we find that they give rise to a new type of finite-size effect.

One source of double-trace structures is the completeness relation \eqref{eq: Ts summed over a}, which appears as colour part of each propagator. 
In double-line notation, it can be written as
\begin{align}
\label{eq: double-line SUN propagator}
\delta^i_l\delta^k_j - \frac{\colors}{N} \delta^i_j\delta^k_l {} &= {} \, %
\settoheight{\eqoff}{$\times$}%
\setlength{\eqoff}{0.5\eqoff}%
\addtolength{\eqoff}{-4.5\unitlength}%
\raisebox{\eqoff}{\fbox{%
\fmfframe(1,2)(1,2){%
 \begin{fmfgraph*}(20,5)
  \fmfstraight
  \fmfpen{10}
  \fmfleft{i}
  \fmfright{o}
  \fmf{plain_n}{i,o}
  \fmffreeze
  \fmfdraw
  \fmfpen{8}
  \fmf{iplain_n,foreground=white}{i,o}
  \fmfdraw
  \fmf{leftrightarrows,foreground=black}{i,o}
  \fmfdraw
  \fmfiv{l=$\scriptstyle i$,l.a=90,l.d=7}{vloc(__i)}
  \fmfiv{l=$\scriptstyle j$,l.a=-90,l.d=7}{vloc(__i)}
  \fmfiv{l=$\scriptstyle l$,l.a=90,l.d=7}{vloc(__o)}
  \fmfiv{l=$\scriptstyle k$,l.a=-90,l.d=7}{vloc(__o)}
  \fmfdraw
 \end{fmfgraph*}%
}}}\,{}%
-{}\frac{\colors}{N}{}\,%
\settoheight{\eqoff}{$\times$}%
\setlength{\eqoff}{0.5\eqoff}%
\addtolength{\eqoff}{-4.5\unitlength}%
\raisebox{\eqoff}{\fbox{%
\fmfframe(1,2)(1,2){%
 \begin{fmfgraph*}(20,5)
  \fmfstraight
  \fmfpen{10}
  \fmfleft{i}
  \fmfright{o}
  \fmf{interrupted_plain_n}{i,o}
  \fmffreeze
  \fmfdraw
  \fmfpen{8}
  \fmf{iinterrupted_plain_n,foreground=white}{i,o}
  \fmfdraw
  \fmf{leftrightarrows_interruptedold,foreground=black}{i,o}
  \fmfdraw
  \fmfiv{l=$\scriptstyle i$,l.a=90,l.d=7}{vloc(__i)}
  \fmfiv{l=$\scriptstyle j$,l.a=-90,l.d=7}{vloc(__i)}
  \fmfiv{l=$\scriptstyle l$,l.a=90,l.d=7}{vloc(__o)}
  \fmfiv{l=$\scriptstyle k$,l.a=-90,l.d=7}{vloc(__o)}
  \fmfdraw
 \end{fmfgraph*}%
}}}{}
\eqncom
\end{align}
where $\colors=1$ for gauge group $\SU{N}$ and $\colors=0$ for gauge group $\U{N}$ as defined in \eqref{eq: color s}.
The double-trace term in \eqref{eq: double-line SUN propagator} subtracts the $\U{1}$ component in the first term in \eqref{eq: double-line SUN propagator}, which is absent for gauge group $\SU{N}$.

In correlation functions of gauge-invariant local composite operators, the fundamental gauge-group indices $i,j,k,l = 1,\dots,N$ in \eqref{eq: double-line SUN propagator} have to be contracted by vertices, composite operators and other propagators.
Two possible cases can occur.
In the first case, $i$ is connected to $l$ and $k$ is connected to $j$:
\begin{equation}
\underbrace{\fbox{
\settoheight{\eqoff}{$\times$}%
\setlength{\eqoff}{0.5\eqoff}%
\addtolength{\eqoff}{-10.5\unitlength}%
\raisebox{\eqoff}{%
\fmfframe(1.5,8)(0,8){
 \begin{fmfgraph*}(20,5)
  \fmfstraight
  \fmfpen{10}
  \fmfleft{i}
  \fmfright{o}
  \fmf{plain_n}{i,o}
  \fmfdraw
  \fmfpen{8}
  \fmf{iplain_n,foreground=white}{i,o}
  \fmfdraw
  \fmf{leftrightarrows,foreground=black}{i,o}
  \fmfdraw
 \fmfpen{1}
 \fmfcmd{z0=(0.5 vloc(__i)+0.5 vloc(__o)+(0,-30)); draw (vloc(__i)+(0,-4.5)){dir 180}...z0...{dir 180}(vloc(__o)+(0,-4.5)) withcolor (0.65,0.65,0.65);}
 \fmfcmd{z1=(0.5 vloc(__i)+0.5 vloc(__o)+(0,+30)); draw (vloc(__i)+(0,+4.5)){dir 180}...z1...{dir 180}(vloc(__o)+(0,+4.5)) withcolor (0.65,0.65,0.65);}
  \fmfiv{l=$\scriptstyle i$,l.a=90,l.d=7}{vloc(__i)}
  \fmfiv{l=$\scriptstyle j$,l.a=-90,l.d=7}{vloc(__i)}
  \fmfiv{l=$\scriptstyle l$,l.a=90,l.d=7}{vloc(__o)}
  \fmfiv{l=$\scriptstyle k$,l.a=-90,l.d=7}{vloc(__o)}
  \fmfdraw
 \end{fmfgraph*}
}
}}
}_{N^2}
\,{}-{}\frac{\colors}{N}{}\,
\underbrace{\fbox{
\settoheight{\eqoff}{$\times$}%
\setlength{\eqoff}{0.5\eqoff}%
\addtolength{\eqoff}{-10.5\unitlength}%
\raisebox{\eqoff}{%
\fmfframe(1.5,8)(0,8){
 \begin{fmfgraph*}(20,5)
  \fmfstraight
  \fmfpen{10}
  \fmfleft{i}
  \fmfright{o}
  \fmf{interrupted_plain_n}{i,o}
  \fmffreeze
  \fmfdraw
  \fmfpen{8}
  \fmf{iinterrupted_plain_n,foreground=white}{i,o}
  \fmfdraw
  \fmf{leftrightarrows_interruptedold,foreground=black}{i,o}
  \fmfdraw 
\fmfpen{1}
 \fmfcmd{z0=(0.5 vloc(__i)+0.5 vloc(__o)+(0,-30)); draw (vloc(__i)+(0,-4.5)){dir 180}...z0...{dir 180}(vloc(__o)+(0,-4.5)) withcolor (0.65,0.65,0.65);}
 \fmfcmd{z1=(0.5 vloc(__i)+0.5 vloc(__o)+(0,+30)); draw (vloc(__i)+(0,+4.5)){dir 180}...z1...{dir 180}(vloc(__o)+(0,+4.5)) withcolor (0.65,0.65,0.65);}
  \fmfiv{l=$\scriptstyle i$,l.a=90,l.d=7}{vloc(__i)}
  \fmfiv{l=$\scriptstyle j$,l.a=-90,l.d=7}{vloc(__i)}
  \fmfiv{l=$\scriptstyle l$,l.a=90,l.d=7}{vloc(__o)}
  \fmfiv{l=$\scriptstyle k$,l.a=-90,l.d=7}{vloc(__o)}
  \fmfdraw
 \end{fmfgraph*}
}
}}
}_{N^1}
\,{}\propto{}\left(1-\frac{\colors}{N^2}\right)N^2 \eqndot
\end{equation}
In this case, the contribution from the double-trace term is suppressed with respect to the one from the single-trace term by a factor of $\frac{1}{N^2}$.
In the second case, $i$ is connected to $j$ and $k$ to $l$:
\begin{equation}
\underbrace{%
\settoheight{\eqoff}{$\times$}%
\setlength{\eqoff}{0.5\eqoff}%
\addtolength{\eqoff}{-4.5\unitlength}%
\raisebox{\eqoff}{\fbox{%
\fmfframe(3,2)(3,2){%
 \begin{fmfgraph*}(20,5)
  \fmfstraight
  \fmfpen{10}
  \fmfleft{i}
  \fmfright{o}
  \fmf{plain_n}{i,o}
  \fmfdraw
  \fmfpen{8}
  \fmf{iplain_n,foreground=white}{i,o}
  \fmfdraw
  \fmf{leftrightarrows,foreground=black}{i,o}
  \fmfdraw
 \fmfpen{1}
 \fmfcmd{z3=(vloc(__i)+(-10,0)); draw (vloc(__i)+(0,-4.5)){dir 180}...z3...(vloc(__i)+(0,4.5)){dir 0} withcolor (0.65,0.65,0.65);}
 \fmfcmd{z4=(vloc(__o)+(10,0)); draw (vloc(__o)+(0,-4.5)){dir 0}...z4...(vloc(__o)+(0,4.5)){dir 180} withcolor (0.65,0.65,0.65);}
  \fmfiv{l=$\scriptstyle i$,l.a=90,l.d=7}{vloc(__i)}
  \fmfiv{l=$\scriptstyle j$,l.a=-90,l.d=7}{vloc(__i)}
  \fmfiv{l=$\scriptstyle l$,l.a=90,l.d=7}{vloc(__o)}
  \fmfiv{l=$\scriptstyle k$,l.a=-90,l.d=7}{vloc(__o)}
  \fmfdraw
 \end{fmfgraph*}%
}%
}}%
}_{N^1}\,%
{}-{}\frac{\colors}{N}\,%
\underbrace{%
\settoheight{\eqoff}{$\times$}%
\setlength{\eqoff}{0.5\eqoff}%
\addtolength{\eqoff}{-4.5\unitlength}%
\raisebox{\eqoff}{\fbox{%
\fmfframe(3,2)(3,2){%
 \begin{fmfgraph*}(20,5)
  \fmfstraight
  \fmfpen{10}
  \fmfleft{i}
  \fmfright{o}
  \fmf{interrupted_plain_n}{i,o}
  \fmffreeze
  \fmfdraw
  \fmfpen{8}
  \fmf{iinterrupted_plain_n,foreground=white}{i,o}
  \fmfdraw
  \fmf{leftrightarrows_interruptedold,foreground=black}{i,o}
  \fmfdraw 
\fmfpen{1}
 \fmfcmd{z3=(vloc(__i)+(-10,0)); draw (vloc(__i)+(0,-4.5)){dir 180}...z3...(vloc(__i)+(0,4.5)){dir 0} withcolor (0.65,0.65,0.65);}
 \fmfcmd{z4=(vloc(__o)+(10,0)); draw (vloc(__o)+(0,-4.5)){dir 0}...z4...(vloc(__o)+(0,4.5)){dir 180} withcolor (0.65,0.65,0.65);}
  \fmfiv{l=$\scriptstyle i$,l.a=90,l.d=7}{vloc(__i)}
  \fmfiv{l=$\scriptstyle j$,l.a=-90,l.d=7}{vloc(__i)}
  \fmfiv{l=$\scriptstyle l$,l.a=90,l.d=7}{vloc(__o)}
  \fmfiv{l=$\scriptstyle k$,l.a=-90,l.d=7}{vloc(__o)}
  \fmfdraw
 \end{fmfgraph*}%
}%
}}%
}_{N^2}\,{}\propto{}\left(1-\colors\right)N\eqndot
\end{equation}
In this case, the double-trace term contributes at the same leading order as the single-trace term.
Moreover, the sum of both contributions vanishes for gauge group $\SU{N}$, where $\colors=1$.
This can be interpreted as follows.
The contribution of the $\U{1}$ component, which is subtracted by the double-trace term, is of leading order only if it is the only contribution.
This is precisely what occurs in the second case, where the connection of the indices projects out all other components. 
In the undeformed $\cN=4$ SYM theory, the $\U{1}$ component is free, as all interactions are of commutator type. Hence, the diagrams of the second case have to vanish due to cancellations between different contributions.  

For two-point functions, a generic diagram corresponding to the second case is
\begin{equation}
\label{eq: generic prewrapping diagram}
\settoheight{\eqoff}{$+$}%
\setlength{\eqoff}{1.5\eqoff}%
\addtolength{\eqoff}{-9\unit}%
\raisebox{\eqoff}{\fbox{%
\rotatebox{90}{%
\begin{pspicture}(0,-1.5)(16,10.5)
\ulinsert{0}{9}
\olvertex{0}{6}
\psline[linestyle=dotted,linewidth=\blacklinew](0,4)(0,2)
\psline[linestyle=dotted,linecolor=ogray,linewidth=\blacklinew](0.75,4)(0.75,2)
\psline[linestyle=dotted,linewidth=\blacklinew](1.5,4)(1.5,2)
\dlinsert{0}{0}
\urinsert{16}{9}
\orvertex{16}{6}
\psline[linestyle=dotted,linewidth=\blacklinew](14.5,4)(14.5,2)
\psline[linestyle=dotted,linecolor=ogray,linewidth=\blacklinew](15.25,4)(15.25,2)
\psline[linestyle=dotted,linewidth=\blacklinew](16,4)(16,2)
\drinsert{16}{0}
\uinex{2}{9}%
\iinex{2}{6}%
\dinex{2}{0}%
\setlength{\ya}{9\unit}
\addtolength{\ya}{0.5\dlinewidth}
\setlength{\yb}{0\unit}
\addtolength{\yb}{-0.5\dlinewidth}
\setlength{\xc}{7.0\unit}
\setlength{\yc}{4.5\unit}
\addtolength{\yc}{-0.5\dlinewidth}
\setlength{\xd}{9.0\unit}
\setlength{\yd}{4.5\unit}
\addtolength{\yd}{0.5\dlinewidth}
\psbezier[linewidth=\blacklinew](3.5,\ya)(5.5,\ya)(5.5,\yd)(\xc,\yd)
\psbezier[linewidth=\blacklinew](3.5,\yb)(5.5,\yb)(5.5,\yc)(\xc,\yc)
\psline[linestyle=dotted,linewidth=\blacklinew](3.5,4)(3.5,2)
\psbezier[linewidth=\blacklinew](12.5,\ya)(10.5,\ya)(10.5,\yd)(\xd,\yd)
\psbezier[linewidth=\blacklinew](12.5,\yb)(10.5,\yb)(10.5,\yc)(\xd,\yc)
\doutex{14}{0}%
\ioutex{14}{6}%
\uoutex{14}{9}%
\psline[linestyle=dotted,linewidth=\blacklinew](12.5,4)(12.5,2)
\setlength{\xa}{3.5\unit}
\addtolength{\xa}{\dlinewidth}
\setlength{\xb}{6.5\unit}
\addtolength{\xb}{-\dlinewidth}
\setlength{\ya}{8\unit}
\addtolength{\ya}{-0.5\dlinewidth}
\setlength{\yb}{1\unit}
\addtolength{\yb}{0.5\dlinewidth}
\psline[linecolor=gray,fillstyle=solid,fillcolor=gray,linearc=\linearc](\xa,\ya)(\xb,4.5)(\xa,\yb)
\setlength{\xa}{12.5\unit}
\addtolength{\xa}{-\dlinewidth}
\setlength{\xb}{9.5\unit}
\addtolength{\xb}{\dlinewidth}
\setlength{\ya}{8\unit}
\addtolength{\ya}{-0.5\dlinewidth}
\setlength{\yb}{1\unit}
\addtolength{\yb}{0.5\dlinewidth}
\psline[linecolor=gray,fillstyle=solid,fillcolor=gray,linearc=\linearc](\xa,\ya)(\xb,4.5)(\xa,\yb)
\psset{linecolor=black,linewidth=\blacklinew}
\psline[doubleline=true,doublesep=4.\blacklinew](7.0,4.5)(9.0,4.5)
\end{pspicture}}}}%
\,{}-{}\displaystyle\frac{\colors}{N}\, %
\settoheight{\eqoff}{$+$}%
\setlength{\eqoff}{1.5\eqoff}%
\addtolength{\eqoff}{-9\unit}%
\raisebox{\eqoff}{\fbox{%
\rotatebox{90}{%
\begin{pspicture}(0,-1.5)(16,10.5)
\ulinsert{0}{9}
\olvertex{0}{6}
\psline[linestyle=dotted,linewidth=\blacklinew](0,4)(0,2)
\psline[linestyle=dotted,linecolor=ogray,linewidth=\blacklinew](0.75,4)(0.75,2)
\psline[linestyle=dotted,linewidth=\blacklinew](1.5,4)(1.5,2)
\dlinsert{0}{0}
\urinsert{16}{9}
\orvertex{16}{6}
\psline[linestyle=dotted,linewidth=\blacklinew](14.5,4)(14.5,2)
\psline[linestyle=dotted,linecolor=ogray,linewidth=\blacklinew](15.25,4)(15.25,2)
\psline[linestyle=dotted,linewidth=\blacklinew](16,4)(16,2)
\drinsert{16}{0}
\uinex{2}{9}%
\iinex{2}{6}%
\dinex{2}{0}%
\setlength{\ya}{9\unit}
\addtolength{\ya}{0.5\dlinewidth}
\setlength{\yb}{0\unit}
\addtolength{\yb}{-0.5\dlinewidth}
\setlength{\xc}{7.0\unit}
\setlength{\yc}{4.5\unit}
\addtolength{\yc}{-0.5\dlinewidth}
\setlength{\xd}{9.0\unit}
\setlength{\yd}{4.5\unit}
\addtolength{\yd}{0.5\dlinewidth}
\psbezier[linewidth=\blacklinew](3.5,\ya)(5.5,\ya)(5.5,\yd)(\xc,\yd)
\psbezier[linewidth=\blacklinew](3.5,\yb)(5.5,\yb)(5.5,\yc)(\xc,\yc)
\psline[linestyle=dotted,linewidth=\blacklinew](3.5,4)(3.5,2)
\psbezier[linewidth=\blacklinew](12.5,\ya)(10.5,\ya)(10.5,\yd)(\xd,\yd)
\psbezier[linewidth=\blacklinew](12.5,\yb)(10.5,\yb)(10.5,\yc)(\xd,\yc)
\doutex{14}{0}%
\ioutex{14}{6}%
\uoutex{14}{9}%
\psline[linestyle=dotted,linewidth=\blacklinew](12.5,4)(12.5,2)
\setlength{\xa}{3.5\unit}
\addtolength{\xa}{\dlinewidth}
\setlength{\xb}{6.5\unit}
\addtolength{\xb}{-\dlinewidth}
\setlength{\ya}{8\unit}
\addtolength{\ya}{-0.5\dlinewidth}
\setlength{\yb}{1\unit}
\addtolength{\yb}{0.5\dlinewidth}
\psline[linecolor=gray,fillstyle=solid,fillcolor=gray,linearc=\linearc](\xa,\ya)(\xb,4.5)(\xa,\yb)
\setlength{\xa}{12.5\unit}
\addtolength{\xa}{-\dlinewidth}
\setlength{\xb}{9.5\unit}
\addtolength{\xb}{\dlinewidth}
\setlength{\ya}{8\unit}
\addtolength{\ya}{-0.5\dlinewidth}
\setlength{\yb}{1\unit}
\addtolength{\yb}{0.5\dlinewidth}
\psline[linecolor=gray,fillstyle=solid,fillcolor=gray,linearc=\linearc](\xa,\ya)(\xb,4.5)(\xa,\yb)
\psset{linecolor=black,linewidth=\blacklinew}
\uoneprop{8}{4.5}{0}
\end{pspicture}}}}
\,{}\propto{}
\,\,\,(1-\colors)N^{2L-1}\eqncom
\end{equation}
where the dark grey area denotes arbitrary planar interactions and both grey-shaded operators are assumed to be of single-trace type and length $L$.%
\footnote{This analysis can be immediately generalised to the length-changing case.}
We denote such diagrams as s-channel type. 
As every cubic vertex comes with a factor of $g_\YM$ and every quartic vertex with a factor $g_\YM^2$, this diagram contains a minimum of $2L-2$ factors of $g_\YM$.
Hence, it can start to contribute to the two-point function and thus to the anomalous dimension at loop order $\ell=L-1$.
As this is one loop order earlier than the critical wrapping order $\ell=L$, we call this new finite-size effect prewrapping.

Another source of double-trace structures are double-trace terms in the action.
While the action of the $\beta$-deformation in $\cN=1$ superspace is free of double-trace terms, they do occur in the component action for gauge group $\SU{N}$ and read
\begin{equation}\label{eq: component action double trace}
\begin{aligned}
\int\de^4x\,\Bigl[-\frac{\colors}{N}
 \frac{g^2_\YM}{2}\tr\bigl(\starcomm{\bar\phi^j}{\bar\phi^k}\bigr)
 \tr\bigl(\starcomm{\phi_j}{\phi_k}\bigr)
\Bigr] \eqncom
\end{aligned}
\end{equation}
where the $\ast$-commutator is defined via the $\ast$-product in complete analogy to the usual commutator.
These double-trace terms arise when integrating out the auxiliary fields and using the completeness relation \eqref{eq: Ts summed over a}, see e.g.\ \cite{Fokken:2013aea} for a detailed derivation.%
\footnote{The double-trace term \eqref{eq: component action double trace} was explicitly written down in \cite{Jin:2012np} but occurred already implicitly in \cite{Freedman:2005cg} several years earlier.} 
Hence, it can be completely understood via the previous analysis by considering the theory in $\cN=1$ superspace or in component space without integrating out the auxiliary fields.
We will encounter other double-trace terms, which do not arise via this mechanism, in the $\gamma_i$-deformation in the next chapters.

Understanding the mechanism of prewrapping, we can also formulate criteria for single-trace operators to be potentially affected by it.
In order for the Feynman diagram \eqref{eq: generic prewrapping diagram} to exist, 
a field in the theory or a trace factor in a multi-trace coupling must have the same  $\SU{4}$ Cartan charges as the composite operator. 
Moreover, in the first case, this field has to have non-trivial $\ast$-products with other fields, i.e.\ it has to have deformed vertices.
If all its vertices are undeformed, the different contributions cancel as in the undeformed theory.
A table of all potentially affected single-trace operators in closed subsectors of the $\beta$-deformation is given in \cite{Fokken:2013mza}. 
As follows from an easy combinatorial analysis,
 this table contains only the operators $\tr(\phi_i\phi_j)$ and $\tr(\bar\phi^i\bar\phi^j)$ with $i\neq j$, which are known to be affected by prewrapping.
In non-compact subsectors or the complete theory, however, large families of potentially affected operators exist, such as 
\begin{equation}\label{eq: candidate}
 \tr\left(\phi_2\phi_3(\phi_1\bar\phi^1)^i(\phi_2\bar\phi^2)^j(\phi_3\bar\phi^3)^k(\psi_1\bar\psi^1)^l(\psi_2\bar\psi^2)^m(\psi_3\bar\psi^3)^n(\psi_4\bar\psi^4)^o\cfstrength^p\cantifstrength^q\right)\eqncom
\end{equation}
where $i,j,k,l,m,n,o,p,q\in \NN_0$ and all spinor indices are suppressed.

Based on the mechanism of prewrapping as well as finiteness theorems in $\cN=1$ superspace, it was furthermore argued in \cite{Fokken:2013mza} that the anomalous dimensions of $\tr(\phi_i\phi_j)$ and $\tr(\bar\phi^i\bar\phi^j)$ with $i\neq j$ as well as those of their superpartners vanish at all orders of planar perturbation theory in the $\beta$-deformation with gauge group $\SU{N}$.%
\footnote{These anomalous dimensions were found to vanish at two-loop order in \cite{Penati:2005hp}.}

\section{Complete one-loop dilatation operator}
\label{sec: one-loop dilatation operator in beta deformation}

Next, we incorporate the finite-size effects of wrapping and prewrapping into the asymptotic dilatation operator density \eqref{eq: deformation of D_2} in order to obtain the complete one-loop dilatation operator of the planar $\beta$-deformation.
Moreover, we comment on its one-loop spectrum.

\subsection{Gauge group \texorpdfstring{$\SU{N}$}{SU(N)}}

For gauge group $\SU{N}$, the finite-size effect of prewrapping has to be incorporated into the asymptotic dilatation operator density \eqref{eq: deformation of D_2}.
Before doing so, let us illustrate why \eqref{eq: deformation of D_2} fails to yield the correct result in the $\beta$-deformation while it does give the correct result in the undeformed theory.
As an example, consider the operator $\cO = \tr (\psi_{\alpha1} \phi_2)$, or, more concretely, the coefficient of $\cO$ in $\loopDila{1} \cO$.
In the undeformed theory, this coefficient is the sum of four matrix elements of the dilatation operator density $\loopdila{1}_{\cN=4}$,  
namely
\begin{equation}
\label{eq: contributions}
 \begin{aligned}
 (\loopdila{1}_{\cN=4})_{\psi_1\phi_2}^{\psi_1\phi_2}= +3\eqncom \quad
 (\loopdila{1}_{\cN=4})_{\psi_1\phi_2}^{\phi_2\psi_1}= -1\eqncom \quad
 (\loopdila{1}_{\cN=4})_{\phi_2\psi_1}^{\phi_2\psi_1}= +3\eqncom \quad
 (\loopdila{1}_{\cN=4})_{\phi_2\psi_1}^{\psi_1\phi_2}= -1 \eqncom
\end{aligned}
\end{equation}
where we are suppressing the index $\alpha$ of the fermion.
The first two matrix elements stem from the following sums of Feynman diagrams
\begin{equation}
\begin{aligned}
 (\loopdila{1}_{\cN=4})_{\psi_1\phi_2}^{\psi_1\phi_2}&=
 \underbrace{\frac{1}{2}\fbox{\FDiagram[labelleftbottom=$\scriptstyle \psi_1$,
 	  labelrightbottom=$\scriptstyle \phi_2$,
 	  labellefttop=$\scriptstyle \psi_1$,
 	  labelrighttop=$\scriptstyle \phi_2$,
	  leftSE,long,longup]{dashes_sarrow}{plain_sarrow}{}{}{}}}_{+2}
\, + \,
  \underbrace{\frac{1}{2}\FDiagram[
 	  labelleftbottom=$\scriptstyle \psi_1$,
 	  labelrightbottom=$\scriptstyle \phi_2$,
 	  labellefttop=$\scriptstyle \psi_1$,
 	  labelrighttop=$\scriptstyle \phi_2$,
	  rightSE,long,longup]{dashes_sarrow}{plain_sarrow}{}{}{}}_{+1}
\, + \,
  \underbrace{\FDiagram[styleleftbottom=dashes,
 	  stylerightbottom=plain,
 	  stylemid=dashes,
 	  stylelefttop=dashes,
 	  stylerighttop=plain,
 	  labelleftbottom=$\scriptstyle \psi_1$,
 	  labelrightbottom=$\scriptstyle \phi_2$,
 	  labellefttop=$\scriptstyle \psi_1$,
 	  labelrighttop=$\scriptstyle \phi_2$,
	  tchannel,long,longup]{dashes_sarrow}{plain_sarrow}{wiggly}{dashes_sarrow}{plain_sarrow}}_{-1}
\, + \,
\underbrace{\FDiagram[styleleftbottom=dashes,
	  stylerightbottom=plain,
	  stylemid=dashes,
	  stylelefttop=dashes,
	  stylerighttop=plain,
	  labelleftbottom=$\scriptstyle \psi_1$,
	  labelrightbottom=$\scriptstyle \phi_2$,
	  labellefttop=$\scriptstyle \psi_1$,
	  labelrighttop=$\scriptstyle \phi_2$,
	  labelmid=$\scriptstyle \psi_3$,
	  schannel,long,longup]{dashes_sarrow}{plain_sarrow}{dashes_srarrow}{dashes_sarrow}{plain_sarrow}}_{+1} \eqncom \\
(\loopdila{1}_{\cN=4})^{\phi_2\psi_1}_{\psi_1\phi_2}&=
\underbrace{\FDiagram[styleleftbottom=dashes,
	  stylerightbottom=plain,
	  stylemid=dashes,
	  stylelefttop=dashes,
	  stylerighttop=plain,
	  labelleftbottom=$\scriptstyle \psi_1$,
	  labelrightbottom=$\scriptstyle \phi_2$,
	  labellefttop=$\scriptstyle \phi_2$,
	  labelrighttop=$\scriptstyle \psi_1$,
	  labelmid=$\scriptstyle \psi_3$,
	  schannel,long,longup]{dashes_sarrow}{plain_sarrow}{dashes_srarrow}{plain_sarrow}{dashes_sarrow}}_{-1}
\eqncom
\end{aligned}
\end{equation}
where complex scalars are depicted by solid lines, fermions by dashed lines, self-energy insertions by black blobs and the composite operators by thick horizontal lines.\footnote{While the whole two-point function is shown, only the part contributing to the off-shell operator renormalisation is depicted in black whereas the rest is depicted in grey.} The fact that only the single-trace part is considered is depicted by the thick horizontal lines' extension beyond the points where the field lines exit them.
The Feynman rules used to calculate these matrix elements can be found in \cite{Fokken:2013aea}.
The last two matrix elements stem from the reflections of the above Feynman diagrams at the vertical axis.
As expected, the sum of all s-channel-type diagrams vanishes due to cancellations among the four asymptotic contributions.

In the $\beta$-deformation, the four matrix elements \eqref{eq: contributions} receive the phase factors $1$, $\e^{i \beta }$, $1$ and $\e^{-i \beta }$, respectively, which follows both from relation \eqref{eq: diagram relation} and the Feynman diagram calculation.
The resulting sum of the s-channel-type diagrams is non-vanishing and given by
\begin{equation}
 1-\e^{i \beta }+1-\e^{-i \beta }=4\sin^2\frac{\beta}{2} 
\eqndot
\end{equation}

In principle, the failure of \eqref{eq: deformation of D_2} to yield the correct result requires the separate calculation of the double-trace contribution to the dilatation operator, e.g.\ by the method of section \ref{sec: one-loop generalised unitarity}.
In the following, however, we will argue that a shortcut is available which avoids any calculation.
The main idea behind this argument is to restore the cancellation mechanism of the undeformed theory in order to set the contributions of all s-channel-type diagrams to zero without altering any other contributions.
In the $\beta$-deformation, only vertices built from matter-type fields, i.e.\ the chiral superfields $\Phi_i$, $\bar\Phi^i$, $i=1,2,3$ and its components $\phi_i$, $\bar\phi^i$, $\psi_{i\alpha}$, $\bar\psi^{i}_{\alphadot}$ in Wess-Zumino gauge, are deformed.
Vertices containing at least one gauge-type field, i.e.\ the vector superfield $V$ and its components $A_\mu$, $\psi_{4\alpha}$, $\bar\psi^{4}_{\alphadot}$, are undeformed.
In s-channel-type diagrams, one undeformed vertex suffices for the cancellation of the undeformed theory to occur.
Hence, we only need to consider s-channel-type diagrams with two deformed vertices.
An exhaustive list of them is shown in table \ref{tab: deformed possibilities}.
Note that these diagrams only occur for matrix elements of the dilatation operator with four matter-type fields or four anti-matter-type fields.
Hence, we can restore the cancellation mechanism of the undeformed theory by setting $\beta=0$ in these matrix elements.
It remains to be shown that this does not alter any non-s-channel-type diagrams, i.e.\ that no other deformed diagrams with these combinations of external fields exist.
It is easy to convince oneself that this is the case.%
\footnote{The only other diagrams with deformed vertices and these combinations of external fields are of self-energy type. Hence, they have range one and we can apply relation \eqref{eq: diagram relation} to show that their planar part is deformation independent.}
Further note that this argument is completely independent of any covariant derivatives that can act on the external fields.

\begin{table}[tbp]
\centering
\begin{tabular}{|c|@{\quad}c@{\quad}|@{\quad}c@{\quad}|}
\hline
 & in components & $\mathcal{N}=1$ \\\hline
\begin{minipage}[c]{2.4cm}
\vspace*{0.4\baselineskip}
\centering {\bf s-channel} \\
\vspace*{0.2\baselineskip}
\begin{tabular}{@{}c@{}@{}l}
 $+${ } & vertical \& \\
 & horizontal \\ 
 & reflections \\
  $+${ } & twists
\end{tabular}%
\vspace*{0.2\baselineskip}
\end{minipage}%
 & %
$\FDiagram[labelleftbottom=$\scriptstyle \psi_i$,
 	  labelrightbottom=$\scriptstyle \psi_j$,
 	  labellefttop=$\scriptstyle \psi_i$,
 	  labelrighttop=$\scriptstyle \psi_j$,
	  labelmid=$\scriptstyle \phi_k$,
	  schannel,long,longup]{dashes_sarrow}{dashes_sarrow}{plain_srarrow}{dashes_sarrow}{dashes_sarrow}$ \,\,\,\,\,\, %
$\FDiagram[labelleftbottom=$\scriptstyle \phi_i$,
 	  labelrightbottom=$\scriptstyle \psi_j$,
 	  labellefttop=$\scriptstyle \phi_i$,
 	  labelrighttop=$\scriptstyle \psi_j$,
	  labelmid=$\scriptstyle \psi_k$,
	  schannel,long,longup]{plain_sarrow}{dashes_sarrow}{dashes_srarrow}{plain_sarrow}{dashes_sarrow}$ \,\,\,\,\,\, %
$\FDiagram[labelleftbottom=$\scriptstyle \phi_i$,
 	  labelrightbottom=$\scriptstyle \phi_j$,
 	  labellefttop=$\scriptstyle \phi_i$,
 	  labelrighttop=$\scriptstyle \phi_j$,
	  xchannel,long,longup]{plain_sarrow}{plain_sarrow}{}{plain_sarrow}{plain_sarrow}%
\!\!=\!\!\FDiagram[
 	  labelleftbottom=$\scriptstyle \phi_i$,
 	  labelrightbottom=$\scriptstyle \phi_j$,
 	  labellefttop=$\scriptstyle \phi_i$,
 	  labelrighttop=$\scriptstyle \phi_j$,
	  labelmid=$\scriptstyle F_k$,
	  schannel,long,longup]{plain_sarrow}{plain_sarrow}{dots_srarrow}{plain_sarrow}{plain_sarrow}$ %
& $\FDiagram[labelleftbottom=$\scriptstyle \varPhi_i$,
 	  labelrightbottom=$\scriptstyle \varPhi_j$,
 	  labellefttop=$\scriptstyle \varPhi_i$,
 	  labelrighttop=$\scriptstyle \varPhi_j$,
	  labelmid=$\scriptstyle \varPhi_k$,
	  schannel,long,longup]{plain_sarrow}{plain_sarrow}{plain_srarrow}{plain_sarrow}{plain_sarrow}$ \\ \hline
\begin{minipage}[c]{2.4cm}
\vspace*{0.4\baselineskip}
\centering {\bf t-channel} \\
\vspace*{0.2\baselineskip}
\begin{tabular}{@{}c@{}@{}l}
 $+${ } & vertical \& \\
 & horizontal \\ 
 & reflections \\
 &
\end{tabular}%
\vspace*{0.2\baselineskip}
\end{minipage}%
 & %
$\FDiagram[labelleftbottom=$\scriptstyle \psi_i$,
 	  labelrightbottom=$\scriptstyle \bar\psi^j$,
 	  labellefttop=$\scriptstyle \bar\psi^j$,
 	  labelrighttop=$\scriptstyle \psi_i$,
	  labelmid=$\scriptstyle \phi_k$,
tchannel,long,longup]{dashes_sarrow}{dashes_srarrow}{plain_srarrow}{dashes_srarrow}{dashes_sarrow}$ \,\,\,\,\,\, %
$\FDiagram[labelleftbottom=$\scriptstyle \phi_i$,
 	  labelrightbottom=$\scriptstyle \bar\psi^j$,
 	  labellefttop=$\scriptstyle \bar\psi^j$,
 	  labelrighttop=$\scriptstyle \phi_i$,
	  labelmid=$\scriptstyle \bar\psi^k$,
	  tchannel,long,longup]{plain_sarrow}{dashes_srarrow}{dashes_srarrow}{dashes_srarrow}{plain_sarrow}$ \,\,\,\,\,\, %
$\FDiagram[labelleftbottom=$\scriptstyle \phi_i$,
 	  labelrightbottom=$\scriptstyle \bar\phi^j$,
 	  labellefttop=$\scriptstyle \bar\phi^j$,
 	  labelrighttop=$\scriptstyle \phi_i$,
	  xchannel,long,longup]{plain_sarrow}{plain_srarrow}{}{plain_srarrow}{plain_sarrow}%
\!\!=\!\!\FDiagram[
 	  labelleftbottom=$\scriptstyle \phi_i$,
 	  labelrightbottom=$\scriptstyle \bar\phi^j$,
 	  labellefttop=$\scriptstyle \bar\phi^j$,
 	  labelrighttop=$\scriptstyle \phi_i$,
	  labelmid=$\scriptstyle F_k$,
	  tchannel,long,longup]{plain_sarrow}{plain_srarrow}{dots_srarrow}{plain_srarrow}{plain_sarrow}$ & %
$\FDiagram[labelleftbottom=$\scriptstyle \varPhi_i$,
  	  labelrightbottom=$\scriptstyle \bar\varPhi^j$,
  	  labellefttop=$\scriptstyle \bar\varPhi^j$,
  	  labelrighttop=$\scriptstyle \varPhi_i$,
 	  labelmid=$\scriptstyle \bar\varPhi^k$,
 	  tchannel,long,longup]{plain_sarrow}{plain_srarrow}{plain_srarrow}{plain_srarrow}{plain_sarrow}$ \\\hline
\end{tabular}
\caption{Feynman diagrams with two deformed vertices that contribute to the dilatation operator at range two. Complex scalars and chiral superfields are depicted by solid lines, fermions by dashed lines, auxiliary fields by dotted lines and the composite operators by thick horizontal lines. `Twist' denotes the reflection of only the upper half of the diagram at a vertical axis.}
\label{tab: deformed possibilities}
\end{table}

We define the alphabets of matter-type fields and anti-matter-type fields
\begin{equation}
\label{eq: alphabets of matter and antimatter fields}
\begin{aligned}
 \cA_\text{matter}&=\{\cder^k\phi_1,\cder^k\phi_2,\cder^k\phi_3,\cder^k\psi_{\alpha1},\cder^k\psi_{\alpha2},\cder^k\psi_{\alpha3} \}
  \eqncom \\
 \bar\cA_\text{matter}&=
 \{\cder^k\bar\phi^1,\cder^k\bar\phi^2,\cder^k\bar\phi^3,\cder^k\bar\psi_{\dot\alpha}^{1},\cder^k\bar\psi_{\dot\alpha}^{2},\cder^k\bar\psi_{\dot\alpha}^{3} \}
\end{aligned} 
\end{equation}
as subsets of the alphabet $\cA$ given in \eqref{eq: alphabet of fields}. We have furthermore abbreviated the index structure of the fields as 
$\cder^k\psi_{\alpha i}\equiv \cder_{(\alpha_1\dot\alpha_1}\cdots \cder_{\alpha_k\dot\alpha_k}\psi_{\alpha)i}$, etc. 
The complete one-loop dilatation operator of the $\beta$-deformation with gauge group $\SU{N}$ is then given by
\begin{equation}
\label{eq: complete one-loop dilatation operator in beta deformation}
 (\loopdila{1}_\beta)_{A_iA_j}^{A_kA_l}= \e^{\frac{\complexi}{2} (\mathbf{q}_{A_k} \wedge \mathbf{q}_{A_l}- \mathbf{q}_{A_i} \wedge \mathbf{q}_{A_j})} \rule[-0.96cm]{0.1mm}{1.415cm}\!{\phantom{|}}_{\substack{\\[0.2cm]
\beta=0\text{ if }L=2\text{ and \phantom{...............}}\\ 
(A_i,A_j,A_k,A_l\in\cA_\text{matter}\text{ or \phantom{)}}\\  
\phantom{{}({}} A_i,A_j,A_k,A_l\in\bar\cA_\text{matter}) \phantom{\text{ or }}}}  (\loopdila{1}_{\cN=4})_{A_iA_j}^{A_kA_l}
\eqndot
\end{equation} 

\subsection{Gauge group \texorpdfstring{$\U{N}$}{U(N)}}

For gauge group $\U{N}$, the prewrapping effect is absent at one-loop order.%
\footnote{This is true if the running double-trace coupling \eqref{eq: component action double trace} is set to zero at tree level, which is the case we are considering here.
}
However, the well known wrapping effect occurs for operators of length $L=1$.
The corresponding wrapping diagrams are one-particle reducible and given entirely by the self-energy diagrams of the $\U{1}$ component.
For the scalar fields, these are calculated in appendix \ref{appsec: one-loop self energies}. 
Via supersymmetry, this also determines the self energies of the matter-type fermions%
\footnote{The self energies of the fermions were also explicitly calculated in \cite{Fokken:2014moa} for the case of the more general $\gamma_i$-deformation.}
 and we find%
\footnote{The anomalous dimensions can be calculated from the self energies \eqref{eq: delta phi U1} using \eqref{ZOL} and \eqref{eq: gamma OL def}.}
\begin{equation}
 \gamma^{(1)}_{\tr\D^k\phi_i}=\gamma^{(1)}_{\tr\D^k\bar\phi^i}=\gamma^{(1)}_{\tr\D^k\psi_{\alpha i}}=\gamma^{(1)}_{\tr\D^k\bar\psi^i_{\dot\alpha}}=4 \sin^2 \tfrac{\beta}{2}  
\end{equation}
in the notation introduced below \eqref{eq: alphabets of matter and antimatter fields}.
The $\U1$ components of the gauge-type fields are free as in the undeformed theory and hence their self energies vanish. 
Thus, we find
\begin{equation}
 \gamma^{(1)}_{\tr\D^k\cfstrength_{\alpha\beta}}=\gamma^{(1)}_{\tr\D^k\cantifstrength_{\dot\alpha\dot\beta}}=\gamma^{(1)}_{\tr\D^k\psi_{\alpha4}}=\gamma^{(1)}_{\tr\D^k\bar\psi^4_{\dot\alpha}}=0 \eqndot
\end{equation}

\subsection{One-loop spectrum}

Using the above results, the spectrum of all primary operators with classical scaling dimension $\Delta_0\leq 4.5$ was calculated in \cite{Fokken:2013mza}.
Interestingly, only the supermultiplets of the operators $\tr(\phi_i\phi_j)$ and $\tr(\bar\phi^i\bar\phi^j)$ with $i\neq j$ were found to be affected by the prewrapping effect at one-loop order, although it affects infinitely many matrix elements, cf.\ \eqref{eq: complete one-loop dilatation operator in beta deformation}.
At the diagrammatic level, this can be understood from cancellations among the contributions of the different single-trace operators whose linear combinations form the primary operators; see \cite{Fokken:2013mza} for a detailed example.
It would be very interesting to find a general principle to explain this, in particular one that can also be applied to constrain the occurrence of prewrapping at higher loop orders.

\chapter{Non-conformality of the \texorpdfstring{$\gamma_i$}{gamma-i}-deformation}
\label{chap: nonconformality}

In this chapter, we show that the non-supersymmetric three-parameter $\gamma_i$-deformation \cite{Frolov:2005dj} is not conformally invariant due to the running of a double-trace coupling without fixed point. Moreover, it cannot be rendered conformally invariant by including further multi-trace couplings that satisfy certain minimal requirements. Via the prewrapping effect, this also affects the planar theory.

In section \ref{sec: multi-trace couplings}, we enlist the aforementioned requirements on multi-trace couplings and give all couplings that satisfy them.
We calculate the one-loop renormalisation of a particular double-trace coupling in section \ref{sec: renormalisation}. In section \ref{sec: beta function}, we give its one-loop beta function and find that it has no zeros for generic deformation parameters.
We provide a short review on renormalisation, which includes the derivation of some of the formulae used in this chapter, in appendix \ref{app: deformed theories}.

The results presented in this chapter were first published in \cite{Fokken:2013aea}.

\section{Multi-trace couplings}
\label{sec: multi-trace couplings}

It is well known that loop corrections in a quantum field theory can introduce UV divergences, which have to be absorbed into counterterms via renormalisation. 
The presence of these counterterms at loop level requires us to consider the corresponding couplings already at tree level.
In principle, the counterterms can depend on all couplings in the theory.
Including further couplings might thus restore conformal invariance which is broken otherwise.
Hence, we consider all multi-trace couplings that fulfil a set of minimal requirements.

The requirements are as follows.
First, the action should be renormalisable by power counting.
Second, the 't Hooft limit should exist, i.e.\ no proliferation of $N$-powers beyond the planar limit is permitted.
Third, the three $\U1$ charges shown in table \ref{tab: su(4) charges} should be preserved by the multi-trace couplings.
Fourth, the theory should reduce to $\mathcal{N}=4$ SYM theory in the limit $\gamma_1=\gamma_2=\gamma_3=0$. 

Via finite-size effects, the planar contraction of each additional trace factor in a coupling with a single-trace operator of the same length increases the $N$-power by one. Hence, it follows from the second requirement that couplings with $n$ traces should have a prefactor of (at least) $\frac{1}{N^{n-1}}$, which we write out explicitly.

For gauge group $\SU{N}$, every trace factor has to contain at least two fields as the generators of $\SU{N}$ are traceless.
This allows for the following set of couplings that satisfy the four requirements:
\begin{equation}\label{eq: double-trace couplings}
-\frac{g_\YM^2}{N}\big[Q^{kl}_{\text{F}\,ij}\tr(\bar\phi^i\bar\phi^j)\tr(\phi_k\phi_l)
+Q^{kl}_{\text{D}\,ij}\tr(\bar\phi^i\phi_k)\tr(\bar\phi^j\phi_l)\big]\eqndot
\end{equation} 
For the action to be real in Euclidean signature, the coupling tensors have to satisfy
\begin{equation}\label{eq: conjugation of QF and QD}
\begin{aligned}
(Q^{kl}_{\text{F}\,ij})^\ast&=Q^{ij}_{\text{F}\,kl}\eqncom \qquad
(Q^{kl}_{\text{D}\,ij})^\ast=Q^{ij}_{\text{D}\,kl}\eqndot
\end{aligned}
\end{equation}

For gauge group $\U{N}$, also trace factors containing a single field can occur, which project to the $\U{1}$ component of that field. This allows for the following cubic multi-trace couplings
\begin{equation}
\begin{aligned}\label{eq: cubic UN multi-trace couplings}
&+\frac{g_\YM}{N}\big[
(\rho_{\psi}^{i})^{BA}\tr(\psi^{\alpha}_{A})\tr(\phi_i\psi_{\alpha B})
+(\rho_{\phi}^{i})^{BA}\tr(\phi_i)\tr(\psi^{\alpha}_{B}\psi_{\alpha A})
\\
&\hphantom{{}+{}\frac{g_\YM}{N}\big[}
+(\rho_{\bar\psi\,i}^{\dagger})_{BA}\tr(\bar\psi^{\dot\alpha A})\tr(\bar\phi^i\bar\psi_{\dot\alpha}^{B})
+(\rho_{\bar\phi\,i}^{\dagger})_{BA}\tr(\bar\phi^i)\tr(\bar\psi^{\dot\alpha B}\bar\psi_{\dot\alpha}^{A})
\\
&\hphantom{{}+{}\frac{g_\YM}{N}\big[}
+(\tilde\rho_{\bar\psi}^{i})_{BA}\tr(\bar\psi^{\dot\alpha A})\tr(\phi_i\bar\psi_{\dot\alpha}^{B})
+(\tilde\rho_{\bar\phi}^{i})_{BA}\tr(\phi_i)\tr(\bar\psi^{\dot\alpha B}\bar\psi_{\dot\alpha}^{A})
\\
&\hphantom{{}+{}\frac{g_\YM}{N}\big[}
+(\tilde\rho_{\psi\,i}^{\dagger})^{BA}\tr(\psi^{\alpha}_{A})\tr(\bar\phi^i\psi_{\alpha B})
+(\tilde\rho_{\bar\phi\,i}^{\dagger})^{BA}\tr(\bar\phi^i)\tr(\psi^{\alpha}_{B}\psi_{\alpha A})
\big]
\\
&+\frac{g_\YM}{N^2}\big[
(\rho_{3}^{i})^{BA}\tr(\psi^{\alpha}_{A})\tr(\phi_i)\tr(\psi_{\alpha B})
 +(\rho_{3\,i}^{\dagger})_{BA}\tr(\bar\psi^{\dot\alpha A})\tr(\bar\phi^i)\tr(\bar\psi_{\dot\alpha}^{B})\\
&\phantom{{}+{}\frac{g_\YM}{N^2}\big[}
+(\tilde\rho_{3}^{i})_{BA}\tr(\bar\psi^{\dot\alpha A})\tr(\phi_i)\tr(\bar\psi_{\dot\alpha}^{B})
+(\tilde\rho_{3\,i}^{\dagger})^{BA}\tr(\psi^{\alpha}_{A})\tr(\bar\phi^i)\tr(\psi_{\alpha B})
\big]
\end{aligned}
\end{equation}
as well as the additional quartic multi-trace couplings 
\begin{equation}
\begin{aligned}\label{qUNc}
&-\frac{g_\YM^2}{N}\big[
Q^{kl}_{\bar\phi\,ij}\tr(\bar\phi^i)\tr(\bar\phi^j\phi_k\phi_l)
+Q^{kl}_{\phi\,ij}\tr(\phi_k)\tr(\bar\phi^i\bar\phi^j\phi_l)\big]\\
&-\frac{g_\YM^2}{N^2}\big[
Q^{kl}_{\bar\phi\bar\phi\,ij}\tr(\bar\phi^i)\tr(\bar\phi^j)\tr(\phi_k\phi_l)
+Q^{kl}_{\phi\phi\,ij}\tr(\phi_k)\tr(\phi_l)\tr(\bar\phi^i\bar\phi^j)\\
&\hphantom{{}-{}\frac{g_\YM^2}{N^2}\big[}+Q^{kl}_{\bar\phi\phi\,ij}\tr(\bar\phi^i)\tr(\phi_k)\tr(\bar\phi^j\phi_l)\big]\\
&-\frac{g_\YM^2}{N^3}
Q^{kl}_{4\,ij}\tr(\bar\phi^i)\tr(\bar\phi^j)\tr(\phi_k)\tr(\phi_l)
\eqndot
\end{aligned}
\end{equation}
The latter have to satisfy the reality conditions
\begin{equation}\label{conjQ}
\begin{aligned}
(Q^{kl}_{\phi\,ij})^\ast&=Q^{ji}_{\bar\phi\,kl}\eqncom\qquad
(Q^{kl}_{\bar\phi\phi\,ij})^\ast=Q^{ij}_{\bar\phi\phi\,kl}\eqncom\qquad
(Q^{kl}_{\phi\phi\,ij})^\ast=Q^{ij}_{\bar\phi\bar\phi\,kl}\eqncom\qquad
(Q^{kl}_{4\,ij})^\ast=Q^{ij}_{4\,kl}\eqndot
\end{aligned}
\end{equation}
For gauge group $\U{N}$, all $\U{1}$ components are understood to be absorbed into the trace factors of length one such that only $\SU{N}$ components occur in the traces containing two or more fields.

\section{Renormalisation}
\label{sec: renormalisation}

In order to find a one-loop conformally invariant theory, one has to renormalise all couplings in the previous section, calculate their beta functions and find a configuration of tree-level values such that the beta functions vanish simultaneously. 
Performing this analysis, one finds, however, that such a configuration does not exist.
To prove this, it is sufficient to identify a single coupling whose beta function does not vanish.
In the following, we show that the three couplings 
\begin{equation}
\label{eq: running couplings}
-\frac{g_\YM^2}{N}Q^{ii}_{\text{F}\,ii}\tr(\bar\phi^i\bar\phi^i)\tr(\phi_i\phi_i)
\eqncom \qquad i=1,2,3 \text{ not summed,}
\end{equation}
have a non-vanishing beta function.

In renormalising the couplings \eqref{eq: running couplings}, we exploit the fact that they are not renormalised in the undeformed $\mathcal{N}=4$ SYM theory.
It is hence sufficient to calculate only deformed contributions to the renormalisation constant 
\begin{equation}
 \cZ_{Q^{ii}_{\text{F}\,ii}}=1+\delta\cZ^{(1)}_{Q^{ii}_{\text{F}\,ii}}+\cO(\g^4)\eqnsem
\end{equation}
the undeformed contributions can be inferred from the requirement that both contributions add up to one in the limit of the undeformed theory. 
In contrast to the first part, in this and the following chapter we work in the DR scheme described in section \ref{sec: subtleties} with the effective planar coupling constant $\g$ defined in \eqref{eq: effective planar coupling constant}.
Moreover, as the $N$-counting is more subtle here, we include a power of $g^{2\ell}$ in the definition of the $\ell$-loop contribution $\delta\cZ^{(\ell)}_{Q^{ii}_{\text{F}\,ii}}$ and similarly for all other quantities.

\renewcommand*{\thesubfigure}{\ensuremath{\mathrm{\Roman{subfigure}}}}
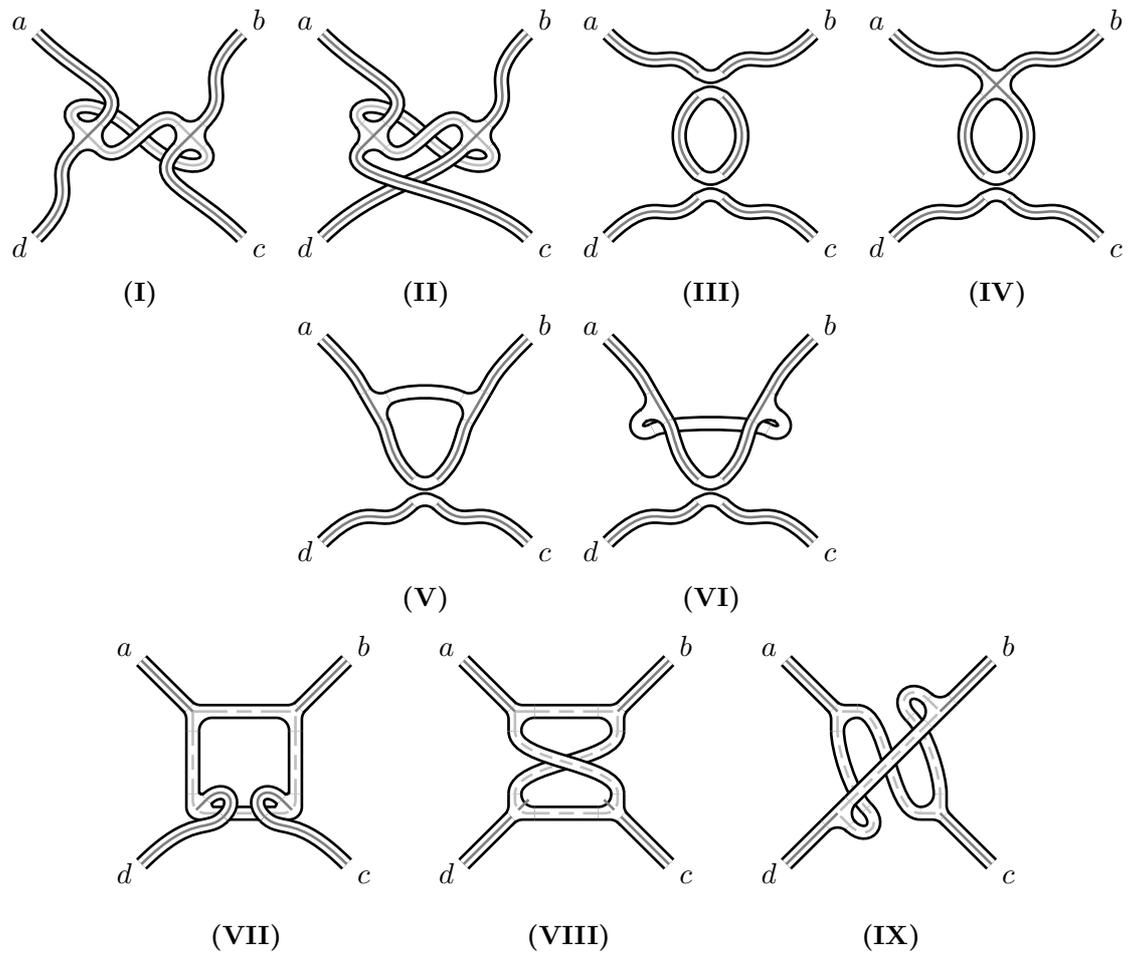
\begin{figure}[htbp]
\begin{center}
\begin{subfigure}[t]{0.240\textwidth}{%
\begin{pspicture}(-7.5,-7.5)(7.5,7.5)
\fourvertex{-3}{0}{45}
\fourvertex{3}{0}{45}
\setlength{\xa}{1.5\unit}
\addtolength{\xa}{-0.5\doublesep}
\addtolength{\xa}{-\linew}
\setlength{\xb}{4.5\unit}
\addtolength{\xb}{0.5\doublesep}
\addtolength{\xb}{\linew}
\setlength{\xc}{7.5\unit}
\addtolength{\xc}{-0.5\doublesep}
\addtolength{\xc}{-\linew}
\setlength{\xd}{10.5\unit}
\addtolength{\xd}{0.5\doublesep}
\addtolength{\xd}{\linew}
\setlength{\ya}{6\unit}
\addtolength{\ya}{0.5\doublesep}
\addtolength{\ya}{\linew}
\setlength{\yb}{3\unit}
\addtolength{\yb}{-0.5\doublesep}
\addtolength{\yb}{-\linew}
\psset{linecolor=black,doubleline=true}
\psbezier(-3.7071,0.7071)(-4.7071,1.7071)(-2.7071,2.7071)(0,0)
\psbezier(0,0)(2.7071,-2.7071)(4.7071,-1.7071)(3.7071,-0.7071)
\psset{linecolor=lightgray,doubleline=false}
\psbezier(-3.7071,0.7071)(-4.7071,1.7071)(-2.7071,2.7071)(0,0)
\psbezier(0,0)(2.7071,-2.7071)(4.7071,-1.7071)(3.7071,-0.7071)
\psset{linecolor=black,doubleline=true}
\psbezier(-2.2921,-0.7071)(-0.87868,-2.2071)(0.87868,2.12132)(2.2929,0.7071)
\psset{linecolor=lightgray,doubleline=false}
\psbezier(2.2929,0.7071)(0.87868,2.12132)(-0.87868,-2.2071)(-2.2921,-0.7071)
\psline(3.7071,-0.7071)(2.2929,0.7071)
\psline(-2.2921,-0.7071)(-3.7071,0.7071)
\psset{linecolor=black,doubleline=true}
\rput[B](-7,6.25){$a$}
\psbezier(-2.2921,0.7071)(-0.2921,2.7071)(-3,3)(-6,6)
\rput[B](7,6.25){$b$}
\psbezier(3.7071,0.7071)(5.7071,2.7071)(3,3)(6,6)
\rput[B](7,-7.125){$c$}
\psbezier(6,-6)(3,-3)(0.2921,-2.7071)(2.2921,-0.7071)
\rput[B](-7,-7.125){$d$}
\psbezier(-6,-6)(-3,-3)(-5.7071,-2.7071)(-3.7071,-0.7071)
\psset{linecolor=gray,doubleline=false}
\psbezier(-6,-6)(-3,-3)(-5.7071,-2.7071)(-3.7071,-0.7071)
\psline(-3.7071,-0.7071)(-2.2921,0.7071)
\psbezier(-2.2921,0.7071)(-0.2921,2.7071)(-3,3)(-6,6)
\psbezier(6,-6)(3,-3)(0.2921,-2.7071)(2.2921,-0.7071)
\psline(2.2921,-0.7071)(3.7071,0.7071)
\psbezier(3.7071,0.7071)(5.7071,2.7071)(3,3)(6,6)
\end{pspicture}
\caption{\phantom{.}}
\label{subfig: ss4s4}
}\end{subfigure}
\begin{subfigure}[t]{0.240\textwidth}{%
\begin{pspicture}(-7.5,-7.5)(7.5,7.5)
\fourvertex{-3}{0}{45}
\fourvertex{3}{0}{45}
\setlength{\xa}{1.5\unit}
\addtolength{\xa}{-0.5\doublesep}
\addtolength{\xa}{-\linew}
\setlength{\xb}{4.5\unit}
\addtolength{\xb}{0.5\doublesep}
\addtolength{\xb}{\linew}
\setlength{\xc}{7.5\unit}
\addtolength{\xc}{-0.5\doublesep}
\addtolength{\xc}{-\linew}
\setlength{\xd}{10.5\unit}
\addtolength{\xd}{0.5\doublesep}
\addtolength{\xd}{\linew}
\setlength{\ya}{6\unit}
\addtolength{\ya}{0.5\doublesep}
\addtolength{\ya}{\linew}
\setlength{\yb}{3\unit}
\addtolength{\yb}{-0.5\doublesep}
\addtolength{\yb}{-\linew}
\psset{linecolor=black,doubleline=true}
\psbezier(-3.7071,0.7071)(-4.7071,1.7071)(-2.7071,2.7071)(0,0)
\psbezier(0,0)(2.7071,-2.7071)(4.7071,-1.7071)(3.7071,-0.7071)
\psset{linecolor=lightgray,doubleline=false}
\psbezier(-3.7071,0.7071)(-4.7071,1.7071)(-2.7071,2.7071)(0,0)
\psbezier(0,0)(2.7071,-2.7071)(4.7071,-1.7071)(3.7071,-0.7071)
\psset{linecolor=black,doubleline=true}
\psbezier(-2.2921,-0.7071)(-0.87868,-2.2071)(0.87868,2.12132)(2.2929,0.7071)
\psset{linecolor=lightgray,doubleline=false}
\psbezier(2.2929,0.7071)(0.87868,2.12132)(-0.87868,-2.2071)(-2.2921,-0.7071)
\psline(3.7071,-0.7071)(2.2929,0.7071)
\psline(-2.2921,-0.7071)(-3.7071,0.7071)
\psset{linecolor=black,doubleline=true}
\rput[B](-7,6.25){$a$}
\psbezier(-2.2921,0.7071)(-0.2921,2.7071)(-3,3)(-6,6)
\rput[B](7,6.25){$b$}
\psbezier(3.7071,0.7071)(5.7071,2.7071)(3,3)(6,6)
\rput[B](-7,-7.125){$d$}
\psbezier(-6,-6)(-3,-3)(0.2921,-2.7071)(2.2921,-0.7071)
\psset{linecolor=gray,doubleline=false}
\psbezier(-6,-6)(-3,-3)(0.2921,-2.7071)(2.2921,-0.7071)
\psline(2.2921,-0.7071)(3.7071,0.7071)
\psbezier(3.7071,0.7071)(5.7071,2.7071)(3,3)(6,6)
\psset{linecolor=black,doubleline=true}
\rput[B](7,-7.125){$c$}
\psbezier(6,-6)(3,-3)(-5.7071,-2.7071)(-3.7071,-0.7071)
\psset{linecolor=gray,doubleline=false}
\psbezier(6,-6)(3,-3)(-5.7071,-2.7071)(-3.7071,-0.7071)
\psline(-3.7071,-0.7071)(-2.2921,0.7071)
\psbezier(-2.2921,0.7071)(-0.2921,2.7071)(-3,3)(-6,6)
\end{pspicture}
\caption{\phantom{.}}
\label{subfig: us4s4}
}\end{subfigure}
\begin{subfigure}[t]{0.240\textwidth}{%
\begin{pspicture}(-7.5,-7.5)(7.5,7.5)
\fourvertexdbltr{0}{-3}{45}
\fourvertexdbltr{0}{3}{45}
\setlength{\xa}{1.5\unit}
\addtolength{\xa}{-0.5\doublesep}
\addtolength{\xa}{-\linew}
\setlength{\xb}{4.5\unit}
\addtolength{\xb}{0.5\doublesep}
\addtolength{\xb}{\linew}
\setlength{\xc}{7.5\unit}
\addtolength{\xc}{-0.5\doublesep}
\addtolength{\xc}{-\linew}
\setlength{\xd}{10.5\unit}
\addtolength{\xd}{0.5\doublesep}
\addtolength{\xd}{\linew}
\setlength{\ya}{6\unit}
\addtolength{\ya}{0.5\doublesep}
\addtolength{\ya}{\linew}
\setlength{\yb}{3\unit}
\addtolength{\yb}{-0.5\doublesep}
\addtolength{\yb}{-\linew}
\psset{doubleline=true}
\psbezier(-0.7071,-2.2921)(-2.2071,-0.87868)(-2.2071,0.87868)(-0.7071,2.2929)
\psbezier(0.7071,2.2929)(2.2071,0.87868)(2.2071,-0.87868)(0.7071,-2.2921)
\rput[B](-7,6.25){$a$}
\psbezier(-0.7071,3.7071)(-2.7071,5.7071)(-3,3)(-6,6)
\rput[B](7,6.25){$b$}
\psbezier(0.7071,3.7071)(2.7071,5.7071)(3,3)(6,6)
\rput[B](7,-7.125){$c$}
\psbezier(6,-6)(3,-3)(2.7071,-5.7071)(0.7071,-3.7071)
\rput[B](-7,-7.125){$d$}
\psbezier(-0.7071,-3.7071)(-2.7071,-5.7071)(-3,-3)(-6,-6)
\psset{linecolor=gray,doubleline=false}
\psbezier(-6,-6)(-3,-3)(-2.7071,-5.7071)(-0.7071,-3.7071)
\psbezier(6,-6)(3,-3)(2.7071,-5.7071)(0.7071,-3.7071)
\psbezier(-0.7071,3.7071)(-2.7071,5.7071)(-3,3)(-6,6)
\psbezier(0.7071,3.7071)(2.7071,5.7071)(3,3)(6,6)
\psbezier(-0.7071,-2.2921)(-2.2071,-0.87868)(-2.2071,0.87868)(-0.7071,2.2929)
\psbezier(0.7071,-2.2921)(2.2071,-0.87868)(2.2071,0.87868)(0.7071,2.2929)
\end{pspicture}
\caption{\phantom{.}}
\label{subfig: ts4s4}
}\end{subfigure}
\begin{subfigure}[t]{0.240\textwidth}{%
\begin{pspicture}(-7.5,-7.5)(7.5,7.5)
\fourvertexdbltr{0}{-3}{45}
\fourvertex{0}{3}{45}
\setlength{\xa}{1.5\unit}
\addtolength{\xa}{-0.5\doublesep}
\addtolength{\xa}{-\linew}
\setlength{\xb}{4.5\unit}
\addtolength{\xb}{0.5\doublesep}
\addtolength{\xb}{\linew}
\setlength{\xc}{7.5\unit}
\addtolength{\xc}{-0.5\doublesep}
\addtolength{\xc}{-\linew}
\setlength{\xd}{10.5\unit}
\addtolength{\xd}{0.5\doublesep}
\addtolength{\xd}{\linew}
\setlength{\ya}{6\unit}
\addtolength{\ya}{0.5\doublesep}
\addtolength{\ya}{\linew}
\setlength{\yb}{3\unit}
\addtolength{\yb}{-0.5\doublesep}
\addtolength{\yb}{-\linew}
\psset{doubleline=true}
\psbezier(-0.7071,-2.2921)(-2.2071,-0.87868)(-2.2071,0.87868)(-0.7071,2.2929)
\psbezier(0.7071,2.2929)(2.2071,0.87868)(2.2071,-0.87868)(0.7071,-2.2921)
\rput[B](-7,6.25){$a$}
\psbezier(-0.7071,3.7071)(-2.7071,5.7071)(-3,3)(-6,6)
\rput[B](7,6.25){$b$}
\psbezier(0.7071,3.7071)(2.7071,5.7071)(3,3)(6,6)
\rput[B](7,-7.125){$c$}
\psbezier(6,-6)(3,-3)(2.7071,-5.7071)(0.7071,-3.7071)
\rput[B](-7,-7.125){$d$}
\psbezier(-0.7071,-3.7071)(-2.7071,-5.7071)(-3,-3)(-6,-6)
\psset{linecolor=gray,doubleline=false}
\psbezier(-6,-6)(-3,-3)(-2.7071,-5.7071)(-0.7071,-3.7071)
\psline(-0.7071,2.2929)(0.7071,3.7071)
\psline(0.7071,2.2929)(-0.7071,3.7071)
\psbezier(6,-6)(3,-3)(2.7071,-5.7071)(0.7071,-3.7071)
\psbezier(-0.7071,3.7071)(-2.7071,5.7071)(-3,3)(-6,6)
\psbezier(0.7071,3.7071)(2.7071,5.7071)(3,3)(6,6)
\psbezier(-0.7071,-2.2921)(-2.2071,-0.87868)(-2.2071,0.87868)(-0.7071,2.2929)
\psbezier(0.7071,-2.2921)(2.2071,-0.87868)(2.2071,0.87868)(0.7071,2.2929)
\end{pspicture}
\caption{\phantom{.}}
\label{subfig: tDs4s4}
}\end{subfigure}

\begin{subfigure}[t]{0.240\textwidth}{%
\begin{pspicture}(-7.5,-7.5)(7.5,7.5)
\fourvertexdbltr{0}{-3}{45}
\threevertex{-3}{2}{30}
\threevertex{3}{2}{150}
\setlength{\xa}{1.5\unit}
\addtolength{\xa}{-0.5\doublesep}
\addtolength{\xa}{-\linew}
\setlength{\xb}{4.5\unit}
\addtolength{\xb}{0.5\doublesep}
\addtolength{\xb}{\linew}
\setlength{\xc}{7.5\unit}
\addtolength{\xc}{-0.5\doublesep}
\addtolength{\xc}{-\linew}
\setlength{\xd}{10.5\unit}
\addtolength{\xd}{0.5\doublesep}
\addtolength{\xd}{\linew}
\setlength{\ya}{6\unit}
\addtolength{\ya}{0.5\doublesep}
\addtolength{\ya}{\linew}
\setlength{\yb}{3\unit}
\addtolength{\yb}{-0.5\doublesep}
\addtolength{\yb}{-\linew}
\psset{doubleline=true}
\psbezier(-2.134,2.5)(-1.268,3)(1.268,3)(2.134,2.5)
\rput[B](-7,6.25){$a$}
\psbezier(-0.7071,-2.2921)(-2.2071,-0.87868)(-2,0.268)(-2.5,1.134)
\psbezier(-3.5,2.866)(-4,3.732)(-4,4)(-6,6)
\rput[B](7,6.25){$b$}
\psbezier(0.7071,-2.2921)(2.2071,-0.87868)(2,0.268)(2.5,1.134)
\psbezier(3.5,2.866)(4,3.732)(4,4)(6,6)
\rput[B](7,-7.125){$c$}
\psbezier(6,-6)(3,-3)(2.7071,-5.7071)(0.7071,-3.7071)
\rput[B](-7,-7.125){$d$}
\psbezier(-0.7071,-3.7071)(-2.7071,-5.7071)(-3,-3)(-6,-6)
\psset{linecolor=gray,doubleline=false}
\psbezier(-6,-6)(-3,-3)(-2.7071,-5.7071)(-0.7071,-3.7071)
\psbezier(6,-6)(3,-3)(2.7071,-5.7071)(0.7071,-3.7071)
\psbezier(-0.7071,-2.2921)(-2.2071,-0.87868)(-2,0.268)(-2.5,1.134)
\psline(-2.5,1.134)(-3.5,2.866)
\psbezier(-3.5,2.866)(-4,3.732)(-4,4)(-6,6)
\psbezier(0.7071,-2.2921)(2.2071,-0.87868)(2,0.268)(2.5,1.134)
\psline(2.5,1.134)(3.5,2.866)
\psbezier(3.5,2.866)(4,3.732)(4,4)(6,6)
\end{pspicture}
\caption{\phantom{.}}
\label{subfig: s4gg2}
}\end{subfigure}
\begin{subfigure}[t]{0.240\textwidth}{%
\begin{pspicture}(-7.5,-7.5)(7.5,7.5)
\psset{doubleline=true}
\psbezier(-3.866,1.5)(-4.732,1)(-4.232,0.134)(-3.366,0.634)
\psbezier(-3.366,0.634)(-2.5,1.134)(2.5,1.134)(3.366,0.634)
\psbezier(3.866,1.5)(4.732,1)(4.232,0.134)(3.366,0.634)
\fourvertexdbltr{0}{-3}{45}
\threevertex{-3}{2}{210}
\threevertex{3}{2}{-30}
\setlength{\xa}{1.5\unit}
\addtolength{\xa}{-0.5\doublesep}
\addtolength{\xa}{-\linew}
\setlength{\xb}{4.5\unit}
\addtolength{\xb}{0.5\doublesep}
\addtolength{\xb}{\linew}
\setlength{\xc}{7.5\unit}
\addtolength{\xc}{-0.5\doublesep}
\addtolength{\xc}{-\linew}
\setlength{\xd}{10.5\unit}
\addtolength{\xd}{0.5\doublesep}
\addtolength{\xd}{\linew}
\setlength{\ya}{6\unit}
\addtolength{\ya}{0.5\doublesep}
\addtolength{\ya}{\linew}
\setlength{\yb}{3\unit}
\addtolength{\yb}{-0.5\doublesep}
\addtolength{\yb}{-\linew}
\psset{doubleline=true}
\rput[B](-7,6.25){$a$}
\psbezier(-0.7071,-2.2921)(-2.2071,-0.87868)(-2,0.268)(-2.5,1.134)
\psbezier(-3.5,2.866)(-4,3.732)(-4,4)(-6,6)
\rput[B](7,6.25){$b$}
\psbezier(0.7071,-2.2921)(2.2071,-0.87868)(2,0.268)(2.5,1.134)
\psbezier(3.5,2.866)(4,3.732)(4,4)(6,6)
\rput[B](7,-7.125){$c$}
\psbezier(6,-6)(3,-3)(2.7071,-5.7071)(0.7071,-3.7071)
\rput[B](-7,-7.125){$d$}
\psbezier(-0.7071,-3.7071)(-2.7071,-5.7071)(-3,-3)(-6,-6)
\psset{linecolor=gray,doubleline=false}
\psbezier(-6,-6)(-3,-3)(-2.7071,-5.7071)(-0.7071,-3.7071)
\psbezier(6,-6)(3,-3)(2.7071,-5.7071)(0.7071,-3.7071)
\psbezier(-0.7071,-2.2921)(-2.2071,-0.87868)(-2,0.268)(-2.5,1.134)
\psline(-2.5,1.134)(-3.5,2.866)
\psbezier(-3.5,2.866)(-4,3.732)(-4,4)(-6,6)
\psbezier(0.7071,-2.2921)(2.2071,-0.87868)(2,0.268)(2.5,1.134)
\psline(2.5,1.134)(3.5,2.866)
\psbezier(3.5,2.866)(4,3.732)(4,4)(6,6)
\end{pspicture}
\caption{\phantom{.}}
\label{subfig: s4gg3}
}\end{subfigure}

\begin{subfigure}[t]{0.272\textwidth}{%
\begin{pspicture}(-8.5,-8.5)(8.5,8.5)
\recthreevertex{-3}{3}{135}
\recthreevertex{3}{3}{45}
\recthreevertexoneside{3}{-3}{-45}
\recthreevertexoneside{-3}{-3}{-135}
\psset{linecolor=black,doubleline=true}
\psline(-1.85,3)(1.85,3)
\psline(3,1.85)(3,-1.85)
\psline(1.85,-3)(-1.85,-3)
\psline(-3,-1.85)(-3,1.85)
\psset{linecolor=lightgray,doubleline=false,linestyle=dashed}
\psline(-3,1.85)(-3,3)
\psline(-3,3)(-1.85,3)
\psline(-1.85,3)(1.85,3)
\psline(1.85,3)(3,3)(3,1.85)
\psline(3,1.85)(3,-1.85)
\psline[linearc=\linearc](3,-1.85)(3,-3)(1.85,-3)
\psline(1.85,-3)(-1.85,-3)
\psline[linearc=\linearc](-1.85,-3)(-3,-3)(-3,-1.85)
\psline(-3,-1.85)(-3,1.85)

\psset{linecolor=black,doubleline=true,linestyle=solid}

\rput[B](-7,6.25){$a$}
\psline(-3.2,3.2)(-6,6)
\rput[B](7,6.25){$b$}
\psline(3.2,3.2)(6,6)
\rput[B](7,-7.125){$c$}
\psbezier(2.2,-2.2)(1.2,-1.2)(0.2,-2.2)(1.2,-3.2)
\psbezier(1.2,-3.2)(2.2,-4.2)(3,-3)(6,-6)
\rput[B](-7,-7.125){$d$}
\psbezier(-2.2,-2.2)(-1.2,-1.2)(-0.2,-2.2)(-1.2,-3.2)
\psbezier(-1.2,-3.2)(-2.2,-4.2)(-3,-3)(-6,-6)
\psset{linecolor=gray,doubleline=false}
\psline(-3,3)(-3.2,3.2)
\psline(-3.2,3.2)(-6,6)
\psline(3,3)(3.2,3.2)
\psline(3.2,3.2)(6,6)
\psline(2.8,-2.8)(2.2,-2.2)
\psbezier(2.2,-2.2)(1.2,-1.2)(0.2,-2.2)(1.2,-3.2)
\psbezier(1.2,-3.2)(2.2,-4.2)(3,-3)(6,-6)
\psline(-2.8,-2.8)(-2.2,-2.2)
\psbezier(-2.2,-2.2)(-1.2,-1.2)(-0.2,-2.2)(-1.2,-3.2)
\psbezier(-1.2,-3.2)(-2.2,-4.2)(-3,-3)(-6,-6)
\end{pspicture}
\caption{\phantom{.}}
\label{subfig: fbox1}
}\end{subfigure}
\begin{subfigure}[t]{0.272\textwidth}{%
\begin{pspicture}(-8.5,-8.5)(8.5,8.5)
\recthreevertex{-3}{3}{135}
\recthreevertex{3}{3}{45}
\recthreevertex{3}{-3}{-45}
\recthreevertex{-3}{-3}{-135}
\psset{linecolor=black,doubleline=true}
\psline(-1.85,3)(1.85,3)
\psbezier(3,1.85)(3,0)(-3,0)(-3,-1.85)
\psline(-1.85,-3)(1.85,-3)
\psset{linecolor=lightgray,doubleline=false,linestyle=dashed}
\psline(-3,1.85)(-3,3)
\psline(-3,3)(-1.85,3)
\psline(-1.85,3)(1.85,3)
\psline(1.85,3)(3,3)(3,1.85)
\psbezier(3,1.85)(3,0)(-3,0)(-3,-1.85)
\psline[linearc=\linearc](-3,-1.85)(-3,-3)(-1.85,-3)
\psline(-1.85,-3)(1.85,-3)
\psline[linearc=\linearc](1.85,-3)(3,-3)(3,-1.85)
\psset{linecolor=black,doubleline=true,linestyle=solid}
\psbezier(3,-1.85)(3,0)(-3,0)(-3,1.85)
\psset{linecolor=lightgray,doubleline=false,linestyle=dashed}
\psbezier(3,-1.85)(3,0)(-3,0)(-3,1.85)
\psset{linecolor=black,doubleline=true,linestyle=solid}

\rput[B](-7,6.25){$a$}
\psline(-3.2,3.2)(-6,6)
\rput[B](7,6.25){$b$}
\psline(3.2,3.2)(6,6)
\rput[B](7,-7.125){$c$}
\psline(-3.2,-3.2)(-6,-6)
\rput[B](-7,-7.125){$d$}
\psline(3.2,-3.2)(6,-6)
\psset{linecolor=gray,doubleline=false}
\psline(-3,3)(-3.2,3.2)
\psline(-3.2,3.2)(-6,6)
\psline(3,3)(3.2,3.2)
\psline(3.2,3.2)(6,6)
\psline(2.8,-2.8)(2.2,-2.2)
\psline(-3.2,-3.2)(-6,-6)
\psline(-2.8,-2.8)(-2.2,-2.2)
\psline(3.2,-3.2)(6,-6)
\end{pspicture}
\caption{\phantom{.}}
\label{subfig: fbox2}
}\end{subfigure}
\begin{subfigure}[t]{0.272\textwidth}{%
\begin{pspicture}(-8.5,-8.5)(8.5,8.5)
\recthreevertex{-3}{3}{135}
\threevertex{3}{3}{135}
\recthreevertex{3}{-3}{-45}
\threevertex{-3}{-3}{-45}
\psset{linecolor=black,doubleline=true}
\psbezier(-1.85,3)(0,3)(0,-3)(1.85,-3)
\psbezier(3,-1.85)(3,-0.85)(2.2929,1.7071)(1.2929,2.7071)
\psbezier(1.2929,2.7071)(0.2929,3.7071)(1.2929,4.7071)(2.2929,3.7071)
\psbezier(-2.2929,-3.7071)(-1.2929,-4.7071)(-0.2929,-3.7071)(-1.2929,-2.7071)
\psbezier(-1.2929,-2.7071)(-2.2929,-1.7071)(-3,0.85)(-3,1.85)

\rput[B](-7,6.25){$a$}
\psline(-3.2,3.2)(-6,6)
\rput[B](7,6.25){$b$}
\psline(3.7071,3.7071)(6,6)
\rput[B](7,-7.125){$c$}
\psline(3.2,-3.2)(6,-6)
\rput[B](-7,-7.125){$d$}
\psline(-3.7071,-3.7071)(-6,-6)
\psset{linecolor=gray,doubleline=false}
\psline(-3,3)(-3.2,3.2)
\psline(-3.2,3.2)(-6,6)
\psline(3,3)(3.7071,3.7071)
\psline(3.7071,3.7071)(6,6)
\psline(3,-3)(3.2,-3.2)
\psline(3.2,-3.2)(6,-6)
\psline(-3,-3)(-3.7071,-3.7071)
\psline(-3.7071,-3.7071)(-6,-6)
\psset{linecolor=lightgray,doubleline=false,linestyle=dashed}
\psline(-3,1.85)(-3,3)(-1.85,3)
\psbezier(-1.85,3)(0,3)(0,-3)(1.85,-3)
\psline(1.85,-3)(3,-3)(3,-1.85)
\psbezier(3,-1.85)(3,-0.85)(2.2929,1.7071)(1.2929,2.7071)
\psbezier(1.2929,2.7071)(0.2929,3.7071)(1.2929,4.7071)(2.2929,3.7071)
\psline(2.2929,3.7071)(3,3)(2.2929,2.2929)
\psline(-2.2929,-2.2929)(-3,-3)(-2.2929,-3.7071)
\psbezier(-2.2929,-3.7071)(-1.2929,-4.7071)(-0.2929,-3.7071)(-1.2929,-2.7071)
\psbezier(-1.2929,-2.7071)(-2.2929,-1.7071)(-3,0.85)(-3,1.85)

\psset{linecolor=black,doubleline=true,linestyle=solid}
\psline(-2.2929,-2.2929)(2.2929,2.2929)
\psset{linecolor=lightgray,doubleline=false,linestyle=dashed}
\psline(-2.2929,-2.2929)(2.2929,2.2929)
\end{pspicture}
\caption{\phantom{.}}
\label{subfig: fbox3}
}\end{subfigure}
\caption{Deformed 1PI diagrams contributing to the renormalisation constant $\delta\cZ^{(1)}_{Q^{ii}_{\text{F}\,ii}}$ at leading order in $N$.
The colour structure of the individual terms is depicted in double-line notation. Scalars are depicted by solid central lines, fermions by dashed central lines and gauge fields by vanishing central lines.}
\label{fig: deformed diagrams contributions to renormalisation}
\end{center}
\end{figure}

Up to reflections, all one-particle-irreducible (1PI) Feynman diagrams that contain deformed interactions and contribute to $\delta\cZ^{(1)}_{Q^{ii}_{\text{F}\,ii}}$ at leading order in $N$ are shown in figure \ref{fig: deformed diagrams contributions to renormalisation}.
Note that the additional couplings present for gauge group $\U{N}$ in comparison to gauge group $\SU{N}$ contribute only at subleading order in $N$; the result is thus independent of considering either $\U{N}$ or $\SU{N}$.
These Feynman diagrams can be evaluated using the Feynman rules given in \cite{Fokken:2013aea}; see \cite{Fokken:2013aea} for the details of this calculation.
In order to regulate the IR divergences, it is sufficient to direct one off-shell momentum $p$ in a suitably chosen way through the diagram, such that only the scale $p^2$ occurs in the integrals.%
\footnote{In particular, we are not calculating the full double-trace contribution to the amplitude here. We only need to extract its UV divergence.} 
The total leading contributions of deformed diagrams with only scalar, gluonic and fermionic intermediate states, respectively, are
\begin{equation}
\label{eq: subsums of deformed contributions}
\begin{aligned}
\Kop\left[(1+\Rop_|)[\,(\subref{subfig: ss4s4})+(\subref{subfig: us4s4})\,]+(\subref{subfig: ts4s4})+2\,(\subref{subfig: tDs4s4})\right]
&=8g_\YM^4 \Kop[I_1]\big(
\cos^22\gamma_i^+\cos^22\gamma_i^-\\
&\hphantom{{}={}8g_\YM^4\Kop[I_1]\big(}
+(Q^{ii}_{\text{F}\,ii})^2+Q^{ii}_{\text{F}\,ii}\big)\big(ab\big)\big(cd\big)
\eqncom\\
2\Kop\left[(\subref{subfig: s4gg2})+(\subref{subfig: s4gg3})\right]
&=8\alpha g_\YM^4\Kop[I_1]Q^{ii}_{\text{F}\,ii}\big(ab\big)\big(cd\big)
\eqncom\\
\Kop\left[(1+\Rop_|)(\subref{subfig: fbox3})\right]
&=-8g_\YM^4\Kop[I_1](\cos2\gamma_i^++\cos2\gamma_i^-)
\big(ab\big)\big(cd\big)
\eqncom
\end{aligned}
\end{equation}
where $\alpha$ is the gauge-fixing parameter, the operator $\Kop$ extracts the poles in $\peps$ and $\Rop_|$ denotes the reflection of the diagram at the vertical 
axis while keeping the original order of the indices.
Moreover, we have abbreviated 
\begin{equation}
 \label{eq: abbreviation for gauge group generators}
 (a_1\cdots a_n)\equiv \tr(\T^{a_1}\cdots \T^{a_n}) \eqndot
\end{equation}
The factors $2$ in front of (\subref{subfig: tDs4s4}), (\subref{subfig: s4gg2}) and (\subref{subfig: s4gg3}) stem from the reflections of these diagrams at the horizontal axis, which yield identical contributions. 
The total contribution of 1PI diagrams to $\delta\cZ^{(1)}_{Q^{ii}_{\text{F}\,ii}}$, which is given by the sum of all terms in \eqref{eq: subsums of deformed contributions} and the undeformed contribution, has to vanish in the limit of $\cN=4$ SYM theory. Hence, we find that the undeformed 1PI diagrams at leading order contribute 
\begin{equation}
 8g_\YM^4\Kop[I_1]\big(ab\big)\big(cd\big)\eqndot
\end{equation}
The occurring integral and its UV divergence are given by
\begin{equation}
\label{eq: def I one}
 I_1=\frac{\e^{-\gamma_{\text{E}}\peps}}{(4\pi)^{2-\peps}}\FDinline[bubble]\eqncom \quad \Kop[I_1]=\frac{1}{(4\pi)^2\peps}\eqncom
\end{equation}
where the bubble integral was defined in \eqref{eq: oneloopint:bubble}. As already mentioned, we use the DR scheme with minimal subtraction in the effective planar coupling constant $\g$ in this and the following chapter; hence the $\peps$-dependent prefactor in \eqref{eq: def I one}.

The counterterm $\delta Q^{ii}_{\text{F}\,ii}$ appears in the action as 
\begin{equation}
\label{eq: counterterm}
-\frac{g_\YM^2}{N}(Q^{ii}_{\text{F}\,ii}+\delta Q^{ii}_{\text{F}\,ii})\tr(\bar\phi^i\bar\phi^i)\tr(\phi_i\phi_i)
\end{equation} 
and has to cancel the UV divergences of the 1PI diagrams.
At one-loop order and leading order in $N$, it hence reads
\begin{equation}\label{eq: counterterm at one-loop order}
\begin{aligned}
\delta Q^{ii\,(1)}_{\text{F}\,ii}
&=-\Kop\left.\left[
\qvertr[
\fmfiv{label=$\scriptstyle i a$,l.dist=2}{vloc(__v1)}
\fmfiv{label=$\scriptstyle i b$,l.dist=2}{vloc(__v2)}
\fmfiv{label=$\scriptstyle i c$,l.dist=2}{vloc(__v3)}
\fmfiv{label=$\scriptstyle i d$,l.dist=2}{vloc(__v4)}
\fmfcmd{fill(fullcircle scaled 10 shifted vloc(__vc1)) withcolor 0.2black;}
]{plain_rar,l.side=left,l.dist=2,l.dist=2}{plain_rar,l.side=left,l.dist=2,l.dist=2}{plain_ar,l.side=left,l.dist=2,l.dist=2}{plain_ar,l.side=left,l.dist=2,l.dist=2}{}{}{}{}
\right]\right|_{-4\frac{g_\YM^2}{N}(ab)(cd)}
\\
&
=\frac{2g^2}{\peps}\big(4\sin^2\gamma_i^+\sin^2\gamma_i^-
+(Q^{ii}_{\text{F}\,ii})^2-(1+\alpha)Q^{ii}_{\text{F}\,ii}\big)
\eqncom
\end{aligned}
\end{equation}
where the vertical bar prescribes to take the coefficient of the specified term while the factor $4$ in this term arises when taking functional derivatives with respect to the fields in order to derive the Feynman rules.

In addition to the 1PI diagrams, also self-energy diagrams contribute to the renormalisation constant $\cZ_{Q^{ii}_{\text{F}\,ii}}$. 
Their total contribution at one-loop order is
\begin{equation}
\label{eq: self-energy contributions}
\begin{aligned}
&-\frac 12\Kop\left.\left[
\qvertr[
\fmfiv{label=$\scriptstyle i a$,l.dist=2}{vloc(__v1)}
\fmfiv{label=$\scriptstyle i b$,l.dist=2}{vloc(__v2)}
\fmfiv{label=$\scriptstyle i c$,l.dist=2}{vloc(__v3)}
\fmfiv{label=$\scriptstyle i d$,l.dist=2}{vloc(__v4)}
\fmfcmd{pair vert[]; vert1 = vloc(__vc1);}
\fmfi{plain_ar}{vert1--vi1}
\fmfi{plain_ar}{vo1--vloc(__v1)}
\fmfcmd{fill(fullcircle scaled 7.5 shifted vm1) withcolor 0.2black;}
]{phantom}{plain_rar}{plain_ar}{plain_ar}{}{}{}{}
+
\qvertr[
\fmfiv{label=$\scriptstyle i a$,l.dist=2}{vloc(__v1)}
\fmfiv{label=$\scriptstyle i b$,l.dist=2}{vloc(__v2)}
\fmfiv{label=$\scriptstyle i c$,l.dist=2}{vloc(__v3)}
\fmfiv{label=$\scriptstyle i d$,l.dist=2}{vloc(__v4)}
\fmfcmd{pair vert[]; vert1 = vloc(__vc1);}
\fmfi{plain_ar}{vert1--vi2}
\fmfi{plain_ar}{vo2--vloc(__v2)}
\fmfcmd{fill(fullcircle scaled 7.5 shifted vm2) withcolor 0.2black;}
]{plain_rar}{phantom}{plain_ar}{plain_ar}{}{}{}{}
+
\qvertr[
\fmfiv{label=$\scriptstyle i a$,l.dist=2}{vloc(__v1)}
\fmfiv{label=$\scriptstyle i b$,l.dist=2}{vloc(__v2)}
\fmfiv{label=$\scriptstyle i c$,l.dist=2}{vloc(__v3)}
\fmfiv{label=$\scriptstyle i d$,l.dist=2}{vloc(__v4)}
\fmfcmd{pair vert[]; vert1 = vloc(__vc1);}
\fmfi{plain_rar}{vert1--vi3}
\fmfi{plain_rar}{vo3--vloc(__v3)}
\fmfcmd{fill(fullcircle scaled 7.5 shifted vm3) withcolor 0.2black;}
]{plain_rar}{plain_rar}{phantom}{plain_ar}{}{}{}{}
+
\qvertr[
\fmfiv{label=$\scriptstyle i a$,l.dist=2}{vloc(__v1)}
\fmfiv{label=$\scriptstyle i b$,l.dist=2}{vloc(__v2)}
\fmfiv{label=$\scriptstyle i c$,l.dist=2}{vloc(__v3)}
\fmfiv{label=$\scriptstyle i d$,l.dist=2}{vloc(__v4)}
\fmfcmd{pair vert[]; vert1 = vloc(__vc1);}
\fmfi{plain_rar}{vert1--vi4}
\fmfi{plain_rar}{vo4--vloc(__v4)}
\fmfcmd{fill(fullcircle scaled 7.5 shifted vm4) withcolor 0.2black;}
]{plain_rar}{plain_rar}{plain_ar}{phantom}{}{}{}{}\right]\right|_{-4\frac{g_\YM^2}{N}(ab)(cd)}\\
&=
-2\Kop\Big[
\settoheight{\eqoff}{$\times$}%
\setlength{\eqoff}{0.5\eqoff}%
\addtolength{\eqoff}{-7.5\unitlength}%
\raisebox{\eqoff}{%
\fmfframe(1,0)(1,0){%
\begin{fmfchar*}(20,15)
\fmfleft{v1}
\fmfright{v2}
\fmf{plain_ar}{v1,vl}
\fmf{phantom}{vl,vr}
\fmf{plain_ar}{vr,v2}
\fmffreeze
\fmfposition
\fmfcmd{pair vert[]; vert1 = vloc(__vl);}
\vacpolp[1]{vert1--vloc(__vr)}
\end{fmfchar*}}}
\Big]\frac{1}{p^2}\!\!\!
\left.
\qvertr[
\fmfiv{label=$\scriptstyle i a$,l.dist=2}{vloc(__v1)}
\fmfiv{label=$\scriptstyle i b$,l.dist=2}{vloc(__v2)}
\fmfiv{label=$\scriptstyle i c$,l.dist=2}{vloc(__v3)}
\fmfiv{label=$\scriptstyle i d$,l.dist=2}{vloc(__v4)}
]{plain_rar}{plain_rar}{plain_ar}{plain_ar}{}{}{}{}\right|_{-4\frac{g_\YM^2}{N}(ab)(cd)}
\!\!\!=-2\delta_{\phi_i}^{\SU{N},(1)} Q^{ii}_{\text{F}\,ii}
=2(1+\alpha)\frac{\g^2}{\peps}Q^{ii}_{\text{F}\,ii}
\eqncom
\end{aligned}
\end{equation}
where $\delta_{\phi_i}^{\SU{N},(1)}$ denotes the one-loop self energy of the $\SU{N}$ components, which is presented in appendix \ref{appsec: one-loop self energies}.\footnote{The self-energy diagrams of the $\U{1}$ component contribute only to couplings in which at least one trace factor of length one occurs.}

Hence, the one-loop renormalisation constant $\delta\cZ_{Q^{ii}_{\text{F}\,ii}}^{(1)}$ is given by 
\begin{equation}
\label{eq: renormalisation constant of the double-trace coupling}
\begin{aligned}
\delta\cZ_{Q^{ii}_{\text{F}\,ii}}^{(1)}&=\frac{\delta Q^{ii\,(1)}_{\text{F}\,ii}}{Q^{ii}_{\text{F}\,ii}}-2\delta_{\phi_i}^{\SU{N},(1)} 
=2\frac{\g^2}{\peps}\frac{1}{Q^{ii}_{\text{F}\,ii}}\big(4\sin^2\gamma_i^+\sin^2\gamma_i^-
+(Q^{ii}_{\text{F}\,ii})^2\big)
\eqncom
\end{aligned}
\end{equation}
cf.\ appendix \ref{appsec: renormalisation group equations}.

\section{Beta function}
\label{sec: beta function}

From the renormalisation constant \eqref{eq: renormalisation constant of the double-trace coupling}, the one-loop beta function is determined as 
\begin{equation}
\label{eq: beta function result}
\begin{aligned}
\beta_{Q_{\text{F}\,ii}^{ii}}^{(1)}
=Q_{\text{F}\,ii}^{ii}\peps g_\YM \frac{\partial}{\partial g_\YM}\delta\cZ_{Q^{ii}_{\text{F}\,ii}}^{(1)}
=4 \g^2
\left(4\sin^2\gamma_i^+\sin^2\gamma_i^-
+(Q^{ii}_{\text{F}\,ii})^2\right)
\eqncom
\end{aligned}
\end{equation}
where the effective planar coupling $\g^2$ is defined in \eqref{eq: effective planar coupling constant}; 
see appendix \ref{appsec: renormalisation group equations} for details.%
\footnote{Recall that, in contrast to the first part, in this as well as the following chapter we are including a factor of $\g^{2\ell}$ in the definition of $\ell$-loop expressions.}
The beta function \eqref{eq: beta function result} is non-zero unless $\gamma_j=\gamma_k\pm n\pi$ with $n\in \ZZ$.
Hence, for generic deformation parameters and generic $N$, the $\gamma_i$-deformation is not conformally invariant.%
\footnote{We have worked in the large $N$ expansion, where different orders in $N$ are assumed to be linearly independent. 
Our arguments cannot exclude cancellations between different orders for some non-generic finite $N$. The fact that $N$ is an integer does, however, severely restrict this possibility.
}
Note that this running double-trace coupling affects the planar spectrum of anomalous dimensions via the finite-size effect of prewrapping, as we will explicitly demonstrate in the next chapter. Thus, conformal invariance is broken even in the planar limit.%
\footnote{The result \eqref{eq: beta function result} agrees with the unpublished result of \cite{DymarskyRoiban}; we thank Radu Roiban for communication on this point. It was also later confirmed in \cite{Jin:2013baa}. However, note that the author of \cite{Jin:2013baa} nevertheless calls the $\gamma_i$-deformation conformally invariant in the planar limit.}%
$^,$%
\footnote{Note that the breakdown of conformal invariance cannot be detected using the analysis based on D-instantons in \cite{Ferrari:2013pq}, neither in the $\gamma_i$-deformation nor in the $\beta$-deformation with gauge group $\U{N}$.
First, the double-trace couplings seem to be discarded by the formalism as they are formally suppressed in $\frac{1}{N}$.
Second, the full geometry is only probed by the instantons at linear order in the deformation parameters $\gamma_i$, while the breakdown of conformal invariance occurs at quadratic order, cf.\ \eqref{eq: beta function result}. 
}
Recall that in our analysis we have been looking for fixed points as functions of the (perturbative) coupling $\g$, i.e.\ fixed lines. We cannot exclude that  isolated Banks-Zaks fixed points \cite{Banks:1981nn} exist as some finite value of $\g$.
Since the beta function \eqref{eq: beta function result} is always positive for generic values of the deformation parameters, the running coupling $Q_{\text{F}\,ii}^{ii}$ moreover has a Landau pole, which makes the theory instable.%
\footnote{In the later article \cite{Jin:2013baa}, also the flow of the deformation parameters $\gamma_i$ was analysed and it was argued that the Landau pole can be avoided for $\gamma_i^-=\cO(1/N^2)$.}

As the AdS/CFT correspondence relates the conformal invariance of the gauge theory to the $\AdS_5$ factor in the string-theory background, several different scenarios are possible.
\begin{enumerate}
 \item The string background is instable due to the emergence of closed string tachyons. In the setup of non-supersymmetric orbifolds, these occur and were related to the running multi-trace couplings in the corresponding gauge theories \cite{Dymarsky:2005nc}. Tachyons were also found in $\gamma_i$-deformed flat space \cite{Spradlin:2005sv}, but could not yet be related to instabilities of the $\gamma_i$-deformation.
 \item String corrections deform the $\AdS_5$ factor in the Frolov background \cite{Frolov:2005dj}.
 \item The $\AdS_5$ factor is exact but the gauge theory dual to this background has not yet been found. All natural candidates are, however, excluded by our analysis. It could be that this theory does not even have a Lagrangian description with the field content of $\mathcal{N}=4$ SYM theory. 
 \item The deformation parameters $\gamma_i$ are functions of the effective planar coupling $\g$ which coincide for $\g=0$. 
 Similar finite functions of the couplings were found in ABJ(M) theory \cite{Aharony:2008ug,Aharony:2008gk} and in the interpolating quiver gauge theory of \cite{Gadde:2009dj}, see \cite{Minahan:2009aq,Minahan:2009wg,Leoni:2010tb} and \cite{Pomoni:2011jj}, respectively.
 This possibility is hard to exclude via perturbation theory as $\gamma_i-\gamma_j$ might always be of one loop-order higher than the one currently analysed. 
\end{enumerate}
It would be very interesting to determine which of these possibilities is the case.%
\footnote{For a recent interpretation of double-trace couplings in the AdS/CFT dictionary, see \cite{Aharony:2015afa}.
}

\chapter{Anomalous dimensions in the \texorpdfstring{$\gamma_i$}{gamma-i}-deformation}
\label{chap: anomalous dimensions}

In this chapter, we calculate the planar anomalous dimensions of the operators $\cO_L=\tr(\phi_i^L)$ 
at $L$-loop order via Feynman diagrams.
For $L\geq3$, we find a perfect match with the predictions of integrability.
For $L=2$, where the integrability-based result diverges, we obtain a finite rational answer. 
Via the prewrapping effect, it depends on the running double-trace coupling $Q_{\text{F}\,ii}^{ii}$ whose non-vanishing beta function we have calculated in the last chapter, and hence on the renormalisation scheme.
This explicitly demonstrates that conformal invariance is broken even in the planar limit.

As we show in section \ref{sec: classification}, the calculation can be vastly simplified by using relation \eqref{eq: diagram relation} between Feynman diagrams in the $\gamma_i$-deformation and the undeformed theory as well as the fact that the operators $\cO_L=\tr(\phi_i^L)$ are protected in the latter. 
In the case of $L\geq3$, only four Feynman diagrams have to be evaluated; the corresponding calculation is shown in section \ref{sec: L geq 3}. 
The calculation for $L=2$ is performed in section \ref{sec: L eq 2}.
We briefly review the foundations of the renormalisation theory used in this chapter in appendix \ref{app: deformed theories}.

This chapter is based on results first published in \cite{Fokken:2014soa}.

\section{Classification of diagrams}
\label{sec: classification}

In the previous chapter, we have exploited the fact that the double-trace coupling \eqref{eq: running couplings} is not renormalised in $\mathcal{N}=4$ SYM theory to simplify the calculation of its renormalisation constant and beta function in the $\gamma_i$-deformation.
Similarly, we can exploit the fact that the operators $\cO_L=\tr(\phi_i^L)$ are protected, i.e.\ not renormalised, in the undeformed theory to simplify the calculation of their renormalisation constant $\cZ_{\cO_L}$ and anomalous dimension $\gamma_{\cO_L}$ in the $\gamma_i$-deformation.
This means we only need to calculate diagrams that are affected by the deformation.

According to the discussion in section \ref{sec: relation}, two classes of diagrams contribute to the operator renormalisation. 
In the first class, the subdiagram of elementary interactions is of single-trace type. It is either a connected diagram with the structure $\tr((\phi_i)^{R}(\bar{\phi}^i)^{R})$, $R\leq L$, or a product of disconnected factors with structure $\tr((\phi_i)^{R_j}(\bar{\phi}^i)^{R_j})$, $\sum_j R_j\leq L$. Applying relation \eqref{eq: diagram relation} to these structures, we find that the diagrams in the first class are independent of the deformation, as the occurring phase factor is $\Phi(\phi_i\ast \phi_i\ast\cdots\ast\phi_i\ast \bar\phi^i\ast \bar\phi^i\cdots \ast \bar\phi^i)=1$.
In the second class of diagrams, the subdiagram of elementary interactions is of double-trace type with structure $\tr((\phi_i)^{L})\tr((\bar{\phi}^i)^{L})$. 
This structure can arise from the finite-size effects of wrapping and prewrapping. 
According to the criteria developed in section \ref{sec: prewrapping}, prewrapping cannot affect the operators $\cO_L=\tr(\phi_i^L)$ for $L\geq 3$.
It starts to affect the operator $\cO_2=\tr(\phi_i\phi_i)$ at the critical prewrapping order $\ell=L-1=1$ and, moreover, stems from the deformation-dependent double-trace coupling \eqref{eq: running couplings} alone.
The wrapping effect, on the other hand, affects all $\cO_L=\tr(\phi_i^L)$ starting at the critical wrapping order $\ell=L$.

We can further decompose the set of wrapping diagrams in two subclasses as
\setlength{\unit}{0.2cm}
\begin{equation}
\label{eq: decomposition of wrapping diagrams}
\settoheight{\eqoff}{$+$}%
\setlength{\eqoff}{0.5\eqoff}%
\addtolength{\eqoff}{-10.5\unit}%
\raisebox{\eqoff}{%
\begin{pspicture}(-0.44444,-5.33333)(19.11111,13.33333)
\rput[r](1.33333,8){$\scriptstyle \phi_i$}
\rput[r](1.33333,0){$\scriptstyle \phi_i$}
\rput[l](12.8888,0){$\scriptstyle \bar\phi^i$}
\rput[l](12.8888,8){$\scriptstyle \bar\phi^i$}
\uinex{2}{9}%
\iinex{2}{6}%
\dinex{2}{0}%
\setlength{\xa}{4\unit}
\addtolength{\xa}{-0.5\dlinewidth}
\setlength{\ya}{9\unit}
\addtolength{\ya}{0.5\dlinewidth}
\setlength{\xb}{6.5\unit}
\addtolength{\xb}{-0.5\dlinewidth}
\setlength{\yb}{14\unit}
\addtolength{\yb}{0.5\dlinewidth}
\setlength{\xc}{9.5\unit}
\addtolength{\xc}{0.5\dlinewidth}
\setlength{\yc}{11\unit}
\addtolength{\yc}{-0.5\dlinewidth}
\setlength{\xd}{12\unit}
\addtolength{\xd}{0.5\dlinewidth}
\setlength{\yd}{0\unit}
\addtolength{\yd}{-0.5\dlinewidth}
\setlength{\xe}{17.5\unit}
\addtolength{\xe}{-0.5\dlinewidth}
\setlength{\ye}{-2\unit}
\addtolength{\ye}{0.5\dlinewidth}
\setlength{\xf}{20.5\unit}
\addtolength{\xf}{0.5\dlinewidth}
\setlength{\yf}{-5\unit}
\addtolength{\yf}{-0.5\dlinewidth}
\setlength{\yg}{8\unit}
\addtolength{\yg}{-0.5\dlinewidth}
\setlength{\yh}{1\unit}
\addtolength{\yh}{0.5\dlinewidth}
\psline[liftpen=1,linearc=\linearc](\xa,\ya)(\xb,\ya)(\xb,\yb)(\xf,\yb)(\xf,\yg)
\psline[liftpen=1,linearc=\linearc](\xf,\yg)(\xf,\yh)
\psline[liftpen=1,linearc=\linearc](\xf,\yh)(\xf,\yf)(\xb,\yf)(\xb,\yd)(\xa,\yd)
\psline[liftpen=1,linearc=\linearc](\xd,\ya)(\xc,\ya)(\xc,\yc)(\xe,\yc)(\xe,\yg)
\psline[liftpen=1,linearc=\linearc](\xe,\yg)(\xe,\yh)
\psline[liftpen=1,linearc=\linearc](\xe,\yh)(\xe,\ye)(\xc,\ye)(\xc,\yd)(\xd,\yd)
\psline[linestyle=dotted](3.11111,4)(3.11111,1.33333)
\doutex{14}{0}%
\ioutex{14}{6}%
\uoutex{14}{9}%
\psline[linestyle=dotted](11.11111,4)(11.11111,1.33333)
\setlength{\xa}{4\unit}
\addtolength{\xa}{0.5\dlinewidth}
\setlength{\ya}{9\unit}
\addtolength{\ya}{-0.5\dlinewidth}
\setlength{\xb}{6.5\unit}
\addtolength{\xb}{0.5\dlinewidth}
\setlength{\yb}{14\unit}
\addtolength{\yb}{-0.5\dlinewidth}
\setlength{\xc}{9.5\unit}
\addtolength{\xc}{-0.5\dlinewidth}
\setlength{\yc}{11\unit}
\addtolength{\yc}{0.5\dlinewidth}
\setlength{\xd}{12\unit}
\addtolength{\xd}{-0.5\dlinewidth}
\setlength{\yd}{0\unit}
\addtolength{\yd}{0.5\dlinewidth}
\setlength{\xe}{17.5\unit}
\addtolength{\xe}{0.5\dlinewidth}
\setlength{\ye}{-2\unit}
\addtolength{\ye}{-0.5\dlinewidth}
\setlength{\xf}{20.5\unit}
\addtolength{\xf}{-0.5\dlinewidth}
\setlength{\yf}{-5\unit}
\addtolength{\yf}{0.5\dlinewidth}
\setlength{\yg}{8\unit}
\addtolength{\yg}{0.5\dlinewidth}
\setlength{\yh}{1\unit}
\addtolength{\yh}{-0.5\dlinewidth}
\pscustom[linecolor=gray,fillstyle=solid,fillcolor=gray,linearc=\linearc]{%
\psline[liftpen=1,linearc=\linearc](\xb,\ya)(\xd,\ya)(\xd,\yd)(\xa,\yd)(\xa,\ya)(\xb,\ya)
\psline[liftpen=2,linearc=\linearc](\xb,\ya)(\xb,\yb)(\xf,\yb)(\xf,\yg)
\psline[liftpen=1](\xf,\yg)(\xe,\yg)
\psline[liftpen=1,linearc=\linearc](\xe,\yg)(\xe,\yc)(\xc,\yc)(\xc,\ya)
\psline[liftpen=2,linearc=\linearc](\xb,\yd)(\xb,\yf)(\xf,\yf)(\xf,\yh)
\psline[liftpen=1](\xf,\yh)(\xe,\yh)
\psline[liftpen=1,linearc=\linearc](\xe,\yh)(\xe,\ye)(\xc,\ye)(\xc,\yd)
}
\pscustom[linecolor=gray,fillstyle=solid,fillcolor=gray]{%
\psline[liftpen=1](\xe,\yg)(\xf,\yg)(\xf,\yh)
\psline[liftpen=2](\xf,\yh)(\xe,\yh)(\xe,\yg)
}
\end{pspicture}}
\,=
\settoheight{\eqoff}{$+$}%
\setlength{\eqoff}{0.5\eqoff}%
\addtolength{\eqoff}{-10.5\unit}%
\raisebox{\eqoff}{%
\begin{pspicture}(-0.44444,-5.33333)(19.11111,13.33333)
\rput[r](1.33333,8){$\scriptstyle \phi_i$}
\rput[r](1.33333,0){$\scriptstyle \phi_i$}
\rput[l](12.8888,0){$\scriptstyle \bar\phi^i$}
\rput[l](12.8888,8){$\scriptstyle \bar\phi^i$}
\uinex{2}{9}%
\iinex{2}{6}%
\dinex{2}{0}%
\setlength{\xa}{4\unit}
\addtolength{\xa}{-0.5\dlinewidth}
\setlength{\ya}{9\unit}
\addtolength{\ya}{0.5\dlinewidth}
\setlength{\xb}{6.5\unit}
\addtolength{\xb}{-0.5\dlinewidth}
\setlength{\yb}{14\unit}
\addtolength{\yb}{0.5\dlinewidth}
\setlength{\xc}{9.5\unit}
\addtolength{\xc}{0.5\dlinewidth}
\setlength{\yc}{11\unit}
\addtolength{\yc}{-0.5\dlinewidth}
\setlength{\xd}{12\unit}
\addtolength{\xd}{0.5\dlinewidth}
\setlength{\yd}{0\unit}
\addtolength{\yd}{-0.5\dlinewidth}
\setlength{\xe}{17.5\unit}
\addtolength{\xe}{-0.5\dlinewidth}
\setlength{\ye}{-2\unit}
\addtolength{\ye}{0.5\dlinewidth}
\setlength{\xf}{20.5\unit}
\addtolength{\xf}{0.5\dlinewidth}
\setlength{\yf}{-5\unit}
\addtolength{\yf}{-0.5\dlinewidth}
\setlength{\yg}{8\unit}
\addtolength{\yg}{-0.5\dlinewidth}
\setlength{\yh}{1\unit}
\addtolength{\yh}{0.5\dlinewidth}
\psline[liftpen=1,linearc=\linearc](\xa,\ya)(\xb,\ya)(\xb,\yb)(\xf,\yb)(\xf,\yg)
\psline[liftpen=1,linearc=\linearc](\xf,\yg)(\xf,\yh)
\psline[liftpen=1,linearc=\linearc](\xf,\yh)(\xf,\yf)(\xb,\yf)(\xb,\yd)(\xa,\yd)
\psline[liftpen=1,linearc=\linearc](\xd,\ya)(\xc,\ya)(\xc,\yc)(\xe,\yc)(\xe,\yg)
\psline[liftpen=1,linearc=\linearc](\xe,\yg)(\xe,\yh)
\psline[liftpen=1,linearc=\linearc](\xe,\yh)(\xe,\ye)(\xc,\ye)(\xc,\yd)(\xd,\yd)
\psline[linestyle=dotted](3.11111,4)(3.11111,1.33333)
\doutex{14}{0}%
\ioutex{14}{6}%
\uoutex{14}{9}%
\psline[linestyle=dotted](11.11111,4)(11.11111,1.33333)
\setlength{\xa}{4\unit}
\addtolength{\xa}{0.5\dlinewidth}
\setlength{\ya}{9\unit}
\addtolength{\ya}{-0.5\dlinewidth}
\setlength{\xb}{6.5\unit}
\addtolength{\xb}{0.5\dlinewidth}
\setlength{\yb}{14\unit}
\addtolength{\yb}{-0.5\dlinewidth}
\setlength{\xc}{9.5\unit}
\addtolength{\xc}{-0.5\dlinewidth}
\setlength{\yc}{11\unit}
\addtolength{\yc}{0.5\dlinewidth}
\setlength{\xd}{12\unit}
\addtolength{\xd}{-0.5\dlinewidth}
\setlength{\yd}{0\unit}
\addtolength{\yd}{0.5\dlinewidth}
\setlength{\xe}{17.5\unit}
\addtolength{\xe}{0.5\dlinewidth}
\setlength{\ye}{-2\unit}
\addtolength{\ye}{-0.5\dlinewidth}
\setlength{\xf}{20.5\unit}
\addtolength{\xf}{-0.5\dlinewidth}
\setlength{\yf}{-5\unit}
\addtolength{\yf}{0.5\dlinewidth}
\setlength{\yg}{8\unit}
\addtolength{\yg}{0.5\dlinewidth}
\setlength{\yh}{1\unit}
\addtolength{\yh}{-0.5\dlinewidth}
\pscustom[linecolor=gray,fillstyle=solid,fillcolor=gray,linearc=\linearc]{%
\psline[liftpen=1,linearc=\linearc](\xb,\ya)(\xd,\ya)(\xd,\yd)(\xa,\yd)(\xa,\ya)(\xb,\ya)
\psline[liftpen=2,linearc=\linearc](\xb,\ya)(\xb,\yb)(\xf,\yb)(\xf,\yg)
\psline[liftpen=1](\xf,\yg)(\xe,\yg)
\psline[liftpen=1,linearc=\linearc](\xe,\yg)(\xe,\yc)(\xc,\yc)(\xc,\ya)
\psline[liftpen=2,linearc=\linearc](\xb,\yd)(\xb,\yf)(\xf,\yf)(\xf,\yh)
\psline[liftpen=1](\xf,\yh)(\xe,\yh)
\psline[liftpen=1,linearc=\linearc](\xe,\yh)(\xe,\ye)(\xc,\ye)(\xc,\yd)
}

\pscustom[linecolor=gray,fillstyle=solid,fillcolor=gray]{%
\psline[liftpen=1](\xe,\yg)(\xf,\yg)(\xf,\yh)
\psline[liftpen=2](\xf,\yh)(\xe,\yh)(\xe,\yg)
}
\setlength{\xb}{8\unit}
\setlength{\yb}{12.5\unit}
\setlength{\xc}{19\unit}
\setlength{\yc}{-3.5\unit}
\setlength{\yg}{8\unit}
\addtolength{\yg}{0.5\dlinewidth}
\setlength{\yh}{1\unit}
\addtolength{\yh}{-0.5\dlinewidth}
\psline[liftpen=2,linearc=\linearc](\xc,\yh)(\xc,\yc)(\xb,\yc)(\xb,\yb)(\xc,\yb)
(\xc,\yg)
\psline[liftpen=2,linearc=\linearc](\xc,\yg)(\xc,\yh)
\end{pspicture}}
\,+
\settoheight{\eqoff}{$+$}%
\setlength{\eqoff}{0.5\eqoff}%
\addtolength{\eqoff}{-10.5\unit}%
\raisebox{\eqoff}{%
\begin{pspicture}(-0.44444,-5.33333)(19.11111,13.33333)
\rput[r](1.33333,8){$\scriptstyle \phi_i$}
\rput[r](1.33333,0){$\scriptstyle \phi_i$}
\rput[l](12.8888,0){$\scriptstyle \bar\phi^i$}
\rput[l](12.8888,8){$\scriptstyle \bar\phi^i$}
\uinex{2}{9}%
\iinex{2}{6}%
\dinex{2}{0}%
\setlength{\xa}{4\unit}
\addtolength{\xa}{-0.5\dlinewidth}
\setlength{\ya}{9\unit}
\addtolength{\ya}{0.5\dlinewidth}
\setlength{\xb}{6.5\unit}
\addtolength{\xb}{-0.5\dlinewidth}
\setlength{\yb}{14\unit}
\addtolength{\yb}{0.5\dlinewidth}
\setlength{\xc}{9.5\unit}
\addtolength{\xc}{0.5\dlinewidth}
\setlength{\yc}{11\unit}
\addtolength{\yc}{-0.5\dlinewidth}
\setlength{\xd}{12\unit}
\addtolength{\xd}{0.5\dlinewidth}
\setlength{\yd}{0\unit}
\addtolength{\yd}{-0.5\dlinewidth}
\setlength{\xe}{17.5\unit}
\addtolength{\xe}{-0.5\dlinewidth}
\setlength{\ye}{-2\unit}
\addtolength{\ye}{0.5\dlinewidth}
\setlength{\xf}{20.5\unit}
\addtolength{\xf}{0.5\dlinewidth}
\setlength{\yf}{-5\unit}
\addtolength{\yf}{-0.5\dlinewidth}
\setlength{\yg}{8\unit}
\addtolength{\yg}{-0.5\dlinewidth}
\setlength{\yh}{1\unit}
\addtolength{\yh}{0.5\dlinewidth}
\psline[liftpen=1,linearc=\linearc](\xa,\ya)(\xb,\ya)(\xb,\yb)(\xf,\yb)(\xf,\yg)
\psline[liftpen=1,linearc=\linearc](\xf,\yg)(\xf,\yh)
\psline[liftpen=1,linearc=\linearc](\xf,\yh)(\xf,\yf)(\xb,\yf)(\xb,\yd)(\xa,\yd)
\psline[liftpen=1,linearc=\linearc](\xd,\ya)(\xc,\ya)(\xc,\yc)(\xe,\yc)(\xe,\yg)
\psline[liftpen=1,linearc=\linearc](\xe,\yg)(\xe,\yh)
\psline[liftpen=1,linearc=\linearc](\xe,\yh)(\xe,\ye)(\xc,\ye)(\xc,\yd)(\xd,\yd)
\psline[linestyle=dotted](3.11111,4)(3.11111,1.33333)
\doutex{14}{0}%
\ioutex{14}{6}%
\uoutex{14}{9}%
\psline[linestyle=dotted](11.11111,4)(11.11111,1.33333)
\setlength{\xa}{4\unit}
\addtolength{\xa}{0.5\dlinewidth}
\setlength{\ya}{9\unit}
\addtolength{\ya}{-0.5\dlinewidth}
\setlength{\xb}{6.5\unit}
\addtolength{\xb}{0.5\dlinewidth}
\setlength{\yb}{14\unit}
\addtolength{\yb}{-0.5\dlinewidth}
\setlength{\xc}{9.5\unit}
\addtolength{\xc}{-0.5\dlinewidth}
\setlength{\yc}{11\unit}
\addtolength{\yc}{0.5\dlinewidth}
\setlength{\xd}{12\unit}
\addtolength{\xd}{-0.5\dlinewidth}
\setlength{\yd}{0\unit}
\addtolength{\yd}{0.5\dlinewidth}
\setlength{\xe}{17.5\unit}
\addtolength{\xe}{0.5\dlinewidth}
\setlength{\ye}{-2\unit}
\addtolength{\ye}{-0.5\dlinewidth}
\setlength{\xf}{20.5\unit}
\addtolength{\xf}{-0.5\dlinewidth}
\setlength{\yf}{-5\unit}
\addtolength{\yf}{0.5\dlinewidth}
\setlength{\yg}{8\unit}
\addtolength{\yg}{0.5\dlinewidth}
\setlength{\yh}{1\unit}
\addtolength{\yh}{-0.5\dlinewidth}
\pscustom[linecolor=gray,fillstyle=solid,fillcolor=gray,linearc=\linearc]{%
\psline[liftpen=1,linearc=\linearc](\xb,\ya)(\xd,\ya)(\xd,\yd)(\xa,\yd)(\xa,\ya)(\xb,\ya)
\psline[liftpen=2,linearc=\linearc](\xb,\ya)(\xb,\yb)(\xf,\yb)(\xf,\yg)
\psline[liftpen=1](\xf,\yg)(\xe,\yg)
\psline[liftpen=1,linearc=\linearc](\xe,\yg)(\xe,\yc)(\xc,\yc)(\xc,\ya)
\psline[liftpen=2,linearc=\linearc](\xb,\yd)(\xb,\yf)(\xf,\yf)(\xf,\yh)
\psline[liftpen=1](\xf,\yh)(\xe,\yh)
\psline[liftpen=1,linearc=\linearc](\xe,\yh)(\xe,\ye)(\xc,\ye)(\xc,\yd)
}
\addtolength{\yh}{0.5\linew}
\addtolength{\yg}{-0.5\linew}
\pssin[periods=8,coilarm=0.1,amplitude=0.2](\xe,\yg)(\xe,\yh)
\pssin[periods=8,coilarm=0.1,amplitude=0.2](\xf,\yg)(\xf,\yh)
\newlength{\yaaa}
\setlength{\yaaa}{0.5\yh}
\addtolength{\yaaa}{0.5\yg}
\addtolength{\xe}{0.7\dlinewidth}
\addtolength{\xf}{-0.7\dlinewidth}
\psline[linestyle=dotted,dotsep=1.5pt](\xe,\yaaa)(\xf,\yaaa)
\newlength{\xaaa}
\setlength{\xaaa}{0.5\xe}
\addtolength{\xaaa}{0.5\xf}
\end{pspicture}}
\eqndot
\end{equation}
\setlength{\unit}{0.225cm}%
Wrapping diagrams in the first subclass contain a closed path around the operator that is built from the propagators of scalars and fermions alone.
In \eqref{eq: decomposition of wrapping diagrams}, this path is depicted as a solid line.
Wrapping diagrams in the second subclass do not contain such a path, i.e.\ every closed path around the operator contains at least one gauge-field propagator, which is depicted by wiggly lines.

We can now prove that every wrapping diagram in the second subclass is independent of the deformation.
Given a wrapping diagram of the second subclass, we can eliminate all gauge fields by the following replacements:
\begin{equation}
\label{eq: vertex replacements} 
\setlength{\unit}{0.4cm}
\psset{xunit=\unit,yunit=\unit,runit=\unit}
\setlength{\dlinewidth}{0.75\unit}
\setlength{\linew}{1pt}
\setlength{\doublesep}{\dlinewidth}
\addtolength{\doublesep}{-\linew}
\psset{doublesep=\doublesep}
\psset{linewidth=\linew}
\setlength{\auxlen}{-0.2929\dlinewidth}
\addtolength{\auxlen}{\unit}
\setlength{\linearc}{0.75\unit}
\settoheight{\eqoff}{$+$}%
\setlength{\eqoff}{0.5\eqoff}%
\addtolength{\eqoff}{-2\unit}%
\raisebox{\eqoff}{%
\begin{pspicture}(-2,-2)(2,2)
\recthreevertex{0}{0}{0}
\psset{doubleline=true}
\psline(-1.4142,1.4142)(-0.7071,0.7071)
\psline(-0.7071,-0.7071)(-1.4142,-1.4142)
\psline(0.7071,0)(1.7071,0)
\psset{doubleline=false}
\psline(-1.4142,1.4142)(0,0)
\psline(0,0)(-1.4142,-1.4142)
\pscoil[coilwidth=0.1666,coilheight=3,coilarm=0,coilaspect=0]{-}(0,0)(1.7071,0)
\end{pspicture}}
\eqncom\,
\settoheight{\eqoff}{$+$}%
\setlength{\eqoff}{0.5\eqoff}%
\addtolength{\eqoff}{-2\unit}%
\raisebox{\eqoff}{%
\begin{pspicture}(-2,-2)(2,2)
\fourvertex{0}{0}{45}
\psset{doubleline=true}
\psline(-1.4142,1.4142)(-0.7071,0.7071)
\psline(1.4142,1.4142)(0.7071,0.7071)
\psline(1.4142,-1.4142)(0.7071,-0.7071)
\psline(-0.7071,-0.7071)(-1.4142,-1.4142)
\psset{doubleline=false}
\psline(-1.4142,1.4142)(0,0)
\psline(0,0)(-1.4142,-1.4142)
\pscoil[coilwidth=0.1666,coilheight=3,coilarm=0,coilaspect=0]{-}(0,0)(1.4142,1.4142)
\pscoil[coilwidth=0.1666,coilheight=3,coilarm=0,coilaspect=0]{-}(0,0)(1.4142,-1.4142)
\end{pspicture}}
\quad\longrightarrow\quad
\settoheight{\eqoff}{$+$}%
\setlength{\eqoff}{0.5\eqoff}%
\addtolength{\eqoff}{-2\unit}%
\raisebox{\eqoff}{%
\begin{pspicture}(-2,-2)(0.2921,2)
\psset{doubleline=true}
\psline[linearc=\linearc](-1.4142,1.4142)(0,0)(-1.4142,-1.4142)
\psset{doubleline=false}
\psline[linearc=\linearc](-1.4142,1.4142)(0,0)(-1.4142,-1.4142)
\end{pspicture}}
\eqncom
\end{equation}
where the solid central line denotes scalars and fermions and the wiggly central line denotes gauge fields. 
By definition, this replacement interrupts every closed path around the operator at least once. 
Thus, the resulting diagram is no longer a wrapping diagram.
Instead, its subdiagram of elementary interactions is of single-trace type, and we can use the above argument to show that it is deformation independent.%
\footnote{In the case that the subdiagram is not connected, we can apply relation \eqref{eq: diagram relation} to each of its connected components.}
However, as the interaction vertices of the gauge field are independent of the deformation parameters, the resulting subdiagram has the same dependence on the deformation parameters as the original diagram, which concludes the proof.

Thus, we have shown that at any loop order only wrapping diagrams of the first subclass in \eqref{eq: decomposition of wrapping diagrams} and prewrapping diagrams that contain the coupling \eqref{eq: running couplings} can be deformation dependent.
Let us now calculate the anomalous dimensions of the operators $\tr(\phi_i^L)$ at $L$-loop order.

\section{Anomalous dimensions for \texorpdfstring{$L\geq3$}{L>=3}}
\label{sec: L geq 3}

As shown in the previous section, the only deformation-dependent diagrams that contribute to the anomalous dimension of $\cO_L=\tr(\phi_i^L)$ with $L\geq3$ are wrapping diagrams of the first class in \eqref{eq: decomposition of wrapping diagrams}. Those are wrapping diagrams which contain a closed path around the operator that is built from scalars and fermions alone. 
In particular, 
\begin{equation}
 \cZ_{\cO_L}=1+\delta\cZ_{\cO_L}^{(L)}+\cO(\g^{2L+2})\eqncom
\end{equation}
and the anomalous dimension vanishes for $\ell<L$.%
\footnote{As in the last chapter, we are including a factor of $\g^{2\ell}$ in the definition of an $\ell$-loop expression throughout this chapter.}
At the critical wrapping order $\ell=L$, only four non-vanishing diagrams in the first class in \eqref{eq: decomposition of wrapping diagrams} exist:
\begin{equation}
\label{eq: wrapping diagrams}
\begin{aligned}
S(L)=
\settoheight{\eqoff}{$\times$}%
\setlength{\eqoff}{0.5\eqoff}%
\addtolength{\eqoff}{-16\unitlength}%
\raisebox{\eqoff}{%
\fmfframe(7,1)(3,1){%
\begin{fmfchar*}(30,30)
  \fmfleft{in}
  \fmfright{out1}
\fmf{phantom}{in,v5}
\fmf{phantom}{out,v2}
\fmf{phantom}{in,va5}
\fmf{phantom}{out,va2}
\fmfforce{(0,0.5h)}{in}
\fmfforce{(w,0.5h)}{out}
\fmfpoly{phantom}{v1,v6,v5,v4,v3,v2}
\fmffixed{(0.75w,0)}{v5,v2}
\fmfpoly{phantom}{va1,va6,va5,va4,va3,va2}
\fmffixed{(w,0)}{va5,va2}
\fmf{phantom}{vc,v1}
\fmf{phantom}{vc,v4}
\fmffreeze
\fmf{plain_ar,left=0}{v1,v2}
\fmf{plain_ar,left=0}{v2,v3}
\fmf{plain_ar,left=0}{v3,v4}
\fmf{dots,left=0}{v4,v5}
\fmf{plain_ar,left=0}{v5,v6}
\fmf{plain_ar,left=0}{v6,v1}
\fmf{plain_rar}{vc,v1}
\fmf{plain_rar}{vc,v2}
\fmf{plain_rar}{vc,v3}
\fmf{dots}{vc,v4}
\fmf{plain_rar}{vc,v5}
\fmf{plain_rar}{vc,v6}
\fmf{plain_ar}{va1,v1}
\fmf{plain_ar}{va2,v2}
\fmf{plain_ar}{va3,v3}
\fmf{dots}{va4,v4}
\fmf{plain_ar}{va5,v5}
\fmf{plain_ar}{va6,v6}
\fmffreeze
\fmfv{l=$\scriptscriptstyle 1$,l.dist=2}{va1}
\fmfv{l=$\scriptscriptstyle 2$,l.dist=2}{va2}
\fmfv{l=$\scriptscriptstyle 3$,l.dist=2}{va3}
\fmfv{l=$\scriptscriptstyle {L-1}$,l.dist=2}{va5}
\fmfv{l=$\scriptscriptstyle L$,l.dist=2}{va6}
\fmfiv{decor.shape=circle,decor.filled=full,decor.size=3}{vloc(__vc)}
\end{fmfchar*}}}
&=g_\YM^{2L}N^L
\Big(2\e^{iL\gamma_i^-}\cos{L\gamma_i^+}+\frac{1}{2^L}\Big)
P_L
\eqncom\\
\bar{S}(L)=
\settoheight{\eqoff}{$\times$}%
\setlength{\eqoff}{0.5\eqoff}%
\addtolength{\eqoff}{-16\unitlength}%
\raisebox{\eqoff}{%
\fmfframe(7,1)(3,1){%
\begin{fmfchar*}(30,30)
  \fmfleft{in}
  \fmfright{out1}
\fmf{phantom}{in,v5}
\fmf{phantom}{out,v2}
\fmf{phantom}{in,va5}
\fmf{phantom}{out,va2}
\fmfforce{(0,0.5h)}{in}
\fmfforce{(w,0.5h)}{out}
\fmfpoly{phantom}{v1,v6,v5,v4,v3,v2}
\fmffixed{(0.75w,0)}{v5,v2}
\fmfpoly{phantom}{va1,va6,va5,va4,va3,va2}
\fmffixed{(w,0)}{va5,va2}
\fmf{phantom}{vc,v1}
\fmf{phantom}{vc,v4}
\fmffreeze
\fmf{plain_rar,left=0}{v1,v2}
\fmf{plain_rar,left=0}{v2,v3}
\fmf{plain_rar,left=0}{v3,v4}
\fmf{dots,left=0}{v4,v5}
\fmf{plain_rar,left=0}{v5,v6}
\fmf{plain_rar,left=0}{v6,v1}
\fmf{plain_rar}{vc,v1}
\fmf{plain_rar}{vc,v2}
\fmf{plain_rar}{vc,v3}
\fmf{dots}{vc,v4}
\fmf{plain_rar}{vc,v5}
\fmf{plain_rar}{vc,v6}
\fmf{plain_ar}{va1,v1}
\fmf{plain_ar}{va2,v2}
\fmf{plain_ar}{va3,v3}
\fmf{dots}{va4,v4}
\fmf{plain_ar}{va5,v5}
\fmf{plain_ar}{va6,v6}
\fmffreeze
\fmfv{l=$\scriptscriptstyle 1$,l.dist=2}{va1}
\fmfv{l=$\scriptscriptstyle 2$,l.dist=2}{va2}
\fmfv{l=$\scriptscriptstyle 3$,l.dist=2}{va3}
\fmfv{l=$\scriptscriptstyle {L-1}$,l.dist=2}{va5}
\fmfv{l=$\scriptscriptstyle L$,l.dist=2}{va6}
\fmfiv{decor.shape=circle,decor.filled=full,decor.size=3}{vloc(__vc)}
\end{fmfchar*}}}
&=g_\YM^{2L}N^L
\Big(2\e^{-iL\gamma_i^-}\cos{L\gamma_i^+}+\frac{1}{2^L}\Big)P_L
\eqncom\\
F(L)=
\settoheight{\eqoff}{$\times$}%
\setlength{\eqoff}{0.5\eqoff}%
\addtolength{\eqoff}{-16\unitlength}%
\raisebox{\eqoff}{%
\fmfframe(7,1)(3,1){%
\begin{fmfchar*}(30,30)
  \fmfleft{in}
  \fmfright{out1}
\fmf{phantom}{in,v10}
\fmf{phantom}{out,v4}
\fmf{phantom}{in,va5}
\fmf{phantom}{out,va2}
\fmfforce{(0,0.5h)}{in}
\fmfforce{(w,0.5h)}{out}
\fmfpoly{phantom}{v1,v12,v11,v10,v9,v8,v7,v6,v5,v4,v3,v2}
\fmffixed{(0.75w,0)}{v10,v4}
\fmfpoly{phantom}{va1,va6,va5,va4,va3,va2}
\fmffixed{(w,0)}{va5,va2}
\fmf{phantom}{vc,v1}
\fmf{phantom}{vc,v7}
\fmffreeze
\fmf{dashes_ar,left=0}{v1,v2}
\fmf{dashes_rar,left=0}{v2,v3}
\fmf{dashes_ar,left=0}{v3,v4}
\fmf{dashes_rar,left=0}{v4,v5}
\fmf{dashes_ar,left=0}{v5,v6}
\fmf{dots,left=0}{v6,v7}
\fmf{dots,left=0}{v7,v8}
\fmf{dashes_rar,left=0}{v8,v9}
\fmf{dashes_ar,left=0}{v9,v10}
\fmf{dashes_rar,left=0}{v10,v11}
\fmf{dashes_ar,left=0}{v11,v12}
\fmf{dashes_rar,left=0}{v12,v1}
\fmf{plain_rar}{vc,v1}
\fmf{plain_rar}{vc,v3}
\fmf{plain_rar}{vc,v5}
\fmf{dots}{vc,v7}
\fmf{plain_rar}{vc,v9}
\fmf{plain_rar}{vc,v11}
\fmf{plain_ar}{va1,v2}
\fmf{plain_ar}{va2,v4}
\fmf{plain_ar}{va3,v6}
\fmf{dots}{va4,v8}
\fmf{plain_ar}{va5,v10}
\fmf{plain_ar}{va6,v12}
\fmffreeze
\fmfv{l=$\scriptscriptstyle 1$,l.dist=2}{va1}
\fmfv{l=$\scriptscriptstyle 2$,l.dist=2}{va2}
\fmfv{l=$\scriptscriptstyle 3$,l.dist=2}{va3}
\fmfv{l=$\scriptscriptstyle {L-1}$,l.dist=2}{va5}
\fmfv{l=$\scriptscriptstyle L$,l.dist=2}{va6}
\fmfiv{decor.shape=circle,decor.filled=full,decor.size=3}{vloc(__vc)}
\end{fmfchar*}}}
&=-4g_\YM^{2L}N^L \cos L\gamma_i^+P_L\vphantom{\Big(\Big)}\eqncom\\
\tilde F(L)
=
\settoheight{\eqoff}{$\times$}%
\setlength{\eqoff}{0.5\eqoff}%
\addtolength{\eqoff}{-16\unitlength}%
\raisebox{\eqoff}{%
\fmfframe(7,1)(3,1){%
\begin{fmfchar*}(30,30)
  \fmfleft{in}
  \fmfright{out1}
\fmf{phantom}{in,v10}
\fmf{phantom}{out,v4}
\fmf{phantom}{in,va5}
\fmf{phantom}{out,va2}
\fmfforce{(0,0.5h)}{in}
\fmfforce{(w,0.5h)}{out}
\fmfpoly{phantom}{v1,v12,v11,v10,v9,v8,v7,v6,v5,v4,v3,v2}
\fmffixed{(0.75w,0)}{v10,v4}
\fmfpoly{phantom}{va1,va6,va5,va4,va3,va2}
\fmffixed{(w,0)}{va5,va2}
\fmf{phantom}{vc,v1}
\fmf{phantom}{vc,v7}
\fmffreeze
\fmf{dashes_rar,left=0}{v1,v2}
\fmf{dashes_ar,left=0}{v2,v3}
\fmf{dashes_rar,left=0}{v3,v4}
\fmf{dashes_ar,left=0}{v4,v5}
\fmf{dashes_rar,left=0}{v5,v6}
\fmf{dots,left=0}{v6,v7}
\fmf{dots,left=0}{v7,v8}
\fmf{dashes_ar,left=0}{v8,v9}
\fmf{dashes_rar,left=0}{v9,v10}
\fmf{dashes_ar,left=0}{v10,v11}
\fmf{dashes_rar,left=0}{v11,v12}
\fmf{dashes_ar,left=0}{v12,v1}
\fmf{plain_rar}{vc,v1}
\fmf{plain_rar}{vc,v3}
\fmf{plain_rar}{vc,v5}
\fmf{dots}{vc,v7}
\fmf{plain_rar}{vc,v9}
\fmf{plain_rar}{vc,v11}
\fmf{plain_ar}{va1,v2}
\fmf{plain_ar}{va2,v4}
\fmf{plain_ar}{va3,v6}
\fmf{dots}{va4,v8}
\fmf{plain_ar}{va5,v10}
\fmf{plain_ar}{va6,v12}
\fmffreeze
\fmfv{l=$\scriptscriptstyle 1$,l.dist=2}{va1}
\fmfv{l=$\scriptscriptstyle 2$,l.dist=2}{va2}
\fmfv{l=$\scriptscriptstyle 3$,l.dist=2}{va3}
\fmfv{l=$\scriptscriptstyle {L-1}$,l.dist=2}{va5}
\fmfv{l=$\scriptscriptstyle L$,l.dist=2}{va6}
\fmfiv{decor.shape=circle,decor.filled=full,decor.size=3}{vloc(__vc)}
\end{fmfchar*}}}
&=-4g_\YM^{2L}N^L \cos L\gamma_i^-P_L\vphantom{\Big(\Big)}
\eqncom
\end{aligned}
\end{equation}
where we have depicted scalars by solid lines, fermions by dashed lines and the operator insertion by the central dot.
The arrows indicate the charge flow.
The four diagrams in \eqref{eq: wrapping diagrams} have been calculated using the Feynman rules in \cite{Fokken:2013aea}; see \cite{Fokken:2014soa} for details of this calculation.
The occurring `cake' integral as well as its divergence, which is only a UV one, read \cite{Broadhurst:1985vq}%
\begin{equation}\label{PL}
P_L=
\settoheight{\eqoff}{$\times$}%
\setlength{\eqoff}{0.5\eqoff}%
\addtolength{\eqoff}{-8\unitlength}%
\raisebox{\eqoff}{%
\fmfframe(3,-2)(0,-2){%
\begin{fmfchar*}(20,20)
  \fmfleft{in}
  \fmfright{out1}
\fmf{phantom}{in,v1}
\fmf{phantom}{out,v2}
\fmfforce{(0,0.5h)}{in}
\fmfforce{(w,0.5h)}{out}
\fmfpoly{phantom}{v1,va4,va3,v2,va2,va1}
\fmffixed{(0.75w,0)}{v1,v2}
\fmf{phantom}{vc,v1}
\fmf{plain}{vc,v2}
\fmffreeze
\fmf{plain,left=0.25}{v1,va1}
\fmf{plain,left=0.25}{va1,va2}
\fmf{plain,left=0.25}{va2,v2}
\fmf{plain,left=0.25}{v2,va3}
\fmf{plain,left=0.25}{va3,va4}
\fmf{dots,left=0.25}{va4,v1}
\fmf{plain}{vc,va1}
\fmf{plain}{vc,va2}
\fmf{plain}{vc,va3}
\fmf{dots}{vc,va4}
\fmf{plain}{vc,v1}
\fmffreeze
\fmfv{l=$\scriptscriptstyle L$,l.dist=2}{va1}
\fmfv{l=$\scriptscriptstyle 1$,l.dist=2}{va2}
\fmfv{l=$\scriptscriptstyle 2$,l.dist=2}{v2}
\fmfv{l=$\scriptscriptstyle 3$,l.dist=2}{va3}
\fmfv{l=$\scriptscriptstyle L-1$,l.dist=2}{v1}
\end{fmfchar*}}}
\eqncom\qquad
\mathcal{P}_L
=\Kop[P_L]
=\frac{1}{(4\pi)^{2L}}\frac{1}{\peps}
\frac{2}{L}\binom{2L-3}{L-1}\zeta_{2L-3}
\eqncom
\end{equation}
where we are working in the DR scheme as in the previous chapter.

The contribution of the four deformed diagrams to the renormalisation constant is 
\begin{equation} 
\label{eq: delta Z OL def}
\begin{aligned}
\delta\mathcal{Z}_{\mathcal{O}_L,\text{def}}^{(L)}
&=
-\Kop[S(L)+\bar{S}(L)+F(L)+\tilde F(L)]\\
&=4g_\YM^{2L}N^L\Big(\cos L\gamma_i^++\cos L\gamma_i^--\cos L\gamma_i^+\cos L\gamma_i^--\frac{1}{2^{L+1}}\Big)\mathcal{P}_L
\eqndot
\end{aligned}
\end{equation}
The contribution of all other diagrams to the renormalisation constant $\delta \mathcal{Z}_{\mathcal{O}_L,\text{non-def}}^{(L)}$ is deformation independent and can be found from the requirement that $\delta \mathcal{Z}_{\mathcal{O}_L}^{(L)}$ vanishes in the undeformed theory:
\begin{equation}
\label{eq: delta Z OL nondef}
\begin{aligned}
\delta\mathcal{Z}_{\mathcal{O}_L,\text{non-def}}^{(L)}
&=-\delta\mathcal{Z}_{\mathcal{O}_L,\text{def}}^{(L)}\,\Big|_{\gamma_i^\pm=0}
=-4g_\YM^{2L}N^L\Big(1-\frac{1}{2^{L+1}}\Big)\mathcal{P}_L
\eqndot
\end{aligned}
\end{equation}

The complete planar $L$-loop renormalisation constant $\delta\mathcal{Z}_{\mathcal{O}_L}^{(L)}$ is thus given by
\begin{equation}
\label{eq: ZL}
\begin{aligned}
\delta\mathcal{Z}_{\mathcal{O}_L}^{(L)}
&=
\delta\mathcal{Z}_{\mathcal{O}_L,\text{def}}^{(L)}
+\delta\mathcal{Z}_{\mathcal{O}_L,\text{non-def}}^{(L)}&=-16g_\YM^{2L}N^L\sin^2\frac{L\gamma_i^+}{2}\sin^2\frac{L\gamma_i^-}{2}\mathcal{P}_L\eqndot
\end{aligned}
\end{equation}
In particular, it vanishes in the limit of the $\beta$-deformation, $\gamma_i^+=\beta$, $\gamma_i^-=0$, as required.

Using \eqref{eq: Z in terms of D}, this yields the planar anomalous dimension 
\begin{equation}
\label{eq: gamma OL}
\begin{aligned}
\gamma_{\mathcal{O}_L}&=-64\g^{2L}\sin^2\frac{L\gamma_i^+}{2}\sin^2\frac{L\gamma_i^-}{2}
\binom{2L-3}{L-1}\zeta_{2L-3}+\cO(g^{2L+2})\eqncom
\end{aligned}
\end{equation}
which perfectly agrees with the integrability-based prediction of \cite{Ahn:2011xq}.%
\footnote{In order to match the definitions of \cite{Ahn:2011xq}, a factor of $2$ has to absorb into the effective planar coupling constant $\g$ defined in \eqref{eq: effective planar coupling constant} and a factor of $L$ has to be absorbed into $\gamma_i^\pm$.}

\section{Anomalous dimension for \texorpdfstring{$L=2$}{L=2}}
\label{sec: L eq 2}

For $\cO_2=\tr(\phi_i\phi_i)$, both wrapping and prewrapping corrections contribute. Moreover, the latter already set in at one-loop order.
Accordingly, we decompose the renormalisation constant $\mathcal{Z}_{\mathcal{O}_2}$ as 
\begin{equation}
\label{eq: Z2}
\mathcal{Z}_{\mathcal{O}_2}
=1+\delta\mathcal{Z}^{(1)}_{\mathcal{O}_2}
+\delta\mathcal{Z}^{(2)}_{\mathcal{O}_2}
+\mathcal{O}(g^6)\eqndot
\end{equation}

At one-loop order, the only deformation-dependent diagram is the prewrapping diagram
\begin{equation}
\label{eq: prewrapping diagram}
\ltwoopren{\fmfforce{(0.5w,0.5h)}{vc}
\fmf{plain_rar,left=0.75}{vo,vc}\fmf{plain_rar,right=0.75}{vo,vc}
\fmf{plain_rar}{vc,vl}\fmf{plain_rar}{vc,vr}
\fmfiv{label=$\scriptstyle Q_{\text{F}}$,l.a=0,l.dist=4}{vloc(__vc)}}
=-2g_\YM^2NQ_{\text{F}\,ii}^{ii}I_1
\eqncom
\end{equation}
where $I_1$ was defined in \eqref{eq: def I one} and the $Q_{\text{F}}$ in the diagram indicates that the quartic vertex next to it is given by the double-trace coupling \eqref{eq: running couplings}.
As $Q_{\text{F}\,ii}^{ii}$ is set to zero when taking the limit of the undeformed theory, all other contributions have to add up to zero as well in order to reproduce the result of $\cN=4$ SYM theory for vanishing deformations.
Hence, the complete one-loop contribution to the renormalisation constant is given by 
\begin{equation}\label{eq: delta Z O2 one-loop}
\delta\mathcal{Z}^{(1)}_{\mathcal{O}_2}=2g_\YM^2NQ_{\text{F}\,ii}^{ii}\Kop[I_1]\eqncom
\end{equation}
where $\Kop[I_1]$ was defined in \eqref{eq: def I one}.

At two-loop order, we need the following additional one-loop diagrams as they can occur as subdiagrams:
\begin{equation}
\label{eq: one-loop subdiagrams}
\begin{gathered}
\ltwoopren{\fmfforce{(0.5w,0.5h)}{vc}
\fmf{plain_rar,left=0.75}{vo,vc}\fmf{plain_rar,right=0.75}{vo,vc}
\fmf{plain_rar}{vc,vl}\fmf{plain_rar}{vc,vr}
\fmffreeze
\fmfposition
}
=g_\YM^2NI_1
\eqncom\qquad
\ltwoopren{%
\fmf{plain_rar}{vo,vcl}\fmf{plain_rar}{vcl,vl}
\fmf{plain_rar}{vo,vcr}\fmf{plain_rar}{vcr,vr}
\fmffreeze
\fmf{photon}{vcl,vcr}
}
=g_\YM^2N\alpha I_1
\eqncom\\
\settoheight{\eqoff}{$\times$}%
\setlength{\eqoff}{0.5\eqoff}%
\addtolength{\eqoff}{-5.625\unitlength}%
\raisebox{\eqoff}{%
\fmfframe(1,0)(1,0){%
\begin{fmfchar*}(15,11.25)
\fmfleft{v1}
\fmfright{v2}
\fmffixed{(0.5w,0)}{vc1,vc2}
\fmf{plain_ar}{v1,vc1}
\fmf{plain_ar}{vc2,v2}
\fmf{phantom,left=1}{vc1,vc2}
\fmf{phantom,left=1}{vc2,vc1}
\fmffreeze
\fmfposition
\fmfipath{p[]}
\fmfiset{p1}{vpath(__vc1,__vc2)}
\fmfiset{p2}{vpath(__vc2,__vc1)}
\fmfcmd{fill(p1--p2--cycle) withcolor 0.2black;}
\end{fmfchar*}}}
=g_\YM^2Np^{2(1-\peps)}
(-(1+\alpha)I_1+2(\alpha-1)I'_1)\eqndot
\end{gathered}
\end{equation}
Here, the gluon is depicted as a wiggly line and the black blob represents all contributions of the self-energy diagrams for the $\SU{N}$ components of the scalar field $\phi_i$ with off-shell momentum $p$.
The finite integral $I'_1$ contains a numerator as well as a doubled propagator and is given by%
\begin{equation}
\label{eq: def I' one}
 I'_1=\frac{\e^{-\gamma_{\text{E}}\peps}}{(4\pi)^{2-\peps}}\frac{(-l^\mu p_\mu)}{l^2}\FDinline[bubble,momentum]=\frac{\e^{-\gamma_{\text{E}}\peps}}{(4\pi)^{2-\peps}}\peps\FDinline[bubble] \eqncom 
\end{equation}
where we have completed the square and used integration-by-parts (IBP) identities in the last step. 
As in the first part, the loop-momentum-dependent prefactor in \eqref{eq: def I' one} is understood to occur inside of the depicted integral, and the arrow depicts the direction of the loop momentum.
The resulting one-loop counterterms read
\begin{equation}
\label{eq: counter term two-point vertex}
\begin{gathered}
\ltwoopren{%
\fmf{plain_rar}{vo,vl}
\fmf{plain_rar}{vo,vr}
\fmffreeze
\fmfiv{d.sh=cross,d.size=8}{vloc(__vo)}
\fmfiv{label=$\scriptstyle Q_{\text{F}}$,l.a=0,l.dist=6}{vloc(__vo)}}
=\delta J^{(1)}_{\mathcal{O}_2,\text{def}}
=2g^2Q_{\text{F}\,ii}^{ii}\frac{1}{\peps}
\eqncom\qquad
\ltwoopren{%
\fmf{plain_rar}{vo,vl}
\fmf{plain_rar}{vo,vr}
\fmffreeze
\fmfiv{d.sh=cross,d.size=8}{vloc(__vo)}
}
=\delta J^{(1)}_{\mathcal{O}_2,\text{non-def}}
=-g^2(1+\alpha)\frac{1}{\peps}
\eqncom\\
\settoheight{\eqoff}{$\times$}%
\setlength{\eqoff}{0.5\eqoff}%
\addtolength{\eqoff}{-5.625\unitlength}%
\raisebox{\eqoff}{%
\fmfframe(1,0)(1,0){%
\begin{fmfchar*}(15,11.25)
\fmfleft{v1}
\fmfright{v2}
\fmf{plain_ar}{v1,vc1}
\fmf{plain_ar}{vc1,v2}
\fmffreeze
\fmfposition
\fmfv{d.sh=cross,d.size=8}{vc1}
\end{fmfchar*}}}
{}={}
-p^2
\delta^{\SU{N},(1)}_{\phi_i}
\eqncom\qquad\delta^{\SU{N},(1)}_{\phi_i}=-g^2(1+\alpha)\frac{1}{\peps}
\eqndot
\end{gathered}
\end{equation}

The contribution of all diagrams that contain only single-trace couplings to the two-loop renormalisation constant $\delta\mathcal{Z}_{\mathcal{O}_2}^{(2)}$ is essentially given by setting $L=2$ in \eqref{eq: ZL}:
\begin{equation}
\label{eq: delta Z O2 single-trace}
\delta\mathcal{Z}_{\mathcal{O}_2,\text{st}}^{(2)}=-16g_\YM^4N^2\sin^2\gamma_i^+\sin^2\gamma_i^-\Kop[I_2]
\eqncom
\end{equation}
where the `cake' integral with two pieces is denoted as 
\begin{equation}\label{eq: I2}
\begin{aligned}
I_2
&= \frac{\e^{-2\gamma_{\text{E}}\peps}}{(4\pi)^{4-2\peps}}\FDinline[fishtop] 
\eqncom\\
\Kop[I_2]&=\frac{1}{(4\pi)^4}\left(\frac{1}{2\peps^2}
+\frac{1}{\peps}\left(\frac{5}{2}-\gamma_{\text{E}}+\log4\pi-\log\frac{p^2}{\mu^2}\right)\right)\eqncom
\end{aligned}
\end{equation}
with the two-loop `fish' integral given in \eqref{eq: two-loop fish integral}.%
\footnote{Recall that, in contrast to the first part of this thesis, we calculate only UV divergent contributions and not the full form factor here. Hence, the two external legs on the right side of the two-loop `fish' integral should be understood as one off-shell momentum $p$ and the third external momentum is set to zero.}
In contrast to the integrals $P_L$ with $L\geq3$, it does have a non-subtracted subdivergence as can be seen from the logarithmic dependence on the square of the off-shell momentum $p$ divided by the 't Hooft mass $\mu$.%
\footnote{In contrast to the first part, we are working in Euclidean signature here. Hence, the positive sign in front of $p^2$.
}
This momentum-dependent divergence cannot be absorbed into $\delta\cZ^{(2)}_{\cO_2}$;
it is required by consistency that this subdivergence is subtracted by other contributions such that only the overall UV divergence 
\begin{equation}
\label{eq: K R I2}
\begin{aligned}
\mathcal{I}_2&=\Kop\Rop[I_2]=\Kop[I_2-\Kop[I_1]I_1]=
\frac{1}{(4\pi)^4}\Big(-\frac{1}{2\peps^2}+\frac{1}{2\peps}\Big)
\end{aligned}
\end{equation}
remains.
Here, the operator $\Rop$ subtracts the subdivergences, cf.\ \cite{Collins:1984xc}.%
\footnote{We trust that the reader does not confuse this operator with the reflection operator $\Rop_|$ of the previous chapter or the $\rr_{ij}$ operators of chapter \ref{chap: tree-level form factors}.}
This also shows that a truncation of the theory to only the single-trace part is inconsistent.%
\footnote{As we will see below, the second term in $\Kop[I_2-\Kop[I_1]I_1]$ of \eqref{eq: K R I2} is provided by the one-loop counterterm \eqref{eq: counterterm at one-loop order} of the running double-trace coupling $Q^{ii}_{\text{F}\,ii}$.}

The following 1PI two-loop diagrams with a double-trace coupling contribute to the renormalisation constant:
\begin{equation}
\begin{aligned}
\ltwoopren{%
\fmfforce{(0.5w,0.721688h)}{vc1}%
\fmf{plain_rar,left=0.75}{vc2,vc1}\fmf{plain_rar,right=0.75}{vc2,vc1}
\fmf{plain_rar,left=0.75}{vo,vc2}\fmf{plain_rar,right=0.75}{vo,vc2}
\fmf{plain_rar}{vc1,vl}\fmf{plain_rar}{vc1,vr}
\fmffreeze
\fmfposition
\fmfiv{label=$\scriptstyle Q_{\text{F}}$,l.a=-15,l.dist=6}{vloc(__vc1)}
}
&=\ltwoopren{%
\fmfforce{(0.5w,0.721688h)}{vc1}%
\fmf{plain_rar,left=0.75}{vc2,vc1}\fmf{plain_rar,right=0.75}{vc2,vc1}
\fmf{plain_rar,left=0.75}{vo,vc2}\fmf{plain_rar,right=0.75}{vo,vc2}
\fmf{plain_rar}{vc1,vl}\fmf{plain_rar}{vc1,vr}
\fmffreeze
\fmfposition
\fmfiv{label=$\scriptstyle Q_{\text{F}}$,l.a=0,l.dist=6}{vloc(__vc2)}
}
=-2g_\YM^4N^2 Q_{\text{F}\,ii}^{ii} I_1^2
\eqncom\qquad
\ltwoopren{%
\fmfforce{(0.5w,0.721688h)}{vc1}%
\fmf{plain_rar,left=0.75}{vc2,vc1}\fmf{plain_rar,right=0.75}{vc2,vc1}
\fmf{plain_rar,left=0.75}{vo,vc2}\fmf{plain_rar,right=0.75}{vo,vc2}
\fmf{plain_rar}{vc1,vl}\fmf{plain_rar}{vc1,vr}
\fmffreeze
\fmfposition
\fmfiv{label=$\scriptstyle Q_{\text{F}}$,l.a=-15,l.dist=6}{vloc(__vc1)}
\fmfiv{label=$\scriptstyle Q_{\text{F}}$,l.a=0,l.dist=6}{vloc(__vc2)}
}
=4g_\YM^4N^2(Q_{\text{F}\,ii}^{ii})^2I_1^2
\eqncom\\
\ltwoopren{%
\fmfforce{(0.5w,0.5h)}{vc}
\fmf{phantom,left=0.75}{vo,vc}\fmf{phantom,left=0.75}{vc,vo}
\fmf{phantom}{vc,vl}\fmf{phantom}{vc,vr}
\fmffreeze\fmfposition
\fmfipath{p[]}
\fmfipair{vm[],vl[],vr[]}
\fmfiset{p1}{vpath(__vo,__vc)}
\fmfiset{p2}{reverse(vpath(__vc,__vo))}
\fmfiset{p3}{vpath(__vc,__vl)}
\fmfiset{p4}{vpath(__vc,__vr)}
\svertex{vm1}{p1}
\svertex{vm2}{p2}
\svertex{vm3}{p3}
\svertex{vm4}{p4}
\fmfi{plain_rar}{subpath (0,length(p3)/2) of p3}
\fmfi{plain_rar}{subpath (length(p3)/2,length(p3)) of p3}
\fmfi{plain_rar}{subpath (0,length(p4)/2) of p4}
\fmfi{plain_rar}{subpath (length(p4)/2,length(p2)) of p4}
\fmfi{plain_rar}{p1}
\fmfi{plain_rar}{p2}
\fmfi{photon}{vm3{dir 60}..{dir -60}vm4}
\fmfiv{label=$\scriptstyle Q_{\text{F}}$,l.a=0,l.dist=6}{vloc(__vc)}
}
&=-2g_\YM^4N^2Q_{\text{F}\,ii}^{ii}\alpha I_1^2
\eqncom\\
\ltwoopren{%
\fmfforce{(0.5w,0.5h)}{vc}
\fmf{phantom,left=0.75}{vo,vc}\fmf{phantom,left=0.75}{vc,vo}
\fmf{phantom}{vc,vl}\fmf{phantom}{vc,vr}
\fmffreeze\fmfposition
\fmfipath{p[]}
\fmfipair{vm[],vl[],vr[]}
\fmfiset{p1}{vpath(__vo,__vc)}
\fmfiset{p2}{reverse(vpath(__vc,__vo))}
\fmfiset{p3}{vpath(__vc,__vl)}
\fmfiset{p4}{vpath(__vc,__vr)}
\svertex{vm1}{p1}
\svertex{vm2}{p2}
\svertex{vm3}{p3}
\svertex{vm4}{p4}
\fmfi{plain_rar}{subpath (0,length(p1)/2) of p1}
\fmfi{plain_rar}{subpath (length(p1)/2,length(p1)) of p1}
\fmfi{plain_rar}{subpath (0,length(p2)/2) of p2}
\fmfi{plain_rar}{subpath (length(p2)/2,length(p2)) of p2}
\fmfi{plain_rar}{p3}
\fmfi{plain_rar}{p4}
\fmfi{photon}{vm1--vm2}
\fmfiv{label=$\scriptstyle Q_{\text{F}}$,l.a=0,l.dist=6}{vloc(__vc)}
}
&=-2g_\YM^4N^2Q_{\text{F}\,ii}^{ii}(2(3-\alpha)I_2-(3-2\alpha)I_1^2)
\eqncom\\
\ltwoopren{%
\fmfforce{(0.5w,0.5h)}{vc}
\fmf{phantom,left=0.75}{vo,vc}\fmf{phantom,left=0.75}{vc,vo}
\fmf{phantom}{vc,vl}\fmf{phantom}{vc,vr}
\fmffreeze\fmfposition
\fmfipath{p[]}
\fmfipair{vm[],vl[],vr[]}
\fmfiset{p1}{vpath(__vo,__vc)}
\fmfiset{p2}{reverse(vpath(__vc,__vo))}
\fmfiset{p3}{vpath(__vc,__vl)}
\fmfiset{p4}{vpath(__vc,__vr)}
\svertex{vm1}{p1}
\svertex{vm2}{p2}
\svertex{vm3}{p3}
\svertex{vm4}{p4}
\fmfi{plain_rar}{subpath (0,length(p1)*0.4) of p1}
\fmfi{plain_rar}{subpath (length(p1)*0.6,length(p1)) of p1}
\fmfi{plain_rar}{p2}
\fmfi{plain_rar}{p3}
\fmfi{plain_rar}{p4}
\vacpolp[0.25]{p1}
\fmfiv{label=$\scriptstyle Q_{\text{F}}$,l.a=0,l.dist=6}{vloc(__vc)}}
&=-4g_\YM^4N^2Q_{\text{F}\,ii}^{ii}(-(1+\alpha)I_2+(\alpha-1)(2I_2-I_1^2))
\eqndot
\end{aligned}
\end{equation}
Moreover, the following 1PI one-loop diagrams with one-loop counterterms contribute:
\begin{equation}
\begin{aligned}
\ltwoopren{%
\fmfforce{(0.5w,0.5h)}{vc}
\fmf{phantom,left=0.75}{vo,vc}\fmf{phantom,left=0.75}{vc,vo}
\fmf{phantom}{vc,vl}\fmf{phantom}{vc,vr}
\fmffreeze\fmfposition
\fmfipath{p[]}
\fmfipair{vm[],vl[],vr[]}
\fmfiset{p1}{vpath(__vo,__vc)}
\fmfiset{p2}{reverse(vpath(__vc,__vo))}
\fmfiset{p3}{vpath(__vc,__vl)}
\fmfiset{p4}{vpath(__vc,__vr)}
\svertex{vm1}{p1}
\svertex{vm2}{p2}
\svertex{vm3}{p3}
\svertex{vm4}{p4}
\fmfi{plain_rar}{subpath (0,length(p1)*0.5) of p1}
\fmfi{plain_rar}{subpath (length(p1)*0.5,length(p1)) of p1}
\fmfi{plain_rar}{p2}
\fmfi{plain_rar}{p3}
\fmfi{plain_rar}{p4}
\fmfi{phantom_cross}{p1}
\fmfiv{label=$\scriptstyle Q_{\text{F}}$,l.a=0,l.dist=6}{vloc(__vc)}}
&=4g_\YM^2N\delta^{\SU{N},(1)}_{\phi_i} Q_{\text{F}\,ii}^{ii}I_1
\eqncom\qquad
&\ltwoopren{\fmfforce{(0.5w,0.5h)}{vc}
\fmf{plain_rar,left=0.75}{vo,vc}\fmf{plain_rar,right=0.75}{vo,vc}
\fmf{plain_rar}{vc,vl}\fmf{plain_rar}{vc,vr}
\fmffreeze
\fmfposition
\fmfcmd{draw (((polycross 4) scaled 8) rotated 45) shifted vloc(__vc);}
\fmfiv{label=$\scriptstyle Q_{\text{F}}$,l.a=0,l.dist=6}{vloc(__vc)}}
&=-2g_\YM^2N\delta Q_{\text{F}\,ii}^{ii\,(1)}I_1
\eqncom\\
\ltwoopren{\fmfforce{(0.5w,0.5h)}{vc}
\fmf{plain_rar,left=0.75}{vo,vc}\fmf{plain_rar,right=0.75}{vo,vc}
\fmf{plain_rar}{vc,vl}\fmf{plain_rar}{vc,vr}
\fmffreeze
\fmfposition
\fmfiv{d.sh=cross,d.size=8}{vloc(__vo)}
\fmfiv{label=$\scriptstyle Q_{\text{F}}$,l.a=0,l.dist=6}{vloc(__vc)}}
&=-2g_\YM^2N\delta J^{(1)}_{\mathcal{O}_2,\text{non-def}}Q_{\text{F}\,ii}^{ii}I_1
\eqncom\qquad
&\ltwoopren{\fmfforce{(0.5w,0.5h)}{vc}
\fmf{plain_rar,left=0.75}{vo,vc}\fmf{plain_rar,right=0.75}{vo,vc}
\fmf{plain_rar}{vc,vl}\fmf{plain_rar}{vc,vr}
\fmffreeze
\fmfposition
\fmfiv{d.sh=cross,d.size=8}{vloc(__vo)}
\fmfiv{label=$\scriptstyle Q_{\text{F}}$,l.a=0,l.dist=6}{vloc(__vo)}}
&=g_\YM^2N\delta J^{(1)}_{\mathcal{O}_2,\text{def}}I_1
\eqncom\\
\ltwoopren{\fmfforce{(0.5w,0.5h)}{vc}
\fmf{plain_rar,left=0.75}{vo,vc}\fmf{plain_rar,right=0.75}{vo,vc}
\fmf{plain_rar}{vc,vl}\fmf{plain_rar}{vc,vr}
\fmffreeze
\fmfposition
\fmfiv{d.sh=cross,d.size=8}{vloc(__vo)}
\fmfiv{label=$\scriptstyle Q_{\text{F}}$,l.a=0,l.dist=6}{vloc(__vc)}
\fmfiv{label=$\scriptstyle Q_{\text{F}}$,l.a=0,l.dist=6}{vloc(__vo)}}
&=-2g_\YM^2N\delta J^{(1)}_{\mathcal{O}_2,\text{def}}Q_{\text{F}\,ii}^{ii}I_1
\eqncom\qquad
&\ltwoopren{%
\fmf{plain_rar}{vo,vcl}\fmf{plain_rar}{vcl,vl}
\fmf{plain_rar}{vo,vcr}\fmf{plain_rar}{vcr,vr}
\fmffreeze
\fmf{photon}{vcl,vcr}
\fmfiv{d.sh=cross,d.size=8}{vloc(__vo)}
\fmfiv{label=$\scriptstyle Q_{\text{F}}$,l.a=0,l.dist=6}{vloc(__vo)}}
&=g_\YM^2N\delta J^{(1)}_{\mathcal{O}_2,\text{def}}\alpha I_1
\eqncom
\end{aligned}
\end{equation}
which are deformation dependent as they contain the double-trace coupling either directly or via a counterterm.

The total contributions of the 1PI diagrams with a double-trace coupling is 
\begin{equation}
\label{eq: delta Z O2 double-trace 1PI}
\begin{aligned}
\delta\mathcal{Z}^{(2)}_{\mathcal{O}_2,\text{dt},1\text{PI}}
&=g_\YM^4N^2
\big(16\sin^2\gamma_i^+\sin^2\gamma_i^-\Kop[\Kop[I_1]I_1]
+2Q_{\text{F}\,ii}^{ii}(\alpha+1
-2Q_{\text{F}\,ii}^{ii})\Kop\Rop[I_1^2]
\big)
\eqncom
\end{aligned}
\end{equation}
where 
\begin{equation}
 \Kop\Rop[I_1^2]=\Kop[I^2_1- 2 \Kop[I_1]I_1]= -\Kop[I_1]^2 \eqndot
\end{equation}
Finally, also one-particle-reducible (non-1PI) diagrams contribute as 
\begin{equation}
\label{eq: delta Z O2 double-trace non-1PI}
\begin{aligned}
\delta\mathcal{Z}^{(2)}_{\mathcal{O}_2,\text{dt},\text{non-$1$PI}}
&=
\frac{1}{2}\Bigg[
\ltwoopren{%
\fmf{phantom}{vo,vl}
\fmf{plain_rar}{vo,vr}
\fmffreeze
\fmfposition
\fmfipath{p[]}
\fmfipair{vm[],vl[],vr[]}
\fmfiset{p1}{vpath(__vo,__vl)}
\fmfiset{p2}{vpath(__vo,__vr)}
\svertex{vm1}{p1}
\svertex{vm2}{p2}
\fmfi{plain_rar}{subpath (0,length(p1)*0.5) of p1}
\fmfi{plain_rar}{subpath (length(p1)*0.5,length(p1)) of p1}
\fmfi{phantom_cross}{p1}
\fmfiv{d.sh=cross,d.size=8}{vloc(__vo)}
\fmfiv{label=$\scriptstyle Q_{\text{F}}$,l.a=0,l.dist=6}{vloc(__vo)}}
+
\ltwoopren{%
\fmf{plain_rar}{vo,vl}
\fmf{phantom}{vo,vr}
\fmffreeze
\fmfposition
\fmfipath{p[]}
\fmfipair{vm[],vl[],vr[]}
\fmfiset{p1}{vpath(__vo,__vl)}
\fmfiset{p2}{vpath(__vo,__vr)}
\svertex{vm1}{p1}
\svertex{vm2}{p2}
\fmfi{plain_rar}{subpath (0,length(p2)*0.5) of p2}
\fmfi{plain_rar}{subpath (length(p2)*0.5,length(p2)) of p2}
\fmfi{phantom_cross}{p2}
\fmfiv{d.sh=cross,d.size=8}{vloc(__vo)}
\fmfiv{label=$\scriptstyle Q_{\text{F}}$,l.a=0,l.dist=6}{vloc(__vo)}}
\Bigg]
=-\delta^{\SU{N},(1)}_{\phi_i}\delta J^{(1)}_{\mathcal{O}_2,\text{def}}
\eqncom
\end{aligned}
\end{equation}
cf.\ \eqref{eq: ZOLexpansion}.

The complete two-loop renormalisation constant $\delta\mathcal{Z}^{(2)}_{\mathcal{O}_2}$ can be obtained by summing \eqref{eq: delta Z O2 single-trace}, \eqref{eq: delta Z O2 double-trace 1PI} and \eqref{eq: delta Z O2 double-trace non-1PI}:
\begin{equation}
\label{eq: delta Z O2 two-loop}
\begin{aligned}
\delta\mathcal{Z}^{(2)}_{\mathcal{O}_2}
&=\delta\mathcal{Z}^{(2)}_{\mathcal{O}_2,\text{st}}+\delta\mathcal{Z}^{(2)}_{\mathcal{O}_2,\text{dt},\text{$1$PI}}+\delta\mathcal{Z}^{(2)}_{\mathcal{O}_2,\text{dt},\text{non-$1$PI}}\\
&=
-g_\YM^4N^2(16\sin^2\gamma_i^+\sin^2\gamma_i^-\,\mathcal{I}_2
+4(Q_{\text{F}\,ii}^{ii})^2\Kop\Rop[I_1^2])
\eqndot
\end{aligned}
\end{equation}

Up to two-loop order, the logarithm of the renormalisation constant is given by 
\begin{equation}
\label{eq: ln Z O2}
\begin{aligned}
\log\mathcal{Z}_{\mathcal{O}_2}
&=\delta\mathcal{Z}^{(1)}_{\mathcal{O}_2}
+\delta\mathcal{Z}^{(2)}_{\mathcal{O}_2}
-\frac{1}{2}\left(\delta\mathcal{Z}^{(1)}_{\mathcal{O}_2}\right)^2+\mathcal{O}(g^6)\\
&=2g^2Q_{\text{F}\,ii}^{ii}\frac{1}{\peps}
+2g^4\left(
8\sin^2\gamma_i^+\sin^2\gamma_i^-\left(\frac{1}{2\peps^2}-\frac{1}{2\peps}\right)
+(Q_{\text{F}\,ii}^{ii})^2\frac{1}{\peps^2}\right)
+\mathcal{O}(g^6)
\eqncom
\end{aligned}
\end{equation}
which still contains double poles. 
In fact, the presence of the double poles is associated with the non-vanishing beta function \eqref{eq: beta function result}.
In appendix \ref{appsec: renormalisation group equations}, we give the more general expression \eqref{eq: gamma OL def} for an anomalous dimension, which  is applicable in this case.
Applying \eqref{eq: gamma OL def}, we find that the double poles cancel with the contribution of \eqref{eq: beta function result}.

The anomalous dimension up to two-loop order is then given by 
\begin{equation}
\label{eq: gamma O2}
\begin{aligned}
\gamma_{\mathcal{O}_2}
&=\Big(\peps \g\frac{\partial}{\partial \g}
-\beta_{Q_{\text{F}\,ii}^{ii}}\frac{\partial}{\partial Q_{\text{F}\,ii}^{ii}}\Big)\log\mathcal{Z}_{\mathcal{O}_2}
=4\g^2Q_{\text{F}\,ii}^{ii}-32\g^4\sin^2\gamma_i^+\sin^2\gamma_i^- +\cO(\g^6)
\eqndot
\end{aligned}
\end{equation}
This result is valid in the DR scheme.
In a different scheme with coupling constant $g_\varrho=\g \e^{\frac{\varrho}{2}\peps}$, 
the anomalous dimension reads
\begin{equation}
\label{eq: gamma O2 varrho}
\begin{aligned}
\gamma_{\mathcal{O}_2}^{\varrho}
&=4g_\varrho^2Q_{\text{F}\,ii}^{ii}-32g_\varrho^4\sin^2\gamma_i^+\sin^2\gamma_i^--2g_\varrho^2\varrho\,\beta_{Q_{\text{F}\,ii}^{ii}}+\cO(g_\varrho^6)\\
&=4g_\varrho^2Q_{\text{F}\,ii}^{ii}-8g_\varrho^4\big(4(1+\varrho)\sin^2\gamma_i^+\sin^2\gamma_i^-+\varrho(Q^{ii}_{\text{F}\,ii})^2\big)+\cO(g_\varrho^6)
\eqndot
\end{aligned}
\end{equation}
In particular, we have $\varrho=-\gamma_{\text{E}}+\log4\pi$ in the $\overline{\text{DR}}$ scheme used in the first part of this thesis.
A general derivation of the scheme change can be found in \cite{Fokken:2014soa}. 
At level of the calculation above, the influence of the scheme change can be seen as follows.
In the presence of double poles in $\log\mathcal{Z}_{\mathcal{O}_2}$, the redefinition of the coupling constant from $\g$ to $\g_\varrho$ does not commute with applying the operator $\Kop$ which extracts the poles. Instead, a multiple of the coefficient of the double pole, which is proportional to the beta function, is added to the single pole, which yields the anomalous dimension. 

The renormalisation-scheme dependence of the planar anomalous dimension \eqref{eq: gamma O2 varrho} explicitly demonstrates that conformal invariance is broken even in the planar theory.

\cleardoublepage
\phantomsection
\addcontentsline{toc}{chapter}{Conclusions}
\chapter*{Conclusions}
\markboth{Conclusions}{Conclusions}
\label{chap: conclusion}

The last one and a half decades have seen tremendous progress in understanding scattering amplitudes and correlation functions in $\cN=4$ SYM theory.
In this thesis, we have addressed the question to which extend the methods developed in this context and the structures found there can be generalised to other quantities in $\cN=4$ SYM theory as well as to other theories.

In the first part of this thesis, we have studied form factors of generic gauge-invariant local composite operators in $\cN=4$ SYM theory.
Form factors form a bridge between the purely on-shell scattering amplitudes and the purely off-shell correlation functions.
They provide an ideal starting point for applying on-shell methods to quantities containing composite operators.
Furthermore, form factors allow us to study the dilatation operator and hence the spectral problem of integrability via powerful on-shell methods.

At tree-level and for a minimal number of external fields, we have found that, up to a momentum-conserving delta function and a normalisation factor, the colour-ordered form factor of an operator is simply given by replacing the oscillators $(\aosc_i^{\dagger\alpha},\bosc_i^{\dagger\alphadot},\dosc_i^{\dagger A})$ in the spin-chain representation of the operator by the super-spinor-helicity variables $(\lambda_i^{\alpha},\lambdat_i^{\alphadot},\etatt_i^{A})$.
Moreover, the generators of $\PSU{2,2|4}$ act on them accordingly.
Hence, minimal form factors realise the spin-chain picture of $\cN=4$ SYM theory in the language of scattering amplitudes. 

At one-loop level, we have calculated the cut-constructible part of the minimal form factor of any operator using generalised unitarity.
While its IR divergence is of the well known universal form, its UV divergence depends on the operator and allows us to read off the complete one-loop dilatation operator of $\cN=4$ SYM theory. 
To the author's knowledge, this is the first derivation of the complete one-loop dilatation operator using field theory alone, i.e.\ without lifting results from a closed subsector of the theory via symmetry.
Moreover, our results
 provide a field-theoretic derivation of the connection between the tree-level four-point amplitude and the complete one-loop dilatation operator derived on the basis of symmetry considerations in \cite{Zwiebel:2011bx}.
Note that our results 
do not rely on the planar limit and are valid for any $N$.

The approach to calculate the dilatation operator via (minimal) form factors and on-shell methods continues to work at higher loop orders, as we have demonstrated using the Konishi primary operator and the operators of the $\SU{2}$ sector at two-loop order as examples.

For the Konishi operator $\cK$, which is the prime example of a non-protected operator, an important subtlety arises in the calculation of its form factors via on-shell methods.
This operator depends on the number of scalars $N_\phi$ in the theory, and, through the relation $N_\phi=10-D=6+2\peps$, also on the dimension of spacetime $D$.
Employing four-dimensional unitarity yields direct results only for $\cK_{N_\phi=6}$, which is 
not the correct analytic continuation of the Konishi primary operator to $D=4-2\peps$ dimensions.
Using a group-theoretic decomposition of the different contributions to its form factor, however, we have given an all-loop prescription how to lift the result for $\cK_{N_\phi=6}$ to $\cK$.
This extends the method of unitarity and solves the long standing puzzle of calculating the Konishi form factor via on-shell methods.
While the difference between the form factors of $\cK_{N_\phi=6}$ and $\cK$ affects the anomalous dimension starting from two-loop order, it is a new source of finite rational terms already at one-loop order.
Moreover, similar subtleties also arise for other dimension-dependent operators and can be solved analogously. 
These subtleties are also not restricted to form factors and to the on-shell method of unitarity;
they equally occur for generalised form factors and correlation functions, as well as for other four-dimensional on-shell methods.
Our results suggest that they can also be solved in these contexts, thus providing the basis to apply the on-shell unitarity method as well as other on-shell methods also to generalised form factors and correlation functions of general operators.

For operators in the $\SU{2}$ sector, a different complication occurs. 
In contrast to the Konishi operator, these operators are not eigenstates under renormalisation but mix among each other. This leads to a non-trivial mixing 
of the universal IR divergences 
and the UV divergences, which are operator-valued due to the operator mixing. 
In order to disentangle the former from the latter, also the exponentiation of the divergences has to be understood in an operatorial form.
Taking the logarithm and subtracting the universal IR divergences, we have read off the dilatation operator from the UV divergences.
Moreover, we have calculated the finite remainder functions for the minimal form factors, which have to be understood as operators as well.
In contrast to the remainders of scattering amplitudes and BPS form factor, they are not of uniform transcendentality.
Their maximally transcendental part, however, is universal and coincides with the result of \cite{Brandhuber:2014ica} for the BPS operators $\tr(\phi_{14}^L)$, thus extending the principle of maximal transcendentality to form factors of non-protected operators.
Due to Ward identities for form factors, the lower transcendental parts can be expressed in terms of one simple function of transcendental degree three and two simple functions of transcendental degree two and less. 
We conjecture that the universality of the maximally transcendental part extends to all operators, also beyond the $\SU{2}$ sector. 

Our results provide a solid stepping stone towards deriving the complete two-loop dilatation operator, which is currently unknown, via on-shell methods.
Moreover, they show that form factors of non-protected operators in $\cN=4$ SYM theory share many features with scattering amplitudes in QCD, such as UV divergences and rational terms. 
Form factors hence allow us to study these features within the simpler $\cN=4$ SYM theory.
Note that our methods do not rely on the planar limit or integrability.
They could also be applied to more general theories.

Aiming to understand the geometry and the integrable structure underlying form factors for a general non-minimal number of external points, 
we have studied tree-level form factors of the chiral part of the stress-tensor supermultiplet for any number of external points $n$.
In particular, we have extended on-shell diagrams to the partially off-shell form factors. 
In addition to the two building blocks and two equivalence moves present for on-shell diagrams of amplitudes, the extension to form factors only requires the minimal form factor as a further building block as well as one further equivalence move. 
We have found a relation between the on-shell diagrams for form factors and those for scattering amplitudes, which allows us to obtain the on-shell diagrams for all tree-level form factors from their counterparts for amplitudes.
In contrast to the case of amplitudes, several top-cell diagrams are required to obtain all BCFW terms for form factors. The different top-cell diagrams are, however, related by cyclic permutations of the on-shell legs.
Furthermore, the permutation associated with the on-shell diagram plays a slightly different role for form factors.
Our results open the path to extend on-shell diagrams 
 to form factors of general operators, generalised form factors and correlation functions.

Following the approach for scattering amplitudes, 
 we have introduced a central-charge deformation for form factors.
This allows their construction via the integrability-based technique of $\rr$ operators, which was initially developed for tree-level amplitudes.
Form factors do not share the Yangian invariance of 
 amplitudes and are hence not eigenstates of the monodromy matrix of the spin chain from the study of 
 amplitudes. However, they are eigenstates of the transfer matrix of this spin chain provided that the corresponding operators are eigenstates of the transfer matrix that appeared in the spin chain of the spectral problem.
This implies the existence of a tower of conserved charges and symmetry under the action of a part of the Yangian.
In particular, form factors embed the integrable spin chain of the spectral problem into the one that appeared for amplitudes.
In addition to $n$-point tree-level form factors of the stress-tensor supermultiplet, we have explicitly shown this transfer-matrix identity for the minimal tree-level form factors of generic operators, but we are confident that it holds for all $n$-point tree-level form factors. 

Turning back to the chiral part of the stress-tensor supermultiplet, we have found that the corresponding form factors can be obtained from a Graßmannian integral representation. As we are using two auxiliary on-shell momenta to parametrise the off-shell momentum of the composite operator, the occurring Graßmannian is $\GrassmannSymbol(n+2,k)$.  We have given the Graßmannian integral representation in spinor-helicity variables, twistors and momentum twistors.
In contrast to the case of planar amplitudes, the on-shell form in this integral contains consecutive as well as non-consecutive minors; the latter also occur in the case of non-planar scattering amplitudes \cite{Arkani-Hamed:2014bca,Chen:2014ara,Franco:2015rma,Chen:2015qna}.

Our results, in addition to other studies, have shown that many structures found in scattering amplitudes have a natural generalisation for quantities containing local composite operators such as form factors and that also the methods developed for amplitudes can be generalised to these cases.
In the end, on-shell methods might turn out to be as powerful for form factors and correlation functions as they are for scattering amplitudes.

Deformations of $\cN=4$ SYM theory provide us with further examples of theories that can be understood using similar methods. Moreover, they can shed some light on the origins and interdependence of the special properties of $\cN=4$ SYM theory. 
In the second part of this thesis, we have studied the $\beta$- and the $\gamma_i$-deformation of $\cN=4$ SYM theory.
They
 were respectively shown to be the most general $\cN=1$ supersymmetric and non-supersymmetric field-theory deformations of $\cN=4$ SYM theory that are integrable at the level of the asymptotic Bethe ansatz \cite{Beisert:2005if}, i.e.\ asymptotically integrable. 
Planar single-trace interactions in these theories are closely related to their undeformed counterparts via Filk's theorem. 
Hence, in the planar limit and in the asymptotic regime, all results for scattering amplitudes, correlation functions, anomalous dimensions and form factors in $\cN=4$ SYM theory remain valid in the deformed theories after some minimal modifications. In particular, this is true for the asymptotic form factor results of chapters \ref{chap: minimal one-loop form factors} and \ref{chap: two-loop su(2) form factors}.
Moreover, we have extended Filk's theorem to composite operators that are neutral with respect to the charges that define the deformation.
Thus, in particular the results for the minimal form factors of the Konishi primary operator, which were obtained in chapter \ref{chap: two-loop Konishi form factor} up to two-loop order, remain valid in the deformed theories to all orders of planar perturbation theory.

For general charged composite operators, however, the operators have to be removed from the diagram before applying Filk's theorem.
In the case of finite-size effects, the resulting subdiagram of elementary interactions is of multi-trace type, and Filk's theorem is not applicable. 
In particular, Filk's theorem cannot be used to prove that integrability is inherited to the deformed theories in the presence of finite-size effects.
This has lead us to investigate multi-trace and especially double-trace terms.
In addition to non-planar combinations of single-trace interactions, they can originate from the double-trace part of the $\SU{N}$ propagator and from double-trace couplings induced by quantum corrections.
In the $\cN=1$ supersymmetric $\beta$-deformation with gauge group $\SU{N}$, such a double-trace coupling is required for conformal invariance.
In contrast, the $\beta$-deformation with gauge group $\U{N}$ is not conformally invariant. 
This shows in particular that the deformed theories distinguish between the gauge groups $\U{N}$ and $\SU{N}$.
In the non-supersymmetric $\gamma_i$-deformation with gauge group $\U{N}$ or $\SU{N}$, we have identified a running double-trace coupling without fixed points, which breaks conformal invariance.
Moreover, conformal invariance cannot be restored by including further multi-trace couplings that satisfy certain minimal requirements.

Double-trace couplings affect planar correlation functions and anomalous dimensions through a new type of finite-size effect.
As it starts to contribute one loop order earlier than the well known wrapping effect, we have called it prewrapping.
We have analysed the mechanism behind prewrapping in detail and given a necessary criterion for composite operators to be affected by it.

In the $\beta$-deformation with gauge groups $\U{N}$ and $\SU{N}$, we have included the respective finite-size correction for wrapping and prewrapping for operators of length one and two into the asymptotic one-loop dilatation operator of \cite{Beisert:2005if} to obtain the complete one-loop dilatation operator of the planar theory.
Interestingly, the only prewrapping-affected supermultiplets in the complete planar one-loop spectrum are those of $\tr(\phi_i\phi_j)$ and $\tr(\bar\phi^i\bar\phi^j)$ with $i\neq j$, which are moreover related by a $\ZZ_3$ symmetry of the theory and charge conjugation; the respective contributions to all other potentially affected operators cancel in the specific combinations that form the eigenstates.

Aiming to test integrability in the $\gamma_i$-deformation and spurred by a puzzling divergence in an integrability-based prediction, we have calculated the planar $L$-loop anomalous dimensions of the operators $\tr(\phi_{14}^L)$ in this theory. 
This calculation can be hugely simplified using the fact that these operators are protected in the undeformed theory.
For generic $L\geq3$, only four diagrams have to be calculated,
which can moreover be evaluated analytically for any $L$. We have found a perfect match between our results and the integrability-based predictions of \cite{Ahn:2011xq}. This is one of the very rare occasions in quantum field theory where quantities can be calculated at generic loop orders. 
Moreover, it yields a highly non-trivial test of integrability in the deformed setting.
For $L=2$, where the integrability-based description diverges, we find a finite rational result. This result depends on the running double-trace coupling and hence on the renormalisation scheme, which explicitly shows that the $\gamma_i$-deformed theory is not conformally invariant --- not even in the planar limit.

\cleardoublepage
\phantomsection
\addcontentsline{toc}{chapter}{Outlook}
\chapter*{Outlook}
\markboth{Outlook}{Outlook}
\label{chap: outlook}

Our results open up many interesting  paths for further investigations, both concerning form factors and deformations.

While we have calculated the cut-constructible part of the one-loop form factor of any operators in chapter \ref{chap: minimal one-loop form factors}, this leaves potential finite rational terms undetermined. In \cite{Wilhelm:2014qua}, we have given an example in which such rational terms indeed occur, and it would be interesting to calculate them in general.
Methods to determine rational terms have been developed for amplitudes in QCD and might also be applicable in our case, see \cite{Bern:2007dw} for a review.

Moreover, it would be very interesting to calculate the minimal form factors of generic operators at two-loop order. 
In particular, this would yield the complete two-loop dilatation operator of $\cN=4$ SYM theory, which is currently unknown.
Apart from further checks of integrability beyond the limitations of closed sectors, the two-loop dilatation operator would also provide the eigenstates that correspond to the anomalous dimensions given by integrability.
Further two-loop results for minimal form factors would also allow for additional checks of our conjecture about the universality of the highest transcendental part of the form factor remainders. 
In addition, it might even be possible to prove this conjecture, cf.\ the discussion in section \ref{sec: remainder}.
A better understanding of the non-trivial behaviour of the minimal form factors under soft and collinear limits would also be desirable. 

In chapter \ref{chap: two-loop su(2) form factors}, we have used the Ward identity \eqref{eq: action on form factor} for the generators of $\SU{2}$ to relate various components in the loop corrections to the minimal form factors.
For general generators, however, these Ward identities are anomalous.
Corresponding corrections to (some of) the generators are known for amplitudes as well as for the spin-chain picture. 
For form factors, both kinds of corrections occur and can hence be studied.

At tree level, we have observed a relation between the top-cell diagrams for the $(n+2)$-point scattering amplitudes and the $n$-point form factors of the stress tensor supermultiplet. It would be interesting to prove this relation and to determine the exact number of top-cell diagrams that are required in the form factor case.
Moreover, it might provide further insights to see whether the above relation is related to the forward limit and the Lagrangian insertion technique.
In addition, a better understanding of the on-shell form that is integrated in the Graßmannian integral representation would be desirable. 
For amplitudes, the on-shell form is completely determined by having logarithmic divergences at all boundaries.
A more refined argument might also determine the on-shell form in the case of form factors.
For amplitudes, the Graßmannian integral representation has been generalised to the amplituhedron \cite{Arkani-Hamed:2013jha,Arkani-Hamed:2013kca,ArkaniHamed:2010gg}, which is also applicable at loop level.
A similar construction for form factors in terms of a ``formfactorhedron'' 
 should also be possible.
Moreover, it would be very interesting to extend on-shell diagrams for form factors to general operators. Different BCFW recursion relations that might serve as a basis of this construction were given for the operators $\tr(\phi_{14}^L)$ as well as for operators in the $\SU{2}$ and $\SL{2}$ sectors in \cite{Penante:2014sza} and \cite{Engelund:2012re}, respectively.
In particular, this would generalise the transfer-matrix identity to these form factors and should also allow for their construction via integrability and Graßmannian integrals.

From the construction of correlation functions via generalised unitarity \cite{Engelund:2012re}, it follows that on-shell diagrams can also be used to describe leading singularities of loop-order correlation functions. It would be interesting to extend this to the full correlation functions. 
Especially correlation functions of the chiral half of the stress-tensor supermultiplet have been intensively studied in the last years, see e.g.\ \cite{Chicherin:2015bza} and references therein.

Finally, form factors can also be calculated using the twistor action \cite{Boels:2006ir}, as will be shown in \cite{KMSW}.%

In the $\beta$- and $\gamma_i$-deformed theories, the important question whether these theories are also integrable beyond the asymptotic regime remains to be answered.
This would require to incorporate the finite-size effect of prewrapping into the integrability-based description.
In the conformal $\beta$-deformation with gauge group $\SU{N}$, a first step would be to obtain a 
 non-divergent and vanishing result for the anomalous dimension of $\tr(\phi_i\phi_j)$ with $1\leq i<j\leq 3$.
Applying the same procedure to the state $\tr(\phi_i\phi_i)$ with $i=1,2,3$ in the $\gamma_i$-deformation, a match with our field-theory result \eqref{eq: gamma O2 varrho} might also be achievable, 
at least for some choice of the tree-level coupling $Q^{ii}_{\text{F}\,ii}$ and the scheme $\rho$.
This choice should then be checked for other states, which might lead to the conclusion that integrability is also valid beyond conformality.
Also for this reason, it would be very interesting to identify further states that are affected by prewrapping and can hence serve as a testing ground. 

In chapter \ref{chap: nonconformality}, we have listed four possible interpretations of the non-conformality of the $\gamma_i$-deformation in the context of the AdS/CFT correspondence. 
It would be highly desirable to see which of them is actually realised.

It would also be very interesting to shed some further light on the role of the gauge group in the AdS/CFT correspondence.
The derivation \cite{Maldacena:1997re} of the AdS/CFT correspondence starts with the brane picture, where the massless part of the open string theory is a $\U{N}$ gauge theory.
In the undeformed setting, however, the $\U{1}$ mode decouples and the theories with gauge group $\U{N}$ and $\SU{N}$ are essentially the same.
In the deformed theories, this is no longer the case; see \cite{Frolov:2005iq} for a discussion. In particular, only the $\beta$-deformation with gauge group $\SU{N}$ is conformally invariant.
In \cite{Frolov:2005iq}, it was also argued that the gauge group in the deformed AdS/CFT correspondence should be $\SU{N}$, as the couplings to the $\U1$ components of the fields are non-vanishing but flow to zero in the IR.
Using the results of chapter \ref{chap: introduction to integrable deformations} -- \ref{chap: anomalous dimensions}, we can make a prediction at all orders of planar perturbation theory.
While the operators in the $20'$ of $\SO{6}$ are all protected in the undeformed theory, this degeneracy lifts in the deformed theories in a specific way.
The operators $\tr(\phi_{i}\phi_{i})$ and $\tr(\bar\phi^{i}\bar\phi^{i})$ with $i=1,2,3$ are protected in the $\beta$-deformation but have non-vanishing anomalous dimensions in the $\gamma_i$-deformation.
The anomalous dimensions of the operators $\tr(\phi_{i}\phi_{j})$ and $\tr(\bar\phi^i\bar\phi^{j})$ with $1\leq i<j\leq 3$ vanish in the $\beta$-deformation with gauge group $\SU{N}$ but are non-vanishing in the $\beta$-deformation with gauge group $\U{N}$.
The anomalous dimensions of $\tr(\bar\phi^{i}\phi_{j})$ with $i,j=1,2,3$, $i\neq j$, are non-vanishing in both deformed theories, while the anomalous dimensions of the traceless combinations $\tr(\bar\phi^{1}\phi_{1}-\bar\phi^{2}\phi_{2})$ and $\tr(\bar\phi^{2}\phi_{2}-\bar\phi^{3}\phi_{3})$ vanish in both deformed theories.
These predictions should be checked at strong coupling in the string theory in order to clarify the question of the gauge group.

\cleardoublepage
\phantomsection
\addcontentsline{toc}{chapter}{Acknowledgements}
\chapter*{Acknowledgements}
\markboth{Acknowledgements}{Acknowledgements}

It is a pleasure to thank my 
advisor Matthias Staudacher
for 
 offering me the opportunity to be part of his wonderful group at the vibrant surrounding of Humboldt-University Berlin,
 for sharing his knowledge and views,
 for inspiring discussions, encouragement, tea, cookies and
 for the opportunity to travel the world through physics.
Further, I am deeply grateful to 
my
coadvisor Christoph Sieg
for 
 sharing his knowledge,
 for stimulating discussions about physics and beyond,
 for countless cups of espresso
 as well as for always having had an open door.

I am indebted to 
 Jan Fokken,
 Rouven Frassek,
 Florian Loebbert,
 David Meidinger,
 Dhitiman Nandan,
 Christoph Sieg and
 Gang Yang 
for the fruitful, inspiring and enjoyable collaborations on the various project that are summarised in this thesis.

I am grateful to Malte Pieper, Christoph Sieg and Matthias Staudacher
for comments on the manuscript and suggestions on how to improve it.
I would like to thank 
 Jan Plefka,
 Matthias Staudacher and 
 Tristan McLoughlin
for acting as referees of this thesis.

I have greatly benefited from discussions and communications with 
 Zoltan Bajnok,
  Benjamin Basso,
  Niklas Beisert,
  David Berenstein,
  Zvi Bern,
  Massimo Bianchi,
  Sophia Borowka,
  Jacob Bourjaily,
  Andreas Brandhuber,
  Johannes Broedel,
  Simon Caron-Huot,
  Lance Dixon,
  Burkhard Eden,
  Oluf Engelund,
  Ewa Felinska,
  Livia Ferro, 
  Jan Fokken,
  Valentina Forini,
  Rouven Frassek,
  Sergey Frolov,
  Philip Hähnel,
  Johannes Henn,
  Paul Heslop,
  Nils Kanning,
  Vladimir Kazakov,
  Minkyoo Kim,
  Thomas Klose,
  Yumi Ko,
  Laura Koster,
  Charlotte Kristjansen,
  Kasper Larsen,
  Marius de Leeuw,
  S\'{e}bastien Leurent,
  Pedro Liendro,
  Georgios Linardopoulos, 
  Florian Loebbert,
  Tomasz \L{}ukowski,
  Christian Marboe,
  Lionel Mason,
  David Meidinger,
  Vladimir Mitev,
  Dhitiman Nandan,
  Erik Panzer,
  Brenda Penante,
  Jan Plefka,
  Elli Pomoni,
  Gregor Richter,
  Radu Roiban,
  Henning Samtleben,
  Volker Schomerus,
  Alessandro Sfondrini,
  Christoph Sieg, 
  Vladimir Smirnov,
  Mads Søgaard,
  Marcus Spradlin, 
  Martin Sprenger,
  Matthias Staudacher,
  Ryo Suzuki,
  Stijn van Tongeren,
  Alessandro Torrielli,
  Gabriele Travaglini,
  Vitaly Velizhanin,
  Christian Vergu,
  Pedro Viera,
  Gang Yang,
  Anastasia Volovich and
  Konstantinos Zoubos
 --- both on various aspects of the topics covered in this work and beyond. 
Moreover, I 
 thank 
Camille Boucher-Veronneau, Lance Dixon and Jeffrey Pennington for sharing their unpublished notes.

I would like to thank everyone in the group at Humboldt University Berlin for the good company, the movie evenings, the cakes, the music, and for having made my time in Berlin a very enjoyable one.

I am grateful to
 the Kavli Institute for the Physics and Mathematics of the Universe in Tokyo,
 the Simons Center for Geometry and Physics, 
 the C.N.\ Yang Institute for Theoretical Physics,
 the Queen Mary University of London,
 the Niels Bohr Institute, 
 the University of Oxford and 
 the ETH Zürich
for warm hospitality during different stages of my PhD studies.

This work was supported by the 
 DFG, SFB 647 \emph{Raum -- Zeit -- Materie. Analytische und Geometrische Strukturen}, the 
 Marie Curie network GATIS (gatis.desy.eu) of the European Union's Seventh Framework Programme FP7/2007-2013/ under REA Grant Agreement No 317089 and the
 Marie Curie International Research Staff Exchange Network UNIFY (FP7-People-2010-IRSES under Grant Agreement No 269217).
 Ich danke der Studienstiftung des deutschen Volkes f\"ur ein Promotionsf\"orderungsstipendium.

Ich danke meinen Freunden für ihre Unterstützung und dafür, dass sie während der Zeit der Doktorarbeit auch für die gelegentlich notwendige Ablenkung gesorgt haben.
Insbesondere danke ich Inka, Malte und Patrick, die mich auf der abenteuerlichen Reise des Physikstudiums von Anfang an begleitet haben.
Ich danke meinen Eltern und meiner Schwester für die konstante Unterstützung, die ich in meinem Studium und Leben durch sie erfahren habe.
Zu guter Letzt danke ich 
 meiner Theresa.

\appendix

\chapter{Feynman integrals}
\label{appchap: integrals}

In this appendix, we summarise our conventions for Feynman integrals and the lifting procedure. Moreover, we provide explicit expressions of Passarino-Veltman reductions as well as several Feynman integrals that occur throughout this thesis.

\section{Conventions and lifting}
\label{appsec: lifting}

Using Feynman diagrams, the following combination of the Yang-Mills coupling $g_\YM$, the number of colours $N$ and the 't Hooft mass $\mu$ occurs at $\ell$-loop order of planar perturbation theory:
\begin{equation}
\label{eq: integral-measure-relation}
(g_\YM \mu^\peps)^{2\ell}N^{\ell} (-i)^\ell \int\frac{\de^Dl_1}{(2\pi)^D}\cdots\frac{\de^Dl_\ell}{(2\pi)^D}
\frac{f(l_1,\dots,l_\ell)}{\prod_j D_j}
=\gmod^{2\ell} I^{(\ell)}[f(l_1,\dots,l_\ell)]\eqncom
\end{equation}
where $\gmod$ is the modified effective planar coupling constant \eqref{eq: modified effective planar coupling constant} and 
\begin{equation}
\label{eq: Ielldef}
I^{(\ell)}[f(l_1,\dots,l_\ell)]=(\e^{\gamma_{\text{E}}}\mu^2)^{\ell\peps}\int\frac{\de^Dl_1}{i\pi^{\frac{D}{2}}}\cdots\frac{\de^Dl_\ell}{i\pi^{\frac{D}{2}}}
\frac{f(l_1,\dots,l_\ell)}{\prod_j D_j}
\eqndot
\end{equation}
In these expressions, $f(l_1,\dots,l_\ell)$ denotes a polynomial in the loop momenta and $D_j=k_j^2$ denote the inverse propagators, where $k_j$ is the combination of external momenta and loop momenta that flow through the propagator.

The lifting procedure of unitarity employed in chapters \ref{chap: minimal one-loop form factors}, \ref{chap: two-loop Konishi form factor} and \ref{chap: two-loop su(2) form factors} consists of two steps.
First, we have to undo the replacements \eqref{eq: cut replacement} by setting 
\begin{equation}
2\pi \delta^+(l_i^2) \to \frac{i}{l_i^2} \eqndot
\end{equation}
Second, we have to change the coupling constant and the measure factor such that the uncut integrals have the form given in \eqref{eq: Ielldef}.
For example, in the one-loop case \eqref{eq: su2 XXXX 2}, we have
\begin{equation}
g_{\YM}^2 N \twopst \frac{ s_{12}}{(l_1-p_1)^2} 
= g_{\YM}^2 N s_{12} \FDinline[triangle, doublecut,cutlabels, twolabels, labelone=p_1, labeltwo=p_2]  
\ \overset{\scriptstyle\text{lifting}}{\longrightarrow} \ - i \gmod ^2 s_{12} \FDinline[triangle, twolabels, labelone=p_1, labeltwo=p_2] \eqndot
\end{equation}
Note that this procedure includes the continuation of the expression from $D=4$ to $D=4-2\peps$, which is not always unique.
For the Lorentz vectors, this issue is discussed in \cite{Bern:1994cg}. 
For the flavour degrees of freedom, a different issue arises, which is discussed in section \ref{sec: subtleties}.

\section{Passarino-Veltman reduction}
\label{appsec: PV reduction}

In loop-level calculations using on-shell methods or Feynman diagrams, Feynman integrals with uncontracted loop momenta $l_i^\mu$ in the numerator can occur.
These can be reduced to scalar integrals or integrals with fully contracted loop momenta in the numerator using 
Passarino-Veltman (PV) reduction \cite{Passarino:1978jh}.

Let us illustrate this reduction for the simple case of the linear bubble integral at one-loop order:
\begin{equation}
 l^\mu\FDinline[bubble,momentum, twolabels, labelone=p_1, labeltwo=p_2]
={(\e^{\gamma_\text{E}}\mu^2)^{\peps}} \int\frac{\de^Dl}{i\pi^{\frac{D}{2}}}\frac{l^\mu}{ l^2 (l+q)^2} \eqncom
\end{equation}
where $q=p_1+p_2$ and the momentum-dependent prefactor is understood to appear inside of the depicted integral as shown.
Lorentz symmetry requires that the resulting expression after integration is proportional to the single external Lorentz vector that occurs in the integral, i.e.\
\begin{equation}
\label{eq: ansatz PV reduction of linear bubble}
 l^\mu \FDinline[bubble,momentum, twolabels, labelone=p_1, labeltwo=p_2]= A \, q^\mu\eqndot
\end{equation}
In order to fix the constant of proportionality $A$, we contract both sides of \eqref{eq: ansatz PV reduction of linear bubble} with $q_\mu$.
We find 
\begin{equation}
\begin{aligned}
 A \,q^2&={(\e^{\gamma_\text{E}}\mu^2)^{\peps}} \int\frac{\de^Dl}{i\pi^{\frac{D}{2}}}\frac{l\cdot q}{ l^2 (l+q)^2}
 =\frac{1}{2}{(\e^{\gamma_\text{E}}\mu^2)^{\peps}} \int\frac{\de^Dl}{i\pi^{\frac{D}{2}}}\frac{(l+q)^2-l^2-q^2}{ l^2 (l+q)^2}\\
 &=-\frac{q^2}{2}\FDinline[bubble, twolabels, labelone=p_1, labeltwo=p_2]\eqncom
\end{aligned}
 \end{equation}
where the two additional terms with $l$-dependent numerators in the second-to-last step drop out as they lead to massless tadpole integrals, which integrate to zero.
Hence,
\begin{equation}
\label{eq: linear bubble PV}
 l^\mu \FDinline[bubble,momentum, twolabels, labelone=p_1, labeltwo=p_2]
=-\frac{q^\mu}{2} \FDinline[bubble, twolabels, labelone=p_1, labeltwo=p_2] \eqndot
\end{equation}

Using a similar procedure, we can also reduce more complicated tensor integrals with higher loop order as well as higher power of the loop momenta in the numerator.
For two powers of the loop momentum $l^\mu l^\nu$ in the numerator, also the four-dimensional metric $g^{\mu\nu}$ occurs in the ansatz for the reduction.
Contracting its inverse with the momenta yields
\begin{equation} 
\label{eq: l_peps}
g_{\mu \nu} l^\mu l^\nu = l_{(4)}^2 = l^2 + l_\peps^2 \eqncom
\end{equation}
where $l_{(4)}$ denotes the four-dimensional part of the loop momentum and $l_\peps$ its $(-2\peps)$-dimensional part.%
\footnote{In the decomposition of $l$ into $l_{(4)}$ and $l_\peps$, we have assumed that $\peps<0$. The sign in \eqref{eq: l_peps} is due to the mostly-minus metric we are using.
}
For instance, the PV reduction of the tensor-two one-mass triangle integral yields
\begin{equation}
\begin{aligned}
 l^\mu l^\nu \FDinline[triangle,momentum, twolabels, labelone=p_1, labeltwo=p_2]
 &=p_1^\mu p_1^\nu\FDinline[triangle, twolabels, labelone=p_1, labeltwo=p_2]+\left(\frac{g^{\mu\nu}}{2}-\frac{p_1^\mu p_2^\nu+p_1^\nu p_2^\mu}{s_{12}}\right)l_\peps^2\FDinline[triangle,momentum, twolabels, labelone=p_1, labeltwo=p_2] \\
 &\phaneq
 +\left(\frac{g^{\mu\nu}}{4}+\frac{3p_1^\mu p_1^\nu}{2s_{12}}-\frac{p_2^\mu p_2^\nu}{2s_{12}}-\frac{p_1^\mu p_2^\nu+p_1^\nu p_2^\mu}{s_{12}}\right)\FDinline[bubble, twolabels, labelone=p_1, labeltwo=p_2] \eqncom
\end{aligned}
\end{equation}
where the one-mass triangle integral with numerator $l_\peps^2$ evaluates to a finite rational function \cite{Bern:1995ix}:
\begin{equation}
 l_\peps^2\FDinline[triangle,momentum, twolabels, labelone=p_1, labeltwo=p_2]=\frac{1}{2} + \cO(\peps) \eqndot
\end{equation}

\section{Selected integrals}
\label{appsec: selected integrals}

In this section, we provide explicit expressions for several one- and two-loop integrals that are required throughout this work.

At one-loop order, we require the bubble integral as well as the one-mass triangle integral. These are given by 
\begin{align}
\label{eq: oneloopint:bubble}
\FDinline[bubble, twolabels, labelone=p_1, labeltwo=p_2]
&={(\e^{\gamma_\text{E}}\mu^2)^{\peps}} \int\frac{\de^Dl}{i\pi^{\frac{D}{2}}}\frac{1}{ l^2 (l+q)^2}
= \e^{\gamma_\text{E} \peps} \, \frac{\Gamma(1-\peps)^2\Gamma(1+\peps)}{\Gamma(1-2\peps)}{1 \over \peps (1-2\peps)} \Big(-\frac{q^2}{\mu^2} \Big)^{-\peps}
\!
\eqncom\\
\label{eq: oneloopint:1masstriangle}
\FDinline[triangle, twolabels, labelone=p_1, labeltwo=p_2] 
&={(\e^{\gamma_\text{E}}\mu^2)^{\peps}} \int\frac{\de^Dl}{i\pi^{\frac{D}{2}}}
\frac{1}{(l+p_1)^2 l^2 (l-p_2)^2} \\
&= - \e^{\gamma_\text{E} \peps} \, \frac{\Gamma(1-\peps)^2\Gamma(1+\peps)}{\Gamma(1-2\peps)}{1 \over \peps^2} {1\over (-q^2)} \Big(-\frac{q^2}{\mu^2} \Big)^{-\peps} 
\eqncom \nonumber
\end{align}
cf.\ e.g.\ \cite{Smirnov:2004ym}.
Both these integrals depend on the single scale $q^2=(p_1+p_2)^2$.

At two-loop order, it is advantageous to reduce the occurring integrals via integration-by-part (IBP) identities as implemented e.g.\ in the \texttt{Mathematica} package \texttt{LiteRed} \cite{Lee:2013mka}.
The two-loop one-scale integrals required in chapter \ref{chap: two-loop Konishi form factor} can be reduced as
\begin{align}
\label{eq: two-loop fish integral}
\FDinline[fishtop,twolabels,labelone=p_1,labeltwo=p_2] 
&= {2 - 3 \peps \over \peps (-q^2)} \FDinline[sunrise,twolabels,labelone=p_1,labeltwo=p_2]\eqncom
\\
(q^2)^2\FDinline[rainbow, twolabels, labelone=p_1,labeltwo=p_2] 
&=  -{3(1-2 \peps) (1-3\peps) (2 - 3 \peps) \over \peps^3 (-q^2)} \FDinline[sunrise,twolabels,labelone=p_1,labeltwo=p_2] \\* \nonumber
& \phantom{{}={}} +  {3(1-2 \peps) (1-3\peps) \over 2\peps^2} \FDinline[itwo,twolabels,labelone=p_1,labeltwo=p_2] 
+  {(1 - 2 \peps)^2 \over \peps^2} \bigg( \FDinline[bubble,twolabels,labelone=p_1,labeltwo=p_2] \bigg)^2\eqncom \\
s_{1l}s_{2l}
\FDinline[rainbow,momentum,twolabels,labelone=p_1,labeltwo=p_2]
&= {(2 - 3 \peps)(2-9\peps+10\peps^2 - 4\peps^3) \over (1-\peps) (1-2\peps) \peps^2 (-q^2)} \FDinline[sunrise,twolabels,labelone=p_1,labeltwo=p_2] \\* \nonumber
& \phantom{{}={}} -  {1-4 \peps + 2\peps^2 \over (1-\peps)\peps} \FDinline[itwo,twolabels,labelone=p_1,labeltwo=p_2] 
-  {2- 3 \peps +2 \peps^2 \over 2 (1-\peps) \peps} \bigg(\FDinline[bubble,twolabels,labelone=p_1,labeltwo=p_2] \bigg)^2\eqncom
\\
s_{1l}s_{2l}
\FDinline[rainbownonplanar,momentum,twolabels,labelone=p_1,labeltwo=p_2]
&= {(1-2 \peps) (2-3\peps) (3 - 5\peps) \over \peps^2(1-4\peps) (-q^2) } \FDinline[sunrise,twolabels,labelone=p_1,labeltwo=p_2] \\* \nonumber
& \phantom{{}={}} - {(1+ \peps) (1-2\peps) \over \peps (1-4\peps)} \FDinline[itwo,twolabels,labelone=p_1,labeltwo=p_2] -  {\peps \, (-q^2)^2 \over (1-4\peps)} \FDinline[rainbownonplanar, twolabels, labelone=p_1,labeltwo=p_2]
\eqndot
\end{align}
The occurring master integrals are \cite{Gehrmann:2005pd}
\begin{equation}
\begin{aligned}
\FDinline[sunrise,twolabels,labelone=p_1,labeltwo=p_2] &= \e^{2\gamma_\text{E}\peps} \frac{\Gamma(1-\peps)^3\Gamma(1+2\peps)}{2\peps (1-2\peps)\Gamma(3-3\peps)}  (-q^2) \left(-{q^2 \over \mu^2} \right)^{-2\peps}  \eqncom
\\
\FDinline[itwo,twolabels,labelone=p_1,labeltwo=p_2] &= \e^{2\gamma_\text{E}\peps} \frac{\Gamma(1-\peps)^2\Gamma(1+\peps)\Gamma(1+2\peps)\Gamma(1-2\peps)}{2\peps^2 (1-2\peps)\Gamma(2-3\peps)}  \left(-{q^2 \over \mu^2} \right)^{-2\peps} \eqncom
\\
\FDinline[rainbownonplanar, twolabels, labelone=p_1,labeltwo=p_2] &= 
\e^{2\gamma_\text{E}\peps} {1\over (-q^2)^2} \left(-{q^2 \over \mu^2} \right)^{-2\peps} \bigg[ \frac{\Gamma(1-2\peps)^4\Gamma(1+2\peps)^3\Gamma(1-\peps)\Gamma(1+\peps)}{\peps^4 (1-4\peps)^2\Gamma(1+4\peps)} \\
& \phaneq+ \frac{4\Gamma(1-\peps)^2\Gamma(1-2\peps)\Gamma(1+2\peps)}{\peps^2 (1+\peps)(1+2\peps)\Gamma(1-4\peps)} \,_3 F_2\big(1,1,1+2\peps; 2+\peps, 2+2\peps; 1 \big)  \\
& \phaneq+ \frac{\Gamma(1-\peps)^2\Gamma(1+\peps)\Gamma(1-2\peps)\Gamma(1+2\peps)}{2\peps^4 \Gamma(1-3\peps)} \,_3 F_2\big(1,-4\peps,-2\peps; 1-3\peps, 1-2\peps; 1 \big)  \\
& \phaneq+ \frac{\Gamma(1-\peps)^3\Gamma(1+2\peps)}{2\peps^4 \Gamma(1-3\peps)} \,_4 F_3\big(1,1-\peps, -4\peps, -2\peps; 1-3\peps, 1-2\peps, 1-2\peps; 1 \big) \bigg] \eqncom
\end{aligned}
\end{equation}
where $_p F_q$ denotes the respective generalised hypergeometric functions. 

In chapter \ref{chap: two-loop su(2) form factors}, also two-loop two- and three-scale integrals occur.
They can be reduced via IBP identities in a similar way. The corresponding master integrals are more lengthy and we hence refrain from showing them; they can be found in \cite{Gehrmann:2000zt}.

\chapter{Scattering amplitudes}
\label{app: scattering amplitudes}

In this appendix, we provide explicit expressions for several colour-ordered tree-level scattering amplitudes that are required throughout the first part of this thesis. We use the notation and conventions introduced in section \ref{sec: generalities about form factors}.

\section{MHV and \texorpdfstring{\MHVb}{MHVbar} amplitudes}
\label{appsec: MHV and MHVb amplitudes}

As explained in section \ref{sec: generalities about form factors}, super amplitudes can be classified according to their degree in the Graßmann variables $\etatt_i^A$. 
Super amplitudes with the minimal Graßmann degree, which is eight, are denoted as maximally-helicity-violating (MHV).
The $n$-point colour-ordered tree-level $\MHV$ super amplitude is given by 
\begin{equation}\label{eq: colour-ordered MHV super amplitude}
 \begin{aligned}
  \ampco^{\MHV\,(0)}_n(1,\dots,n)=\frac{i(2\pi)^4\delta^4(P)\delta^8(Q)}{\ab{12}\ab{23}\cdots\ab{n1}} \eqncom
 \end{aligned}
\end{equation}
where
\begin{equation}
 \delta^8(Q)=\frac{1}{2^4}\prod_{A=1}^4 \sum_{i,j=1}^{n} \epsilon_{\alpha\beta}{Q}_i^{\alpha A}{Q}^{\beta A}_j=\prod_{A=1}^4 \sum_{1\leq i< j\leq n} \ab{ij}\etatt_i^A\etatt_j^A\eqndot
\end{equation}

Super amplitudes of the maximal degree in the Graßmann variables $\etatt_i^A$, which is $4n-8$, are denoted as $\MHVb$.
They can be obtained from the MHV super amplitudes by conjugation, i.e.\ by replacing $\lambda_i \to \lambdat_i$, $\lambdat_i \to \lambda_i$, $\etatt_i \to \etatt_i^*$ and applying the fermionic Fourier transformation 
$\prod_{i=1}^n \int \de^4 \etatt_i^*  \e^{\etatt_i^C \etatt_{i,C}^*}$.%
\footnote{Here, we have assumed all on-shell fields 
 to have positive energy.
For negative energy, we have $\lambda_i \to -\lambdat_i$ and $\lambdat_i \to -\lambda_i$ as discussed in section \ref{sec: generalities about form factors}.
}
For example, the three-point colour-ordered tree-level $\MHVb$ super amplitude is given by  
\begin{equation}
 \begin{aligned}
  \ampco^{\MHVb\,(0)}_3(1,2,3)=\frac{(-i)(2\pi)^4\delta^4(P)\delta^4(\bar{Q})}{\sb{12}\sb{23}\sb{31}} \eqncom
 \end{aligned}
\end{equation}
where
\begin{equation}
 \delta^4(\bar{Q})=\prod_{A=1}^4  \left(\sb{12}\etatt_3^A+\sb{23}\etatt_1^A+\sb{31}\etatt_2^A\right) \eqndot
\end{equation}

\section{Scalar NMHV six-point amplitudes}
\label{appsec: scalar nmhv six-point amplitudes}

In this section, we collect some scalar NMHV six-point amplitudes at tree level, which are required in the unitarity calculation of chapter \ref{chap: two-loop su(2) form factors}.
We abbreviate the component amplitudes as 
\begin{equation}
\atreef{Z_6 Z_5 Z_4}{Z_1 Z_2 Z_3} = \ampco_6^{(0)}(1^{Z_1}, 2^{Z_2}, 3^{Z_3}, 4^{\bar Z_4}, 5^{\bar Z_5}, 6^{\bar Z_6}) \eqndot
\end{equation}
Using MHV or BCFW recursion relation and simplifying the result such that only Mandelstam variables occur in it, we find
\begin{equation}
\label{eq: six- point amplitudes}
\begin{aligned}
\atreef{YXX}{XXY} = & - {1\over s_{234}} \eqncom \\
\atreef{XXY}{YXX} = & - {1\over s_{345}} \eqncom \\
\atreef{XXX}{XXX} = & {s_{23} s_{56} \over s_{16} s_{34} s_{234}} + {s_{12} s_{45} \over s_{16} s_{34} s_{345}} - {s_{123} \over s_{16} s_{34}} \eqncom \\
\atreef{XXY}{XYX} = & {s_{12} \over s_{16} s_{345}} + {s_{56} \over s_{16} s_{234}} - {1\over s_{16}} + {1\over s_{345}} \eqncom \\
\atreef{YXX}{XYX} = & {s_{23} \over s_{34} s_{234}} + {s_{45} \over s_{34} s_{345}} - {1\over s_{34}} + {1\over s_{234}} \eqncom \\
\atreef{XYX}{XXY} = & {s_{12} \over s_{16} s_{345}} + {s_{56} \over s_{16} s_{234}} - {1\over s_{16}} + {1\over s_{234}} \eqncom \\
\atreef{XYX}{YXX} = & {s_{23} \over s_{34} s_{234}} + {s_{45} \over s_{34} s_{345}} - {1\over s_{34}} + {1\over s_{345}} \eqncom \\
\atreef{XXY}{XXY} = & - {s_{23} s_{56} \over s_{16} s_{34} s_{234}} - {s_{12} s_{45} \over s_{16} s_{34} s_{345}} +  {s_{123} \over s_{16} s_{34}} - {s_{12} \over s_{16} s_{345}} - {s_{56} \over s_{16} s_{234}} + {1\over s_{16}} \eqncom \\
\atreef{YXX}{YXX} = & - {s_{23} s_{56} \over s_{16} s_{34} s_{234}} - {s_{12} s_{45} \over s_{16} s_{34} s_{345}} +  {s_{123} \over s_{16} s_{34}} - {s_{23} \over s_{34} s_{234}} - {s_{45} \over s_{34} s_{345}} + {1\over s_{34}} \eqncom \\
\atreef{XYX}{XYX} = & - {s_{23} s_{56} \over s_{16} s_{34} s_{234}} - {s_{12} s_{45} \over s_{16} s_{34} s_{345}} +  {s_{123} \over s_{16} s_{34}} \\ 
& - {s_{12} \over s_{16} s_{345}} - {s_{56} \over s_{16} s_{234}} + {1\over s_{16}} - {1\over s_{345}} - {s_{23} \over s_{34} s_{234}} - {s_{45} \over s_{34} s_{345}} + {1\over s_{34}} - {1\over s_{234} } \eqncom
\end{aligned}
\end{equation}
where we have suppressed a factor of $i(2\pi)^4\delta^4(\sum_{i=1}^6p_i)$ in each amplitude.

\chapter{Deformed theories}
\label{app: deformed theories}

In this appendix, we provide some details on the renormalisation of fields, couplings and composite operators in the deformed theories. 
Moreover, we give the self energies of the scalar fields.
The presentation in this appendix is based on \cite{Fokken:2013aea,Fokken:2014soa}.

\section{Renormalisation}
\label{appsec: renormalisation group equations}

In this section, we discuss the renormalisation of fields, couplings and composite operators. In particular, we discuss the coefficients occurring in the renormalisation group equations (RGEs), i.e.\ the beta functions and anomalous dimensions. We focus on the cases required for chapters \ref{chap: prewrapping}--\ref{chap: anomalous dimensions} and refer the reader to \cite{ZinnJustin:2002ru,Collins:1984xc} for general treatments. 

Some of the most basic quantities that in general require renormalisation are the elementary fields themselves.
Although the corresponding self energies vanish in the supersymmetric formulation employed in the first part of this thesis, they are  non-vanishing in a component formulation, in particular in the one used in the second part of this work.
For our purpose, it is sufficient to look at the scalar fields $\phi_i$.
We define the renormalised field $\phi_i$ in terms of the bare field $\phi_{i\,0}$ and the renormalisation constant $\mathcal{Z}_{\phi_i}$ as 
\begin{equation}
 \begin{aligned}
  \phi_i&=\mathcal{Z}_{\phi_i}^{-\frac{1}{2}}\phi_{i\,0}
\eqndot
 \end{aligned}
\end{equation}
The latter can be expanded in terms of its counterterm as
\begin{equation}
\mathcal{Z}_{\phi_i}=1+\delta_{\phi_i}\eqndot
\end{equation}
We calculate $\delta_{\phi_i}$ up to one-loop order in the next section.
Note that in theories with gauge group $\U{N}$, we also have to distinguish between the self energies of the $\SU{N}$ and the $\U{1}$ components.
In our conventions, traces of more than one field are always understood to contain only the $\SU{N}$ components of the fields, cf.\ section \ref{sec: multi-trace couplings}. Hence, only the $\SU{N}$ self energies occur in the context of such traces.

The renormalised Yang-Mills coupling $g_{\YM}$ is defined in terms of its bare counterpart by a rescaling with the 't Hooft mass $\mu$, which sets the renormalisation scale:
\begin{equation}
 g_{\YM}=\mu^{-\peps} g_{\YM\,0} \eqndot
\end{equation}
Its beta function is defined as 
\begin{equation}
 \beta_{g_\YM}=\mu\deriv{\mu}g_{\YM} \eqndot
\end{equation}
Using that the bare Yang-Mills coupling $g_{\YM\,0}$ is independent of $\mu$, we find 
\begin{equation}
\begin{aligned}
0&=\mu\deriv{\mu}g_{\YM\,0}=\Big(\mu\parderiv{\mu}+\beta_{g_\YM}\parderiv{g_\YM}\Big)\mu^\peps g_\YM=
\mu^\peps(\peps g_\YM+\beta_{g_\YM})\eqncom\\
\end{aligned}
\end{equation}
and hence 
\begin{equation}
\label{eq: beta g YM}
\beta_{g_\YM}=-\peps g_{\YM}
\eqndot
\end{equation}
Note that $\beta_{g_\YM}$ vanishes for $D=4$ as is required for conformal invariance.

Next, we turn to the renormalisation of the running coupling \eqref{eq: running couplings}.
Writing it once in terms of bare quantities and once in terms of renormalised quantities and counterterms, we have
\begin{equation}\label{eq: zombiekillerrel}
-\frac{g_{\YM\,0}^2}{N}Q_{0\,\text{F}\,ii}^{ii}\tr(\bar\phi_{0}^{i}\bar\phi_{0}^{i})\tr(\phi_{i\,0}\phi_{i\,0})
=-\frac{\mu^{2\peps}g_\YM^2}{N}(Q_{\text{F}\,ii}^{ii}+\delta Q_{\text{F}\,ii}^{ii})\tr(\bar\phi^i\bar\phi^i)\tr(\phi_i\phi_i)\eqncom
\end{equation}
where $i=1,2,3$ is not summed over.
The renormalised coupling is related to the bare coupling as 
\begin{equation}
\begin{aligned}\label{eq: Q  renormalisation}
Q_{\text{F}\,ii}^{ii}&=\mathcal{Z}_{Q_{\text{F}\,ii}^{ii}}^{-1}Q_{0\,\text{F}\,ii}^{ii}
\eqndot
\end{aligned}
\end{equation}
The renormalisation constant $\mathcal{Z}_{Q_{\text{F}\,ii}^{ii}}$ can explicitly be written as
\begin{equation}
\mathcal{Z}_{Q_{\text{F}\,ii}^{ii}}
=\left(1+(Q_{\text{F}\,ii}^{ii})^{-1}\delta Q_{\text{F}\,ii}^{ii}\right)\mathcal{Z}_{\phi_i}^{-2}
\eqndot
\end{equation}

The beta function of this coupling is defined as 
\begin{equation}
\beta_{Q_{\text{F}\,ii}^{ii}}=\mu\deriv{\mu}Q_{\text{F}\,ii}^{ii}
\eqncom
\end{equation}
and, in analogy to the case of $g_\YM$, we have
\begin{equation}
\begin{aligned}
0&=\mu\deriv{\mu}Q_{0\,\text{F}\,ii}^{ii}
=Q_{\text{F}\,ii}^{ii}\Big(\beta_{g_\YM}\parderiv{g_\YM}+\beta_{Q_{\text{F}\,ii}^{ii}}\parderiv{Q_{\text{F}\,ii}^{ii}}\Big)\mathcal{Z}_{Q_{\text{F}\,ii}^{ii}}+\mathcal{Z}_{Q_{\text{F}\,ii}^{ii}}\beta_{Q_{\text{F}\,ii}^{ii}}
\eqndot
\end{aligned}
\end{equation}
Inserting \eqref{eq: beta g YM}, we find
\begin{equation}
\begin{aligned}
0
&=Q_{\text{F}\,ii}^{ii}\Big(-\peps g_\YM\parderiv{g_\YM}+\beta_{Q_{\text{F}\,ii}^{ii}}\parderiv{Q_{\text{F}\,ii}^{ii}}\Big)\log\mathcal{Z}_{Q_{\text{F}\,ii}^{ii}}+\beta_{Q_{\text{F}\,ii}^{ii}}
\eqndot
\end{aligned}
\end{equation}
This equation can be solved order by order. At lowest order, we have%
\footnote{As in chapters \ref{chap: nonconformality} and \ref{chap: anomalous dimensions}, we include an appropriate factor of the coupling constant in the definition of $\ell$-loop quantities.}
\begin{equation}
\label{eq: beta Q lowest order}
\begin{aligned}
\beta_{Q_{\text{F}\,ii}^{ii}}
&=Q_{\text{F}\,ii}^{ii}\peps g_\YM\parderiv{g_\YM}\delta\mathcal{Z}_{Q_{\text{F}\,ii}^{ii}}^{(1)} + \cO(g_\YM^4)
\eqndot
\end{aligned}
\end{equation}

Now, let us come to the renormalisation of gauge-invariant local composite operators. For the sake of concreteness, we focus on $\cO_L=\tr(\phi_i^L)$. Composite operators can be regarded as external states and can be added to the action with appropriate source terms $J_{\mathcal{O}_L}$. Writing this once in terms of bare quantities and once in terms of renormalised quantities and counterterms, we have
\begin{equation}
\label{eq: S OL}
\delta S_{\mathcal{O}_L}
=\int\de^Dx\,J_{\mathcal{O}_L,0}\,\mathcal{O}_{L,0}(\phi_{i\,0})
=\int\de^Dx\,J_{\mathcal{O}_L}\big[\mathcal{O}_L(\phi_i)+\delta J_{\mathcal{O}_L}\mathcal{O}_L(\phi_i)\big]
\eqndot
\end{equation}
The renormalisation constant of the source term is defined via
\begin{equation}
\begin{aligned}\label{eq: source renormalisation}
J_{\mathcal{O}_L}=\mathcal{Z}_{J_{\mathcal{O}_L}}^{-1}J_{\mathcal{O}_L,0}\eqncom \qquad \mathcal{Z}_{J_{\mathcal{O}_L}}=\mathcal{Z}_{\mathcal{O}_L,1\text{PI}}\mathcal{Z}_{\phi_i}^{-\frac{L}{2}}
\eqndot
\end{aligned}
\end{equation}
Its one-particle-irreducible (1PI) contribution is related to the counterterm as
\begin{equation}
\begin{aligned}\label{Zsource}
\mathcal{Z}_{\mathcal{O}_L,1\text{PI}}=1+\delta J_{\mathcal{O}_L}\eqndot
\end{aligned}
\end{equation}

Note that the renormalisation of the source term is equivalent to the renormalisation of the operator
\begin{equation}
\mathcal{O}_L(\phi_i)=\mathcal{Z}_{\mathcal{O}_L}\mathcal{O}_{L,0}(\phi_{i\,0})
\eqncom
\end{equation}
which we have been using in the first part of this work.
Furthermore, from
\begin{equation}\label{eq: JO}
J_{\mathcal{O}_L,0}\,\mathcal{O}_{L,0}(\phi_{i\,0})=J_{\mathcal{O}_L}\mathcal{O}_L(\phi_i)
\eqncom
\end{equation}
we find 
\begin{equation}\label{ZOL}
\mathcal{Z}_{\mathcal{O}_L}=\mathcal{Z}_{J_{\mathcal{O}_L}} =\mathcal{Z}_{\mathcal{O}_L,1\text{PI}}\mathcal{Z}_{\phi_i}^{-\frac{L}{2}}
\eqndot
\end{equation}
At the first two loop orders, we have 
\begin{equation}\label{eq: ZOLexpansion}
\begin{aligned}
\delta\mathcal{Z}^{(1)}_{\mathcal{O}_L}&=\delta J^{(1)}_{\mathcal{O}_L}-\frac{L}{2}\delta^{(1)}_{\phi^i}\eqncom\\
\delta\mathcal{Z}^{(2)}_{\mathcal{O}_L}&=\delta J^{(2)}_{\mathcal{O}_L}
-\frac{L}{2}\delta^{(2)}_{\phi^i}-\frac{L}{2}\delta^{(1)}_{\phi^i}\Big(\delta J^{(1)}_{\mathcal{O}_L}-\frac{L+2}{4}\delta^{(1)}_{\phi^i}\Big)
\eqndot
\end{aligned}
\end{equation}

The anomalous dimension $\gamma_{\cO_L}$ is defined as 
\begin{equation}
\label{eq: definition of gamma}
\gamma_{\mathcal{O}_L}=-\mu\deriv{\mu}\log\mathcal{Z}_{\mathcal{O}_L}
\eqndot
\end{equation}
Using \eqref{eq: beta g YM}, we find
\begin{equation}
\label{eq: gamma OL def}
\gamma_{\mathcal{O}_L}=\Big(\peps g_\YM\parderiv{g_\YM}-\beta_{Q_{\text{F}\,ii}^{ii}}\parderiv{{Q_{\text{F}\,ii}^{ii}}}\Big)\log\mathcal{Z}_{\mathcal{O}_L}
=\Big(\peps \g\parderiv{\g}-\beta_{Q_{\text{F}\,ii}^{ii}}\parderiv{{Q_{\text{F}\,ii}^{ii}}}\Big)\log\mathcal{Z}_{\mathcal{O}_L}
\eqncom
\end{equation}
which generalises the relation \eqref{eq: Z in terms of D} used in the first part of this thesis and becomes important in section \ref{sec: L eq 2}, where $\cZ_{\mathcal{O}_2}$ depends on $Q_{\text{F}\,ii}^{ii}$ in addition to $g_\YM$.

\section{One-loop self energies}
\label{appsec: one-loop self energies}

In the following, we compute the one-loop self energies of the scalar fields.

The UV divergences of the corresponding Feynman diagrams are
\begin{equation}
\label{eq: one- loop scalar self-energy diagrams}
\begin{aligned}
\Kop\Big[\swfone[
\fmfiv{label=$\scriptstyle i a$,l.dist=2}{vloc(__v1)}
\fmfiv{label=$\scriptstyle j b$,l.dist=2}{vloc(__v2)}
]{plain_ar}{dashes_ar,left=1}{dashes_rar,left=1}
\Big]
&=-2p^2\frac{\g^2}{\peps}\delta_i^j\big[\big(ab\big)-\cos \gamma_i^-
\frac{1}{N}\big(a\big)\big(b\big)\big]
\eqncom\\
\Kop\Big[\swfone[
\fmfiv{label=$\scriptstyle i a$,l.dist=2}{vloc(__v1)}
\fmfiv{label=$\scriptstyle j b$,l.dist=2}{vloc(__v2)}
]{plain_ar}{dashes_rar,left=1}{dashes_ar,left=1}
\Big]
&=-2p^2\frac{\g^2}{\peps}\delta_i^j\big[\big(ab\big)
-\cos\gamma_i^+
\frac{1}{N}\big(a\big)\big(b\big)\big]
\eqncom\\
\Kop\Big[\swfone[
\fmfiv{label=$\scriptstyle i a$,l.dist=2}{vloc(__v1)}
\fmfiv{label=$\scriptstyle j b$,l.dist=2}{vloc(__v2)}
]{plain_ar}{photon,left=1}{plain_rar}
\Big]
&=p^2\frac{\g^2}{\peps}\delta_i^j(3-\alpha)\big[\big(ab\big)-\frac{1}{N}\big(a\big)\big(b\big)\big]
\eqncom
\end{aligned}
\end{equation}
where we have depicted the scalars by solid lines, the fermions by dashed lines and the gauge fields by wiggly lines. The charge flow is indicated by arrows, $\alpha$ is the gauge-fixing parameter and we have used the abbreviation \eqref{eq: abbreviation for gauge group generators}. We have separated the diagrams with exchanged fermions into those which contain a $\psi_4$ in the first line and those which do not contain a  $\psi_4$ in the second line.
The diagrams have been calculated using the Feynman rules of \cite{Fokken:2013aea}.

The one-loop counterterm for the $\SU{N}$ components of the scalar field is
\begin{equation}
\begin{aligned}
\label{eq: delta phi SUN}
\delta_{\phi_i}^{\SU{N},(1)}&=\frac{1}{p^2}\Kop\Big[
\settoheight{\eqoff}{$\times$}%
\setlength{\eqoff}{0.5\eqoff}%
\addtolength{\eqoff}{-7.5\unitlength}%
\raisebox{\eqoff}{%
\fmfframe(3,0)(3,0){%
\begin{fmfchar*}(20,15)
\fmfleft{v1}
\fmfright{v2}
\fmf{plain_ar}{v1,vl}
\fmf{phantom}{vl,vr}
\fmf{plain_ar}{vr,v2}
\fmffreeze
\fmfposition
\fmfiv{label=$\scriptstyle i a$,l.dist=2}{vloc(__v1)}
\fmfiv{label=$\scriptstyle j b$,l.dist=2}{vloc(__v2)}
\fmfcmd{pair vert[]; vert1 = vloc(__vl);}
\vacpolp[1]{vert1--vloc(__vr)}
\end{fmfchar*}}}
\Big]\Big|_{\delta_i^j(ab)}
&=-\frac{\g^2}{\peps}(1+\alpha)
\eqncom
\end{aligned}
\end{equation}
where the vertical bar prescribes to take the coefficient of the specified expression.

The one-loop counterterm for the $\U1$ component reads
\begin{equation}
\begin{aligned}
\label{eq: delta phi U1}
\delta_{\phi_i}^{\U1,(1)}&=\frac{1}{p^2}\Kop\Big[
\settoheight{\eqoff}{$\times$}%
\setlength{\eqoff}{0.5\eqoff}%
\addtolength{\eqoff}{-7.5\unitlength}%
\raisebox{\eqoff}{%
\fmfframe(3,0)(3,0){%
\begin{fmfchar*}(20,15)
\fmfleft{v1}
\fmfright{v2}
\fmf{plain_ar}{v1,vl}
\fmf{phantom}{vl,vr}
\fmf{plain_ar}{vr,v2}
\fmffreeze
\fmfposition
\fmfiv{label=$\scriptstyle i a$,l.dist=2}{vloc(__v1)}
\fmfiv{label=$\scriptstyle j b$,l.dist=2}{vloc(__v2)}
\fmfcmd{pair vert[]; vert1 = vloc(__vl);}
\vacpolp[1]{vert1--vloc(__vr)}
\end{fmfchar*}}}
\Big]\Big|_{\delta_i^j(ab)}
+\frac{1}{p^2}\Kop\Big[
\settoheight{\eqoff}{$\times$}%
\setlength{\eqoff}{0.5\eqoff}%
\addtolength{\eqoff}{-7.5\unitlength}%
\raisebox{\eqoff}{%
\fmfframe(3,0)(3,0){%
\begin{fmfchar*}(20,15)
\fmfleft{v1}
\fmfright{v2}
\fmf{plain_ar}{v1,vl}
\fmf{phantom}{vl,vr}
\fmf{plain_ar}{vr,v2}
\fmffreeze
\fmfposition
\fmfiv{label=$\scriptstyle i a$,l.dist=2}{vloc(__v1)}
\fmfiv{label=$\scriptstyle j b$,l.dist=2}{vloc(__v2)}
\fmfcmd{pair vert[]; vert1 = vloc(__vl);}
\vacpolp[1]{vert1--vloc(__vr)}
\end{fmfchar*}}}
\Big]\Big|_{\delta_i^j\frac{1}{N}(a)(b)}\\
&=-4\frac{\g^2}{\peps}\left(\sin\frac{\gamma_i^+}{2}+\sin\frac{\gamma_i^-}{2}\right)
\eqncom
\end{aligned}
\end{equation}
which vanishes for $\gamma^\pm_i=0$ as required.

\cleardoublepage
\phantomsection
\addcontentsline{toc}{chapter}{Bibliography}
\bibliographystyle{utcaps}
\bibliography{ThesisINSPIRE}

\end{fmffile}
\end{document}

%% file: deformationpicturemacro.tex
\newcommand{\svertex}[2]{%
\fmfiequ{#1}{point length(#2)/2 of (#2)}
}
\newcommand{\dvertex}[4][0.33]{%
\fmfiequ{#2}{point (0.5-(#1)/2)*length(#4) of (#4)}
\fmfiequ{#3}{point (0.5+(#1)/2)*length(#4) of (#4)}
}

\newgray{ogray}{0.85}
\newgray{hatchgray}{1}
\newgray{sgray}{0.8}
\newcommand{\olcolor}{ogray}
\newcommand{\olfillstyle}{crosshatch*}
\newcommand{\olhatchcolor}{hatchgray}
\newcommand{\orcolor}{ogray}
\newcommand{\orfillstyle}{crosshatch*}
\newcommand{\orhatchcolor}{hatchgray}
\newcommand{\sfillstyle}{solid}

\newlength{\unit}
\newlength{\rad}
\newlength{\roff}
\newlength{\ri}
\psset{xunit=\unit,yunit=\unit,runit=\unit}
\newlength{\dlinewidth}
\newlength{\linew}
\newlength{\blacklinew}
\newlength{\doublesep}
\psset{doublesep=\doublesep}
\psset{linewidth=\linew}
\newlength{\auxlen}
\newlength{\linearc}
\newlength{\xa}
\newlength{\ya}
\newlength{\xb}
\newlength{\yb}
\newlength{\xc}
\newlength{\yc}
\newlength{\xd}
\newlength{\yd}
\newlength{\xe}
\newlength{\ye}
\newlength{\xf}
\newlength{\yf}

\newlength{\yg}

\newlength{\yh}
\newcommand{\ulinsert}[3][white]{%
\setlength{\xa}{#2\unit}
\addtolength{\xa}{0\unit}
\setlength{\xb}{#2\unit}
\addtolength{\xb}{1.5\unit}
\setlength{\xc}{#2\unit}
\addtolength{\xc}{3\unit}
\setlength{\ya}{#3\unit}
\addtolength{\ya}{0.5\dlinewidth}
\setlength{\yb}{#3\unit}
\addtolength{\yb}{-0.5\dlinewidth}
\setlength{\yc}{#3\unit}
\addtolength{\yc}{-1.5\unit}
\psset{doubleline=false}
\pscustom[fillstyle=\sfillstyle,fillcolor=#1,linecolor=#1,linewidth=0pt]{%
\psline[liftpen=1,linearc=\linearc](\xc,\yb)(\xb,\yb)(\xb,\yc)
\psline[liftpen=1](\xb,\ya)(\xc,\ya)}
\pscustom[fillstyle=\olfillstyle,fillcolor=\olcolor,hatchcolor=\olhatchcolor,
linecolor=\olcolor,linewidth=\linew]{%
\psline[linearc=\linearc](\xb,\yc)(\xb,\ya)
\psline[liftpen=1,linearc=2\linearc](\xb,\ya)(\xa,\ya)(\xa,\yc)}
\psline[linearc=\linearc](\xb,\yc)(\xb,\yb)(\xc,\yb)
\psline[linearc=2\linearc](\xa,\yc)(\xa,\ya)(\xb,\ya)
\psline(\xb,\ya)(\xc,\ya)
}
\newcommand{\dlinsert}[3][white]{%
\setlength{\xa}{#2\unit}
\addtolength{\xa}{0\unit}
\setlength{\xb}{#2\unit}
\addtolength{\xb}{1.5\unit}
\setlength{\xc}{#2\unit}
\addtolength{\xc}{3\unit}
\setlength{\ya}{#3\unit}
\addtolength{\ya}{-0.5\dlinewidth}
\setlength{\yb}{#3\unit}
\addtolength{\yb}{0.5\dlinewidth}
\setlength{\yc}{#3\unit}
\addtolength{\yc}{1.5\unit}
\psset{doubleline=false}
\pscustom[fillstyle=\sfillstyle,fillcolor=#1,linecolor=#1,linewidth=0pt]{%
\psline[linearc=\linearc](\xc,\yb)(\xb,\yb)(\xb,\yc)
\psline(\xb,\ya)(\xc,\ya)}
\pscustom[fillstyle=\olfillstyle,fillcolor=\olcolor,hatchcolor=\olhatchcolor,
linecolor=\olcolor,linewidth=\linew]{%
\psline[linearc=\linearc](\xb,\yc)(\xb,\ya)
\psline[liftpen=1,linearc=2\linearc](\xb,\ya)(\xa,\ya)(\xa,\yc)}
\psline[linearc=\linearc](\xb,\yc)(\xb,\yb)(\xc,\yb)
\psline[linearc=2\linearc](\xa,\yc)(\xa,\ya)(\xb,\ya)
\psline(\xb,\ya)(\xc,\ya)
}
\newcommand{\drinsert}[3][white]{%
\setlength{\xa}{#2\unit}
\addtolength{\xa}{0\unit}
\setlength{\xb}{#2\unit}
\addtolength{\xb}{-1.5\unit}
\setlength{\xc}{#2\unit}
\addtolength{\xc}{-3\unit}
\setlength{\ya}{#3\unit}
\addtolength{\ya}{-0.5\dlinewidth}
\setlength{\yb}{#3\unit}
\addtolength{\yb}{0.5\dlinewidth}
\setlength{\yc}{#3\unit}
\addtolength{\yc}{1.5\unit}
\psset{doubleline=false}
\pscustom[fillstyle=\sfillstyle,fillcolor=#1,linecolor=#1,linewidth=0pt]{%
\psline[linearc=\linearc](\xc,\yb)(\xb,\yb)(\xb,\yc)
\psline(\xb,\ya)(\xc,\ya)}
\pscustom[fillstyle=\orfillstyle,fillcolor=\orcolor,hatchcolor=\orhatchcolor,
linecolor=\orcolor,linewidth=\linew]{%
\psline[linearc=\linearc](\xb,\yc)(\xb,\ya)
\psline[liftpen=1,linearc=2\linearc](\xb,\ya)(\xa,\ya)(\xa,\yc)}
\psline[linearc=\linearc](\xb,\yc)(\xb,\yb)(\xc,\yb)
\psline[linearc=2\linearc](\xa,\yc)(\xa,\ya)(\xb,\ya)
\psline(\xb,\ya)(\xc,\ya)
}
\newcommand{\urinsert}[3][white]{%
\setlength{\xa}{#2\unit}
\addtolength{\xa}{0\unit}
\setlength{\xb}{#2\unit}
\addtolength{\xb}{-1.5\unit}
\setlength{\xc}{#2\unit}
\addtolength{\xc}{-3\unit}
\setlength{\ya}{#3\unit}
\addtolength{\ya}{0.5\dlinewidth}
\setlength{\yb}{#3\unit}
\addtolength{\yb}{-0.5\dlinewidth}
\setlength{\yc}{#3\unit}
\addtolength{\yc}{-1.5\unit}
\psset{doubleline=false}
\pscustom[fillstyle=\sfillstyle,fillcolor=#1,linecolor=#1,linewidth=0pt]{%
\psline[linearc=\linearc](\xc,\yb)(\xb,\yb)(\xb,\yc)
\psline[liftpen=1](\xb,\ya)(\xc,\ya)}
\pscustom[fillstyle=\orfillstyle,fillcolor=\orcolor,hatchcolor=\orhatchcolor,
linecolor=\orcolor,linewidth=\linew]{%
\psline[linearc=\linearc](\xb,\yc)(\xb,\ya)
\psline[liftpen=1,linearc=2\linearc](\xb,\ya)(\xa,\ya)(\xa,\yc)}
\psline[linearc=\linearc](\xb,\yc)(\xb,\yb)(\xc,\yb)
\psline[linearc=2\linearc](\xa,\yc)(\xa,\ya)(\xb,\ya)
\psline(\xb,\ya)(\xc,\ya)
}
\newcommand{\olvertex}[3][white]{%
\setlength{\xa}{#2\unit}
\addtolength{\xa}{0\unit}
\setlength{\xb}{#2\unit}
\addtolength{\xb}{1.5\unit}
\setlength{\xc}{#2\unit}
\addtolength{\xc}{3\unit}
\setlength{\ya}{#3\unit}
\addtolength{\ya}{1.5\unit}
\setlength{\yb}{#3\unit}
\addtolength{\yb}{0.5\dlinewidth}
\setlength{\yc}{#3\unit}
\addtolength{\yc}{-0.5\dlinewidth}
\setlength{\yd}{#3\unit}
\addtolength{\yd}{-1.5\unit}
\psset{doubleline=false}
\pscustom[fillstyle=\sfillstyle,fillcolor=#1,linecolor=#1,linewidth=0pt]{%
\psline[linearc=\linearc](\xc,\yb)(\xb,\yb)(\xb,\ya)
\psline[liftpen=1,linearc=\linearc](\xb,\yd)(\xb,\yc)(\xc,\yc)}
\pscustom[fillstyle=\olfillstyle,fillcolor=\olcolor,hatchcolor=\olhatchcolor,
linecolor=\olcolor,linewidth=0pt]{%
\psline[liftpen=0](\xa,\yd)(\xb,\yd)
\psline[liftpen=0](\xb,\ya)(\xa,\ya)
}
\psline[linecolor=\olcolor,linewidth=\linew](\xb,\ya)(\xb,\yd)
\psline[linearc=\linearc]{-C}(\xc,\yb)(\xb,\yb)(\xb,\ya)
\psline[liftpen=1,linearc=\linearc](\xb,\yd)(\xb,\yc)(\xc,\yc)
\psline{C-}(\xa,\ya)(\xa,\yd)
}
\newcommand{\orvertex}[3][white]{%
\setlength{\xa}{#2\unit}
\addtolength{\xa}{0\unit}
\setlength{\xb}{#2\unit}
\addtolength{\xb}{-1.5\unit}
\setlength{\xc}{#2\unit}
\addtolength{\xc}{-3\unit}
\setlength{\ya}{#3\unit}
\addtolength{\ya}{1.5\unit}
\setlength{\yb}{#3\unit}
\addtolength{\yb}{0.5\dlinewidth}
\setlength{\yc}{#3\unit}
\addtolength{\yc}{-0.5\dlinewidth}
\setlength{\yd}{#3\unit}
\addtolength{\yd}{-1.5\unit}
\psset{doubleline=false}
\pscustom[fillstyle=\sfillstyle,fillcolor=#1,linecolor=#1,linewidth=0pt]{%
\psline[linearc=\linearc](\xc,\yb)(\xb,\yb)(\xb,\ya)
\psline[liftpen=1,linearc=\linearc](\xb,\yd)(\xb,\yc)(\xc,\yc)}
\pscustom[fillstyle=\orfillstyle,fillcolor=\orcolor,hatchcolor=\orhatchcolor,
linecolor=\orcolor,linewidth=0pt]{%
\psline[liftpen=0](\xb,\ya)(\xa,\ya)
\psline[liftpen=0](\xa,\yd)(\xb,\yd)
\psline[liftpen=0](\xb,\ya)(\xa,\ya)
}
\psline[linecolor=\orcolor,linewidth=\linew](\xb,\ya)(\xb,\yd)
\psline[linearc=\linearc]{-C}(\xc,\yb)(\xb,\yb)(\xb,\ya)
\psline[liftpen=1,linearc=\linearc](\xb,\yd)(\xb,\yc)(\xc,\yc)
\psline{C-}(\xa,\ya)(\xa,\yd)
}

\newcommand{\threevertex}[4][white]{%
\setlength{\xa}{0\unit}
\addtolength{\xa}{-0.5\dlinewidth}
\setlength{\xb}{0\unit}
\addtolength{\xb}{0.5\dlinewidth}
\setlength{\xc}{0\unit}
\addtolength{\xc}{1\unit}
\setlength{\ya}{0\unit}
\addtolength{\ya}{1\unit}
\setlength{\yb}{0\unit}
\addtolength{\yb}{0.5\dlinewidth}
\setlength{\yc}{0\unit}
\addtolength{\yc}{-0.5\dlinewidth}
\setlength{\yd}{0\unit}
\addtolength{\yd}{-1\unit}
\psset{doubleline=false}
\rput{0}(#2\unit,#3\unit){%
\pscustom[fillstyle=\sfillstyle,fillcolor=#1,linecolor=#1,linewidth=0pt]{%
\rotate{#4}
\psline[liftpen=1,linearc=\linearc](\xb,\ya)(\xb,\yb)(\xc,\yb)
\psline[liftpen=1,linearc=\linearc](\xc,\yc)(\xb,\yc)(\xb,\yd)
\psline[liftpen=1](\xa,\yd)(\xa,\ya)}
\pscustom{%
\rotate{#4}
\psline[liftpen=2,linearc=\linearc](\xb,\ya)(\xb,\yb)(\xc,\yb)
\psline[liftpen=2,linearc=\linearc](\xc,\yc)(\xb,\yc)(\xb,\yd)
\psline[liftpen=2](\xa,\yd)(\xa,\ya)
}
}
}

\newcommand{\recthreevertex}[4][white]{%
\setlength{\xa}{-0.7071\auxlen}
\addtolength{\xa}{-0.7071\dlinewidth}
\setlength{\xb}{0\unit}
\addtolength{\xb}{-0.7071\auxlen}
\setlength{\xc}{-0.7071\auxlen}
\addtolength{\xc}{0.7071\unit}
\setlength{\xd}{-0.7071\auxlen}
\addtolength{\xd}{1.2071\unit}

\setlength{\ya}{-0.7071\auxlen}
\addtolength{\ya}{-0.7071\dlinewidth}
\setlength{\yb}{0\unit}
\addtolength{\yb}{-0.7071\auxlen}
\setlength{\yc}{-0.7071\auxlen}
\addtolength{\yc}{-0.7071\dlinewidth}
\addtolength{\yc}{0.7071\unit}
\setlength{\yd}{0.7071\auxlen}
\addtolength{\yd}{0.7071\dlinewidth}
\addtolength{\yd}{-0.7071\unit}
\setlength{\ye}{0\unit}
\addtolength{\ye}{0.7071\auxlen}
\setlength{\yf}{0.7071\auxlen}
\addtolength{\yf}{0.7071\dlinewidth}
\psset{doubleline=false}
\rput{0}(#2\unit,#3\unit){%
\pscustom[fillstyle=\sfillstyle,fillcolor=#1,linecolor=#1,linewidth=0pt]{%
\rotate{#4}
\psline[liftpen=1,linearc=\linearc](\xa,\yb)(-0.7071\dlinewidth,0)(\xa,\ye)
\psline[liftpen=1,linearc=\linearc](\xb,\yf)(\xc,\yd)(\xd,\yd)
\psline[liftpen=1,linearc=\linearc](\xd,\yc)(\xc,\yc)(\xb,\ya)}
\pscustom{%
\rotate{#4}
\psline[liftpen=1,linearc=\linearc](\xa,\yb)(-0.7071\dlinewidth,0)(\xa,\ye)
\psline[liftpen=2,linearc=\linearc](\xb,\yf)(\xc,\yd)(\xd,\yd)
\psline[liftpen=2,linearc=\linearc](\xd,\yc)(\xc,\yc)(\xb,\ya)
}
}
}

\newcommand{\recthreevertexoneside}[4][white]{%
\setlength{\xa}{-0.7071\auxlen}
\addtolength{\xa}{-0.7071\dlinewidth}
\setlength{\xb}{0\unit}
\addtolength{\xb}{-0.7071\auxlen}
\setlength{\xc}{-0.7071\auxlen}
\addtolength{\xc}{-0.353505\unit}
\setlength{\xd}{-\auxlen}
\addtolength{\xd}{-0.5\dlinewidth}
\setlength{\ya}{-0.7071\auxlen}
\addtolength{\ya}{-0.7071\dlinewidth}
\setlength{\yb}{0\unit}
\addtolength{\yb}{-0.7071\auxlen}
\setlength{\yc}{-0.7071\auxlen}
\addtolength{\yc}{-0.7071\dlinewidth}
\addtolength{\yc}{0.7071\unit}
\setlength{\yd}{0.7071\auxlen}
\addtolength{\yd}{0.7071\dlinewidth}
\addtolength{\yd}{-0.7071\unit}
\setlength{\ye}{0\unit}
\addtolength{\ye}{0.7071\auxlen}
\setlength{\yf}{0.7071\auxlen}
\addtolength{\yf}{0.7071\dlinewidth}
\psset{doubleline=false}
\rput{0}(#2\unit,#3\unit){%
\pscustom[fillstyle=\sfillstyle,fillcolor=#1,linecolor=#1,linewidth=0pt]{%
\rotate{#4}
\psline[liftpen=1,linearc=\linearc](\xb,\yf)(0.7071\dlinewidth,0)(\xb,\ya)
\psline[liftpen=1,linearc=0.2\linearc](\xa,\yb)(\xc,\yc)(\xd,\yc)
\psline[liftpen=1,linearc=0.2\linearc](\xd,\yd)(\xc,\yd)(\xa,\ye)}
\pscustom{%
\rotate{#4}
\psline[liftpen=1,linearc=\linearc](\xb,\yf)(0.7071\dlinewidth,0)(\xb,\ya)
\psline[liftpen=2,linearc=0.2\linearc](\xa,\yb)(\xc,\yc)(\xd,\yc)
\psline[liftpen=2,linearc=0.2\linearc](\xd,\yd)(\xc,\yd)(\xa,\ye)
}
}
}
\newcommand{\fourvertex}[4][white]{%
\setlength{\xa}{0\unit}
\addtolength{\xa}{-1\unit}
\setlength{\xb}{0\unit}
\addtolength{\xb}{-0.5\dlinewidth}
\setlength{\xc}{0\unit}
\addtolength{\xc}{0.5\dlinewidth}
\setlength{\xd}{0\unit}
\addtolength{\xd}{1\unit}
\setlength{\ya}{0\unit}
\addtolength{\ya}{-1\unit}
\setlength{\yb}{0\unit}
\addtolength{\yb}{-0.5\dlinewidth}
\setlength{\yc}{0\unit}
\addtolength{\yc}{0.5\dlinewidth}
\setlength{\yd}{0\unit}
\addtolength{\yd}{1\unit}
\psset{doubleline=false}
\rput{0}(#2\unit,#3\unit){%
\pscustom[fillstyle=\sfillstyle,fillcolor=#1,linecolor=#1,linewidth=0pt]{%
\rotate{#4}
\psline[liftpen=1,linearc=\linearc](\xc,\ya)(\xc,\yb)(\xd,\yb)
\psline[liftpen=1,linearc=\linearc](\xd,\yc)(\xc,\yc)(\xc,\yd)
\psline[liftpen=1,linearc=\linearc](\xb,\yd)(\xb,\yc)(\xa,\yc)
\psline[liftpen=1,linearc=\linearc](\xa,\yb)(\xb,\yb)(\xb,\ya)}
\pscustom{%
\rotate{#4}
\psline[liftpen=1,linearc=\linearc](\xc,\ya)(\xc,\yb)(\xd,\yb)
\psline[liftpen=2,linearc=\linearc](\xd,\yc)(\xc,\yc)(\xc,\yd)
\psline[liftpen=2,linearc=\linearc](\xb,\yd)(\xb,\yc)(\xa,\yc)
\psline[liftpen=2,linearc=\linearc](\xa,\yb)(\xb,\yb)(\xb,\ya)
}
}
}
\newcommand{\fourvertexdbltr}[4][white]{%
\setlength{\xa}{0\unit}
\addtolength{\xa}{-1\unit}
\setlength{\xb}{0\unit}
\addtolength{\xb}{-0.5\dlinewidth}
\setlength{\xc}{0\unit}
\addtolength{\xc}{0.5\dlinewidth}
\setlength{\xd}{0\unit}
\addtolength{\xd}{1\unit}
\setlength{\ya}{0\unit}
\addtolength{\ya}{-1\unit}
\setlength{\yb}{0\unit}
\addtolength{\yb}{-0.5\dlinewidth}
\setlength{\yc}{0\unit}
\addtolength{\yc}{0.5\dlinewidth}
\setlength{\yd}{0\unit}
\addtolength{\yd}{1\unit}
\psset{doubleline=false}
\rput{0}(#2\unit,#3\unit){%
\pscustom[fillstyle=\sfillstyle,fillcolor=#1,linecolor=#1,linewidth=0pt]{%
\rotate{#4}
\psline[liftpen=1,linearc=1.5\linearc](\xc,\ya)(0,0)(\xa,\yc)
\psline[liftpen=1,linearc=\linearc](\xd,\yc)(\xc,\yc)(\xc,\yd)
\psline[liftpen=1,linearc=1.5\linearc](\xb,\yd)(0,0)(\xa,\yc)
\psline[liftpen=1,linearc=\linearc](\xa,\yb)(\xb,\yb)(\xb,\ya)}
\pscustom{%
\rotate{#4}
\psline[liftpen=1,linearc=1.5\linearc](\xc,\ya)(0,0)(\xa,\yc)
\psline[liftpen=2,linearc=\linearc](\xd,\yc)(\xc,\yc)(\xc,\yd)
\psline[liftpen=2,linearc=1.5\linearc](\xb,\yd)(0,0)(\xd,\yb)
\psline[liftpen=2,linearc=\linearc](\xa,\yb)(\xb,\yb)(\xb,\ya)
}
}
}

\newcommand{\uoutex}[3][white]{%
\setlength{\xb}{#2\unit}
\addtolength{\xb}{-1.5\unit}
\setlength{\xc}{#2\unit}
\setlength{\ya}{#3\unit}
\addtolength{\ya}{0.5\dlinewidth}
\setlength{\yb}{#3\unit}
\addtolength{\yb}{-0.5\dlinewidth}
\setlength{\yc}{#3\unit}
\addtolength{\yc}{-1.5\unit}
\psset{doubleline=false}
\pscustom[fillstyle=\sfillstyle,fillcolor=#1,linecolor=#1]{%
\psline[liftpen=1,linearc=\linearc](\xc,\yb)(\xb,\yb)(\xb,\yc)
\psline[liftpen=1](\xb,\ya)(\xc,\ya)}
\psline[linearc=\linearc](\xb,\yc)(\xb,\yb)(\xc,\yb)
\psline(\xb,\ya)(\xc,\ya)
}
\newcommand{\doutex}[3][white]{%
\setlength{\xb}{#2\unit}
\addtolength{\xb}{-1.5\unit}
\setlength{\xc}{#2\unit}
\setlength{\ya}{#3\unit}
\addtolength{\ya}{-0.5\dlinewidth}
\setlength{\yb}{#3\unit}
\addtolength{\yb}{0.5\dlinewidth}
\setlength{\yc}{#3\unit}
\addtolength{\yc}{1.5\unit}
\psset{doubleline=false}
\pscustom[fillstyle=\sfillstyle,fillcolor=#1,linecolor=#1]{%
\psline[linearc=\linearc](\xc,\yb)(\xb,\yb)(\xb,\yc)
\psline(\xb,\ya)(\xc,\ya)}
\psline[linearc=\linearc](\xb,\yc)(\xb,\yb)(\xc,\yb)
\psline(\xb,\ya)(\xc,\ya)
}
\newcommand{\dinex}[3][white]{%
\setlength{\xb}{#2\unit}
\addtolength{\xb}{1.5\unit}
\setlength{\xc}{#2\unit}
\setlength{\ya}{#3\unit}
\addtolength{\ya}{-0.5\dlinewidth}
\setlength{\yb}{#3\unit}
\addtolength{\yb}{0.5\dlinewidth}
\setlength{\yc}{#3\unit}
\addtolength{\yc}{1.5\unit}
\psset{doubleline=false}
\pscustom[fillstyle=\sfillstyle,fillcolor=#1,linecolor=#1]{%
\psline[linearc=\linearc](\xc,\yb)(\xb,\yb)(\xb,\yc)
\psline(\xb,\ya)(\xc,\ya)}
\psline[linearc=\linearc](\xb,\yc)(\xb,\yb)(\xc,\yb)
\psline(\xb,\ya)(\xc,\ya)
}
\newcommand{\uinex}[3][white]{%
\setlength{\xb}{#2\unit}
\addtolength{\xb}{1.5\unit}
\setlength{\xc}{#2\unit}
\setlength{\ya}{#3\unit}
\addtolength{\ya}{0.5\dlinewidth}
\setlength{\yb}{#3\unit}
\addtolength{\yb}{-0.5\dlinewidth}
\setlength{\yc}{#3\unit}
\addtolength{\yc}{-1.5\unit}
\psset{doubleline=false}
\pscustom[fillstyle=\sfillstyle,fillcolor=#1,linecolor=#1]{%
\psline[linearc=\linearc](\xc,\yb)(\xb,\yb)(\xb,\yc)
\psline[liftpen=1](\xb,\ya)(\xc,\ya)}
\psline[linearc=\linearc](\xb,\yc)(\xb,\yb)(\xc,\yb)
\psline(\xb,\ya)(\xc,\ya)
}
\newcommand{\ioutex}[3][white]{%
\setlength{\xb}{#2\unit}
\addtolength{\xb}{-1.5\unit}
\setlength{\xc}{#2\unit}
\setlength{\ya}{#3\unit}
\addtolength{\ya}{1.5\unit}
\setlength{\yb}{#3\unit}
\addtolength{\yb}{0.5\dlinewidth}
\setlength{\yc}{#3\unit}
\addtolength{\yc}{-0.5\dlinewidth}
\setlength{\yd}{#3\unit}
\addtolength{\yd}{-1.5\unit}
\psset{doubleline=false}
\pscustom[fillstyle=\sfillstyle,fillcolor=#1,linecolor=#1]{%
\psline[linearc=\linearc](\xc,\yb)(\xb,\yb)(\xb,\ya)
\psline[liftpen=1,linearc=\linearc](\xb,\yd)(\xb,\yc)(\xc,\yc)}
\psline[linecolor=#1,linewidth=\linew](\xb,\ya)(\xb,\yd)
\psline[linearc=\linearc]{-C}(\xc,\yb)(\xb,\yb)(\xb,\ya)
\psline[liftpen=1,linearc=\linearc](\xb,\yd)(\xb,\yc)(\xc,\yc)
}
\newcommand{\iinex}[3][white]{%
\setlength{\xb}{#2\unit}
\addtolength{\xb}{1.5\unit}
\setlength{\xc}{#2\unit}
\setlength{\ya}{#3\unit}
\addtolength{\ya}{1.5\unit}
\setlength{\yb}{#3\unit}
\addtolength{\yb}{0.5\dlinewidth}
\setlength{\yc}{#3\unit}
\addtolength{\yc}{-0.5\dlinewidth}
\setlength{\yd}{#3\unit}
\addtolength{\yd}{-1.5\unit}
\psset{doubleline=false}
\pscustom[fillstyle=\sfillstyle,fillcolor=#1,linecolor=#1,linewidth=0pt]{%
\psline[linearc=\linearc](\xc,\yb)(\xb,\yb)(\xb,\ya)
\psline[liftpen=1,linearc=\linearc](\xb,\yd)(\xb,\yc)(\xc,\yc)}
\psline[linecolor=#1,linewidth=\linew](\xb,\ya)(\xb,\yd)
\psline[linearc=\linearc]{-C}(\xc,\yb)(\xb,\yb)(\xb,\ya)
\psline[liftpen=1,linearc=\linearc](\xb,\yd)(\xb,\yc)(\xc,\yc)
}

\newcommand{\qvertr}[5][]{%
\settoheight{\eqoff}{$\times$}%
\setlength{\eqoff}{0.5\eqoff}%
\addtolength{\eqoff}{-12\unitlength}%
\raisebox{\eqoff}{%
\fmfframe(4,6)(4,6){%
\begin{fmfchar*}(12,12)
\fmfleft{v1,v4}
\fmfright{v2,v3}
\fmfforce{(0,h)}{v1}
\fmfforce{(0,0)}{v4}
\fmfforce{(w,0)}{v3}
\fmfforce{(w,h)}{v2}
\fmf{#2}{v1,vc1}
\fmf{#3}{v2,vc1}
\fmf{#4}{v3,vc1}
\fmf{#5}{v4,vc1}
\fmffreeze
\fmfposition
\fmfipath{p[]}
\fmfipair{vm[],vo[],vi[]}
\fmfiset{p1}{vpath(__v1,__vc1)}
\fmfiset{p2}{vpath(__v2,__vc1)}
\fmfiset{p3}{vpath(__v3,__vc1)}
\fmfiset{p4}{vpath(__v4,__vc1)}
\svertex{vm1}{p1}
\dvertex{vo1}{vi1}{p1}
\svertex{vm2}{p2}
\dvertex{vo2}{vi2}{p2}
\svertex{vm3}{p3}
\dvertex{vo3}{vi3}{p3}
\svertex{vm4}{p4}
\dvertex{vo4}{vi4}{p4}
{#1}
\end{fmfchar*}}}
}

\newcommand{\vacpolp}[2][0.5]{%
\fmfcmd{
begingroup;
save t,v,tv,do,di,ppol,pstr,dia;
path ppol,pstr;
pair v[],tv[],do[],di[];
ppol=#2;
t3=arctime (0.5*arclength ppol) of ppol;
v3=point t3 of ppol;
dia=#1 arclength ppol; 
fill(fullcircle scaled dia shifted v3) withcolor 0.2black;
endgroup;
}
}
\newcommand{\swfone}[4][]{%
\settoheight{\eqoff}{$\times$}%
\setlength{\eqoff}{0.5\eqoff}%
\addtolength{\eqoff}{-7.5\unitlength}%
\raisebox{\eqoff}{%
\fmfframe(4,0)(4,0){%
\begin{fmfchar*}(20,15)
\fmfleft{v1}
\fmfright{v2}
\fmffixed{(0.5w,0)}{vc1,vc2}
\fmf{#2}{v1,vc1}
\fmf{#2}{vc2,v2}
\fmf{#3}{vc1,vc2}
\fmf{#4}{vc2,vc1}
\fmffreeze
\fmfposition
\fmfipath{p[]}
\fmfipair{vm[]}
\fmfiset{p1}{vpath(__v1,__vc1)}
\fmfiset{p2}{vpath(__vc1,__vc2)}
\fmfiset{p3}{reverse vpath(__vc2,__vc1) }
\fmfiset{p4}{vpath(__vc2,__v2)}
\svertex{vm1}{p2}
\svertex{vm2}{p3}
{#1}
\end{fmfchar*}}}}

\newcommand{\ltwoopren}[1]{%
\settoheight{\eqoff}{$\times$}%
\setlength{\eqoff}{0.5\eqoff}%
\addtolength{\eqoff}{-7.5\unitlength}%
\raisebox{\eqoff}{%
\fmfframe(1,-1.25)(1,1.25){%
\begin{fmfchar*}(15,15)
\fmfleft{vl}
\fmfright{vr}
\fmfforce{(0.5w,0)}{vo}
\fmfforce{(0,0.866025h)}{vl}
\fmfforce{(w,0.866025h)}{vr}
\fmffreeze
\fmfposition
{#1}
\fmfiv{decor.shape=circle,decor.filled=full,decor.size=3}{vloc(__vo)}
\end{fmfchar*}}}}

\newcommand{\uoneprop}[4][white]{%
\setlength{\xa}{0\unit}
\addtolength{\xa}{-1\unit}
\setlength{\xb}{-.6\unit}
\addtolength{\xb}{-0.5\dlinewidth}
\setlength{\xc}{0.6\unit}
\addtolength{\xc}{0.5\dlinewidth}
\setlength{\xd}{0\unit}
\addtolength{\xd}{1\unit}
\setlength{\ya}{0\unit}
\addtolength{\ya}{-1\unit}
\setlength{\yb}{0\unit}
\addtolength{\yb}{-0.5\dlinewidth}
\setlength{\yc}{0\unit}
\addtolength{\yc}{0.5\dlinewidth}
\setlength{\yd}{0\unit}
\addtolength{\yd}{1\unit}
\psset{doubleline=false}
\rput{0}(#2\unit,#3\unit){%
\pscustom[fillstyle=\sfillstyle,fillcolor=#1,linecolor=#1,linewidth=0pt]{%
\rotate{#4}
\psbezier[liftpen=1,linearc=\linearc](\xb,\yc)(0,\yc)(0,\yb)(\xb,\yb)
}
\pscustom[fillstyle=\sfillstyle,fillcolor=#1,linecolor=#1,linewidth=0pt]{%
\rotate{#4}
\psbezier[liftpen=2,linearc=\linearc](\xc,\yb)(0,\yb)(0,\yc)(\xc,\yc)
}
\pscustom{%
\rotate{#4}
\psbezier[linearc=\linearc](\xb,\yc)(-0.1,\yc)(-0.1,\yb)(\xb,\yb)
\psbezier[liftpen=2,linearc=\linearc](\xc,\yb)(0.1,\yb)(0.1,\yc)(\xc,\yc)
}
}
}

%% file: DoubleLineMakro.tex
\newcommand{\feynmpdefinitionscaps}{

\fmfcmd{%
linejoin:=rounded;
}

\fmfcmd{%
style_def plain_n expr p =
 linecap:=butt;
 cdraw p;
 linecap:=rounded;
enddef;
style_def plain_h expr p =
 linecap:=butt;
 cdraw subpath (0, 1length p ) of p;
 linecap:=rounded;
 cdraw subpath (0.95length p ,length p ) of p;
enddef;
style_def plain_t expr p =
 linecap:=rounded;
 cdraw subpath (0, 0.05length p ) of p;
 linecap:=butt;
 cdraw subpath (0,length p ) of p;
 linecap:=rounded;
enddef;
style_def plain_ht expr p =
 linecap:=rounded;
 cdraw p;
enddef;
style_def iplain_n expr p =
 linecap:=squared;
 cdraw p;
 linecap:=rounded;
enddef;
style_def iplain_h expr p =
 linecap:=squared;
 cdraw subpath (0, 0.5length p ) of p;
 linecap:=rounded;
 cdraw subpath (0.5length p ,length p ) of p;
enddef;
style_def iplain_t expr p =
 linecap:=rounded;
 cdraw subpath (0, 0.5length p ) of p;
 linecap:=squared;
 cdraw subpath (0.5length p,length p ) of p;
 linecap:=rounded;
enddef;
style_def iplain_ht expr p =
 linecap:=rounded;
 cdraw p;
enddef;
style_def interrupted_plain_n expr p =
 draw_plain_h subpath (0, 0.25length p ) of p;
 draw_plain_t subpath (0.75length p ,length p ) of p;
enddef;
style_def interrupted_plain_h expr p =
 draw_plain_h subpath (0, 0.25length p ) of p;
 draw_plain_ht subpath (0.75length p ,length p ) of p;
enddef;
style_def interrupted_plain_t expr p =
 draw_plain_ht subpath (0, 0.25length p ) of p;
 draw_plain_t subpath (0.75length p ,length p ) of p;
enddef;
style_def interrupted_plain_ht expr p =
 draw_plain_ht subpath (0, 0.25length p ) of p;
 draw_plain_ht subpath (0.75length p ,length p ) of p;
enddef;
style_def iinterrupted_plain_n expr p =
 draw_iplain_h subpath (0, 0.25length p ) of p;
 draw_iplain_t subpath (0.75length p ,length p ) of p;
enddef;
style_def iinterrupted_plain_h expr p =
 draw_iplain_h subpath (0, 0.25length p ) of p;
 draw_iplain_ht subpath (0.75length p ,length p ) of p;
enddef;
style_def iinterrupted_plain_t expr p =
 draw_iplain_ht subpath (0, 0.25length p ) of p;
 draw_iplain_t subpath (0.75length p ,length p ) of p;
enddef;
style_def iinterrupted_plain_ht expr p =
 draw_iplain_ht subpath (0, 0.25length p ) of p;
 draw_iplain_ht subpath (0.75length p ,length p ) of p;
enddef;
vardef shift_p expr p =  
p  shifted 4.5 (unitvector direction (length (p)/2) of p rotated +90)
enddef;
style_def leftrightarrows expr p =
  shrink 0.5;
  cfill arrow (shift_p (p));
  cfill arrow (shift_p (reverse p));
  endshrink;
enddef;%
style_def leftrightarrows_interrupted expr p =
  shrink 0.5;
  cfill arrow (shift_p (subpath (0, 0.4length p ) of p));
  cfill arrow (shift_p (reverse (subpath (0, 0.4length p ) of p)));
  cfill arrow (shift_p (subpath (0.6length p ,length p ) of p));
  cfill arrow (shift_p (reverse (subpath (0.6length p ,length p ) of p)));
  endshrink;
enddef;%
style_def leftrightarrows_interruptedold expr p =
  shrink 0.5;
  cfill arrow (shift_p (subpath (0, 0.25length p ) of p));
  cfill arrow (shift_p (reverse (subpath (0, 0.25length p ) of p)));
  cfill arrow (shift_p (subpath (0.75length p ,length p ) of p));
  cfill arrow (shift_p (reverse (subpath (0.75length p ,length p ) of p)));
  endshrink;
enddef;%
style_def leftrightarrows_start expr p =
  shrink 0.5;
  cfill arrow (shift_p (subpath (0, 0.1length p ) of p));
  cfill arrow (shift_p (reverse (subpath (0, 0.1length p ) of p)));
  endshrink;
enddef;}
}

%% file: Macros2.tex
\newcommand{\la}{\langle}
\newcommand{\ra}{\rangle}
\newcommand{\ab}[1]{\la #1 \ra}
\renewcommand{\sb}[1]{[ #1 ]}
\newcommand{\asb}[3]{\la #1 | #2 | #3 ]}
\newcommand{\sab}[3]{[ #1 | #2 | #3 \ra}
\newcommand{\MHV}{{\ensuremath{\text{MHV}} \xspace}}
\newcommand{\MHVb}{{\ensuremath{\overline{\text{MHV}}} \xspace}}

\newcommand{\lambdat}{\tilde{\lambda}}
\newcommand{\etat}{\tilde{\eta}}

\newcommand{\CC}{\ensuremath{\mathbb{C}}}
\newcommand{\ZZ}{\ensuremath{\mathbb{Z}}}
\renewcommand{\SS}{\ensuremath{\mathbb{S}}}
\newcommand{\NN}{\ensuremath{\mathbb{N}}}
\newcommand{\PP}{\ensuremath{\mathbb{P}}}

\newcommand{\U}[1]{\ensuremath{U(#1)}\xspace}
\newcommand{\SU}[1]{\ensuremath{SU(#1)}\xspace}
\newcommand{\SO}[1]{\ensuremath{SO(#1)}\xspace}

\newcommand{\complexi}{i}

\DeclareMathOperator{\idm}{\mathds{1}}

\newcommand{\comm}[3][]{{}[{}#2{}\,{\overset{#1}{,}}\,{}#3{}]{}}
\newcommand{\acomm}[3][]{{}\{{}#2{}\,{\overset{#1}{,}}\,{}#3{}\}{}}

\newcommand{\YM}{{\mathrm{\scriptscriptstyle YM}}}

\DeclareMathOperator{\tr}{tr}
\newcommand{\Tr}{\tr}

\DeclareMathOperator{\phaneq}{\phantom{{}=}}

\newcommand{\colors}{s}

\newcommand{\e}{\operatorname{e}}
\newcommand{\cstar}{\ensuremath{\ast}}

\newcommand{\ev}[1]{\langle #1 \rangle}

\newcommand{\ket}[1]{\mathopen{\mid}{}#1{}\mathclose{\rangle}}

\newcommand{\bra}[1]{\mathopen{\langle}{}#1{}\mathclose{\mid}}

\newcommand{\starcomm}[2]{\comm{#1}{#2}_\cstar}
\newcommand{\staracomm}[2]{\acomm{#1}{#2}_\cstar}

\newcommand{\vac}{\operatorname{\mid} 0 \, \operatorname{\rangle}}

\DeclareMathOperator{\Kop}{K}
\DeclareMathOperator{\Rop}{R}
\DeclareMathOperator{\Top}{T}

\newcommand{\eqndot}{\, .}
\newcommand{\eqncom}{\, ,}
\newcommand{\eqnsem}{\, ;}

\newcommand{\de}{\operatorname{d}\!}

\newcommand{\cA}{\mathcal{A}}
\newcommand{\cB}{\mathcal{B}}
\newcommand{\cC}{\mathcal{C}}
\newcommand{\cD}{\mathcal{D}}

\newcommand{\cF}{\mathcal{F}}
\newcommand{\cG}{\mathcal{G}}

\newcommand{\cI}{\mathcal{I}}

\newcommand{\cK}{\mathcal{K}}
\newcommand{\cL}{\mathcal{L}}

\newcommand{\cN}{\mathcal{N}}
\newcommand{\cO}{\mathcal{O}}

\newcommand{\cV}{\mathcal{V}}

\newcommand{\cZ}{\mathcal{Z}}

\newcommand{\ba}{\mathbf{a}}
\newcommand{\bb}{\mathbf{b}}

\newcommand{\bd}{\mathbf{d}}

\newcommand{\bA}{\mathbf{A}}

\newcommand{\alphadot}{{\dot{\alpha}}}
\newcommand{\betadot}{{\dot{\beta}}}
\newcommand{\gammadot}{{\dot{\gamma}}}

\newcommand{\ferm}{\psi}
\newcommand{\antiferm}{\bar{\psi}}

\newcommand{\cfstrength}{F}
\newcommand{\cantifstrength}{\bar{\cfstrength}}

\newcommand{\cder}{\D}

\newcommand{\aosc}{\ba}
\newcommand{\aoscdag}{{\aosc^\dagger}}
\newcommand{\bosc}{\bb}
\newcommand{\boscdag}{{\bosc^\dagger}}

\newcommand{\dosc}{\bd}

\newcommand{\Aosc}{\bA}

\newcommand{\akindsite}[2][ ]{a^{#1}_{#2}}
\newcommand{\bkindsite}[2][ ]{b^{#1}_{#2}}

\newcommand{\dkindsite}[2][ ]{d^{#1}_{#2}}

\DeclareMathOperator{\T}{T}

\newlength{\eqoff}

\DeclareMathOperator{\D}{D}

%% file: PhDThesis.bbl
\providecommand{\href}[2]{#2}\begingroup\raggedright\begin{thebibliography}{100}

\bibitem{Fokken:2013aea}
J.~Fokken, C.~Sieg, and M.~Wilhelm, ``{Non-conformality of $\gamma_i$-deformed
  $\mathcal{N}=4$ SYM theory},''
  \href{http://dx.doi.org/10.1088/1751-8113/47/45/455401}{{\em J. Phys. A:
  Math. Theor.} {\bfseries 47} (2014) },
\href{http://arxiv.org/abs/1308.4420}{{\ttfamily arXiv:1308.4420 [hep-th]}}.

\bibitem{Fokken:2013mza}
J.~Fokken, C.~Sieg, and M.~Wilhelm, ``{The complete one-loop dilatation
  operator of planar real $\beta$-deformed $ \mathcal{N} $ = 4 SYM theory},''
  \href{http://dx.doi.org/10.1007/JHEP07(2014)150}{{\em JHEP} {\bfseries 1407}
  (2014) 150},
\href{http://arxiv.org/abs/1312.2959}{{\ttfamily arXiv:1312.2959 [hep-th]}}.

\bibitem{Fokken:2014soa}
J.~Fokken, C.~Sieg, and M.~Wilhelm, ``{A piece of cake: the ground-state
  energies in $\gamma_{i}$-deformed $\mathcal{N} = 4$ SYM theory at leading
  wrapping order},'' \href{http://dx.doi.org/10.1007/JHEP09(2014)078}{{\em
  JHEP} {\bfseries 09} (2014) 78},
\href{http://arxiv.org/abs/1405.6712}{{\ttfamily arXiv:1405.6712 [hep-th]}}.

\bibitem{Wilhelm:2014qua}
M.~Wilhelm, ``{Amplitudes, Form Factors and the Dilatation Operator in
  $\mathcal{N}=4$ SYM Theory},''
  \href{http://dx.doi.org/10.1007/JHEP02(2015)149}{{\em JHEP} {\bfseries 02}
  (2015) 149},
\href{http://arxiv.org/abs/1410.6309}{{\ttfamily arXiv:1410.6309 [hep-th]}}.

\bibitem{Nandan:2014oga}
D.~Nandan, C.~Sieg, M.~Wilhelm, and G.~Yang, ``{Cutting through form factors
  and cross sections of non-protected operators in $ \mathcal{N}=4 $ SYM},''
  \href{http://dx.doi.org/10.1007/JHEP06(2015)156}{{\em JHEP} {\bfseries 1506}
  (2015) 156},
\href{http://arxiv.org/abs/1410.8485}{{\ttfamily arXiv:1410.8485 [hep-th]}}.

\bibitem{Loebbert:2015ova}
F.~Loebbert, D.~Nandan, C.~Sieg, M.~Wilhelm, and G.~Yang, ``{On-Shell Methods
  for the Two-Loop Dilatation Operator and Finite Remainders},''
  \href{http://dx.doi.org/10.1007/JHEP10(2015)012}{{\em JHEP} {\bfseries 10}
  (2015) 012},
\href{http://arxiv.org/abs/1504.06323}{{\ttfamily arXiv:1504.06323 [hep-th]}}.

\bibitem{Frassek:2015rka}
R.~Frassek, D.~Meidinger, D.~Nandan, and M.~Wilhelm, ``{On-shell diagrams,
  Gra{\ss}mannians and integrability for form factors},''
  \href{http://dx.doi.org/10.1007/JHEP01(2016)182}{{\em JHEP} {\bfseries 01}
  (2016) 182},
\href{http://arxiv.org/abs/1506.08192}{{\ttfamily arXiv:1506.08192 [hep-th]}}.

\bibitem{Schroers:2014dua}
B.~Schroers and M.~Wilhelm, ``{Towards Non-Commutative Deformations of
  Relativistic Wave Equations in 2+1 Dimensions},''
  \href{http://dx.doi.org/10.3842/SIGMA.2014.053}{{\em SIGMA} {\bfseries 10}
  (2014) 053},
\href{http://arxiv.org/abs/1402.7039}{{\ttfamily arXiv:1402.7039 [hep-th]}}.

\bibitem{Fokken:2014moa}
J.~Fokken and M.~Wilhelm, ``{One-Loop Partition Functions in Deformed
  $\mathcal{N}=4$ SYM Theory},''
  \href{http://dx.doi.org/10.1007/JHEP03(2015)018}{{\em JHEP} {\bfseries 1503}
  (2015) 018},
\href{http://arxiv.org/abs/1411.7695}{{\ttfamily arXiv:1411.7695 [hep-th]}}.

\bibitem{Chatrchyan:2012xdj}
{\bfseries CMS} Collaboration, S.~Chatrchyan {\em et~al.}, ``{Observation of a
  new boson at a mass of 125 GeV with the CMS experiment at the LHC},''
  \href{http://dx.doi.org/10.1016/j.physletb.2012.08.021}{{\em Phys. Lett.}
  {\bfseries B716} (2012) 30--61},
\href{http://arxiv.org/abs/1207.7235}{{\ttfamily arXiv:1207.7235 [hep-ex]}}.

\bibitem{Aad:2012tfa}
{\bfseries ATLAS} Collaboration, G.~Aad {\em et~al.}, ``{Observation of a new
  particle in the search for the Standard Model Higgs boson with the ATLAS
  detector at the LHC},''
  \href{http://dx.doi.org/10.1016/j.physletb.2012.08.020}{{\em Phys. Lett.}
  {\bfseries B716} (2012) 1--29},
\href{http://arxiv.org/abs/1207.7214}{{\ttfamily arXiv:1207.7214 [hep-ex]}}.

\bibitem{Anastasiou:2015ema}
C.~Anastasiou, C.~Duhr, F.~Dulat, F.~Herzog, and B.~Mistlberger, ``{Higgs Boson
  Gluon-Fusion Production in QCD at Three Loops},''
  \href{http://dx.doi.org/10.1103/PhysRevLett.114.212001}{{\em Phys. Rev.
  Lett.} {\bfseries 114} no.~21, (2015) 212001},
\href{http://arxiv.org/abs/1503.06056}{{\ttfamily arXiv:1503.06056 [hep-ph]}}.

\bibitem{Brink:1976bc}
L.~Brink, J.~H. Schwarz, and J.~Scherk, ``{Supersymmetric Yang-Mills
  Theories},''
\href{http://dx.doi.org/10.1016/0550-3213(77)90328-5}{{\em Nucl.Phys.}
  {\bfseries B121} (1977) 77}.

\bibitem{Maldacena:1997re}
J.~M. Maldacena, ``The Large $N$ limit of superconformal field theories and
  supergravity,'' {\em Adv.Theor.Math.Phys.} {\bfseries 2} (1998) 231--252,
\href{http://arxiv.org/abs/hep-th/9711200}{{\ttfamily arXiv:hep-th/9711200
  [hep-th]}}.

\bibitem{Gubser:1998bc}
S.~Gubser, I.~R. Klebanov, and A.~M. Polyakov, ``{Gauge theory correlators from
  noncritical string theory},''
  \href{http://dx.doi.org/10.1016/S0370-2693(98)00377-3}{{\em Phys.Lett.}
  {\bfseries B428} (1998) 105--114},
\href{http://arxiv.org/abs/hep-th/9802109}{{\ttfamily arXiv:hep-th/9802109
  [hep-th]}}.

\bibitem{Witten:1998qj}
E.~Witten, ``{Anti-de Sitter space and holography},'' {\em
  Adv.Theor.Math.Phys.} {\bfseries 2} (1998) 253--291,
\href{http://arxiv.org/abs/hep-th/9802150}{{\ttfamily arXiv:hep-th/9802150
  [hep-th]}}.

\bibitem{Aharony:1999ti}
O.~Aharony, S.~S. Gubser, J.~M. Maldacena, H.~Ooguri, and Y.~Oz, ``{Large N
  field theories, string theory and gravity},''
  \href{http://dx.doi.org/10.1016/S0370-1573(99)00083-6}{{\em Phys.Rept.}
  {\bfseries 323} (2000) 183--386},
\href{http://arxiv.org/abs/hep-th/9905111}{{\ttfamily arXiv:hep-th/9905111
  [hep-th]}}.

\bibitem{D'Hoker:2002aw}
E.~D'Hoker and D.~Z. Freedman, ``{Supersymmetric gauge theories and the AdS /
  CFT correspondence},''
\href{http://arxiv.org/abs/hep-th/0201253}{{\ttfamily arXiv:hep-th/0201253
  [hep-th]}}.

\bibitem{'tHooft:1973jz}
G.~'t~Hooft, ``A Planar Diagram Theory for Strong Interactions,''
\href{http://dx.doi.org/10.1016/0550-3213(74)90154-0}{{\em Nucl.Phys.}
  {\bfseries B72} (1974) 461}.

\bibitem{Beisert:2010jr}
N.~Beisert {\em et~al.}, ``{Review of AdS/CFT Integrability: An Overview},''
  \href{http://dx.doi.org/10.1007/s11005-011-0529-2}{{\em Lett.Math.Phys.}
  {\bfseries 99} (2012) 3--32},
\href{http://arxiv.org/abs/1012.3982}{{\ttfamily arXiv:1012.3982 [hep-th]}}.

\bibitem{Bethe:1931}
H.~Bethe, ``{Zur Theorie der Metalle},''
  \href{http://dx.doi.org/10.1007/BF01341708}{{\em Zeitschrift für Physik A:
  Hadrons and Nuclei} {\bfseries 71} (1931) 205--226}.

\bibitem{Lipatov:1993yb}
L.~Lipatov, ``{High-energy asymptotics of multicolor QCD and exactly solvable
  lattice models},''
\href{http://arxiv.org/abs/hep-th/9311037}{{\ttfamily arXiv:hep-th/9311037
  [hep-th]}}.

\bibitem{Bogomolny:1975de}
E.~B. Bogomolny, ``{Stability of Classical Solutions},'' {\em Sov. J. Nucl.
  Phys.} {\bfseries 24} (1976) 449.
[Yad. Fiz.24,861(1976)].

\bibitem{Prasad:1975kr}
M.~K. Prasad and C.~M. Sommerfield, ``{An Exact Classical Solution for the 't
  Hooft Monopole and the Julia-Zee Dyon},''
\href{http://dx.doi.org/10.1103/PhysRevLett.35.760}{{\em Phys. Rev. Lett.}
  {\bfseries 35} (1975) 760--762}.

\bibitem{Minahan:2002ve}
J.~Minahan and K.~Zarembo, ``{The Bethe ansatz for $\cN=4$ superYang-Mills},''
  {\em JHEP} {\bfseries 0303} (2003) 013,
\href{http://arxiv.org/abs/hep-th/0212208}{{\ttfamily arXiv:hep-th/0212208
  [hep-th]}}.

\bibitem{Beisert:2003yb}
N.~Beisert and M.~Staudacher, ``{The $\cN=4$ SYM integrable super spin
  chain},'' \href{http://dx.doi.org/10.1016/j.nuclphysb.2003.08.015}{{\em
  Nucl.Phys.} {\bfseries B670} (2003) 439--463},
\href{http://arxiv.org/abs/hep-th/0307042}{{\ttfamily arXiv:hep-th/0307042
  [hep-th]}}.

\bibitem{Beisert:2003jj}
N.~Beisert, ``{The complete one-loop dilatation operator of $\cN=4$
  superYang-Mills theory},''
  \href{http://dx.doi.org/10.1016/j.nuclphysb.2003.10.019}{{\em Nucl.Phys.}
  {\bfseries B676} (2004) 3--42},
\href{http://arxiv.org/abs/hep-th/0307015}{{\ttfamily arXiv:hep-th/0307015
  [hep-th]}}.

\bibitem{Beisert:2005fw}
N.~Beisert and M.~Staudacher, ``{Long-range $\mathfrak{psu}(2,2|4)$ Bethe
  Ansatze for gauge theory and strings},''
  \href{http://dx.doi.org/10.1016/j.nuclphysb.2005.06.038}{{\em Nucl.Phys.}
  {\bfseries B727} (2005) 1--62},
\href{http://arxiv.org/abs/hep-th/0504190}{{\ttfamily arXiv:hep-th/0504190
  [hep-th]}}.

\bibitem{Sieg:2005kd}
C.~Sieg and A.~Torrielli, ``{Wrapping interactions and the genus expansion of
  the 2-point function of composite operators},''
  \href{http://dx.doi.org/10.1016/j.nuclphysb.2005.06.011}{{\em Nucl.Phys.}
  {\bfseries B723} (2005) 3--32},
\href{http://arxiv.org/abs/hep-th/0505071}{{\ttfamily arXiv:hep-th/0505071
  [hep-th]}}.

\bibitem{Beisert:2004ry}
N.~Beisert, ``{The Dilatation operator of $\cN=4$ super Yang-Mills theory and
  integrability},'' \href{http://dx.doi.org/10.1016/j.physrep.2004.09.007}{{\em
  Phys.Rept.} {\bfseries 405} (2005) 1--202},
  \href{http://arxiv.org/abs/hep-th/0407277}{{\ttfamily arXiv:hep-th/0407277
  [hep-th]}}.
Ph.D. Thesis.

\bibitem{Beisert:2004hm}
N.~Beisert, V.~Dippel, and M.~Staudacher, ``{A Novel long range spin chain and
  planar $\mathcal{N}=4$ super Yang-Mills},''
  \href{http://dx.doi.org/10.1088/1126-6708/2004/07/075}{{\em JHEP} {\bfseries
  07} (2004) 075},
\href{http://arxiv.org/abs/hep-th/0405001}{{\ttfamily arXiv:hep-th/0405001
  [hep-th]}}.

\bibitem{Fiamberti:2007rj}
F.~Fiamberti, A.~Santambrogio, C.~Sieg, and D.~Zanon, ``{Wrapping at four loops
  in $\mathcal{N}=4$ SYM},''
  \href{http://dx.doi.org/10.1016/j.physletb.2008.06.061}{{\em Phys.Lett.}
  {\bfseries B666} (2008) 100--105},
\href{http://arxiv.org/abs/0712.3522}{{\ttfamily arXiv:0712.3522 [hep-th]}}.

\bibitem{Fiamberti:2008sh}
F.~Fiamberti, A.~Santambrogio, C.~Sieg, and D.~Zanon, ``{Anomalous dimension
  with wrapping at four loops in $\mathcal{N}=4$ SYM},''
  \href{http://dx.doi.org/10.1016/j.nuclphysb.2008.07.014}{{\em Nucl.Phys.}
  {\bfseries B805} (2008) 231--266},
\href{http://arxiv.org/abs/0806.2095}{{\ttfamily arXiv:0806.2095 [hep-th]}}.

\bibitem{Velizhanin:2008jd}
V.~Velizhanin, ``{The four-loop anomalous dimension of the Konishi operator in
  $\cN=4$ supersymmetric Yang-Mills theory},''
  \href{http://dx.doi.org/10.1134/S0021364009010020}{{\em JETP Lett.}
  {\bfseries 89} (2009) 6--9},
\href{http://arxiv.org/abs/0808.3832}{{\ttfamily arXiv:0808.3832 [hep-th]}}.

\bibitem{Bajnok:2008bm}
Z.~Bajnok and R.~A. Janik, ``{Four-loop perturbative Konishi from strings and
  finite size effects for multiparticle states},''
  \href{http://dx.doi.org/10.1016/j.nuclphysb.2008.08.020}{{\em Nucl. Phys.}
  {\bfseries B807} (2009) 625--650},
\href{http://arxiv.org/abs/0807.0399}{{\ttfamily arXiv:0807.0399 [hep-th]}}.

\bibitem{Ambjorn:2005wa}
J.~Ambjorn, R.~A. Janik, and C.~Kristjansen, ``{Wrapping interactions and a new
  source of corrections to the spin-chain/string duality},''
  \href{http://dx.doi.org/10.1016/j.nuclphysb.2005.12.007}{{\em Nucl.Phys.}
  {\bfseries B736} (2006) 288--301},
\href{http://arxiv.org/abs/hep-th/0510171}{{\ttfamily arXiv:hep-th/0510171
  [hep-th]}}.

\bibitem{Arutyunov:2007tc}
G.~Arutyunov and S.~Frolov, ``{On String S-matrix, Bound States and TBA},''
  \href{http://dx.doi.org/10.1088/1126-6708/2007/12/024}{{\em JHEP} {\bfseries
  12} (2007) 024},
\href{http://arxiv.org/abs/0710.1568}{{\ttfamily arXiv:0710.1568 [hep-th]}}.

\bibitem{Arutyunov:2009zu}
G.~Arutyunov and S.~Frolov, ``{String hypothesis for the $\text{AdS}_5\times
  \text{S}^5$ mirror},''
  \href{http://dx.doi.org/10.1088/1126-6708/2009/03/152}{{\em JHEP} {\bfseries
  03} (2009) 152},
\href{http://arxiv.org/abs/0901.1417}{{\ttfamily arXiv:0901.1417 [hep-th]}}.

\bibitem{Bombardelli:2009ns}
D.~Bombardelli, D.~Fioravanti, and R.~Tateo, ``{Thermodynamic Bethe Ansatz for
  planar AdS/CFT: A Proposal},''
  \href{http://dx.doi.org/10.1088/1751-8113/42/37/375401}{{\em J. Phys.}
  {\bfseries A42} (2009) 375401},
\href{http://arxiv.org/abs/0902.3930}{{\ttfamily arXiv:0902.3930 [hep-th]}}.

\bibitem{Gromov:2009bc}
N.~Gromov, V.~Kazakov, A.~Kozak, and P.~Vieira, ``{Exact Spectrum of Anomalous
  Dimensions of Planar $\mathcal{N} = 4$ Supersymmetric Yang-Mills Theory: TBA
  and excited states},''
  \href{http://dx.doi.org/10.1007/s11005-010-0374-8}{{\em Lett. Math. Phys.}
  {\bfseries 91} (2010) 265--287},
\href{http://arxiv.org/abs/0902.4458}{{\ttfamily arXiv:0902.4458 [hep-th]}}.

\bibitem{Arutyunov:2009ur}
G.~Arutyunov and S.~Frolov, ``{Thermodynamic Bethe Ansatz for the
  AdS$_5\times$S$^5$ Mirror Model},''
  \href{http://dx.doi.org/10.1088/1126-6708/2009/05/068}{{\em JHEP} {\bfseries
  05} (2009) 068},
\href{http://arxiv.org/abs/0903.0141}{{\ttfamily arXiv:0903.0141 [hep-th]}}.

\bibitem{Gromov:2009tv}
N.~Gromov, V.~Kazakov, and P.~Vieira, ``{Exact Spectrum of Anomalous Dimensions
  of Planar $\mathcal{N}=4$ Supersymmetric Yang-Mills Theory},''
  \href{http://dx.doi.org/10.1103/PhysRevLett.103.131601}{{\em Phys. Rev.
  Lett.} {\bfseries 103} (2009) 131601},
\href{http://arxiv.org/abs/0901.3753}{{\ttfamily arXiv:0901.3753 [hep-th]}}.

\bibitem{Gromov:2011cx}
N.~Gromov, V.~Kazakov, S.~Leurent, and D.~Volin, ``{Solving the AdS/CFT
  Y-system},'' \href{http://dx.doi.org/10.1007/JHEP07(2012)023}{{\em JHEP}
  {\bfseries 1207} (2012) 023},
\href{http://arxiv.org/abs/1110.0562}{{\ttfamily arXiv:1110.0562 [hep-th]}}.

\bibitem{Gromov:2013pga}
N.~Gromov, V.~Kazakov, S.~Leurent, and D.~Volin, ``{Quantum Spectral Curve for
  Planar $\mathcal{N} =$ Super-Yang-Mills Theory},''
  \href{http://dx.doi.org/10.1103/PhysRevLett.112.011602}{{\em Phys.Rev.Lett.}
  {\bfseries 112} no.~1, (2014) 011602},
\href{http://arxiv.org/abs/1305.1939}{{\ttfamily arXiv:1305.1939 [hep-th]}}.

\bibitem{Marboe:2014gma}
C.~Marboe and D.~Volin, ``{Quantum spectral curve as a tool for a perturbative
  quantum field theory},''
  \href{http://dx.doi.org/10.1016/j.nuclphysb.2015.08.021}{{\em Nucl. Phys.}
  {\bfseries B899} (2015) 810--847},
\href{http://arxiv.org/abs/1411.4758}{{\ttfamily arXiv:1411.4758 [hep-th]}}.

\bibitem{Gromov:2015wca}
N.~Gromov, F.~Levkovich-Maslyuk, and G.~Sizov, ``{Quantum Spectral Curve and
  the Numerical Solution of the Spectral Problem in AdS$_5$/CFT$_4$},''
\href{http://arxiv.org/abs/1504.06640}{{\ttfamily arXiv:1504.06640 [hep-th]}}.

\bibitem{Kotikov:2001sc}
A.~Kotikov and L.~Lipatov, ``{DGLAP and BFKL evolution equations in the
  $\mathcal{N}=4$ supersymmetric gauge theory},''
\href{http://arxiv.org/abs/hep-ph/0112346}{{\ttfamily arXiv:hep-ph/0112346
  [hep-ph]}}.

\bibitem{Kotikov:2004er}
A.~Kotikov, L.~Lipatov, A.~Onishchenko, and V.~Velizhanin, ``{Three loop
  universal anomalous dimension of the Wilson operators in $\mathcal{N}=4$ SUSY
  Yang-Mills model},''
  \href{http://dx.doi.org/10.1016/j.physletb.2004.05.078}{{\em Phys.Lett.}
  {\bfseries B595} (2004) 521--529},
\href{http://arxiv.org/abs/hep-th/0404092}{{\ttfamily arXiv:hep-th/0404092
  [hep-th]}}.

\bibitem{Kotikov:2006ts}
A.~Kotikov and L.~Lipatov, ``{On the highest transcendentality in
  $\mathcal{N}=4$ SUSY},''
  \href{http://dx.doi.org/10.1016/j.nuclphysb.2007.01.020}{{\em Nucl.Phys.}
  {\bfseries B769} (2007) 217--255},
\href{http://arxiv.org/abs/hep-th/0611204}{{\ttfamily arXiv:hep-th/0611204
  [hep-th]}}.

\bibitem{Gehrmann:2011xn}
T.~Gehrmann, J.~M. Henn, and T.~Huber, ``{The three-loop form factor in $\cN=4$
  super Yang-Mills},'' \href{http://dx.doi.org/10.1007/JHEP03(2012)101}{{\em
  JHEP} {\bfseries 1203} (2012) 101},
\href{http://arxiv.org/abs/1112.4524}{{\ttfamily arXiv:1112.4524 [hep-th]}}.

\bibitem{Li:2014afw}
Y.~Li, A.~von Manteuffel, R.~M. Schabinger, and H.~X. Zhu, ``{Soft-virtual
  corrections to Higgs production at N$^3$LO},''
  \href{http://dx.doi.org/10.1103/PhysRevD.91.036008}{{\em Phys. Rev.}
  {\bfseries D91} (2015) 036008},
\href{http://arxiv.org/abs/1412.2771}{{\ttfamily arXiv:1412.2771 [hep-ph]}}.

\bibitem{Elvang:2013cua}
H.~Elvang and Y.-t. Huang, ``{Scattering Amplitudes},''
\href{http://arxiv.org/abs/1308.1697}{{\ttfamily arXiv:1308.1697 [hep-th]}}.

\bibitem{Henn:2014yza}
J.~M. Henn and J.~C. Plefka, ``{Scattering Amplitudes in Gauge Theories},''
\href{http://dx.doi.org/978-3-642-54021-9, 10.1007/978-3-642-54022-6}{{\em
  Lect.Notes Phys.} {\bfseries 883} (2014) 1--195}.

\bibitem{Parke:1986gb}
S.~J. Parke and T.~Taylor, ``{An Amplitude for $n$ Gluon Scattering},''
\href{http://dx.doi.org/10.1103/PhysRevLett.56.2459}{{\em Phys.Rev.Lett.}
  {\bfseries 56} (1986) 2459}.

\bibitem{Nair:1988bq}
V.~Nair, ``{A Current Algebra for Some Gauge Theory Amplitudes},''
\href{http://dx.doi.org/10.1016/0370-2693(88)91471-2}{{\em Phys.Lett.}
  {\bfseries B214} (1988) 215}.

\bibitem{Bern:1994zx}
Z.~Bern, L.~J. Dixon, D.~C. Dunbar, and D.~A. Kosower, ``{One-loop $n$-point
  gauge theory amplitudes, unitarity and collinear limits},''
  \href{http://dx.doi.org/10.1016/0550-3213(94)90179-1}{{\em Nucl.Phys.}
  {\bfseries B425} (1994) 217--260},
\href{http://arxiv.org/abs/hep-ph/9403226}{{\ttfamily arXiv:hep-ph/9403226
  [hep-ph]}}.

\bibitem{Bern:1994cg}
Z.~Bern, L.~J. Dixon, D.~C. Dunbar, and D.~A. Kosower, ``{Fusing gauge theory
  tree amplitudes into loop amplitudes},''
  \href{http://dx.doi.org/10.1016/0550-3213(94)00488-Z}{{\em Nucl.Phys.}
  {\bfseries B435} (1995) 59--101},
\href{http://arxiv.org/abs/hep-ph/9409265}{{\ttfamily arXiv:hep-ph/9409265
  [hep-ph]}}.

\bibitem{Britto:2004nc}
R.~Britto, F.~Cachazo, and B.~Feng, ``{Generalized unitarity and one-loop
  amplitudes in $\cN=4$ super-Yang-Mills},''
  \href{http://dx.doi.org/10.1016/j.nuclphysb.2005.07.014}{{\em Nucl.Phys.}
  {\bfseries B725} (2005) 275--305},
\href{http://arxiv.org/abs/hep-th/0412103}{{\ttfamily arXiv:hep-th/0412103
  [hep-th]}}.

\bibitem{Bern:2005iz}
Z.~Bern, L.~J. Dixon, and V.~A. Smirnov, ``{Iteration of planar amplitudes in
  maximally supersymmetric Yang-Mills theory at three loops and beyond},''
  \href{http://dx.doi.org/10.1103/PhysRevD.72.085001}{{\em Phys.Rev.}
  {\bfseries D72} (2005) 085001},
\href{http://arxiv.org/abs/hep-th/0505205}{{\ttfamily arXiv:hep-th/0505205
  [hep-th]}}.

\bibitem{Anastasiou:2003kj}
C.~Anastasiou, Z.~Bern, L.~J. Dixon, and D.~Kosower, ``{Planar amplitudes in
  maximally supersymmetric Yang-Mills theory},''
  \href{http://dx.doi.org/10.1103/PhysRevLett.91.251602}{{\em Phys.Rev.Lett.}
  {\bfseries 91} (2003) 251602},
\href{http://arxiv.org/abs/hep-th/0309040}{{\ttfamily arXiv:hep-th/0309040
  [hep-th]}}.

\bibitem{Catani:1998bh}
S.~Catani, ``{The Singular behavior of QCD amplitudes at two loop order},''
  \href{http://dx.doi.org/10.1016/S0370-2693(98)00332-3}{{\em Phys.Lett.}
  {\bfseries B427} (1998) 161--171},
\href{http://arxiv.org/abs/hep-ph/9802439}{{\ttfamily arXiv:hep-ph/9802439
  [hep-ph]}}.

\bibitem{Sterman:2002qn}
G.~F. Sterman and M.~E. Tejeda-Yeomans, ``{Multiloop amplitudes and
  resummation},'' \href{http://dx.doi.org/10.1016/S0370-2693(02)03100-3}{{\em
  Phys.Lett.} {\bfseries B552} (2003) 48--56},
\href{http://arxiv.org/abs/hep-ph/0210130}{{\ttfamily arXiv:hep-ph/0210130
  [hep-ph]}}.

\bibitem{Beisert:2006ez}
N.~Beisert, B.~Eden, and M.~Staudacher, ``{Transcendentality and Crossing},''
  \href{http://dx.doi.org/10.1088/1742-5468/2007/01/P01021}{{\em J.Stat.Mech.}
  {\bfseries 0701} (2007) P01021},
\href{http://arxiv.org/abs/hep-th/0610251}{{\ttfamily arXiv:hep-th/0610251
  [hep-th]}}.

\bibitem{Alday:2007he}
L.~F. Alday and J.~Maldacena, ``{Comments on gluon scattering amplitudes via
  AdS/CFT},'' \href{http://dx.doi.org/10.1088/1126-6708/2007/11/068}{{\em JHEP}
  {\bfseries 0711} (2007) 068},
\href{http://arxiv.org/abs/0710.1060}{{\ttfamily arXiv:0710.1060 [hep-th]}}.

\bibitem{Bartels:2008ce}
J.~Bartels, L.~N. Lipatov, and A.~Sabio~Vera, ``{BFKL Pomeron, Reggeized gluons
  and Bern-Dixon-Smirnov amplitudes},''
  \href{http://dx.doi.org/10.1103/PhysRevD.80.045002}{{\em Phys. Rev.}
  {\bfseries D80} (2009) 045002},
\href{http://arxiv.org/abs/0802.2065}{{\ttfamily arXiv:0802.2065 [hep-th]}}.

\bibitem{Bern:2008ap}
Z.~Bern, L.~Dixon, D.~Kosower, R.~Roiban, M.~Spradlin, C.~Vergu, and
  A.~Volovich, ``{The Two-Loop Six-Gluon MHV Amplitude in Maximally
  Supersymmetric Yang-Mills Theory},''
  \href{http://dx.doi.org/10.1103/PhysRevD.78.045007}{{\em Phys.Rev.}
  {\bfseries D78} (2008) 045007},
\href{http://arxiv.org/abs/0803.1465}{{\ttfamily arXiv:0803.1465 [hep-th]}}.

\bibitem{Drummond:2008aq}
J.~Drummond, J.~Henn, G.~Korchemsky, and E.~Sokatchev, ``{Hexagon Wilson loop =
  six-gluon MHV amplitude},''
  \href{http://dx.doi.org/10.1016/j.nuclphysb.2009.02.015}{{\em Nucl.Phys.}
  {\bfseries B815} (2009) 142--173},
\href{http://arxiv.org/abs/0803.1466}{{\ttfamily arXiv:0803.1466 [hep-th]}}.

\bibitem{Goncharov09}
B.~Goncharov, A, ``{A simple construction of Grassmannian polylogarithms},''
\href{http://arxiv.org/abs/0908.2238}{{\ttfamily arXiv:0908.2238 [math.AG]}}.

\bibitem{Goncharov:2010jf}
A.~B. Goncharov, M.~Spradlin, C.~Vergu, and A.~Volovich, ``{Classical
  Polylogarithms for Amplitudes and Wilson Loops},''
  \href{http://dx.doi.org/10.1103/PhysRevLett.105.151605}{{\em Phys.Rev.Lett.}
  {\bfseries 105} (2010) 151605},
\href{http://arxiv.org/abs/1006.5703}{{\ttfamily arXiv:1006.5703 [hep-th]}}.

\bibitem{Duhr:2014woa}
C.~Duhr, ``{Mathematical aspects of scattering amplitudes},'' in {\em
  {Theoretical Advanced Study Institute in Elementary Particle Physics:
  Journeys Through the Precision Frontier: Amplitudes for Colliders (TASI 2014)
  Boulder, Colorado, June 2-27, 2014}}.
\newblock 2014.
\newblock
\href{http://arxiv.org/abs/1411.7538}{{\ttfamily arXiv:1411.7538 [hep-ph]}}.
\newblock

\bibitem{Dixon:2011pw}
L.~J. Dixon, J.~M. Drummond, and J.~M. Henn, ``{Bootstrapping the three-loop
  hexagon},'' \href{http://dx.doi.org/10.1007/JHEP11(2011)023}{{\em JHEP}
  {\bfseries 11} (2011) 023},
\href{http://arxiv.org/abs/1108.4461}{{\ttfamily arXiv:1108.4461 [hep-th]}}.

\bibitem{Dixon:2014xca}
L.~J. Dixon, J.~M. Drummond, C.~Duhr, M.~von Hippel, and J.~Pennington,
  ``{Bootstrapping six-gluon scattering in planar $\cN=4$ super-Yang-Mills
  theory},'' {\em PoS} {\bfseries LL2014} (2014) 077,
\href{http://arxiv.org/abs/1407.4724}{{\ttfamily arXiv:1407.4724 [hep-th]}}.

\bibitem{CaronHuot:2012ab}
S.~Caron-Huot and K.~J. Larsen, ``{Uniqueness of two-loop master contours},''
  \href{http://dx.doi.org/10.1007/JHEP10(2012)026}{{\em JHEP} {\bfseries 1210}
  (2012) 026},
\href{http://arxiv.org/abs/1205.0801}{{\ttfamily arXiv:1205.0801 [hep-ph]}}.

\bibitem{ArkaniHamed:2012nw}
N.~Arkani-Hamed, J.~L. Bourjaily, F.~Cachazo, A.~B. Goncharov, A.~Postnikov,
  and J.~Trnka, ``{Scattering Amplitudes and the Positive Grassmannian},''
\href{http://arxiv.org/abs/1212.5605}{{\ttfamily arXiv:1212.5605 [hep-th]}}.

\bibitem{Nandan:2013ip}
D.~Nandan, M.~F. Paulos, M.~Spradlin, and A.~Volovich, ``{Star Integrals,
  Convolutions and Simplices},''
  \href{http://dx.doi.org/10.1007/JHEP05(2013)105}{{\em JHEP} {\bfseries 05}
  (2013) 105},
\href{http://arxiv.org/abs/1301.2500}{{\ttfamily arXiv:1301.2500 [hep-th]}}.

\bibitem{Cachazo:2004kj}
F.~Cachazo, P.~Svrcek, and E.~Witten, ``{MHV vertices and tree amplitudes in
  gauge theory},'' \href{http://dx.doi.org/10.1088/1126-6708/2004/09/006}{{\em
  JHEP} {\bfseries 0409} (2004) 006},
\href{http://arxiv.org/abs/hep-th/0403047}{{\ttfamily arXiv:hep-th/0403047
  [hep-th]}}.

\bibitem{Britto:2004ap}
R.~Britto, F.~Cachazo, and B.~Feng, ``{New recursion relations for tree
  amplitudes of gluons},''
  \href{http://dx.doi.org/10.1016/j.nuclphysb.2005.02.030}{{\em Nucl.Phys.}
  {\bfseries B715} (2005) 499--522},
\href{http://arxiv.org/abs/hep-th/0412308}{{\ttfamily arXiv:hep-th/0412308
  [hep-th]}}.

\bibitem{Britto:2005fq}
R.~Britto, F.~Cachazo, B.~Feng, and E.~Witten, ``{Direct proof of tree-level
  recursion relation in Yang-Mills theory},''
  \href{http://dx.doi.org/10.1103/PhysRevLett.94.181602}{{\em Phys.Rev.Lett.}
  {\bfseries 94} (2005) 181602},
\href{http://arxiv.org/abs/hep-th/0501052}{{\ttfamily arXiv:hep-th/0501052
  [hep-th]}}.

\bibitem{Drummond:2008cr}
J.~Drummond and J.~Henn, ``{All tree-level amplitudes in $\cN=4$ SYM},''
  \href{http://dx.doi.org/10.1088/1126-6708/2009/04/018}{{\em JHEP} {\bfseries
  0904} (2009) 018},
\href{http://arxiv.org/abs/0808.2475}{{\ttfamily arXiv:0808.2475 [hep-th]}}.

\bibitem{ArkaniHamed:2010kv}
N.~Arkani-Hamed, J.~L. Bourjaily, F.~Cachazo, S.~Caron-Huot, and J.~Trnka,
  ``{The All-Loop Integrand For Scattering Amplitudes in Planar $\cN=4$ SYM},''
  \href{http://dx.doi.org/10.1007/JHEP01(2011)041}{{\em JHEP} {\bfseries 1101}
  (2011) 041},
\href{http://arxiv.org/abs/1008.2958}{{\ttfamily arXiv:1008.2958 [hep-th]}}.

\bibitem{Penrose:1967wn}
R.~Penrose, ``{Twistor algebra},''
\href{http://dx.doi.org/10.1063/1.1705200}{{\em J.Math.Phys.} {\bfseries 8}
  (1967) 345}.

\bibitem{Hodges:2009hk}
A.~Hodges, ``{Eliminating spurious poles from gauge-theoretic amplitudes},''
  \href{http://dx.doi.org/10.1007/JHEP05(2013)135}{{\em JHEP} {\bfseries 1305}
  (2013) 135},
\href{http://arxiv.org/abs/0905.1473}{{\ttfamily arXiv:0905.1473 [hep-th]}}.

\bibitem{Arkani-Hamed:2014via}
N.~Arkani-Hamed, J.~L. Bourjaily, F.~Cachazo, and J.~Trnka, ``{Singularity
  Structure of Maximally Supersymmetric Scattering Amplitudes},''
  \href{http://dx.doi.org/10.1103/PhysRevLett.113.261603}{{\em Phys. Rev.
  Lett.} {\bfseries 113} no.~26, (2014) 261603},
\href{http://arxiv.org/abs/1410.0354}{{\ttfamily arXiv:1410.0354 [hep-th]}}.

\bibitem{Chen:2014ara}
B.~Chen, G.~Chen, Y.-K.~E. Cheung, Y.~Li, R.~Xie, and Y.~Xin, ``{Nonplanar
  On-shell Diagrams and Leading Singularities of Scattering Amplitudes},''
\href{http://arxiv.org/abs/1411.3889}{{\ttfamily arXiv:1411.3889 [hep-th]}}.

\bibitem{Arkani-Hamed:2014bca}
N.~Arkani-Hamed, J.~L. Bourjaily, F.~Cachazo, A.~Postnikov, and J.~Trnka,
  ``{On-Shell Structures of MHV Amplitudes Beyond the Planar Limit},''
  \href{http://dx.doi.org/10.1007/JHEP06(2015)179}{{\em JHEP} {\bfseries 06}
  (2015) 179},
\href{http://arxiv.org/abs/1412.8475}{{\ttfamily arXiv:1412.8475 [hep-th]}}.

\bibitem{Bern:2014kca}
Z.~Bern, E.~Herrmann, S.~Litsey, J.~Stankowicz, and J.~Trnka, ``{Logarithmic
  Singularities and Maximally Supersymmetric Amplitudes},''
  \href{http://dx.doi.org/10.1007/JHEP06(2015)202}{{\em JHEP} {\bfseries 06}
  (2015) 202},
\href{http://arxiv.org/abs/1412.8584}{{\ttfamily arXiv:1412.8584 [hep-th]}}.

\bibitem{Franco:2015rma}
S.~Franco, D.~Galloni, B.~Penante, and C.~Wen, ``{Non-Planar On-Shell
  Diagrams},'' \href{http://dx.doi.org/10.1007/JHEP06(2015)199}{{\em JHEP}
  {\bfseries 06} (2015) 199},
\href{http://arxiv.org/abs/1502.02034}{{\ttfamily arXiv:1502.02034 [hep-th]}}.

\bibitem{Chen:2015qna}
B.~Chen, G.~Chen, Y.-K.~E. Cheung, R.~Xie, and Y.~Xin, ``{Top-forms of Leading
  Singularities in Nonplanar Multi-loop Amplitudes},''
\href{http://arxiv.org/abs/1506.02880}{{\ttfamily arXiv:1506.02880 [hep-th]}}.

\bibitem{Benincasa:2015zna}
P.~Benincasa, ``{On-shell diagrammatics and the perturbative structure of
  planar gauge theories},''
\href{http://arxiv.org/abs/1510.03642}{{\ttfamily arXiv:1510.03642 [hep-th]}}.

\bibitem{ArkaniHamed:2009dn}
N.~Arkani-Hamed, F.~Cachazo, C.~Cheung, and J.~Kaplan, ``{A Duality For The S
  Matrix},'' \href{http://dx.doi.org/10.1007/JHEP03(2010)020}{{\em JHEP}
  {\bfseries 1003} (2010) 020},
\href{http://arxiv.org/abs/0907.5418}{{\ttfamily arXiv:0907.5418 [hep-th]}}.

\bibitem{Mason:2009qx}
L.~Mason and D.~Skinner, ``{Dual Superconformal Invariance, Momentum Twistors
  and Grassmannians},''
  \href{http://dx.doi.org/10.1088/1126-6708/2009/11/045}{{\em JHEP} {\bfseries
  0911} (2009) 045},
\href{http://arxiv.org/abs/0909.0250}{{\ttfamily arXiv:0909.0250 [hep-th]}}.

\bibitem{ArkaniHamed:2009vw}
N.~Arkani-Hamed, F.~Cachazo, and C.~Cheung, ``{The Grassmannian Origin Of Dual
  Superconformal Invariance},''
  \href{http://dx.doi.org/10.1007/JHEP03(2010)036}{{\em JHEP} {\bfseries 1003}
  (2010) 036},
\href{http://arxiv.org/abs/0909.0483}{{\ttfamily arXiv:0909.0483 [hep-th]}}.

\bibitem{Arkani-Hamed:2013jha}
N.~Arkani-Hamed and J.~Trnka, ``{The Amplituhedron},''
  \href{http://dx.doi.org/10.1007/JHEP10(2014)030}{{\em JHEP} {\bfseries 1410}
  (2014) 30},
\href{http://arxiv.org/abs/1312.2007}{{\ttfamily arXiv:1312.2007 [hep-th]}}.

\bibitem{Arkani-Hamed:2013kca}
N.~Arkani-Hamed and J.~Trnka, ``{Into the Amplituhedron},''
  \href{http://dx.doi.org/10.1007/JHEP12(2014)182}{{\em JHEP} {\bfseries 1412}
  (2014) 182},
\href{http://arxiv.org/abs/1312.7878}{{\ttfamily arXiv:1312.7878 [hep-th]}}.

\bibitem{ArkaniHamed:2010gg}
N.~Arkani-Hamed, J.~L. Bourjaily, F.~Cachazo, A.~Hodges, and J.~Trnka, ``{A
  Note on Polytopes for Scattering Amplitudes},''
  \href{http://dx.doi.org/10.1007/JHEP04(2012)081}{{\em JHEP} {\bfseries 1204}
  (2012) 081},
\href{http://arxiv.org/abs/1012.6030}{{\ttfamily arXiv:1012.6030 [hep-th]}}.

\bibitem{Dixon:2010ik}
L.~J. Dixon, J.~M. Henn, J.~Plefka, and T.~Schuster, ``{All tree-level
  amplitudes in massless QCD},''
  \href{http://dx.doi.org/10.1007/JHEP01(2011)035}{{\em JHEP} {\bfseries 01}
  (2011) 035},
\href{http://arxiv.org/abs/1010.3991}{{\ttfamily arXiv:1010.3991 [hep-ph]}}.

\bibitem{Drummond:2009fd}
J.~M. Drummond, J.~M. Henn, and J.~Plefka, ``{Yangian symmetry of scattering
  amplitudes in $\mathcal{N}=4$ super Yang-Mills theory},''
  \href{http://dx.doi.org/10.1088/1126-6708/2009/05/046}{{\em JHEP} {\bfseries
  0905} (2009) 046},
\href{http://arxiv.org/abs/0902.2987}{{\ttfamily arXiv:0902.2987 [hep-th]}}.

\bibitem{Drummond:2008vq}
J.~Drummond, J.~Henn, G.~Korchemsky, and E.~Sokatchev, ``{Dual superconformal
  symmetry of scattering amplitudes in $\mathcal{N}=4$ super-Yang-Mills
  theory},'' \href{http://dx.doi.org/10.1016/j.nuclphysb.2009.11.022}{{\em
  Nucl.Phys.} {\bfseries B828} (2010) 317--374},
\href{http://arxiv.org/abs/0807.1095}{{\ttfamily arXiv:0807.1095 [hep-th]}}.

\bibitem{Zwiebel:2011bx}
B.~I. Zwiebel, ``{From Scattering Amplitudes to the Dilatation Generator in
  $\cN=4$ SYM},'' \href{http://dx.doi.org/10.1088/1751-8113/45/11/115401}{{\em
  J.Phys.} {\bfseries A45} (2012) 115401},
\href{http://arxiv.org/abs/1111.0083}{{\ttfamily arXiv:1111.0083 [hep-th]}}.

\bibitem{Ferro:2012xw}
L.~Ferro, T.~Łukowski, C.~Meneghelli, J.~Plefka, and M.~Staudacher,
  ``{Harmonic R-matrices for Scattering Amplitudes and Spectral
  Regularization},''
  \href{http://dx.doi.org/10.1103/PhysRevLett.110.121602}{{\em Phys.Rev.Lett.}
  {\bfseries 110} no.~12, (2013) 121602},
\href{http://arxiv.org/abs/1212.0850}{{\ttfamily arXiv:1212.0850 [hep-th]}}.

\bibitem{Ferro:2013dga}
L.~Ferro, T.~Łukowski, C.~Meneghelli, J.~Plefka, and M.~Staudacher,
  ``{Spectral Parameters for Scattering Amplitudes in $\cN=4$ Super Yang-Mills
  Theory},'' \href{http://dx.doi.org/10.1007/JHEP01(2014)094}{{\em JHEP}
  {\bfseries 1401} (2014) 094},
\href{http://arxiv.org/abs/1308.3494}{{\ttfamily arXiv:1308.3494 [hep-th]}}.

\bibitem{Chicherin:2013ora}
D.~Chicherin, S.~Derkachov, and R.~Kirschner, ``{Yang-Baxter operators and
  scattering amplitudes in $\cN=4$ super-Yang-Mills theory},''
  \href{http://dx.doi.org/10.1016/j.nuclphysb.2014.02.016}{{\em Nucl.Phys.}
  {\bfseries B881} (2014) 467--501},
\href{http://arxiv.org/abs/1309.5748}{{\ttfamily arXiv:1309.5748 [hep-th]}}.

\bibitem{Frassek:2013xza}
R.~Frassek, N.~Kanning, Y.~Ko, and M.~Staudacher, ``{Bethe Ansatz for Yangian
  Invariants: Towards Super Yang-Mills Scattering Amplitudes},''
  \href{http://dx.doi.org/10.1016/j.nuclphysb.2014.03.015}{{\em Nucl.Phys.}
  {\bfseries B883} (2014) 373--424},
\href{http://arxiv.org/abs/1312.1693}{{\ttfamily arXiv:1312.1693 [math-ph]}}.

\bibitem{Beisert:2014qba}
N.~Beisert, J.~Broedel, and M.~Rosso, ``{On Yangian-invariant regularization of
  deformed on-shell diagrams in $\mathcal{N}=4$ super-Yang-Mills theory},''
  \href{http://dx.doi.org/10.1088/1751-8113/47/36/365402}{{\em J.Phys.}
  {\bfseries A47} (2014) 365402},
\href{http://arxiv.org/abs/1401.7274}{{\ttfamily arXiv:1401.7274 [hep-th]}}.

\bibitem{Kanning:2014maa}
N.~Kanning, T.~Łukowski, and M.~Staudacher, ``{A shortcut to general
  tree-level scattering amplitudes in $\mathcal{N} = 4$ SYM via
  integrability},'' \href{http://dx.doi.org/10.1002/prop.201400017}{{\em
  Fortsch.Phys.} {\bfseries 62} (2014) 556--572},
\href{http://arxiv.org/abs/1403.3382}{{\ttfamily arXiv:1403.3382 [hep-th]}}.

\bibitem{Broedel:2014pia}
J.~Broedel, M.~de~Leeuw, and M.~Rosso, ``{A dictionary between R-operators,
  on-shell graphs and Yangian algebras},''
  \href{http://dx.doi.org/10.1007/JHEP06(2014)170}{{\em JHEP} {\bfseries 1406}
  (2014) 170},
\href{http://arxiv.org/abs/1403.3670}{{\ttfamily arXiv:1403.3670 [hep-th]}}.

\bibitem{Broedel:2014hca}
J.~Broedel, M.~de~Leeuw, and M.~Rosso, ``{Deformed one-loop amplitudes in $
  \mathcal{N}=4 $ super-Yang-Mills theory},''
  \href{http://dx.doi.org/10.1007/JHEP11(2014)091}{{\em JHEP} {\bfseries 11}
  (2014) 091},
\href{http://arxiv.org/abs/1406.4024}{{\ttfamily arXiv:1406.4024 [hep-th]}}.

\bibitem{Bargheer:2014mxa}
T.~Bargheer, Y.-t. Huang, F.~Loebbert, and M.~Yamazaki, ``{Integrable Amplitude
  Deformations for $\cN=4$ Super Yang-Mills and ABJM Theory},''
  \href{http://dx.doi.org/10.1103/PhysRevD.91.026004}{{\em Phys. Rev.}
  {\bfseries D91} no.~2, (2015) 026004},
\href{http://arxiv.org/abs/1407.4449}{{\ttfamily arXiv:1407.4449 [hep-th]}}.

\bibitem{Ferro:2014gca}
L.~Ferro, T.~Łukowski, and M.~Staudacher, ``{$\mathcal N=4$ scattering
  amplitudes and the deformed Graßmannian},''
  \href{http://dx.doi.org/10.1016/j.nuclphysb.2014.10.012}{{\em Nucl.Phys.}
  {\bfseries B889} (2014) 192--206},
\href{http://arxiv.org/abs/1407.6736}{{\ttfamily arXiv:1407.6736 [hep-th]}}.

\bibitem{Kanning:2014cca}
N.~Kanning, Y.~Ko, and M.~Staudacher, ``{Graßmannian integrals as matrix
  models for non-compact Yangian invariants},''
  \href{http://dx.doi.org/10.1016/j.nuclphysb.2015.03.011}{{\em Nucl. Phys.}
  {\bfseries B894} (2015) 407--421},
\href{http://arxiv.org/abs/1412.8476}{{\ttfamily arXiv:1412.8476 [hep-th]}}.

\bibitem{Alday:2007hr}
L.~F. Alday and J.~M. Maldacena, ``{Gluon scattering amplitudes at strong
  coupling},'' \href{http://dx.doi.org/10.1088/1126-6708/2007/06/064}{{\em
  JHEP} {\bfseries 0706} (2007) 064},
\href{http://arxiv.org/abs/0705.0303}{{\ttfamily arXiv:0705.0303 [hep-th]}}.

\bibitem{Alday:2009dv}
L.~F. Alday, D.~Gaiotto, and J.~Maldacena, ``{Thermodynamic Bubble Ansatz},''
  \href{http://dx.doi.org/10.1007/JHEP09(2011)032}{{\em JHEP} {\bfseries 09}
  (2011) 032},
\href{http://arxiv.org/abs/0911.4708}{{\ttfamily arXiv:0911.4708 [hep-th]}}.

\bibitem{Alday:2010vh}
L.~F. Alday, J.~Maldacena, A.~Sever, and P.~Vieira, ``{Y-system for Scattering
  Amplitudes},'' \href{http://dx.doi.org/10.1088/1751-8113/43/48/485401}{{\em
  J.Phys.} {\bfseries A43} (2010) 485401},
\href{http://arxiv.org/abs/1002.2459}{{\ttfamily arXiv:1002.2459 [hep-th]}}.

\bibitem{Basso:2013vsa}
B.~Basso, A.~Sever, and P.~Vieira, ``{Spacetime and Flux Tube S-Matrices at
  Finite Coupling for $\cN=4$ Supersymmetric Yang-Mills Theory},''
  \href{http://dx.doi.org/10.1103/PhysRevLett.111.091602}{{\em Phys.Rev.Lett.}
  {\bfseries 111} no.~9, (2013) 091602},
\href{http://arxiv.org/abs/1303.1396}{{\ttfamily arXiv:1303.1396 [hep-th]}}.

\bibitem{Basso:2014hfa}
B.~Basso, J.~Caetano, L.~Cordova, A.~Sever, and P.~Vieira, ``{OPE for all
  Helicity Amplitudes},'' \href{http://dx.doi.org/10.1007/JHEP08(2015)018}{{\em
  JHEP} {\bfseries 08} (2015) 018},
\href{http://arxiv.org/abs/1412.1132}{{\ttfamily arXiv:1412.1132 [hep-th]}}.

\bibitem{Schmidt:1997wr}
C.~R. Schmidt, ``{H $\longrightarrow$ g g g (g q anti-q) at two loops in the
  large M(t) limit},''
  \href{http://dx.doi.org/10.1016/S0370-2693(97)01102-7}{{\em Phys.Lett.}
  {\bfseries B413} (1997) 391--395},
\href{http://arxiv.org/abs/hep-ph/9707448}{{\ttfamily arXiv:hep-ph/9707448
  [hep-ph]}}.

\bibitem{Hofman:2008ar}
D.~M. Hofman and J.~Maldacena, ``{Conformal collider physics: Energy and charge
  correlations},'' \href{http://dx.doi.org/10.1088/1126-6708/2008/05/012}{{\em
  JHEP} {\bfseries 05} (2008) 012},
\href{http://arxiv.org/abs/0803.1467}{{\ttfamily arXiv:0803.1467 [hep-th]}}.

\bibitem{Engelund:2012re}
O.~T. Engelund and R.~Roiban, ``{Correlation functions of local composite
  operators from generalized unitarity},''
  \href{http://dx.doi.org/10.1007/JHEP03(2013)172}{{\em JHEP} {\bfseries 1303}
  (2013) 172},
\href{http://arxiv.org/abs/1209.0227}{{\ttfamily arXiv:1209.0227 [hep-th]}}.

\bibitem{Belitsky:2013xxa}
A.~V. Belitsky, S.~Hohenegger, G.~P. Korchemsky, E.~Sokatchev, and
  A.~Zhiboedov, ``{From correlation functions to event shapes},''
  \href{http://dx.doi.org/10.1016/j.nuclphysb.2014.04.020}{{\em Nucl. Phys.}
  {\bfseries B884} (2014) 305--343},
\href{http://arxiv.org/abs/1309.0769}{{\ttfamily arXiv:1309.0769 [hep-th]}}.

\bibitem{Belitsky:2013bja}
A.~V. Belitsky, S.~Hohenegger, G.~P. Korchemsky, E.~Sokatchev, and
  A.~Zhiboedov, ``{Event shapes in $\mathcal{N} = 4$ super-Yang-Mills
  theory},'' \href{http://dx.doi.org/10.1016/j.nuclphysb.2014.04.019}{{\em
  Nucl. Phys.} {\bfseries B884} (2014) 206--256},
\href{http://arxiv.org/abs/1309.1424}{{\ttfamily arXiv:1309.1424 [hep-th]}}.

\bibitem{Bianchi:2013sta}
L.~Bianchi, V.~Forini, and A.~V. Kotikov, ``{On DIS Wilson coefficients in
  $\mathcal{N}=4$ super Yang-Mills theory},''
  \href{http://dx.doi.org/10.1016/j.physletb.2013.07.013}{{\em Phys.Lett.}
  {\bfseries B725} (2013) 394--401},
\href{http://arxiv.org/abs/1304.7252}{{\ttfamily arXiv:1304.7252 [hep-th]}}.

\bibitem{Mueller:1979ih}
A.~H. Mueller, ``{On the Asymptotic Behavior of the Sudakov Form-factor},''
\href{http://dx.doi.org/10.1103/PhysRevD.20.2037}{{\em Phys.Rev.} {\bfseries
  D20} (1979) 2037}.

\bibitem{Collins:1980ih}
J.~C. Collins, ``{Algorithm to Compute Corrections to the Sudakov
  Form-factor},''
\href{http://dx.doi.org/10.1103/PhysRevD.22.1478}{{\em Phys.Rev.} {\bfseries
  D22} (1980) 1478}.

\bibitem{Sen:1981sd}
A.~Sen, ``{Asymptotic Behavior of the Sudakov Form-Factor in QCD},''
\href{http://dx.doi.org/10.1103/PhysRevD.24.3281}{{\em Phys.Rev.} {\bfseries
  D24} (1981) 3281}.

\bibitem{Magnea:1990zb}
L.~Magnea and G.~F. Sterman, ``{Analytic continuation of the Sudakov
  form-factor in QCD},''
\href{http://dx.doi.org/10.1103/PhysRevD.42.4222}{{\em Phys.Rev.} {\bfseries
  D42} (1990) 4222--4227}.

\bibitem{vanNeerven:1985ja}
W.~van Neerven, ``{Infrared Behavior of On-shell Form-factors in a $\cN=4$
  Supersymmetric {Yang-Mills} Field Theory},''
\href{http://dx.doi.org/10.1007/BF01571808}{{\em Z.Phys.} {\bfseries C30}
  (1986) 595}.

\bibitem{Maldacena:2010kp}
J.~Maldacena and A.~Zhiboedov, ``{Form factors at strong coupling via a
  Y-system},'' \href{http://dx.doi.org/10.1007/JHEP11(2010)104}{{\em JHEP}
  {\bfseries 1011} (2010) 104},
\href{http://arxiv.org/abs/1009.1139}{{\ttfamily arXiv:1009.1139 [hep-th]}}.

\bibitem{Gao:2013dza}
Z.~Gao and G.~Yang, ``{Y-system for form factors at strong coupling in $AdS_5$
  and with multi-operator insertions in $AdS_3$},''
  \href{http://dx.doi.org/10.1007/JHEP06(2013)105}{{\em JHEP} {\bfseries 1306}
  (2013) 105},
\href{http://arxiv.org/abs/1303.2668}{{\ttfamily arXiv:1303.2668 [hep-th]}}.

\bibitem{Brandhuber:2010ad}
A.~Brandhuber, B.~Spence, G.~Travaglini, and G.~Yang, ``{Form Factors in
  $\cN=4$ Super Yang-Mills and Periodic Wilson Loops},''
  \href{http://dx.doi.org/10.1007/JHEP01(2011)134}{{\em JHEP} {\bfseries 1101}
  (2011) 134},
\href{http://arxiv.org/abs/1011.1899}{{\ttfamily arXiv:1011.1899 [hep-th]}}.

\bibitem{Bork:2010wf}
L.~Bork, D.~Kazakov, and G.~Vartanov, ``{On form factors in $\cN=4$ SYM},''
  \href{http://dx.doi.org/10.1007/JHEP02(2011)063}{{\em JHEP} {\bfseries 1102}
  (2011) 063},
\href{http://arxiv.org/abs/1011.2440}{{\ttfamily arXiv:1011.2440 [hep-th]}}.

\bibitem{Brandhuber:2011tv}
A.~Brandhuber, O.~Gurdogan, R.~Mooney, G.~Travaglini, and G.~Yang, ``{Harmony
  of Super Form Factors},''
  \href{http://dx.doi.org/10.1007/JHEP10(2011)046}{{\em JHEP} {\bfseries 1110}
  (2011) 046},
\href{http://arxiv.org/abs/1107.5067}{{\ttfamily arXiv:1107.5067 [hep-th]}}.

\bibitem{Bork:2011cj}
L.~Bork, D.~Kazakov, and G.~Vartanov, ``{On MHV Form Factors in Superspace for
  $\mathcal{N}=4$ SYM Theory},''
  \href{http://dx.doi.org/10.1007/JHEP10(2011)133}{{\em JHEP} {\bfseries 1110}
  (2011) 133},
\href{http://arxiv.org/abs/1107.5551}{{\ttfamily arXiv:1107.5551 [hep-th]}}.

\bibitem{Henn:2011by}
J.~M. Henn, S.~Moch, and S.~G. Naculich, ``{Form factors and scattering
  amplitudes in $\cN=4$ SYM in dimensional and massive regularizations},''
  \href{http://dx.doi.org/10.1007/JHEP12(2011)024}{{\em JHEP} {\bfseries 1112}
  (2011) 024},
\href{http://arxiv.org/abs/1109.5057}{{\ttfamily arXiv:1109.5057 [hep-th]}}.

\bibitem{Brandhuber:2012vm}
A.~Brandhuber, G.~Travaglini, and G.~Yang, ``{Analytic two-loop form factors in
  $\cN=4$ SYM},'' \href{http://dx.doi.org/10.1007/JHEP05(2012)082}{{\em JHEP}
  {\bfseries 1205} (2012) 082},
\href{http://arxiv.org/abs/1201.4170}{{\ttfamily arXiv:1201.4170 [hep-th]}}.

\bibitem{Bork:2012tt}
L.~Bork, ``{On NMHV form factors in $\cN=4$ SYM theory from generalized
  unitarity},'' \href{http://dx.doi.org/10.1007/JHEP01(2013)049}{{\em JHEP}
  {\bfseries 1301} (2013) 049},
\href{http://arxiv.org/abs/1203.2596}{{\ttfamily arXiv:1203.2596 [hep-th]}}.

\bibitem{Johansson:2012zv}
H.~Johansson, D.~A. Kosower, and K.~J. Larsen, ``{Two-Loop Maximal Unitarity
  with External Masses},''
  \href{http://dx.doi.org/10.1103/PhysRevD.87.025030}{{\em Phys.Rev.}
  {\bfseries D87} (2013) 025030},
\href{http://arxiv.org/abs/1208.1754}{{\ttfamily arXiv:1208.1754 [hep-th]}}.

\bibitem{Boels:2012ew}
R.~H. Boels, B.~A. Kniehl, O.~V. Tarasov, and G.~Yang, ``{Color-kinematic
  Duality for Form Factors},''
  \href{http://dx.doi.org/10.1007/JHEP02(2013)063}{{\em JHEP} {\bfseries 1302}
  (2013) 063},
\href{http://arxiv.org/abs/1211.7028}{{\ttfamily arXiv:1211.7028 [hep-th]}}.

\bibitem{Penante:2014sza}
B.~Penante, B.~Spence, G.~Travaglini, and C.~Wen, ``{On super form factors of
  half-BPS operators in $\mathcal{N}=4$ super Yang-Mills},''
  \href{http://dx.doi.org/10.1007/JHEP04(2014)083}{{\em JHEP} {\bfseries 1404}
  (2014) 083},
\href{http://arxiv.org/abs/1402.1300}{{\ttfamily arXiv:1402.1300 [hep-th]}}.

\bibitem{Brandhuber:2014ica}
A.~Brandhuber, B.~Penante, G.~Travaglini, and C.~Wen, ``{The last of the simple
  remainders},'' \href{http://dx.doi.org/10.1007/JHEP08(2014)100}{{\em JHEP}
  {\bfseries 1408} (2014) 100},
\href{http://arxiv.org/abs/1406.1443}{{\ttfamily arXiv:1406.1443 [hep-th]}}.

\bibitem{Bork:2014eqa}
L.~V. Bork, ``{On form factors in $ \mathcal{N}=4 $ SYM theory and
  polytopes},'' \href{http://dx.doi.org/10.1007/JHEP12(2014)111}{{\em JHEP}
  {\bfseries 12} (2014) 111},
\href{http://arxiv.org/abs/1407.5568}{{\ttfamily arXiv:1407.5568 [hep-th]}}.

\bibitem{Huang:2016bmv}
R.~Huang, Q.~Jin, and B.~Feng, ``{Form Factor and Boundary Contribution of
  Amplitude},''
\href{http://arxiv.org/abs/1601.06612}{{\ttfamily arXiv:1601.06612 [hep-th]}}.

\bibitem{Bern:2008qj}
Z.~Bern, J.~Carrasco, and H.~Johansson, ``{New Relations for Gauge-Theory
  Amplitudes},'' \href{http://dx.doi.org/10.1103/PhysRevD.78.085011}{{\em
  Phys.Rev.} {\bfseries D78} (2008) 085011},
\href{http://arxiv.org/abs/0805.3993}{{\ttfamily arXiv:0805.3993 [hep-ph]}}.

\bibitem{Gehrmann:2011aa}
T.~Gehrmann, M.~Jaquier, E.~Glover, and A.~Koukoutsakis, ``{Two-Loop QCD
  Corrections to the Helicity Amplitudes for $H \to$ 3 partons},''
  \href{http://dx.doi.org/10.1007/JHEP02(2012)056}{{\em JHEP} {\bfseries 1202}
  (2012) 056},
\href{http://arxiv.org/abs/1112.3554}{{\ttfamily arXiv:1112.3554 [hep-ph]}}.

\bibitem{Belitsky:2013ofa}
A.~V. Belitsky, S.~Hohenegger, G.~P. Korchemsky, E.~Sokatchev, and
  A.~Zhiboedov, ``{Energy-Energy Correlations in $\cN=4$ Supersymmetric
  Yang-Mills Theory},''
  \href{http://dx.doi.org/10.1103/PhysRevLett.112.071601}{{\em Phys. Rev.
  Lett.} {\bfseries 112} no.~7, (2014) 071601},
\href{http://arxiv.org/abs/1311.6800}{{\ttfamily arXiv:1311.6800 [hep-th]}}.

\bibitem{Boels:2015yna}
R.~Boels, B.~A. Kniehl, and G.~Yang, ``{Master integrals for the four-loop
  Sudakov form factor},''
\href{http://arxiv.org/abs/1508.03717}{{\ttfamily arXiv:1508.03717 [hep-th]}}.

\bibitem{Gunaydin:1981yq}
M.~Gunaydin and C.~Saclioglu, ``{Oscillator Like Unitary Representations of
  Noncompact Groups With a Jordan Structure and the Noncompact Groups of
  Supergravity},''
\href{http://dx.doi.org/10.1007/BF01218560}{{\em Commun. Math. Phys.}
  {\bfseries 87} (1982) 159}.

\bibitem{Gunaydin:1998sw}
M.~Gunaydin, D.~Minic, and M.~Zagermann, ``{4-D doubleton conformal theories,
  CPT and IIB string on $\text{AdS}_5\times \text{S}^5$},''
  \href{http://dx.doi.org/10.1016/S0550-3213(98)00543-4}{{\em Nucl. Phys.}
  {\bfseries B534} (1998) 96--120},
  \href{http://arxiv.org/abs/hep-th/9806042}{{\ttfamily arXiv:hep-th/9806042
  [hep-th]}}.
[Erratum: Nucl. Phys.B538,531(1999)].

\bibitem{Beisert:2010jq}
N.~Beisert, ``{On Yangian Symmetry in Planar $\cN=4$ SYM},''
\href{http://arxiv.org/abs/1004.5423}{{\ttfamily arXiv:1004.5423 [hep-th]}}.

\bibitem{Koster:2014fva}
L.~Koster, V.~Mitev, and M.~Staudacher, ``{A Twistorial Approach to
  Integrability in $\mathcal N=4$ SYM},''
  \href{http://dx.doi.org/10.1002/prop.201400085}{{\em Fortsch. Phys.}
  {\bfseries 63} no.~2, (2015) 142--147},
\href{http://arxiv.org/abs/1410.6310}{{\ttfamily arXiv:1410.6310 [hep-th]}}.

\bibitem{Brandhuber:2014pta}
A.~Brandhuber, B.~Penante, G.~Travaglini, and D.~Young, ``{Integrability and
  MHV diagrams in $\cN=4$ supersymmetric Yang-Mills theory},''
  \href{http://dx.doi.org/10.1103/PhysRevLett.114.071602}{{\em Phys. Rev.
  Lett.} {\bfseries 114} (2015) 071602},
\href{http://arxiv.org/abs/1412.1019}{{\ttfamily arXiv:1412.1019 [hep-th]}}.

\bibitem{Brandhuber:2015boa}
A.~Brandhuber, B.~Penante, G.~Travaglini, and D.~Young, ``{Integrability and
  unitarity},'' \href{http://dx.doi.org/10.1007/JHEP05(2015)005}{{\em JHEP}
  {\bfseries 05} (2015) 005},
\href{http://arxiv.org/abs/1502.06627}{{\ttfamily arXiv:1502.06627 [hep-th]}}.

\bibitem{Engelund:2015cfa}
O.~T. Engelund, ``{Lagrangian Insertion in the Light-Like Limit and the
  Super-Correlators/Super-Amplitudes Duality},''
\href{http://arxiv.org/abs/1502.01934}{{\ttfamily arXiv:1502.01934 [hep-th]}}.

\bibitem{Laenen:2015jia}
E.~Laenen, K.~J. Larsen, and R.~Rietkerk, ``{Position-space cuts for Wilson
  line correlators},'' \href{http://dx.doi.org/10.1007/JHEP07(2015)083}{{\em
  JHEP} {\bfseries 07} (2015) 083},
\href{http://arxiv.org/abs/1505.02555}{{\ttfamily arXiv:1505.02555 [hep-th]}}.

\bibitem{Chicherin:2014uca}
D.~Chicherin, R.~Doobary, B.~Eden, P.~Heslop, G.~P. Korchemsky, L.~Mason, and
  E.~Sokatchev, ``{Correlation functions of the chiral stress-tensor multiplet
  in $ \mathcal{N}=4 $ SYM},''
  \href{http://dx.doi.org/10.1007/JHEP06(2015)198}{{\em JHEP} {\bfseries 06}
  (2015) 198},
\href{http://arxiv.org/abs/1412.8718}{{\ttfamily arXiv:1412.8718 [hep-th]}}.

\bibitem{Bork:2015fla}
L.~V. Bork and A.~I. Onishchenko, ``{On Soft Theorems And Form Factors In
  $\cN=4$ SYM Theory},''
\href{http://arxiv.org/abs/1506.07551}{{\ttfamily arXiv:1506.07551 [hep-th]}}.

\bibitem{Zoubos:2010kh}
K.~Zoubos, ``{Review of AdS/CFT Integrability, Chapter IV.2: Deformations,
  Orbifolds and Open Boundaries},''
  \href{http://dx.doi.org/10.1007/s11005-011-0515-8}{{\em Lett.Math.Phys.}
  {\bfseries 99} (2012) 375--400},
\href{http://arxiv.org/abs/1012.3998}{{\ttfamily arXiv:1012.3998 [hep-th]}}.

\bibitem{vanTongeren:2013gva}
S.~J. van Tongeren, ``{Integrability of the ${\rm Ad}{{{\rm S}}_{5}}\times
  {{{\rm S}}^{5}}$ superstring and its deformations},''
  \href{http://dx.doi.org/10.1088/1751-8113/47/43/433001}{{\em J.Phys.}
  {\bfseries A47} (2014) 433001},
\href{http://arxiv.org/abs/1310.4854}{{\ttfamily arXiv:1310.4854 [hep-th]}}.

\bibitem{Leigh:1995ep}
R.~G. Leigh and M.~J. Strassler, ``{Exactly marginal operators and duality in
  four-dimensional $\cN=1$ supersymmetric gauge theory},''
  \href{http://dx.doi.org/10.1016/0550-3213(95)00261-P}{{\em Nucl.Phys.}
  {\bfseries B447} (1995) 95--136},
\href{http://arxiv.org/abs/hep-th/9503121}{{\ttfamily arXiv:hep-th/9503121
  [hep-th]}}.

\bibitem{Lunin:2005jy}
O.~Lunin and J.~M. Maldacena, ``{Deforming field theories with $\U1 \times \U1$
  global symmetry and their gravity duals},''
  \href{http://dx.doi.org/10.1088/1126-6708/2005/05/033}{{\em JHEP} {\bfseries
  0505} (2005) 033},
\href{http://arxiv.org/abs/hep-th/0502086}{{\ttfamily arXiv:hep-th/0502086
  [hep-th]}}.

\bibitem{Frolov:2005dj}
S.~Frolov, ``{Lax pair for strings in Lunin-Maldacena background},''
  \href{http://dx.doi.org/10.1088/1126-6708/2005/05/069}{{\em JHEP} {\bfseries
  0505} (2005) 069},
\href{http://arxiv.org/abs/hep-th/0503201}{{\ttfamily arXiv:hep-th/0503201
  [hep-th]}}.

\bibitem{Szabo:2001kg}
R.~J. Szabo, ``{Quantum field theory on noncommutative spaces},''
  \href{http://dx.doi.org/10.1016/S0370-1573(03)00059-0}{{\em Phys.Rept.}
  {\bfseries 378} (2003) 207--299},
\href{http://arxiv.org/abs/hep-th/0109162}{{\ttfamily arXiv:hep-th/0109162
  [hep-th]}}.

\bibitem{Filk:1996dm}
T.~Filk, ``{Divergencies in a field theory on quantum space},''
\href{http://dx.doi.org/10.1016/0370-2693(96)00024-X}{{\em Phys.Lett.}
  {\bfseries B376} (1996) 53--58}.

\bibitem{Bershadsky:1998mb}
M.~Bershadsky, Z.~Kakushadze, and C.~Vafa, ``{String expansion as large N
  expansion of gauge theories},''
  \href{http://dx.doi.org/10.1016/S0550-3213(98)00272-7}{{\em Nucl. Phys.}
  {\bfseries B523} (1998) 59--72},
\href{http://arxiv.org/abs/hep-th/9803076}{{\ttfamily arXiv:hep-th/9803076
  [hep-th]}}.

\bibitem{Bershadsky:1998cb}
M.~Bershadsky and A.~Johansen, ``{Large N limit of orbifold field theories},''
  \href{http://dx.doi.org/10.1016/S0550-3213(98)00526-4}{{\em Nucl. Phys.}
  {\bfseries B536} (1998) 141--148},
\href{http://arxiv.org/abs/hep-th/9803249}{{\ttfamily arXiv:hep-th/9803249
  [hep-th]}}.

\bibitem{Beisert:2005if}
N.~Beisert and R.~Roiban, ``{Beauty and the twist: The Bethe ansatz for twisted
  $\cN=4$ SYM},'' \href{http://dx.doi.org/10.1088/1126-6708/2005/08/039}{{\em
  JHEP} {\bfseries 0508} (2005) 039},
\href{http://arxiv.org/abs/hep-th/0505187}{{\ttfamily arXiv:hep-th/0505187
  [hep-th]}}.

\bibitem{Fiamberti:2008sn}
F.~Fiamberti, A.~Santambrogio, C.~Sieg, and D.~Zanon, ``{Single impurity
  operators at critical wrapping order in the $\beta$-deformed $\cN=4$ SYM},''
  \href{http://dx.doi.org/10.1088/1126-6708/2009/08/034}{{\em JHEP} {\bfseries
  0908} (2009) 034},
\href{http://arxiv.org/abs/0811.4594}{{\ttfamily arXiv:0811.4594 [hep-th]}}.

\bibitem{Gunnesson:2009nn}
J.~Gunnesson, ``{Wrapping in maximally supersymmetric and marginally deformed
  $\cN=4$ Yang-Mills},''
  \href{http://dx.doi.org/10.1088/1126-6708/2009/04/130}{{\em JHEP} {\bfseries
  0904} (2009) 130},
\href{http://arxiv.org/abs/0902.1427}{{\ttfamily arXiv:0902.1427 [hep-th]}}.

\bibitem{Gromov:2010dy}
N.~Gromov and F.~Levkovich-Maslyuk, ``{Y-system and $\beta$-deformed $\cN=4$
  Super-Yang-Mills},''
  \href{http://dx.doi.org/10.1088/1751-8113/44/1/015402}{{\em J.Phys.}
  {\bfseries A44} (2011) 015402},
\href{http://arxiv.org/abs/1006.5438}{{\ttfamily arXiv:1006.5438 [hep-th]}}.

\bibitem{Arutyunov:2010gu}
G.~Arutyunov, M.~de~Leeuw, and S.~J. van Tongeren, ``{Twisting the Mirror
  TBA},'' \href{http://dx.doi.org/10.1007/JHEP02(2011)025}{{\em JHEP}
  {\bfseries 1102} (2011) 025},
\href{http://arxiv.org/abs/1009.4118}{{\ttfamily arXiv:1009.4118 [hep-th]}}.

\bibitem{Ahn:2011xq}
C.~Ahn, Z.~Bajnok, D.~Bombardelli, and R.~I. Nepomechie, ``{TBA, NLO Luscher
  correction, and double wrapping in twisted AdS/CFT},''
  \href{http://dx.doi.org/10.1007/JHEP12(2011)059}{{\em JHEP} {\bfseries 1112}
  (2011) 059},
\href{http://arxiv.org/abs/1108.4914}{{\ttfamily arXiv:1108.4914 [hep-th]}}.

\bibitem{Kazakov:2015efa}
V.~Kazakov, S.~Leurent, and D.~Volin, ``{T-system on T-hook: Grassmannian
  Solution and Twisted Quantum Spectral Curve},''
\href{http://arxiv.org/abs/1510.02100}{{\ttfamily arXiv:1510.02100 [hep-th]}}.

\bibitem{Frolov:2009in}
S.~Frolov and R.~Suzuki, ``{Temperature quantization from the TBA equations},''
  \href{http://dx.doi.org/10.1016/j.physletb.2009.06.069}{{\em Phys.Lett.}
  {\bfseries B679} (2009) 60--64},
\href{http://arxiv.org/abs/0906.0499}{{\ttfamily arXiv:0906.0499 [hep-th]}}.

\bibitem{deLeeuw:2011rw}
M.~de~Leeuw and S.~J. van Tongeren, ``{Orbifolded Konishi from the Mirror
  TBA},'' \href{http://dx.doi.org/10.1088/1751-8113/44/32/325404}{{\em J.Phys.}
  {\bfseries A44} (2011) 325404},
\href{http://arxiv.org/abs/1103.5853}{{\ttfamily arXiv:1103.5853 [hep-th]}}.

\bibitem{deLeeuw:2012hp}
M.~de~Leeuw and S.~J. van Tongeren, ``{The spectral problem for strings on
  twisted $\text{AdS}_5\times\text{S}^5$},''
  \href{http://dx.doi.org/10.1016/j.nuclphysb.2012.03.004}{{\em Nucl.Phys.}
  {\bfseries B860} (2012) 339--376},
\href{http://arxiv.org/abs/1201.1451}{{\ttfamily arXiv:1201.1451 [hep-th]}}.

\bibitem{FrolovPC}
S.~Frolov.
\newblock \!\!\!\!, private communication.

\bibitem{Freedman:2005cg}
D.~Z. Freedman and U.~Gursoy, ``{Comments on the $\beta$-deformed $\cN=4$ SYM
  theory},'' \href{http://dx.doi.org/10.1088/1126-6708/2005/11/042}{{\em JHEP}
  {\bfseries 0511} (2005) 042},
\href{http://arxiv.org/abs/hep-th/0506128}{{\ttfamily arXiv:hep-th/0506128
  [hep-th]}}.

\bibitem{Frolov:2005iq}
S.~Frolov, R.~Roiban, and A.~A. Tseytlin, ``{Gauge-string duality for
  (non)supersymmetric deformations of $\cN=4$ super Yang-Mills theory},''
  \href{http://dx.doi.org/10.1016/j.nuclphysb.2005.10.004}{{\em Nucl.Phys.}
  {\bfseries B731} (2005) 1--44},
\href{http://arxiv.org/abs/hep-th/0507021}{{\ttfamily arXiv:hep-th/0507021
  [hep-th]}}.

\bibitem{Hollowood:2004ek}
T.~J. Hollowood and S.~P. Kumar, ``{An $\cN=1$ duality cascade from a
  deformation of $\cN=4$ SUSY Yang-Mills theory},''
  \href{http://dx.doi.org/10.1088/1126-6708/2004/12/034}{{\em JHEP} {\bfseries
  0412} (2004) 034},
\href{http://arxiv.org/abs/hep-th/0407029}{{\ttfamily arXiv:hep-th/0407029
  [hep-th]}}.

\bibitem{Ananth:2006ac}
S.~Ananth, S.~Kovacs, and H.~Shimada, ``{Proof of all-order finiteness for
  planar $\beta$-deformed Yang-Mills},''
  \href{http://dx.doi.org/10.1088/1126-6708/2007/01/046}{{\em JHEP} {\bfseries
  0701} (2007) 046},
\href{http://arxiv.org/abs/hep-th/0609149}{{\ttfamily arXiv:hep-th/0609149
  [hep-th]}}.

\bibitem{Ananth:2007px}
S.~Ananth, S.~Kovacs, and H.~Shimada, ``{Proof of ultra-violet finiteness for a
  planar non-supersymmetric Yang-Mills theory},''
  \href{http://dx.doi.org/10.1016/j.nuclphysb.2007.04.005}{{\em Nucl.Phys.}
  {\bfseries B783} (2007) 227--237},
\href{http://arxiv.org/abs/hep-th/0702020}{{\ttfamily arXiv:hep-th/0702020
  [HEP-TH]}}.

\bibitem{Dymarsky:2005uh}
A.~Dymarsky, I.~Klebanov, and R.~Roiban, ``{Perturbative search for fixed lines
  in large N gauge theories},''
  \href{http://dx.doi.org/10.1088/1126-6708/2005/08/011}{{\em JHEP} {\bfseries
  0508} (2005) 011},
\href{http://arxiv.org/abs/hep-th/0505099}{{\ttfamily arXiv:hep-th/0505099
  [hep-th]}}.

\bibitem{Dymarsky:2005nc}
A.~Dymarsky, I.~Klebanov, and R.~Roiban, ``{Perturbative gauge theory and
  closed string tachyons},''
  \href{http://dx.doi.org/10.1088/1126-6708/2005/11/038}{{\em JHEP} {\bfseries
  0511} (2005) 038},
\href{http://arxiv.org/abs/hep-th/0509132}{{\ttfamily arXiv:hep-th/0509132
  [hep-th]}}.

\bibitem{Banks:1981nn}
T.~Banks and A.~Zaks, ``{On the Phase Structure of Vector-Like Gauge Theories
  with Massless Fermions},''
\href{http://dx.doi.org/10.1016/0550-3213(82)90035-9}{{\em Nucl. Phys.}
  {\bfseries B196} (1982) 189}.

\bibitem{Armoni:2003va}
A.~Armoni, E.~Lopez, and A.~M. Uranga, ``{Closed strings tachyons and
  noncommutative instabilities},''
  \href{http://dx.doi.org/10.1088/1126-6708/2003/02/020}{{\em JHEP} {\bfseries
  02} (2003) 020},
\href{http://arxiv.org/abs/hep-th/0301099}{{\ttfamily arXiv:hep-th/0301099
  [hep-th]}}.

\bibitem{Blumenhagen:1999ns}
R.~Blumenhagen, A.~Font, and D.~Lust, ``{Tachyon free orientifolds of type 0B
  strings in various dimensions},''
  \href{http://dx.doi.org/10.1016/S0550-3213(99)00381-8}{{\em Nucl. Phys.}
  {\bfseries B558} (1999) 159--177},
\href{http://arxiv.org/abs/hep-th/9904069}{{\ttfamily arXiv:hep-th/9904069
  [hep-th]}}.

\bibitem{Blumenhagen:1999uy}
R.~Blumenhagen, A.~Font, and D.~Lust, ``{Nonsupersymmetric gauge theories from
  D-branes in type 0 string theory},''
  \href{http://dx.doi.org/10.1016/S0550-3213(99)00443-5}{{\em Nucl. Phys.}
  {\bfseries B560} (1999) 66--92},
\href{http://arxiv.org/abs/hep-th/9906101}{{\ttfamily arXiv:hep-th/9906101
  [hep-th]}}.

\bibitem{Angelantonj:1999qg}
C.~Angelantonj and A.~Armoni, ``{Nontachyonic type 0B orientifolds,
  nonsupersymmetric gauge theories and cosmological RG flow},''
  \href{http://dx.doi.org/10.1016/S0550-3213(00)00136-X}{{\em Nucl. Phys.}
  {\bfseries B578} (2000) 239--258},
\href{http://arxiv.org/abs/hep-th/9912257}{{\ttfamily arXiv:hep-th/9912257
  [hep-th]}}.

\bibitem{Liendo:2011da}
P.~Liendo, ``{Orientifold daughter of $\cN=4$ SYM and double-trace running},''
  \href{http://dx.doi.org/10.1103/PhysRevD.86.105032}{{\em Phys. Rev.}
  {\bfseries D86} (2012) 105032},
\href{http://arxiv.org/abs/1107.3125}{{\ttfamily arXiv:1107.3125 [hep-th]}}.

\bibitem{Minahan:2010js}
J.~A. Minahan, ``{Review of AdS/CFT Integrability, Chapter I.1: Spin Chains in
  $\cN=4$ Super Yang-Mills},''
  \href{http://dx.doi.org/10.1007/s11005-011-0522-9}{{\em Lett.Math.Phys.}
  {\bfseries 99} (2012) 33--58},
\href{http://arxiv.org/abs/1012.3983}{{\ttfamily arXiv:1012.3983 [hep-th]}}.

\bibitem{ArkaniHamed:2008gz}
N.~Arkani-Hamed, F.~Cachazo, and J.~Kaplan, ``{What is the Simplest Quantum
  Field Theory?},'' \href{http://dx.doi.org/10.1007/JHEP09(2010)016}{{\em JHEP}
  {\bfseries 1009} (2010) 016},
\href{http://arxiv.org/abs/0808.1446}{{\ttfamily arXiv:0808.1446 [hep-th]}}.

\bibitem{Zwiebel:2007cpa}
B.~I. Zwiebel, ``{The $\mathfrak{psu}(1,1|2)$ Spin Chain of $\cN=4$
  Supersymmetric Yang-Mills Theory},'' 2007.
\newblock \url{http://search.proquest.com/docview/304839945?accountid=38978}.
  Ph.D. Thesis.

\bibitem{Lehmann:1954rq}
H.~Lehmann, K.~Symanzik, and W.~Zimmermann, ``{On the formulation of quantized
  field theories},''
\href{http://dx.doi.org/10.1007/BF02731765}{{\em Nuovo Cim.} {\bfseries 1}
  (1955) 205--225}.

\bibitem{Witten:2003nn}
E.~Witten, ``{Perturbative gauge theory as a string theory in twistor space},''
  \href{http://dx.doi.org/10.1007/s00220-004-1187-3}{{\em Commun.Math.Phys.}
  {\bfseries 252} (2004) 189--258},
\href{http://arxiv.org/abs/hep-th/0312171}{{\ttfamily arXiv:hep-th/0312171
  [hep-th]}}.

\bibitem{Bargheer:2011mm}
T.~Bargheer, N.~Beisert, and F.~Loebbert, ``{Exact Superconformal and Yangian
  Symmetry of Scattering Amplitudes},''
  \href{http://dx.doi.org/10.1088/1751-8113/44/45/454012}{{\em J.Phys.}
  {\bfseries A44} (2011) 454012},
\href{http://arxiv.org/abs/1104.0700}{{\ttfamily arXiv:1104.0700 [hep-th]}}.

\bibitem{KMSW}
L.~Koster, V.~Mitev, M.~Staudacher, and M.~Wilhelm., ``{Form Factors from the
  Twistor Action}.'' In preparation.

\bibitem{Kinoshita:1962ur}
T.~Kinoshita, ``{Mass singularities of Feynman amplitudes},''
\href{http://dx.doi.org/10.1063/1.1724268}{{\em J.Math.Phys.} {\bfseries 3}
  (1962) 650--677}.

\bibitem{Lee:1964is}
T.~Lee and M.~Nauenberg, ``{Degenerate Systems and Mass Singularities},''
\href{http://dx.doi.org/10.1103/PhysRev.133.B1549}{{\em Phys.Rev.} {\bfseries
  133} (1964) B1549--B1562}.

\bibitem{'tHooft:1973mm}
G.~'t~Hooft, ``{Dimensional regularization and the renormalization group},''
\href{http://dx.doi.org/10.1016/0550-3213(73)90376-3}{{\em Nucl.Phys.}
  {\bfseries B61} (1973) 455--468}.

\bibitem{Dixon:1996wi}
L.~J. Dixon, ``{Calculating scattering amplitudes efficiently},'' in {\em {QCD
  and beyond. Proceedings, Theoretical Advanced Study Institute in Elementary
  Particle Physics, TASI-95, Boulder, USA, June 4-30, 1995}}.
\newblock 1996.
\newblock
\href{http://arxiv.org/abs/hep-ph/9601359}{{\ttfamily arXiv:hep-ph/9601359
  [hep-ph]}}.
\newblock

\bibitem{Bern:2007dw}
Z.~Bern, L.~J. Dixon, and D.~A. Kosower, ``{On-Shell Methods in Perturbative
  QCD},'' \href{http://dx.doi.org/10.1016/j.aop.2007.04.014}{{\em Annals Phys.}
  {\bfseries 322} (2007) 1587--1634},
\href{http://arxiv.org/abs/0704.2798}{{\ttfamily arXiv:0704.2798 [hep-ph]}}.

\bibitem{Abreu:2014cla}
S.~Abreu, R.~Britto, C.~Duhr, and E.~Gardi, ``{From multiple unitarity cuts to
  the coproduct of Feynman integrals},''
  \href{http://dx.doi.org/10.1007/JHEP10(2014)125}{{\em JHEP} {\bfseries 1410}
  (2014) 125},
\href{http://arxiv.org/abs/1401.3546}{{\ttfamily arXiv:1401.3546 [hep-th]}}.

\bibitem{Britto:2010xq}
R.~Britto, ``{Loop Amplitudes in Gauge Theories: Modern Analytic Approaches},''
  \href{http://dx.doi.org/10.1088/1751-8113/44/45/454006}{{\em J. Phys.}
  {\bfseries A44} (2011) 454006},
\href{http://arxiv.org/abs/1012.4493}{{\ttfamily arXiv:1012.4493 [hep-th]}}.

\bibitem{Passarino:1978jh}
G.~Passarino and M.~Veltman, ``{One-Loop Corrections for $e^+ e^-$ Annihilation
  Into $\mu^+ \mu^-$ in the Weinberg Model},''
\href{http://dx.doi.org/10.1016/0550-3213(79)90234-7}{{\em Nucl.Phys.}
  {\bfseries B160} (1979) 151}.

\bibitem{Ossola:2006us}
G.~Ossola, C.~G. Papadopoulos, and R.~Pittau, ``{Reducing full one-loop
  amplitudes to scalar integrals at the integrand level},''
  \href{http://dx.doi.org/10.1016/j.nuclphysb.2006.11.012}{{\em Nucl.Phys.}
  {\bfseries B763} (2007) 147--169},
\href{http://arxiv.org/abs/hep-ph/0609007}{{\ttfamily arXiv:hep-ph/0609007
  [hep-ph]}}.

\bibitem{Forde:2007mi}
D.~Forde, ``{Direct extraction of one-loop integral coefficients},''
  \href{http://dx.doi.org/10.1103/PhysRevD.75.125019}{{\em Phys.Rev.}
  {\bfseries D75} (2007) 125019},
\href{http://arxiv.org/abs/0704.1835}{{\ttfamily arXiv:0704.1835 [hep-ph]}}.

\bibitem{Kosower:2011ty}
D.~A. Kosower and K.~J. Larsen, ``{Maximal Unitarity at Two Loops},''
  \href{http://dx.doi.org/10.1103/PhysRevD.85.045017}{{\em Phys.Rev.}
  {\bfseries D85} (2012) 045017},
\href{http://arxiv.org/abs/1108.1180}{{\ttfamily arXiv:1108.1180 [hep-th]}}.

\bibitem{Gluza:2010ws}
J.~Gluza, K.~Kajda, and D.~A. Kosower, ``{Towards a Basis for Planar Two-Loop
  Integrals},'' \href{http://dx.doi.org/10.1103/PhysRevD.83.045012}{{\em
  Phys.Rev.} {\bfseries D83} (2011) 045012},
\href{http://arxiv.org/abs/1009.0472}{{\ttfamily arXiv:1009.0472 [hep-th]}}.

\bibitem{Velizhanin:2009gv}
V.~N. Velizhanin, ``{The Non-planar contribution to the four-loop universal
  anomalous dimension in $\cN=4$ Supersymmetric Yang-Mills theory},''
  \href{http://dx.doi.org/10.1134/S0021364009120017}{{\em JETP Lett.}
  {\bfseries 89} (2009) 593--596},
\href{http://arxiv.org/abs/0902.4646}{{\ttfamily arXiv:0902.4646 [hep-th]}}.

\bibitem{Anselmi:1996mq}
D.~Anselmi, M.~T. Grisaru, and A.~Johansen, ``{A Critical behavior of anomalous
  currents, electric - magnetic universality and CFT in four-dimensions},''
  \href{http://dx.doi.org/10.1016/S0550-3213(97)00108-9}{{\em Nucl.Phys.}
  {\bfseries B491} (1997) 221--248},
\href{http://arxiv.org/abs/hep-th/9601023}{{\ttfamily arXiv:hep-th/9601023
  [hep-th]}}.

\bibitem{Anselmi:1996dd}
D.~Anselmi, D.~Freedman, M.~T. Grisaru, and A.~Johansen, ``{Universality of the
  operator product expansions of SCFT in four-dimensions},''
  \href{http://dx.doi.org/10.1016/S0370-2693(97)00007-5}{{\em Phys.Lett.}
  {\bfseries B394} (1997) 329--336},
\href{http://arxiv.org/abs/hep-th/9608125}{{\ttfamily arXiv:hep-th/9608125
  [hep-th]}}.

\bibitem{Bianchi:1999ge}
M.~Bianchi, S.~Kovacs, G.~Rossi, and Y.~S. Stanev, ``{On the logarithmic
  behavior in $\mathcal{N}=4$ SYM theory},''
  \href{http://dx.doi.org/10.1088/1126-6708/1999/08/020}{{\em JHEP} {\bfseries
  9908} (1999) 020},
\href{http://arxiv.org/abs/hep-th/9906188}{{\ttfamily arXiv:hep-th/9906188
  [hep-th]}}.

\bibitem{Bianchi:2000hn}
M.~Bianchi, S.~Kovacs, G.~Rossi, and Y.~S. Stanev, ``{Anomalous dimensions in
  $\cN=4$ SYM theory at order $g^4$},''
  \href{http://dx.doi.org/10.1016/S0550-3213(00)00312-6}{{\em Nucl.Phys.}
  {\bfseries B584} (2000) 216--232},
\href{http://arxiv.org/abs/hep-th/0003203}{{\ttfamily arXiv:hep-th/0003203
  [hep-th]}}.

\bibitem{Eden:2000mv}
B.~Eden, C.~Schubert, and E.~Sokatchev, ``{Three loop four point correlator in
  $\cN=4$ SYM},'' \href{http://dx.doi.org/10.1016/S0370-2693(00)00515-3}{{\em
  Phys.Lett.} {\bfseries B482} (2000) 309--314},
\href{http://arxiv.org/abs/hep-th/0003096}{{\ttfamily arXiv:hep-th/0003096
  [hep-th]}}.

\bibitem{Eden:2004ua}
B.~Eden, C.~Jarczak, and E.~Sokatchev, ``{A Three-loop test of the dilatation
  operator in $\cN = 4$ SYM},''
  \href{http://dx.doi.org/10.1016/j.nuclphysb.2005.01.036}{{\em Nucl.Phys.}
  {\bfseries B712} (2005) 157--195},
\href{http://arxiv.org/abs/hep-th/0409009}{{\ttfamily arXiv:hep-th/0409009
  [hep-th]}}.

\bibitem{Sieg:2010tz}
C.~Sieg, ``{Superspace computation of the three-loop dilatation operator of
  $\cN=4$ SYM theory},''
  \href{http://dx.doi.org/10.1103/PhysRevD.84.045014}{{\em Phys.Rev.}
  {\bfseries D84} (2011) 045014},
\href{http://arxiv.org/abs/1008.3351}{{\ttfamily arXiv:1008.3351 [hep-th]}}.

\bibitem{Eden:2012fe}
B.~Eden, P.~Heslop, G.~P. Korchemsky, V.~A. Smirnov, and E.~Sokatchev,
  ``{Five-loop Konishi in $\mathcal{N}=4$ SYM},''
  \href{http://dx.doi.org/10.1016/j.nuclphysb.2012.04.015}{{\em Nucl.Phys.}
  {\bfseries B862} (2012) 123--166},
\href{http://arxiv.org/abs/1202.5733}{{\ttfamily arXiv:1202.5733 [hep-th]}}.

\bibitem{Beisert:2003tq}
N.~Beisert, C.~Kristjansen, and M.~Staudacher, ``{The Dilatation operator of
  conformal $\mathcal{N}=4$ superYang-Mills theory},''
  \href{http://dx.doi.org/10.1016/S0550-3213(03)00406-1}{{\em Nucl.Phys.}
  {\bfseries B664} (2003) 131--184},
\href{http://arxiv.org/abs/hep-th/0303060}{{\ttfamily arXiv:hep-th/0303060
  [hep-th]}}.

\bibitem{Bajnok:2009vm}
Z.~Bajnok, A.~Hegedus, R.~A. Janik, and T.~Łukowski, ``{Five loop Konishi from
  AdS/CFT},'' \href{http://dx.doi.org/10.1016/j.nuclphysb.2009.10.015}{{\em
  Nucl.Phys.} {\bfseries B827} (2010) 426--456},
\href{http://arxiv.org/abs/0906.4062}{{\ttfamily arXiv:0906.4062 [hep-th]}}.

\bibitem{Arutyunov:2010gb}
G.~Arutyunov, S.~Frolov, and R.~Suzuki, ``{Five-loop Konishi from the Mirror
  TBA},'' \href{http://dx.doi.org/10.1007/JHEP04(2010)069}{{\em JHEP}
  {\bfseries 1004} (2010) 069},
\href{http://arxiv.org/abs/1002.1711}{{\ttfamily arXiv:1002.1711 [hep-th]}}.

\bibitem{Balog:2010xa}
J.~Balog and A.~Hegedus, ``{5-loop Konishi from linearized TBA and the XXX
  magnet},'' \href{http://dx.doi.org/10.1007/JHEP06(2010)080}{{\em JHEP}
  {\bfseries 1006} (2010) 080},
\href{http://arxiv.org/abs/1002.4142}{{\ttfamily arXiv:1002.4142 [hep-th]}}.

\bibitem{Bajnok:2012bz}
Z.~Bajnok and R.~A. Janik, ``{Six and seven loop Konishi from Luscher
  corrections},'' \href{http://dx.doi.org/10.1007/JHEP11(2012)002}{{\em JHEP}
  {\bfseries 1211} (2012) 002},
\href{http://arxiv.org/abs/1209.0791}{{\ttfamily arXiv:1209.0791 [hep-th]}}.

\bibitem{Leurent:2013mr}
S.~Leurent and D.~Volin, ``{Multiple zeta functions and double wrapping in
  planar $\mathcal{N}=4$ SYM},''
  \href{http://dx.doi.org/10.1016/j.nuclphysb.2013.07.020}{{\em Nucl.Phys.}
  {\bfseries B875} (2013) 757--789},
\href{http://arxiv.org/abs/1302.1135}{{\ttfamily arXiv:1302.1135 [hep-th]}}.

\bibitem{Volin:IGST}
D.~Volin, ``{Quantum spectral curve for AdS$_5$/CFT$_4$ spectral problem}.''
  Talk given at Integrability in Gauge and String Theory (IGST) 2013.

\bibitem{Bern:1996je}
Z.~Bern, L.~J. Dixon, and D.~A. Kosower, ``{Progress in one loop QCD
  computations},'' \href{http://dx.doi.org/10.1146/annurev.nucl.46.1.109}{{\em
  Ann. Rev. Nucl. Part. Sci.} {\bfseries 46} (1996) 109--148},
\href{http://arxiv.org/abs/hep-ph/9602280}{{\ttfamily arXiv:hep-ph/9602280
  [hep-ph]}}.

\bibitem{BDP-notes}
C.~Boucher-Veronneau, L.~Dixon, and J.~Pennington \!\!, Unpublished notes.

\bibitem{Collins:1984xc}
J.~C. Collins, {\em {Renormalization}}, vol.~26 of {\em Cambridge Monographs on
  Mathematical Physics}.
\newblock Cambridge University Press, Cambridge,
1986.
\newblock

\bibitem{'tHooft:1972fi}
G.~'t~Hooft and M.~Veltman, ``{Regularization and Renormalization of Gauge
  Fields},''
\href{http://dx.doi.org/10.1016/0550-3213(72)90279-9}{{\em Nucl.Phys.}
  {\bfseries B44} (1972) 189--213}.

\bibitem{Siegel:1979wq}
W.~Siegel, ``{Supersymmetric Dimensional Regularization via Dimensional
  Reduction},''
\href{http://dx.doi.org/10.1016/0370-2693(79)90282-X}{{\em Phys.Lett.}
  {\bfseries B84} (1979) 193}.

\bibitem{Capper:1979ns}
D.~Capper, D.~Jones, and P.~van Nieuwenhuizen, ``{Regularization by Dimensional
  Reduction of Supersymmetric and Nonsupersymmetric Gauge Theories},''
\href{http://dx.doi.org/10.1016/0550-3213(80)90244-8}{{\em Nucl.Phys.}
  {\bfseries B167} (1980) 479}.

\bibitem{Bern:1991aq}
Z.~Bern and D.~A. Kosower, ``{The Computation of loop amplitudes in gauge
  theories},''
\href{http://dx.doi.org/10.1016/0550-3213(92)90134-W}{{\em Nucl.Phys.}
  {\bfseries B379} (1992) 451--561}.

\bibitem{Bern:2002zk}
Z.~Bern, A.~De~Freitas, L.~J. Dixon, and H.~Wong, ``{Supersymmetric
  regularization, two loop QCD amplitudes and coupling shifts},''
  \href{http://dx.doi.org/10.1103/PhysRevD.66.085002}{{\em Phys.Rev.}
  {\bfseries D66} (2002) 085002},
\href{http://arxiv.org/abs/hep-ph/0202271}{{\ttfamily arXiv:hep-ph/0202271
  [hep-ph]}}.

\bibitem{Buras:1989xd}
A.~J. Buras and P.~H. Weisz, ``{QCD Nonleading Corrections to Weak Decays in
  Dimensional Regularization and 't Hooft-Veltman Schemes},''
\href{http://dx.doi.org/10.1016/0550-3213(90)90223-Z}{{\em Nucl.Phys.}
  {\bfseries B333} (1990) 66}.

\bibitem{Siegel:1980qs}
W.~Siegel, ``{Inconsistency of Supersymmetric Dimensional Regularization},''
\href{http://dx.doi.org/10.1016/0370-2693(80)90819-9}{{\em Phys.Lett.}
  {\bfseries B94} (1980) 37}.

\bibitem{Avdeev:1981vf}
L.~Avdeev, G.~Chochia, and A.~Vladimirov, ``{On the Scope of Supersymmetric
  Dimensional Regularization},''
\href{http://dx.doi.org/10.1016/0370-2693(81)90886-8}{{\em Phys.Lett.}
  {\bfseries B105} (1981) 272}.

\bibitem{Avdeev:1982np}
L.~Avdeev, ``{Noninvariance of Regularization by Dimensional Reduction: An
  Explicit Example of Supersymmetry Breaking},''
\href{http://dx.doi.org/10.1016/0370-2693(82)90726-2}{{\em Phys.Lett.}
  {\bfseries B117} (1982) 317}.

\bibitem{Avdeev:1982xy}
L.~Avdeev and A.~Vladimirov, ``{Dimensional Regularization and
  Supersymmetry},''
\href{http://dx.doi.org/10.1016/0550-3213(83)90437-6}{{\em Nucl.Phys.}
  {\bfseries B219} (1983) 262}.

\bibitem{Lee:2013mka}
R.~N. Lee, ``{LiteRed 1.4: a powerful tool for reduction of multiloop
  integrals},'' \href{http://dx.doi.org/10.1088/1742-6596/523/1/012059}{{\em
  J.Phys.Conf.Ser.} {\bfseries 523} (2014) 012059},
\href{http://arxiv.org/abs/1310.1145}{{\ttfamily arXiv:1310.1145 [hep-ph]}}.

\bibitem{Gehrmann:2000zt}
T.~Gehrmann and E.~Remiddi, ``{Two loop master integrals for $\gamma^*$
  $\longrightarrow$ 3 jets: The Planar topologies},''
  \href{http://dx.doi.org/10.1016/S0550-3213(01)00057-8}{{\em Nucl.Phys.}
  {\bfseries B601} (2001) 248--286},
\href{http://arxiv.org/abs/hep-ph/0008287}{{\ttfamily arXiv:hep-ph/0008287
  [hep-ph]}}.

\bibitem{VerguCode}
C.~Vergu, ``Lecture Notes for the Mathematica Summer School on Theoretical
  Physics.'' \url{http://msstp.org/sites/default/files/Demo.nb}, 2011.

\bibitem{Hartwell:1994rp}
G.~Hartwell and P.~S. Howe, ``{(N, p, q) harmonic superspace},''
  \href{http://dx.doi.org/10.1142/S0217751X95001820}{{\em Int.J.Mod.Phys.}
  {\bfseries A10} (1995) 3901--3920},
\href{http://arxiv.org/abs/hep-th/9412147}{{\ttfamily arXiv:hep-th/9412147
  [hep-th]}}.

\bibitem{Eden:2011yp}
B.~Eden, P.~Heslop, G.~P. Korchemsky, and E.~Sokatchev, ``{The
  super-correlator/super-amplitude duality: Part I},''
  \href{http://dx.doi.org/10.1016/j.nuclphysb.2012.12.015}{{\em Nucl.Phys.}
  {\bfseries B869} (2013) 329--377},
\href{http://arxiv.org/abs/1103.3714}{{\ttfamily arXiv:1103.3714 [hep-th]}}.

\bibitem{ArkaniHamed:2009si}
N.~Arkani-Hamed, F.~Cachazo, C.~Cheung, and J.~Kaplan, ``{The S-Matrix in
  Twistor Space},'' \href{http://dx.doi.org/10.1007/JHEP03(2010)110}{{\em JHEP}
  {\bfseries 1003} (2010) 110},
\href{http://arxiv.org/abs/0903.2110}{{\ttfamily arXiv:0903.2110 [hep-th]}}.

\bibitem{Bullimore:2010pa}
M.~Bullimore, ``{Inverse Soft Factors and Grassmannian Residues},''
  \href{http://dx.doi.org/10.1007/JHEP01(2011)055}{{\em JHEP} {\bfseries 01}
  (2011) 055},
\href{http://arxiv.org/abs/1008.3110}{{\ttfamily arXiv:1008.3110 [hep-th]}}.

\bibitem{Nandan:2012rk}
D.~Nandan and C.~Wen, ``{Generating All Tree Amplitudes in $\mathcal{N}=4$ SYM
  by Inverse Soft Limit},''
  \href{http://dx.doi.org/10.1007/JHEP08(2012)040}{{\em JHEP} {\bfseries 1208}
  (2012) 040},
\href{http://arxiv.org/abs/1204.4841}{{\ttfamily arXiv:1204.4841 [hep-th]}}.

\bibitem{Postnikov:2006kva}
A.~Postnikov, ``{Total positivity, Grassmannians, and networks},''
\href{http://arxiv.org/abs/math/0609764}{{\ttfamily arXiv:math/0609764
  [math.CO]}}.

\bibitem{Bourjaily:2012gy}
J.~L. Bourjaily, ``{Positroids, Plabic Graphs, and Scattering Amplitudes in
  Mathematica},''
\href{http://arxiv.org/abs/1212.6974}{{\ttfamily arXiv:1212.6974 [hep-th]}}.

\bibitem{Faddeev:1996iy}
L.~D. Faddeev, ``{How algebraic Bethe ansatz works for integrable model},'' in
  {\em {Relativistic gravitation and gravitational radiation. Proceedings,
  School of Physics, Les Houches, France, September 26-October 6, 1995}},
  pp.~149--219.
\newblock 1996.
\newblock
\href{http://arxiv.org/abs/hep-th/9605187}{{\ttfamily arXiv:hep-th/9605187
  [hep-th]}}.
\newblock

\bibitem{PhysRevLett.19.103}
B.~Sutherland, ``Exact Solution of a Two-Dimensional Model for Hydrogen-Bonded
  Crystals,'' \href{http://dx.doi.org/10.1103/PhysRevLett.19.103}{{\em Phys.
  Rev. Lett.} {\bfseries 19} (Jul, 1967) 103--104}.

\bibitem{Sklyanin:1991ss}
E.~K. Sklyanin, ``{Quantum inverse scattering method. Selected topics},''
\href{http://arxiv.org/abs/hep-th/9211111}{{\ttfamily arXiv:hep-th/9211111
  [hep-th]}}.

\bibitem{Kazama:2015iua}
Y.~Kazama, S.~Komatsu, and T.~Nishimura, ``{On the singlet projector and the
  monodromy relation for psu$(2,2|4)$ spin chains and reduction to
  subsectors},'' \href{http://dx.doi.org/10.1007/JHEP09(2015)183}{{\em JHEP}
  {\bfseries 09} (2015) 183},
\href{http://arxiv.org/abs/1506.03203}{{\ttfamily arXiv:1506.03203 [hep-th]}}.

\bibitem{Elvang:2014fja}
H.~Elvang, Y.-t. Huang, C.~Keeler, T.~Lam, T.~M. Olson, S.~B. Roland, and D.~E.
  Speyer, ``{Grassmannians for scattering amplitudes in 4d $\mathcal{N}=4$ SYM
  and 3d ABJM},'' \href{http://dx.doi.org/10.1007/JHEP12(2014)181}{{\em JHEP}
  {\bfseries 1412} (2014) 181},
\href{http://arxiv.org/abs/1410.0621}{{\ttfamily arXiv:1410.0621 [hep-th]}}.

\bibitem{FokkenPhDthesis}
J.~Fokken \!\!, Ph.D.\ thesis, to appear.

\bibitem{Khoze:2005nd}
V.~V. Khoze, ``{Amplitudes in the $\beta$-deformed conformal Yang-Mills},''
  \href{http://dx.doi.org/10.1088/1126-6708/2006/02/040}{{\em JHEP} {\bfseries
  0602} (2006) 040},
\href{http://arxiv.org/abs/hep-th/0512194}{{\ttfamily arXiv:hep-th/0512194
  [hep-th]}}.

\bibitem{Berenstein:2000ux}
D.~Berenstein, V.~Jejjala, and R.~G. Leigh, ``{Marginal and relevant
  deformations of $\cN=4$ field theories and noncommutative moduli spaces of
  vacua},'' \href{http://dx.doi.org/10.1016/S0550-3213(00)00394-1}{{\em
  Nucl.Phys.} {\bfseries B589} (2000) 196--248},
\href{http://arxiv.org/abs/hep-th/0005087}{{\ttfamily arXiv:hep-th/0005087
  [hep-th]}}.

\bibitem{Berenstein:2000hy}
D.~Berenstein and R.~G. Leigh, ``{Discrete torsion, AdS/CFT and duality},''
  \href{http://dx.doi.org/10.1088/1126-6708/2000/01/038}{{\em JHEP} {\bfseries
  0001} (2000) 038},
\href{http://arxiv.org/abs/hep-th/0001055}{{\ttfamily arXiv:hep-th/0001055
  [hep-th]}}.

\bibitem{David:2013oha}
J.~R. David and A.~Sadhukhan, ``{Structure constants of $\beta$ deformed super
  Yang-Mills},'' \href{http://dx.doi.org/10.1007/JHEP10(2013)206}{{\em JHEP}
  {\bfseries 10} (2013) 206},
\href{http://arxiv.org/abs/1307.3909}{{\ttfamily arXiv:1307.3909 [hep-th]}}.

\bibitem{Jin:2012np}
Q.~Jin and R.~Roiban, ``{On the non-planar $\beta$-deformed $\mathcal{N}=4$
  super-Yang-Mills theory},''
  \href{http://dx.doi.org/10.1088/1751-8113/45/29/295401}{{\em J.Phys.}
  {\bfseries A45} (2012) 295401},
\href{http://arxiv.org/abs/1201.5012}{{\ttfamily arXiv:1201.5012 [hep-th]}}.

\bibitem{Penati:2005hp}
S.~Penati, A.~Santambrogio, and D.~Zanon, ``{Two-point correlators in the
  $\beta$-deformed $\cN=4$ SYM at the next-to-leading order},''
  \href{http://dx.doi.org/10.1088/1126-6708/2005/10/023}{{\em JHEP} {\bfseries
  0510} (2005) 023},
\href{http://arxiv.org/abs/hep-th/0506150}{{\ttfamily arXiv:hep-th/0506150
  [hep-th]}}.

\bibitem{DymarskyRoiban}
A.~Dymarsky and R.~Roiban. \!\!\!\!, unpublished.

\bibitem{Jin:2013baa}
Q.~Jin, ``{The Emergence of Supersymmetry in $\gamma_i$-deformed ${\cal N}=4$
  super-Yang-Mills theory},''
\href{http://arxiv.org/abs/1311.7391}{{\ttfamily arXiv:1311.7391 [hep-th]}}.

\bibitem{Ferrari:2013pq}
F.~Ferrari, M.~Moskovic, and A.~Rovai, ``{Examples of Emergent Type IIB
  Backgrounds from Matrices},''
  \href{http://dx.doi.org/10.1016/j.nuclphysb.2013.03.010}{{\em Nucl. Phys.}
  {\bfseries B872} (2013) 184--212},
\href{http://arxiv.org/abs/1301.3738}{{\ttfamily arXiv:1301.3738 [hep-th]}}.

\bibitem{Spradlin:2005sv}
M.~Spradlin, T.~Takayanagi, and A.~Volovich, ``{String theory in beta deformed
  spacetimes},'' \href{http://dx.doi.org/10.1088/1126-6708/2005/11/039}{{\em
  JHEP} {\bfseries 11} (2005) 039},
\href{http://arxiv.org/abs/hep-th/0509036}{{\ttfamily arXiv:hep-th/0509036
  [hep-th]}}.

\bibitem{Aharony:2008ug}
O.~Aharony, O.~Bergman, D.~L. Jafferis, and J.~Maldacena, ``{$\cN=6$
  superconformal Chern-Simons-matter theories, M2-branes and their gravity
  duals},'' \href{http://dx.doi.org/10.1088/1126-6708/2008/10/091}{{\em JHEP}
  {\bfseries 10} (2008) 091},
\href{http://arxiv.org/abs/0806.1218}{{\ttfamily arXiv:0806.1218 [hep-th]}}.

\bibitem{Aharony:2008gk}
O.~Aharony, O.~Bergman, and D.~L. Jafferis, ``{Fractional M2-branes},''
  \href{http://dx.doi.org/10.1088/1126-6708/2008/11/043}{{\em JHEP} {\bfseries
  11} (2008) 043},
\href{http://arxiv.org/abs/0807.4924}{{\ttfamily arXiv:0807.4924 [hep-th]}}.

\bibitem{Gadde:2009dj}
A.~Gadde, E.~Pomoni, and L.~Rastelli, ``{The Veneziano Limit of $\cN = 2$
  Superconformal QCD: Towards the String Dual of $\cN = 2$ SU(N$_c$) SYM with
  N$_f$ = 2 N$_c$},''
\href{http://arxiv.org/abs/0912.4918}{{\ttfamily arXiv:0912.4918 [hep-th]}}.

\bibitem{Minahan:2009aq}
J.~A. Minahan, O.~Ohlsson~Sax, and C.~Sieg, ``{Magnon dispersion to four loops
  in the ABJM and ABJ models},''
  \href{http://dx.doi.org/10.1088/1751-8113/43/27/275402}{{\em J. Phys.}
  {\bfseries A43} (2010) 275402},
\href{http://arxiv.org/abs/0908.2463}{{\ttfamily arXiv:0908.2463 [hep-th]}}.

\bibitem{Minahan:2009wg}
J.~A. Minahan, O.~Ohlsson~Sax, and C.~Sieg, ``{Anomalous dimensions at four
  loops in $\cN=6$ superconformal Chern-Simons theories},''
  \href{http://dx.doi.org/10.1016/j.nuclphysb.2011.01.016}{{\em Nucl. Phys.}
  {\bfseries B846} (2011) 542--606},
\href{http://arxiv.org/abs/0912.3460}{{\ttfamily arXiv:0912.3460 [hep-th]}}.

\bibitem{Leoni:2010tb}
M.~Leoni, A.~Mauri, J.~A. Minahan, O.~Ohlsson~Sax, A.~Santambrogio, C.~Sieg,
  and G.~Tartaglino-Mazzucchelli, ``{Superspace calculation of the four-loop
  spectrum in $\cN=6$ supersymmetric Chern-Simons theories},''
  \href{http://dx.doi.org/10.1007/JHEP12(2010)074}{{\em JHEP} {\bfseries 12}
  (2010) 074},
\href{http://arxiv.org/abs/1010.1756}{{\ttfamily arXiv:1010.1756 [hep-th]}}.

\bibitem{Pomoni:2011jj}
E.~Pomoni and C.~Sieg, ``{From $\cN=4$ gauge theory to $\cN=2$ conformal QCD:
  three-loop mixing of scalar composite operators},''
\href{http://arxiv.org/abs/1105.3487}{{\ttfamily arXiv:1105.3487 [hep-th]}}.

\bibitem{Aharony:2015afa}
O.~Aharony, G.~Gur-Ari, and N.~Klinghoffer, ``{The Holographic Dictionary for
  Beta Functions of Multi-trace Coupling Constants},''
  \href{http://dx.doi.org/10.1007/JHEP05(2015)031}{{\em JHEP} {\bfseries 05}
  (2015) 031},
\href{http://arxiv.org/abs/1501.06664}{{\ttfamily arXiv:1501.06664 [hep-th]}}.

\bibitem{Broadhurst:1985vq}
D.~J. Broadhurst, ``{Evaluation of a Class of Feynman Diagrams for All Numbers
  of Loops and Dimensions},''
\href{http://dx.doi.org/10.1016/0370-2693(85)90340-5}{{\em Phys.Lett.}
  {\bfseries B164} (1985) 356}.

\bibitem{Chicherin:2015bza}
D.~Chicherin, R.~Doobary, B.~Eden, P.~Heslop, G.~P. Korchemsky, and
  E.~Sokatchev, ``{Bootstrapping correlation functions in $\cN=4$ SYM},''
\href{http://arxiv.org/abs/1506.04983}{{\ttfamily arXiv:1506.04983 [hep-th]}}.

\bibitem{Boels:2006ir}
R.~Boels, L.~J. Mason, and D.~Skinner, ``{Supersymmetric Gauge Theories in
  Twistor Space},'' \href{http://dx.doi.org/10.1088/1126-6708/2007/02/014}{{\em
  JHEP} {\bfseries 02} (2007) 014},
\href{http://arxiv.org/abs/hep-th/0604040}{{\ttfamily arXiv:hep-th/0604040
  [hep-th]}}.

\bibitem{Bern:1995ix}
Z.~Bern and G.~Chalmers, ``{Factorization in one-loop gauge theory},''
  \href{http://dx.doi.org/10.1016/0550-3213(95)00226-I}{{\em Nucl.Phys.}
  {\bfseries B447} (1995) 465--518},
\href{http://arxiv.org/abs/hep-ph/9503236}{{\ttfamily arXiv:hep-ph/9503236
  [hep-ph]}}.

\bibitem{Smirnov:2004ym}
V.~A. Smirnov, ``{Evaluating Feynman integrals},''
{\em Springer Tracts Mod.Phys.} {\bfseries 211} (2004) 1--244.

\bibitem{Gehrmann:2005pd}
T.~Gehrmann, T.~Huber, and D.~Maitre, ``{Two-loop quark and gluon form-factors
  in dimensional regularisation},''
  \href{http://dx.doi.org/10.1016/j.physletb.2005.07.019}{{\em Phys.Lett.}
  {\bfseries B622} (2005) 295--302},
\href{http://arxiv.org/abs/hep-ph/0507061}{{\ttfamily arXiv:hep-ph/0507061
  [hep-ph]}}.

\bibitem{ZinnJustin:2002ru}
J.~Zinn-Justin, ``{Quantum field theory and critical phenomena},''
{\em Int.Ser.Monogr.Phys.} {\bfseries 113} (2002) 1--1054.

\end{thebibliography}\endgroup
